\documentclass[PhD]{dukethesis2006}

\usepackage{amsmath}
\usepackage{amssymb}
\usepackage{amsthm}
\usepackage{array}
\usepackage{epsfig}
\usepackage[hidelinks=true]{hyperref}
\usepackage{graphicx}
\usepackage{xy}
\usepackage{mathtools}
\usepackage{slashed}
\usepackage{amssymb}
\graphicspath{{images/}}
\usepackage{subcaption}
\usepackage{cite}
\newcommand{\trento}{T\raisebox{-0.5ex}{R}ENTo}

\newcommand{\fig}{\begin{figure}[htbp]\centering}

\newcommand{\efig}{\end{figure}} 

\newcommand{\eq}{\begin{equation}}
\newcommand{\eeq}{\end{equation}}
\newcommand{\eqa}{\begin{eqnarray}}
\newcommand{\eeqa}{\end{eqnarray}}

\newcommand*\diff{\mathop{}\!\mathrm{d}}
\newcommand*\Diff[1]{\mathop{}\!\mathrm{d^#1}}
\newcommand{\nn}{\nonumber}

\newcommand{\be}{\begin{eqnarray}}
\newcommand{\ee}{\end{eqnarray}}
\newcommand{\ma}{\mathrm}
\newcommand{\ml}{\mathcal}
\newcommand{\bs}{\boldsymbol}
\newcommand{\Tr}{\mathrm{Tr}}

\DeclareMathOperator{\sign}{sign}
\DeclareMathOperator\Arg{Arg}

\newcommand{\singlespacing}{%
  \let\CS=\small\renewcommand{\baselinestretch}{1.0}\CS}
\newcommand{\doublespacing}{%
  \let\CS=\small\renewcommand{\baselinestretch}{1.6}\CS}

\author{Xiaojun Yao}
\advisor{Berndt O. M\"uller}
\member{Thomas C. Mehen}
\member{Steffen A. Bass}
\member{Mark C. Kruse}
\member{Harold U. Baranger}

\department{Physics}
\title{A\MakeLowercase{pplication of }E\MakeLowercase{ffective} F\MakeLowercase{ield} T\MakeLowercase{heory in }N\MakeLowercase{uclear} P\MakeLowercase{hysics}}

\begin{document}

\maketitle

\makeabstract

\Copyright

\abstract
The production of heavy quarkonium in heavy ion collisions has been used as an important probe of the quark-gluon plasma (QGP). Due to the plasma screening effect, the color attraction between the heavy quark antiquark pair inside a quarkonium is significantly suppressed at high temperature and thus no bound states can exist, i.e., they ``melt". In addition, a bound heavy quark antiquark pair can dissociate if enough energy is transferred to it in a dynamical process inside the plasma. So one would expect the production of quarkonium to be considerably suppressed in heavy ion collisions. However, experimental measurements have shown that a large amount of quarkonia survive the evolution inside the high temperature plasma. It is realized that the in-medium recombination of unbound heavy quark pairs into quarkonium is as crucial as the melting and dissociation. Thus, phenomenological studies have to account for static screening, dissociation and recombination in a consistent way. But recombination is less understood theoretically than the melting and dissociation. Many studies using semi-classical transport equations model the recombination effect from the consideration of detailed balance at thermal equilibrium. However, these studies cannot explain how the system of quarkonium reaches equilibrium and estimate the time scale of the thermalization. Recently, another approach based on the open quantum system formalism started being used. In this framework, one solves a quantum evolution for in-medium quarkonium. Dissociation and recombination are accounted for consistently. However, the connection between the semi-classical transport equation and the quantum evolution is not clear.

In this dissertation, I will try to address the issues raised above. As a warm-up project, I will first study a similar problem: $\alpha$-$\alpha$ scattering at the $^8$Be resonance inside an $e^-e^+\gamma$ plasma. By applying pionless effective field theory and thermal field theory, I will show how the plasma screening effect modifies the $^8$Be resonance energy and width. I will discuss the need to use the open quantum system formalism when studying the time evolution of a system embedded inside a plasma. Then I will use effective field theory of QCD and the open quantum system formalism to derive a Lindblad equation for bound and unbound heavy quark antiquark pairs inside a weakly-coupled QGP. Under the Markovian approximation and the assumption of weak coupling between the system and the environment, the Lindblad equation will be shown to turn to a Boltzmann transport equation if a Wigner transform is applied to the open system density matrix. These assumptions will be justified by using the separation of scales, which is assumed in the construction of effective field theory. I will show the scattering amplitudes that contribute to the collision terms in the Boltzmann equation are gauge invariant and infrared safe. By coupling the transport equation of quarkonium with those of open heavy flavors and solving them using Monte Carlo simulations, I will demonstrate how the system of bound and unbound heavy quark antiquark pairs reaches detailed balance and equilibrium inside the QGP. Phenomenologically, my calculations can describe the experimental data on bottomonium production. Finally I will extend the framework to study the in-medium evolution of heavy diquarks and estimate the production rate of the doubly charmed baryon $\Xi_{cc}^{++}$ in heavy ion collisions.

\acknowledgements
Coming to the stage of today, I have so many persons to thank. The first person I want to thank, who is also the one I want to thank the most, is my advisor Berndt M\"uller. I am extremely lucky to have Berndt as my advisor. The story can be dated back to my senior year as an undergraduate student, when I was studying at Duke University as an exchange student. In the summer just before the exchange program, I was looking for an advisor for my undergraduate research project. So I searched the website of Duke physics department. When I read through the department webpage of Berndt, I had the direct feeling that it would be great if I could work with him. Indeed, under Berndt's guidance, I quickly expanded my knowledge of modern physics in my senior year. And I quickly realized one simple fact: I was very excited and happy about what I learnt from the discussion after each meeting with Berndt. I believe I benefited a lot from Berndt's methodology to inspire and motivate young students. After entering the graduate school, I would like to have Berndt as my dissertation advisor. But he served as a vice-director at Brookhaven National Laboratory (BNL) and was not at Duke most of the time. I am very grateful that Berndt makes it possible for me to visit BNL frequently so that I can meet with him regularly to discuss physics. The procedure is not always easy, but Berndt managed it. He also managed to leave some time for me to discuss physics with him, usually after the office hour each day. The more discussions I have with him, the more admired I am by his broad knowledge and deep understanding of physics in a wide range to topics. He not only instructed me on physics but also inspired me often spiritually. I benefited significantly from working with him. Without him, I could not have made the accomplishments, many of which are covered in this dissertation, during my graduate study and research. 

The second person I want to thank is Thomas Mehen. I began to know Tom as I took his class on quantum mechanics when I was an exchange undergraduate student at Duke. I really enjoyed his lectures. They were challenging and fun. I was very grateful for Tom when he introduced the idea of using effective field theory to do a quantum mechanics calculation that I presented in my preliminary exam. That conversation totally changed the trajectory that I would follow in my graduate research. Without him, I would probably not be motivated to learn effective field theory and use it extensively in my research. Tom often drew my attention to interesting new developments in our fields and I really benefited a lot from discussions with him. One of my important contributions to the understanding of quarkonium in-medium evolution, was motivated from a discussion with him. He introduced the concept of open effective field theory and the papers written by Eric Braaten, et. al., to me, which motivated me to think about using it to study quarkonium in-medium evolution. Later, we successfully derived the semi-classical in-medium transport equation of quarkonium, by applying effective field theory and the open quantum system method. Quite recently, I had another inspiring discussion with Tom and we began to think about interesting physics questions in the jet in-medium evolution. I can see we will have another successful publication and even more probably.

I also want to thank Steffen Bass. I feel very grateful for him because during my first few years as a graduate student, it was Steffen who taught me the basic concepts and current understandings of heavy ion collisions, which helped me to quickly catch up with the cutting edge of research in heavy ion collisions. In my later research studies, I benefited a lot from his expertise in numerical computations and code designing. The Monte Carlo method that will be intensively used to solve transport equations in Chapter 4 of this dissertation is largely motivated from his work and improved by discussions with him. I also learnt a lot from his advice and suggestions on how to advertise one's research work and how to prepare poster and oral presentations.

Furthermore, I want to thank the former chair of the physics department Haiyan Gao. I am extremely grateful for her because she initiated the student exchange program in physics between Duke University and Shandong University. I was lucky to be among the first round of students in the program when I was a senior. Without her, the whole story told above would never exist and I would probably not be in such a good stage of my scientific career today. 

In addition, I want to thank my colleagues Weiyao Ke and Yingru Xu for the useful discussions and collaborations. I thank Jean-Francois Paquet for useful discussions and advice. I thank Johann Rafelski for teaching me the basic concept of Debye screening in the beginning of my graduate research. I would also like to thank my (former) Duke theory colleagues Jussi Auvinen, Reggie Bain, Harold Baranger, Jonah Bernhard, Shanshan Cao, Shailesh Chandrasekharan, Leo Fang, Emilie Huffman, Hanqing Liu, Jian-Guo Liu, Yiannis Makris, Scott Moreland, Marlene Nahrgang, Arya Roy, Hersh Singh, Roxanne Springer, Di-Lun Yang, Gu Zhang and Xin Zhang for teaching me interesting physics and making the second floor of the physics building a nice place to work. Since I spent a significant amount of my graduate time at BNL, I would also like to thank my colleagues who are working at or once visited the nuclear theory group there: Abhay Deshpande, Adrian Dumitru, Yoshitaka Hatta, Luchang Jin, Frithjof Karsch, Dmitri Kharzeev, Yacine Mehtar-Tani, Swagato Mukherjee, Peter Petreczky, Rob Pisarski, Bj\"orn Schenke, Chun Shen, Derek Teaney and Raju Venugopalan for inspiring discussions.

Finally, I would like to thank my parents for their love, spiritual support and encouragements during these years. I thank my friends for the relaxing chatting. I thank JJ Lin and Aimer for their songs accompanying me in my bright and dark days over the years.

\tableofcontents

\listoffigures

\listoftables


\chapter{Introduction}
\pagenumbering{arabic}
\vspace{0.2in}
\doublespacing
\section{Heavy Ion Collisions}

The central aim of heavy ion collision experiments is to search for the quark-gluon plasma (QGP) and study its properties. The QGP is a deconfined phase of nuclear matter that can be achieved at high temperature and high density. QGP is also believed to be the main content of the universe up to microseconds after the Big Bang. The phase transition and the properties of QGP are governed by the underlying microscopic theory, the Quantum Chromodynamics (QCD). QCD describes the dynamics of quarks and gluons and the formations of hadrons such as protons and neutrons. It is part of the Standard Model of particle physics and explains most of the observed mass in everyday life.

\begin{figure}[h]
\begin{center}
\includegraphics[height=3.in]{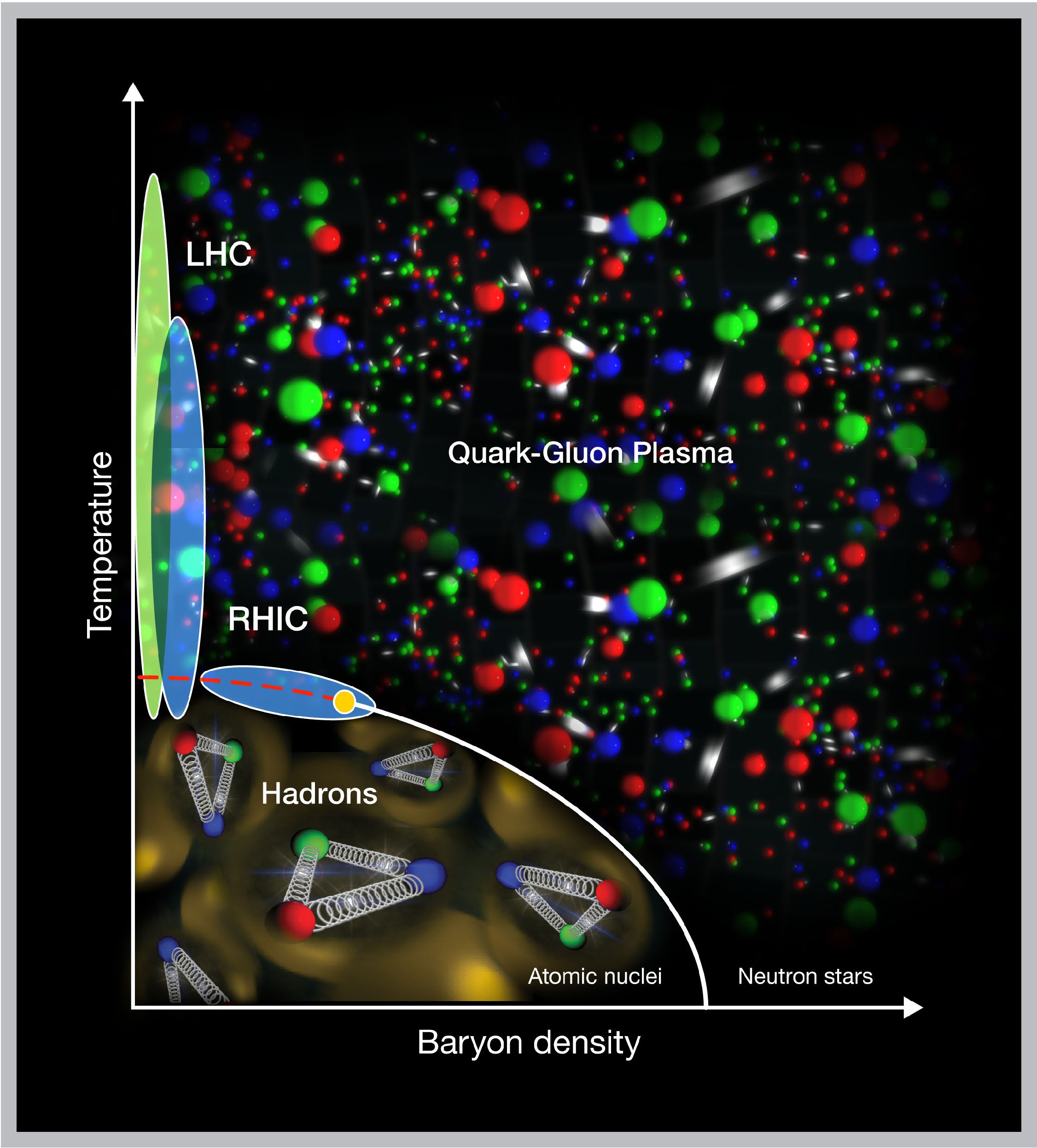}
\caption[Sketch of the phase diagram of QCD matter.]{Sketch of the phase diagram of QCD matter. At zero baryon chemical potential, the transition between the hadronic phase and the QGP is a smooth cross-over around the temperature $T\simeq 154$ MeV, indicated by the dashed red line. At low temperatures and large baryon chemical potentials, there is likely a first-order phase transition, marked by the white line. A critical point may exist between the smooth cross-over and the first-order transition, denoted by the yellow point. Credit: Brookhaven National Laboratory.}
\label{fig_chap1:phase}
\end{center}
\end{figure}

A sketch of the QCD matter phase structure at different temperatures and baryon chemical potentials is shown in Fig.~\ref{fig_chap1:phase}. At zero baryon chemical potential, lattice QCD calculations at the physical pion mass have shown that the transition between the hadronic phase and the QGP is a smooth crossover \cite{Aoki:2006we}. The transition occurs around the temperature $T\simeq 154$ MeV \cite{Aoki:2006br,Bazavov:2011nk} (we will use the natural unit system where $c=\hbar=k_B=1$). At low temperatures, as the baryon chemical potential increases, the hadronic phase is likely to turn into a color superconducting phase with quark Cooper pairs, i.e., diquark condensates. Many studies have indicated that this transition is likely of first order \cite{Alford:1997zt,Rapp:1997zu}. Therefore, there probably exists a critical point between the smooth crossover and the first-order phase transition. Lattice QCD calculations at non-vanishing chemical potentials have the notorious sign problem and thus it is extremely difficult to calculate the position of the critical point via lattice QCD. A recent development uses a Taylor expansion in $\mu_B/T$ to study the smooth crossover at non-zero baryon chemical potential \cite{Bazavov:2018mes}. So far, the first-order phase transition and the critical point are conjectures based on model calculations.

QCD has the property called ``asymptotic freedom" \cite{Gross:1973id,Politzer:1973fx}, which means the interaction among quarks and gluons becomes weaker at shorter distances or larger energy scales. So naively one expects that the QGP is a weakly-coupled plasma at high temperature. However, it was not until the 21st century, when the Relativistic Heavy Ion Collider (RHIC) at Brookhaven National Laboratory started running, that detailed and systematic experimental studies of QGP became available. It was shown for the first time that the QGP produced at the current collider energies is a strongly-coupled fluid \cite{Arsene:2004fa,Adcox:2004mh,Back:2004je,Adams:2005dq}.

At RHIC, gold nuclei (Au) are accelerated circularly to the speed of $0.99996c$ roughly ($c$ is the speed of light and the kinetic energy of each nucleon contained in the accelerated nuclei is $100$ GeV in the laboratory frame) and are made to collide with each other. A large amount of particles are produced from the collision. By measuring and analyzing these final-state particles, one can determine whether a QGP is created and study its properties. In the transverse plane, most of the observed particles have thermal spectra, blue-shifted by the radial expansions of the plasma. Particle distributions also exhibit an azimuthal angular anisotropy, known as the ``elliptic flow". At more peripheral collisions with larger impact parameters, the signal of the elliptic flow is stronger. The physical understanding is that during the initial collision, a large amount of energies are deposited into the collision region. Shortly after the initial collision (about $0.5-1$ fm$/c$), a QGP close to thermal equilibrium is formed due to interactions within the system. Then the QGP expands collectively and cools down. When the QGP temperature drops to about $154$ MeV, hadrons consisting of confined quarks and gluons are formed. These hadrons still interact with each other for a period of time and eventually freeze-out (no more interactions) and move to the detectors. The ``elliptic flow" comes from the geometric eccentricity and fluctuations in the initial collisions and is a feature of the collective expansion. 

With data collected at RHIC, it was discovered that ideal hydrodynamics with no viscosity \cite{Kolb:2000sd} and simple initial conditions and hadronization models \cite{Cooper:1974mv} can approximately describe the spectra of observed particles. This fact indicates that the QGP is a plasma with low viscosity, i.e., an almost ``perfect" fluid. This contradicts the naive expectation that the QGP is weakly-coupled at high temperature because perturbative calculations estimated the shear viscosity as \cite{Baym:1990uj,Arnold:2000dr}
\be 
\eta \sim \frac{T^3}{g^4\ln{(1/g)}}\,.
\ee
In other words, perturbative QCD states that the viscosity is large when the plasma is weakly-coupled.

The smallness of the shear viscosity is constrained by the uncertainty principle of quantum mechanics \cite{Danielewicz:1984ww}. An explicit calculation by mapping the strongly-coupled gauge theory to a weakly-coupled Einsteinian gravity theory via the AdS/CFT correspondence \cite{Maldacena:1997re} showed a lower bound of the viscosity-to-entropy ratio in a strongly-coupled plasma \cite{Kovtun:2004de}
\be
\frac{\eta}{s} \geq \frac{1}{4\pi}\,.
\ee
Recent analyses based on viscous hydrodynamics obtain values of viscosity consistent with the lower bound \cite{Song:2010mg}, or slightly larger (see Ref.~\cite{Bernhard:2016tnd} for a quantitative estimate of uncertainties). In a nutshell, the QGP is a strongly-coupled plasma with small shear viscosity. 

Around 2010, the Large Hadron Collider (LHC) at the European Organization for Nuclear Research started its heavy ion collision experiments. The LHC heavy ion experiments collide lead nuclei (Pb) at much higher center-of-mass energies ($\sqrt{s} = 2.76$, $5.02$ TeV) than the RHIC energy. The LHC experiments not only confirmed the ``perfectness" of the QGP fluid \cite{Muller:2012zq}, but also provided more statistics and precision in the data. Using these data sets, one can construct the multi-particle correlations and constrain parameters of the hydrodynamics by a theory-experiment comparison. Furthermore, these high quality data allow more detailed analyzes of other observables that may contain signatures of the QGP such as strange particles, jets, open and hidden heavy flavors, photons and di-leptons. The last three observables are called ``hard" probes because they are not part of the collective hydrodynamics, which is dominated by soft modes (whose energies are $\lesssim T$). Jets and heavy flavors have large energy scales given by the jet energy or the heavy quark mass. Photons and di-leptions only interact electroweakly with the QGP, which is negligible compared with the strong interaction.

\vspace{0.2in}
\subsection{Strange Particles}
Inside a high temperature QGP, strange quarks reach chemical and kinetic equilibrium with the gluons, up and down quarks within the QGP lifetime due to the conversion process $gg\leftrightarrow q\bar{q}$ where $q=u,d,s$. The strange quark can be produced thermally inside the QGP because the strange quark mass is smaller than the QGP temperature. The strange quark abundance at the freeze-out saturates and the production of strange hadrons is enhanced with respect to the same collision without the QGP formation. Therefore, the strange enhancement is a signature of the QGP formation in heavy ion collisions \cite{Rafelski:1982pu}.

\vspace{0.2in}
\subsection{Jets}
A jet is a group of collimated particles with large transverse momentum. The leading partons (quarks or gluons) of jets are produced in the initial hard scattering when the two nuclei collide. Subsequently the leading parton radiates out other partons and a shower of hadrons is produced. Most of the time, two jets are produced back-to-back in the transverse plane. When they travel through the QGP, they lose energy and change directions due to scattering with medium constituents and in-medium radiations. As a result, the jet production at the same transverse momentum is suppressed in heavy ion collisions with respect to that in proton-proton collisions, normalized properly by the number of binary nucleon-nucleon collisions in heavy ion collisions. Furthermore, the initial angular correlation disappears in the final measured jets. These phenomena are called jet quenching.

\vspace{0.2in}
\subsection{Open and Hidden Heavy Flavors}
Heavy quarks are produced early in the initial hard scattering when the two nuclei collide and subsequently lose energy and change momentum inside the QGP medium. The initial spectra typically have a power-law tail. So the in-medium evolution is an approach to thermalization. If the QGP lasts long enough, eventually heavy quarks will thermalize and become part of the QGP. But since QGP has a finite lifetime, the thermalization is incomplete. Observables that can probe the degree of thermalization are the suppression of D-mesons, B-mesons and singly heavy baryons and their azimuthal angular anisotropy accumulated during the in-medium evolution. By analyzing these observables, one can learn how the QGP influences the evolution of heavy quarks.

Quarkonium is a bound state of heavy quark-antiquark pair. The name charmonium refers to the charm-anticharm mesons and the name bottomonium to the bottom-antibottom mesons. Bound states of top-antitop do not exist because the top quark decays fast via weak interactions (the lifetime of a top quark is roughly $5\times10^{-25}s$ while the typical strong interaction time scale is about $10^{-23}s$). The modification of quarkonium production in heavy ion collisions can also be used as a probe of the QGP. This will be discussed in more detail in the next section.

\vspace{0.2in}
\subsection{Electromagnetic Probes}
When a quark scatters inelastically in the medium and changes its momentum, it may radiate electromagnetically and emit a photon or a pair of leptons. The produced photons and leptons then escape the QGP with almost no further modifications on their energy and momentum. The radiation rate depends on the temperature of the plasma. Therefore measurements of the electromagnetic radiations can tell us the temperature profile inside the QGP. In practice, systematic background subtractions are required to remove contributions from the electromagnetic decays of hadrons, which occurs during the hadronic gas stage outside of the QGP.

\vspace{0.2in}
\subsection{Future Experimental Developments}
In the forthcoming years, RHIC will conduct the program ``Phase 2 of Beam Energy Scan" by running Au-Au collisions at several low center-of-mass energies: $\sqrt{s}=7.7,9.1,11.5,14.5$, $19.6,27,39,62.4$ GeV. The program aims at measuring particle multiplicities with higher precision and more statistics than ``Phase 1", which occurred in 2010, 2011 and 2014. The purpose of running these low-energy collisions is to scan the QCD phase structure and search for the critical point \cite{Yang:2019bjr}, as shown in Fig.~\ref{fig_chap1:phase}. Since the searching is relied on measuring event-by-event fluctuations of conserved quantities such as the baryon number (for an overview of the critical point search, see Ref.~\cite{Luo:2017faz} and references therein), one needs a detector with high performance. For this purpose, the STAR collaboration upgraded the STAR detector by upgrading its inner tracking projection chamber, endcap time of flight detector and event plane detector. These upgrades will improve the detector energy and momentum resolution and the particle identification capability so that the STAR detector can serve better for the critical point search. 

At the same time, the sPHENIX collaboration at RHIC is constructing a new detector with high efficiency and resolution \cite{Aidala:2012nz}. The main purpose of the sPHENIX detector is to measure hard probe observables at the RHIC energy with higher precision and statistics. These observables include suppression factors of bottomonium, jets and open heavy flavors. These measurements are important for our understanding of the hard probes in heavy ion collisions because they can provide experimental data as precise as those at the LHC energies. So one can use these measurements at the RHIC energy to further constrain theoretical models. The sPHENIX detector will start collecting experimental data within the next few years. 

The LHC heavy ion experiments will upgrade their detectors in the near future, for example, the ALICE detector \cite{Abelevetal:2014cna}. Designing a new detector is also under consideration \cite{Adamova:2019vkf}. The main motivation is to increase the precision and statistics in the measurements of hard probe observables such as the azimuthal angular anisotropy of quarkonium. Measurements of new observables such as the quarkonium polarization in heavy ion collisions and jet substructure modifications may become possible. It is likely that future detector upgrades at LHC will make it possible to measure the production rates of doubly heavy baryons and doubly heavy tetraquarks in heavy ion collisions. An estimate of the production rate will be presented in Chapter 5 of this dissertation.

Data on hard probes with more statistics and higher precision will be available in the coming years. Theoretical descriptions also need improvements to match the data precision so that we can learn properties of the QGP from the theory-experiment comparisons.

\vspace{0.2in}
\singlespacing
\section{Quarkonium Production in Heavy Ion Collisions}
\doublespacing

\vspace{0.2in}
\subsection{$Q\bar{Q}$ Phenomenology}
\begin{figure}
    \centering
    \begin{subfigure}[t]{0.48\textwidth}
        \centering
        \includegraphics[height=2.1in]{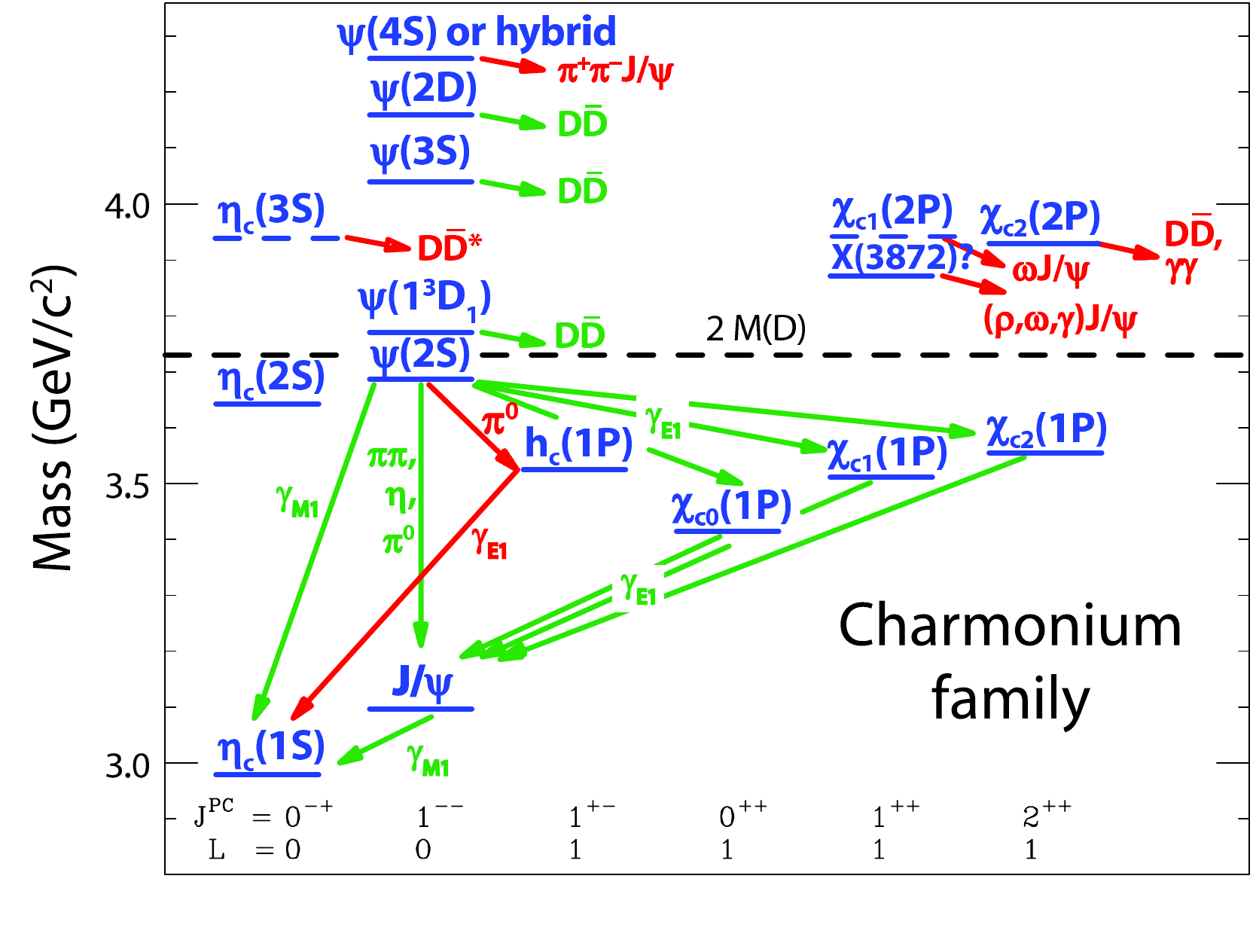}
        \caption{Charmonium family.}\label{}
    \end{subfigure}%
    ~~
    \begin{subfigure}[t]{0.48\textwidth}
        \centering
        \includegraphics[height=2.1in]{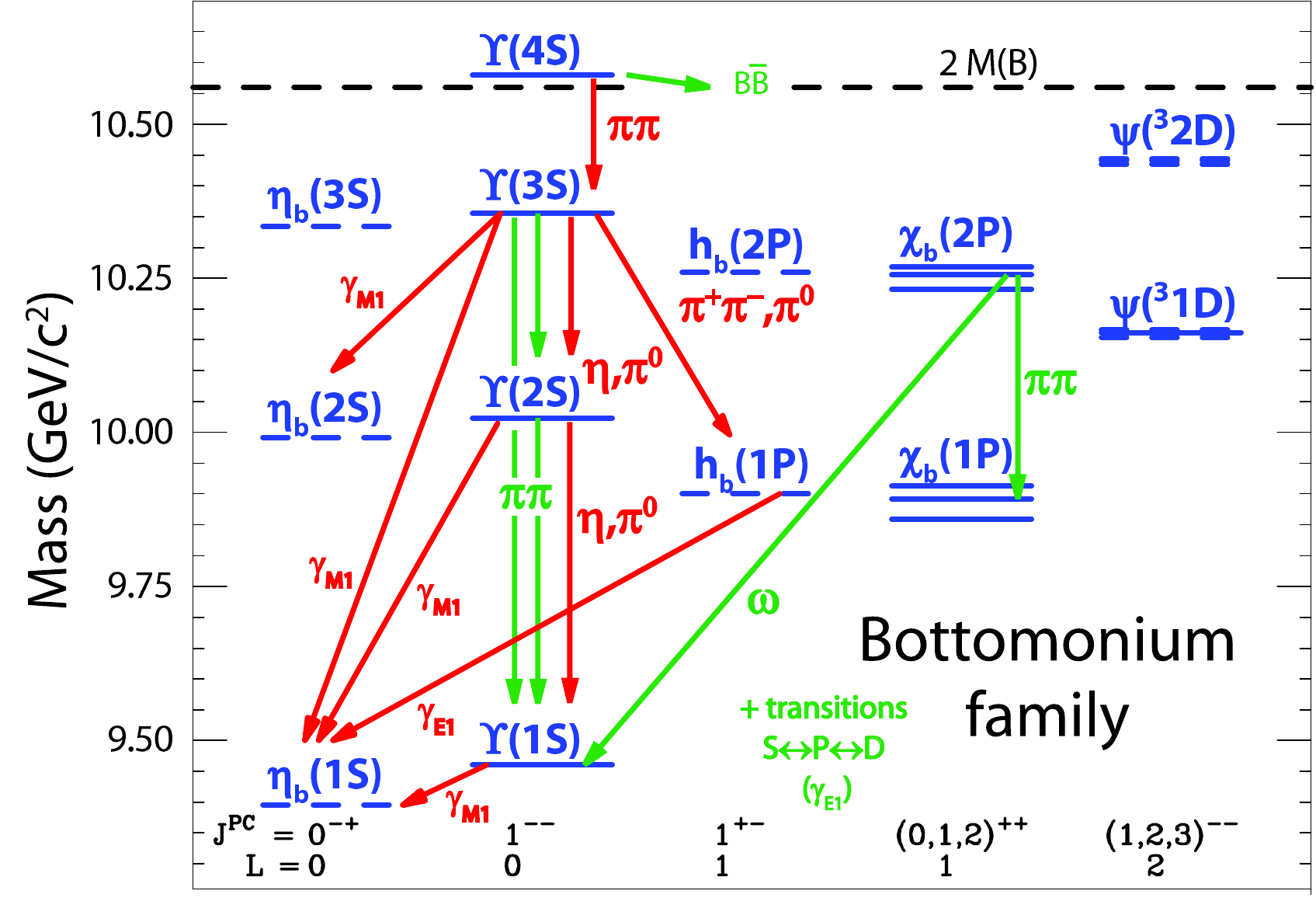}
        \caption{Bottomonium family.}\label{}
    \end{subfigure}%
\caption[Spectra and transitions of different quarkonium states.]{Spectra and transitions of different quarkonium states. The plots are taken from Ref.~\cite{Eichten:2007qx}. The states are labeled by a name and the corresponding quantum numbers $nL$ and $J^{PC}$. The dashed line is the open heavy meson threshold. Quarkonium states above the threshold can decay into a pair of open heavy mesons. $\gamma_{E1}$ indicates the electric dipole transition while $\gamma_{M1}$ indicates the magnetic dipole transition.}
\label{fig_chap1:QQbar_family}
\end{figure}

The first quarkonium, the charmonium state $J/\psi$, was discovered in 1974 \cite{Aubert:1974js,Augustin:1974xw}. The discovery confirmed the existence of a fourth quark, the charm quark, in addition to the ``light" up, down and strange quarks, and stimulated a sequence of discoveries of new particle states, which include the bottomonium states. Most quarkonium states can be thought of as a bound state of heavy quark-antiquark pair $Q\bar{Q}$ (we will give an argument in Section 1.4). Since the charm quark and bottom quark masses are large $M_c \approx 1.3$ GeV, $M_b \approx 4.5$ GeV, most quarkonium states can be approximately described by a nonrelativistic Schr\"odinger equation with a Cornell potential between the $Q\bar{Q}$
\be
V(r) = - \frac{a}{r} + br \,.
\ee
The part $- a/r$ is an attractive Coulomb potential just as in the hydrogen atom. The linear part $br$ includes the effect of confinement of QCD. Solving the Schr\"odinger equation one can obtain a spectrum of states labeled by the quantum numbers $n$ (radial excitation, starting from $n=1$) and $L$ (orbital angular momentum). In addition, the spin-spin interactions between the $Q\bar{Q}$ and the spin-orbital coupling can be included as perturbation just like the hyperfine and fine structure splittings in the hydrogen atom spectrum. Therefore, we need two more quantum numbers, the spin $S$ and the total angular momentum $J$ to label a quarkonium state. We will use the spectroscopic notation $n^{2S+1}L_J$. With the quantum numbers given, one can also specify the parity $P$ and charge conjugation $C$ of the state. The notation $J^{PC}$ is also used to label different states. The $J/\psi$ mentioned above has $n=1$, $S=1$, $L=0$ and $J^{PC} = 1^{--}$. A similar charmonium state with $n=2$ is called $\psi$(2S). The corresponding bottomonium states with $S=1$, $L=0$ and $J^{PC} = 1^{--}$ are named $\Upsilon$(nS). 
States that do not fit into the pattern of $n^{2S+1}L_J$ are called exotics. The spectra of the charmonium and the bottomonium families are shown in Fig.~\ref{fig_chap1:QQbar_family}, in which different quarkonium states are labeled by the quantum numbers $nL$ and $J^{PC}$. Each combination of the quantum numbers corresponds to a specific name such as $\eta_c$, $\psi$ or $\chi$.

Quarkonium states with masses larger than the open heavy meson threshold will decay into two open heavy mesons. The threshold is given by twice the mass of D-mesons for charmonium and twice the mass of B-mesons for bottomonium. In fact, there are more than one threshold. For charmonium, possible thresholds can be given by the masses of $D\bar{D}$, $D^*\bar{D}^*$ ($D^*$ is an radially excited state of $D$) or $D_s \bar{D}_s$. For those quarkonium states whose masses are close to the thresholds, whether these states are $Q\bar{Q}$ bound states or molecules of an open heavy meson pair is still under investigation. A good discussion of these states and the exotics can be found in Ref.~\cite{Guo:2017jvc}.

Below the open heavy meson threshold, higher excited quarkonium states can decay into lower excited ones via radiating out photons, pions, omegas or etas. Selection rules can be constructed based on the quantum numbers of the initial and final quarkonium states and the types of the transitions. Some examples of the possible transitions among different quarkonium states are also shown in Fig.~\ref{fig_chap1:QQbar_family}.

\begin{figure}
    \centering
    \begin{subfigure}[t]{0.48\textwidth}
        \centering
        \includegraphics[height=2.3in]{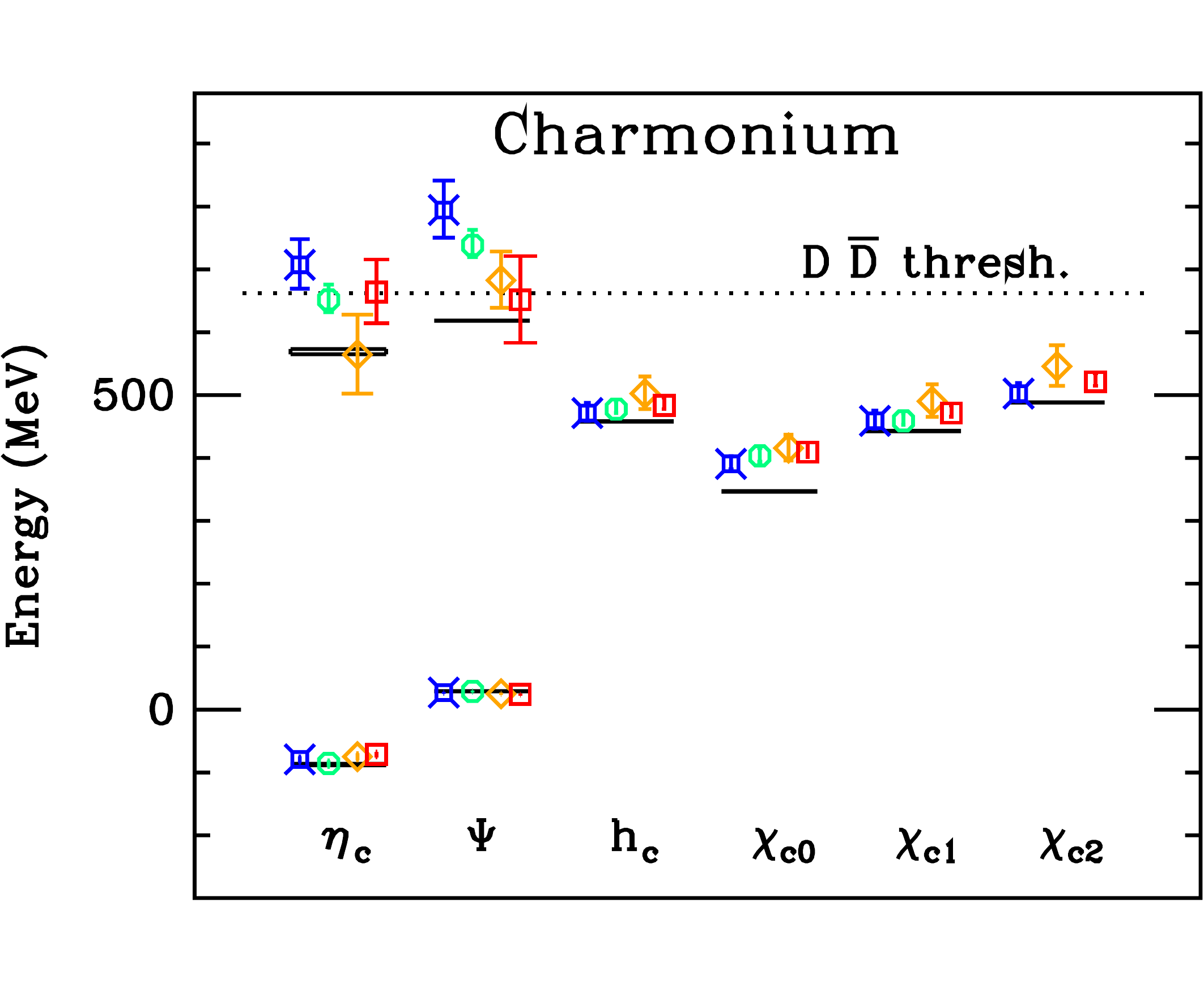}
        \caption{Charmonium family. The plot is taken from Ref.~\cite{Burch:2009az}. The calculation includes dynamical $u,d,s$ quarks. The spin averaged 1S state mass $\frac{3m_{J/\psi}+m_{\eta_c}}{4}$ is used as the reference point. Different colors represent different lattice spacings.}\label{}
    \end{subfigure}%
    ~~
    \begin{subfigure}[t]{0.48\textwidth}
        \centering
        \includegraphics[height=2.3in]{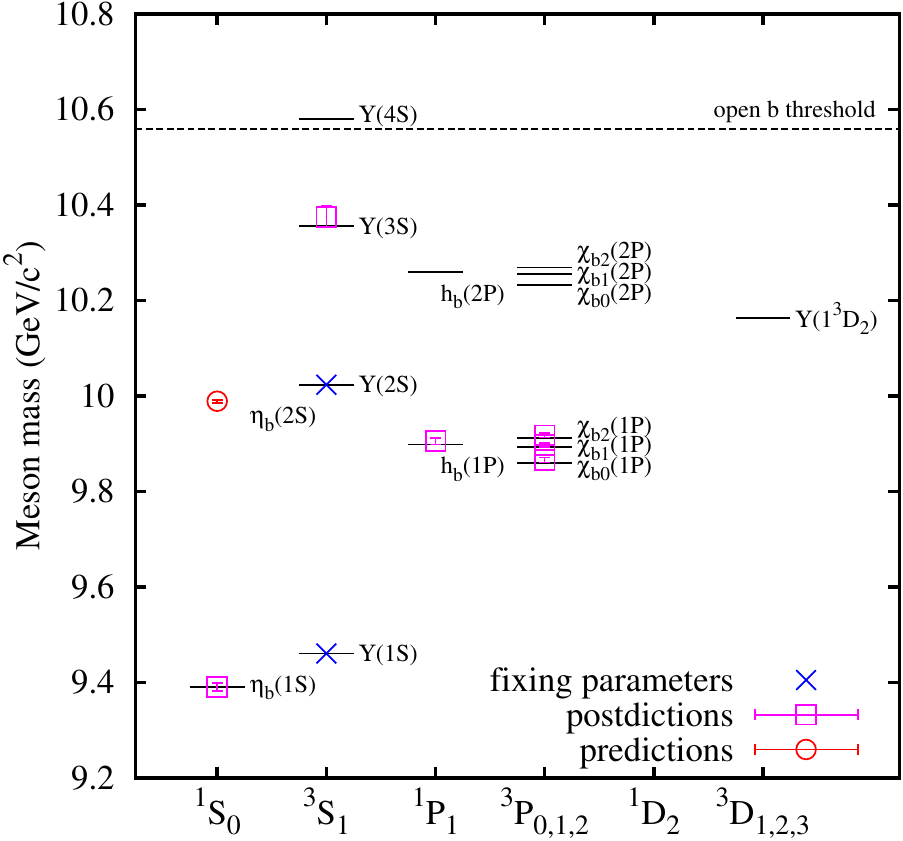}
        \caption{Bottomonium family. The plot is taken from Ref.~\cite{Dowdall:2011wh}. The calculation uses nonrelativistic QCD action (which will be explained in Section 1.4.3.) with dynamical $u,d,s,c$ quarks. The blue crosses are used as inputs of the calculations.}\label{}
    \end{subfigure}%
\caption[Lattice calculations of quarkonium spectra.]{Lattice calculations of quarkonium spectra. Black solid lines indicate the experimental measurement results.}
\label{fig_chap1:QQbar_lattice}
\end{figure}

Most lower excited states are ordinary states and their mass spectra are well-understood within the Cornell potential description. This picture is also confirmed by lattice QCD calculations. Some lattice calculation results of the quarkonium spectra are shown in Fig.~\ref{fig_chap1:QQbar_lattice}. The spectra of ground states and lower excited states can be well-described by the lattice calculations. For excited states close to the open heavy meson threshold, it may be necessary to include their mixing with $D\bar{D}$ (for charmonium) or $B\bar{B}$ (for bottomonium) in the calculation to accurately describe their spectra.

\vspace{0.2in}
\subsection{$Q\bar{Q}$ Suppression}

Not long after the discovery of $J/\psi$, its properties inside a hot QCD medium was investigated. It was found that due to the plasma screening effect, the attractive potential between the heavy quark pair $Q\bar{Q}$ is significantly suppressed at high temperature. As a result, quarkonium bound states no longer exist or ``melt" \cite{Matsui:1986dk,Karsch:1987pv}. It was speculated that due to this screening effect, quarkonium production in heavy ion collisions will be suppressed \cite{Matsui:1986dk}. 

\begin{figure}[h]
\begin{center}
\includegraphics[height=2.8in]{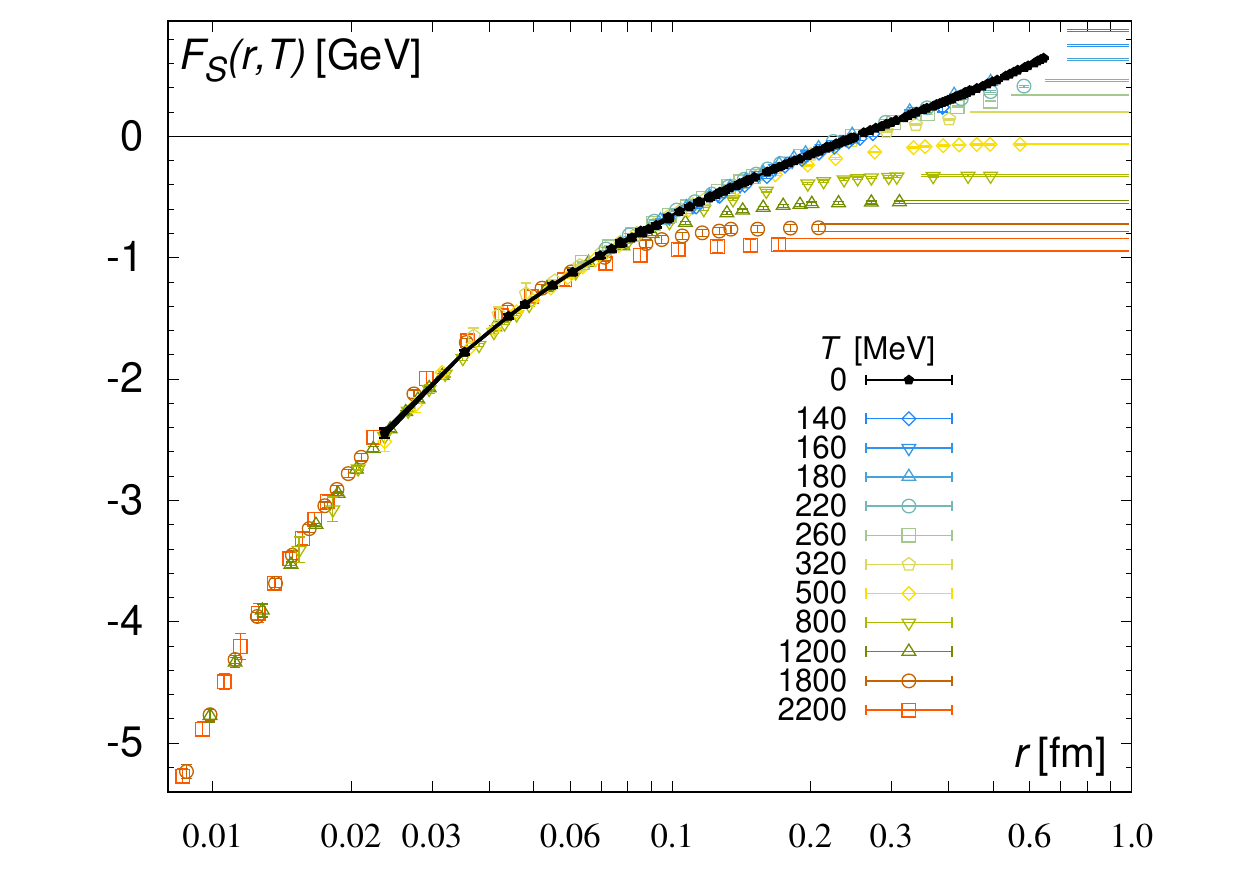}
\caption[Free energy of a static $Q\bar{Q}$ color singlet at different temperatures.]{Free energy of a static $Q\bar{Q}$ color singlet at different temperatures. The black solid line indicates the zero temperature potential between the $Q\bar{Q}$ pair. At high temperatures, the confining part of the potential is flattened out. The plot is taken from Ref.~\cite{Bazavov:2018wmo}.}
\label{fig_chap1:free_energy}
\end{center}
\end{figure}

It was proposed later that the screening of the potential can be studied on a lattice by computing the free energy of a static $Q\bar{Q}$ pair \cite{McLerran:1981pb}. A recent (2+1)-flavor lattice calculation of the free energy of a static $Q\bar{Q}$ color singlet is shown in Fig.~\ref{fig_chap1:free_energy}. The black solid line indicates the zero temperature Cornell potential between the $Q\bar{Q}$. As shown in the plot, the confining part of the potential is screened at high temperatures. The distance where the potential starts to be flat becomes smaller as temperature increases.  

\begin{figure}[h]
\begin{center}
\includegraphics[height=2.2in]{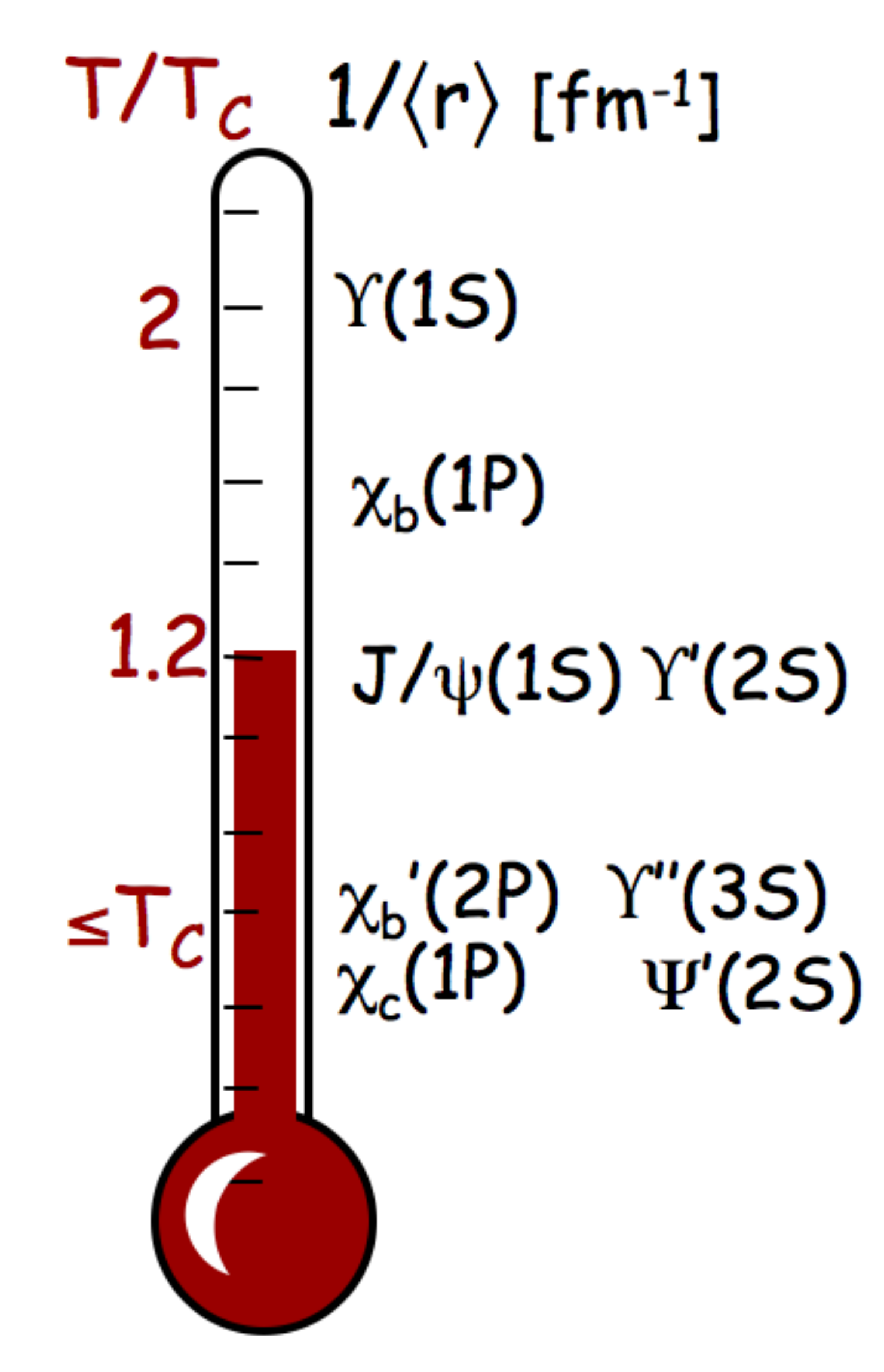}
\caption[Different quarkonium states as a thermometer of the QGP.]{Different quarkonium states as a thermometer of the QGP. The plot is taken from Ref.~\cite{Mocsy:2008eg}.}
\label{fig_chap1:thermometer}
\end{center}
\end{figure}

Many studies solve the nonrelativisitic Schr\"odinger equation for a $Q\bar{Q}$ pair with a temperature dependent attractive potential to obtain the melting temperatures of different quarkonium states \cite{Wong:2004zr,Alberico:2005xw,Mocsy:2007jz}. Some use the singlet free energy $F_1(r,T)$ calculated on the lattice as the potential while others use the internal energy given by the thermodynamic relation
\be
U_1(r,T) = F_1(r,T) - T\frac{\partial F_1(r,T)}{\partial T}\,.
\ee
The qualitative conclusions are the same: since different quarkonium states have different binding energies and typical sizes, they start to melt at different temperatures. More deeply bound states can survive at higher temperatures and thus are expected to be less suppressed in heavy ion collisions. Therefore the quarkonium suppression is sensitive to the temperature profile of the QGP. In this sense, quarkonium can be thought of as a thermometer of QGP \cite{Mocsy:2008eg}, as shown in Fig.~\ref{fig_chap1:thermometer}. 

However, the use of quarkonium as a QGP thermometer is more complicated because screening of the attractive potential is not the only suppression mechanism. A quarkonium state inside a QGP below its melting temperature, may break up after absorbing enough energy from the medium. More precisely, when an amount of energy exceeding its binding energy is transferred to the quarkonium state, it dissociates to an unbound $Q\bar{Q}$ pair. The thermal breakup may happen when quarkonium scatters with medium constituents such as quarks and gluons in a weak coupling picture of the QGP. As a result of the interaction, the quarkonium state acquires thermal width. The thermal width is also a type of plasma screening effect. To distinguish it from the screening of the attractive potential, we will call the thermal width a {\it dynamical} plasma screening effect since it is generated from dynamical scattering processes. The screening of the attractive potential is called the {\it static} plasma screening effect.

The plasma screening effect is temperature dependent and becomes stronger at higher temperatures. Therefore one would expect that quarkonium production is more suppressed at LHC than at RHIC because higher collision energies lead to a hotter QGP. However, this simple expectation was proven wrong by the LHC data (which will be shown in the next section). The production of $J/\psi$ is surprisingly less suppressed. Actually, this was predicted long before the experimental observation \cite{Thews:2000rj,Andronic:2007bi}. It was noted that the (re)combination of unbound heavy quark-antiquark pairs becomes gradually more important as the collision energy increases and more heavy quarks are produced. Here by recombination, I indicate both the recombination of correlated $Q\bar{Q}$ pairs from previous quarkonium dissociations and the combination of uncorrelated $Q\bar{Q}$ pairs from different initial hard scattering vertices.

In addition to the hot medium effect, other factors can also contribute to the quarkonium suppression in heavy ion collisions such as the ``cold nuclear matter effect" affecting the initial hard scattering. The parton distribution function (PDF) of a large nucleus is generally different from that of a proton due to the nuclear interactions among the nucleons. The nuclear PDF can be extracted from measurements in proton-nucleus collisions and used as inputs in nucleus-nucleus collisions. This is because in proton-nucleus collisions, the modification of quarkonium production due to the hot plasma is believed to be small.

In the hadronic phase, higher excited quarkonium states can decay to lower excited quarkonium states. Open bottom mesons can also decay to charmonium states. Since higher excited quarkonium states are expected to be more suppressed, fewer feed-down contributions can lead to the suppressions of the lower excited states. Feed-down contributions can be measured in $e^+e^-$ and proton-proton collisions.

Since the cold nuclear matter effect and feed-down contributions can be measured in other experiments, the key to understanding the quarkonium suppression in heavy ion collisions is to understand the quarkonium evolution inside the QGP. The evolution includes both static and dynamical plasma screening effects and the recombination inside the medium or near the phase crossover. There are mainly three approaches: the statistical hadronization model, transport equations, and the open quantum system formalism.

\subsubsection{Statistical Hadronization Model}
The statistical hadronization models have been used to describe charmonium production \cite{Andronic:2003zv,Andronic:2007bi}. In these models it is assumed that the charm quark evolves as an unbound state inside the hot medium due to the Debye screening. During the evolution, the charm quark equilibrates kinematically but not chemically, because the annihilation of charm quarks is negligible during the lifetime of the QGP and the total number of charm and anticharm quarks is fixed by the initial hard scattering. Thermal production is also negligible because of the large charm quark mass, compared with the QGP temperature. Charmonium is assumed to be produced from coalescence of charm-anticharm pairs at the crossover hypersurface from the QGP phase to a hadron gas phase. At the crossover, the momentum spectra of charm quarks are assumed to be thermal. Although the model has some phenomenological success, it is limited to the study of charmonium with low transverse momentum. The kinematic thermalization assumption is never justified for charmonium at large transverse momentum and for bottomonium because we expect that the thermalization time increases with the particle's energy.

\subsubsection{Transport Equations}
A more popular approach is to use a transport equation \cite{Grandchamp:2003uw, Grandchamp:2005yw, Yan:2006ve, Liu:2009nb, Song:2011xi, Song:2011nu, Sharma:2012dy, Nendzig:2014qka, Krouppa:2015yoa, Chen:2017duy, Zhao:2017yan, Du:2017qkv, Aronson:2017ymv, Ferreiro:2018wbd}. In this approach, a rate equation is used to describe the dissociation and recombination of quarkonium inside the hot medium. The dissociation and recombination rates depend on the bound state wave function at finite temperature. Debye screening of the potential is taken into account when one solves the bound state wave function from the Schr\"odinger equation.

In most studies, the dissociation rate is calculated from perturbative QCD. The dissociation process at leading order (LO) in the coupling constant is the gluon absorption process $g+H\rightarrow Q+\bar{Q}$ where $H$ indicates a quarkonium state. It was first investigated by using large-$N_c$ expansions \cite{Peskin:1979va,Bhanot:1979vb}. At next-leading order (NLO), inelastic scattering between quarkonium and medium constituents contributes to the dissociation $l(\bar{l},g)+H\rightarrow l(\bar{l},g)+Q+\bar{Q}$ where $l$ indicates a light quark. The inelastic scattering was first studied in the quasi-free limit where the $Q$ and $\bar{Q}$ are treated as free particles and each of them scatters independently with medium constituents \cite{Grandchamp:2001pf}. Later, the interference effect was taken into account. This leads to a dependence of the dissociation rate on the relative distance of the heavy quark-antiquark pair \cite{Laine:2006ns,Beraudo:2007ky}, which maps into a dependence of the inelastic scattering on the bound state size or wave function \cite{Park:2007zza}, as in the case of gluon absorption.
More recently, these dissociation rates were studied in potential nonrelativistic QCD (pNRQCD) \cite{Brambilla:2010vq,Brambilla:2011sg,Brambilla:2013dpa} by systematic weak coupling and nonrelativistic expansions. Anisotropic corrections to dissociation rates have also been considered \cite{Dumitru:2007hy, Dumitru:2009fy, Du:2016wdx}.

The recombination process has been analyzed in the framework of perturbative QCD with parametrized non-thermal heavy quark momentum distributions \cite{Song:2012at}. But in phenomenology, one needs to use momentum distributions from real heavy quark in-medium evolutions, which start with a power law tail. In most phenomenological studies, recombination is modeled from detailed balance with an extra suppression factor accounting for the incomplete thermalization of heavy quarks. However, these studies cannot explain how the system approaches detailed balance and thermalization. The functional form of the extra suppression factor lacks a theoretical foundation.

In this dissertation, I will address the issues raised above. In Chapter 4, I will construct a set of coupled Boltzmann transport equations of both open heavy quarks and quarkonia, in which the heavy quark momentum distribution is not from an assumed parametrization but rather calculated from real-time dynamics, and quarkonium dissociation and recombination are calculated in the same theoretical framework of pNRQCD \cite{Yao:2017fuc, Yao:2018zrg, Yao:2018sgn, Yao:2018dap}. By using the coupled Boltzmann transport equations, detailed balance and thermalization of heavy quarks and quarkonium can be demonstrated from the real-time dynamics of heavy quark diffusion and energy changes and the interplay between quarkonium dissociation and recombination. This framework will be justified theoretically in Chapter 3, by using separation of scales in pNRQCD \cite{Yao:2018nmy}.

\subsubsection{Open Quantum System}
\begin{figure}[h]
\begin{center}
\includegraphics[height=3.0in]{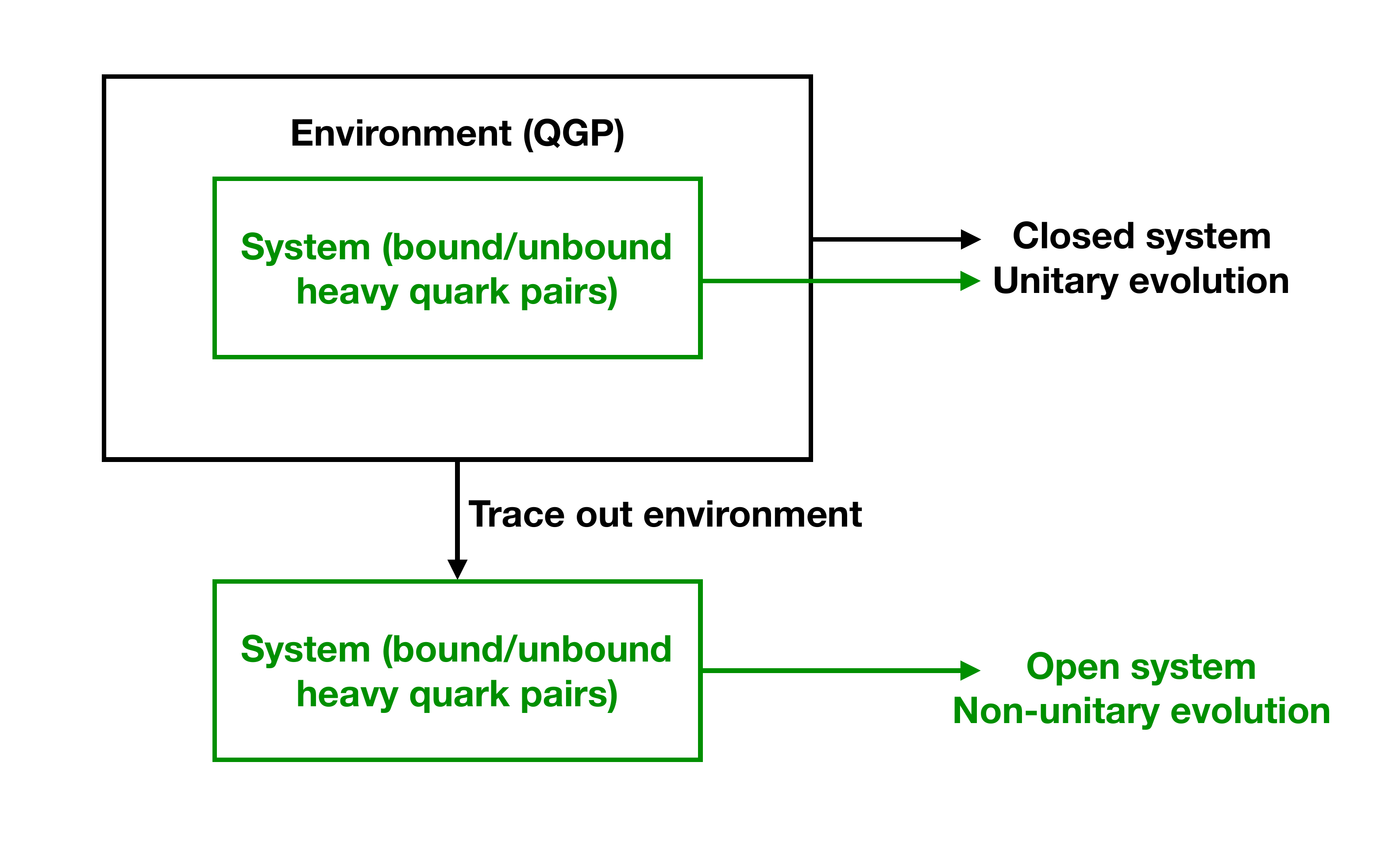}
\caption[Schematic diagram of the open quantum system.]{Schematic diagram of the open quantum system. The system and the environment evolve unitarily together as a closed system. When the environment is traced out, the system is open and evolves non-unitarily.}
\label{fig_chap1:opensystem}
\end{center}
\end{figure}

More recently, an approach based on the theory of open quantum systems has been studied widely \cite{Young:2010jq, Borghini:2011ms, Akamatsu:2011se, Akamatsu:2014qsa, Blaizot:2015hya, Kajimoto:2017rel, DeBoni:2017ocl, Blaizot:2017ypk, Brambilla:2017zei, Akamatsu:2018xim}. This approach is a quantum description in terms of the density matrix rather than a semi-classical transport equation. In this approach, we call bound and unbound heavy quark-antiquark pairs as the system, and the QGP as the environment. The system and the environment together evolve unitarily as a closed system. When we integrate out the environment degrees of freedom and focus on the dynamics of the system, we find the system evolves non-unitarily as an open system and stochastic interactions can appear. A schematic diagram of the open quantum system is shown in Fig.~\ref{fig_chap1:opensystem}. The non-unitary evolution of the system density matrix can be written in the form of a Lindblad equation \cite{Lindblad:1975ef}. Quarkonium dissociation occurs as a result of the wave function decoherence during the non-unitary evolution. At the same time, the Lindblad equation preserves the trace of the system density matrix. Physically this means the total number of heavy quarks is conserved during the evolution (unless one introduces annihilation of $Q\bar{Q}$, which is negligible in current heavy ion collisions). The unbound heavy quark pair from the quarkonium dissociation never disappears from the system. They stay as active degrees of freedom and may recombine in the later evolution. The advantage of this framework is that the recombination effect is included systematically. This feature is never easily achieved in transport models.

The difficulty of applying the open quantum system framework to the phenomenology resides in how to couple it to hydrodynamics. The quantum evolution equation depends on the temperature of the QGP, which can only be obtained from hydrodynamics simulations for realistic heavy ion collisions. But the hydrodynamics is a semi-classical description. The quantum evolution of the open quantum system may have some semi-classical correspondences. This will be explored in Chapter 3 of this dissertation, where a deep connection between the open quantum system formalism and the Boltzmann transport equation will be illuminated.

\vspace{0.2in}
\section{Experimental Measurements}

In collider experiments, the production cross section of a vector quarkonium state is measured by detecting di-lepton pairs ($e^-e^+$ or $\mu^-\mu^+$) from the leptonic decay. Usually the di-muon channel has lower background than the di-electron channel. The invariant mass of the all detected di-lepton pairs is reconstructed and fitted with models of signal and background. From the fit, the number $N_{H\to l\bar{l}}$ of a specific quarkonium state $H$ produced and decaying into the di-lepton pairs can be deduced. Then the total number $N_H$ of this specific quarkonium state produced can be obtained 
\be
N_H = \frac{N_{H\to l\bar{l}}}{Br(H\rightarrow l\bar{l}) \times a \times \epsilon} \,,
\ee
where $a$ and $\epsilon $ denote the detector acceptance and efficiency and $Br(H\rightarrow l\bar{l})$ is the di-lepton branching ratio of the quarkonium $H$. The cross section can be calculated from the integrated luminosity $L = \int \diff t \ml{L}$
\be
\sigma_H = \frac{N_H}{L}\,.
\ee
More differential cross sections can be obtained in a similar procedure. This is the standard procedure in proton-proton collisions. In heavy ion collisions, instead of the cross section, one construct the nuclear modification factor $R_{AB}$ after fitting the yield,
\be
R_{AB}(\bs b) \equiv \frac{\sigma_{AB \rightarrow H+X}(\bs b)}{ \sigma_{pp \rightarrow H+X}   \langle N_\ma{coll} \rangle(\bs b)  } = \frac{N_{AB \rightarrow H+X}(\bs b)}{ T_{AB}(\bs b) \sigma_{pp \rightarrow H+X} }\,,
\ee
where $\bs b$ is the impact parameter of the collision and $A$ and $B$ indicate the two colliding nuclei, i.e., proton, Au or Pb. The $+X$ indicates the inclusive process: all particles other than $H$ are integrated over. $\langle N_\ma{coll} \rangle(\bs b)$ is the averaged number of binary nucleon-nucleon collisions in one nucleus-nucleus collision at a given impact parameter. It is given by
\be
\langle N_\ma{coll} \rangle(\bs b) = \sigma_{NN} T_{AB}(\bs b)\,,
\ee
where $\sigma_{NN}$ is the inelastic nucleon-nucleon scattering cross section and $T_{AB}(\bs b)$ is the nuclear overlap function. The nuclear overlap function is an effective nucleon-nucleon density in the transverse plane
\be
\label{chap1_eqn_overlap}
T_{AB}(\bs b) &\equiv& \int \diff^2r T_A(\bs r) T_B({\bs b} - {\bs r})\\
T_A(\bs r) &\equiv& \int \diff z \rho_A({\bs r}, z)\,,
\ee
where $T_A(\bs r)$ is the nuclear thickness function at the transverse position $\bs r$ and $\rho_A({\bs r}, z)$ is the nuclear density function at the transverse position $\bs r$ and longitudinal position $z$. The normalization condition is $\int\diff^2b T_A(\bs b) = A$. The nuclear density function $\rho_A$ can be parametrized by a Woods-Saxon distribution function. Finally, 
\be
\label{chap1_eqn_N_AB}
N_{AB \rightarrow H+X}(\bs b) \equiv \frac{\sigma_{AB \rightarrow H+X}(\bs b)}{\sigma_{NN}}\,,
\ee
is the averaged number of quarkonium state $H$ produced in one nucleus-nucleus collision. In the expression of (\ref{chap1_eqn_N_AB}), $1/\sigma_{NN}$ can be thought of as the integrated luminosity that is needed to have one nucleon-nucleon inelastic scattering, which is the same integrated luminosity needed to have a $A$-$B$ collision with $\langle N_\ma{coll} \rangle$ binary nucleon-nucleon collisions. If $R_{AB}=1$, there is no suppression due to the cold and hot nuclear matter effects.

\begin{figure}
    \centering
    \begin{subfigure}[t]{0.48\textwidth}
        \centering
        \includegraphics[height=1.8in]{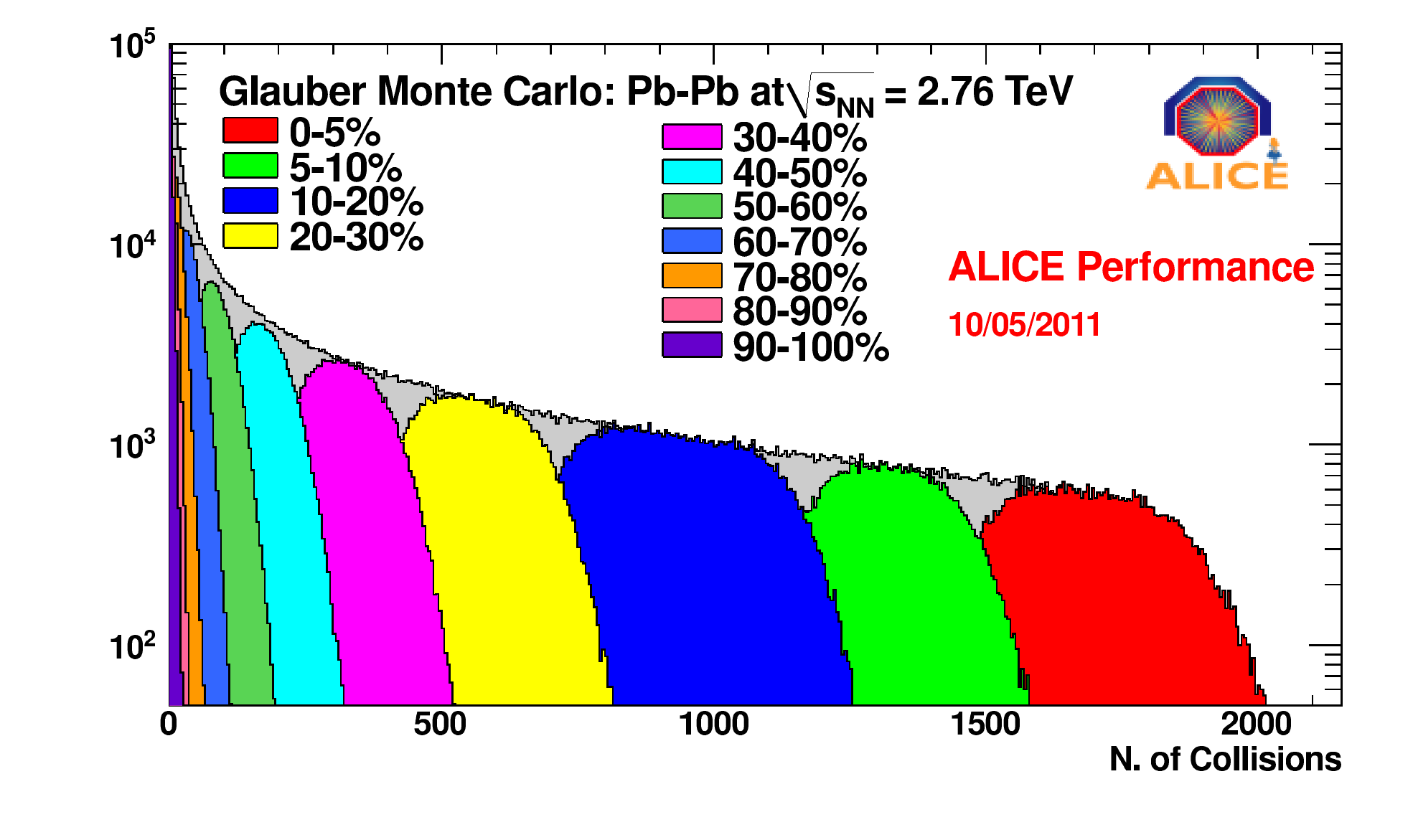}
        \caption{Relation between the centrality and $\langle N_{\ma{coll}} \rangle $.}\label{}
    \end{subfigure}%
    ~
    \begin{subfigure}[t]{0.48\textwidth}
        \centering
        \includegraphics[height=1.8in]{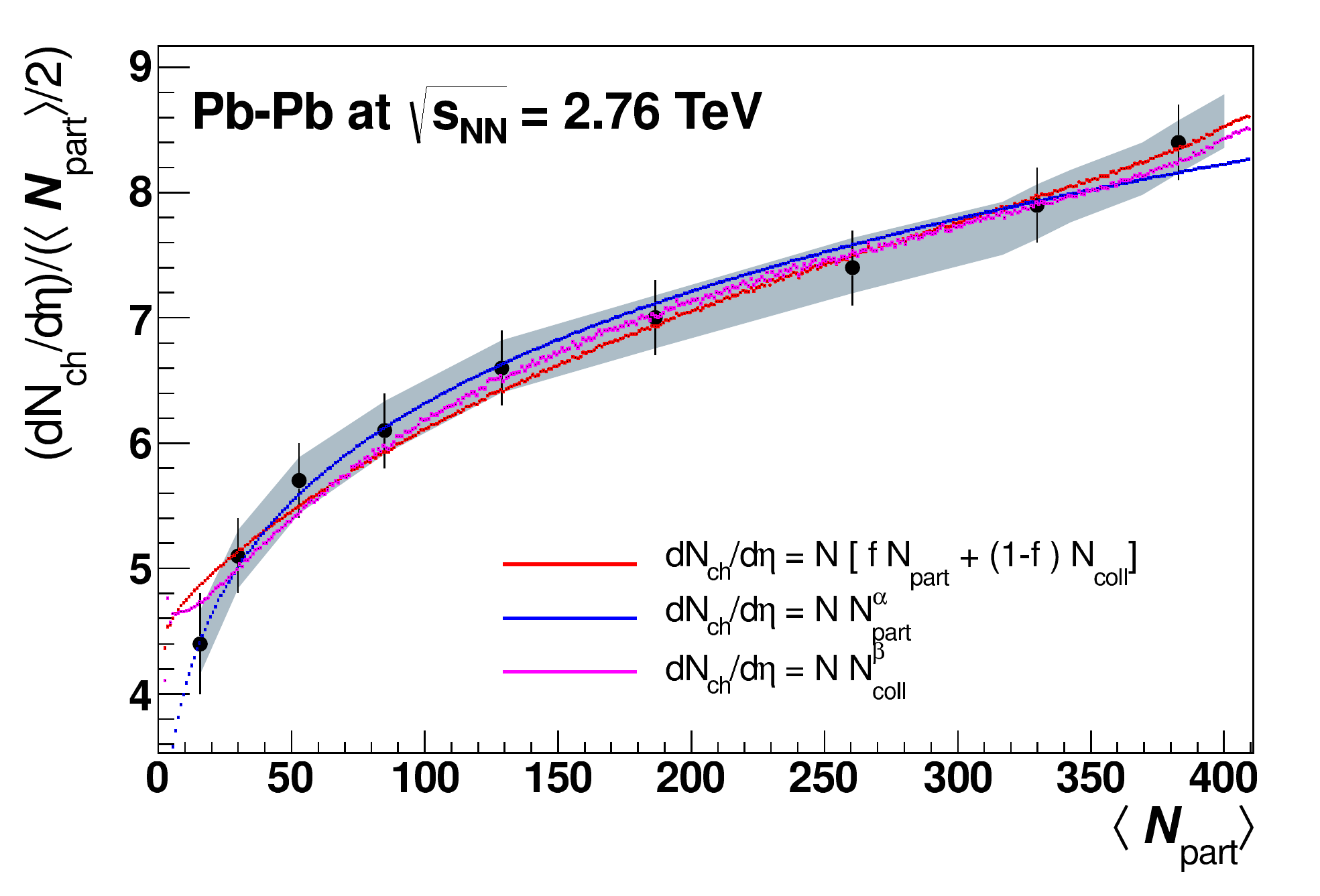}
        \caption{Relation between the number of charged particles per rapidity per participant pair and $\langle N_{\ma{part}} \rangle $.}\label{}
    \end{subfigure}%
\caption[Relations between the centrality, $\langle N_{\ma{coll}} \rangle $, $\langle N_{\ma{part}} \rangle $ and the number of charged particles per rapidity per participant pair.]{Relations between the centrality, $\langle N_{\ma{coll}} \rangle $, $\langle N_{\ma{part}} \rangle $ and the number of charged particles per rapidity per participant pair. The plots are taken from Ref.~\cite{Abelev:2013qoq}, which is based on Glauber Monte Carlo simulations and the measurements of the ALICE collaboration.}
\label{fig_chap1:centrality_Npart}
\end{figure}

Usually experimental results of $R_{AA}$ are plotted as a function of {\it centrality} rather than the impact parameter. The centrality is defined by binning all $AA$ collision events according to the number of charged particles $N_{\ma{ch}}$ produced in each collision event. For example, a $0$-$10\%$ centrality corresponds to the top $10\%$ of all the $AA$ collision events, if they are listed from high to low, based on the number of charged particles produced in each collision event. Alternatively, one can use some binary collision models to estimate the averaged number of binary collisions $\langle N_{\ma{coll}} \rangle $ and the averaged number of participants of the collision $\langle N_{\ma{part}} \rangle $ at a given impact parameter. (We use the averaged number because fluctuations can happen.) One can also use $\langle N_{\ma{part}} \rangle $ to represent the centrality since we expect the number of charged particles produced in one $AA$ collision, to be positively correlated with the number of participant nucleons in the $AA$ collision event. Therefore experimental results of $R_{AA}$ are often plotted as a function of $\langle N_{\ma{part}} \rangle $. Using a binary collision model, one can connect the impact parameter, the averaged number of participant nucleons, the averaged number of binary collisions and the centrality with each other. In Fig.~\ref{fig_chap1:centrality_Npart}, the relation between the centrality and $\langle N_{\ma{coll}} \rangle $ and the relation between the number of charged particles per rapidity per participant pair and $\langle N_{\ma{part}} \rangle $ are depicted. The plots show results from Monte Carlo simulations that are based on the Glauber model \cite{Glauber:1955qq,Miller:2007ri}.

\begin{figure}
    \centering
    \begin{subfigure}[b]{0.48\textwidth}
        \centering
        \includegraphics[height=2.8in]{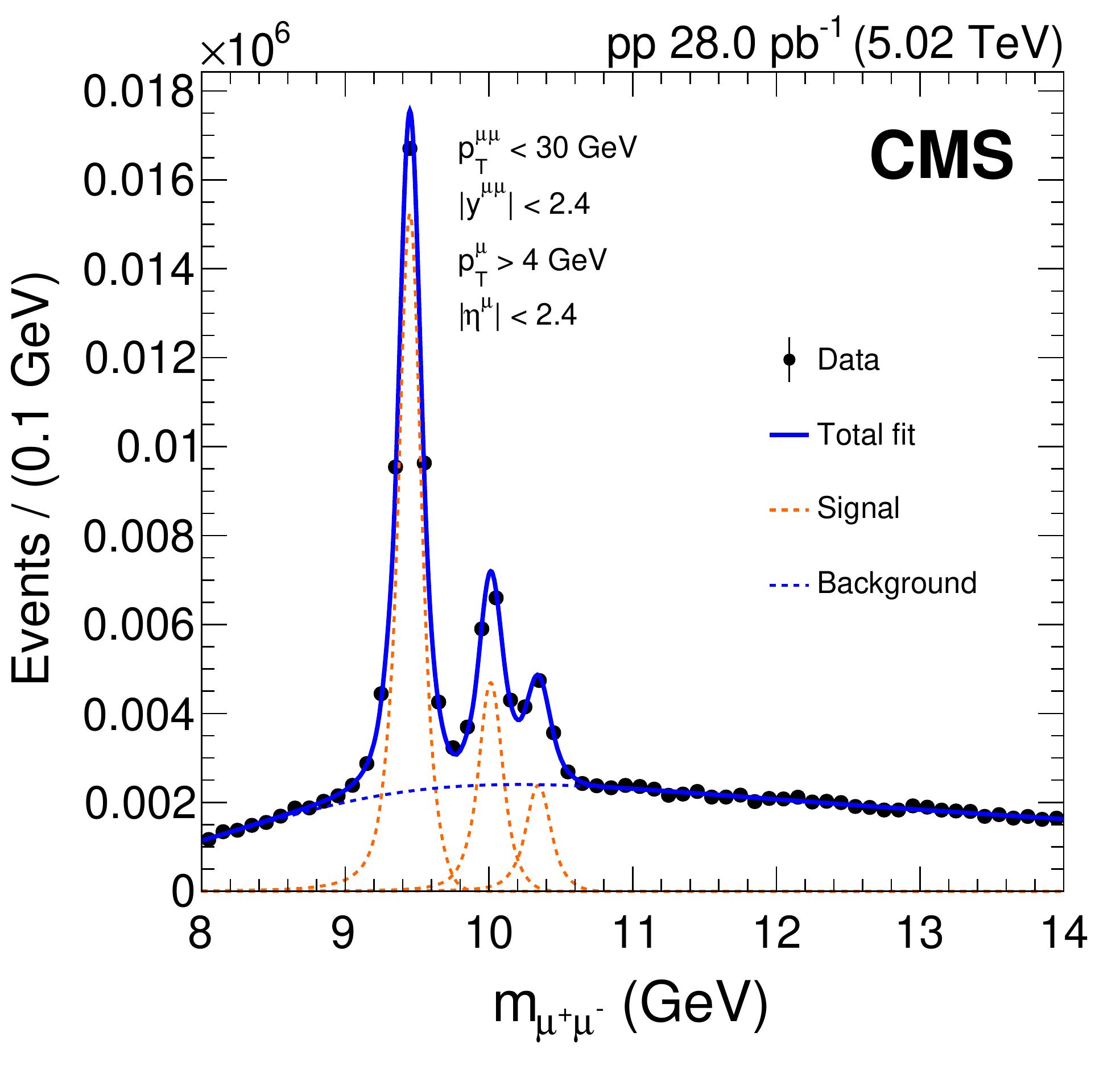}
        \caption{}\label{}
    \end{subfigure}%
    ~~
    \begin{subfigure}[b]{0.48\textwidth}
        \centering
        \includegraphics[height=2.8in]{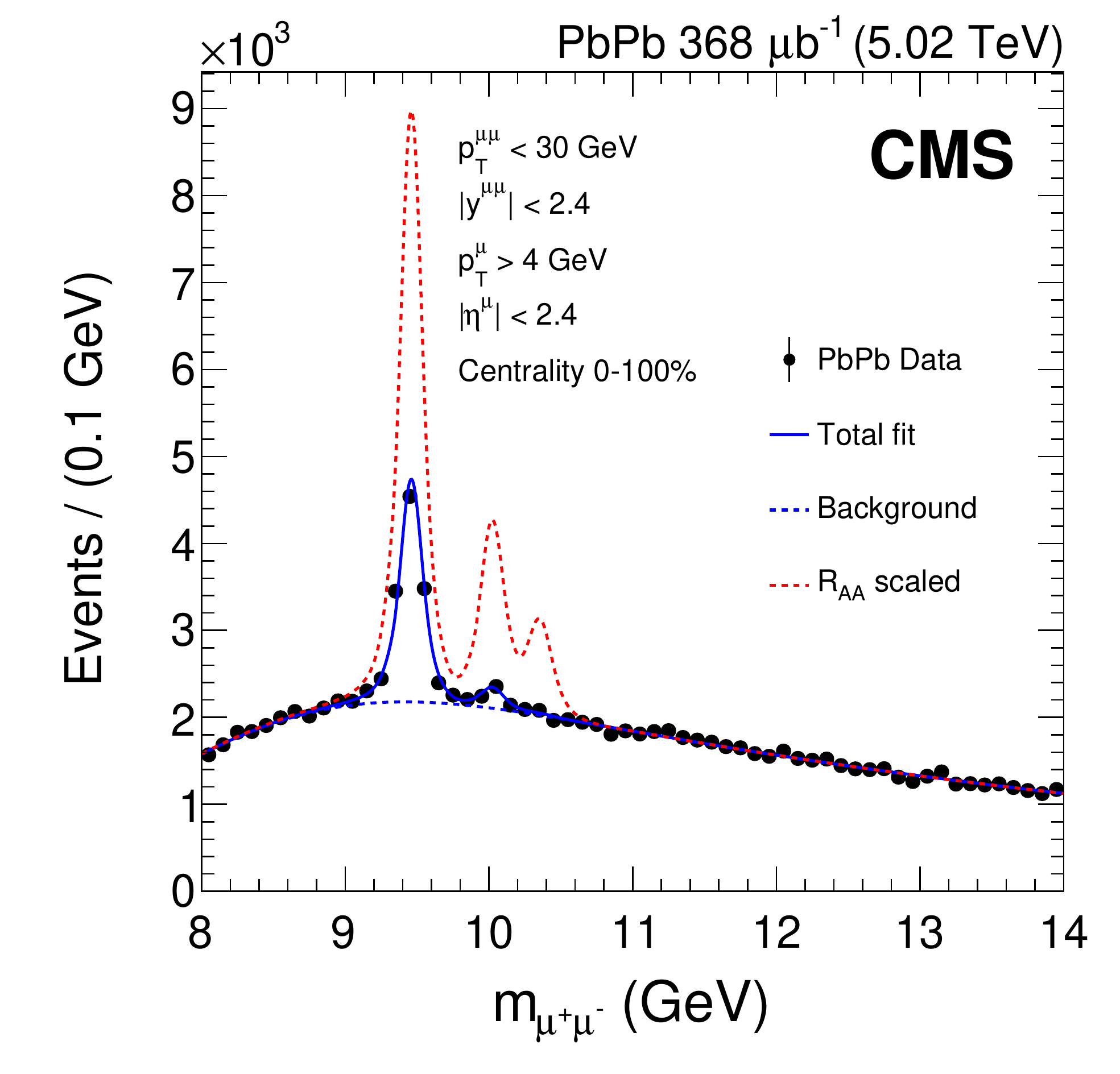}
        \caption{}\label{}
    \end{subfigure}%
\caption[Spectra of the $\mu^-\mu^+$ invariant mass and fitted number of $\Upsilon$(nS) produced.]{Spectra of the $\mu^-\mu^+$ invariant mass and fitted number of $\Upsilon$(nS) produced. The red dashed line is the spectra of $\Upsilon$(nS) in an equivalent proton-proton collision without cold and hot medium effects. Clearly the bottomonium production in heavy ion collisions is suppressed. The plots are taken from Ref.~\cite{Sirunyan:2018nsz}.}
\label{fig_chap1:cms_yield}
\end{figure}

In Fig.~\ref{fig_chap1:cms_yield}, the spectra of the $\mu^-\mu^+$ invariant mass measured in $5.02$ TeV Pb-Pb collisions at LHC by the CMS collaboration and the fitted number of $\Upsilon$(nS) produced are shown \cite{Sirunyan:2018nsz}. In the right subplot, the measured heights of all the three peaks are significantly lowered than those in an equivalent proton-proton collision (marked in red dashed line). Thus it is clear that quarkonium production in heavy ion collisions is suppressed.

\begin{figure}
    \centering
    \begin{subfigure}[b]{0.48\textwidth}
        \centering
        \includegraphics[height=2.1in]{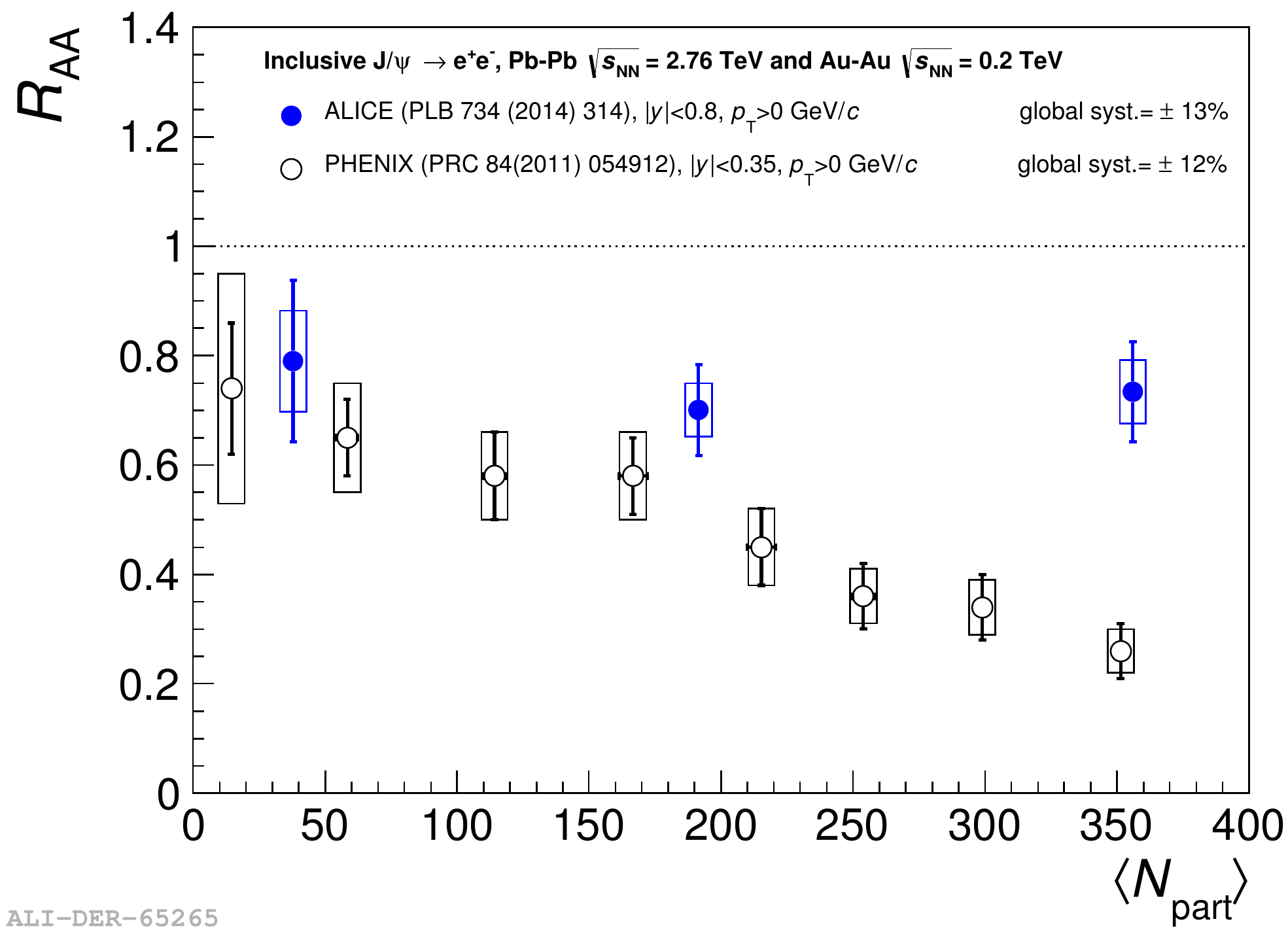}
        \caption{Mid rapidity}\label{}
    \end{subfigure}%
    ~~
    \begin{subfigure}[b]{0.48\textwidth}
        \centering
        \includegraphics[height=2.1in]{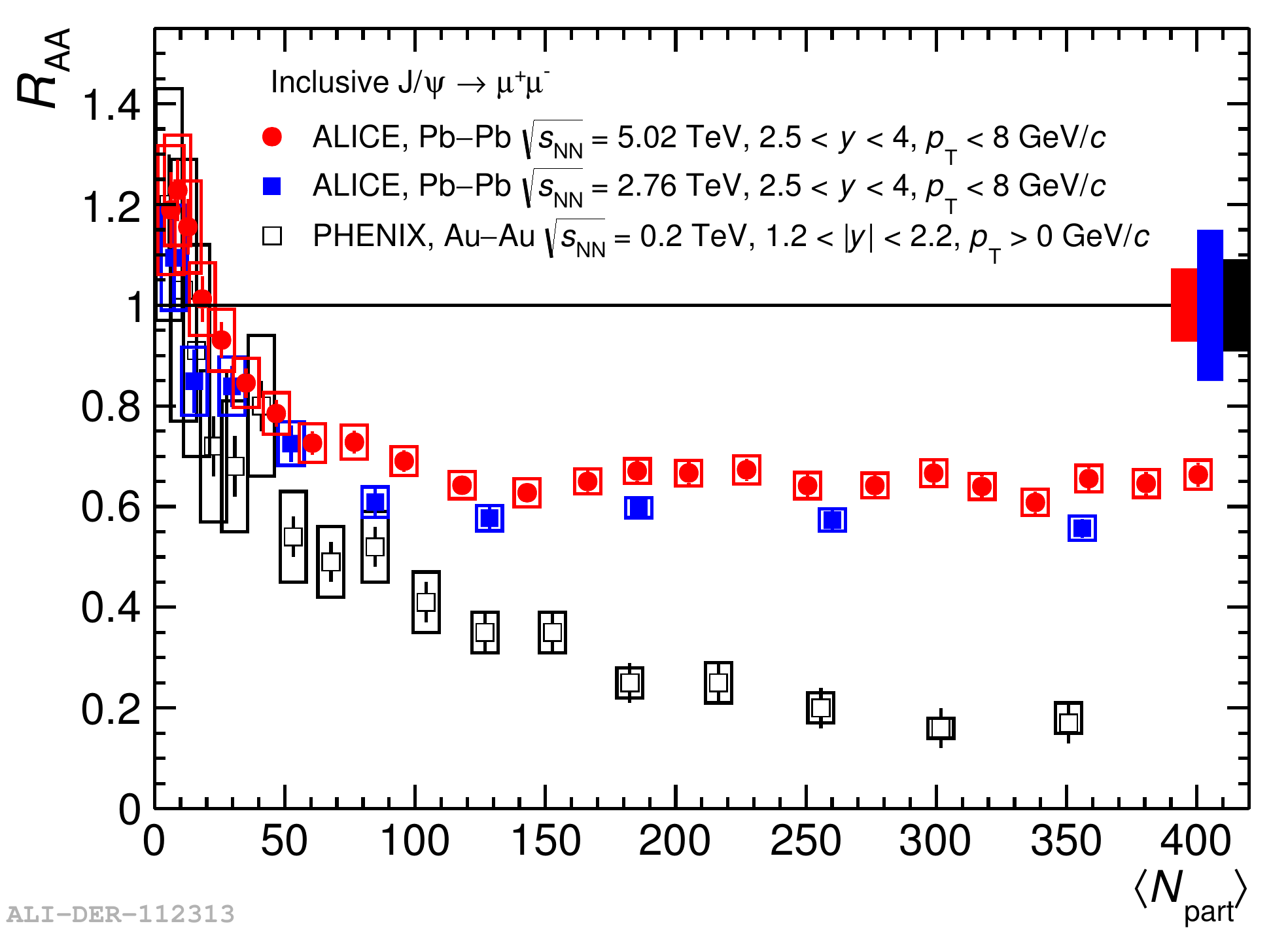}
        \caption{Forward rapidity}\label{}
    \end{subfigure}%
\caption[$R_{AA}$ of inclusive $J/\psi$ in mid and forward rapidities.]{$R_{AA}$ of inclusive $J/\psi$ in mid and forward rapidities. The data are measured by the PHENIX collaboration at RHIC \cite{Adare:2011yf} and the ALICE collaboration at LHC \cite{Abelev:2012rv,Abelev:2013ila,Adam:2016rdg}. The left plot is taken from Ref.~\cite{Andronic:2015wma} while the right plot is taken from Ref.~\cite{Scomparin:2017pno}. The $J/\psi$ $R_{AA}$ measured by ALICE increases slightly when the rapidity changes from $2.5 < y <4$ to $|y|<0.8$. If the ALICE data is constrained to the same rapidity range $|y|<0.35$ as in the PHENIX measurements, we would expect the $J/\psi$ $R_{AA}$ to further increase slightly at the LHC energy. Therefore, the comparison of $J/\psi$ $R_{AA}$ in the mid rapidity clearly shows that $J/\psi$ production is less suppressed at LHC than that at RHIC. This implies significant recombination of $Q\bar{Q}$ at the LHC energies. Another evidence is the slight increase in $J/\psi$ $R_{AA}$ at $5.02$ TeV above that at $2.76$ TeV.}
\label{fig_chap1:alice_vs_phenix}
\end{figure}

The $R_{AA}$ results of inclusive $J/\psi$ as a function of centrality at both the RHIC $200$ GeV and LHC $2.76$ and $5.02$ TeV heavy ion collisions are shown in Fig.~\ref{fig_chap1:alice_vs_phenix}. Though the measurements at RHIC and LHC have different rapidity cuts, we can still compare the results because the rapidity dependence at the LHC energy is weak. The $J/\psi$ $R_{AA}$ measured by ALICE increases slightly when the rapidity changes from $2.5 < y <4$ to $|y|<0.8$. If the ALICE data is constrained to the same rapidity range $|y|<0.35$ as in the PHENIX measurements, we would expect the $J/\psi$ $R_{AA}$ to further increase slightly at the LHC energy. Therefore, the comparison of $J/\psi$ $R_{AA}$ in the mid rapidity clearly shows that $J/\psi$ production is less suppressed at LHC than that at RHIC. As discussed in the last subsection, the reason is the enhanced contribution from recombination of unbound charm-anticharm pairs. At the higher LHC energies, more charm-anticharm quarks are produced in the initial hard scattering and the charm (anticharm) quark density (number per rapidity) increases significantly. Therefore, the probability for recombination is higher for them inside the QGP. The fact that the $J/\psi$ $R_{AA}$ at the LHC energy is larger at mid rapidity than that at forward/backward rapidity also indicates the importance of the recombination effect. Another evidence of significant recombination at the LHC energies is the slight increase in $J/\psi$ $R_{AA}$ at $5.02$ TeV above that at $2.76$ TeV.

\begin{figure}
\begin{center}
\includegraphics[height=2.2in]{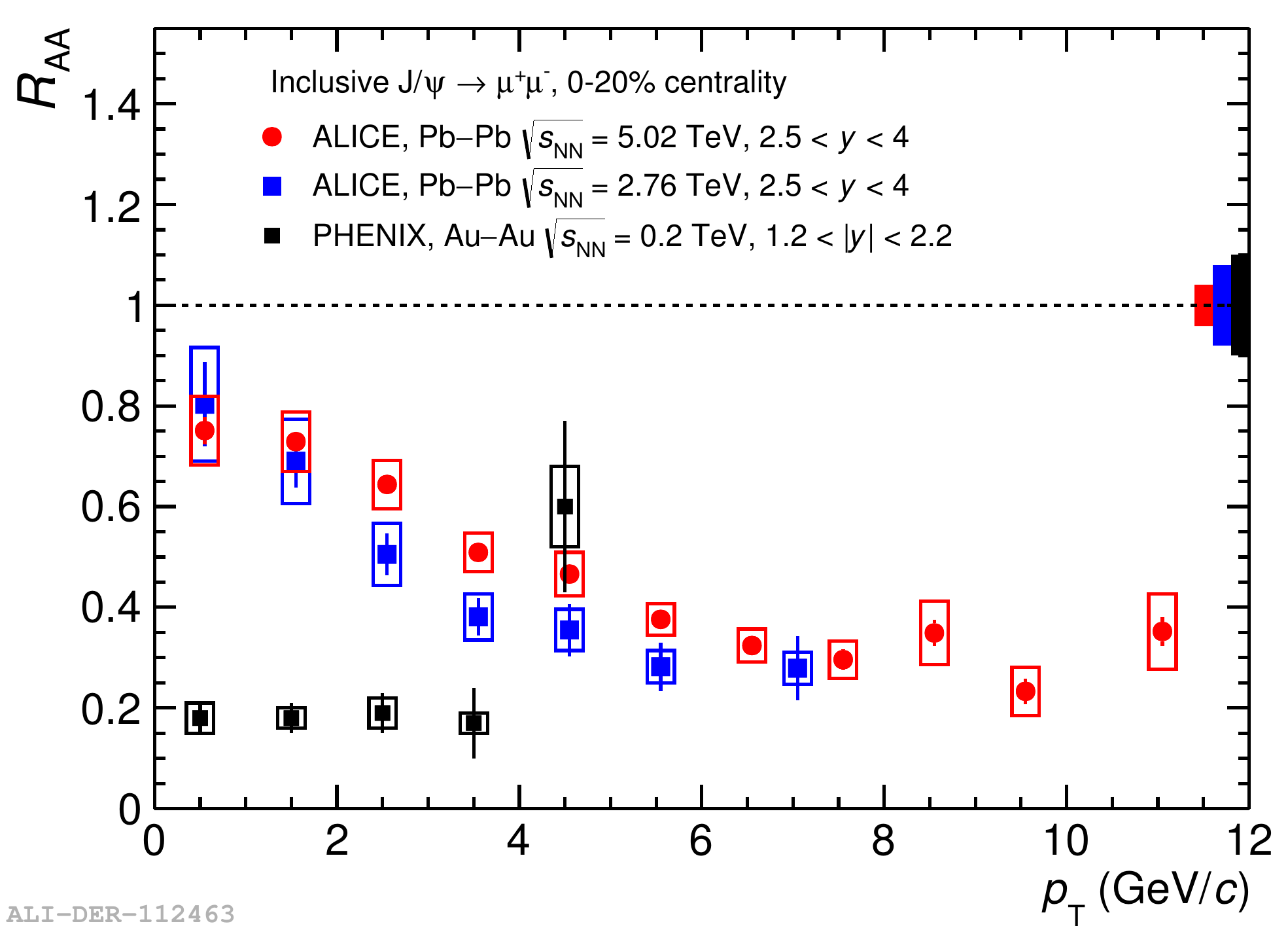}
\caption[$R_{AA}$ of inclusive $J/\psi$ as a function of transverse momentum.]{$R_{AA}$ of inclusive $J/\psi$ as a function of transverse momentum. The data are measured by the PHENIX collaboration at RHIC \cite{Adare:2011yf} and the ALICE collaboration at LHC \cite{Adam:2016rdg,Adam:2015isa}. Low-$p_T$ $J/\psi$ is more suppressed at the RHIC energy than that at the LHC energies. This implies the importance of recombination effect at LHC energies.}
\label{fig_chap1:alice_vs_phenix_pT}
\end{center}
\end{figure}

\begin{figure}
    \centering
    \begin{subfigure}[b]{0.48\textwidth}
        \centering
        \includegraphics[height=2.6in]{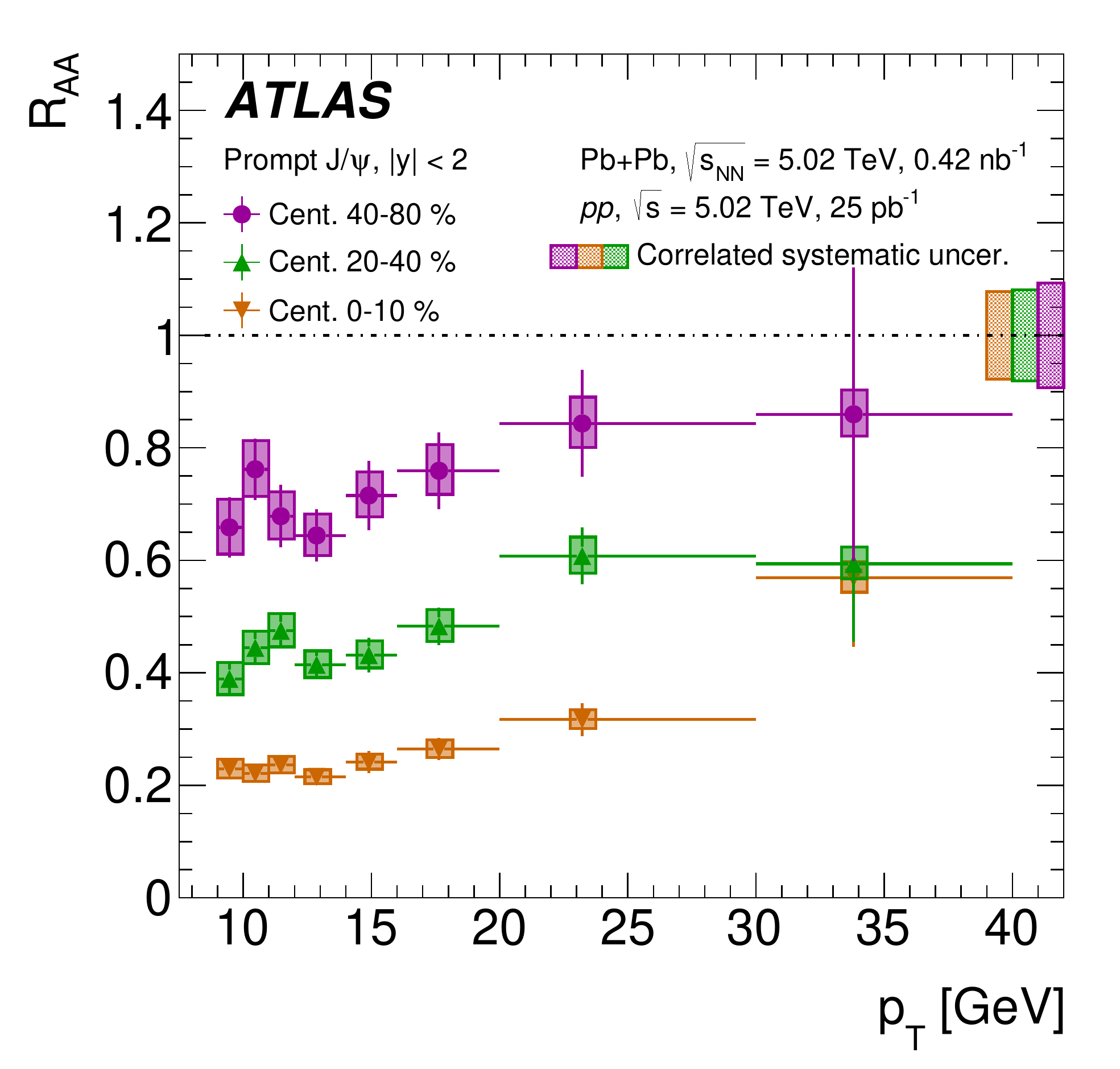}
        \caption{Prompt $J/\psi$.}\label{atlas:pT_p}
    \end{subfigure}%
    ~~
    \begin{subfigure}[b]{0.48\textwidth}
        \centering
        \includegraphics[height=2.6in]{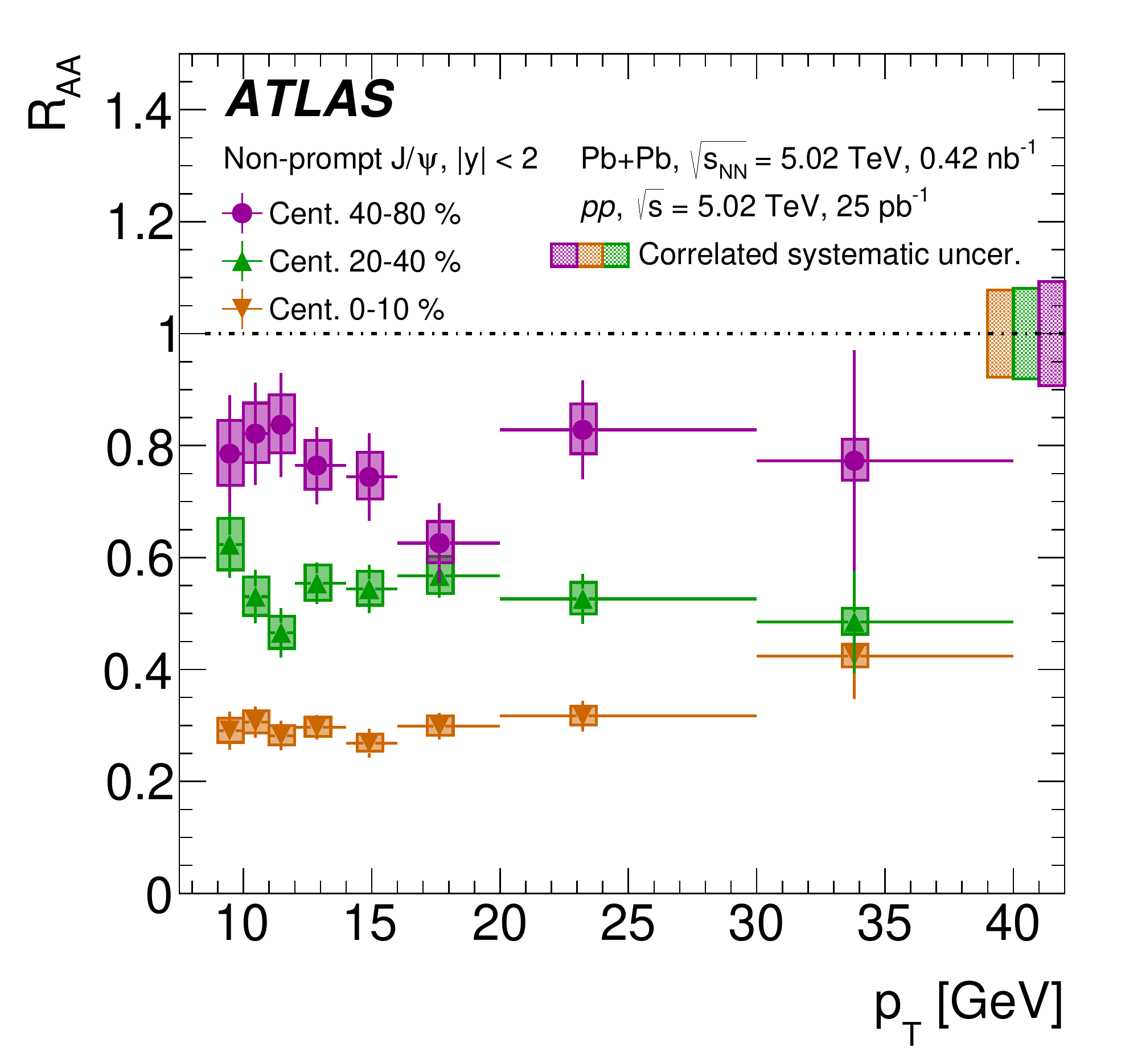}
        \caption{Non-prompt $J/\psi$.}\label{atlas:pT_np}
    \end{subfigure}%
    ~~
    
    \begin{subfigure}[b]{0.48\textwidth}
        \centering
        \includegraphics[height=2.6in]{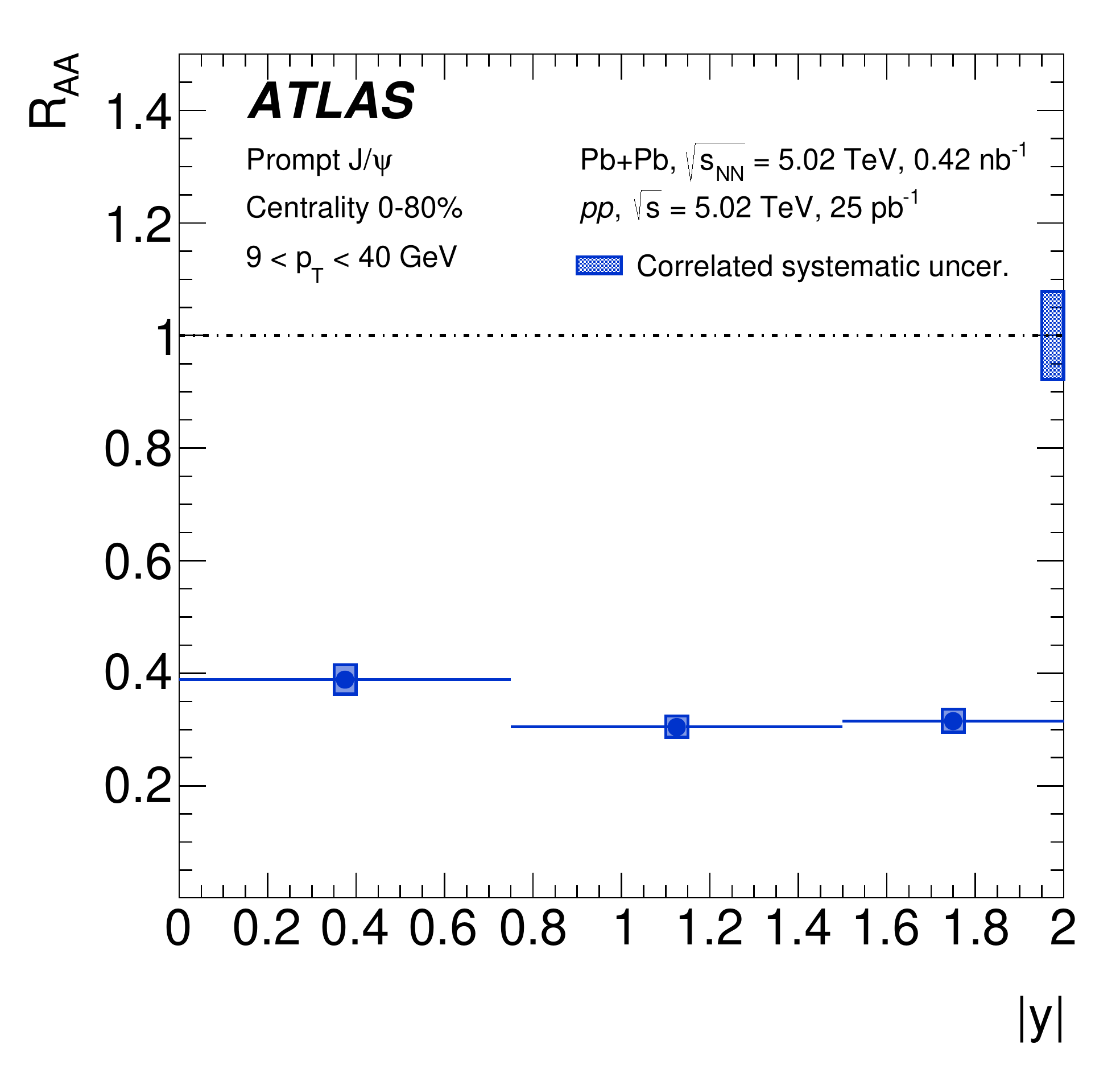}
        \caption{Prompt $J/\psi$.}\label{atlas:rap_p}
    \end{subfigure}%
    ~~ 
    \begin{subfigure}[b]{0.48\textwidth}
        \centering
        \includegraphics[height=2.6in]{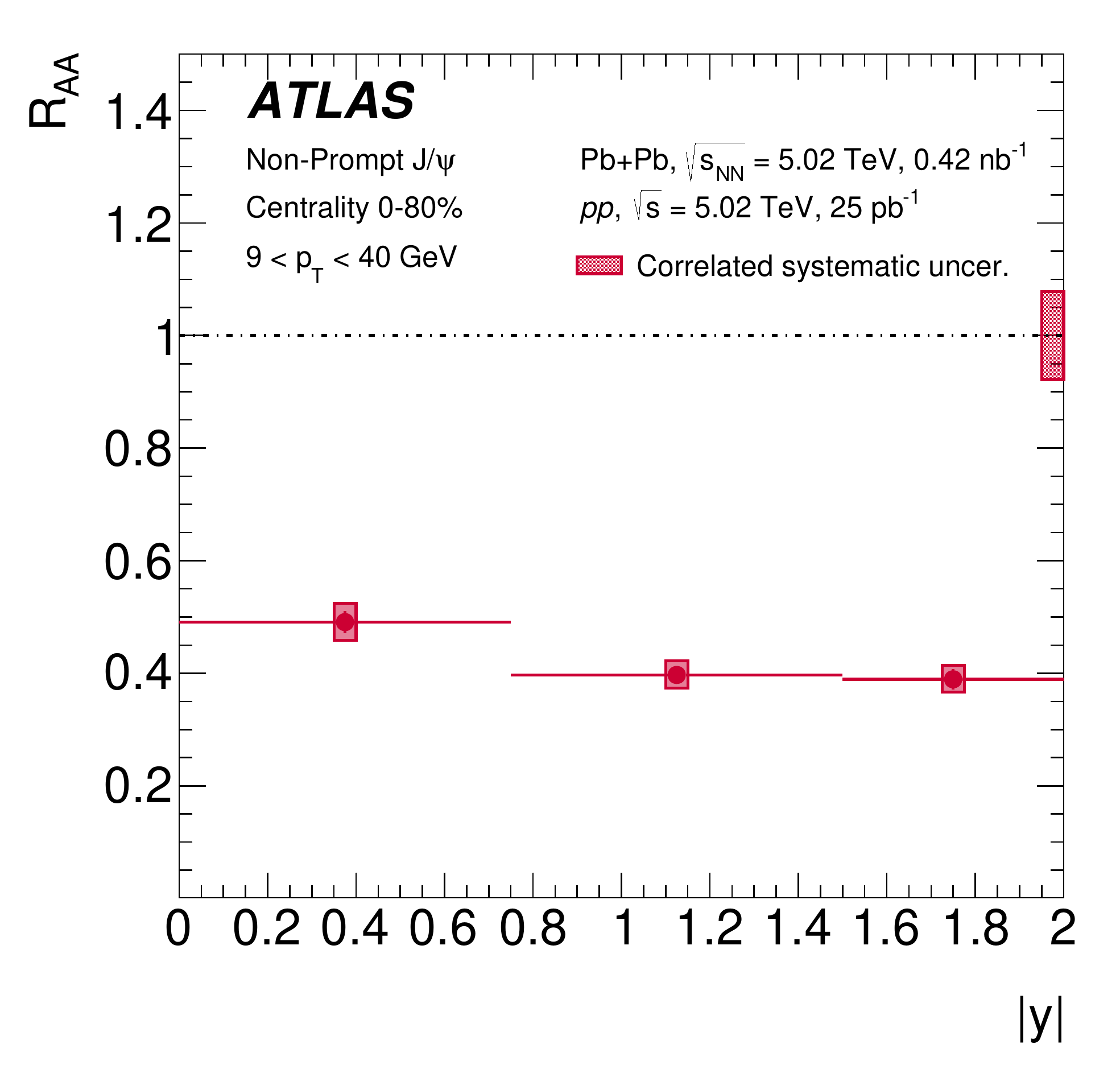}
        \caption{Non-prompt $J/\psi$.}\label{atlas:rap_np}
    \end{subfigure}%
    ~~
    
\caption[$R_{AA}$ of high-$p_T$ prompt and non-prompt $J/\psi$ as functions of transverse momentum and rapidity, measured by the ATLAS collaboration.]{$R_{AA}$ of high-$p_T$ prompt and non-prompt $J/\psi$ as functions of transverse momentum and rapidity, measured by the ATLAS collaboration \cite{Aaboud:2018quy}. Prompt $J/\psi$ is formed from $c\bar{c}$ that are produced in the initial hard scattering of heavy ion collisions while non-prompt $J/\psi$ is produced from $b$-hadron decays. We can see from (a) and (b) that both the prompt and non-prompt $J/\psi$ are more suppressed in more central collisions. The dependences of suppression on the transverse momentum and the rapidity are weak.}
\label{fig_chap1:atlas}
\end{figure}

With more statistics in experimental measurements, more differential observables can be measured such as the transverse momentum ($p_T$) dependent $R_{AA}$ and rapidity dependent $R_{AA}$. These measurements contain information of how the static screening, dissociation and recombination depend on the relative velocity between the quarkonium and the medium. The results at both the RHIC and LHC energies are shown in Fig.~\ref{fig_chap1:alice_vs_phenix_pT}. It is clear that low-$p_T$ $J/\psi$ is more suppressed at the RHIC energy than that at the LHC energies. Again, this implies the importance of recombination effect at LHC energies. The $R_{AA}$ of high-$p_T$ $J/\psi$ at the LHC $5.02$ TeV is measured by the ATLAS collaboration and the result is shown in Fig.~\ref{fig_chap1:atlas} as functions of transverse momentum and rapidity. Both the prompt and non-prompt $J/\psi$ $R_{AA}$ are measured. Prompt $J/\psi$ is formed from $c\bar{c}$ that are produced in the initial hard scattering of heavy ion collisions while non-prompt $J/\psi$ is produced from $b$-hadron decays. From Figs.~\ref{atlas:pT_p} and \ref{atlas:pT_np}, we can see that both the prompt and non-prompt $J/\psi$ are more suppressed in more central collisions. The dependences of suppression on the transverse momentum and the rapidity are weak.

\begin{figure}
    \centering
    \begin{subfigure}[b]{0.48\textwidth}
        \centering
        \includegraphics[height=3in]{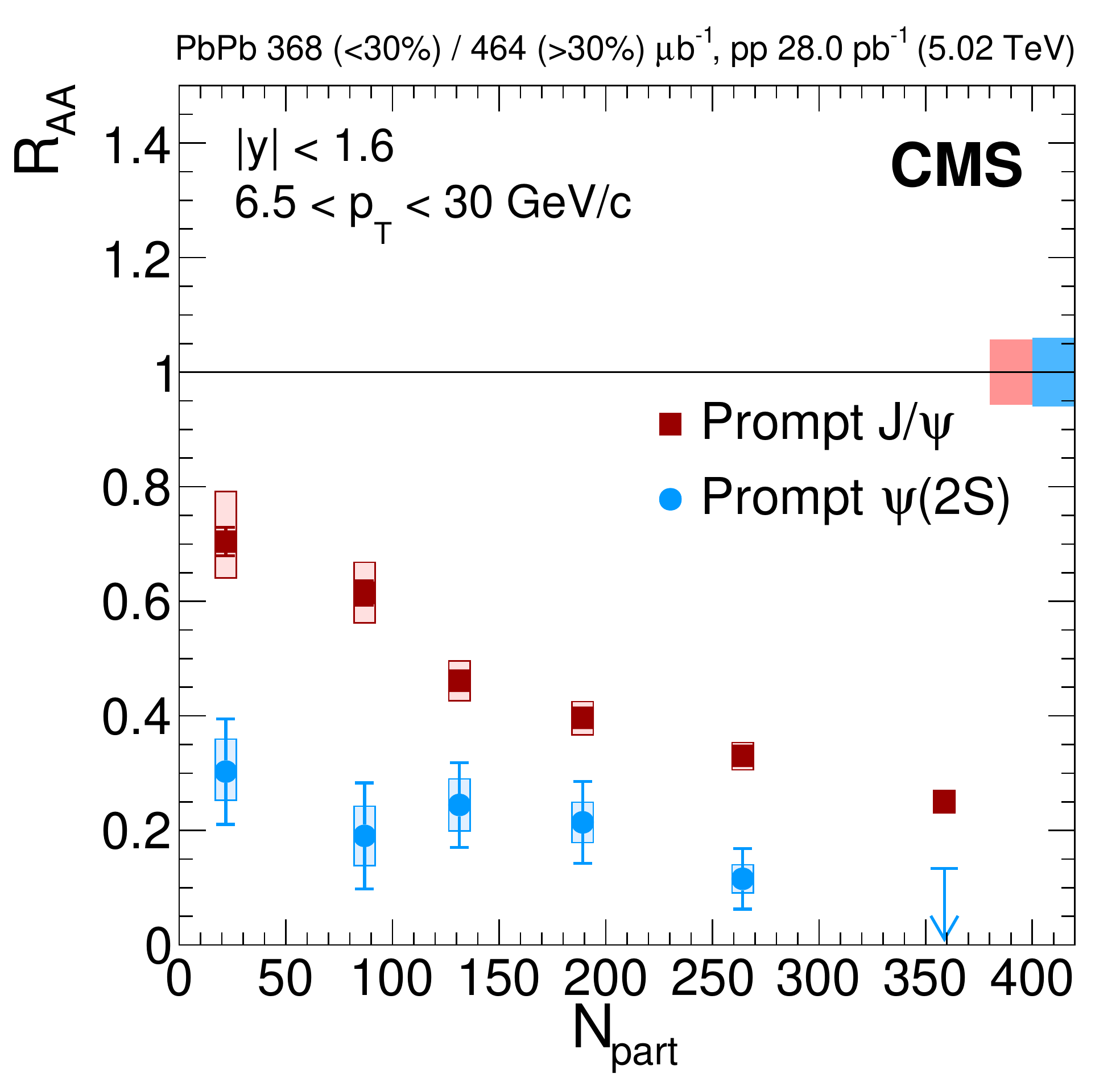}
        \caption{}\label{cms:psi2S_centrality}
    \end{subfigure}%
    ~~
    \begin{subfigure}[b]{0.48\textwidth}
        \centering
        \includegraphics[height=3in]{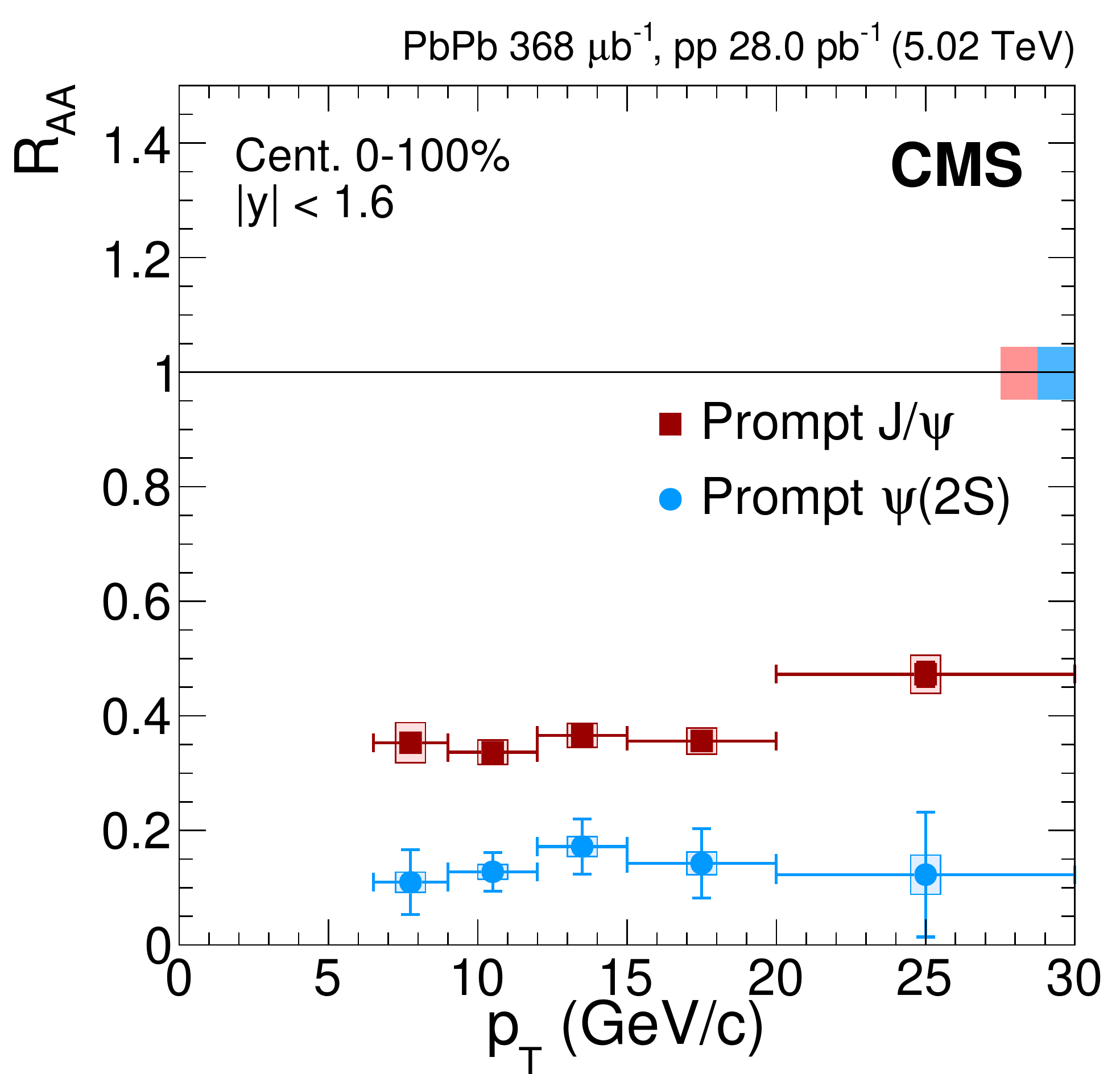}
        \caption{}\label{cms:psi2S_pT}
    \end{subfigure}%
    ~~
\caption[$R_{AA}$ of prompt $J/\psi$ and $\psi$(2S) as functions of centrality and transverse momentum, measured by the CMS collaboration.]{$R_{AA}$ of prompt $J/\psi$ and $\psi$(2S) as functions of centrality and transverse momentum, measured by the CMS collaboration \cite{Sirunyan:2017isk}. The excited charmonium state $\psi$(2S) is more suppressed than the ground state $J/\psi$ because the size of $\psi$(2S) is bigger and is more strongly affected by the medium.}
\label{fig_chap1:cms_psi2S}
\end{figure}

The nuclear modification factor $R_{AA}$ of excited charmonium states can also be measured with enough experimental statistics. This kind of measurement is interesting because we can learn how the plasma screening effects and the recombination depend on the size of the quarkonium state, or equivalently, the binding energy. The CMS collaboration results of prompt $\psi$(2S) $R_{AA}$ as functions of centrality and transverse momentum at the LHC energy are shown in Fig.~\ref{fig_chap1:cms_psi2S}. Obviously $\psi$(2S) has bigger size and is thus more suppressed than $J/\psi$. Theoretically it would be interesting to understand how the suppression depends on the size of quarkonium, which is the combined result of static screening, dissociation and recombination. 

\begin{figure}
    \centering
    \begin{subfigure}[b]{0.5\textwidth}
        \centering
        \includegraphics[height=2.3in]{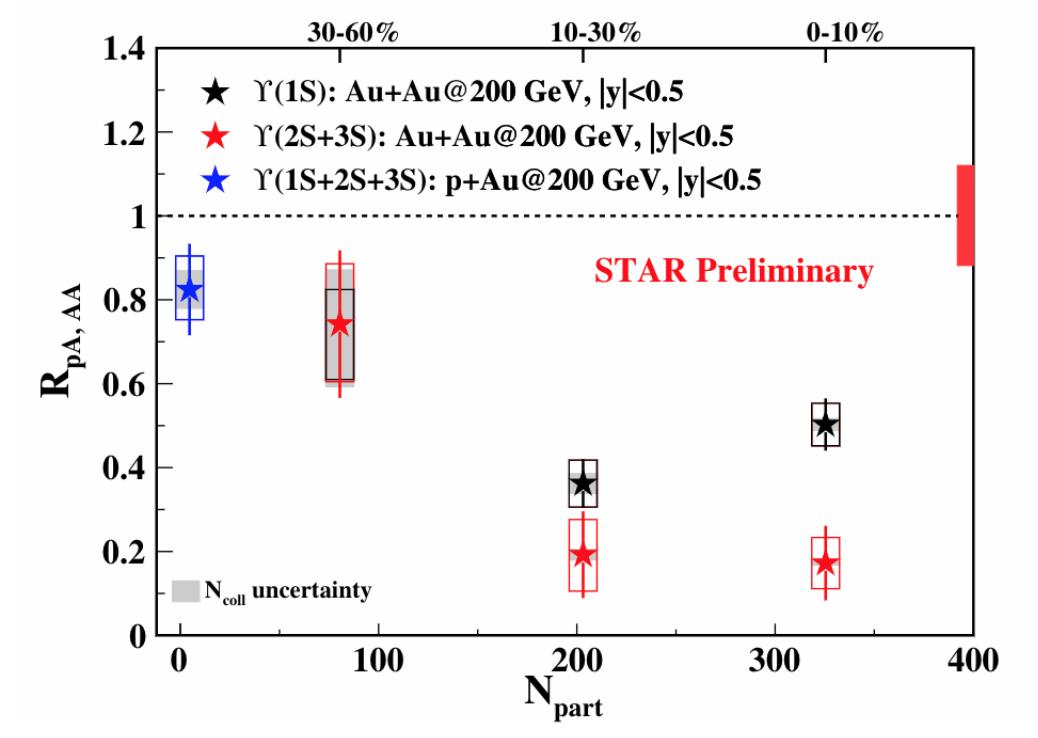}
        \caption{}\label{}
    \end{subfigure}%
    ~
    \begin{subfigure}[b]{0.4\textwidth}
        \centering
        \includegraphics[height=2.3in]{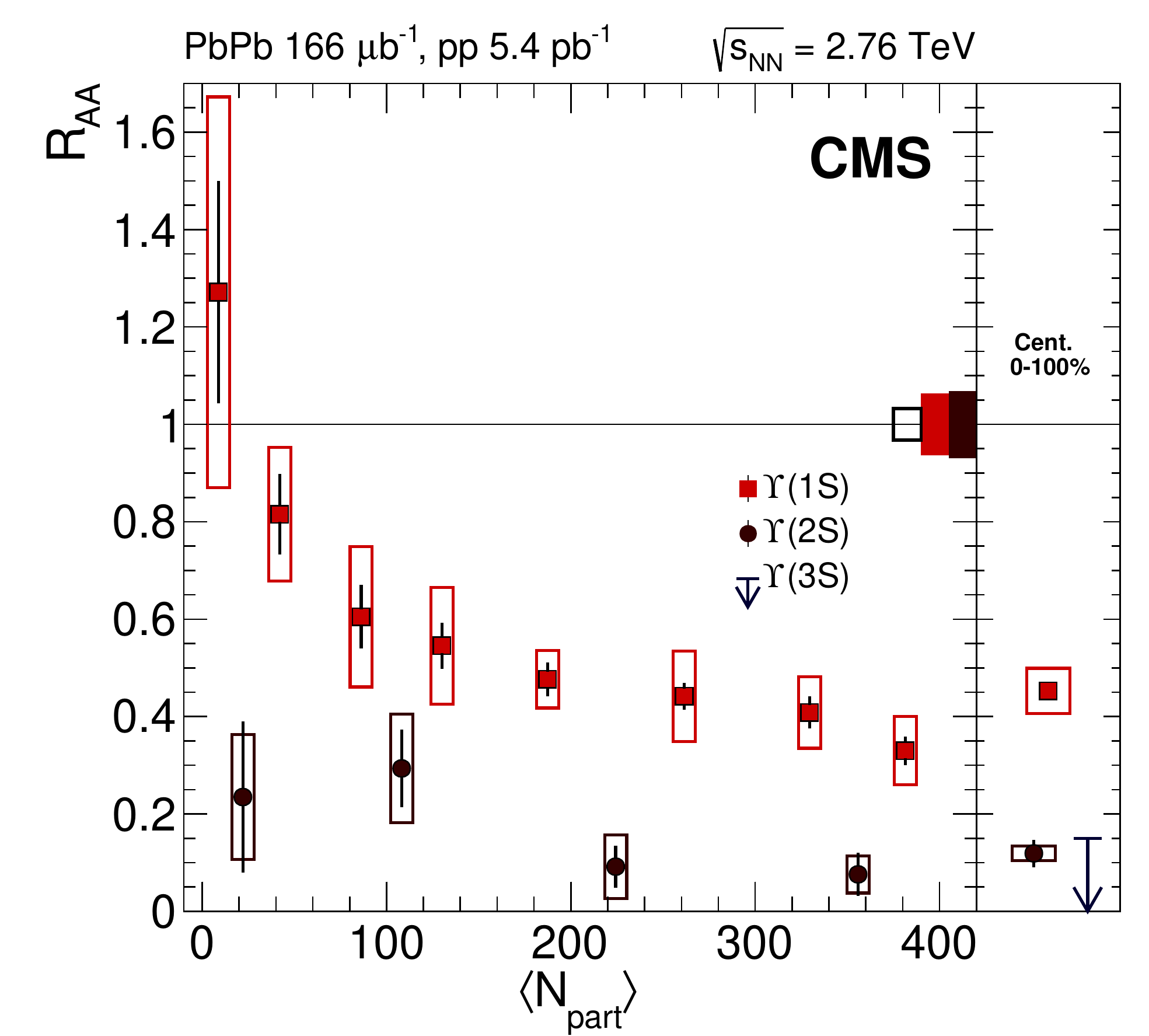}
        \caption{}\label{}
    \end{subfigure}%

    \begin{subfigure}[b]{0.5\textwidth}
        \centering
        \includegraphics[height=2.3in]{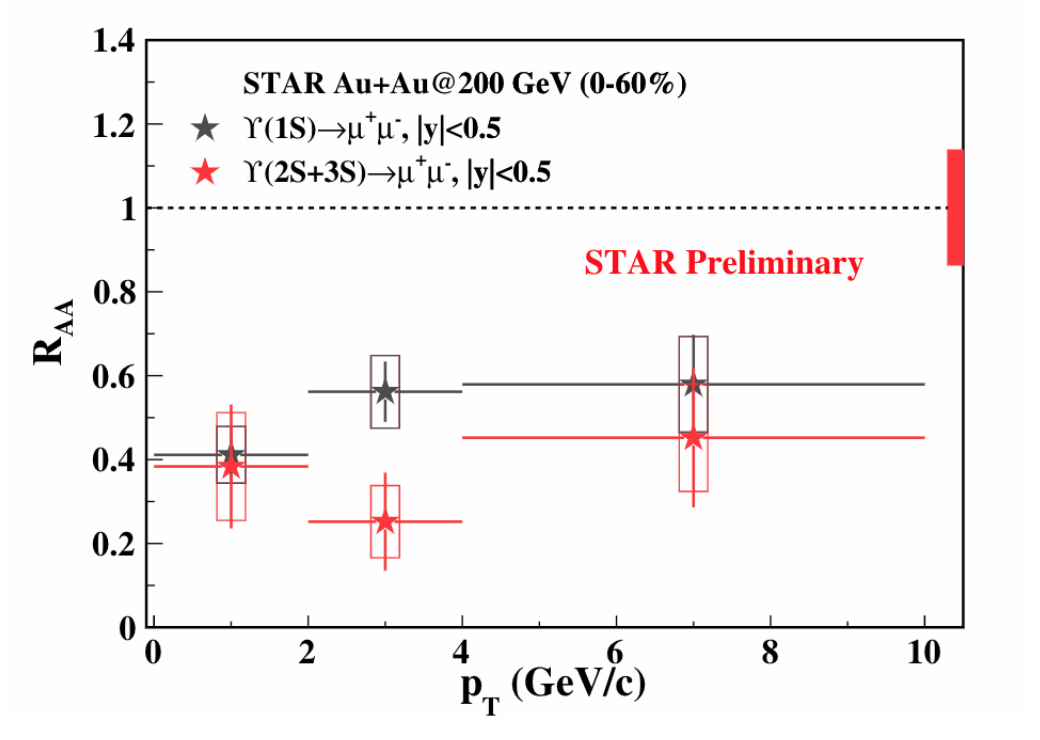}
        \caption{}\label{}
    \end{subfigure}%
    ~~~~
    \begin{subfigure}[b]{0.4\textwidth}
        \centering
        \includegraphics[height=2.3in]{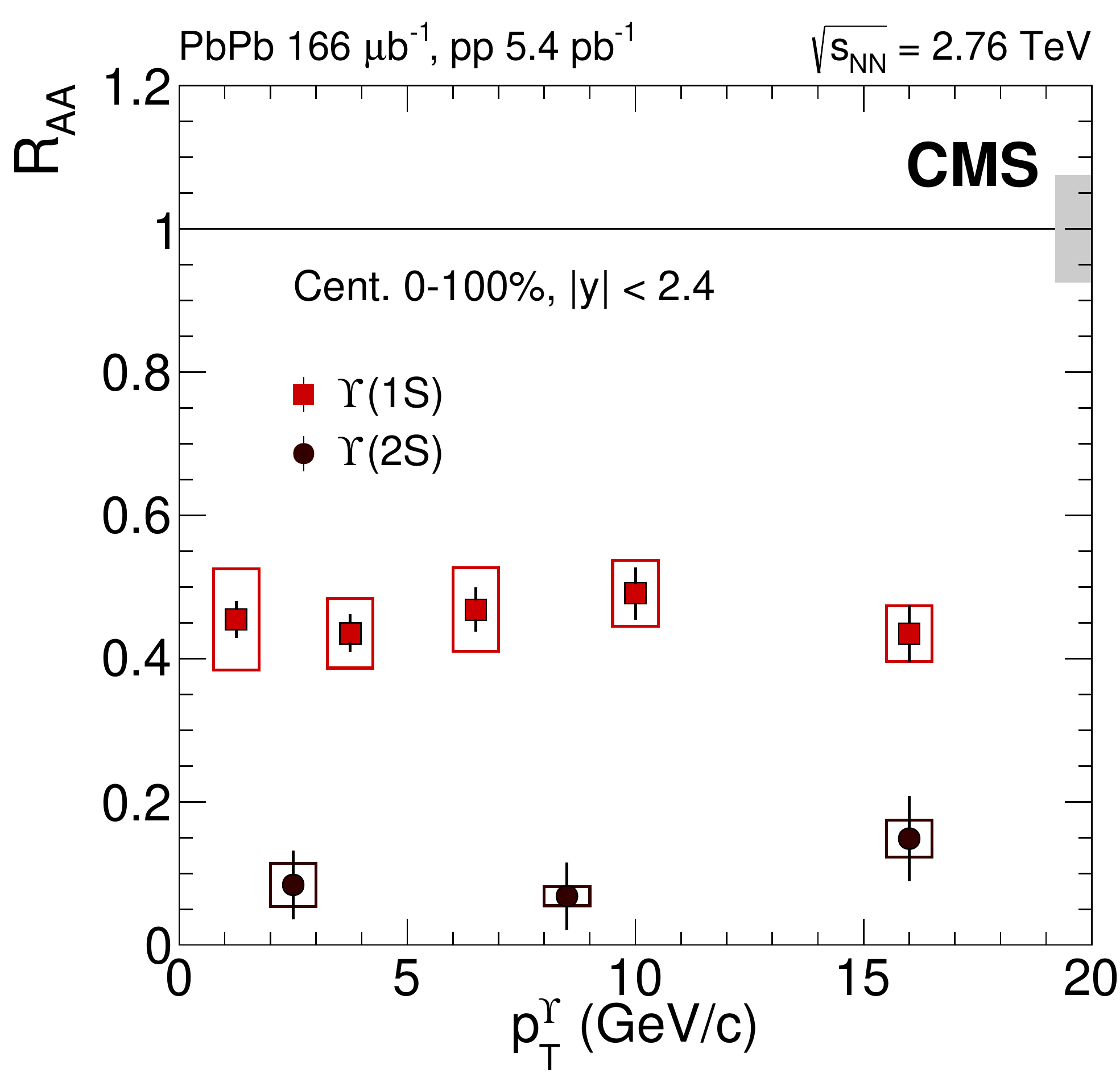}
        \caption{}\label{}
    \end{subfigure}%
    
\caption[$R_{AA}$ of $\Upsilon$(nS) as functions of centrality and transverse momentum measured at the RHIC and LHC energies.]{$R_{AA}$ of $\Upsilon$(nS) as functions of centrality and transverse momentum measured at the RHIC and LHC energies. The RHIC data is measured by the STAR collaboration \cite{Wang:2019vau} while the LHC data is measured by the CMS collaboration \cite{Khachatryan:2016xxp}. The suppression factors of $\Upsilon$(nS) are ordered by the sizes of these states. The smallest state $\Upsilon$(1S) is the least suppressed. The CMS measurements only obtain an upper limit of the $R_{AA}$ of $\Upsilon$(3S), which may imply a complete suppression of $\Upsilon$(3S) at the LHC energies. The $R_{AA}$'s of $\Upsilon$(1S) at the RHIC and LHC energies are approximately equal to each other. This implies recombination from uncorrelated $b\bar{b}$'s for bottomonium at the LHC energies is weaker than that for charmonium because the number of open bottom quarks produced is much smaller than the number of open charm quarks.} 
\label{fig_chap1:cms_vs_star}
\end{figure}

For bottomonium, similar measurements can be conducted. The results of $R_{AA}$ of $\Upsilon$(nS) with $n=1,2,3$ at both the RHIC and LHC energies are shown in Fig.~\ref{fig_chap1:cms_vs_star}. The suppression factors of $\Upsilon$(nS) are ordered by the sizes of these states. The smallest state $\Upsilon$(1S) is the least suppressed. The CMS measurements only obtain an upper limit of the $R_{AA}$ of $\Upsilon$(3S), which may imply a complete suppression of $\Upsilon$(3S) at the LHC energies. The $R_{AA}$'s of $\Upsilon$(1S) at the RHIC and LHC energies are approximately equal to each other. This implies recombination from uncorrelated $b\bar{b}$'s for bottomonium at the LHC energies is weaker than that for charmonium because the number of open bottom quarks produced is much smaller than the number of open charm quarks. More experimental results on $\Upsilon$(nS) $R_{AA}$ will be shown later in Chapter 4 when we compare the calculated results with measurements.

\begin{figure}
\begin{center}
\includegraphics[height=2.2in]{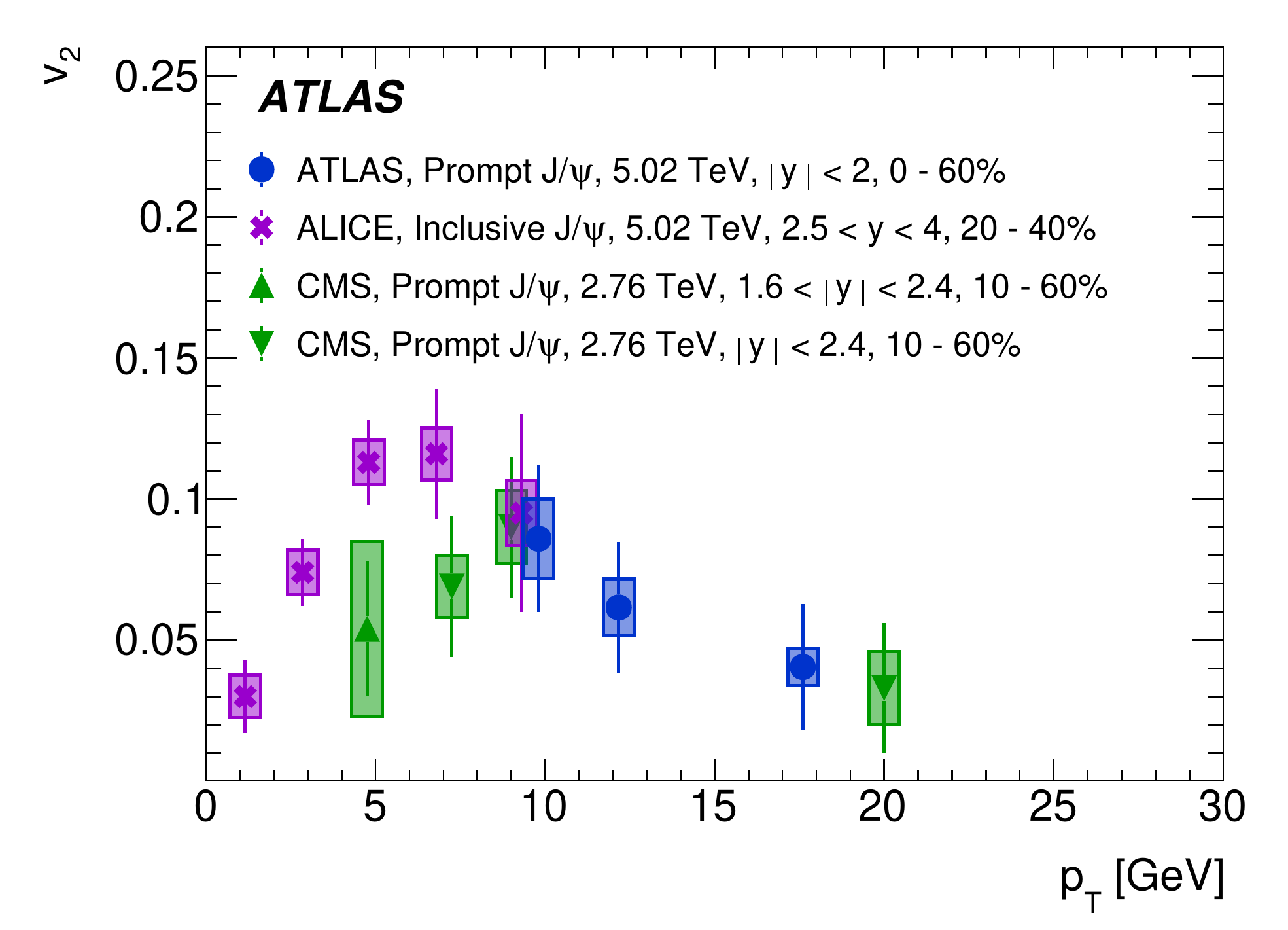}
\caption[$v_2$ of $J/\psi$ measured by the ALICE, ATLAS and CMS collaborations.]{$v_2$ of $J/\psi$ measured by the ALICE \cite{Acharya:2017tgv}, ATLAS \cite{Aaboud:2018ttm} and CMS \cite{Khachatryan:2016ypw} collaborations. The plot is taken from Ref.~\cite{Aaboud:2018ttm}. The non-vanishing values of $J/\psi$ $v_2$ at low $p_T$ imply that charm quarks accumulate the azimuthal angular anisotropy from interactions with the medium. When they recombine as $J/\psi$, the anisotropy is inherited by the bound state.}
\label{fig_chap1:v2}
\end{center}
\end{figure}

Another important experimental observable is the azimuthal angular distribution of the produced quarkonium state $H$
\be
E\frac{\diff^3N_H}{\diff p^3} =\frac{1}{2\pi} \frac{\diff^2N_H}{p_T\diff p_T\diff y}\Big(1+2 \sum_{n=1}v_n({\bs b})\cos[n (\phi - \Psi_\ma{RP} ) ]\Big)\,,
\ee
where $\Psi_\ma{RP}$ is the angle of the reaction plane. For light particles such as pions, kaons or protons, the parameters $v_1$, $v_2$ and $v_3$ are called the directed, elliptic and triangular flows. The flow is a signature of the collective expansion of QGP. In general, the flow $v_n({\bs b})$ depends on the impact parameter or centrality. For heavy quarks and quarkonium, we would rather just call them parameters in the azimuthal angular anisotropy. This is because most light quarks are thought of as part of the medium and participate in the collective expansion while heavy quarks are ``external" currents or hard probes of the QGP. The azimuthal angular anisotropies of heavy quarks and quarkonium are generated when they travel through and interact with the expanding medium. The experimental measurements of $J/\psi$ $v_2$ are shown in Fig.~\ref{fig_chap1:v2}. One explanation of the non-zero $v_2$ is that $J/\psi$ inherits the azimuthal angular anisotropy from recombining charm quarks. The open charm quarks gradually accumulate the anisotropy by interacting with the medium when travel through the medium. The $v_2$ parameter is an important observable for our theoretical understanding of heavy quarks and quarkonium evolution inside the hot medium.

As both RHIC and LHC upgrade their detectors in the forthcoming years, more quarkonium measurements with high statistics and precision will come. Therefore it is time to improve and deepen our theoretical understanding of quarkonium in-medium dynamics.

\vspace{0.2in}
\section{Theoretical Tools}

\vspace{0.2in}
\subsection{Thermal Field Theory}
\subsubsection{Imaginary-time Formalism}
The central theme of thermal field theory is to use the techniques of quantum field theory to study statistical properties of systems at thermal equilibrium and their dynamical properties out of thermal equilibrium. The construction starts from the formal analogy between the partition function in statistical mechanics and the generating functional in field theory. In statistical mechanics, the partition function is defined as
\be
Z \equiv \Tr{e^{-\beta H}} \,,
\ee
where $\beta = 1/T$ is the inverse of the temperature and $H$ is the system Hamiltonian. In field theory, the generating functional is defined as a path integral over all possible field configurations
\be
Z[j] \equiv \int \ml{D} \phi  \exp \Big\{ i \int \diff^4x \big( \ml{L}(x) + j(x)\phi(x) \big) \Big\}\,, 
\ee
where $\ml{L}(x)$ is the Lagrangian density of the system of $\phi$ fields and $j(x)$ is an external current that couples with the $\phi $ field. It can be shown that by analytically continuing the real time $t$ to an imaginary time $\tau = it$, the partition function in statistical mechanics can be written as
\be
Z[j] = \int_{b.c.} \ml{D} \phi \exp \Big\{ - \int_0^\beta \diff \tau \int \diff^3x  \big(  \ml{L}_E(\tau, {\bs x}) -j(\tau, {\bs x})\phi(\tau, {\bs x})   \big)    \Big\} \,,
\ee
where the boundary condition is periodic $\phi(0,{\bs x}) = \phi(\tau, {\bs x})$ if $\phi$ is bosonic and anti-periodic $\phi(0,{\bs x}) = - \phi(\tau, {\bs x})$ if $\phi$ is fermionic. $\ml{L}_E$ is the Lagrangian density in the Euclidean space defined by
\be
\ml{L}_E (\tau, {\bs x}) = - \ml{L}(t = -i\tau , {\bs x}) \,.
\ee
For any operator $O$, its expectation value at thermal equilibrium is defined as
\be
\label{chap1_eqn_expectation}
\langle O \rangle_T \equiv \frac{1}{Z} \Tr \{ e^{-\beta H} O  \} \,.
\ee
The operator expectation value can also be calculated by using the field theory technique,
\be
\langle \ml{T}_{\tau} \phi(x_1) \phi(x_2) \cdots \phi(x_n) \rangle = \frac{1}{Z} \frac{\delta^n Z}{\delta j(x_1) \delta j(x_2) \cdots \delta j(x_n) } \Big|_{j=0} \,.
\ee
The connected Green's functions are defined as
\be
G^{(n)}(x_1,x_2,\cdots,x_n) \equiv \frac{\delta^n \ln{Z} }{\delta j(x_1) \delta j(x_2) \cdots \delta j(x_n) } \Big|_{j=0} \,.
\ee
This formalism of thermal field theory is called the imaginary-time formalism. Feynman rules for perturbative calculations can be derived similarly as in the zero-temperature field theory. The only difference is the discrete energy spectrum in the imaginary-time formalism, which is due to the (anti-)periodicity of the field. To show this, let us consider the propagator $G^{(2)}(x_1,x_2) = G^{(2)}(x_1-x_2) \equiv G(x)$. We have
\be
G(\tau - \beta, {\bs x} ) = \pm G(\tau , {\bs x} )\ \ \ \ \ \ \ \ma{for}\ 0\leq\tau\leq\beta \,,
\ee
where $+$ is for bosonic fields while $-$ is for fermionic fields. Then when we decompose the propagator in the energy-momentum space, we have a Fourier series with a discrete energy spectrum rather than a Fourier transform with a continuum,
\be
G(\tau) = T\sum_n e^{- i\omega_n \tau} G(i\omega_n) \,,
\ee
in which $\omega_n$ is the Matsubara frequency. For bosons $\omega_n=2 n \pi T$ while for fermions $\omega_n= (2n+1)\pi T$. In loop calculations, instead of integrating over the loop energy, we will do a summation over the Matsubara frequencies in the imaginary-time formalism. Methods to do the summation can be found in Ref.~\cite{Pisarski:1987wc}. 

\subsubsection{Real-time Formalism}
If we are only interested in the properties of the system at thermal equilibrium, the imaginary-time formalism is enough. But if we want to know the dynamical properties of the system out of thermal equilibrium, we need to extend the imaginary-time formalism to obtain an explicit time dependence in the construction. This can be done in the real-time formalism, in which one can define different 2-point Green's functions (propagators)
\be 
D_{>}(t,\bs x)&=& \big\langle \phi (t,\bs x)  \phi(0,0)   \big\rangle _T \\ 
D_{<}(t,\bs x) &=& \big\langle  \phi (0,0) \phi (t,\bs x) \big\rangle_T\\ 
D_{R} (t,\bs x)&=& \big\langle  \theta(t)[   \phi(t,\bs x), \phi(0,0)  ] \big\rangle_T\\ 
D_{A} (t,\bs x)&=& - \big\langle  \theta(-t) [\phi (t,\bs x), \phi(0,0)  ] \big\rangle_T\\
D_{\ml{T}} (t,\bs x) &=& \big\langle  \ml{T}( \phi(t,\bs x)\phi(0,0)  ) \big\rangle_T\,,
\ee
where we list the definitions for a real scalar field $\phi$. We follow the standard convention to use $D$ rather than $G$ to label the Green's functions. Generalizations to fermionic fields or gauge fields can be easily made. The expectation value $\langle \cdots \rangle_T$ is defined in Eq.~(\ref{chap1_eqn_expectation}). The subscripts $R$, $A$ and $\ml{T}$ stand for the retarded, advanced and time-ordered Greens's functions. Their Fourier transform is defined as
\be
D_{X} (q_0,\bs q) = \int \diff^4x e^{i(q_0t-\bs q\cdot \bs x)}D_{X} (t,\bs x)\,,
\ee
where $X$ could be $>$, $<$, $R$, $A$ or $\ml{T}$. The $>$ and $<$ propagators satisfy the Kubo-Martin-Schwinger relation, which can be derived from the boundary conditions
\be
D_{<} (q_0,\bs q) = \pm e^{q_0/T} D_{>} (q_0,\bs q) \,,
\ee
in which the $+$ sign is for bosons while the $-$ sign is for fermions. The spectral function is defined by
\be
\rho  (q_0,\bs q) \equiv D_{>} (q_0,\bs q) - D_{<} (q_0,\bs q) = D_{R}(q_0,\bs q) - D_{A} (q_0,\bs q) \,.
\ee
By using the definitions, we can write the time-ordered propagator as
\be
\label{chap1_eqn_timeordered}
D_{\ml{T}} (q_0,\bs q) = D_{R} (q_0,\bs q) + D_{<}(q_0,\bs q) = D_{A} (q_0,\bs q) + D_{>}(q_0,\bs q)\,.
\ee
In the energy-momentum space, the real-time propagators can be obtained from the analytic continuation of the imaginary-time propagators (This is standard and explained in textbooks such as Ref.~\cite{Bellac:2011kqa}.)
\be
D_R (q_0,\bs q) &=& -i D_E ( q_0 = q_0 + i\epsilon,\bs q) \\
D_A (q_0,\bs q) &=& -i D_E ( q_0 = q_0 - i\epsilon,\bs q) \,,
\ee
where the subscript $E$ is a shorthand of Euclidean and means the Green's function in the imaginary-time formalism. 
Using the definitions and relations shown above, we can write out explicitly the propagators of a free scalar field
\be 
\label{eqn:>}
D_{>}(q) &=& (1+n_B(q_0))  \rho_F(q)\\ 
\label{eqn:<}
D_{<}(q) &=& n_B(q_0)  \rho_F(q)\\ 
\rho_F(q) &=&  (2\pi)\sign(q_0) \delta(q_0^2-{\bs q}^2)\\ 
D_{R} (q) &=& \frac{i }{q_0^2-{\bs q}^2+i\sign(q_0)\epsilon} \\ 
D_{A} (q)&=& \frac{i }{q_0^2-{\bs q}^2-i\sign(q_0)\epsilon} \\ 
D_{\ml{T}} (q) &=& \frac{i }{q_0^2-{\bs q}^2+i\epsilon} + n_B(|q_0|)(2\pi)\delta(q_0^2-{\bs q}^2) \,,
\ee
where $n_B(x) \equiv \frac{1}{e^{x/T}-1}$ is the Bose-Einstein distribution.

For an interacting theory at finite temperature, one can use the imaginary-time formalism to calculate loop-corrected Euclidean propagators and then obtain the retarded and advanced propagators via analytic continuations. We can then calculate the spectral function and obtain all the other propagators by using the Kubo-Martin-Schwinger relation and Eq.~(\ref{chap1_eqn_timeordered}).

The real-time thermal field theory can also be formulated in terms of a path integral. To incorporate both the real time dynamics and the thermal properties, one has to double the degrees of freedom and use the Keldysh-Schwinger contour. Since we will not use the Keldysh-Schwinger path integral throughout this dissertation, we will not explain it here.

\subsubsection{Hard Thermal Loops}
Now we will use the imaginary-time formalism to calculate the photon polarization tensor $\Pi^{\mu\nu}$ at one loop. We will use this as an example to explain the hard thermal loops and their power counting. We will focus on the illumination of the power counting rather than a complete calculation here. In the imaginary-time formalism, we can write (we only consider the contribution from the electron and positron and neglect their masses)
\be
\Pi^{\mu\nu}(q_0, {\bs q})=e^2 T\sum_n \int\frac{\diff^3k}{(2\pi)^3} \frac{\Tr \big({\gamma^{\mu}  \slashed{K} \gamma^{\nu} (\slashed{K}-\slashed{Q}) } \big)}{ [\omega_n^2+ {\bs k} ^2 ] [(\omega_n+iq_0)^2 + ({\bs k} - {\bs q})^2]} \,,
\ee
where $K_\mu = ({\bs k}, \omega_n)$ and $Q_\mu = ({\bs q}, -iq_0)$ for $\mu=1,2,3,4$. For the gamma matrices, $\gamma^4 = i\gamma^0$ such that $\{\gamma^\mu, \gamma^\nu \} = -2\delta^{\mu\nu}$. Simplifying the trace we obtain
\be
\Tr \big({\gamma^{\mu}  \slashed{K} \gamma^{\nu} (\slashed{K}-\slashed{Q}) } \big) = 8K^\mu K^\nu - 4\delta^{\mu\nu} K^2 + 4\delta^{\mu\nu} K \cdot Q - 4K^\mu Q^\nu - 4 Q^\mu K^\nu \,.
\ee
We now take the term ${\bs k}^2 = k^2$ in the trace as an example and consider the integral
\be
I \equiv T\sum_n \int\frac{\diff^3k}{(2\pi)^3} \frac{k^2}{  [\omega_n^2+ {\bs k} ^2 ] [(\omega_n+iq_0)^2 + ({\bs k} - {\bs q})^2]  } \,.
\ee
The summation can be done \cite{Braaten:1989mz}
\be \nn
I &=&  \int\frac{\diff^3k}{(2\pi)^3} \frac{k^2}{2 k |{\bs k} - {\bs q} | } \\ \nn
&& \Big[  \big( 1 + n_E(k) + n_E(|{\bs k} - {\bs q} |) \big) 
\Big(  \frac{1}{iq_0 + k + |{\bs k} - {\bs q} |}  - \frac{1}{iq_0 - k - |{\bs k} - {\bs q} |} \Big)  \\
&& - \big( n_E(k) - n_E(|{\bs k} - {\bs q} |) \big) 
\Big(\frac{1}{iq_0 + k - |{\bs k} - {\bs q} |}  - \frac{1}{iq_0 - k + |{\bs k} - {\bs q} |} \Big)
\Big]\,.
\ee
Now let us consider soft external momenta $q \sim eT$ ($q$ here means the four-momentum rather than the magnitude of $|{\bs q}|$). The term ``$1$" in $1 + n_E(k) + n_E(|{\bs k} - {\bs q} |)$ corresponds to the vacuum contribution. It has a quadratic power divergence. After renormalization, we expect $I \sim q^2 \sim e^2T^2$. Then the polarization tensor $\Pi \sim e^4 T^2$, which is $e^2$ suppressed compared with $q^2$. So this loop correction is a perturbation and resummation in this case is not crucial. Then we consider the other terms with the Fermi-Dirac distributions. Due to the Fermi-Dirac distribution, we expect the typical value of $k$ to be $\lesssim T$. For soft loop momenta $k\sim eT$, we expect $I \sim n_E(eT) e^2T^2$ just by counting the power of $k$ of the integral. Since $eT \ll T$, we approximate the distribution by $n_E(eT) \sim 1$ (For the Bose-Einstein distribution of gluons in thermal QCD, we would have $n_B(gT) \sim \frac{T}{gT} \sim 1/g$.). So for soft loop momenta, $I\sim e^2T^2$, $\Pi \sim e^4 T^2$ and resummation is not crucial.

Finally we analyze the case of hard loop momenta $k\sim T$. We can neglect the $q_0\sim |{\bs q}| \sim eT$ in the denominator and approximate $n_E(T)$ by $1$. Then the integral scales as $I\sim T^2$ and $\Pi \sim e^2 T^2$. In the case of hard loop momenta, the polarization tensor scales on the same order as $q^2$. So resummation is crucial in the case of hard loop momenta. In general, if a loop correction has to be resummed when the external momenta are soft $\sim eT$ and the loop momenta are hard $\sim T$, this loop is identified to have a hard thermal loop contribution. The general power counting rule to identify hard thermal loops has been analyzed \cite{Braaten:1989mz}. The complete result of $\Pi^{00}$ in the hard thermal loop approximation (keeping only the contributions from hard thermal loops) at one loop is given by
\be
\label{chap1_eqn_Pi00}
\Pi^{00} (q_0, {\bs q}) = -\frac{4e^2}{\pi^2} \int_0^\infty k \diff k \, n_E(k) \Big [ 1 + \frac{q_0}{|{\bs q}|} \ln\Big( \frac{ q_0 - |{\bs q}| }{q_0 + |{\bs q}| } \Big) \Big] \,.
\ee
If we set $q_0=0$ and take the limit $|{\bs q}| \rightarrow 0$, we obtain the Debye mass
\be
\label{chap1_eqn_Debye}
m_D^2 = - \Pi^{00} (q_0=0, |{\bs q}| \rightarrow 0) = \frac{1}{3}e^2T^2\,.
\ee
which leads to the screening of the electric interaction. The screening is due to the massive force carrier and the interaction becomes short-range with a typical length scale $\sim 1/m_D$. 
In both thermal QED and QCD, the hard thermal loops of the photon and gluon self energies (polarization tensors) satisfy the Ward identity and thus are gauge invariant \cite{Kalashnikov:1979cy,Weldon:1982bn}. More specifically, the Debye mass is gauge invariant and thus is a physical quantity. As a result, the photon and gluon propagators with the hard thermal loop corrections are gauge covariant. The hard thermal loop approximation can be formulated as an effective field theory with an effective Lagrangian, which is manifestly gauge invariant \cite{Braaten:1991gm}.

We can analytically continue the Euclidean $\Pi^{00} $ to real time and obtain the retarded and advanced polarization tensors 
\be
\label{chap1_eqn_Landau_damp}
\Pi^{00}_{R(A)} (q_0, {\bs q}) = \Pi^{00} (q_0 \pm i\epsilon, {\bs q})\,,
\ee
where the $+(-)$ sign is for the retarded (advanced) one. Due to the logarithmic term in the expression (\ref{chap1_eqn_Pi00}), $\Pi^{00}_{R(A)}$ will have an imaginary part when the external momentum is spatial, $q_0<|{\bs q}|$. This is the origin of the Landau damping phenomenon in thermal QED and QCD plasmas. For the quarkonium dynamics inside the QGP, the appearance of the imaginary part corresponds to the quarkonium dissociation caused by inelastic scattering with medium constituents.

\vspace{0.2in}
Thermal field theory is an important theoretical tool in the study of quarkonium inside the QGP. It will be widely used in the following chapters.

\vspace{0.2in}
\subsection{Effective Field Theory}
As mentioned in Section 1.4.1, the hard thermal loop approximation can be formulated as an effective field theory (EFT). EFT is based on separation of scales, symmetries and systematic expansions. It is a theory of effective degrees of freedom. Different terms in the theory are organized by a certain power counting rule: lower-order terms in the power counting have more important contributions to a certain physical process. Thus different terms in the EFT are well-organized by their importance in the process. We have two ways to construct an EFT in general: ``bottom-up" and ``top-down" approaches. In the ``bottom-up" approach, one writes down the most general form of a Lagrangian consistent with all the symmetry properties of the system under study. The parameters of each term will be fitted from experimental data. Chiral perturbation theory is an example of ``bottom-up" construction. In the ``top-down" approach, one integrates out degrees of freedom with a highf energy scale and derives an effective Lagrangian for the low-energy modes. Examples of the ``top-down" construction are soft-collinear effective theory and nonrelativistic QCD which will be introduced below.

EFT is a powerful tool when the system exhibits a well-separated hierarchy of scales. We will use the formulation of EFT extensively throughout this dissertation.

\vspace{0.2in}
\subsection{Nonrelativistic QCD}
Another crucial theoretical tool is the effective field theory of heavy quarks. We will explain the construction of nonrelativistic QCD (NRQCD) and its power counting in this section. It was first constructed in Refs.~\cite{Lepage:1992tx,Bodwin:1994jh}. The construction relies on the following separation of scales
\be
M \gg Mv \gg Mv^2\,,
\ee
where $M$ is the heavy quark mass and $v$ is the magnitude of the relative velocity between the $Q\bar{Q}$ inside the bound state. $M$ is a scale for the quarkonium annihilation and the creation of $Q\bar{Q}$ pairs from gluons. Since $M$ is large for heavy quarks, the process of creating $Q\bar{Q}$ pairs from the initial hard partonic scattering in proton-proton or heavy ion collisions is thought of as a short-distance process. $1/(Mv)$ controls the typical size of the quarkonium. For charmonium, $v^2\sim0.3$ and for bottomonium, $v^2\sim 0.1$. $Mv^2$ is the radial or orbital-angular-momentum excitation energy of the same quarkonium family. $Mv^2\sim 500$ MeV for both charmonium and bottomonium. Thus $Mv^2\sim \Lambda_{QCD}$. 

The idea of NRQCD is to separate the scale $M$ from the scales $Mv$, $Mv^2$, $\Lambda_{QCD}$ by integrating out the degrees of freedom of momenta $\gtrsim M$. These degrees of freedom include relativistic heavy quarks, light quarks and gluons with momenta $\gtrsim M$. The removal of these degrees of freedom is compensated by a set of new local operators and their coefficients in the Lagrangian. As will be discussed below, for NRQCD they are the four-fermion operators. Since $M$ is assumed to be a perturbative scale, the coefficients of the new local operators can be calculated in perturbation theory. Non-perturbative physics in a certain physical process is written in terms of matrix elements of the new local operators. In this way, one factorizes the perturbative and non-perturbative physics. The factorization allows us to extract the information of non-perturbative physics from experimental measurements and perturbative calculations. Usually, the set of new local operators can be organized by their importance to a certain process, encoded in a power counting rule. In NRQCD, the parameter of the power counting is the velocity $v$. So a typical NRQCD calculation is a double expansion in $\alpha_s(M)$ and $v$.

Consider the Lagrangian density of the heavy quark sector of QCD
\be
\ml{L} = \bar{\Psi} ( i \slashed{D} - M ) \Psi \,,
\ee
where $\Psi = \Psi(x)$ is the Dirac spinor, which contains both the quark and antiquark. The first step in the construction is to do a Foldy-Wouthuysen-Tani transformation
\be
\Psi(x) = e^{-iMt} \psi(x) \,,
\ee
then we have
\be
\ml{L} = \bar{\psi} (   i \slashed{D} - M + \gamma_0 M ) \psi \,.
\ee
We will use the $\gamma$ matrices in the Dirac representation
\be
\gamma^0 = \begin{pmatrix} 
\mathbb{I}_{2\times2} & 0 \\
0  & - \mathbb{I}_{2\times2} 
\end{pmatrix}\,, \ \ \ \ \ \ \ \ \ 
\gamma^i = \begin{pmatrix} 
0 & \sigma^i \\
-\sigma^i  & 0
\end{pmatrix}\,,
\ee
where $\sigma^i$ is the Pauli matrices. Under the Dirac representation, we can decompose the Dirac spinor into a large component $\psi_L $ and a small component $\psi_S$
\be
\psi(x) = \begin{pmatrix} 
\psi_L \\
\psi_S
\end{pmatrix}\,.
\ee
Then the Lagrangian can be written as
\be
\ml{L}_\psi = \psi_L^\dagger  iD_0  \psi_L +  \psi_L^\dagger  i {\bs \sigma} \cdot {\bs D}   \psi_S + \psi_S^\dagger  i {\bs \sigma} \cdot {\bs D}   \psi_L + \psi_S^\dagger  (iD_0+2M)  \psi_S \,.
\ee
The equation of motion of $\psi_S$ is obtained by the Euler-Lagrangian equation and differentiating with respect to $\psi_S^\dagger $
\be
\psi_S = -(iD_0+2M)^{-1}   i {\bs \sigma} \cdot {\bs D} \psi_L \,. 
\ee
By replacing $\psi_S $ with the above expression in the Lagrangian gives
\be
\ml{L}_\psi = \psi_L^\dagger  iD_0  \psi_L -  \psi_L^\dagger  i {\bs \sigma} \cdot {\bs D} (iD_0+2M)^{-1}   i {\bs \sigma} \cdot {\bs D} \psi_L \,.
\ee
Then we can expand the Lagrangian in terms of $1/M$
\be
(iD_0+2M)^{-1} = \frac{1}{2M} \Big( 1 - \frac{iD_0}{2M} + \frac{(iD_0)^2}{4M^2} + \cdots \Big)\,.
\ee
We can then organize the Lagrangian by powers of $1/M$ and write 
\be
\label{chap1_eqn_L_psi}
\ml{L}_\psi &=& \ml{L}^{(0)}_\psi + \ml{L}^{(1)}_\psi + \ml{L}^{(2)}_\psi + \cdots \\
\ml{L}^{(0)}_\psi &=& \psi_L^\dagger iD_0  \psi_L \\
\ml{L}^{(1)}_\psi &=& \psi_L^\dagger \Big(  \frac{{\bs D}^2}{2M} + \frac{g}{2M} {\bs \sigma} \cdot {\bs B}   \Big)  \psi_L  \\
\ml{L}^{(2)}_\psi &=& \frac{g}{8M^2} \psi_L^\dagger \Big( [D_i, E_i] + i \epsilon_{ijk} \sigma_i (D_j E_k - E_j D_k) \Big)  \psi_L \,.
\ee
For the antiquark part, we let
\be
\Psi(x) &=& e^{iMt} \chi(x) \,
\ee
and decompose $\chi(x)$ into a small component and a large component
\be
\chi(x) = \begin{pmatrix} 
\chi_S \\
\chi_L
\end{pmatrix}\,.
\ee
Repeating the similar procedure we obtain
\be
\label{chap1_eqn_L_chi}
\ml{L}_\chi &=& \ml{L}^{(0)}_\chi + \ml{L}^{(1)}_\chi + \ml{L}^{(2)}_\chi + \cdots \\
\ml{L}^{(0)}_\chi &=& - \chi_L^\dagger iD_0  \chi_L \\
\ml{L}^{(1)}_\chi &=& \chi_L^\dagger \Big(  \frac{{\bs D}^2}{2M} + \frac{g}{2M} {\bs \sigma} \cdot {\bs B}   \Big)  \chi_L  \\
\ml{L}^{(2)}_\chi &=& -\frac{g}{8M^2} \chi_L^\dagger \Big( [D_i, E_i] + i \epsilon_{ijk} \sigma_i (D_j E_k - E_j D_k) \Big)  \chi_L  \,.
\ee
From now on we will omit the subscript ``$L$" and only write $\psi$ and $\chi$. They are the annihilation operator of heavy quarks and the creation operator of heavy antiquarks respectively. Now we will explain the power counting rule. 

Since the typical momentum of the heavy quark and heavy antiquark is $\sim Mv$, from dimensional analysis we get $\psi,\chi\sim (Mv)^{3/2}$. 
We also expect ${\bs D} \sim Mv$ and then from the equation of motion of the heavy quark we get $D_0\sim Mv^2$. From the equations of motion of the gauge fields, we find $gA_0\sim Mv^2$ and $g{\bs A} \sim Mv^3$. The power counting rule will help us to organize operators according to their importance. 

Let us first apply the power counting to understand the structure of quarkonium. The Fock space decomposition of a quarkonium state $H$ can generally be written as
\be
| H \rangle = | Q\bar{Q} \rangle + | Q\bar{Q}g \rangle + \cdots \,,
\ee
where $g$ indicates a dynamical gluon and the dots indicate higher Fock states with more gluons and light quarks. The lowest Fock state $| Q\bar{Q} \rangle$ is just a heavy quark antiquark pair interacting with each other. In Coulomb gauge, $A_0$ gauge field is not dynamical. So the lowest order operator that connects $ | Q\bar{Q} \rangle$ and $ | Q\bar{Q} g \rangle$ is 
\be
\frac{{\bs D}^2}{2M} = \frac{\nabla^2}{2M} - \frac{ ig{\bs A}\cdot \nabla }{M} - \frac{g^2{\bs A}^2}{2M} \,,
\ee
where we have used the Coulomb gauge condition $\nabla \cdot {\bs A} = 0$. The energy correction to the quarkonium due to the existence of the dynamical gluon is given by
\be
\Delta E = -\frac{ig}{M} \langle H |   \int \diff^3x ( \psi^\dagger  {\bs A}\cdot \nabla  \psi + \chi^\dagger  {\bs A}\cdot \nabla \chi )  | H \rangle \,.
\ee
Using our power counting and $\langle H | H \rangle \sim v^0$, we find $\Delta E \sim Mv^4$. We can write $\Delta E$ in another way: $P_{| Q\bar{Q}g \rangle} \times E_{| Q\bar{Q}g \rangle}$ where $P_{| Q\bar{Q}g \rangle}$ is the probability of the quarkonium in this Fock state and $E_{| Q\bar{Q}g \rangle}$ is the energy of this state. If the dynamical gluon has an energy $\sim Mv$, then $E_{| Q\bar{Q}g \rangle} \sim Mv^2 + Mv \sim Mv$ and $P_{| Q\bar{Q}g \rangle} \sim v^3$; On the other hand, if the dynamical gluon has an energy $\sim Mv^2$, then $E_{| Q\bar{Q}g \rangle} \sim Mv^2$ and $P_{| Q\bar{Q}g \rangle} \sim v^2$. In either case, the Fock state with one dynamical gluon is suppressed at least by $v^2$. So up to higher order corrections in $v^2$, quarkonium can be thought of as a bound state of $Q\bar{Q}$. This power counting analysis also explains the phenomenological success of potential models to describe the quarkonium spectra. 

In addition to the terms in (\ref{chap1_eqn_L_psi}) and (\ref{chap1_eqn_L_chi}), we also need to add four-fermion operators to describe the annihilation and creation processes for quarkonium, to compensate the degrees of freedom integrated out. For the annihilation processes, the dimension-$6$ operators are
\be \nn
(\delta\ml{L})_{d=6} &=& \frac{f_1(^1S_0)}{M^2} \ml{O}_1(^1S_0) + \frac{f_1(^3S_1)}{M^2} \ml{O}_1(^3S_1)  \\
&&+ \frac{f_8(^1S_0)}{M^2} \ml{O}_8(^1S_0)  + \frac{f_8(^3S_1)}{M^2} \ml{O}_8(^3S_1) \\
\ml{O}_1(^1S_0) &=& \psi^\dagger \chi \chi^\dagger \psi \\
\ml{O}_1(^3S_1) &=& \psi^\dagger \sigma_i \chi \chi^\dagger  \sigma_i  \psi  \\
\ml{O}_8(^1S_0) &=& \psi^\dagger T^a  \chi \chi^\dagger  T^a  \psi \\
\ml{O}_8(^3S_1) &=& \psi^\dagger \sigma_iT^a \chi \chi^\dagger \sigma_iT^a \psi \,.
\ee 
The operators are specified by $\ml{O}_i(^{2S+1}L_J)$ where $i=1$ for a color singlet and $i=8$ for a color octet and $S$, $L$ and $J$ are the spin, orbital and total angular momentum quantum numbers. P-wave operators are of the form $\psi^\dagger \overset\leftrightarrow{D}_i \chi \chi^\dagger \overset\leftrightarrow{D}_i \psi$ (one can also insert spin $\sigma_i$ and color $T^a$ operators between the quark antiquark fields), which are dimension-8 operators. When calculating the annihilation rates, matrix elements such as $\langle H | \ml{O}_i(^{2S+1}L_J) | H \rangle $ will appear. One can use our power counting to organize different matrix elements based on their $v^2$-scaling. For quarkonium creation processes, matrix elements of the form $\langle 0 | \psi^\dagger \Sigma \chi | H \rangle \langle H | \chi^\dagger \Sigma \psi  | 0 \rangle$ will show up and $|0\rangle$ is the vacuum state. For dimension-6 operators, $\Sigma$ can be $\mathbb{I}$, $\sigma_i$, $T^a$ or $\sigma_iT^a$. We will omit further details on how to use the four-fermion operators in the calculations of quarkonium annihilation and creation. The NRQCD factorization has been applied widely in the phenomenological studies of quarkonium production in proton-proton collisions \cite{Butenschoen:2011yh,Chao:2012iv,Gong:2012ug,Butenschoen:2012qr,Bodwin:2014gia,Butenschoen:2014dra,Han:2014jya,Bodwin:2015iua,Bain:2016clc,Bain:2017wvk}.

\vspace{0.2in}
\subsection{Potential Nonrelativistic QCD}

The effective field theory potential nonrelativistic QCD (pNRQCD) in vacuum can be systematically constructed from NRQCD by further integrating out the scale $Mv$ \cite{Brambilla:1999xf,Brambilla:2004jw,Fleming:2005pd}. In a hot medium, the construction may be complicated due to the existence of extra scales such as the temperature $T$ and the Debye mass $m_D$. Depending on where $T$ and $m_D$ fit into the scale hierarchy $M \gg Mv \gg Mv^2$, one can obtain different versions of pNRQCD. Since in current heavy ion experiments, the highest temperature achieved is $\sim 500 $ MeV, which is on the same order as $Mv^2$ for both charmonium and bottomonium. We will assume $M \gg Mv \gg Mv^2 \gtrsim T \gtrsim m_D$ and integrate out the scale $Mv$ without worrying about the thermal scales. Thermal medium effects will start to modify the theory if we consider physical processes at the scale $Mv^2$. It is worthing noticing that the typical size of quarkonium is given by $r \sim 1/Mv$. Under our assumed separation of scales, the Debye screening effect is not too strong: $r m_D \lesssim v$. If $r m_D\sim1$, Debye screening of the attractive potential would be too strong to support bound quarkonium states. We will assume $Mv \gg \Lambda_{QCD}$ and perform a perturbative construction of pNRQCD here.

The construction of pNRQCD is as follows: we start with the annihilation operators of a heavy quark antiquark pair $ \psi_i({\bs x}_1, t)  \chi^\dagger_j({\bs x}_2, t)  $ where $i$ and $j$ are color indexes. We want to map it onto composite fields: a color singlet $S({\bs R}, {\bs r}, t)$ and a color octet $O^a({\bs R}, {\bs r}, t)$. The center-of-mass (c.m.) and relative positions of the $Q\bar{Q}$ pair are ${\bs R} = \frac{{\bs x}_1 + {\bs x}_2}{2}$ and ${\bs r} = {\bs x}_1 - {\bs x}_2$ respectively. We will assume the medium is translationally invariant so the existence of the medium does not break the separation into the c.m.~and relative motions. Under a gauge transformation $U$, we expect the color singlet $S$ to be invariant and the color octet $O^a$ to transform like a gluon at $({\bs R}, t)$. Since the operator $\Phi_{ij}({\bs x}_1, {\bs x}_2, t) \equiv \psi_i({\bs x}_1, t)  \chi^\dagger_j({\bs x}_2, t) $ is not gauge invariant and transforms as $U_{ii'} ({\bs x}_1, t) \psi_{i'}({\bs x}_1, t)  \chi^\dagger_{j'}({\bs x}_2, t)  U^\dagger_{j'j} ({\bs x}_2, t)  $, we need to add Wilson lines to make sure the two sides of the mapping have the same gauge transformation laws. We make the following construction
\be
\psi_i({\bs x}_1, t)  \chi^\dagger_j({\bs x}_2, t) &=& W_{ii'}({\bs x}_1, {\bs R}, t) (  \ma{S}_{i'j'} +   \ma{O}_{i'j'}  ) W_{j'j}({\bs R}, {\bs x}_2, t) \\
\ma{S}_{ij} &\equiv& \frac{\delta_{ij}}{\sqrt{N_c}} S({\bs R}, {\bs r}, t)\\
\ma{O}_{ij} &\equiv& \frac{1}{\sqrt{T_F}} T^a_{ij}O^a({\bs R}, {\bs r}, t) \,,
\ee
where $N_c=3$ and $T_F = \frac{1}{2}$ is defined by $\Tr (T^aT^b) = T_F\delta^{ab}$. The pre-factors in front of $S$ and $O^a$ are our normalization conditions. We define the quadratic Casimir of the fundamental representation $C_F \equiv \frac{T_F}{N_c}(N_c^2-1)$ for later use.
The Wilson line is defined as
\be
W_{ij}({\bs x}, {\bs y}, t) \equiv \Big[  \exp\Big\{ig\int_{{\bs y}}^{{\bs x}} \diff{\bs r} \cdot A({\bs r}, t) \Big\} \Big]_{ij} \,,
\ee 
where the gauge field is $A({\bs r}, t) = T^aA^a({\bs r}, t)$.

Then the Lagrangian density of the composite fields can be derived from the Lagrangian density of NRQCD shown in expressions (\ref{chap1_eqn_L_psi}) and (\ref{chap1_eqn_L_chi})
\be
\label{chap1_eqn_L_com}
\ml{L} = \Tr\Big\{\Phi^\dagger({\bs x}_1, {\bs x}_2, t) \Big(  iD_0 + \frac{{\bs D}^2_{{\bs x}_1}}{2M} + \frac{{\bs D}^2_{{\bs x}_2}}{2M} +\cdots \Big) \Phi({\bs x}_1, {\bs x}_2, t) \Big\} \,,
\ee
where $iD_0\Phi({\bs x}_1, {\bs x}_2, t) = i\partial_0 \Phi({\bs x}_1, {\bs x}_2, t) + gA_0({\bs x}_1,t)\Phi({\bs x}_1, {\bs x}_2, t) - \Phi({\bs x}_1, {\bs x}_2, t)gA_0({\bs x}_2,t)$.
Finally we expand the Lagrangian in (\ref{chap1_eqn_L_com}) in powers of $g$ (weak coupling expansion) and ${\bs r}$ (multipole expansion) and obtain the Lagrangian density of pNRQCD
\be\nn
\ml{L}_\ma{pNRQCD} &=& \int \diff^3r \Tr\Big(  \ma{S}^{\dagger}(i\partial_0-H_s)\ma{S} +\ma{O}^{\dagger}( iD_0-H_o )\ma{O} + V_A( \ma{O}^{\dagger}\bs r \cdot g{\bs E} \ma{S} + \ma{h.c.})  \\
\label{eq:lagr}
&&+ \frac{V_B}{2}\ma{O}^{\dagger}\{ \bs r\cdot g\bs E, \ma{O}  \} +\cdots \Big) + \ml{L}_\ma{light\ quark} + \ml{L}_\ma{gluon} \, ,
\ee
where ${\bs E}$ represents the chromoelectric field and $D_0\ma{O} = \partial_0\ma{O} -ig [A_0, \ma{O}]$. The gluon and light quark parts are just QCD with momenta $\lesssim Mv$. The degrees of freedom are the color singlet $\ma{S}(\bs R, \bs r, t)$ and color octet $\ma{O}(\bs R, \bs r, t)$. The color singlet and octet Hamiltonians are expanded in powers of $1/M$ or equivalently, $v$:
\be
H_{s} &=& \frac{(i\bs \nabla_\ma{cm})^2}{4M} + \frac{(i\bs \nabla_\ma{rel})^2}{M} + V_{s}^{(0)} + \frac{V_{s}^{(1)}}{M} + \frac{V_{s}^{(2)}}{M^2} + \cdots\\
H_{o} &=& \frac{(i\bs D_\ma{cm})^2}{4M} + \frac{(i\bs \nabla_\ma{rel})^2}{M} + V_{o}^{(0)} + \frac{V_{o}^{(1)}}{M} + \frac{V_{o}^{(2)}}{M^2} + \cdots\,.
\ee
We will work to the lowest order in the expansion of $v$. By the virial theorem, $\bs p_\ma{rel}^2/M \sim V_{s,o}^{(0)}\sim Mv^2$. Higher-order terms of the potentials including the relativistic corrections and spin-orbital and spin-spin interactions are suppressed by extra powers of $v$. In the $Q\bar{Q}$ pair (bound or unbound) rest frame, the initial c.m.~momentum is zero. If the medium is static with respect to the $Q\bar{Q}$ pair, the final c.m.~momentum after a scattering is of order $\sim T$. Since in our power counting, $T\lesssim Mv^2$, the c.m.~kinetic energy is of order $\lesssim Mv^4$ and thus suppressed by $v^2$.\footnote{If the medium is moving with respect to the $Q\bar{Q}$ pair at a velocity $v_\ma{med}$, the c.m.~kinetic energy is still suppressed at least by one power of $v$ if $v_\ma{med}\lesssim\sqrt{1-v}$. We assume the medium is static with respect to the $Q\bar{Q}$ pair here. Generalization to the case of moving medium with $v_\ma{med}\lesssim\sqrt{1-v}$ will be worked out in Chapter 4.} Therefore, at the lowest order in the nonrelativistic expansion
\be
\label{eqn:hamiltonian}
H_{s,o}  =  \frac{(i\bs \nabla_\ma{rel})^2}{M} + V_{s,o}^{(0)}\,.
\ee
The potentials and Wilson coefficients $V_{A,B}$ in the chromoelectric dipole vertices can be obtained in the construction. Up to order $g^2r$, 
\be
\label{eqn:match}
V_{s}^{(0)} = -C_F\frac{\alpha_s}{r}\,,\ \ \ \ \ \ \ V_{o}^{(0)} = \frac{1}{2N_c}\frac{\alpha_s}{r}\,,\ \ \ \ \ \ \ V_A=V_B=1\,.
\ee
The chromomagnetic vertices are suppressed by powers of $v$. The potential is Coulomb, which is approximately valid inside QGP since the confining part is flattened. One can improve the potentials by using a non-perturbative construction of pNRQCD. 

Under a gauge transformation $U(\bs R, t)$,
\be
\ma{S}(\bs R, \bs r, t) &\rightarrow& \ma{S}(\bs R, \bs r, t)\\
\ma{O}(\bs R, \bs r, t) &\rightarrow& U(\bs R, t)\ma{O}(\bs R, \bs r, t)U^\dagger(\bs R, t)\\
D_\ma{cm} ^\mu &\rightarrow&    U(\bs R, t)D_\ma{cm} ^\mu U^\dagger(\bs R, t)\,,
\ee
therefore the Lagrangian density is invariant under a gauge transformation associated with the c.m.~motion. It is worth noting that the relative motion is not gauged due to the multipole expansion.

\begin{figure}
	\centering
	\begin{tabular}{  c  c  l  }
	\hspace*{-0.26in}{ \raisebox{-0.26in}{\includegraphics[height=0.35in]{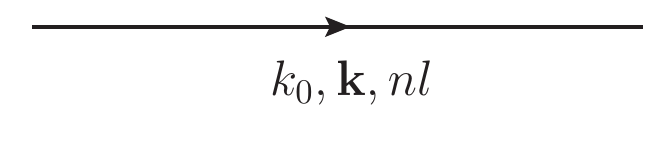}} } &=\ \ \ \ & $\frac{i| \psi_{nl} \rangle \langle \psi_{nl} | }{k_0-\frac{{\bs k}^2}{4M} + |E_{nl}| +i\epsilon } $\\
	\raisebox{-0.26in}{\includegraphics[height=0.5in]{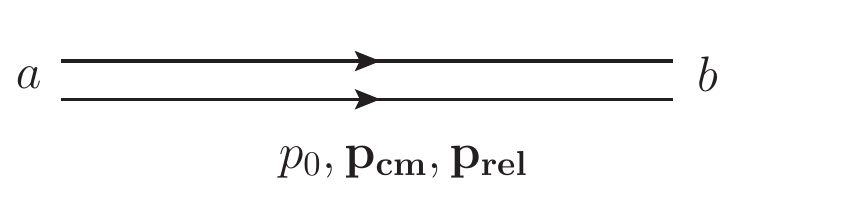}}   &=\ \ \ \ & $\frac{i| \Psi_{{\bs p}_{\ma{rel}}} \rangle \langle \Psi_{{\bs p}_{\ma{rel}}} |}{p_0-\frac{{\bs p^2_{\ma{cm}}}}{4M} - \frac{{\bs p^2_{\ma{rel}}}}{M} +i\epsilon } \delta^{ab}$ \\
	\raisebox{-0.26in}{ \includegraphics[height=1.3in]{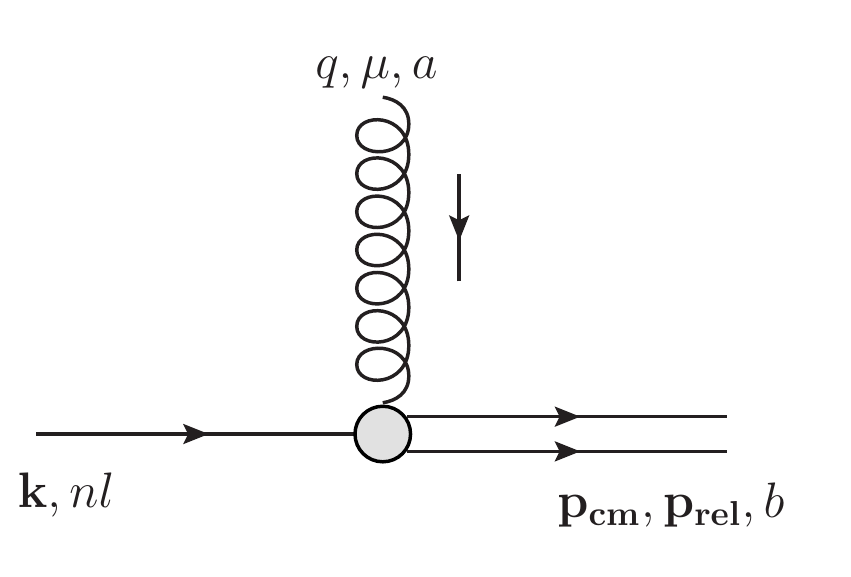}}    &=\ \ \ \ &  $g\sqrt{\frac{T_F}{N_c}}\delta^{ab}(q^0g^{\mu i}  - q^i g^{\mu0}) \langle \Psi_{{\bs p}_\ma{rel}} | r^i | \psi_{nl} \rangle$\\
	\raisebox{-0.26in}{  \includegraphics[height=1.3in]{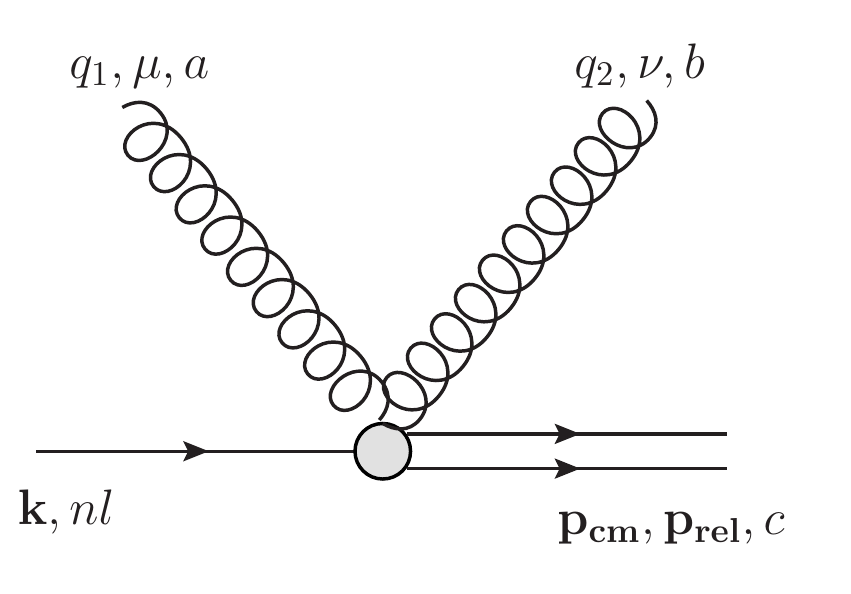}}   &=\ \ \ \ & $ig^2\sqrt{\frac{T_F}{N_c}}f^{abc}(g^{\mu 0}g^{\nu i}-g^{\mu i}g^{\nu 0}) \langle  \Psi_{{\bs p}_\ma{rel}} | r^i | \psi_{nl} \rangle $ \\
	 \raisebox{-0.26in}{ \includegraphics[height=1.3in]{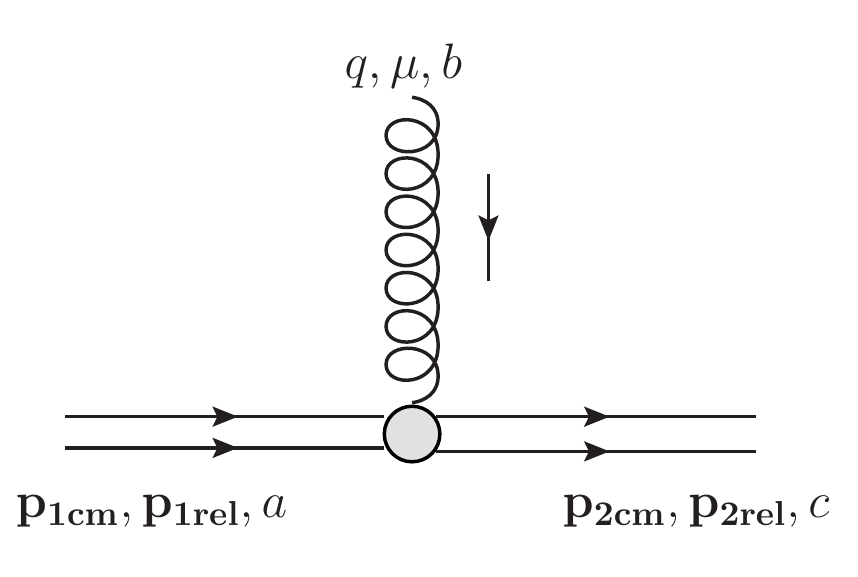} } &=\ \ \ \ &  $gd^{abc}(q^0g^{\mu i} - q^i g^{\mu0}) \langle  \Psi_{{\bs p}_\ma{2rel}}  | r^i | \Psi_{{\bs p}_\ma{1rel}} \rangle$ \\
	 \raisebox{-0.26in}{\includegraphics[height=1.3in]{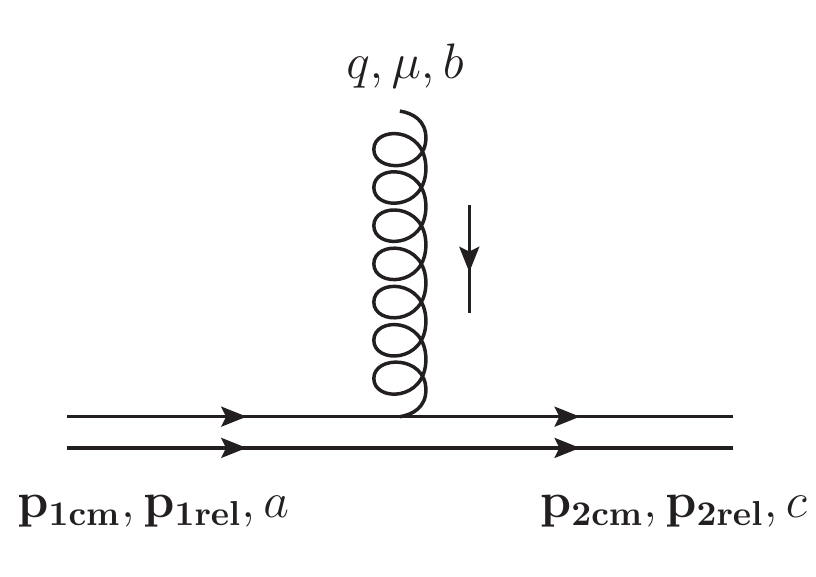} }   &=\ \ \ \ & $gf^{abc} g^{\mu0} (2\pi)^3\delta^3({\bs p}_\ma{rel1} - {\bs p}_\ma{rel2}) $
	\end{tabular}
\caption[Feynman rules in pNRQCD.]{Feynman rules in pNRQCD. Single solid line represents the bound color singlet while double solid lines represent the unbound color octet. The grey blob indicates the dipole interaction. The vertex with no grey blob means the gauge coupling in the c.m.~motion. Unbound singlet propagator will not be used throughout the dissertation and not shown here. The octet wave function $| \Psi_{{\bs p}_{\ma{rel}}} \rangle $ is a Coulomb scattering wave and thus the effect of the octet potential has been resummed in the octet propagator.}
\label{fig:rules}
\end{figure}

To make the wave function associated with the relative motion explicit, we do a change of basis in the relative motion by defining
\be
\ma{S}(\bs R, \bs r, t) &=& \frac{1}{\sqrt{N_c}}S(\bs R, \bs r, t) \equiv \frac{1}{\sqrt{N_c}}\langle \bs r | S(\bs R, t) \rangle \\
\ma{O}(\bs R, \bs r, t) &=& \frac{1}{\sqrt{T_F}}O^a(\bs R, \bs r, t)T^a  \equiv  \frac{1}{\sqrt{T_F}} \langle  \bs r | O^a(\bs R, t)\rangle T^a \,.
\ee
Then the Lagrangian density of the singlet and octet part can be written as \cite{Fleming:2005pd}
\be
\label{eq:lagr2}
\ml{L}_\ma{pNRQCD}({\bs R}, t) &=&  \ml{L}_{\ma{kin},s} + \ml{L}_{\ma{kin},o} + \ml{L}_{\ma{int},so} + \ml{L}_{\ma{int},oo} +\cdots \\
\ml{L}_{\ma{kin},s}  &=&  \langle S(\bs R, t) | (i\partial_0-H_s) | S(\bs R, t)\rangle \\
\ml{L}_{\ma{kin},o}  &=&  \langle O^a(\bs R, t) | ( i\partial_0-H_o )  |O^a(\bs R, t)\rangle \\
\ml{L}_{\ma{int},so} &=&  \sqrt{\frac{T_F}{N_C}}\Big( \langle O^a(\bs R, t) | \bs r \cdot g{\bs E}^a(\bs R, t) | S(\bs R, t)\rangle + \ma{h.c.} \Big) \\
\ml{L}_{\ma{int},oo} &=& if^{abc} \langle O^a(\bs R, t) | g A_0^b(\bs R, t)   |O^c(\bs R, t)\rangle \\\nn
&&  + d^{abc}  \langle O^{a}(\bs R, t)| g \bs r \cdot \bs E^b(\bs R, t) | O^{c}(\bs R, t)\rangle +\cdots  \, ,
\ee
The bra-ket notation saves us from writing the integral over the relative position explicitly. The singlet and octet composite fields are quantized by
\be \nn
|S(\bs R, t) \rangle &=& \int\frac{\diff^3 p_\ma{cm}}{(2\pi)^3}  e^{-i(Et-\bs p_\ma{cm} \cdot \bs R)} \bigg( \sum_{nl}  | \psi_{nl} \rangle a_{nl}(\bs p_\ma{cm})   + \int\frac{\diff^3 p_{\ma{rel}}}{(2\pi)^3}   | \psi_{{\bs p}_{\ma{rel}}} \rangle b_{{\bs p}_{\ma{rel}}}(\bs p_\ma{cm})  \bigg) \\
\\
|O^a(\bs R, t) \rangle &=&  \int\frac{\diff^3 p_\ma{cm}}{(2\pi)^3} e^{-i(Et-\bs p_\ma{cm}\cdot \bs R)}  \int\frac{\diff^3 p_{\ma{rel}}}{(2\pi)^3} 
| \Psi_{{\bs p}_{\ma{rel}}} \rangle c^a_{{\bs p}_{\ma{rel}}}(\bs p_\ma{cm})  \,,
\ee
where $E$ is the eigenenergy of the state under the Hamiltonians, Eq.~(\ref{eqn:hamiltonian}). The whole Hilbert space factorizes into two parts: one for the c.m.~motion and the other for the relative motion. 
The wave functions of the relative motion can be obtained by solving Schr\"odinger equations, which are part of the equations of motion of the free composite fields. 
They can be hydrogen-like wave functions $| \psi_{nl} \rangle$ for bound singlets with the eigenenergy $-|E_{nl}|$, or Coulomb scattering waves $| \psi_{{\bs p}_{\ma{rel}}} \rangle$ and $ | \Psi_{{\bs p}_{\ma{rel}}} \rangle $ for unbound singlets and octets with the eigenenergy ${\bs p}_\ma{rel}^2/M$. No bound state exists in the octet channel due to the repulsive potential. 
We will average over the polarizations of non-$S$ wave quarkonium states when computing scattering amplitudes squared (this will be explained in Chapter 3). So we omit the quantum number $m$ of the bound singlet state.
The operators $a^{(\dagger)}_{nl}(\bs p_\ma{cm})$, $b^{(\dagger)}_{{\bs p}_{\ma{rel}}}(\bs p_\ma{cm})$ and $c^{a(\dagger)}_{{\bs p}_{\ma{rel}}}(\bs p_\ma{cm})$ act on the Fock space to annihilate (create) a composite particle with a c.m.~momentum ${\bs p_\ma{cm}}$ and corresponding quantum numbers in the relative motion. 
These annihilation and creation operators satisfy the following commutation rules:
\be
[a_{n_1l_1}({\bs p}_{\ma{cm}1}),\ a^{\dagger}_{n_2l_2}({\bs p}_{\ma{cm}2})] &=& (2\pi)^3 \delta^3({\bs p}_{\ma{cm}1} - {\bs p}_{\ma{cm}2}) \delta_{n_1n_2}\delta_{l_1l_2} \\
{[b_{{\bs p}_{\ma{rel}1}}({\bs p}_{\ma{cm}1}),\ b^{\dagger}_{{\bs p}_{\ma{rel}2}}({\bs p}_{\ma{cm}2})]} & =& (2\pi)^6 \delta^3({\bs p}_{\ma{cm}1} - {\bs p}_{\ma{cm}2})\delta^3({\bs p}_{\ma{rel}1}-{\bs p}_{\ma{rel}2}) \\
{[c^{a_1}_{{\bs p}_{\ma{rel}1}}({\bs p}_{\ma{cm}1}),\ c^{a_2\dagger}_{{\bs p}_{\ma{rel}2}}({\bs p}_{\ma{cm}2})]} &=& (2\pi)^6 \delta^3({\bs p}_{\ma{cm}1} - {\bs p}_{\ma{cm}2}) \delta^3({\bs p}_{\ma{rel}1}-{\bs p}_{\ma{rel}2}) \delta^{a_1a_2}\,.
\ee
All other commutators are zero. The Feynman rules can be derived and are summarized in Fig.~\ref{fig:rules}. We use the notation $a^ib^i \equiv \sum_i a^ib^i = a_ib_i$ to denote Euclidean summation and ${\bs r}^{\mu} = 0$ when $\mu=0$. These Feynman rules are the starting point of the calculations shown in the following chapters.

\singlespacing
\chapter{Low Energy Scattering of $\alpha$-Particles in an $e^-e^+\gamma$ Plasma}
\doublespacing
\vspace{0.2in}

This chapter seems like a digression. But it is not. In fact, the physical process discussed in this chapter has many similarities with the QGP screeining effect on quarkonium. Yet, the physical process studied here is much simpler both conceptually and technically. Therefore it is a good starting point to understand the physics, develop intuition, and get familiar with some of the theoretical tools that will be used to study quarkonium in-medium transport in later chapters. The work presented in this chapter was done in collaboration with Thomas Mehen and Berndt M\"uller and published in Refs~\cite{Yao:2016gny,Yao:2016cjs}.

\vspace{0.2in}

\section{Physical Motivation: Plasma Screening Effect}
This research was inspired by the following intuitive idea: Suppose a nuclear scattering process is described by short-range nuclear attraction and long-range Coulomb repulsion. The interplay between the two competing forces results in a resonant state and determines its energy and width. If the system is embedded inside an $e^-e^+\gamma$ plasma and the Coulomb repulsion is screened due to the plasma, the resonance energy will be lowered. This is because first, part of the Hamiltonian is lowered and second, the wave function is more centered around the origin where the potential is negative. If the screening is strong enough, the resonance will become a bound state. This screening effect will be most prominent when the resonance energy and the plasma temperature are on the same order. The production of $^8$Be from S-wave resonant $\alpha$ particle scattering in the primordial Big Bang nucleosynthesis serves as a good physical example. The $\alpha$ particle, $^4\ma{He}$, is a tightly bound nucleus with charge $+2e$ and isospin $I=0$.
The $^8$Be resonance energy between two $\alpha$ particles in vacuum is about $91.8$ keV. The temperature of the $e^-e^+\gamma$ plasma during the  Big Bang nucleosynthesis is roughly below $1$ MeV.

The traditional approach of this problem is to use potential models. The short-range nuclear interaction can be modeled by a parametrization such as the Woods-Saxon potential and the long-range potential is just Coulombic. The $e^-e^+\gamma$ plasma screening effect on the Coulomb potential can be studied by solving the Thomas-Fermi model:
\be
-\nabla^2 V_C(\bs x) &=& 4\pi (\rho_1(\bs x) + \rho_2(\bs x) + \rho_e(\bs x))\\
\rho_1(\bs x)   &=& Z_1e\delta(\bs x)\\
\rho_2(\bs x)   &=& Z_2e\delta(\bs x + \bs r)\\
\rho_e(\bs x)   &=& eg_s \int\frac{\diff^3 p}{(2\pi)^3} \bigg(  \frac{1}{ 1+e^{(\epsilon(p)+eV_C)/T}} -  \frac{1}{ 1+e^{(\epsilon(p)-eV_C)/T} } \bigg)\,,
\ee
where $V_C(\bs x)$ is the (screened) Coulomb interaction between the two nuclei and $\rho_1(\bs x)$, $\rho_2(\bs x)$ and $\rho_e(\bs x)$ are the charge densities of the first nucleus, the second nucleus and the $e^-e^+$ in the plasma respectively. Here $\bs r$ is the relative position between the two nuclei and $g_s=2$ is the spin degeneracy of the lepton. After solving the screened Coulomb potential, one can solve the Schr\"odinger equation with the nuclear plus Comlomb potential to obtain the resonance properties.

However, the nuclear potential model is just a parametrization and there is no theoretical guiding principle that can determine the functional form of the potential. Furthermore, when the system exhibits a well separated hierarchy of scales, the separation of scales is not generally built in potential models. On the contrary, the EFT framework is constructed on well-separated scales, which can greatly simplify the calculation. Parameters of the EFT Lagrangian exhibit a simple power counting law. This feature is not easily captured by potential models. Here we will use an effective field theory approach to describe the interactions between $\alpha$ particles.

\vspace{0.2in}

\section{Pionless Effective Field Theory}
Since the $^8$Be resonance energy is much smaller than the $\alpha$ particle mass $M\approx3.7$ GeV, a nonrelativistic description is valid. The typical momentum transferred at resonance is about $18.5$ MeV, much smaller than the pion mass $m_{\pi} \approx 135$ MeV. Thus the internal structure of the $\alpha$ particle can be neglected because the internal dynamics is governed by pion exchanges. The finer details of the $\alpha$ particle structure is not probed at the resonant scattering. The long-range interaction between two $\alpha$ particles is the Coulomb repulsion. To model the short-range nuclear interaction, we use the separation of scales in the process. The scale of the nuclear interaction between $\alpha$ particles is set by twice the pion mass because one pion exchange between $\alpha$ particles is forbidden by isospin symmetry, which is again much larger than the momentum at resonance. Thus, a contact nuclear interaction description between $\alpha$ particles is a reasonable approximation. The effective field theory corresponding to this case is the pionless effective field theory. It was first developed to describe low-energy scattering between nucleons and achieved appealing phenomenological success \cite{Kaplan:1996xu,Kaplan:1998we}. It has been extended to include Coulomb effects \cite{Kong:1999sf} and has been used to study the $\alpha$-$\alpha$ low-energy scattering \cite{Higa:2008dn}.

In the pionless EFT, the $\ma{\alpha}$ particle is described by a scalar field $N$ with a mass $M$. The only nuclear interactions involved in the effective Lagrangian are contact interactions which are organized by derivative expansions. The effective Lagrangian is
\be
\ml{L}=N^{\dagger}\bigg(iD_0+\frac{{\bs D}^2}{2M}\bigg)N -  \frac{C_0}{4}N^{\dagger}N^{\dagger} N N + \frac{C_2}{32}\Big(N^{\dagger} \overleftrightarrow{\nabla}^2 N^\dagger N  N  + h.c. \Big) +\cdots \,,
\ee
where $D_\mu$ is the covariant derivative for the gauge coupling to photons. The propagator of the scalar field with an energy-momentum $(E, {\bs p})$ is 
\be
\frac{i}{E-\frac{{\bs p}^2}{2M}+i\epsilon} \,.
\ee
The four-point vertex for $\alpha$ particles with incoming momentum ${\bs p}$ in the c.m.~frame is
\be
-i\sum_{n=0}^{\infty}C_{2n}p^{2n} \,.
\ee
It is worth noticing that the $C_0$ term corresponds to a delta potential in quantum mechanics. The parameters $C_{2n}$ are bare parameters here. Later we will use the same notation $C_{2n}(\mu)$ for renormalized parameters at the scale $\mu$. We will explain the power counting of $C_{2n}(\mu)$ in the next subsection.

\vspace{0.2in}
\subsection{Scattering Amplitude without Coulomb Interactions}

\begin{figure}
\centering
    \begin{subfigure}[b]{0.3\textwidth}
        \centering
        \includegraphics[height=1.25in]{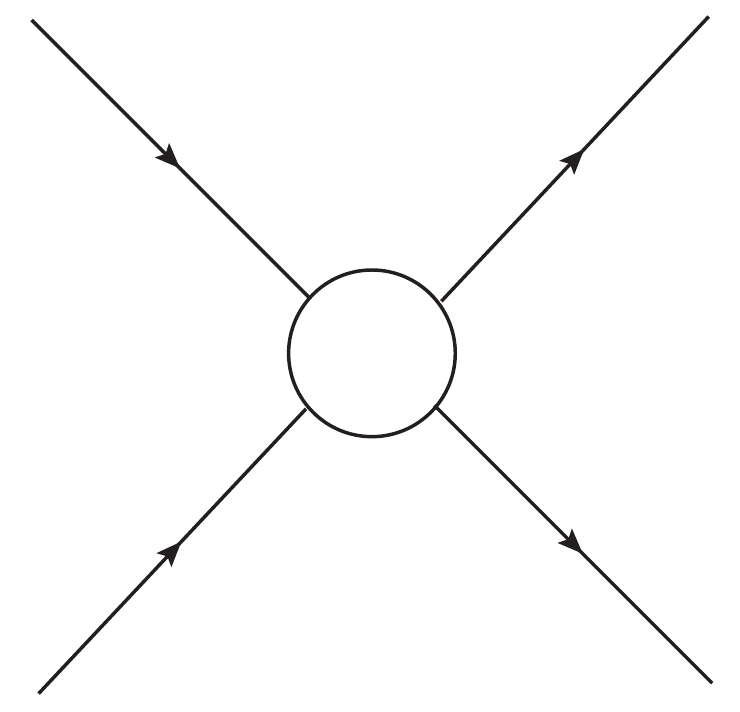}
        \caption{}\label{subfig:NN1}
    \end{subfigure}%
    ~
    \begin{subfigure}[b]{0.6\textwidth}
        \centering
        \includegraphics[height=1.25in]{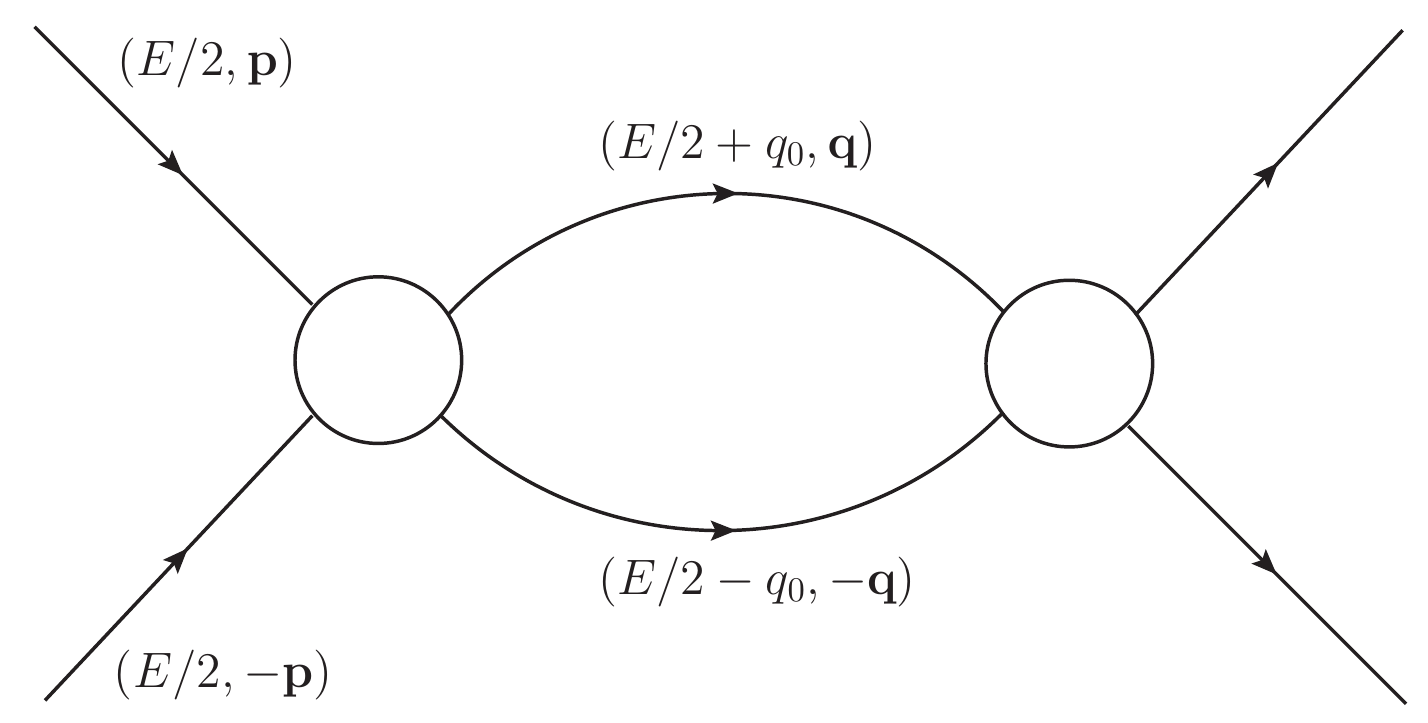}
        \caption{}\label{subfig:NN2}
    \end{subfigure}%
    
    \begin{subfigure}[b]{0.8\textwidth}
        \centering
        \includegraphics[height=1.25in]{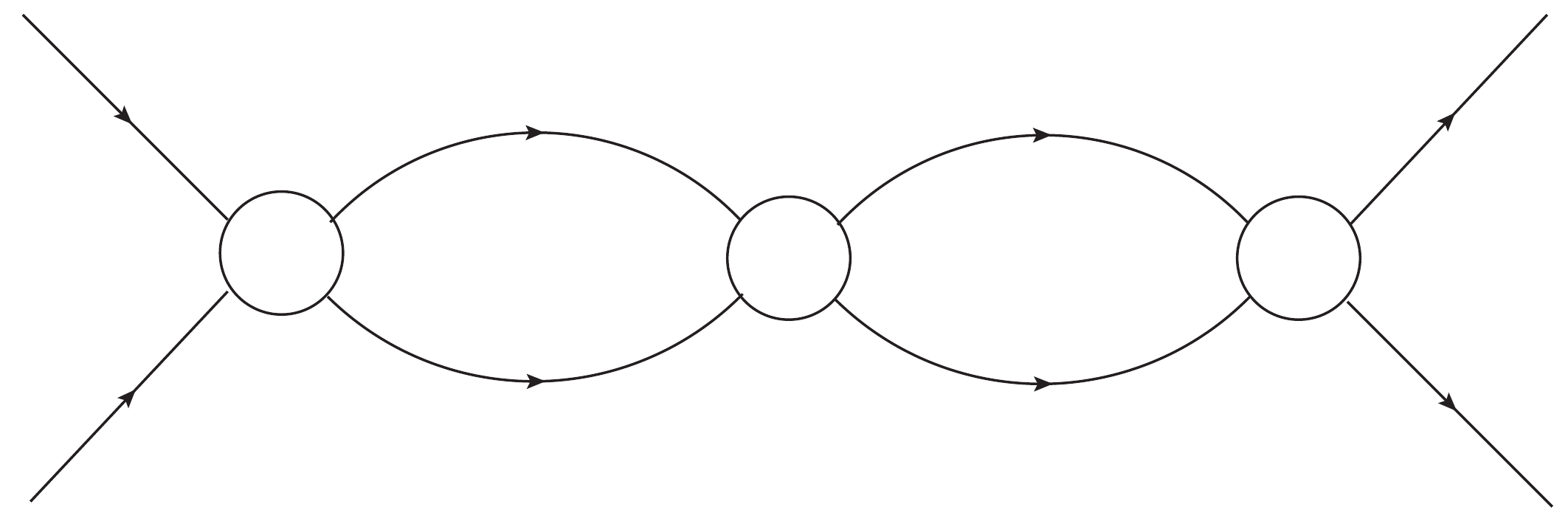}
        \caption{}\label{subfig:NN3}
    \end{subfigure}%
\caption[First three Feynman diagrams contributing to the $\alpha$-$\alpha$ scattering in the geometric series.]{First three Feynman diagrams contributing to the $\alpha$-$\alpha$ scattering in the geometric series. The singlet solid line indicates the $\alpha$ particle and the white blob denotes the vertex $-i\sum_{n=0}^{\infty}C_{2n}p^{2n}$.}
\label{fig_chap2:NN}
\end{figure}
We will focus on the S-wave scattering between two $\alpha$ particles. The tree-level Feynman diagram is shown in Fig.~\ref{subfig:NN1} and the scattering amplitude is
\be
i\ml{T}_\ma{tree} = -i\sum_{n=0}^{\infty}C_{2n}p^{2n}\,.
\ee
The one-loop Feynman diagram with the definition of kinetic variables is shown in Fig.~\ref{subfig:NN2} and the amplitude is given by
\be
i\ml{T}_\ma{1loop} = i\ml{T}_\ma{tree} \sum_n C_{2n} I_n\,,
\ee
where $I_n$ is the loop integral
\be
I_n &=&-i \int\frac{\diff^4q}{(2\pi)^4}{\bs q}^{2n} \frac{i}{E/2+q_0-{\bs q}^2/2M +i\epsilon} \frac{i}{E/2-q_0-{\bs q}^2/2M +i\epsilon}\,,
\ee
which is divergent. So we use dimensional regularization and analytically continue it to an arbitrary dimension $d$,
\be
I_n 
&=& \mu^{4-d} \int \frac{\diff^{d-1}q}{(2\pi)^{d-1}}{\bs q}^{2n}\frac{1}{E-{\bs q}^2/M +i\epsilon}\,,
\ee
where in the last line we have integrated over $q_0$. The renormalization scale $\mu$ is introduced so that the mass dimensions of the parameters $C_{2n}(\mu)$ are the same as in the bare theory. One can find
\be
I_n = -M(ME)^n  \mu^{4-d} \frac{\Gamma(\frac{3-d}{2})}{(4\pi)^{(d-1)/2}} (-ME-i\epsilon)^{(d-3)/2}\,.
\ee
The divergences in $I_n$ will be renormalized (absorbed into the definitions of $C_{2n}$) and in general $C_{2n}$ will be $\mu$-dependent. In the minimal subtraction scheme (MS scheme),
\be
I_n^\ma{MS} = \frac{M}{4\pi}(ME)^n \sqrt{-ME-i\epsilon} = -i \frac{M}{4\pi} p^{2n+1}\,.
\ee
Similarly the two-loop amplitude shown in Fig.~\ref{subfig:NN3} is 
\be
i\ml{T}_\ma{2loop} = i\ml{T}_\ma{tree} (\sum_n C_{2n} I_n)^2\,.
\ee
The n-loop corrections form a geometric series and one can resum all the loop corrections to obtain the total scattering amplitude
\be
\label{chap2:amplitude_eft}
i\ml{T} &=& \frac{ -i\sum_{n=0}^{\infty}C_{2n}p^{2n} }{1-\sum_{n=0}^{\infty}C_{2n}I_n} \,.
\ee
Once we have the scattering amplitude, we can calculate the phase shift defined by\footnote{The factor $\frac{4\pi}{M}$ is a convention when we compare the quantum field theory scattering amplitude and that in quantum mechanics.}
\be
\label{chap2:amplitude_qm}
\ml{T} = \frac{4\pi}{M} \frac{1}{p\cot\delta_0 - ip}\,,
\ee
where $\cot\delta_0$ is the S-wave phase shift and it has an effective range expansion for low-energy scatterings
\be
\label{chap2:eqn_phaseshift_qm_pure_strong}
p\cot\delta_0 = -\frac{1}{a} + \frac{1}{2}r_0p^2 - \frac{1}{4}P_0 p^4 + \cdots\,, 
\ee
where $a$ is the scattering length, $r_0$ the effective range, and $P_0$ the shape parameter. In the EFT framework, the effective range expansion can be generalized to be
\be
p\cot{\delta_0} = -\frac{1}{a} + \frac{1}{2}\Lambda^2\sum_{n=0}^{\infty}r_n\Big( \frac{p^2}{\Lambda^2}  \Big)^{n+1}\,,
\ee
where $1/\Lambda$ is the length scale of the nuclear interaction and low-energy scattering means $p\ll\Lambda$. In our case, the length scale is set by twice the pion mass $\Lambda\sim2m_{\pi}$. We expect $r_n\sim 1/\Lambda$ based on dimensional analysis.

\subsubsection{Small Scattering Length}
When the scattering length $a$ is small, i.e., $p\ll 1/|a| \sim \Lambda$. One can expand the scattering amplitude written in the EFT language (\ref{chap2:amplitude_eft}) and the amplitude in the effective range expansion (\ref{chap2:amplitude_qm}) in powers of $p$ and match the two. Expansion of (\ref{chap2:amplitude_eft}) in the MS scheme gives
\be
\ml{T} =  -C_0
+ iC_0^2\frac{M}{4\pi} p 
+ \Big[  C_0^3\Big(\frac{M}{4\pi}\Big)^2-C_2  \Big]p^2 + \ml{O}(p^3) \,.
\ee
Expansion of (\ref{chap2:amplitude_qm}) leads to
\be
\label{eq:ere_small_a}
\ml{T} =  -\frac{4\pi a}{M} \Big[  1-iap+ \Big(\frac{1}{2}ar_0 - a^2\Big)p^2  + \ml{O}(p^3) \Big] \,.
\ee
Matching these two expressions we obtain
\be
C_0(\mu) &=& \frac{4\pi a}{M} \\
C_2(\mu) &=& C_0\frac{ar_0}{2}\sim \frac{4\pi}{M\Lambda}\Big( \frac{1}{\Lambda^2} \Big),
\ee
where we have used $|a|\sim r_n\sim 1/\Lambda$. The right-hand-sides are $\mu$-independent because in the MS scheme, there is no pole at $d=4$ in $I_n$.

Generally, the coefficients of $p^{2n}$ terms in the scattering amplitude are given by $C_{2n}$ in the EFT and
\be
C_{2n}(\mu)\sim \frac{4\pi}{M\Lambda}\Big( \frac{1}{\Lambda^2} \Big)^n\,.
\ee
So the EFT is indeed an expansion in powers of $p/\Lambda$.

\subsubsection{Large Scattering Length}
When the scattering length $a$ is large, i.e., $|a|\gg 1/\Lambda$, which is a more realistic case because a resonance exists, the naive expansion in powers of $p$ described above fails. In fact, $pa \ll 1$ breaks down and one must reorganize the series of $p^{n}$ in (\ref{eq:ere_small_a}) and absorb all the terms with $(pa)^n$ into the leading-order term. In this case, we need to expand in powers of $p/\Lambda$ while retain $pa$ to all orders. This new expansion of (\ref{chap2:amplitude_qm}) gives
\be
\label{eq:ere_large_a}
\ml{T} = -\frac{4\pi}{M}\frac{1}{1/a+ip}\Big(   1+\frac{r_0/2}{1/a+ip}p^2 + \frac{r_1/2\Lambda^2}{1/a+ip}p^4 + \frac{(r_0/2)^2}{(1/a+ip)^2}p^4+\cdots                             \Big).
\ee
In the EFT method, each Feynman diagram has a positive power of $p$. So to get a negative power of $p$ as in the leading-order term of (\ref{eq:ere_large_a}), an infinite number of Feynman diagrams must be resummed. If we still use the $\ma{MS}$ subtraction scheme and conduct a similar expansion
\be
\ml{T} = -\frac{C_0}{1+\frac{ipM}{4\pi}C_0 }
- \frac{C_2p^2}{ ( 1+\frac{ipM}{4\pi}C_0 )^2}
- \frac{C_4p^4}{ ( 1+\frac{ipM}{4\pi}C_0 )^2} + \frac{\frac{ipM}{4\pi} C_2^2p^4 }{{(1+\frac{ipM}{4\pi}C_0  )^3}} + \cdots\,.
\ee
Matching these two expansions leads to
\be
C_0 &=& \frac{4\pi }{M}a  \\
C_2 &=& \frac{4\pi }{M}\frac{r_0}{2} a^2 \\
C_4 &=& \frac{4\pi }{M} \Big(  \frac{r_1}{2\Lambda^2}a^2 + \frac{r_0^2}{4}a^3 \Big) \,.
\ee
In general $C_{2n}\sim a^{n+1}$. This ruins the power counting because $a$ is large and the expansion parameter in the EFT $pa$ can be $\ml{O}(1)$. 

There are two ways to fix this. One way is to expand the phase shift rather than the scattering amplitude. This is the method we will follow in practical calculations shown later. The other way is to change the renormalization scheme to the power divergence subtraction (PDS) scheme, first introduced in Ref.~\cite{Kaplan:1998we}. PDS subtracts from the dimensionally regularized loop integrals not only the $d\rightarrow4$ poles as in MS scheme, but also poles in lower dimensions which correspond to power law divergences (for example, poles at $d\rightarrow3$),
\be
I_n^\ma{PDS} = -\frac{M}{4\pi}(\mu+ip)p^{2n} \,.
\ee
Then repeating the above matching procedure one finds
\be
C_0(\mu)&=&\frac{4\pi}{M}\Big( \frac{1}{-\mu+1/a} \Big)  \\
C_2(\mu)&=&\frac{4\pi}{M}\Big( \frac{1}{-\mu+1/a} \Big)^2 \frac{r_0}{2}  \\
C_4(\mu)&=&\frac{4\pi}{M}\Big( \frac{1}{-\mu+1/a} \Big)^3 \bigg[ \frac{1}{4}r_0^2 + \frac{1}{2}\frac{r_1}{\Lambda^2}\Big( -\mu+\frac{1}{a} \Big)  \bigg]\,.
\ee
For $\mu \gg 1/|a|$, using $r_n\sim 1/\Lambda$ we see that 
\be
C_0(\mu) &\sim&-\frac{4\pi}{M}\frac{1}{\mu} \\
C_2(\mu) &\sim&-\frac{4\pi}{M}\frac{1}{\mu^2\Lambda} \\
C_4(\mu) &\sim&-\frac{4\pi}{M}\frac{1}{\mu^3\Lambda^2}\,.
\ee
In general
\be
C_{2n}(\mu)&\sim&-\frac{4\pi}{M}\frac{1}{\mu^{n+1}\Lambda^n}\,,
\ee
which makes the power counting manifest. The EFT expands in terms of $\frac{p^2}{\mu\Lambda}$.

\vspace{0.2in} 
\subsection{Scattering Amplitude with Coulomb Interactions}
\begin{figure}
\centering
\includegraphics[width=1.7in]{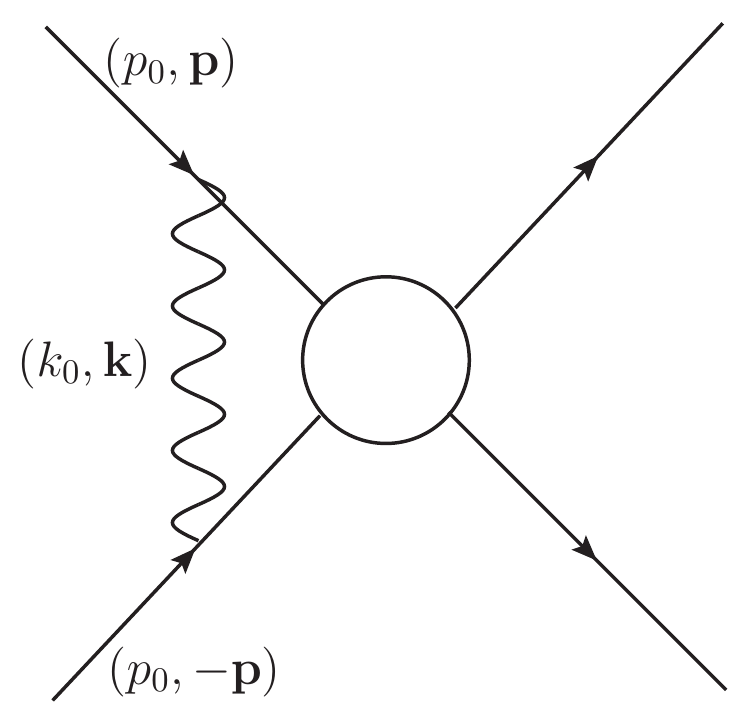}
\caption[One photon exchange correction on the tree-level amplitude with only the $C_0$ term.]{One photon exchange correction on the tree-level amplitude with only the $C_0$ term. Here $p_0 = {\bs p}^2/M$.}
\label{fig_chap2:1gamma}
\end{figure}

\begin{figure}
\centering
    \begin{subfigure}[b]{0.3\textwidth}
        \centering
        \includegraphics[height=1.25in]{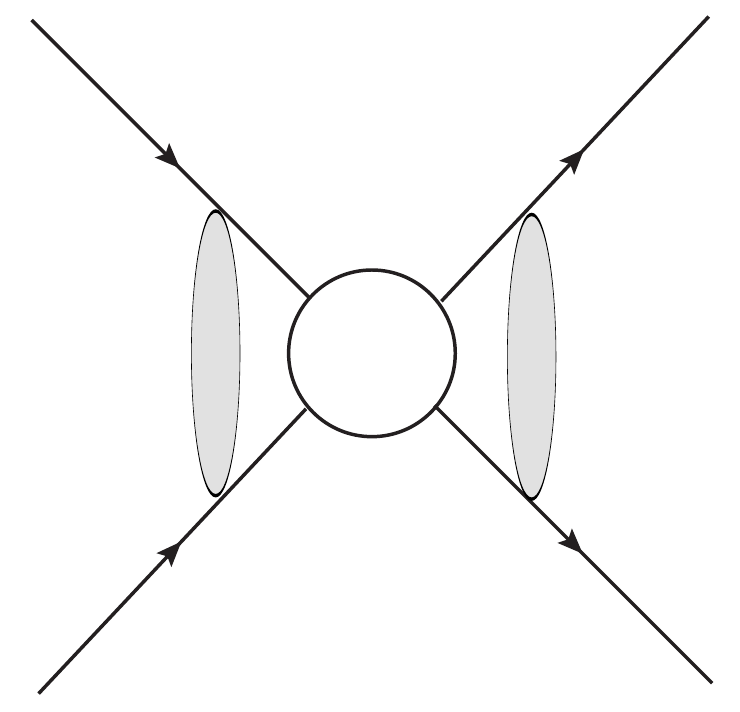}
        \caption{}\label{subfig:NNC1}
    \end{subfigure}%
    ~
    \begin{subfigure}[b]{0.6\textwidth}
        \centering
        \includegraphics[height=1.25in]{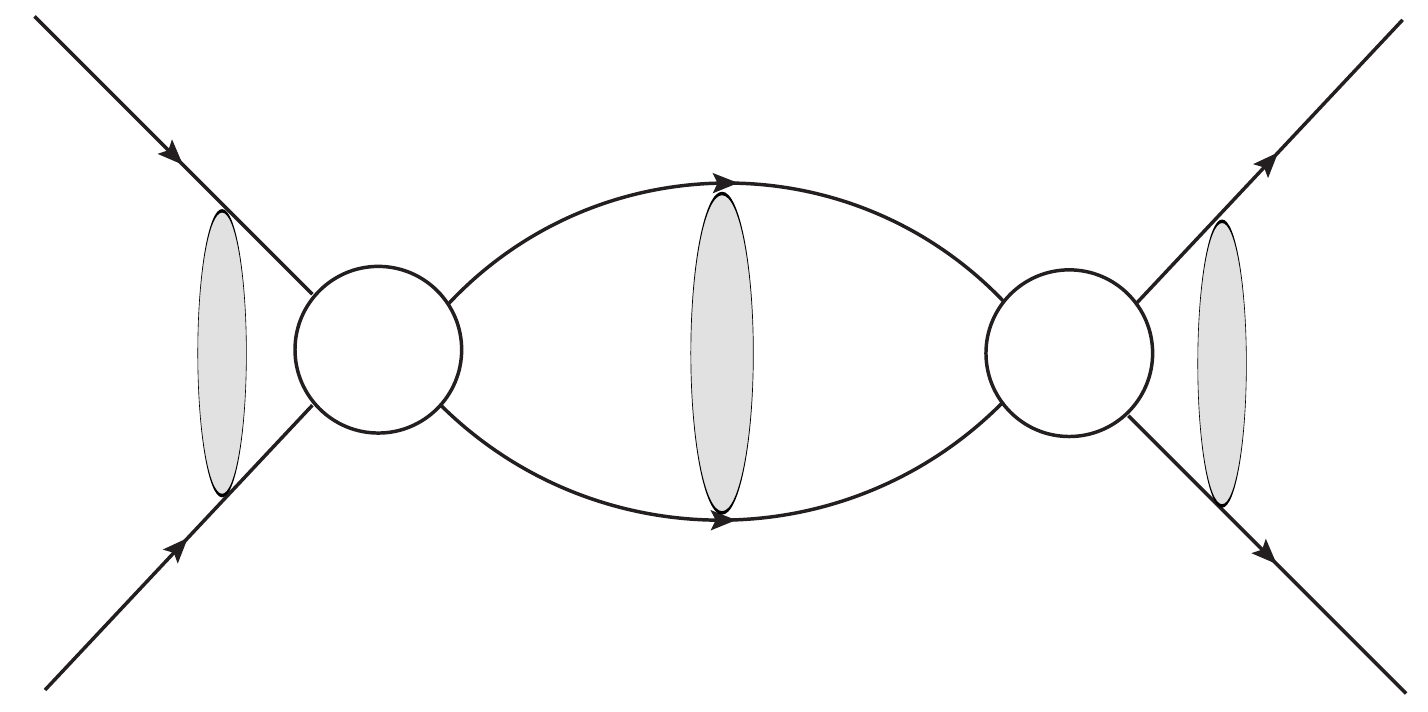}
        \caption{}\label{subfig:NNC2}
    \end{subfigure}%
    
    \begin{subfigure}[b]{0.8\textwidth}
        \centering
        \includegraphics[height=1.25in]{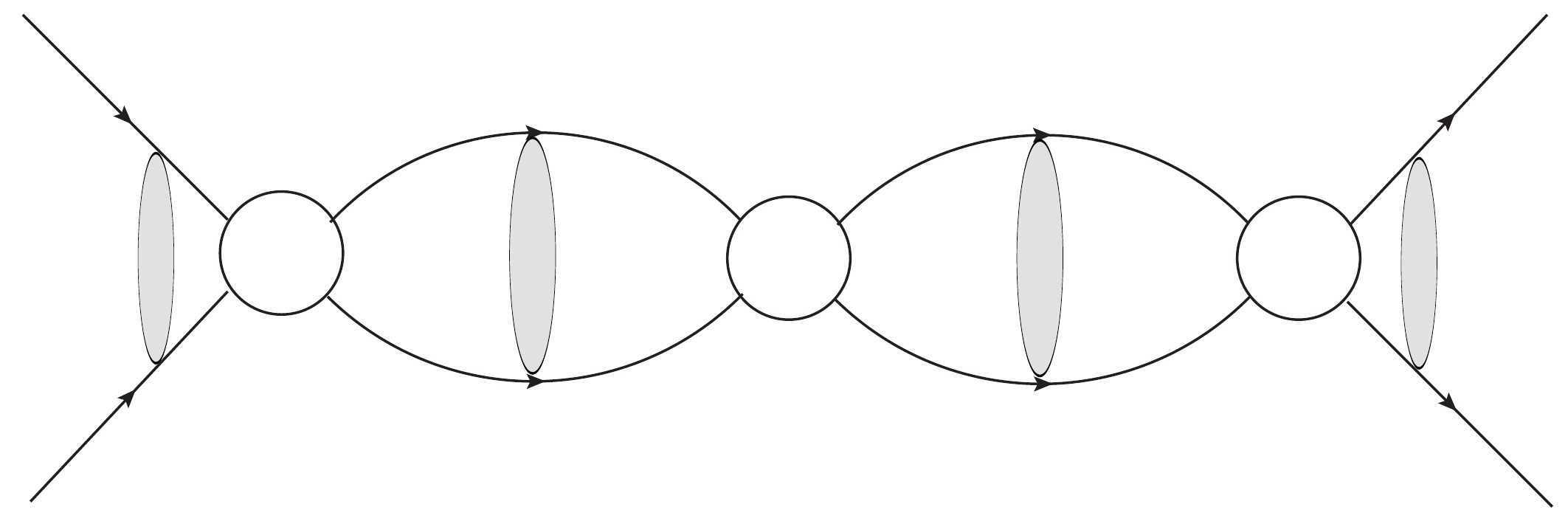}
        \caption{}\label{subfig:NNC3}
    \end{subfigure}%
\caption[First three Feynman diagrams contributing to the $\alpha$-$\alpha$ scattering in the geometric series with all the Coulomb exchanges resummed.]{First three Feynman diagrams contributing to the $\alpha$-$\alpha$ scattering in the geometric series with all the Coulomb exchanges resummed. The grey blob indicates the Green's function (propagator) with all the photon exchanges, shown in Fig.~\ref{fig_chap2:coulomb}}
\label{fig_chap2:NNC}
\end{figure}

\begin{figure}
\centering
\includegraphics[width=5in]{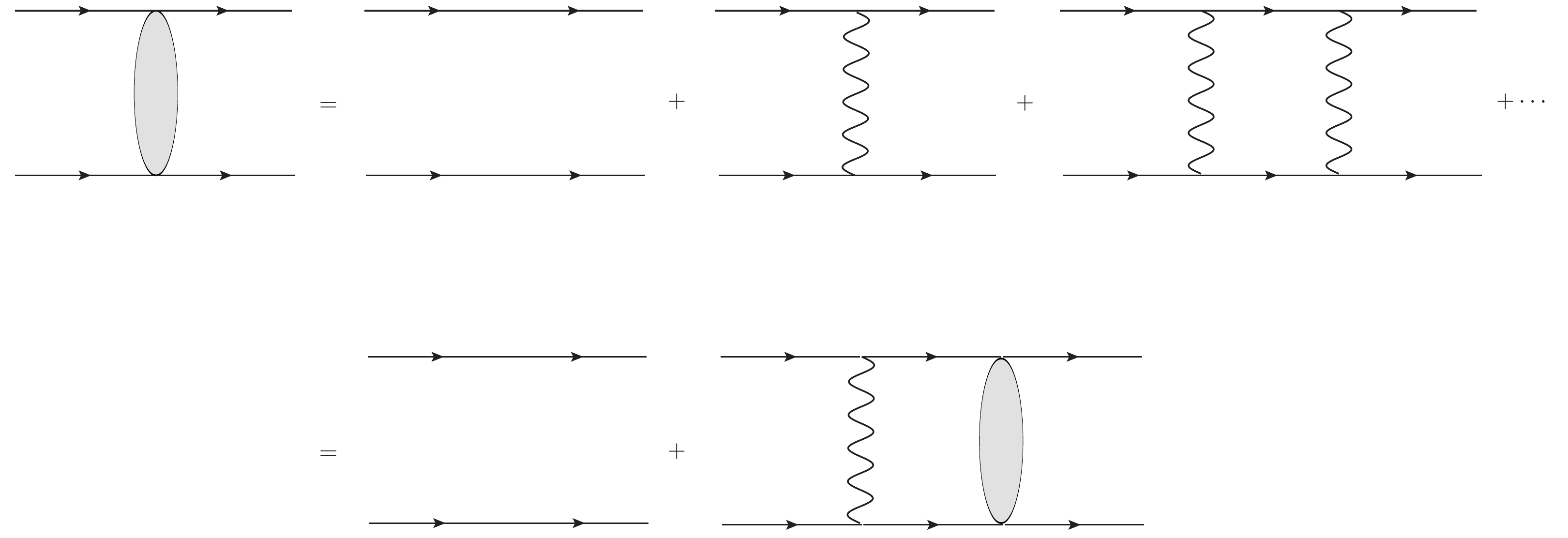}
\caption{Green's function of two nucleons/nuclei with all the photon exchanges resummed.}
\label{fig_chap2:coulomb}
\end{figure}

We first need to check whether the Coulomb effect is perturbative. Consider the one photon exchange correction on the incoming external lines of the $C_0$ vertex shown in Figure~\ref{fig_chap2:1gamma}.
Here we will conduct dimensional analysis. After integrating over the loop momentum $k=(k_0,{\bs k})$, the amplitude is independent of $k$. The nucleon/nucleus propagator scales as $M/p^2$, the photon propagator scales as $1/p^2$ and the loop integration $\int \diff^4k$ scales as $p^5/M$ nonrelativistically. Combining all these, we find the one photon exchange correction on the $C_0$ vertex scales as
\be
Z_1Z_2\alpha_\ma{em} \Big(\frac{M}{p^2}\Big)^2 \frac{1}{p^2} \frac{p^5}{M}= Z_1Z_2 \frac{\alpha_\ma{em} M}{p}\,,
\ee
where $Z_1$ and $Z_2$ are the atomic numbers of the two nuclei. We see that if the scattering energy is low enough, i.e., $p< Z_1Z_2 \alpha_\ma{em} M$, the Coulomb correction is non-perturbative. This scaling can also be shown by an explicit calculation \cite{Kong:1999sf}. The Coulomb effect is non-perturbative in the low-energy $\alpha$ scattering. We need to apply non-perturbative method to sum over all photon exchanges, in addition to resumming the geometric series in the nuclear strong interaction. The associated Feynman diagrams are shown in Fig.~\ref{fig_chap2:NNC} and the summation over all photon exchanges is plotted in Fig.~\ref{fig_chap2:coulomb}. Since we are doing a nonrelativistic calculation, diagrams with crossed photon lines are suppressed. 

The summation can be done by solving the Schr$\ma{\ddot{o}}$dinger equation to obtain the Green's function. In the free theory, we define the retarded and advanced Green's function
\be
\hat{G}_0^{(\pm)}(E)=\frac{1}{E-H_0\pm i\epsilon}\,,
\ee
where $E={\bs p}^2/M$ and $H_0= \hat{\bs p}^2/M$. Inserting a complete set of momentum eigenstates $|{\bs q}\rangle$. We obtain a representation of the free Green's function
\be
\hat{G}_0^{(\pm)}(E)=M\int\frac{\diff^3q}{(2\pi)^3}\frac{|{\bs q}\rangle\langle{\bs q}|}{{\bs p}^2-{\bs q}^2\pm i\epsilon}\,.
\ee

In the case of pure Coulomb repulsion, the Green's functions are given by
\be
\hat{G}_C^{(\pm)}(E)=\frac{1}{E-H_0-V_C\pm i\epsilon},
\ee
where $V_C=\frac{Z_1Z_2e^2}{4\pi r}=\frac{Z_1Z_2\alpha_\ma{em}}{r}$ and 
\be
\hat{G}_C^{(\pm)} = \hat{G}_0^{(\pm)}+\hat{G}_0^{(\pm)}V_C\hat{G}_C^{(\pm)} = \sum_{n=0}^{\infty}\hat{G}_{0}^{(\pm)} ( \hat{V}_C\hat{G}_{0}^{(\pm)} )^n \,.
\ee
This formula comes from the diagrammatic calculation of the Green's function literally, as shown in Figure~\ref{fig_chap2:coulomb}.

Now we add the short-range nuclear strong interaction labeled by the potential $V_S$,
\be
\hat{G}_{SC}^{(\pm)}(E)=\frac{1}{E-H_0-V_C-V_S\pm i\epsilon}\,.
\ee
For operators $A$ and $B$ we have a general relation $A^{-1}-B^{-1}=B^{-1}(B-A)A^{-1}$. Setting $A^{-1}=\hat{G}_{SC}^{(\pm)}$ and $B^{-1}=\hat{G}_{C}^{(\pm)}$ leads to
\be
\label{chap2:eqn_green_sc}
\hat{G}_{SC}^{(\pm)} = \hat{G}_{C}^{(\pm)}+\hat{G}_{C}^{(\pm)}\hat{V}_S\hat{G}_{SC}^{(\pm)} = \sum_{n=0}^{\infty}\hat{G}_{C}^{(\pm)} ( \hat{V}_S\hat{G}_{C}^{(\pm)} )^n \,.
\ee

To define the scattering amplitude, we need the asymptotic states in the far past and far future. Let $|{\bs p}\rangle$ be the free scattering wave function (plane wave) and $|\psi_{\bs p}^{(\pm)}\rangle$ be the incoming $(+)$ and outgoing $(-)$ scattering waves, which are solutions to the Schr$\ma{\ddot{o}}$dinger equation with the Coulomb repulsion and correct boundary conditions at $t=\pm \infty$. Then the Lippmann-Schwinger equation gives
\be
|\psi_{\bs p}^{(\pm)}\rangle = (  1+\hat{G}_C^{(\pm)}V_C )|{\bs p}\rangle = \hat{G}_C^{(\pm)}\hat{G}_0^{-1}|{\bs p}\rangle\,.
\ee
The normalization condition is
\be
\langle \psi_{\bs q}^{(\pm)} | \psi_{\bs p}^{(\pm)}\rangle = (2\pi)^3\delta^{(3)}({\bs q} - {\bs p}).
\ee
The solution to the Schr$\ma{\ddot{o}}$dinger equation under this normalization condition is the Coulomb scattering wave. Their spatial representations are
\be
\psi_{\bs p}^{(+)}({\bs r}) &=& e^{-\pi\eta/2}\Gamma(1+i\eta)M(-i\eta,1;ipr-i{\bs p}\cdot{\bs r})e^{i{\bs p}\cdot{\bs r}}\\
\psi_{\bs p}^{(-)}({\bs r}) &=& e^{-\pi\eta/2}\Gamma(1-i\eta)M(i\eta,1;-ipr-i{\bs p}\cdot{\bs r})e^{i{\bs p}\cdot{\bs r}}\,,
\ee
where $M(a,b;z)$ is the confluent hypergeometric function and $\eta=Z_1Z_2\alpha_\ma{em} M/2p$. 

Similarly, the incoming and outgoing scattering waves $|\Psi_{\bs p}^{(\pm)}\rangle$ with both the Coulomb and nuclear strong interactions are
\be
|\Psi_{\bs p}^{(\pm)}\rangle &=& \big[  1+\hat{G}_{SC}^{(\pm)}(V_C+V_S) \big] |{\bs p}\rangle \,.
\ee
Substituting the relation (\ref{chap2:eqn_green_sc}) into $|\Psi_{\bs p}^{(\pm)}\rangle$ gives
\be \nn 
|\Psi_{\bs p}^{(\pm)}\rangle&=&\Big[ 1+\sum_{n=0}^{\infty}\hat{G}_{C}^{(\pm)} ( \hat{V}_S\hat{G}_{C}^{(\pm)} )^n  ( \hat{V}_S+\hat{V}_C ) \Big]|{\bs p}\rangle \\ \nn
&=&\Big[ 1+\sum_{n=1}^{\infty}( \hat{G}_{C}^{(\pm)}\hat{V}_S )^n\Big] (  1+ \hat{G}_{C}^{(\pm)}\hat{V}_C  ) |{\bs p}\rangle \\
&=&\Big[ 1+\sum_{n=1}^{\infty}(\hat{G}_CV_S)^n  \Big] |\psi_{\bs p}^{(\pm)}\rangle \,.
\ee

\begin{figure}
\centering
\includegraphics[width=1.0in]{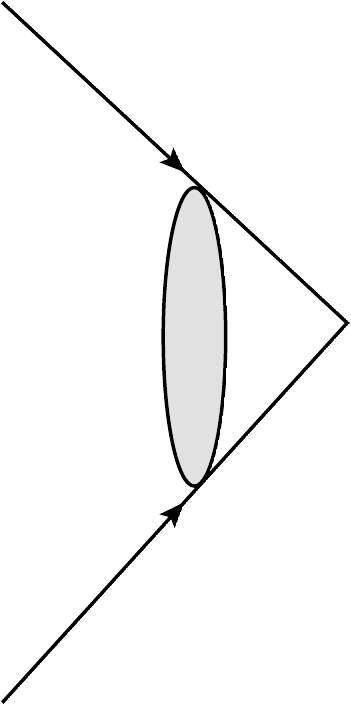}
\caption[Diagrammatic representation of $\psi_{\bs p}^{(+)}({\bs r}=0)$.]{Diagrammatic representation of $\psi_{\bs p}^{(+)}({\bs r}=0)$. Its magnitude squared gives the Sommerfeld factor $C_{\eta}^2$.}
\label{fig_chap2:sommerfeld}
\end{figure}

The scattering amplitude $\ml{T}$ due to both the Coulomb and nuclear strong interactions is defined as
\be
\ml{T} ({\bs p}', {\bs p}) \equiv  - \langle{\bs p'}| V_C+V_S | \Psi_{\bs p}^{(+)}\rangle \,,
\ee
where the negative sign is just a convention. 
Using the Lippmann-Schwinger equations derived above, one can easily show that the scattering amplitude $\ml{T}$ can be separated into two parts: one comes from the pure Coulomb interaction and the other from the nuclear strong interaction modified by the Coulomb corrections,
\be
\ml{T} ({\bs p}', {\bs p}) &=& \ml{T}_C({\bs p}',{\bs p}) + \ml{T}_{SC}({\bs p}',{\bs p}) \\
\ml{T}_{C}({\bs p'},{\bs p})&=& - \langle{\bs p'}|V_C|\psi_{\bs p}^{(+)}\rangle \\ \nn
\ml{T}_{SC}({\bs p'},{\bs p})&=& - \langle\psi_{\bs p'}^{(-)}|V_S|\Psi_{\bs p}^{(+)}\rangle \\
\label{chap2:eqn_T_SC}
&=& - \sum_{n=0}^{\infty}\langle\psi_{\bs p'}^{(-)}|V_S(\hat{G}_C^{(+)}V_S)^n|\psi_{\bs p}^{(+)}\rangle \,.
\ee
It should be noted that the second term $\ml{T}_{SC}({\bs p'},{\bs p})$ is not purely caused by the nuclear strong interaction. Coulomb corrections appear in both the incoming/outgoing waves and the Green's functions. The formula (\ref{chap2:eqn_T_SC}) justifies the summation of the Feynman diagrams in Fig.~\ref{fig_chap2:NNC}. The two grey blobs at the left and right ends of each diagram are represented by the Coulomb wave functions $\langle\psi_{\bs p'}^{(-)}|$ and $|\psi_{\bs p}^{(+)}\rangle$. The grey blobs in the middle correspond to the $(\hat{G}_C^{(+)}V_S)^n$ term. The process we are interested in is an elastic scattering, so we set $|{\bs p}| = |{\bs p}'| \equiv p$ from now on and write the amplitude as $\ml{T}(p)$.
If we only consider the contact interaction $\langle{\bs q'}|V_S|{\bs q}\rangle=C_0$, the diagram in Fig.~\ref{subfig:NNC1} corresponds to the contribution that is first order in the nuclear strong interaction,
\be \nn
\ml{T}_{SC}^{(1)}(p)&=& - \int\frac{\diff^3q}{(2\pi)^3}\int\frac{\diff^3q'}{(2\pi)^3}\langle\psi_{\bs p}^{(-)}|{\bs q'}\rangle\langle{\bs q'}|V_S|{\bs q}\rangle \langle{\bs q}|\psi_{\bs p}^{(+)}\rangle \\
&=& - C_0\psi_{\bs p}^{(-)*}({\bs r} = 0)\psi_{\bs p}^{(+)}({\bs r} =  0) = - C_0C_{\eta}^2e^{2i\sigma_0} \,,
\ee
where $\sigma_\ell=\Arg \Gamma(1+\ell+i\eta)$ is the phase shift of the $\ell$-wave caused purely by the Coulomb interaction and $C_{\eta}^2$ is the Sommerfeld factor defined as
\be
C_{\eta}^2\equiv | \psi_{\bs p}^{(\pm)}({\bs r}=0) |^2=e^{-\pi\eta}\Gamma(1+i\eta)\Gamma(1-i\eta)=\frac{2\pi\eta}{e^{2\pi\eta}-1} \,.
\ee
It is the probability of the two $\alpha$-particles being overlapped, or at the zero-separation. Its diagrammatic interpretation is shown in Fig.~\ref{fig_chap2:sommerfeld}.

The second order amplitude corresponds to the diagram in Fig.~\ref{subfig:NNC2},
\be
\ml{T}_{SC}^{(2)}(p)&=& - C_0^2C_{\eta}^2e^{2i\sigma_0}G_C^{(+)}(E;0,0),
\ee
where $E=p^2/M$ and 
\be
G_C^{(+)}(E;0,0) \equiv \langle {\bs r}' = 0 | \hat{G}_C^{(+)}(E) |  {\bs r} = 0 \rangle \,,
\ee
is the zero-separation to zero-separation propagator,
\be \nn
G_C^{(+)}(E;{\bs r}'=0,{\bs r}=0)&=&\int\frac{\diff^3q}{(2\pi)^3}\int\frac{\diff^3q'}{(2\pi)^3}\langle {\bs q'} | \hat{G}_C^{(+)}(E)| {\bs q} \rangle \\ 
\label{chap2:eqn_GC00}
&=& M\int\frac{\diff^3q}{(2\pi)^3} \frac{\psi_{\bs q}^{(+)}(0)\psi_{\bs q}^{(+)*}(0)}{{\bs p}^2-{\bs q}^2+i\epsilon} \,.
\ee
Summing over all diagrams leads to
\be
\ml{T}_{SC}(p)= - \frac{C_{\eta}^2 C_0e^{2i\sigma_0}}{1-C_0G_C^{(+)}(E;0,0)}= - \frac{C_{\eta}^2e^{2i\sigma_0}}{ C_0^{-1}-G_C^{(+)}(E;0,0)}.
\ee
Generally, the potential includes all the contact interaction terms in the effective Lagrangian\footnote{One may worry that inside loops the loop momentum will show up in the contact interaction and thus powers of $|{\bs q}'|$ and $|{\bs q}|$ will appear in the integrand of (\ref{chap2:eqn_GC00}) and ruin the simple interpretation as the zero-separation to zero-separation propagator. But it has been shown by using the free field equation of motion that only the total relative energy $ME = p^2$ appears in the contact interaction provided one makes a choice of the operator basis \cite{Beane:2000fi}.} $\langle{\bs q'}|V_S|{\bs q}\rangle=\sum_{n=0}^{\infty}C_{2n}p^{2n}$. Therefore the scattering amplitude of the Coulomb modified nuclear strong interaction has a generic form
\be
\label{chap2:eqn_amplitude_SC}
\ml{T} _{SC}(p)= - \frac{C_{\eta}^2e^{2i\sigma_0}}{(\sum_{n=0}^{\infty}C_{2n}p^{2n} )^{-1}-G_C^{(+)}(E;0,0)} \,.
\ee

\begin{figure}
\centering
\includegraphics[width=3.5in]{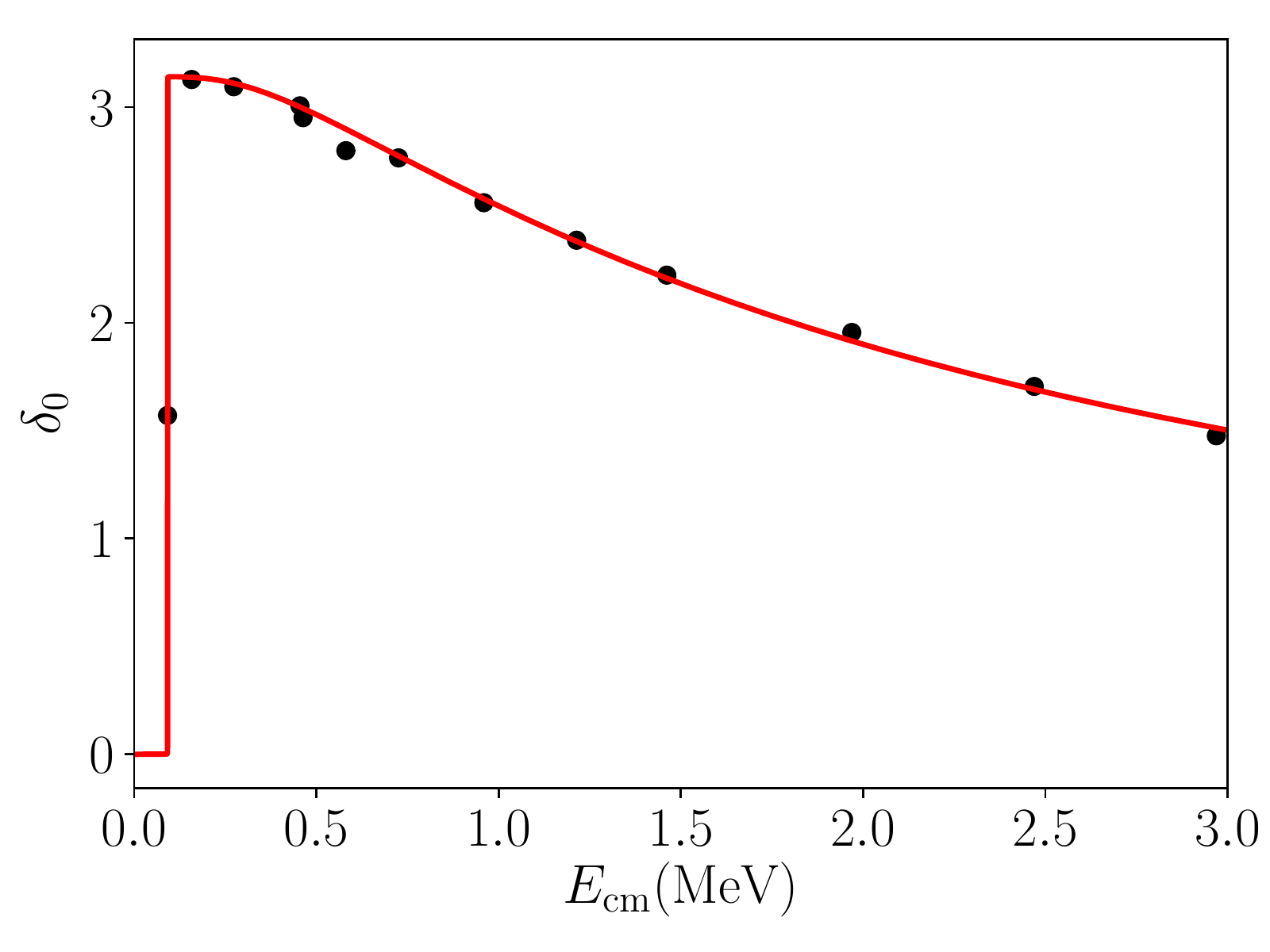}
\caption[The phase shift of the nuclear strong interaction modified by the Coulomb interaction.]{The phase shift of the nuclear strong interaction modified by the Coulomb interaction. The red solid line is our calculation result while the black dots are experimental data from Ref.~\cite{Russell:1956zz}. The experimental uncertainties are small and not shown here.}
\label{fig_chap2:phase_shift}
\end{figure}

Once we have the scattering amplitude, we can calculate the phase shift $\delta_0$ of the nuclear strong interaction modified by the Coulomb repulsion via
\be
\label{chap2_eqn_SC_ERS}
\ml{T} = \frac{4\pi}{M}\frac{e^{2i\sigma_0}}{p\cot\delta_0 -ip} \,.
\ee
As we discussed earlier, the naive expansion of the scattering amplitude in powers of $p$ in the MS scheme does not give a manifest power counting. Instead, we will expand the inverse of the scattering amplitude in powers of $p$, which corresponds to the effective range expansion. By doing so we can also preserve exact unitarity. Expanding the effective range has been shown to be better than expanding the scattering amplitude in powers of $p$ at reproducing the phase shift \cite{Higa:2008dn}. We expand the sum $(\sum_nC_{2n}p^{2n})^{-1}$ to the order $p^4$ and obtain
\be
\label{chap2_eqn_SC_EFT}
\ml{T} = -\frac{C_\eta^2e^{2i\sigma_0}}{{\frac{1}{C_0}-\frac{C_2}{C_0^2}p^2+\big( \frac{C_2^2}{C_0^3}-\frac{C_4}{C_0^2} \big)p^4}-G_C^{(+)}(E,0,0)} \,,
\ee
Comparing the two expressions (\ref{chap2_eqn_SC_ERS}) and (\ref{chap2_eqn_SC_EFT}) leads to a formula for the Coulomb modified nuclear phase shift in terms of the EFT parameters:
\be
\label{chap2:eqn_phase_shift_EFT}
p\cot\delta_0 -ip &=&-\frac{4\pi}{MC_\eta^2}\bigg( \frac{1}{C_0}-\frac{C_2}{C_0^2}p^2+\Big( \frac{C_2^2}{C_0^3}-\frac{C_4}{C_0^2} \Big) p^4 -G_C^{(+)}(E,0,0)  \bigg).
\ee
We need to compute $G_C^{(+)}(E;0,0)$. Using expression (\ref{chap2:eqn_GC00}),
\be
G_C^{(+)}(E;0,0) = M\int\frac{\diff^3q}{(2\pi)^3}\frac{2\pi\eta(q)}{e^{2\pi\eta(q)}-1}\frac{1}{{\bs p}^2-{\bs q}^2+i\epsilon} \,,
\ee
where $\eta(q)=\frac{Z_1Z_2\alpha_\ma{em} M}{2q}$. The integral is ultra-violet (UV) singular and needs regularizing. We separate into a finite part and a divergent part,
\be
G_C^{(+)}(E;0,0)&=&G_C^\ma{fin}+G_C^\ma{div} \\
G_C^\ma{fin}&=&M\int\frac{\diff^3q}{(2\pi)^3}\frac{2\pi\eta(q)}{e^{2\pi\eta(q)}-1}\frac{1}{{\bs q}^2}\frac{{\bs p}^2}{{\bs p}^2-{\bs q}^2+i\epsilon} \\
G_C^\ma{div}&=&-M\int\frac{\diff^3q}{(2\pi)^3}\frac{2\pi\eta(q)}{e^{2\pi\eta(q)}-1}\frac{1}{{\bs q}^2}.
\ee
Let $x=2\pi\eta(q)=\frac{Z_1Z_2\alpha_\ma{em}\pi M}{q}$, then $\diff q=-\frac{q^2}{Z_1Z_2\alpha_\ma{em}\pi M} \diff x$ and we obtain
\be \nn
G_C^\ma{fin}&=&M\int_0^{\infty}\frac{4\pi q^2 \diff q}{(2\pi)^3}\frac{x}{e^x-1}\frac{1}{q^2}\frac{p^2}{p^2-q^2+i\epsilon} \\ \nn
&=&\frac{M}{2\pi^2}\int_{\infty}^0\frac{x}{e^x-1}\Big( -\frac{q^2}{Z_1Z_2\alpha_\ma{em}\pi M} \diff x \Big)\frac{p^2}{p^2-q^2+i\epsilon} \\ \nn
&=&-\frac{Z_1Z_2\alpha_\ma{em}\pi M^2}{2\pi^2}\int_{\infty}^0\frac{x\diff x}{e^x-1}\frac{1}{\frac{(Z_1Z_2\alpha_\ma{em}\pi M)^2}{q^2}-\frac{(Z_1Z_2\alpha_\ma{em}\pi M)^2}{p^2}+i\epsilon} \\ \nn
&=&\frac{Z_1Z_2\alpha_\ma{em} M^2}{2\pi}\int^{\infty}_0\frac{x\diff x}{e^x-1}\frac{1}{x^2+x_0^2} \\
&=&\frac{Z_1Z_2\alpha_\ma{em} M^2}{4\pi}\Big[ \ln\big(\frac{x_0}{2\pi}\big) -\frac{\pi}{x_0}-\psi\big(\frac{x_0}{2\pi}\big)   \Big]\,,
\ee
where $x_0=2\pi i\eta(p)$ and $\psi$ is the Digamma function, i.e., logarithmic derivative of the Gamma function. We evaluate $G_C^\ma{div}$ by using the dimensional regularization. In $d = 3 - \epsilon$ dimension,
\be \nn
G_C^\ma{div}&=&-M\mu^{\epsilon}\frac{\Omega_d}{(2\pi)^d}\int_0^{\infty} \diff qq^{d-3}\frac{2\pi\eta(q)}{e^{2\pi\eta(q)}-1} \\ \nn
&=&-M\mu^{\epsilon}\frac{2\pi^{d/2}}{\Gamma(d/2)(2\pi)^d}(Z_1Z_2\alpha_\ma{em}\pi M)^{d-2}\int_0^{\infty} \diff x\frac{x^{\epsilon-1}}{e^x-1} \\
&=&-\frac{Z_1Z_2\alpha_\ma{em} M^2}{4\sqrt{\pi}}\Big( \frac{2\mu}{Z_1Z_2\alpha_\ma{em} M \sqrt{\pi} } \Big)^{\epsilon}\frac{\Gamma(\epsilon)\zeta(\epsilon)}{\Gamma(\frac{3-\epsilon}{2})} \,,
\ee
where $\mu$ is the renormalization scale. Since we are expanding the phase shift rather than the scattering amplitude, we do not have to use the PDS scheme. The calculation using the PDS scheme has been done in Ref.~\cite{Kong:1999sf}. Here we will only consider poles at $ d \rightarrow 3$, which corresponds to $\epsilon\rightarrow0$. Noticing $\Gamma(\epsilon)$ has a pole as $\epsilon\rightarrow0$, we obtain the regularized integral
\be
G_C^\ma{div}=\frac{Z_1Z_2\alpha_\ma{em} M^2}{4\pi}\Big[   \frac{1}{\epsilon}+\ln\frac{2\mu\sqrt{\pi}}{Z_1Z_2\alpha_\ma{em} M}+1-\frac{3}{2}\gamma  \Big] \,.
\ee
Putting the two terms together we obtain
\be
G_C^{(+)}(E,0,0) &=& \frac{Z_1Z_2\alpha_\ma{em} M^2}{4\pi}\bigg( \frac{1}{\epsilon}-H(\eta)+\ln{\frac{2\mu\sqrt{\pi}}{Z_1Z_2\alpha_\ma{em} M}}+1-\frac{3}{2}\gamma \bigg)\\
H(\eta)&=&\psi(i\eta)+\frac{1}{2i\eta}-\ln{i\eta}=\Re\psi(1+i\eta)+\frac{iC_{\eta}^2}{2\eta}-\ln{\eta} \,.
\ee 
We renormalize the theory in a scheme that the $\epsilon$-pole and all the other $\eta$-independent terms are absorbed into the definition of $C_{2n}$ so they are $\mu$-dependent. Now we can match the renormalized expression (\ref{chap2:eqn_phase_shift_EFT}) with the effective range expansion of the phase shift. The effective range expansion is given by
\be
\label{chap2:eqn_phaseshift_qm}
C_\eta^2 p \cot\delta_0 + Z_1Z_2\alpha_\ma{em} M \Re H(\eta) = -\frac{1}{a} + \frac{1}{2}r_0p^2 - \frac{1}{4}P_0p^4+\cdots \,,
\ee
in which the difference from expression (\ref{chap2:eqn_phaseshift_qm_pure_strong}) is due to the Coulomb correction. We find
\be
-\frac{1}{a}&=&-\frac{4\pi}{M}\frac{1}{C_0}\equiv A \\
\frac{1}{2}r_0&=&\frac{4\pi}{M}\frac{C_2}{C_0^2}\equiv B \\
-\frac{1}{4}P_0&=&-\frac{4\pi}{M}\Big(  \frac{C_2^2}{C_0^3}-\frac{C_4}{C_0^2} \Big)\equiv C \,.
\ee
Then we fit these three parameters by using the experimentally determined $^8$Be resonance energy, width and the phase shift up to $E_\ma{cm}=3\ \ma{MeV}$. The resonance at $E_\ma{cm}\equiv E_0 = 91.8$ keV corresponds to a S-wave phase shift $\delta_0=\frac{\pi}{2}$, i.e., $\cot\delta_0=0$. The width of the resonance $\Gamma$ is given by 
\be
\label{chap2:eqn_width}
\frac{\diff \cot\delta_0(E)}{\diff E}\Big|_{E=E_0}\equiv-\frac{2}{\Gamma} \,.
\ee
We use Eq.~(\ref{chap2:eqn_phaseshift_qm}) and Eq.~(\ref{chap2:eqn_width}) to calculate the resonance energy, width and phase shift and apply a least square fit of the parameters. The best fit result is shown in Table~\ref{table:1}. The best fit resonance energy and width are summarized in Table~\ref{table:2} with a comparison to experimental measurements. The calculated phase shift is plotted in Fig.~\ref{fig_chap2:phase_shift} along with the experimental measurements from Ref.~\cite{Russell:1956zz}. The agreement with experimental data is excellent. Our values for the extracted scattering length, effective range, and shape parameter are consistent with a similar fit in Ref.~\cite{Higa:2008dn}. Our numerical values differ slightly because we fit up to $E_\ma{cm} = 3$ MeV, while Ref.~\cite{Higa:2008dn} fits up to $E_\ma{lab}=3$ MeV, which corresponds to $E_\ma{cm} =1.5$ MeV. We will use these best fit parameters in the study of plasma screening effect on the resonance in the next section.

\begin{table*}
\caption{\label{table:1}Best fit parameters}
\begin{center}
\begin{tabular}{|c|c|c|c|}
\hline
Parameter & $a\ (10^3\ \ma{fm})$ & $r_0\ (\ma{fm})$ & $P_0\ (\ma{fm}^3)$\\
\hline
Best fit value (accurate to $10^{-3}$) &  -2.029 & 1.104 &  -1.824  \\
\hline
\end{tabular}
\end{center}
\end{table*}

\begin{table*}
\caption{\label{table:2}Best fit resonance energy and width}
\begin{center}
\begin{tabular}{|c|c|c|}
\hline
Physical quantity & Resonance energy $E_0$ (keV) & Width $\Gamma$ (eV)\\
\hline
Best fit value (accurate to $10^{-3}$) &   91.838   &   5.715   \\
\hline
Experimental value \cite{Wustenbeckeretal.1992} & 91.84 $\pm$ 0.04   &  5.57 $\pm$ 0.25 \\
\hline
\end{tabular}
\end{center}
\end{table*}

\vspace{0.2in} 
\section{Plasma Screening Effects on the $^8$Be Resonance}
\subsection{Static Plasma Screening Effect}
In this subsection, we will focus on the static plasma screening effect on the $^8$Be resonance. Inside an $e^-e^+\gamma$ plasma, the Coulomb repulsion will be suppressed by
\be
\label{chap2:eqn_yukawa}
V_C = \frac{Z_1Z_2e^2}{4\pi r }e^{-m_Dr}\,,
\ee
where $m_D$ is the Debye mass of the plasma. In the hard thermal loop approximation with a massless electron, we have shown in Eq.~(\ref{chap1_eqn_Debye}) that $m_D^2 = \frac{1}{3}e^2T^2$. Here we leave the Debye mass as a variable and demonstrate the screening effect as a function of the Debye mass. One can connect the Debye mass to the temperature and/or the chemical potential of the plasma via thermal field theory calculations.

\begin{figure}
\centering
    \begin{subfigure}[b]{0.5\textwidth}
        \centering
        \includegraphics[height=2.2in]{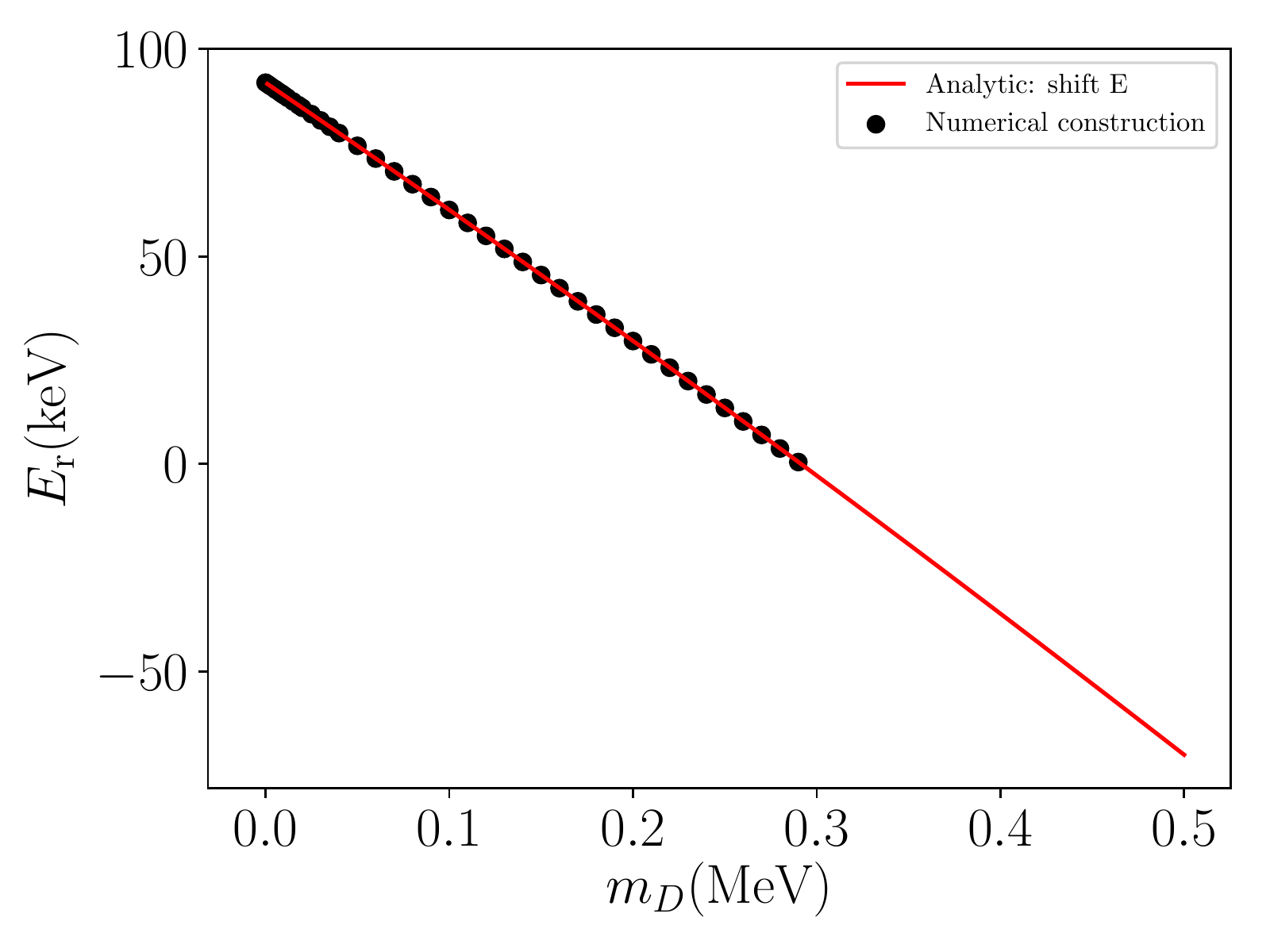}
        \caption{}\label{}
    \end{subfigure}%
    ~
    \begin{subfigure}[b]{0.5\textwidth}
        \centering
        \includegraphics[height=2.2in]{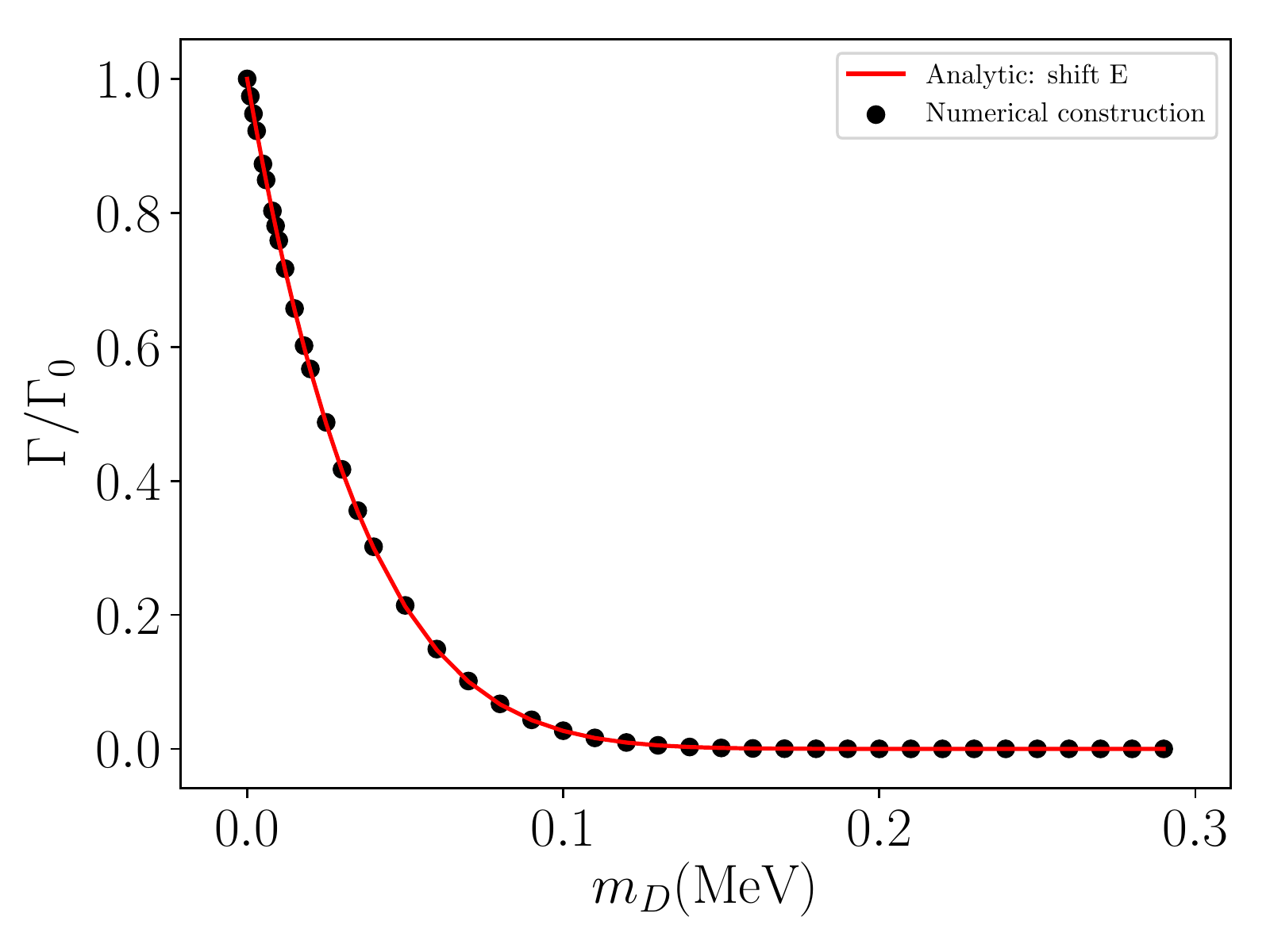}
        \caption{}\label{}
    \end{subfigure}%
\caption[Resonance energy and width as functions of $m_D$.]{Resonance energy and width as functions of $m_D$. The results based on the fully numerical construction are taken from Ref.~\cite{Yao:2016gny}. The fact that the resonance energy turns negative implies the formation of a bound state.}\label{fig_chap2:resonance}
\end{figure}

The expression (\ref{chap2:eqn_amplitude_SC}) is still valid. We only need to modify the Coulomb Green's function by replacing the Coulomb potential in (\ref{chap2:eqn_GC00}) with a Yukawa potential (\ref{chap2:eqn_yukawa}). But the scattering wave associated with a Yukawa potential has no analytic solution. One has to solve the Green's function numerically. We have done the numerical construction of the Green's function in Ref.~\cite{Yao:2016gny} by following the procedure proposed in Ref.~\cite{Kaplan:1996xu}. We also noticed a simple trick later that can greatly simplify the calculation~\cite{Yao:2016cjs}. Here we describe the trick. In the physical process we are considering here, the typical relative momentum between the two $\alpha$ particles at resonance is around $|{\bs p}|=18.5$ MeV. If we consider Debye mass $m_D \ll |{\bs p}|$, we can expand the Yukawa potential in terms of $m_Dr\sim m_D/|{\bs p}|$. The expansion can also be justified from the perspective that we only need the zero-separation to zero-separation propagators. We find
\be \nn
&&G_C^{(+)}(E;0,0; m_D ) \equiv \Big\langle {\bs r}' = 0 \Big|  \frac{1}{ E-H_0 - \frac{ Z_1Z_2\alpha_\ma{em} }{r} e^{-m_Dr}  + i\epsilon }  \Big|  {\bs r} = 0 \Big\rangle \,\,\,\,\,\,\,\,\,\,\, \\ \nn
&=& \Big\langle  0 \Big|  \frac{1}{ E-H_0 - \frac{ Z_1Z_2\alpha_\ma{em} }{r} + Z_1Z_2\alpha_\ma{em} m_D + \ml{O}(m_Dr) + i\epsilon }  \Big|   0  \Big\rangle \\
&\approx& G_C^{(+)}(E+Z_1Z_2\alpha_\ma{em} m_D;0,0; m_D=0) \,.
\ee
So the Coulomb Green's function inside the plasma is approximately equal to that in vacuum, with the energy shifted by $Z_1Z_2\alpha_\ma{em} m_D$. At the same time, we also need to shift the energy argument in the Sommerfeld factor $C_\eta^2(p)$ because it represents $|\psi^{(\pm)}_{\bs p}({\bs r}=0)|^2$, the magnitude squared of the Coulomb scattering wave at the origin. In vacuum, $p^2=ME$ while inside the plasma $p^2 = M(E+Z_1Z_2\alpha_\ma{em} m_D)$. So we modify Eqs.~(\ref{chap2:eqn_phase_shift_EFT}) and (\ref{chap2:eqn_phaseshift_qm}) as
\be
C_\eta^2(\widetilde{p}) p\cot\delta_0 + Z_1Z_2\alpha_\ma{em} M \Re H(\eta(\widetilde{p}))=  -\frac{1}{a} + \frac{1}{2}r_0 p^2  -\frac{1}{4}P_0 p^4 \,,
\ee
where the phase shift $\delta_0 = \delta_0(E,m_D)$ depends on the scattering energy and the Debye mass. $\widetilde{E} = E+Z_1Z_2\alpha_\ma{em} m_D$ and $E$ is the energy in the c.m.~frame. The arguments of both $C_\eta^2$ and $\eta$ are changed accordingly $\widetilde{p} = \sqrt{M\widetilde{E}}$. The effective range expansion parameters have been fitted to the vacuum low-energy scattering data, as shown in the last section. Using them we can compute the resonance energy $E_\ma{r}$ (by setting $\cot\delta_0 = 0$) and width (defined in Eq.~(\ref{chap2:eqn_width})) as functions of the Debye mass $m_D$. The results are shown in Fig.~\ref{fig_chap2:resonance}. The results agree with those from the fully numerical construction of the Green's function with a screened Coulomb potential. We scaled the in-medium width by the vacuum width $\Gamma_0$. As the Debye mass increases, both the resonance energy and the width decrease. This implies that the $^8$Be resonance becomes more stable due to the plasma screening effect. We also analytically continue the equation $\cot\delta_0 = 0$ to the negative energy region, where the solution gives the binding energy of the system. As we can see from the plot, the $^8$Be resonance becomes a bound state when $m_D \gtrsim 0.3$ MeV. A bound $^8$Be does not decay via tunneling and is thus long-lived. 

\subsection{Dynamical Plasma Screening Effect}
\begin{figure}
\centering
\includegraphics[width=4.5in]{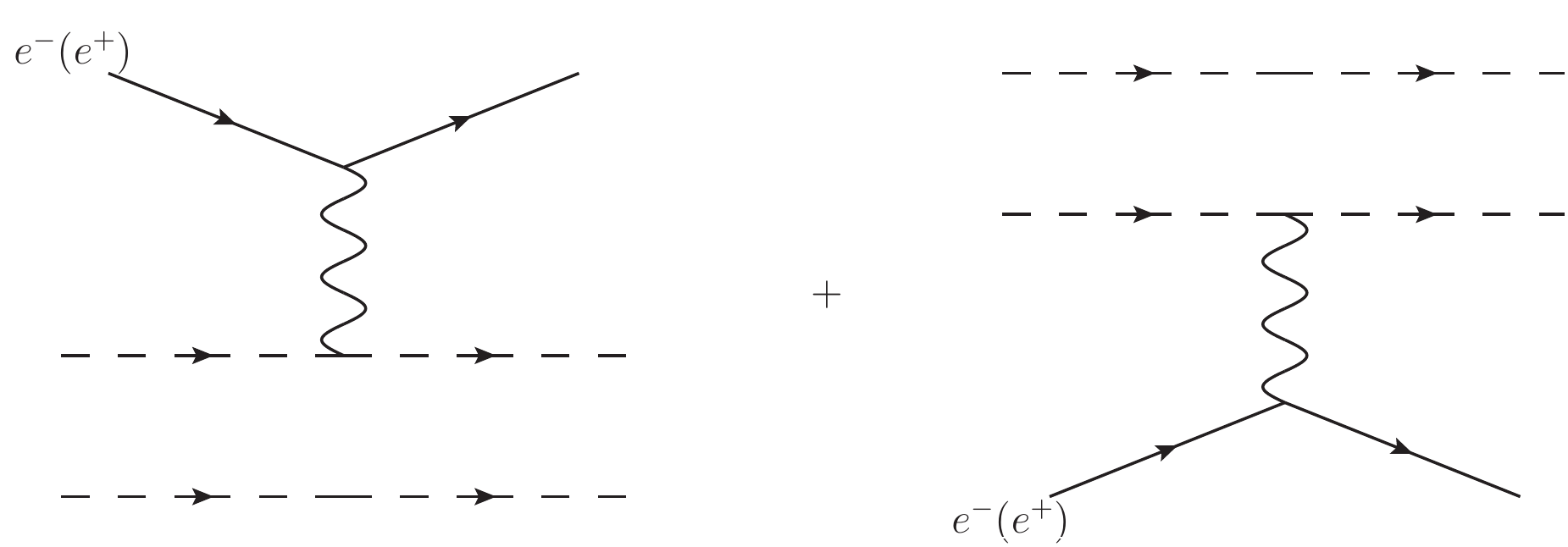}
\caption[Leading order Feynman diagrams contributing to the damping rate of an $\alpha$-pair.]{Leading order Feynman diagrams contributing to the damping rate of an $\alpha$-pair. The dashed line indicates an $\alpha$ particle.}
\label{chap2:fig:damping}
\end{figure}

The suppression of the Coulomb potential is not the only plasma screening effect inside an $e^-e^+\gamma$ plasma. Generally, when a charged particle moves inside a plasma, its momentum will no longer be a constant because it will scatter constantly with medium particles (electrons or positrons). This elastic scattering can change the relative momentum of an $\alpha$ pair but not their total kinetic energy. This leads to an imaginary part in the potential, which describes the damping rate of charged particles with given momenta. Leading order diagrams are shown in Fig.~\ref{chap2:fig:damping}. The photon transferred in the scattering has spatial momenta. Therefore this imaginary potential arises from the imaginary part of the photon polarization tensor (\ref{chap1_eqn_Landau_damp}) with spatial external momenta.

\begin{figure}
\centering
\includegraphics[width=6.0in]{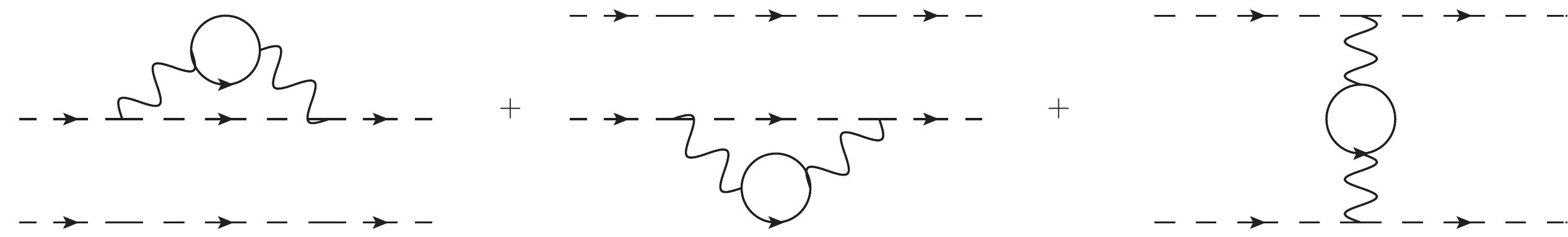}
\caption[Loop corrected $\alpha$-$\alpha$ forward scattering amplitude.]{Loop corrected $\alpha$-$\alpha$ forward scattering amplitude, which contains the lowest order contribution to the imaginary potential. The solid line in loops indicates electron or positron. These diagrams can also represent the loop corrections to the $\alpha$ propagators and the Coulomb exchange interaction.}
\label{chap2:fig:imaginary}
\end{figure}

We can see this connection more clearly in an alternative way. According to unitarity and the cutting rules, the square of the scattering amplitude in Fig.~\ref{chap2:fig:damping}, with all final state summed over, corresponds to the imaginary part of the $\alpha$-$\alpha$ forward scattering amplitude, shown in Fig.~\ref{chap2:fig:imaginary}. The first two diagrams in Fig.~\ref{chap2:fig:imaginary} are the loop corrections of the $\alpha$ propagators and the third diagram is the loop correction of the Coulomb exchange. Inside a plasma, the dominant thermal contributions of these loops come from the hard thermal loops. We have shown in Chapter 1 that the hard thermal loops will generate an imaginary part in the photon self energies when the photon momentum is spatial. This imaginary part in the self-energy will finally result in an imaginary part in the potential between the $\alpha$ pair via the diagrams in Fig.~\ref{chap2:fig:imaginary}. We will extract the imaginary potential by computing the hard thermal loops in these diagrams.

Since we are interested in the temperature range $T<1$ MeV, the electron mass $m_e\sim0.5$ MeV is not small. We have to compute hard thermal loops with an finite electron mass. The time-ordered thermal photon propagator in this case has been calculated~\cite{Escobedo:2008sy},
\be
D_{00}(q_0=0,\bs{q})=\frac{i}{\bs{q}^2+m_D^2} + \frac{16\alpha_\ma{em} g(m_e\beta)}{|\bs{q}|(\bs{q}^2+m_D^2)^2\beta^3} \, .
\ee
The Debye mass is given by 
\be
m_D^2 &=&\frac{8m_e^2}{(2\pi)^2}e^2(2f(m_e\beta)+h(m_e\beta)) \, ,
\ee
where the functions $f$, $g$ and $h$ are defined as
\be
f(m_e\beta) &=& \frac{1}{m_e^2}\int_0^{\infty} \diff k \frac{k^2}{\sqrt{k^2+m_e^2}(e^{\beta\sqrt{k^2+m_e^2}}+1)}=-\sum_{n=1}^{\infty}(-1)^n\frac{K_1(n\beta m_e)}{n\beta m_e} \,\,\,\,\,\,\,\,\,\, \\
h(m_e\beta) &=& \int_0^{\infty} \diff k \frac{1}{\sqrt{k^2+m_e^2}(e^{\beta\sqrt{k^2+m_e^2}}+1)}=-\sum_{n=1}^{\infty}(-1)^nK_0(n\beta m_e)\\
g(m_e\beta) &=& \beta^2\int_0^{\infty} \diff k \frac{k}{e^{\beta\sqrt{k^2+m_e^2}}+1} = m_e\beta\ln{(1+e^{-m_e\beta})} -Li_2(-e^{-m_e\beta}) \, ,
\ee
where $K_0(x)$ and $K_1(x)$ are the modified Bessel functions and $Li_2(x)$ is the dilogarithmic function. For low temperatures $m_e\beta \gg 1$,
these functions are approximated by
\be
m_D^2  &=& 8\alpha_\ma{em} \sqrt{\frac{m_e^3}{2\pi\beta}}e^{-m_e\beta}\left[1+ \ml{O} \left(\frac{1}{m_e\beta}\right) \right],\\
g(m_e\beta) &=&  (m_e\beta +1) e^{-m_e\beta} +\ml{O}(m_e \beta e^{-2m_e \beta}) \, .
\ee
In the limit $m_e \beta\to 0$ we recover the standard hard thermal loop result with a massless electron, $m_D^2 = 4 \pi \alpha_\ma{em} T^2/3$, $g(0) = \pi^2/12$, and the second term in $D_{00}(q_0=0,\bs{q})$ becomes $\frac{\pi m_D^2T}{q(q^2+m_D^2)^2}$.

In the infinite $\alpha$ particle mass approximation, we neglect the kinetic energy term in the $\alpha$ particle propagator. Then each of the first two diagrams in Fig.~\ref{chap2:fig:imaginary} contributes to the $\alpha$ particle (time-ordered) self-energy
\be
\label{eqn:sigma}
i\Sigma_{1(2)} &=& (iZ_{1(2)}e)^2\int \frac{\diff^4q}{(2\pi)^4}\frac{i}{q_0+i\epsilon}D_{00}(q_0,{\bs q})  \nn \\
&=& i(iZ_{1(2)}e)^2\int \frac{\diff^4q}{(2\pi)^4} \bigg[\ml{P}\frac{1}{q_0}-i\pi\delta(q_0)\bigg]D_{00}(q_0,\bs{q}) \nn \\ \nn
&=& -\frac{1}{2}(Z_{1(2)}e)^2\int \frac{\diff^3q}{(2\pi)^3} D_{00}(q_0=0,\bs{q}) \\
&=& iZ_{1(2)}^2\bigg(\frac{1}{2}\alpha_\ma{em} m_D + i\frac{8\alpha_\ma{em}^2g(m_e\beta)T^3}{\pi m_D^2}   \bigg) \, ,
\ee
where in the second line the principle value vanishes because $D_{00}(q_0,\bs{q})=D_{00}(-q_0,\bs{q})$ and the integrand is odd in $q_0$. In the last line a linear divergence that exists in vacuum has been absorbed into the renormalization of the $\alpha$ particle mass. Then we can write the propagator of a single $\alpha$ particle as
\be
\frac{i}{E-\frac{{\bs p}^2}{2M}+i\epsilon + \Sigma_{1(2)}}= \frac{i}{E-\frac{{\bs p}^2}{2M}+i\epsilon +Z_{1(2)}^2\big(\frac{1}{2}\alpha_\ma{em} m_D + i\frac{8\alpha_\ma{em}^2g(m_e\beta)T^3}{\pi m_D^2}   \big)} \, .
\ee
The third diagram modifies the Coulomb exchange potential 
\be
\label{eqn:coulomb}
V_C({\bs r}) &=& i(iZ_1e)(iZ_2e)\int\frac{\diff^3q}{(2\pi)^3}e^{i{\bs q}\cdot{\bs r}} D_{00}(q_0=0,\bs{q})
\nn \\
&=& \frac{Z_1Z_2\alpha_\ma{em}}{ r}e^{-m_Dr} -iZ_1Z_2e^2\int\frac{\diff^3q}{(2\pi)^3}e^{i{\bs q}\cdot{\bs r}}\frac{16\alpha_\ma{em} g(m_e\beta)T^3}{q(q^2+m_D^2)^2} \, ,
\ee
where the first term is the statically screened Coulomb potential and the second term is the dynamical screening contribution. 

The Coulomb Green's function can be constructed from the Lippmann-Schwinger equation as in the previous sections,
\be
\hat{G}^{(+)}_C(E) = \frac{1}{E-\hat{H}_0-\frac{Z_1 Z_2\alpha_\ma{em}}{r} e^{-m_Dr} + \frac{1}{2}( Z_1^2+Z_2^2)\alpha_\ma{em} m_D + iW({\bs r})   +i\epsilon} \, ,
\ee
where
\be
\label{chap2:Wr}
W(r) 
&=&\frac{16\alpha_\ma{em}^2g(m_e\beta)T^3}{\pi m_D^2}\phi(m_Dr, Z_1,Z_2) \\
\label{chap2:phi}
\phi(m_Dr,Z_1,Z_2) &=& 2\int_0^{\infty}\frac{x\diff x}{(1+x^2)^2}\Big(\frac{1}{2}(Z_1^2+Z_2^2) + Z_1 Z_2\frac{\sin{(xm_Dr)}}{xm_Dr} \Big) \, . \,\,\,\,\,\,
\ee
If we recall the definition of the Green's function
\be
\hat{G}^{(+)}_C(E) \frac{1}{E - \hat{H}_0 - V_C + i\epsilon}\,,
\ee
we can see an imaginary part appears in the potential. 

Before we move on to calculations, we consider the behavior of the real and imaginary parts of the potential in the limit $ r\to 0$. In this limit, we can write
\be
\frac{Z_1 Z_2\alpha_\ma{em}}{r} e^{-m_Dr} = \frac{Z_1 Z_2\alpha_\ma{em}}{r} - Z_1 Z_2\alpha_\ma{em} m_D + \ml{O}(r) \,.
\ee
Except for the unscreened Coulomb interaction $Z_1Z_2\alpha/r$, the real contribution from the self energies of $\alpha$ particles combines with the contribution from the static screening of the Coulomb potential to give a negative shift of the potential of $(Z_1+Z_2)^2 \alpha_\ma{em} m_D/2$. It is also easy to see from Eqs.~(\ref{chap2:Wr}) and (\ref{chap2:phi}) that 
$W(0) \propto (Z_1+Z_2)^2$ in the limit $ r\to 0$. This shows that both the real and imaginary parts of the potential vanish at the origin when the two particles have equal and opposite charges. Two oppositely charged particles 
placed at the same point  appear to the plasma like a neutral particle, in which case the plasma will have no effect on their energy.

\begin{figure}
\centering
\includegraphics[width=3in]{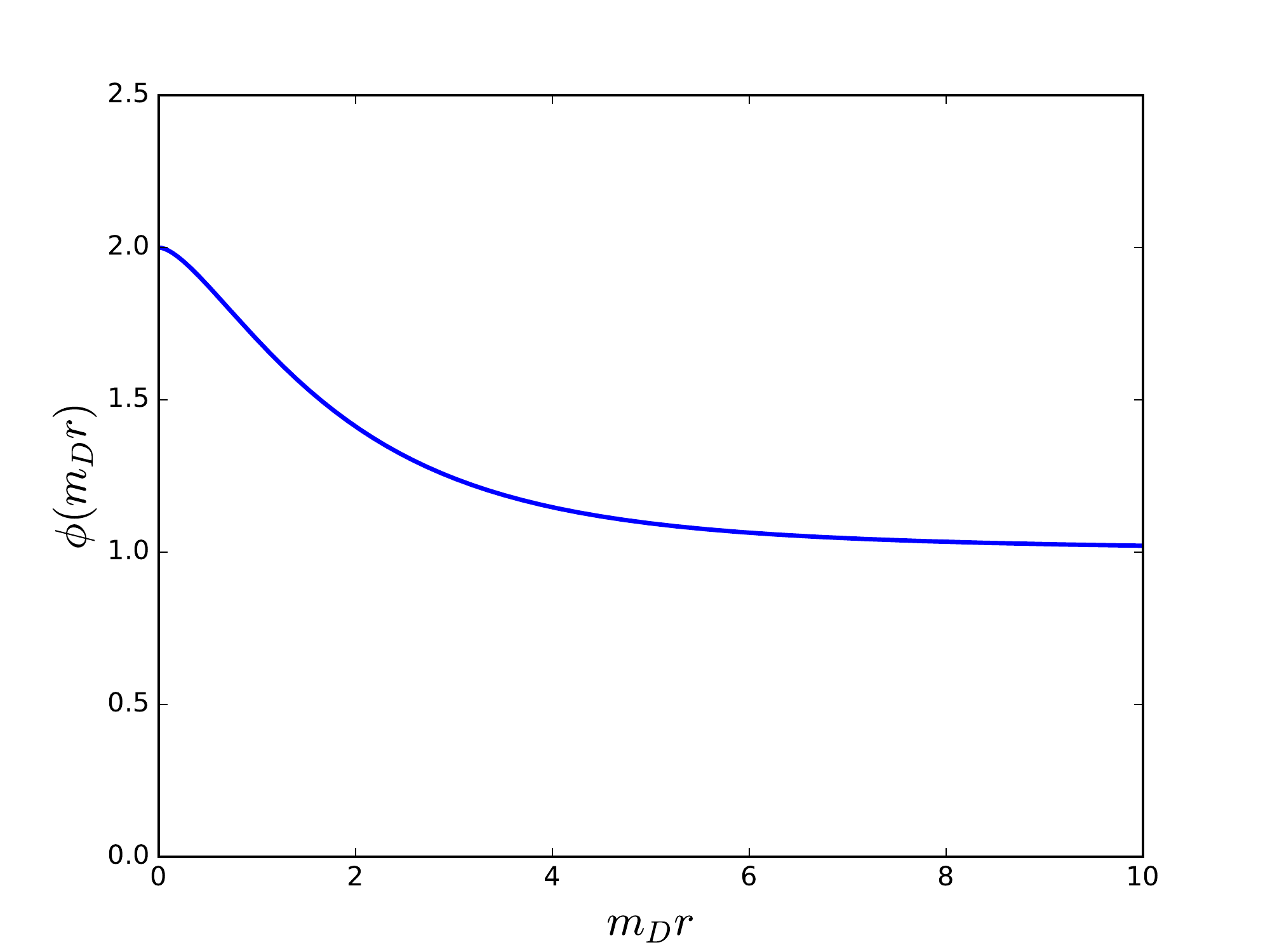}
\caption[The $r$-dependence of $\phi(m_Dr)$.]{The $r$-dependence of $\phi(m_Dr)$, which appears in the imaginary potential. The function is defined by $Z^2\phi(m_Dr) = \phi(m_Dr,Z_1,Z_2)$ for $Z_1=Z_2=Z$.}
\label{fig:phi}
\end{figure}

Henceforth we restrict ourselves to the case $Z_1= Z_2 = Z$, then $\phi(m_Dr,Z,Z) = Z^2 \phi(m_D r)$ and
the function $\phi(m_Dr)$ is plotted in Fig.~\ref{fig:phi}. It can be seen that $\phi(0)=2$ and $\phi(\infty)=1$. When the two 
$\alpha$ particles are far separated, the total damping rate is the sum of the individual damping rate of each $\alpha$ particle while when they are close, the damping rate is doubled due
 to their interactions. 

Finally we compare the relative importance of static versus dynamical screening effects. As argued in the study of static screening effect, the real correction of the unscreened Coulomb potential is about $-Z^2 \alpha_\ma{em} m_D$. When $T\ll m_e$, this scales as $Z^2 \alpha_\ma{em} m_D \sim Z^2 \alpha_\ma{em}^{3/2} (m_e^3 T)^{1/4} e^{-m_e/2T}$. In the same limit,
the coefficient of $\phi(m_Dr)$ in the imaginary part of the potential scales as $ Z^2\alpha_\ma{em}\sqrt{\frac{T^3}{m_e}}$. We see that the static screening is suppressed relative to dynamical screening by $\alpha_\ma{em}^{1/2}(m_e/T)^{5/4} e^{-m_e/2T}$ for $T \ll m_e$. In the opposite limit, $T \gg m_e$, the coefficient of $\phi(m_Dr)$ in $W(r)$ is $Z^2 \alpha_\ma{em} T$, while static screening correction to the potential is $Z^2\alpha_\ma{em} m_D \sim Z^2 \alpha_\ma{em}^{3/2} T$, so static screening is again suppressed relative to dynamical screening by a factor of $\sqrt{\alpha_\ma{em}}$. Thus, in either limit dynamical screening should be expected to be more important. As will be seen in our calculations below, the dynamical screening effect dominates. 

With both the real and imaginary corrections to the Coulomb potential included, the zero-separation to zero-separation propagator is written as
\be \nn
&& G(E,0,0;T) \\ 
\label{eq:green}
&=& \Big\langle {\bs 0} \Big|  \frac{1}{E-\hat{H}_0-\frac{Z^2\alpha_\ma{em}}{r}+2 Z^2\alpha_\ma{em} m_D +iZ^2\frac{32\alpha_\ma{em}^2g(m_e\beta)T^3}{\pi m_D^2} + \ml{O}(r) +i\epsilon}  \Big|  {\bs 0}  \Big\rangle \,,
\ee
where we have expanded the potential to the order $r^0$. The expansion was justified in the previous subsection when we study the static screening effect. We now use the same trick introduced in the previous subsection to calculate the Green's function at finite temperature $G(E,0,0;T)$, which contains both the real and imaginary corrections to the potential. Using Eq.~(\ref{eq:green}), we can write
\be
G(E,0,0;T) = G(\widetilde{E},0,0;T=0) \,,
\ee
where
\be\label{shift}
\widetilde{E} = E + 2 Z^2\alpha_\ma{em} m_D +iZ^2\frac{32\alpha_\ma{em}^2g(m_e\beta)T^3}{\pi m_D^2} \,.
\ee
As in the previous subsection, we obtain the screened Coulomb Green's function in the plasma by analytically continuing the vacuum Coulomb Green's function from $E$ to $\widetilde{E}$. We also need to analytically continue $C_{\eta}^2$ in the same way since the Coulomb wave function is the solution to an analogous analytic continuation of the Schr\"odinger equation. The scattering amplitude in the plasma can be written as
\be
T_{SC}&=&\frac{4\pi}{M}\frac{C_{\widetilde{\eta}}^2e^{2i\sigma_0}}{-\frac{1}{a}+\frac{r_0}{2}p^2- \frac{P_0}{4}p^4-Z^2\alpha_\ma{em} MH(\widetilde{\eta})} \, ,
\ee
where $\widetilde{\eta}$ is computed from $\widetilde{E}$ via
\be
\widetilde{\eta} = \frac{Z_1Z_2\alpha_\ma{em} M}{2 \sqrt{M\widetilde{E}}} \,.
\ee
Then the scattering amplitude squared can be computed at different c.m.~energies and fitted to a Breit-Wigner formula:
\be
\bigg(  \frac{4\pi}{M} \bigg)^2 \bigg| \frac{C_{\widetilde{\eta}}^2e^{2i\sigma_0}}{-\frac{1}{a}+ \frac{r_0}{2}ME-\frac{P_0}{4} M^2E^2-Z^2\alpha_\ma{em} MH(\widetilde{\eta})}  \bigg|^2 =\frac{1}{p^2}\frac{A_0}{(E-E_r)^2+\Gamma^2/4}\,,
\ee
where $E_r$ is the resonance energy and $A_0$ is a constant.  An arbitrary constant $A_0$ shows up in the numerator of our Breit-Wigner parametrization because the potential has an imaginary part which violates unitarity. So the maximum of the amplitude squared no longer corresponds to the unitary limit. The total width $\Gamma$ is the sum of the thermal width, $\Gamma_{\ma{thermal}}$, caused by collisions with medium particles and the intrinsic width, $\Gamma_{\ma{intrinsic}}$, due to the spontaneous decay into two $\alpha$ particles.  $\Gamma_{\ma{intrinsic}}$ 
is defined as the width when only the static screening has been included, which can be calculated similarly by using the Green's function with only a real shift in the energy: $G(E,0,0;T) = G(E+Z^2\alpha_\ma{em} m_D,0,0;T=0)$. 

\begin{figure}
\centering
\includegraphics[height=2.5in]{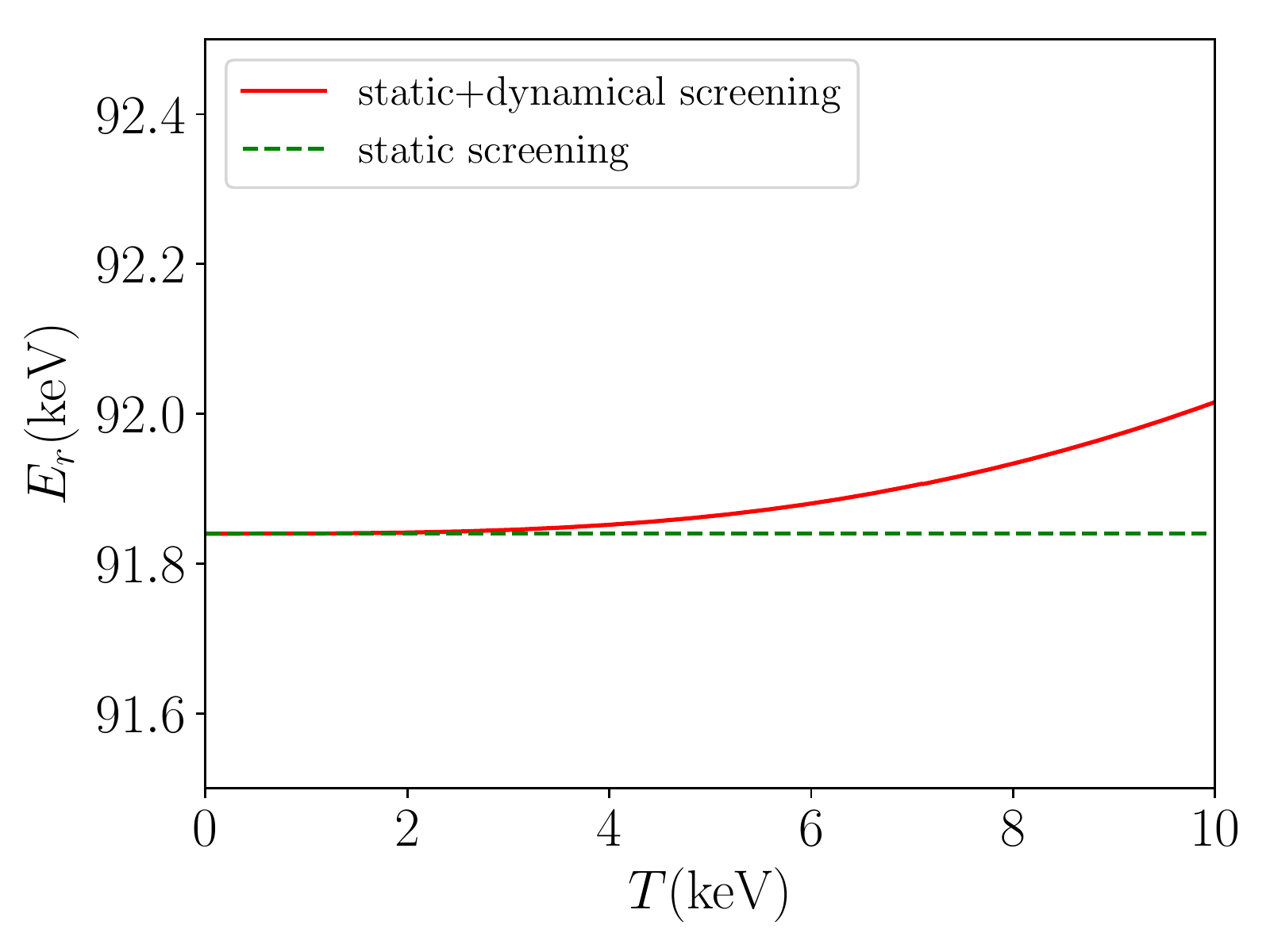}
\caption[The resonance energy, $E_r$, as a function of the temperature, $T$.]{The resonance energy, $E_r$, as a function of the temperature, $T$. The solid red line is the resonance energy with both the static and dynamical screening included and the dotted green line is the energy when only static screening is included.}
\label{fig:resonance_energy}
\end{figure}

\begin{figure}
\centering
    \begin{subfigure}[b]{0.48\textwidth}
        \centering
        \includegraphics[height=2.2in]{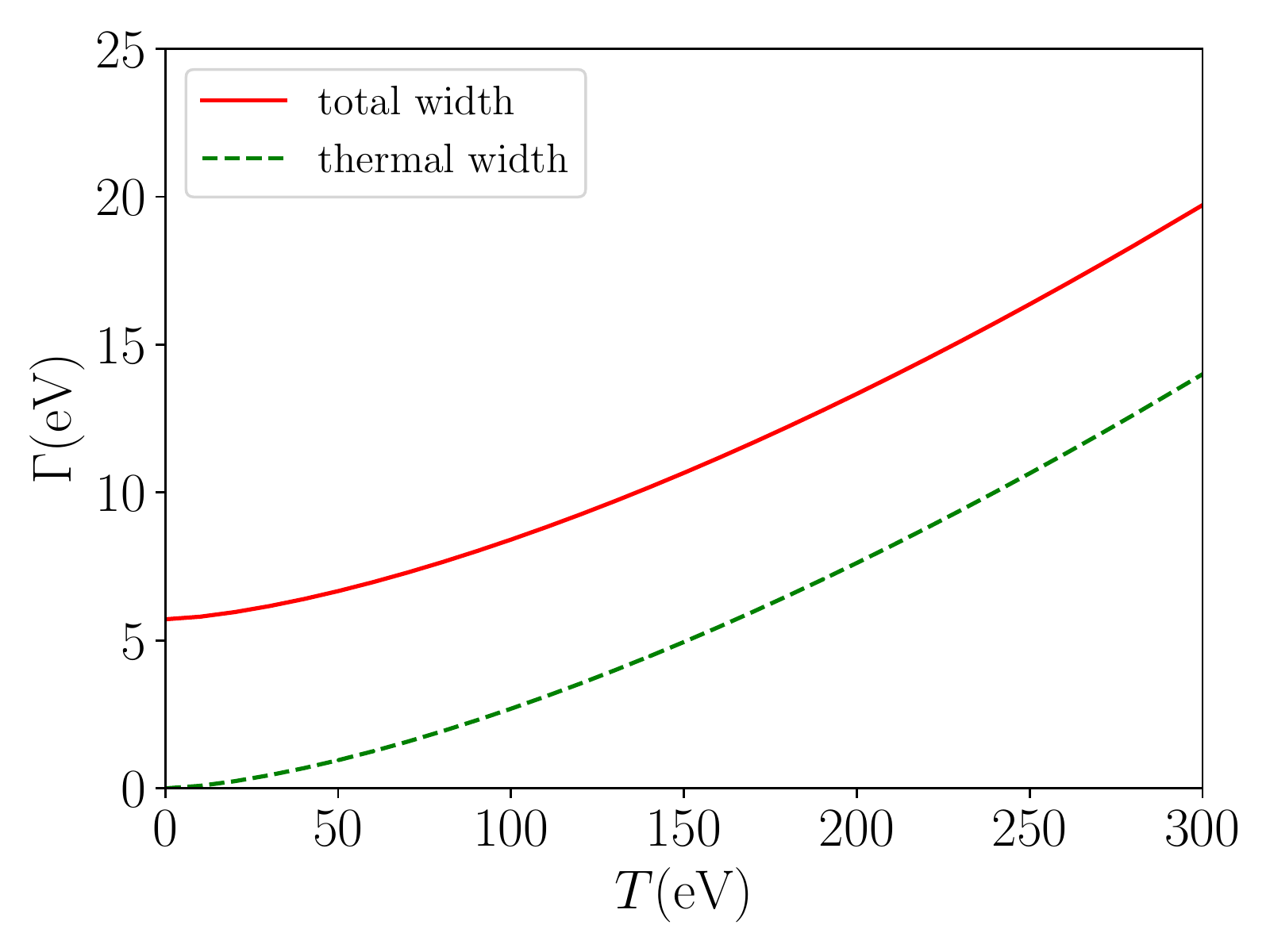}
        \caption{$0 < T < 300$ eV.}\label{}
    \end{subfigure}%
    ~
    \begin{subfigure}[b]{0.48\textwidth}
        \centering
        \includegraphics[height=2.2in]{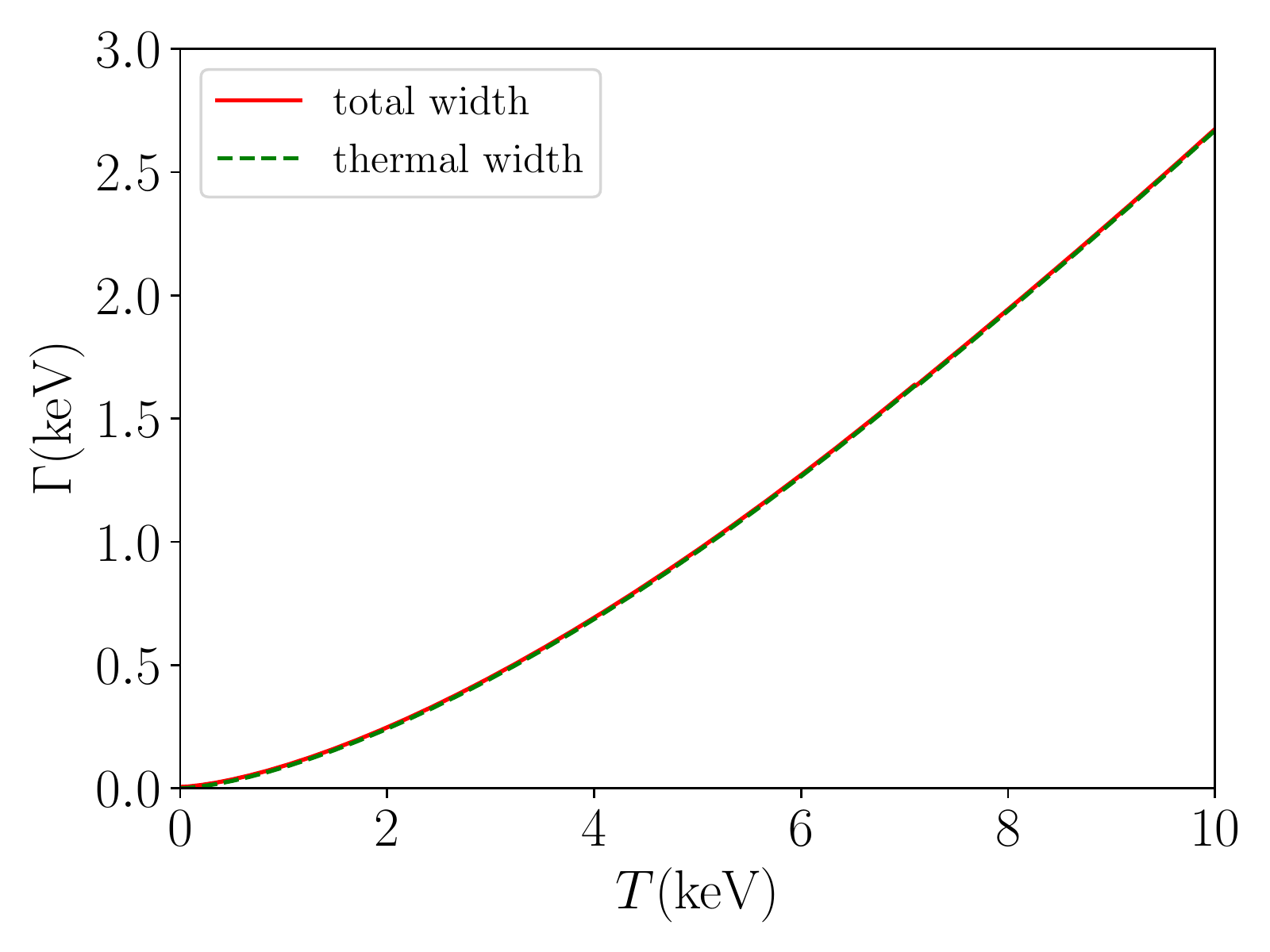}
        \caption{$0 < T < 10$ keV.}\label{}
    \end{subfigure}%
\caption[The total width and the thermal width as a function of plasma temperature, $T$.]{The total width (red solid line) and the thermal width (green dotted line) as a function of plasma temperature, $T$.}
\label{fig:width}
\end{figure}

In Fig.~\ref{fig:resonance_energy} we plot the resonance energy up to $T=10$ keV since in this temperature range the resonance is well described by a Breit-Wigner form. 
The resonance energy with both the static and dynamical screening accounted is marked by the red line and the green dotted line shows the resonance energy when only static screening is considered. The resonance energy increases with the plasma temperature due to the dynamical screening effect. When only static screening is included, the resonance energy decreases with temperature, but only very slightly in the temperature range shown. In Fig.~\ref{fig:width} the solid red line shows the total width of the resonance as a function of temperature and the green dotted line is just the thermal width. 
The total width is an increasing function of the temperature. For temperatures $\sim$ keV, the total width is dominated by the thermal width due to the dynamical screening. As we argued earlier, the dynamical screening dominates over the static screening. 

That the resonance energy increases with the plasma temperature after taking into account the dynamical screening can be understood as follows: The imaginary potential describes the probability loss in the elastic channel as the two $\alpha$ particles move toward each other, which leads to the suppression of the wave function. This suppression is similar to that caused by a repulsive real potential. The imaginary potential and the associated suppression effect increase with the plasma temperature. As a result, it requires a higher kinetic energy to bring the two $\alpha$ particles into the nuclear interaction range. Therefore the resonance energy increases. The resonance also becomes wider. This effect obviously vanishes when $T=0$ and increases with the plasma temperature. Its value is comparable to the intrinsic width $\Gamma_0 = 5.57$ eV when $T \approx 160$ eV, and for temperatures on the order of keV, the thermal width completely dominates the total width of the resonance.

\vspace{0.2in} 
\subsection{Conclusion}
In this section, we calculated the $e^-e^+\gamma$ plasma screening effects on the $^8$Be resonance energy and width. We demonstrate that the dynamical screening effect dominates over the static screening effect. When only static screening is considered, we see the $^8$Be becomes a bound state when the temperature is large enough. After including dynamical screening, the resonance energy and width increase with the plasma temperature, in the opposite way as in the case with only static screening considered. However, this does not contradict with the existence of a bound $^8$Be at high temperature. The explanation is rooted in how we ask questions. We consider a low-energy $\alpha$-$\alpha$ scattering experiment inside an $e^-e^+\gamma$ plasma: We prepare $\alpha$ particles at a given energy when they are far away from each other and shoot them towards each other. In the collision region, an $e^-e^+\gamma$ plasma exists. We measure the final state $\alpha$ particles far away from the collision region. From our measurements, we will see a Breit-Wigner peak in the cross section. We will find that both the peak location and the width increase with the plasma temperature as in the case with both static and dynamical screening considered. This is what we will observe if we prepare and measure only asymptotic states.

But if we ask what happens in the middle time steps, a bound $^8$Be may be generated if the plasma temperature is high enough. On its way out towards our detectors, the bound $^8$Be just gets destroyed by scattering with the plasma constituents. The hotter the plasma is, the easier the bound $^8$Be gets destroyed. After the dissociation, the $\alpha$ pair emitted is correlated and moves back-to-back in the rest frame of the $^8$Be. We could reconstruct the bound $^8$Be if we could measure the in- and out-states of the scatterer that leads to the dissociation and the $\alpha$ pair immediately after the dissociation. But in practice we cannot. There is no way to identify that scatterer in the plasma. Furthermore, on their ways towards our detectors, the momenta of the $\alpha$ pair change significantly due to scattering with the medium constituents and the original correlation is lost. This is why we will never be able to directly observe or easily reconstruct a bound $^8$Be using our detectors.

If we have a plasma whose temperature is changing with time and spatial positions, for a reliable calculation of what we will observe in our detectors, we would need a framework that allows us to keep track of what is happening in the middle time steps. So we need a dynamical evolution equation or a transport equation to describe a system embedded in a medium. As we will discuss in the next chapter, this can be achieved by using the open quantum system framework and the Lindblad equation of the system density matrix.

\singlespacing
\chapter{Quarkonium Transport inside QGP: Theory}
\doublespacing
\vspace{0.2in}
\section{Open Quantum System}
As discussed in the end of last chapter, a natural framework to describe dynamics of particles embedded in a medium is the open quantum system formalism. To elaborate this point, we first consider a simple two-level system in quantum mechanics. We will develop important intuitions on the quarkonium dynamics inside the QGP from this simple example.

\vspace{0.2in}
\subsection{Two-Level System}
We consider a two-level system with a Hamiltonian in the Schr\"odinger picture given by
\be
H &=& H_0 + V(t) \\
H_0 &=& E_0 | 0 \rangle \langle 0| + E_1 | 1 \rangle \langle 1 | \\
V(t) &=& A e^{ i\omega t } | 0 \rangle \langle 1 | + A e^{ - i\omega t } | 1 \rangle \langle 0 | \,,
\ee
in which $A$ is real and controls the interaction strength between the two levels. In the interaction picture
\be
V^{(\ma{int})}(t) \equiv e^{ iH_0t } V(t) e^{ -iH_0t } = A e^{ i (\omega+E_0 - E_1) t } | 0 \rangle \langle 1 | + A e^{ - i (\omega+E_0 - E_1) t } | 1 \rangle \langle 0 | \,.
\ee
For simplicity, we consider the case of resonant absorption and radiation, $\omega = E_1 - E_0$. Let the wave function in the interaction picture be
\be
| \psi(t) \rangle^{(\ma{int})} = c_0(t) |0\rangle + c_1(t) |1\rangle \,,
\ee
where the normalization condition is $|c_0|^2 + |c_1|^2 = 1$. Solving the Schr\"odinger equation in the interaction picture
\be
i\partial_t | \psi(t) \rangle^{(\ma{int})} = V^{(\ma{int})}(t) | \psi(t) \rangle^{(\ma{int})} \,,
\ee
we obtain
\be
\dot{c}_0 &=& -i A c_1 \\
\dot{c}_1 &=& -i A c_0 \,.
\ee
We assume the initial condition is $c_0(0) = 1$, $c_1(0) = 0$. Then the solution is
\be
c_0(t) &=& \cos{At} \\
c_1(t) &=& \sin{At} \,.
\ee
We can obtain the probability of the system being in the ground or the excited state as a function of time, $|c_i(t)|^2$.

Alternatively, we can consider the semi-classical master equation of the system
\be
\dot{P}_0(t) &=& -A^2 P_0(t) + A^2 P_1(t) \\
\dot{P}_1(t) &=& A^2 P_0(t) - A^2 P_1(t) \,.
\ee
The physical meaning is very simple, in an infinitesimal time step $\diff t$, if a particle in the ground state goes to the excited state, the number of ground states will decrease while that of excited states will increase, and similarly if a particle in the excited state decays to the ground state. The transition probability in the time step is given by $A^2\diff t$. We can also solve the master equation under the initial condition $P_0(0)=1$, $P_1(0)=0$.

\begin{figure}
\centering
    \begin{subfigure}[t]{0.5\textwidth}
        \centering
        \includegraphics[height=2.2in]{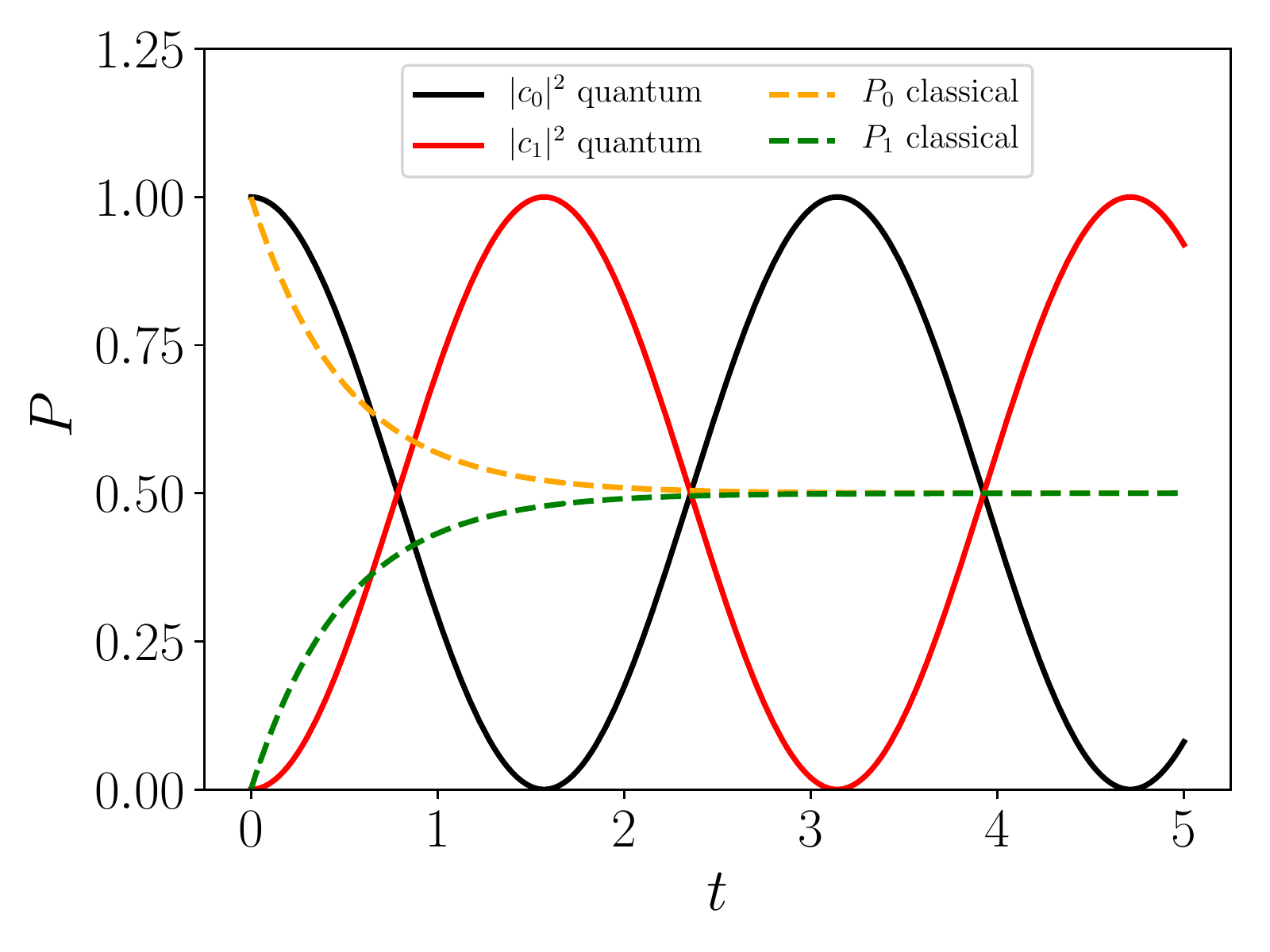}
        \caption{$A=1$.}\label{fig_chap3:P0P1_coherent}
    \end{subfigure}%
    ~
    \begin{subfigure}[t]{0.5\textwidth}
        \centering
        \includegraphics[height=2.2in]{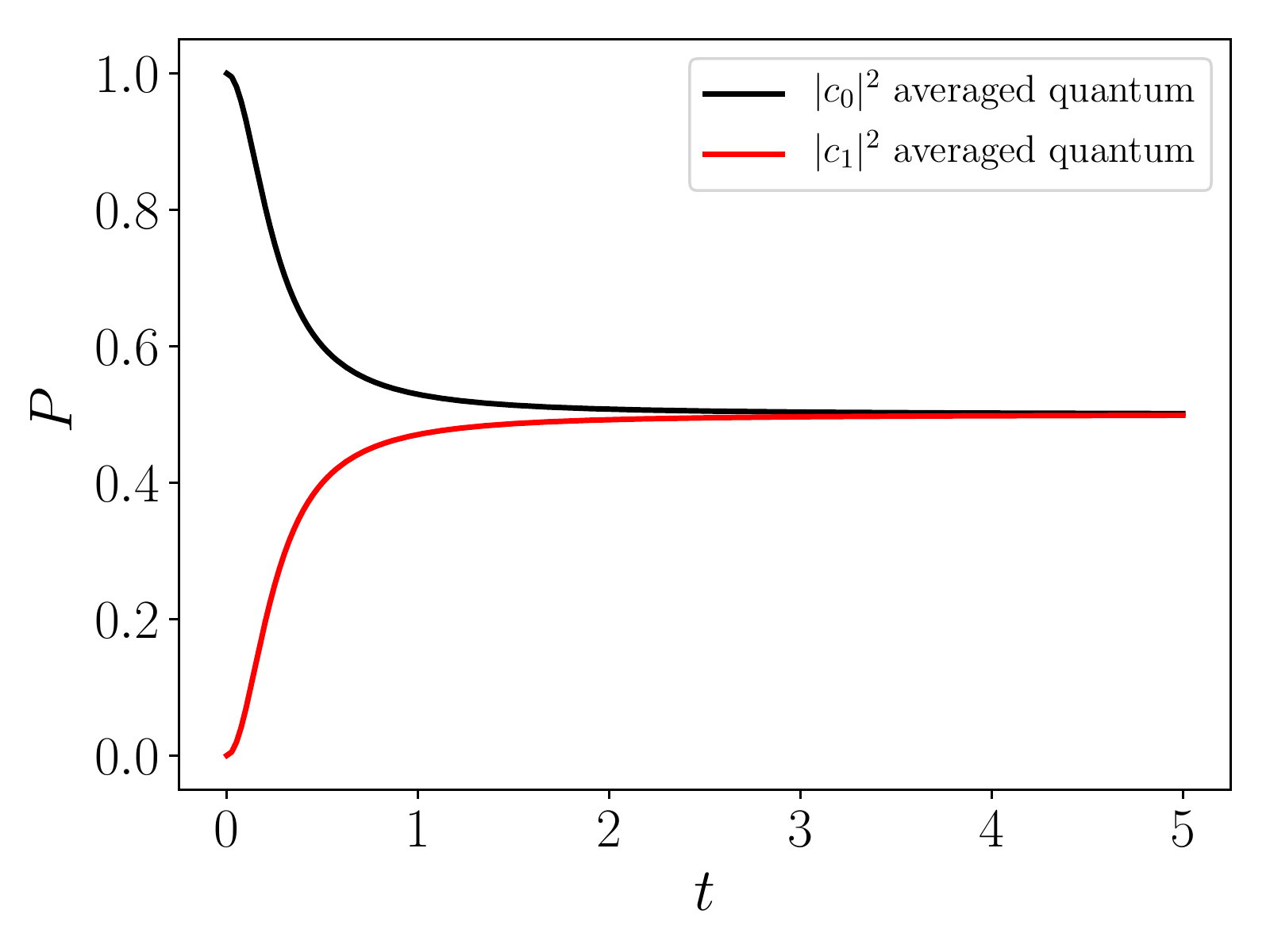}
        \caption{Averaged over $A$.}\label{fig_chap3:P0P1_decoherent}
    \end{subfigure}%
\caption{Probabilities in the ground and excited states.}
\label{fig_chap3:P0P1}
\end{figure}

The results of $|c_i(t)|^2$ in the quantum evolution and $P_i(t)$ in the master equation when $A=1$ (everything is assumed unitless for simplicity) are shown in Fig.~\ref{fig_chap3:P0P1_coherent}. The total probability is conserved for both the quantum evolution $|c_0(t)|^2 + |c_1(t)|^2 = 1$ and the classical master equation $P_0+P_1=1$. Now we find a problem: the probability given by the quantum evolution is oscillating between $0$ and $1$ while that in the classical master equation reaches a plateau after a short time, i.e., the system described by the master equation thermalizes in the general sense. The question we need to answer is how the quantum evolution reduces to the classical evolution. The key to the answer is decoherence. The quantum evolution is based on a closed system so it is unitary without any dissipation or damping. The state $| \psi(t) \rangle^{(\ma{int})}$ is a pure state throughout the evolution and the wave function maintains coherent (the off-diagonal element of the state density matrix is non-zero and imprints crucial information of the state). To destroy the coherence in the state, we have to introduce some randomness into the system. For example, in the Random Matrix Theory (RMT), the system Hamiltonian is represented by a matrix in a fixed basis and the elements of the matrix are random numbers taken from a certain distributions. The eigenstates of the random matrices are random orthogonal vectors. We consider a Hermitian operator
\be
\hat{O} = \sum_i  O_i  | i \rangle \langle i |\,,
\ee
where $| i \rangle$ is the eigenstate of the operator $\hat{O}$: $\hat{O} | i \rangle = O_i  | i \rangle$. Its matrix elements are given by
\be
\langle m | \hat{O} | n \rangle = \sum_i  O_i  \langle m | i \rangle \langle i | n \rangle\,,
\ee
where $| m  \rangle$ and $|n \rangle$ are the eigenstates of random matrices. Since $| m  \rangle$ and $|n \rangle$ are random orthogonal vectors, we expect the averaged matrix elements satisfy
\be
\overline{  \langle m | i \rangle \langle j | n \rangle  } = \frac{1}{D} \delta_{mn}\delta_{ij} \,,
\ee
where $D$ is the dimension of the Hilbert space. So we have
\be
\overline{\langle m | \hat{O} | n \rangle} =  \begin{cases}
      \frac{1}{D}\sum_i O_i, & \text{if}\ m=n \\
      0, & \text{if}\ m\neq n
    \end{cases} \,.
\ee
We see that after the average, the off-diagonal elements vanish. Only diagonal elements survive and become constant, i.e., the system behaves classically and reaches thermalization in the general sense. This is the essential ingredient of the RMT or the Eigenstate Thermalization Hypothesis which can explain thermalization in the general sense. A good review of this topics is given in Ref.~\cite{DAlessio:2016rwt}.

We now will apply this idea of RMT to our two-level system. We assume the interaction strength $A$ satisfies a certain distribution $f(A)$. For each value of $A$, the dynamics of the system is coherent. But what we observe classically is an ensemble average,
\be
\int \diff A f(A) |c_i(t,A)|^2 \,.
\ee
We take a simple distribution
\be
f(A) = \frac{1}{A_0} e^{-A/A_0} \,,
\ee
with $A_0=2$, normalized to unity and calculate the ensemble average. The results of the averaged probabilities are shown in Fig.~\ref{fig_chap3:P0P1_decoherent}. We see that the ensemble average of the coherent evolutions results in a classical evolution.

For the quarkonium evolution inside the QGP, the transition amplitude between the quarkonium and the unbound continuum, depends on the energy of the medium constituents that the quarkonium interacts with. The energy of the medium constituents is given by some distribution such as the thermal distribution. Thus the transition between the bound and unbound $Q\bar{Q}$ has some randomness. The randomness will lead to the decoherence of the quarkonium wave function and thus resulting in the dissociation of quarkonium. The randomness appears because we neglect degrees of freedom in the medium and focus on the system of heavy quark-antiquark pairs. In other words, the system is an open quantum system. In the next subsection, we will briefly review the formalism of open quantum system, which will be applied to study the quarkonium dynamics inside the QGP later.

\vspace{0.2in}
\subsection{General Theory of Open Quantum System}
In this section we briefly review the standard results in open quantum systems, which are covered in many textbooks, see, for example, Ref.~\cite{oqs_book1}. We assume the Hamiltonians of the system and the environment (thermal bath) are given by
\be
H = H_S +H_B + H_I\,,
\ee
where $H_S$ is the system Hamiltonian, $H_B$ is the environment Hamiltonian, and $H_I$ contains the interactions between the system and the environment. The interaction Hamiltonian is assumed to be factorized as follows: $H_I = \sum_{\alpha} O^{(S)}_{\alpha} \otimes O^{(B)}_{\alpha}$ where $\alpha$ denotes all quantum numbers. $O^{(S)}_{\alpha}$ are the system operators while $O^{(B)}_{\alpha}$ are the environment operators. We can assume $\langle O^{(B)}_{\alpha}\rangle \equiv \Tr_B(O^{(B)}_{\alpha}\rho_B) = 0$ because we can redefine $O^{(B)}_{\alpha}$ and $H_S$ by $O^{(B)}_{\alpha} - \langle O^{(B)}_{\alpha}\rangle$ and $H_S + \sum_{\alpha} O^{(S)}_{\alpha} \langle O^{(B)}_{\alpha}\rangle $ respectively. Here $\rho_B$ is the density matrix of the environment. Each part of the Hamiltonian is assumed to be Hermitian.

The von-Neumann equation for the time evolution of the density matrix in the interaction picture is given by
\be
\frac{\diff \rho^{(\ma{int})}(t)}{\diff t} = -i [H^{(\ma{int})}_I(t), \rho^{(\ma{int})}(t)] \,.
\ee 
We will omit the superscript ``(int)" in the following. The symbolic solution is given by
\be
\rho(t) = U(t) \rho(0) U^{\dagger}(t)\,,
\ee
where the evolution operator is
\be
U(t) = \ml{T} \exp\Big\{-i \int_0^t H_I(t') \diff t'  \Big\}\,,
\ee
and $\ml{T}$ is the time-ordering operator.
We assume the interaction is a weak perturbation and expand the evolution operator to the second order in $H_I$.
\be
\label{eqn:2order}
\rho(t) &=& \rho(0) - i \int_0^t  \diff t' [H_I(t'), \rho(0) ] + \int_0^t \diff t_1 \int_0^t \diff t_2 \Big(H_I(t_1)\rho(0)H_I(t_2) \\ \nn
&& - \theta(t_1-t_2)H_I(t_1)H_I(t_2)\rho(0) - \theta(t_2-t_1)\rho(0)H_I(t_1)H_I(t_2)\  \Big) + \ml{O}(H_I^3)\,.
\ee
We shall assume the initial condition is given by
\be
\rho(0) = \rho_S(0) \otimes \rho_B\,,
\ee
where the environment density matrix is assumed to be time-independent. We define 
\be
\label{eqn:corr}
C_{\alpha\beta}(t_1, t_2) \equiv \Tr_B(O^{(B)}_{\alpha}(t_1)O^{(B)}_{\beta}(t_2)\rho_B)\,.
\ee
Then by taking the partial trace over the environment we can obtain the evolution equation of the system $\rho_S(t) \equiv \Tr_B(\rho(t))$
\be\nn
\rho_S(t) &=& \rho_S(0) - i\int_0^t\diff t'\sum_{\alpha} [O^{(S)}_{\alpha}(t'), \rho_S(0)] \Tr_B(O^{(B)}_{\alpha}(t')\rho_B) \\ \nn
&&  + \sum_{\alpha, \beta} \int_0^t \diff t_1 \int_0^t \diff t_2 C_{\alpha\beta}(t_1, t_2)  \Big(  O^{(S)}_{\beta}(t_2)\rho_S(0) O^{(S)}_{\alpha}(t_1)  \\  \nn
&& - \theta(t_1-t_2)O^{(S)}_{\alpha}(t_1)O^{(S)}_{\beta}(t_2)\rho_S(0) \\
&& - \theta(t_2-t_1)\rho_S(0) O^{(S)}_{\alpha}(t_1)O^{(S)}_{\beta}(t_2)    \Big) + \ml{O}(H_I^3)\,.
\ee
Using $\langle O^{(B)}_{\alpha}\rangle = 0$ and inserting four complete sets of the system $ |a\rangle \langle a |$ where $ |a\rangle$ is an eigenstate of $H_S$,  we obtain
\be \nn
\rho_S(t) &=& \rho_S(0) + \sum_{\alpha, \beta} \int_0^t \diff t_1 \int_0^t \diff t_2 C_{\alpha\beta}(t_1, t_2) \sum_{a,b,c,d} \langle a | O^{(S)}_{\beta}(t_2) | b\rangle \langle c | O^{(S)}_{\alpha}(t_1) | d \rangle ^* \\ \nn
&&\Big( |a\rangle\langle b| \rho_S(0) ( |c\rangle\langle d|)^{\dagger}  - \theta(t_1-t_2) ( |c\rangle\langle d|)^{\dagger}|a\rangle\langle b|\rho_S(0) \\
\label{chap2_eqn_pre_lindblad}
&& - \theta(t_2-t_1) \rho_S(0) ( |c\rangle\langle d|)^{\dagger}|a\rangle\langle b| \Big) + \ml{O}(H_I^3)\,.
\ee
Finally defining the Lindblad operator $L_{ab} \equiv |a\rangle \langle b|$ and
\be
\gamma_{ab,cd} (t) &\equiv& \sum_{\alpha, \beta} \int_0^t \diff t_1 \int_0^t \diff t_2 C_{\alpha\beta}(t_1, t_2) \langle a | O^{(S)}_{\beta}(t_2) | b\rangle \langle c | O^{(S)}_{\alpha}(t_1) | d \rangle ^*  \\
\sigma_{ab}(t) &\equiv& \frac{-i}{2} \sum_{\alpha, \beta} \int_0^t \diff t_1 \int_0^t \diff t_2 C_{\alpha\beta}(t_1, t_2) \sign(t_1-t_2) \langle a | O^{(S)}_{\alpha}(t_1) O^{(S)}_{\beta}(t_2) | b\rangle \,,   \,\,\,\,\,\,\,\,\,\,\,\,\,\,
\ee
we obtain the Lindblad equation up to second order in perturbation theory
\be \nn
\rho_S(t) &=& \rho_S(0) -i\sum_{a,b} \sigma_{ab}(t) [L_{ab}, \rho_S(0)]  \\
\label{eqn:lindblad}
&& + \sum_{a,b,c,d} \gamma_{ab,cd} (t) \Big( L_{ab}\rho_S(0)L^{\dagger}_{cd} - \frac{1}{2}\{ L^{\dagger}_{cd}L_{ab}, \rho_S(0)\}  \Big) 
 + \ml{O}(H_I^3)\,.
\ee
The relation $\theta(t)=(1+\sign(t))/2$ has been used when we derive Eq.~(\ref{eqn:lindblad}) from Eq.~(\ref{chap2_eqn_pre_lindblad}). It will be shown in the next section that for quarkonium, the commutator term is a loop correction of the real part of the Hamiltonian (more specifically, the real part of the potential). The anticommutator term describes the dissociation of quarkonium, which can also be thought of as an imaginary part of the potential. The second term on the right hand side of Eq.~(\ref{eqn:lindblad}) represents the recombination contribution.  Using the cyclic property of trace, we can see from Eq.~(\ref{eqn:lindblad}) that the probability is conserved during the evolution: $\Tr{\rho_S(t)} = \Tr{\rho_S(0)}$. This implies that the unbound heavy quark-antiquark pair from the quarkonium dissociation stays as active degrees of freedom of the system and may recombine later in the evolution.

The form of the Lindblad equation is valid up to all orders in the perturbative expansion \cite{Breuer:2002pc}. So the higher-order terms neglected here can also be written in the form of the Lindblad equation. The Lindblad equation cannot be written in the form of a von-Neumann equation because the evolution is non-unitary. Another property of the Lindblad equation is its time-irreversibility. This can be shown from the monotonicity of the relative entropy \cite{Breuer:2002pc}. The relative entropy between two states $\rho$ and $\sigma$ of the whole system (including the system and the environment) is defined as
\be
S( \rho || \sigma ) \equiv \Tr (\rho \ln\rho) - \Tr( \rho \ln \sigma ) \,.
\ee
We assume both states can be written as
\be
\rho &=& \rho_S \otimes \rho_B \\
\sigma &=& \sigma_S \otimes \sigma_B \,.
\ee
The relative entropy has the property of monotonically decreasing under partial trace
\be
S( \rho_S || \sigma_S )  \leq S( \rho || \sigma ) \,.
\ee
Now we assume the bath is in thermal equilibrium and define a steady system state $\rho_S^{\ma{steady}}$ by
\be
\rho_S^{\ma{steady}} = \Tr_B \big\{ U(t) ( \rho_S^{\ma{steady}}\otimes \rho_B^\ma{eq}  )  U^\dagger(t)  \big\} \,.
\ee
If we consider an arbitrary system state $\rho_S(t)$, we find
\be \nn
S( \rho_S(t) || \rho_S^{\ma{steady}} )  &=&  S(   \Tr_B\{ U(t) ( \rho_S(0) \otimes \rho_B^\ma{eq}) U^\dagger(t) \} ||   \Tr_B\{ U(t) ( \rho_S^{\ma{steady}} \otimes \rho_B^\ma{eq}) U^\dagger(t) \}   ) \\ \nn
& \leq &  S(  U(t) ( \rho_S(0) \otimes \rho_B^\ma{eq}) U^\dagger(t)  ||    U(t) ( \rho_S^{\ma{steady}} \otimes \rho_B^\ma{eq}) U^\dagger(t)   ) \\ \nn
&=&  S(   \rho_S(0) \otimes \rho_B^\ma{eq}  ||     \rho_S^{\ma{steady}} \otimes \rho_B^\ma{eq}   ) \\ 
&=& S(   \rho_S(0)   ||     \rho_S^{\ma{steady}}   ) \,.
\ee
So in general the evolution of $\rho_S(t)$ is time-irreversible even when the Hamiltonian of the underlying theory $H$ is time-reversible: $[H, T]=0$ where $T$ is the anti-unitary time reversal operator. The equal sign is achieved when $\rho_S(t) = \rho_S(0)$, which means the system is a steady state. 
The partial trace over the environment can be thought of as an average over different environment configurations. Though the dynamics involving each configuration is governed by a time-reversible theory and a unitary evolution, after averaging, the dynamics becomes time-irreversible and non-unitary. A general discussion of the transition from time-reversible microdynamics to time-irreversible transport has been given in Ref.~\cite{Rau:1995ea} using the projection method.

\vspace{0.2in}
\section{Derivation of Boltzmann Transport Equation}
In this section, we will apply the Lindblad equation to the theory of pNRQCD and derive the in-medium transport equation of quarkonium.
The interaction part of the Hamiltonian of the theory is given in Eq.~(\ref{eq:lagr2}) but only the singlet-octet transition is relevant for the dissociation and recombination of quarkonium at lowest order in the coupling constant. The octet-octet interaction governs the dynamical evolution of unbound heavy quarks and antiquarks and thus will only be relevant to the transport of open heavy quarks at lowest order in the coupling constant. We will neglect the octet-octet interaction when deriving the quarkonium transport equation. The minus sign in the Hamiltonian is of no importance at the order $\ml{O}(H_I^2)$. The weak coupling expansion in $H_I$ is valid because the quarkonium size is small $rT \sim \frac{T}{Mv} \lesssim v$ in our power counting. This is true in both perturbative and non-perturbative constructions of pNRQCD under the assumed hierarchy of scales. 

To use the Lindblad equation derived in the last section, we write $H_I$ as $\sum_{\alpha} O^{(S)}_{\alpha} \otimes O^{(B)}_{\alpha}$ where
\be \nn
O^{(S)}_{\alpha} &\rightarrow& \langle S(\bs R, t) | r_i | O^a(\bs R, t)\rangle  + \langle O^{a}(\bs R, t) | r_i | S(\bs R, t)\rangle \\
O^{(B)}_{\alpha} &\rightarrow& \sqrt{\frac{T_F}{N_C}}g E_i^{a}(\bs R, t)\,.
\ee
The sum over $\alpha$ in Eq.~(\ref{chap2_eqn_pre_lindblad}) means
\be
\sum_{\alpha} \rightarrow \int \diff^3 R \sum_i \sum_a \,.
\ee
The complete set $|a \rangle $ used to construct the Lindblad operators $|a\rangle \langle b|$ are
\be \nn
|\bs k, nl, 1\rangle &=& a^{\dagger}_{nl}(\bs k) | 0 \rangle \\ \nn
 | {\bs p}_{\ma{cm}}, {\bs p}_{\ma{rel}} ,1\rangle  &=& b^{\dagger}_{{\bs p}_{\ma{rel}}}({\bs p}_{\ma{cm}}) | 0 \rangle\\
 | {\bs p}_{\ma{cm}}, {\bs p}_{\ma{rel}}, a\rangle  &=& c^{a\dagger}_{{\bs p}_{\ma{rel}}}({\bs p}_{\ma{cm}})  | 0 \rangle\,,
\ee
where $1$ denotes a color singlet and $a$ is the color index of an octet. These states are a bound color singlet (quarkonium) with momentum $\bs k$ and quantum number $nl$, an unbound color singlet with c.m.~momentum ${\bs p}_{\ma{cm}}$ and relative momentum ${\bs p}_{\ma{rel}}$ and an unbound color octet with c.m.~momentum ${\bs p}_{\ma{cm}}$, relative momentum ${\bs p}_{\ma{rel}}$ and color $a$ respectively. The unbound singlet state will not be used in our current calculation because at the order we are working, an unbound singlet cannot form a bound singlet by radiating out one gluon, only unbound octet can do so.

We are interested in the bound state evolution. Therefore our basic strategy is to study the time evolution of $\langle {\bs k}_1, n_1l_1, 1 | \rho_S(t) | {\bs k}_2, n_2l_2, 1\rangle $ by sandwiching Eq.~(\ref{eqn:lindblad}) between $\langle {\bs k}_1, n_1l_1, 1 |$ and $ | {\bs k}_2, n_2l_2, 1\rangle $. To obtain the evolution equation of the semi-classical phase space distribution function, we will take the Wigner transform of the density matrix
\be
\label{eqn:wigner}
f_{nl}({\bs x}, {\bs k}, t) \equiv \int\frac{\diff^3k'}{(2\pi)^3} e^{i {\bs k}'\cdot {\bs x} } \langle  {\bs k}+\frac{{\bs k}'}{2}, nl,1   | \rho_S(t)  |   {\bs k}-\frac{{\bs k}'}{2} , nl, 1\rangle \,.
\ee
We will extract the linear dependence on $t$ of $\gamma_{ab,cd}(t)$ and $\sigma_{ab}(t)$ terms in Eq.~(\ref{eqn:lindblad}) and then take time derivative at $t=0$ on both sides of Eq.~(\ref{eqn:lindblad}). The double time integrals are simplified by assuming the Markovian approximation, i.e., the upper limit of the time integrals is large and can be taken to be infinity, $t\rightarrow\infty$. The Markovian approximation is valid when the environment correlation time is much smaller than the relaxation time of the system. The former is roughly given by $1/T$ while the latter can be estimated by the inverse of the dissociation rate. The dissociation rate is $\sim(grT)^2T \lesssim \alpha_s v^2T$ in our power counting and $\alpha_s$ is at the scale $Mv$. So our assumed hierarchy of scales $M\gg Mv \gg Mv^2 \gtrsim T \gtrsim m_D$, translates into a separation of time scales of the environment and the system, and thus justifies the Markovian approximation. The Markovian approximation leads to a coarse grained evolution, in which dynamics with finer time scales than the system relaxation time is not resolved. Therefore the $t\rightarrow0$ limit in the time derivative we will do later and the $t\rightarrow\infty$ limit in the integral are not contradictory. The Markovian approximation also means that there is no memory effect \cite{Breuer:2002pc}. The absence of memory effects is reflected in the Boltzmann transport equation in the assumption of molecular chaos, namely that the correlation between particles generated from their previous collisions are completely forgotten in the next collision. We will use our assumed hierarchy of scales to justify the assumption of molecular chaos later. Under the Markovian assumption, $t\rightarrow\infty$, the double time integrals give two delta functions in energy. When the two delta functions correspond to the same energy conservation, one can write them as one delta function multiplied by the time length $t$. This is how we extract the linear dependence on $t$. This trick is also used in the derivation of Fermi's golden rule.

We will consider the three terms on the right hand side of Eq.~(\ref{eqn:lindblad}), when they are sandwiched by $\langle {\bs k}_1, n_1l_1, 1 |$ and $ | {\bs k}_2, n_2l_2, 1\rangle $. We will show detailed derivations in the following three subsections.

\vspace{0.2in}
\subsection{$-i\sigma_{ab} [L_{ab},\rho_S]$ Term}
To compute $\sum_{a,b}\sigma_{ab} L_{ab}$, we first note that 
\be \nn
 && \bigg( \sum_{a,b}\frac{-i}{2}\int_0^t \diff t_1 \int_0^t  \diff t_2 \sum_{\alpha,\beta} C_{\alpha\beta}(t_1,t_2) \theta(t_1-t_2) \langle a | O^{(S)}_{\alpha}(t_1) O^{(S)}_\beta(t_2) | b \rangle L_{ab}\bigg)^\dagger\\ \nn
 &=& \sum_{a,b}\frac{i}{2}\int_0^t \diff t_1 \int_0^t  \diff t_2 \sum_{\alpha,\beta} C_{\beta\alpha}(t_2,t_1) \theta(t_1-t_2) \langle b |  O^{(S)}_\beta(t_2) O^{(S)}_{\alpha}(t_1) | a \rangle L_{ba}\\ 
 \label{eqn:sigma_ab_theta}
  &=& \sum_{a,b}\frac{-i}{2}\int_0^t \diff t_1 \int_0^t  \diff t_2 \sum_{\alpha,\beta} C_{\alpha\beta}(t_1,t_2) \big(-\theta(t_2-t_1)\big) \langle a | O^{(S)}_{\alpha}(t_1) O^{(S)}_\beta(t_2) | b \rangle L_{ab}\,, \,\,\,\,\,\,\,\,\,\,\,\,\,\,\,
\ee
where in the last line we flipped $\alpha\leftrightarrow\beta$, $\ t_1\leftrightarrow t_2$ and $\ |a\rangle \leftrightarrow |b\rangle$. We can split $\sign(t_1-t_2)$ into $\theta(t_1-t_2)$ and $-\theta(t_2-t_1)$ in $\sum_{a,b}\sigma_{ab} L_{ab}$ and just need to compute the $\theta(t_1-t_2)$ term. The $-\theta(t_2-t_1)$ term is given by the Hermitian conjugate of the $\theta(t_1-t_2)$ term, as shown in Eq.~(\ref{eqn:sigma_ab_theta}). Therefore $\sum_{a,b}\sigma_{ab}L_{ab}$ is Hermitian and this term can be thought of as a correction to the system Hamiltonian. To carry out the calculation explicitly, we use the definitions of different Green's functions in the real-time formalism of thermal field theory, introduced in Chapter 1, and write
\be  \nn
&&D^{>\,ab}_{\mu\nu}({\bs R}_1,t_1; {\bs R}_2, t_2)\theta(t_1-t_2) \\
&=& \big( D^{R\,ab}_{\mu\nu}({\bs R}_1,t_1; {\bs R}_2, t_2) + D^{<\,ab}_{\mu\nu}({\bs R}_1,t_1; {\bs R}_2, t_2)  \big) \theta(t_1-t_2)\,.
\ee
Then we can replace $C_{\alpha\beta}(t_1,t_2)$ in the first line of Eq.~(\ref{eqn:sigma_ab_theta}), due to the $\theta(t_1-t_2)$, with 
\be \nn
&& C_{\alpha\beta}(t_1,t_2) = C_{{\bs R}_1i_1a_1,{\bs R}_2i_2a_2}(t_1,t_2) = \frac{T_F}{N_C}g^2\langle  E_{i_1}^{a_1}({\bs R}_1, t_1)  E_{i_2}^{a_2}({\bs R}_2, t_2)  \rangle_T \\ \nn
&\rightarrow & \frac{T_F}{N_C}g^2 \delta^{a_1a_2} \int\frac{\diff^4q}{(2\pi)^4}  e^{-iq_0(t_1-t_2) + i\bs q\cdot({\bs R}_1-{\bs R}_2)}  \\ \nn
&& (q_0^2\delta_{i_1i_2} -q_{i_1}q_{i_2}) \bigg[ \frac{i}{q_0^2 -{\bs q}^2 + i\sign(q_0)\epsilon} +   n_B(q_0)(2\pi) \sign(q_0) \delta(q_0^2-{\bs q}^2)  \bigg]\\ \nn
&=&  \frac{T_F}{N_C}g^2 \delta^{a_1a_2} \int\frac{\diff^4q}{(2\pi)^4}  e^{-iq_0(t_1-t_2) + i\bs q\cdot({\bs R}_1-{\bs R}_2)}  \\ 
\label{eqn:real1}
&& (q_0^2\delta_{i_1i_2} -q_{i_1}q_{i_2})  \bigg[ \frac{i}{q_0^2 -{\bs q}^2 + i\epsilon} +   n_B(|q_0|)(2\pi)  \delta(q_0^2-{\bs q}^2)  \bigg]\,.
\ee
The term inside the square brackets in the last line is the time-ordered thermal propagator in momentum space.

We consider $|a\rangle = |{\bs k}_1, n_1l_1, 1\rangle$ and $|b\rangle = | {\bs k}_2, n_2l_2, 1 \rangle$. We omit other cases here because we are interested in the correction in the Hamiltonian relevant to the bound singlet. Then we can compute
\be  \nn
&& \theta(t_1-t_2) \langle a | O^{(S)}_{\alpha}(t_1) O^{(S)}_\beta(t_2) | b \rangle  \\ \nn
&=&  \theta(t_1-t_2) \langle {\bs k}_1, n_1l_1, 1|   \langle S({\bs R}_1, t_1) | r_{i_1} | O^{a_1}({\bs R}_1, t_1) \rangle 
 \langle O^{a_2}({\bs R}_2, t_2) | r_{i_2} | S({\bs R}_2, t_2) \rangle |  {\bs k}_2, n_2l_2, 1 \rangle \\ \nn
&=&  \delta^{a_1a_2}  \int \frac{\diff^3p_\ma{cm} }{(2\pi)^3}\int \frac{\diff^3p_\ma{rel} }{(2\pi)^3} \langle \psi_{n_1l_1} | r_{i_1} | \Psi_{{\bs p}_\ma{rel}} \rangle
\langle \Psi_{{\bs p}_\ma{rel}} | r_{i_2} |  \psi_{n_2l_2} \rangle \\ 
&&  \theta(t_1-t_2) e^{iE_{p}(t_2-t_1) - i {\bs p}_\ma{cm}\cdot ({\bs R}_2- {\bs R}_1)}  e^{-iE_{k_2}t_2 + i{\bs k}_2\cdot {\bs R}_2} e^{iE_{k_1}t_1 - i{\bs k}_1\cdot {\bs R}_1}  \,,
\ee
where $E_p = \frac{{\bs p}_{\ma{rel}}^2}{M}$. This can be written as 
\be \nn
&& \theta(t_1-t_2) \langle a | O^{(S)}_{\alpha}(t_1) O^{(S)}_\beta(t_2) | b \rangle \\ \nn
 &=&   \langle \psi_{n_1l_1} | r_{i_1}    \int \frac{\diff^4p_\ma{cm} }{(2\pi)^4}\int \frac{\diff^3p_\ma{rel} }{(2\pi)^3}  \frac{i  | \Psi_{{\bs p}_\ma{rel}} \rangle  \langle \Psi_{{\bs p}_\ma{rel}} |}{p_{\ma{cm}}^0-E_{p}+i\epsilon}   r_{i_2} |  \psi_{n_2l_2} \rangle \\ 
 \label{eqn:real2}
&&  \delta^{a_1a_2} e^{ip_\ma{cm}^0(t_2-t_1) - i {\bs p}_\ma{cm}\cdot ({\bs R}_2- {\bs R}_1)}  e^{-iE_{k_2}t_2 + i{\bs k}_2\cdot {\bs R}_2} e^{iE_{k_1}t_1 - i{\bs k}_1\cdot {\bs R}_1}\,.
\ee
It should be noted that $p_{\ma{cm}}^0$ here does not represent the c.m.~energy of the octet. In fact, it is the total energy of the composite octet particle, $p_{\ma{cm}}^0 = \frac{{\bs p}_{\ma{cm}}^2}{4M} + \frac{{\bs p}_{\ma{rel}}^2}{M} = \frac{{\bs p}_{\ma{rel}}^2}{M} + \ml{O}(Mv^4)$.

\begin{figure}
\centering
\includegraphics[width=0.6\linewidth]{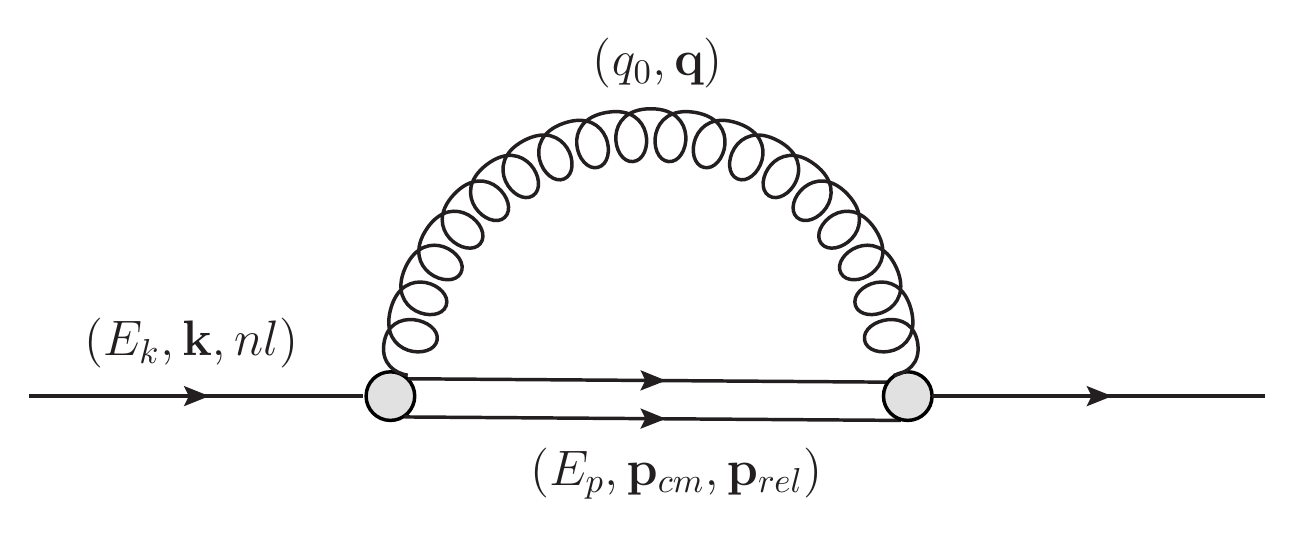}
\caption[Loop correction (self-energy) of the singlet field.]{Loop correction (self-energy) of the singlet field. Single solid line indicates the bound singlet state while the double solid lines represent the unbound octet state.}
\label{fig:loop_singlet}
\end{figure}

To simplify the expression, we make the Markovian approximation $t\rightarrow\infty$. Then integrating $t_1$ and $t_2$ will give two $\delta$-functions in energy. Plugging Eqs.~(\ref{eqn:real1}) and (\ref{eqn:real2}) into $\sum_{a,b}\sigma_{ab}L_{ab}$ and integrating over $t_1$, $t_2$, ${\bs R}_1$ and ${\bs R}_2$ we find 
\be  \nn
&&\sum_{a,b}\sigma_{ab} L_{ab} =\frac{1}{2}    \bigg\{ -i \sum_{n_1,l_1}\sum_{n_2,l_2} \sum_{i_1,i_2}\int\frac{\diff^3k_1}{(2\pi)^3}\int\frac{\diff^3k_2}{(2\pi)^3}\int\frac{\diff^4q}{(2\pi)^4} 
\int\frac{\diff^4 p_\ma{cm}}{(2\pi)^4} \int\frac{\diff^3 p_\ma{rel}}{(2\pi)^3} \\ \nn
&&  \frac{T_F}{N_C}(N_C^2-1) g^2  (q_0^2 \delta_{i_1i_2} -q_{i_1}q_{i_2}) \Big( \frac{i}{q_0^2 - {\bs q}^2 + i\epsilon} +   n_B(|q_0|)(2\pi)  \delta(q_0^2-{\bs q}^2)  \Big) \\ \nn
&& (2\pi)^3\delta^3 ({\bs k}_1-{\bs p}_\ma{cm} - {\bs q}) (2\pi)^3\delta^3 ({\bs k}_2-{\bs p}_\ma{cm} - {\bs q}) \\\nn
&& (2\pi)\delta(E_{k_1}-p_\ma{cm}^0-q^0) (2\pi)\delta(E_{k_2}-p_\ma{cm}^0-q^0)\\
&& \langle \psi_{n_1l_1} | r_{i_1}  \frac{i  | \Psi_{{\bs p}_\ma{rel}} \rangle  \langle \Psi_{{\bs p}_\ma{rel}} |}{p_{\ma{cm}}^0-E_{p}+i\epsilon}   r_{i_2} |  \psi_{n_2l_2} \rangle 
L_{| {\bs k}_1,n_1l_1,1 \rangle \langle  {\bs k}_2,n_2l_2,1   |} +\ma{h.c.}\bigg\} + \cdots\,,
\ee
where contributions from the case where $|a\rangle$ and $|b\rangle$ are the unbound singlets or octets are omitted here. 
The two time integrals give a product of two delta functions in energy $\delta(\omega_1)\delta(\omega_2)$, where $\omega_i=E_{k_i}-p_\ma{cm}^0-q^0 = -|E_{n_il_i}|-p_\ma{cm}^0-q^0$ for $i=1,2$. The $\delta$-functions in energy and momentum lead to ${\bs k}_1={\bs k}_2={\bs k}$, $n_1=n_2=n$ and $l_1=l_2=l$ (we assume no degeneracy in the bound state eigenenergy beyond that implied by rotational invariance). So we have $\omega_1=\omega_2 =\omega$ (because the contributions from terms with $n_1\neq n_2$ or $l_1 \neq l_2$ vanish). We interpret one factor of $2\pi\delta(\omega)$ to be the time interval, so the double time integral is interpreted as follows
\be
\int_0^t\diff t_1\int_0^t \diff t_2 e^{i\omega t_1}e^{-i\omega t_2} = \frac{4\sin^2(\omega t/2)}{\omega^2} \xrightarrow{t\to\infty} t2\pi\delta(\omega)\,.
\ee
This argument also applies in the next two subsections. So for the part relevant to the bound singlet, we have
\be \nn
\sum_{a,b}\sigma_{ab}L_{ab} &\rightarrow& t \sum_{n,l} \int\frac{\diff^3k}{(2\pi)^3} \Re  \bigg\{ -i g^2 C_F  \sum_{i_1,i_2} \int\frac{\diff^4q}{(2\pi)^4} 
\int\frac{\diff^4 p_\ma{cm}}{(2\pi)^4} \int\frac{\diff^3 p_\ma{rel}}{(2\pi)^3} \\ \nn
&& (q_0^2 \delta_{i_1i_2} -q_{i_1}q_{i_2}) \Big( \frac{i}{q_0^2 - {\bs q}^2 + i\epsilon} +   n_B(|q_0|)(2\pi)  \delta(q_0^2-{\bs q}^2)  \Big) \\ \nn
&& \langle \psi_{nl} | r_{i_1}  \frac{i  | \Psi_{{\bs p}_\ma{rel}} \rangle  \langle \Psi_{{\bs p}_\ma{rel}} |}{p_{\ma{cm}}^0-E_{p}+i\epsilon}   r_{i_2} |  \psi_{nl} \rangle\\ 
\label{chap3_eqn_sigmaL}
&& (2\pi)^3\delta^3 ({\bs k}-{\bs p}_\ma{cm} - {\bs q}) (2\pi)\delta(E_{k}-p_\ma{cm}^0-q^0) 
\bigg\} L_{| {\bs k},nl,1 \rangle \langle  {\bs k},nl,1   |} \,. \,\,\,\,
\ee
The part inside the curly bracket gives the one-loop self-energy of the bound color singlet (loop correction of the propagator), which can be calculated as usual by the standard quantum field theory perturbative technique. The one-loop Feynman diagram of the self-energy of the bound color singlet is shown in Fig.~\ref{fig:loop_singlet}. Only the real part of the correction shows up in the $\sum_{a,b}\sigma_{ab}L_{ab}$ term.

We see that the expression (\ref{chap3_eqn_sigmaL}) is an operator $L_{| {\bs k},nl,1 \rangle \langle  {\bs k},nl,1   |} = | {\bs k},nl,1 \rangle \langle  {\bs k},nl,1   |$, which is diagonal in the bound state space. In addition to the term, $\sum_{a,b}\sigma_{ab}L_{ab}$ contains another two terms, which are operators of the form $ | {\bs p}_{\ma{cm}}, {\bs p}_{\ma{rel}} ,1 \rangle \langle  {\bs p}_{\ma{cm}}, {\bs p}_{\ma{rel}} ,1  |$ and $ | {\bs p}_{\ma{cm}}, {\bs p}_{\ma{rel}} , a \rangle \langle  {\bs p}_{\ma{cm}}, {\bs p}_{\ma{rel}} , a  |$. These two terms are not shown here because we focus on the medium modification on the bound color singlet (quarkonium) here. Since $\sum_{a,b}\sigma_{ab}L_{ab}$ is an operator diagonal in the eigenstate of $H_S$, we have
\be
H_S | {\bs k},nl,1 \rangle = ( -|E_{nl}| + \ml{O}(v^3) ) | {\bs k},nl,1 \rangle \\
\sum_{a,b}\sigma_{ab}   L_{ab}  | {\bs k},nl,1 \rangle   = t \Delta E | {\bs k},nl,1 \rangle \,,
\ee
where $ \Delta E$ is real and is a function of ${\bs k}$ and $nl$. Thus $\frac{1}{t}\sum_{a,b}\sigma_{ab}L_{ab}$ is a loop correction of the real part of the system Hamiltonian, or the $Q\bar{Q}$ potential. We can write $\sum_{a,b}\sigma_{ab}L_{ab}\equiv t H_{\ma{1-loop}}$. If we go back to the Schr\"odinger picture, the Lindblad equation turns to be
\be
\rho_S(t) = \rho_S(0)-it [H_S, \rho_S(0)] - i [\sum_{a,b}\sigma_{ab}L_{ab},  \rho_S(0)] + \cdots
\ee
where other terms in the Lindblad Eq.~(\ref{eqn:lindblad}) have been omitted temporarily. Now we define an effective Hamiltonian $H_\ma{eff} = H_S + H_{\ma{1-loop}} $. For the pNRQCD under our assumed hierarchy of scales, we can start with a potential in $H_S$, calculate wave functions of relative motions and the one-loop correction of the real part of the potential to obtain $H_\ma{eff}$. At lowest order perturbation, $H_\ma{eff}$ leads to a correction in the bound state eigenenergy but not in its wave function. Since the one-loop correction contains two dipole vertices and a thermal gluon propagator, we expect the real thermal correction scales as
\be
\Delta E({\bs k}, nl) \sim g^2\frac{T^4}{(Mv)^2Mv^2} \lesssim \alpha_sMv^4\,,
\ee
which is suppressed by $\alpha_sv^2$. So the thermal loop correction of the real part of the potential is perturbative. 
In some cases, it may be necessary to resum all-order loop corrections of the real part of the potential into $H_\ma{eff}$ and then use $H_\ma{eff}$ to calculate the wave function of the relative motion. Explicit calculations in this case are difficult. One may model the real part of the potential in $H_\ma{eff}$ by using recent high statistics lattice studies of the color singlet free energy at finite temperature \cite{Bazavov:2018wmo}. In this dissertation, we will just use a Coulomb potential for simplicity and neglect the loop correction because it is perturbative in our power counting. 

Now if we do a Wigner transform of the form Eq.~(\ref{eqn:wigner}) on
\be
\rho_S(t) = \rho_S(0) -it(H_\ma{eff}\rho_S(0) - \rho_S(0)H_\ma{eff}) + \cdots \,,
\ee
we obtain 
\be \nn
f_{nl}({\bs x}, {\bs k}, t) &=&  f_{nl}({\bs x}, {\bs k}, 0)  -it\int\frac{\diff^3k'}{(2\pi)^3} e^{i {\bs k}'\cdot \bs x}  \\ \nn
&&  \langle \bs k+\frac{{\bs k}'}{2}, nl, 1| (H_\ma{eff}\rho_S -\rho_SH_\ma{eff}) | \bs k-\frac{{\bs k}'}{2}, nl, 1 \rangle + \cdots\\ \nn
&=& f_{nl}({\bs x}, {\bs k}, 0)  - it\int\frac{\diff^3k'}{(2\pi)^3} e^{i {\bs k}'\cdot \bs x} \\
&&  (E_{\bs k +\frac{{\bs k}'}{2}} - E_{\bs k -\frac{{\bs k}'}{2}}) \langle \bs k+\frac{{\bs k}'}{2}, nl, 1| \rho_S(0) | \bs k-\frac{{\bs k}'}{2}, nl, 1 \rangle + \cdots \,.
\ee
Here if we restore the c.m.~kinetic energy temporarily,
\be
E_{\bs k \pm \frac{{\bs k}'}{2}}  = -|E_{nl}| + \frac{(\bs k \pm \frac{{\bs k}'}{2})^2}{4M}\,,
\ee
we can write
\be \nn
f_{nl}({\bs x}, {\bs k}, t) &=& f_{nl}({\bs x}, {\bs k}, 0)  -it\int\frac{\diff^3k'}{(2\pi)^3} \frac{\bs k}{2Mi}\cdot\nabla_{\bs x}e^{i {\bs k}'\cdot \bs x} \\ \nn
&&  \langle \bs k+\frac{{\bs k}'}{2}, nl, 1| \rho_S(0) | \bs k-\frac{{\bs k}'}{2}, nl, 1 \rangle + \cdots \\ 
\label{eqn:stream_f}
&=& f_{nl}({\bs x}, {\bs k}, 0)  - t {\bs v} \cdot\nabla_{\bs x} f_{nl}(\bs x, \bs k, 0) + \cdots \,,
\ee
where the c.m.~velocity of the quarkonium is defined as $\bs v = \frac{\bs k}{2M}$. Under a time derivative, we can see that Eq.~(\ref{eqn:stream_f}) leads to the free streaming of a Boltzmann equation. In the next two subsections, we will show how the collision terms in the Boltzmann equation are derived from the remaining terms in the Lindblad equation. The imaginary part of the diagram in Fig.~\ref{fig:loop_singlet} corresponds to the process of quarkonium dissociation, by the cutting rule. This will appear in the $-\frac{1}{2}\gamma_{ab,cd}\{L^{\dagger}_{cd}L_{ab}, \rho_S(0)\}$ term. We now discuss the calculation of this term in pNRQCD.

\vspace{0.2in}
\subsection{$-\frac{1}{2}\gamma_{ab,cd}\{L^{\dagger}_{cd}L_{ab}, \rho_S(0)\}$ Term}
We will show this term gives the dissociation term in the Boltzmann equation. We first compute the term $-\frac{1}{2}\gamma_{ab,cd}L^{\dagger}_{cd}L_{ab} \rho_S(0)$\,. As explained previously, we are interested in the bound state part of the density matrix and take $\langle {\bs k}_1, n_1l_1, 1 | \rho_S(t) | {\bs k}_2, n_2l_2, 1\rangle $, so we set $|d\rangle=|{\bs k}_1, n_1l_1, 1 \rangle$. Since at lowest order of the expansion, the transition between a bound state and an unbound $Q\bar{Q}$ pair only occurs via the singlet-octet dipole transition, we have $|a\rangle=|c\rangle=|{\bs p}_{\ma{cm}}, {\bs p}_{\ma{rel}}, a_1\rangle$ and double summations over $|a\rangle$ and $|c\rangle$ become just one summation. Similarly we have $|b\rangle = |{\bs k}_3, n_3l_3, 1\rangle$. We need to compute
\be \nn
\gamma_{ab,cd} &=& \int \diff^3R_1 \int \diff^3R_2 \sum_{i_1,i_2,b_1,b_2}\int_0^t \diff t_1 \int_0^t \diff t_2 C_{{\bs R}_1i_1b_1,{\bs R}_2i_2b_2}(t_1, t_2) \\ \nn
&&\langle {\bs k}_1, n_1l_1, 1 | \langle S({\bs R}_1, t_1) | r_{i_1} | O^{b_1}({\bs R}_1, t_1) \rangle | {\bs p}_{\ma{cm}}, {\bs p}_{\ma{rel}},a_1\rangle 
\\ 
&&\langle {\bs p}_{\ma{cm}}, {\bs p}_{\ma{rel}},a_1 |  \langle O^{b_2}({\bs R}_2, t_2) | r_{i_2} | S({\bs R}_2, t_2) \rangle | {\bs k}_3, n_3l_3, 1\rangle \,.
\ee
We can start with
\be \nn
&&\langle {\bs k}_1, n_1l_1, 1 | \langle S({\bs R}_1, t_1) | r_{i_1} | O^{b_1}({\bs R}_1, t_1) \rangle | {\bs p}_{\ma{cm}}, {\bs p}_{\ma{rel}},a_1\rangle \\
  &=&  \langle \psi_{n_1l_1} | r_{i_1}| \Psi_{{\bs p}_{\ma{rel}}} \rangle \delta^{a_1b_1}e^{-i(E_{{\bs p}}t_1 - {\bs p}_{\ma{cm}} \cdot {\bs R}_1)} e^{i(E_{{\bs k}_1}t_1 - {\bs k}_1\cdot{\bs R}_1)} \\ \nn
&&\langle {\bs p}_{\ma{cm}}, {\bs p}_{\ma{rel}},a_1 |  \langle O^{b_2}({\bs R}_2, t_2) | r_{i_2} | S({\bs R}_2, t_2) \rangle | {\bs k}_3, n_3l_3, 1\rangle \\
 &=&  \langle \Psi_{{\bs p}_{\ma{rel}}}   | r_{i_2}|  \psi_{n_3l_3} \rangle \delta^{a_1b_2}e^{ - i(E_{{\bs k}_3}t_2 - {\bs k}_3\cdot{\bs R}_2 )} e^{ i(E_{{\bs p}}t_2 - {\bs p}_{\ma{cm}} \cdot {\bs R}_2)} \,,
\ee
where $E_{{\bs p}} = \frac{{\bs p}_{\ma{rel}}^2}{M}$ and $E_{{\bs k}_i} = -|E_{n_il_i}|$ up to $v^2$-corrections. The correlation needed is 
\be\nn
&&C_{{\bs R}_1i_1b_1,{\bs R}_2i_2b_2}(t_1, t_2) = \frac{T_F}{N_C}g^2\langle  E_{i_1}^{b_1}({\bs R}_1, t_1)  E_{i_2}^{b_2}({\bs R}_2, t_2)  \rangle_T \\ \nn
&=&\frac{T_F}{N_C}g^2 \delta^{b_1b_2} \int\frac{\diff^4q}{(2\pi)^4}  e^{iq_0(t_1-t_2)-i\bs q\cdot({\bs R}_1-{\bs R}_2)} \\
&& (q_0^2\delta_{i_1i_2} -q_{i_1}q_{i_2}) n_B(q_0) (2\pi) \sign(q_0) \delta(q_0^2-{\bs q}^2) + \ml{O}(g^3) \,,
\ee
where we have used the real-time propagator $D^{<\,ab}_{\mu\nu}(q_0, {\bs q})$ of a free gauge field, which is a generalization (by adding the polarization and the color) of the real-time propagator (\ref{eqn:<}) introduced in Chapter 1 . It should be pointed out that one can also use the propagator $D^{>\,ab}_{\mu\nu}(q_0, {\bs q})$ and will obtain the same result due to the relation $1+n_B(q_0)+n_B(-q_0) = 0$. Now we can write the term $\langle {\bs k}_1, n_1l_1, 1 | \gamma_{ab,cd}L^{\dagger}_{cd}L_{ab} \rho_S(0)  | {\bs k}_2, n_2l_2, 1\rangle$ out explicitly as
\be\nn
&&\int\frac{\diff^3p_{\ma{cm}}}{(2\pi)^3} \frac{\diff^3p_{\ma{rel}}}{(2\pi)^3} \frac{\diff^3k_{3}}{(2\pi)^3} \frac{\diff^4q}{(2\pi)^4} \int \diff^3R_1 \int \diff^3R_2 \int_0^t\diff t_1 \int_0^t\diff t_2 \sum_{n_3,l_3,a_1,b_1,b_2,i_1,i_2}  \\ \nn
&& \frac{T_F}{N_C}g^2 (q_0^2\delta_{i_1i_2} -q_{i_1}q_{i_2})  n_B(q_0)\delta^{b_1b_2}(2\pi) \sign(q_0) \delta(q_0^2-{\bs q}^2) e^{iq_0(t_1-t_2)-i\bs q\cdot({\bs R}_1-{\bs R}_2)} \\ \nn
&& \langle \psi_{n_1l_1} | r_{i_1}| \Psi_{{\bs p}_{\ma{rel}}} \rangle \delta^{a_1b_1}  e^{-i(E_{{\bs p}}t_1 - {\bs p}_{\ma{cm}} \cdot {\bs R}_1)} e^{i(E_{{\bs k}_1}t_1 - {\bs k}_1\cdot{\bs R}_1)}  \\ \nn
&&  \langle \Psi_{{\bs p}_{\ma{rel}}}   | r_{i_2}|  \psi_{n_3l_3} \rangle \delta^{a_1b_2}  e^{ -i(E_{{\bs k}_3}t_2 - {\bs k}_3\cdot{\bs R}_2 )} e^{ i(E_{{\bs p}}t_2 - {\bs p}_{\ma{cm}} \cdot {\bs R}_2)}  \\ 
&&  \langle {\bs k}_3, n_3l_3, 1| \rho_S(0) | {\bs k}_2, n_2l_2, 1\rangle \,.
\ee
Integrating over ${\bs R}_1$ and ${\bs R}_2$ gives two delta functions in momenta, $\delta^3({\bs k}_1-{\bs p}_{\ma{cm}}+\bs q)\delta^3({\bs k}_3-{\bs p}_{\ma{cm}}+\bs q)$. Under the Markovian approximation, $t\rightarrow\infty$, integrating over $t_1$ and $t_2$ will give another two delta functions $\delta(E_{k_1}-E_{p}+q_0)\delta(E_{k_3}-E_{p}+q_0)$. Since $E_{k_i} = -|E_{n_il_i} | <0$ and $E_{p} = \frac{{\bs p}_\ma{rel}^2}{M} >0$, some energy has to be transferred to the bound state to break it up to an unbound state, and thus $q_0$ has to be positive. (If we use $D^{>\,ab}_{\mu\nu}(q_0, {\bs q})$, here we would have $\delta(E_{k_1}-E_{p}-q_0)\delta(E_{k_3}-E_{p}-q_0)$ and $q_0$ is negative.) After integrating ${\bs R}_1$, ${\bs R}_2$, $t_1$, $t_2$ and ${\bs k}_3$ we obtain
\be \nn
&& \sum_{n_3,l_3,a_1,i_1}\int\frac{\diff^3p_{\ma{cm}}}{(2\pi)^3} \frac{\diff^3p_{\ma{rel}}}{(2\pi)^3} \frac{\diff^3q}{(2\pi)^32q} n_B(q) \\ \nn
&& (2\pi)^5\delta(E_{k_1}-E_{p}+q)\delta(E_{k_3}-E_{p}+q) \delta^3({\bs k}_1 - {\bs p}_{\ma{cm}} +{\bs q} )\\
&&  \frac{2T_F}{3N_C}q^2g^2  \langle \psi_{n_1l_1} | r_{i_1} | \Psi_{{\bs p}_{\ma{rel}}} \rangle   \langle \Psi_{{\bs p}_{\ma{rel}}}   | r_{i_1} |  \psi_{n_3l_3} \rangle   \langle {\bs k}_1, n_3l_3, 1| \rho_S(0) | {\bs k}_2, n_2l_2, 1\rangle \,,
\ee
where we have used for any smooth function $f(q)$
\be
\int \frac{\diff^3q}{(2\pi)^3} (q^2\delta_{i_1i_2} -q_{i_1}q_{i_2})  f(q) = \frac{2}{3} \delta_{i_1i_2} \int \frac{\diff^3q}{(2\pi)^3} q^2  f(q) \,.
\ee
Due to the energy $\delta$-functions, the sum over $n_3$ and $l_3$ gives $n_3=n_1$, $l_3=l_1$ (we assume no degeneracy in the bound state eigenenergy beyond that implied by rotational invariance). Then one of the energy $\delta$-functions multiplied by $2\pi$ can be interpreted as the time interval $t$, as explained in the last subsection in detail. So we have
\be \nn
&&t\int\frac{\diff^3p_{\ma{cm}}}{(2\pi)^3} \frac{\diff^3p_{\ma{rel}}}{(2\pi)^3} \frac{\diff^3q}{(2\pi)^32q} n_B(q)  (2\pi)^4\delta^3({\bs k}_1 - {\bs p}_{\ma{cm}}  +{\bs q} )  \delta(E_{k_1}-E_{p}+q) \\
\label{eqn:disso_rho}
 && \frac{2T_F}{3N_C}(N_C^2-1)q^2g^2 | \langle \psi_{n_1l_1} | \bs r | \Psi_{{\bs p}_{\ma{rel}}} \rangle|^2  \langle {\bs k}_1, n_1l_1, 1| \rho_S(0) | {\bs k}_2, n_2l_2, 1\rangle \,.
\ee
Under a Wigner transform of the form Eq.~(\ref{eqn:wigner}) (where we set ${\bs k}_1 = {\bs k} + \frac{{\bs k}'}{2}$, ${\bs k}_2 = {\bs k}-\frac{{\bs k}'}{2}$,  $n_1=n_2=n$ and $l_1=l_2=l$ and then do a shift in c.m.~momentum ${\bs p}_{\ma{cm}} \rightarrow {\bs p}_{\ma{cm}} + \frac{{\bs k}'}{2}$), Eq.~(\ref{eqn:disso_rho}) finally leads to
\be \nn
 &&t\int\frac{\diff^3p_{\ma{cm}}}{(2\pi)^3} \frac{\diff^3p_{\ma{rel}}}{(2\pi)^3} \frac{\diff^3q}{(2\pi)^32q} n_B(q)  (2\pi)^4\delta^3({\bs k} - {\bs p}_{\ma{cm}} +{\bs q} ) \delta(E_{k}-E_{p}+q)  \\
 \label{chap3_eqn:disso_boltz}
 && \frac{2}{3}C_Fq^2g^2  | \langle \psi_{nl} | \bs r | \Psi_{{\bs p}_{\ma{rel}}} \rangle|^2  f_{nl}(\bs x, \bs k ,t=0)  \equiv t \ml{C}_{nl}^{-}({\bs x}, {\bs k}, t=0)  \,.
\ee
The other term in the anti-commutator gives the same result. So applying the Wigner transform to the $-\frac{1}{2} \gamma_{ab,cd} \langle {\bs k}_1, n_1l_1, 1 |\{L^{\dagger}_{cd}L_{ab}, \rho_S(0)\} | {\bs k}_2, n_2l_2, 1\rangle$ term in the Lindblad equation,
yields the negative of Eq.~(\ref{chap3_eqn:disso_boltz}). In Eq.~(\ref{chap3_eqn:disso_boltz}), we define a collision term $\ml{C}_{nl}^{-}({\bs x}, {\bs k}, t=0)$ for quarkonium dissociation. We will show later that it corresponds to a collision term in a Boltzmann equation. The structure of $\ml{C}_{nl}^{-}$ is clear and consists of three parts: the phase space integral, the delta functions for the energy-momentum conservation and the square of the amplitude of the dissociation process shown in Fig.~\ref{subfig:a}. We will calculate the amplitude directly in the next section using the Feynman rules of pNRQCD.

\vspace{0.2in}
\subsection{$\gamma_{ab,cd}L_{ab}\rho_S(0)L_{cd}^{\dagger}$ Term}
Finally we consider the $\gamma_{ab,cd}L_{ab}\rho_S(0)L_{cd}^{\dagger}$ term. Since we just want to compute 
\be
\langle {\bs k}_1, n_1l_1, 1 | \rho_S(t) | {\bs k}_2, n_2l_2, 1\rangle \,,
\ee 
at lowest order, we set $|a\rangle = |{\bs k}_1, n_1l_1,1 \rangle$, $|c\rangle=|{\bs k}_2, n_2l_2,1\rangle$, $|b\rangle=|{\bs p}_{1\ma{cm}}, {\bs p}_{1\ma{rel}},a_1\rangle$ and $|d\rangle=|{\bs p}_{2\ma{cm}}, {\bs p}_{2\ma{rel}},a_2\rangle$ where $a_1$ and $a_2$ are color indexes. At lowest order in perturbation, the transition between a bound state and an unbound $Q\bar{Q}$ pair only occurs via the singlet-octet dipole transition. We need to evaluate
\be \nn
\gamma_{ab,cd} &=& \int \diff^3R_1 \int \diff^3R_2 \sum_{i_1,i_2,b_1,b_2}\int_0^t \diff t_1 \int_0^t \diff t_2 C_{{\bs R}_2i_2b_2,{\bs R}_1i_1b_1}(t_2, t_1) \\ \nn
&&\langle {\bs k}_1, n_1l_1, 1 | \langle S({\bs R}_1, t_1) | r_{i_1} | O^{b_1}({\bs R}_1, t_1) \rangle  | {\bs p}_{1\ma{cm}}, {\bs p}_{1\ma{rel}},a_1\rangle \\
&&\langle {\bs p}_{2\ma{cm}}, {\bs p}_{2\ma{rel}},a_2 |  \langle O^{b_2}({\bs R}_2, t_2) | r_{i_2} | S({\bs R}_2, t_2) \rangle | {\bs k}_2, n_2l_2, 1\rangle \,.
\ee
We first compute the singlet-octet transition term,
\be \nn
&&  \langle {\bs k}_1, n_1l_1, 1 | \langle S({\bs R}_1, t_1) | r_{i_1} | O^{b_1}({\bs R}_1, t_1) \rangle  | {\bs p}_{1\ma{cm}}, {\bs p}_{1\ma{rel}},a_1\rangle \\
  &=&  \langle \psi_{n_1l_1} | r_{i_1}| \Psi_{{\bs p}_{1\ma{rel}}} \rangle \delta^{a_1b_1}e^{-i(E_{{\bs p}_1}t_1 - {\bs p}_{1\ma{cm}} \cdot {\bs R}_1)} e^{i(E_{{\bs k}_1}t_1 - {\bs k}_1\cdot{\bs R}_1)} \\ \nn
 &&\langle {\bs p}_{2\ma{cm}}, {\bs p}_{2\ma{rel}},a_2 |  \langle O^{b_2}({\bs R}_2, t_2) | r_{i_2} | S({\bs R}_2, t_2) \rangle | {\bs k}_2, n_2l_2, 1\rangle \\
 &=&  \langle \Psi_{{\bs p}_{2\ma{rel}}}   | r_{i_2}|  \psi_{n_2l_2} \rangle \delta^{a_2b_2}e^{ - i(E_{{\bs k}_2}t_2 - {\bs k}_2\cdot{\bs R}_2 )} e^{ i(E_{{\bs p}_2}t_2 - {\bs p}_{2\ma{cm}} \cdot {\bs R}_2)} 
\ee
The correlation of the thermal chromo-electric fields in real-time thermal field theory is
\be \nn
&&C_{{\bs R}_2i_2b_2,{\bs R}_1i_1b_1}(t_2, t_1) = \frac{T_F}{N_C}g^2\langle  E_{i_2}^{b_2}({\bs R}_2, t_2)   E_{i_1}^{b_1}({\bs R}_1, t_1) \rangle_T \\ \nn
&=&\frac{T_F}{N_C}g^2 \delta^{b_1b_2} \int\frac{\diff^4q}{(2\pi)^4}  e^{iq_0(t_1-t_2)-i\bs q\cdot({\bs R}_1-{\bs R}_2)} \\
&&(q_0^2\delta_{i_1i_2} -q_{i_1}q_{i_2}) (1+n_B(q_0))   (2\pi) \sign(q_0) \delta(q_0^2-{\bs q}^2) +\ml{O}(g^3)\,,
\ee
where we have used the real-time propagator $D^{>\,ab}_{\mu\nu}(q_0, {\bs q})$ of free gluon fields, which is a simple generalization of the real-time propagator (\ref{eqn:>}) introduced in Chapter 1.

Now we can combine everything and write $  \langle {\bs k}_1, n_1l_1, 1 | \gamma_{ab,cd}L_{ab}\rho_S(0)L_{cd}^{\dagger} | {\bs k}_2, n_2l_2, 1\rangle $ out explicitly
\be\nn
&&   \int\frac{\diff^4q}{(2\pi)^4}\frac{\diff^3 p_{1\ma{cm}}}{(2\pi)^3} \frac{\diff^3 p_{1\ma{rel}}}{(2\pi)^3} \frac{\diff^3 p_{2\ma{cm}}}{(2\pi)^3} \frac{\diff^3 p_{2\ma{rel}}}{(2\pi)^3}  \int \diff^3R_1 \int \diff^3R_2 \int_0^t \diff t_1 \int_0^t \diff t_2  \sum_{a_1,a_2, b_1,b_2, i_1,i_2} \\ \nn
&&   \frac{T_F}{N_C}g^2 \delta^{b_1b_2}    (q_0^2\delta_{i_1i_2}-q_{i_1}q_{i_2})(1+n_B(q_0)) (2\pi)\sign(q_0)\delta(q_0^2-{\bs q}^2)   e^{iq_0(t_1-t_2)-i\bs q\cdot({\bs R}_1-{\bs R}_2)}  \\ \nn
&&   \langle \psi_{n_1l_1} | r_{i_1}| \Psi_{{\bs p}_{1\ma{rel}}} \rangle \delta^{a_1b_1} e^{-i(E_{{\bs p}_1}t_1 - {\bs p}_{1\ma{cm}} \cdot {\bs R}_1)} e^{i(E_{{\bs k}_1}t_1 - {\bs k}_1\cdot{\bs R}_1)} \\ \nn
&& \langle \Psi_{{\bs p}_{2\ma{rel}}}   | r_{i_2}|  \psi_{n_2l_2} \rangle \delta^{a_2b_2}  e^{ - i(E_{{\bs k}_2}t_2 - {\bs k}_2\cdot{\bs R}_2 )} e^{ i(E_{{\bs p}_2}t_2 - {\bs p}_{2\ma{cm}} \cdot {\bs R}_2)} \\ 
&&\langle {\bs p}_{1\ma{cm}}, {\bs p}_{1\ma{rel}},a_1  |\rho_S(0) | {\bs p}_{2\ma{cm}}, {\bs p}_{2\ma{rel}},a_2 \rangle \,.
\ee
Integrating over ${\bs R}_1$ and ${\bs R}_2$ gives two delta functions in momenta $\delta^3({\bs k}_1-{\bs p}_{1\ma{cm}}+\bs q)\delta^3({\bs k}_2-{\bs p}_{2\ma{cm}}+\bs q)$. Under the Markovian approximation, $t\rightarrow\infty$, integrating over $t_1$ and $t_2$ gives another two delta functions $\delta(E_{k_1}-E_{p_1}+q_0)\delta(E_{k_2}-E_{p_2}+q_0)$. Then $q_0$ has to be positive because $E_{k_i} = -| E_{n_il_i} | <0$ and $E_{p_i} = \frac{{\bs p}^2_{i,\ma{rel}}}{M}>0$. (Using the propagator $D^{<\,ab}_{\mu\nu}(q_0, {\bs q})$ will give $q_0<0$ but lead to the same result due to $1+n_B(q_0)+n_B(-q_0) = 0$). Again we can set $q_0^2\delta_{i_1i_2}-q_{i_1}q_{i_2} \rightarrow \frac{2}{3}q^2\delta_{i_1i_2}$ since the gluon is on shell $q_0 = |\bs q| =q $ and we will integrate over the angles of ${\bs q}$. Now we have
\be \nn
\label{chap3_eqn:reco}
&&\int \frac{\diff^3 p_{1\ma{cm}}}{(2\pi)^3} \frac{\diff^3 p_{1\ma{rel}}}{(2\pi)^3} \frac{\diff^3 p_{2\ma{cm}}}{(2\pi)^3} \frac{\diff^3 p_{2\ma{rel}}}{(2\pi)^3}  \frac{\diff^3q}{(2\pi)^32q} (1+n_B(q)) \sum_{a, i} (2\pi)^8 \\ \nn
&&\delta^3({\bs k}_1-{\bs p}_{1\ma{cm}}+\bs q)\delta^3({\bs k}_2-{\bs p}_{2\ma{cm}}+\bs q)\delta(E_{k_1}-E_{p_1}+q)\delta(E_{k_2}-E_{p_2}+q) \\
&& \frac{2T_F}{3N_C}q^2g^2 \langle \psi_{n_1l_1} | r_{i}| \Psi_{{\bs p}_{1\ma{rel}}} \rangle\langle \Psi_{{\bs p}_{2\ma{rel}}}   | r_{i}|  \psi_{n_2l_2} \rangle \langle {\bs p}_{1\ma{cm}}, {\bs p}_{1\ma{rel}},a  |\rho_S(0) | {\bs p}_{2\ma{cm}}, {\bs p}_{2\ma{rel}},a \rangle \,.\,\,\,\,\,\,\,\,
\ee
Before integrating the $\delta$-functions, we first apply the Wigner transform on Eq.~(\ref{chap3_eqn:reco}) (by setting ${\bs k}_1 = {\bs k} + \frac{{\bs k}'}{2}$, ${\bs k}_2 = {\bs k}-\frac{{\bs k}'}{2}$,  $n_1=n_2=n$ and $l_1=l_2=l$):
\be \nn
&& \int\frac{\diff^3k'}{(2\pi)^3} e^{i {\bs k}'\cdot {\bs x} } 
\frac{\diff^3 p_{1\ma{cm}}}{(2\pi)^3} \frac{\diff^3 p_{1\ma{rel}}}{(2\pi)^3} \frac{\diff^3 p_{2\ma{cm}}}{(2\pi)^3} \frac{\diff^3 p_{2\ma{rel}}}{(2\pi)^3}  \frac{\diff^3q}{(2\pi)^32q} (1+n_B(q)) \sum_{a, i} (2\pi)^8 \\ \nn
&&\delta^3({\bs k}+\frac{{\bs k}'}{2}-{\bs p}_{1\ma{cm}}+\bs q)\delta^3({\bs k}-\frac{{\bs k}'}{2}-{\bs p}_{2\ma{cm}}+\bs q)\delta(E_{k_1}-E_{p_1}+q)\delta(E_{k_2}-E_{p_2}+q) \\ 
&& \frac{2T_F}{3N_C}q^2g^2 \langle \psi_{nl} | r_{i}| \Psi_{{\bs p}_{1\ma{rel}}} \rangle\langle \Psi_{{\bs p}_{2\ma{rel}}}   | r_{i}|  \psi_{nl} \rangle \langle {\bs p}_{1\ma{cm}}, {\bs p}_{1\ma{rel}},a  |\rho_S(0) | {\bs p}_{2\ma{cm}}, {\bs p}_{2\ma{rel}},a \rangle \,.
\ee
At order $Mv^2$, the c.m.~momentum does not enter the energy: $E_{k_i} = -|E_{nl}|$ and  $E_{p_i} =\frac{{\bs p}^2_{i,\ma{rel}}}{M}$. If we shift the momentum
\be \nn
{\bs p}_{1\ma{cm}} &=& {\bs p}'_{1\ma{cm}} + \frac{{\bs k}'}{2} \\
{\bs p}_{2\ma{cm}} &=& {\bs p}'_{2\ma{cm}} -  \frac{{\bs k}'}{2} \,,
\ee
then the two momentum $\delta$-functions become $\delta^3({\bs k}-{\bs p}'_{1\ma{cm}}+\bs q)\delta^3({\bs k}-{\bs p}'_{2\ma{cm}}+\bs q)$. So we can integrate over ${\bs p}'_{2\ma{cm}}$ and set ${\bs p}'_{2\ma{cm}}={\bs p}'_{1\ma{cm}} = {\bs p}_{\ma{cm}}$. Due to the two energy $\delta$-functions, we have $p_{1\ma{rel}} = p_{2\ma{rel}} $. To further simplify the expression, we assume the octet scattering wave function can be factorized,
\be
\langle \bs r | \Psi_{{\bs p}_{\ma{rel}}} \rangle = e^{i   {\bs p}_{\ma{rel}}   \cdot   \bs r} f(r, p_{\ma{rel}})\,,
\ee
which is true for the plane wave solution. 
If we further let
\be \nn
{\bs p}_{1\ma{rel}} &=& {\bs p}_{\ma{rel}} \\
{\bs p}_{2\ma{rel}} &=& {\bs p}_{\ma{rel}} + {\bs p}'_{\ma{rel}} \,,
\ee
(remember that we have shown $p_{1\ma{rel}}=p_{2\ma{rel}}$,) we obtain
\be\nn
&& t\int\frac{\diff^3 p_{\ma{cm}}}{(2\pi)^3} \frac{\diff^3 p_{\ma{rel}}}{(2\pi)^3} \frac{\diff^3q}{(2\pi)^32q} (1+n_B(q)) \sum_{a,i}
(2\pi)^4\delta^3({\bs k}-{\bs p}_{\ma{cm}}+\bs q) \delta(-|E_{nl}|+q-\frac{{\bs p}_{\ma{rel}}^2}{M})\\ \nn
&& \frac{2T_F}{3N_C}q^2g^2 \langle  \psi_{nl}  | r_i |  \Psi_{{\bs p}_{\ma{rel}}}  \rangle \int \diff^3r \,  \psi_{nl}(\bs r) r_i \Psi^*_{{\bs p}_{\ma{rel}}}(\bs r) \\
\label{eqn:reco_contribute}
&& \int \frac{\diff^3p'_{\ma{rel}}}{(2\pi)^3} e^{-i  {\bs p}'_{\ma{rel}}  \cdot  \bs r } \int \frac{\diff^3k'}{(2\pi)^3} e^{i {\bs k}'\cdot {\bs x} }  
 \langle {\bs p}_{\ma{cm}}+\frac{{\bs k}'}{2}, {\bs p}_{\ma{rel}},a  |\rho_S(0) | {\bs p}_{\ma{cm}}-\frac{{\bs k}'}{2}, {\bs p}_{\ma{rel}}+{\bs p}'_{\ma{rel}},a \rangle \,.\,\,\,\,\,\,\,\,\,\,\,\,\,\,\,\,
\ee
The last line is just the phase space distribution function of a heavy quark-antiquark pair whose c.m.~position is located at $\bs x$ and relative position is $\bs r$:
\be 
&& f_{Q\bar{Q}}(\bs x, {\bs p}_{\ma{cm}}, {\bs r}, {\bs p}_{\ma{rel}}, a,t=0)  \\\nn
 &=& \int \frac{\diff^3k'}{(2\pi)^3} e^{i {\bs k}'\cdot {\bs x} } \int \frac{\diff^3p'_{\ma{rel}}}{(2\pi)^3} e^{-i  {\bs p}'_{\ma{rel}}    \cdot   \bs r } 
 \langle {\bs p}_{\ma{cm}}+\frac{{\bs k}'}{2}, {\bs p}_{\ma{rel}}, a  |\rho_S(0) | {\bs p}_{\ma{cm}}-\frac{{\bs k}'}{2}, {\bs p}_{\ma{rel}}-{\bs p}'_{\ma{rel}},a \rangle \,.
\ee
So the $ \langle {\bs k} + \frac{{\bs k}'}{2} , nl, 1 |  \gamma_{ab,cd}L_{ab}\rho_S(0)L_{cd}^{\dagger}  | {\bs k} - \frac{{\bs k}'}{2} , nl, 1\rangle$ term in the Lindblad equation under the Wigner transform leads to
\be\nn
&& t\int\frac{\diff^3 p_{\ma{cm}}}{(2\pi)^3} \frac{\diff^3 p_{\ma{rel}}}{(2\pi)^3} \frac{\diff^3q}{(2\pi)^32q} (1+n_B(q)) \sum_{a,i} 
 (2\pi)^4\delta^3({\bs k}-{\bs p}_{\ma{cm}}+\bs q) \delta(-|E_{nl}|+q-\frac{{\bs p}_{\ma{rel}}^2}{M}) \\ \nn
&& \frac{2T_F}{3N_C}q^2g^2 \langle  \psi_{nl}  | r_i |  \Psi_{{\bs p}_{\ma{rel}}}  \rangle \int \diff^3r \,  \psi_{nl}(\bs r) r_i \Psi^*_{{\bs p}_{\ma{rel}}}(\bs r) 
 f_{Q\bar{Q}}(\bs x, {\bs p}_{\ma{cm}}, {\bs r}, {\bs p}_{\ma{rel}}, a,t=0) \\
\label{chap3_eqn:reco_boltz}
&\equiv &  t\ml{C}_{nl}^{+}({\bs x}, {\bs k}, t=0) \,,
\ee
where we define a collision term $\ml{C}_{nl}^{+}({\bs x}, {\bs k}, t=0)$ for quarkonium recombination. As we will show below, it will appear as a collision term in a Boltzmann equation. The structure of $\ml{C}_{nl}^{+}$ is slightly more complicated than $\ml{C}_{nl}^{-}$. But it still consists of three parts: the phase space integral, the delta functions for the energy-momentum conservation and the produce of an amplitude with the complex conjugate of a different amplitude. We will discuss later under what conditions the third part reduces to a simple square of amplitude magnitude, as in the case of $\ml{C}_{nl}^{-}$.

\vspace{0.2in}
Now we can put Eqs.~(\ref{eqn:stream_f}), (\ref{chap3_eqn:disso_boltz}) and (\ref{chap3_eqn:reco_boltz}) together and finally infer the Boltzmann transport equations
\be
\label{chap3_eqn_boltz_transport}
\frac{\partial}{\partial t} f_{nl}({\bs x}, {\bs k}, t) + {\bs v}\cdot \nabla_{\bs x}f_{nl}({\bs x}, {\bs k}, t) = \ml{C}_{nl}^{+}({\bs x}, {\bs k}, t) - \ml{C}_{nl}^{-}({\bs x}, {\bs k}, t)\,,
\ee
where the dissociation term $\ml{C}_{nl}^{-}({\bs x}, {\bs k}, t)$ and the recombination term $\ml{C}_{nl}^{+}({\bs x}, {\bs k}, t)$ terms are defined in Eqs.~(\ref{chap3_eqn:disso_boltz}) and (\ref{chap3_eqn:reco_boltz}).

The integral over $\diff^3 r$ in $\ml{C}_{nl}^{(+)} $ is non-trivial: not only the wave function but also the distribution function are involved. We now consider under what conditions the integral can be further simplified. We note that the support (the region with non-zero function value) of the integrand is on the order of the Bohr radius $a_B$ of the bound state. So if the distribution function is almost uniform in $\bs r$ for $r \lesssim a_B$, one can take the distribution function out of the integral. This is true when the diffusion length scale $\sqrt{Dt}$ is much larger than $r\sim a_B\sim \frac{1}{Mv}$ where $D$ is the diffusion constant of open heavy flavors. The distribution function in ${\bs r}$ caused solely by diffusion is a Gaussian with a variance $\sim Dt$. In other words, the distribution function varies significantly at a length scale $\sqrt{Dt}$ and when one focuses on a region with a much smaller length scale, one can treat the distribution function as uniform. 
A perturbative estimate gives $D\sim\frac{1}{\alpha_s^2T}$ \cite{Moore:2004tg}. The time period for the $Q\bar{Q}$ to be close within the bound state formation range is roughly $t\sim \frac{a_B}{v_\ma{rel}} \sim \frac{1/Mv}{p_\ma{rel}/M}\sim \frac{1}{p_\ma{rel}v}$. So $\sqrt{Dt} \gg \frac{1}{Mv}$ gives
\be
\frac{1}{\alpha_s^2  p_\ma{rel} T}\gg \frac{1}{M^2v}\,.
\ee
Taking $T \lesssim Mv^2$ as previously assumed we find we must have $p_\ma{rel} \ll M v/(v^2 \alpha_s^2)$ which is clearly satisfied for $p_\ma{rel}\sim Mv$. For  $p_\ma{rel}$ large enough that this condition is not satisfied, the contribution to the integral involving $\langle \psi_{nl}| {\bs r}|\Psi_{{\bs p}_{\ma{rel}}}\rangle$ is negligible for such large $p_\ma{rel}$. This agrees with the intuition: a heavy quark-antiquark pair with a large relative momentum cannot form a bound state. So we can take $ f_{Q\bar{Q}}(\bs x, {\bs p}_{\ma{cm}}, {\bs r}, {\bs p}_{\ma{rel}}, a,t)$ out of the wave function integral with the awareness that the contribution from $r\gg a_B$ should vanish.

Furthermore, we make the molecular chaos assumption and write
\be
 f_{Q\bar{Q}}(\bs x, {\bs p}_{\ma{cm}}, {\bs r}, {\bs p}_{\ma{rel}}, a,t) = \frac{1}{9}f_Q({\bs x}_1, {\bs p}_1, t) f_{\bar{Q}}({\bs x}_2, {\bs p}_2, t)\,,
\ee
where ${\bs x}, {\bs r}, {\bs p}_{\ma{cm}}, {\bs p}_{\ma{rel}}$ are the c.m.~and relative positions and momenta of the heavy quark-antiquark pair with positions ${\bs x}_1, {\bs x}_2$ and momenta ${\bs p}_1$, ${\bs p}_2$. The factor $\frac{1}{9}$ accounts the probability of the color state of $Q\bar{Q}$ being in a specific octet state $a$. The molecular chaos assumption is valid when the rate of decorrelation between the heavy quark and antiquark is much larger than the relaxation rate of the system. The former is given by $D^{-1}\sim \alpha_s^2T$ with $\alpha_s$ at the scale $T$ or $m_D$ while the later has been estimated above and is $\sim \alpha_s v^2T$ with $\alpha_s$ at the scale $Mv$. In NRQCD, $v\sim \alpha_s(Mv)$, so the molecular chaos assumption is valid.

Combining these two assumptions gives
\be\nn
C_{nl}^{+}({\bs x}, {\bs k}, t)&=&\frac{1}{9}\int\frac{\diff^3 p_{\ma{cm}}}{(2\pi)^3} \frac{\diff^3 p_{\ma{rel}}}{(2\pi)^3} \frac{\diff^3q}{(2\pi)^32q} (1+n_B(q)) f_Q({\bs x}_1, {\bs p}_1, t) f_{\bar{Q}}({\bs x}_2, {\bs p}_2, t) \\ \nn
&& (2\pi)^4 \delta^3({\bs k}-{\bs p}_{\ma{cm}}+\bs q) \delta(-|E_{nl}|+q-\frac{{\bs p}_{\ma{rel}}^2}{M}) \frac{2}{3} C_F q^2g^2 |\langle  \psi_{nl}  | {\bs r} |  \Psi_{{\bs p}_{\ma{rel}}}  \rangle|^2  \,, \\
\ee
where the sum over color index $a$ has been carried out. This is the recombination term used in the Boltzmann equation in Ref.~\cite{Yao:2017fuc}. 

So far our pNRQCD calculations of the dissociation and recombination terms are independent of the quarkonium spin. The dissociation rate is the same for two quarkonium states with all quantum numbers the same except the spin: one has $S=0$ and the other has $S=1$. For recombination, there is a spin statistics factor. The unbound $Q\bar{Q}$ that recombines has four possible spin configurations. Three of them correspond to the $S=1$ state while one corresponds to $S=0$. In order to take the spin multiplicity into account, one must further insert a factor $g_s$ into $C_{nl}^{+}$ where $g_s=\frac{3}{4}$ for $S=1$ quarkonium and $g_s=\frac{1}{4}$ for $S=0$ quarkonium. After including the spin degrees of freedom, the Boltzmann equation (\ref{chap3_eqn_boltz_transport}) becomes
\be
\label{chap3_eqn_boltz_transport_nls}
\frac{\partial}{\partial t} f_{nls}({\bs x}, {\bs p}, t) + {\bs v}\cdot \nabla_{\bs x}f_{nls}({\bs x}, {\bs p}, t) = \ml{C}_{nls}^{+}({\bs x}, {\bs p}, t) - \ml{C}_{nls}^{-}({\bs x}, {\bs p}, t)\,,
\ee
where 
\be
\label{chap3_eqn_Cnls+}
\ml{C}_{nls}^{+}({\bs x}, {\bs p}, t) &=& g_s \ml{C}_{nl}^{+}({\bs x}, {\bs p}, t) \\
\label{chap3_eqn_Cnls-}
\ml{C}_{nls}^{-}({\bs x}, {\bs p}, t)  &=& \ml{C}_{nl}^{-}({\bs x}, {\bs p}, t) \,.
\ee
The derivation of the quarkonium transport equation inside the QGP is one of the main results of the dissertation. During the derivation, the connections among our assumed hierarchy of scales, the Markovian approximation and the molecular chaos will be illustrated.

\vspace{0.2in}
\section{Dissociation and Recombination}

In this section, we will use pNRQCD and compute directly the Feynman diagrams that contribute to the dissociation and recombination terms in the Boltzmann equation. By doing so, we will have a more clear understanding of the dependence of dissociation and recombination on the choice of gauge. At the order $gr$, we will reproduce the results shown in the last section. Contributions at the order $g^2r$ will also be studied, which correspond to inelastic scattering with light quarks and gluons in the QGP. All contributing Feynman diagrams are shown in Fig.~\ref{chap3_fig:diagrams1}.

\begin{figure}
    \centering
    \begin{subfigure}[b]{0.33\textwidth}
        \centering
        \includegraphics[height=1.05in]{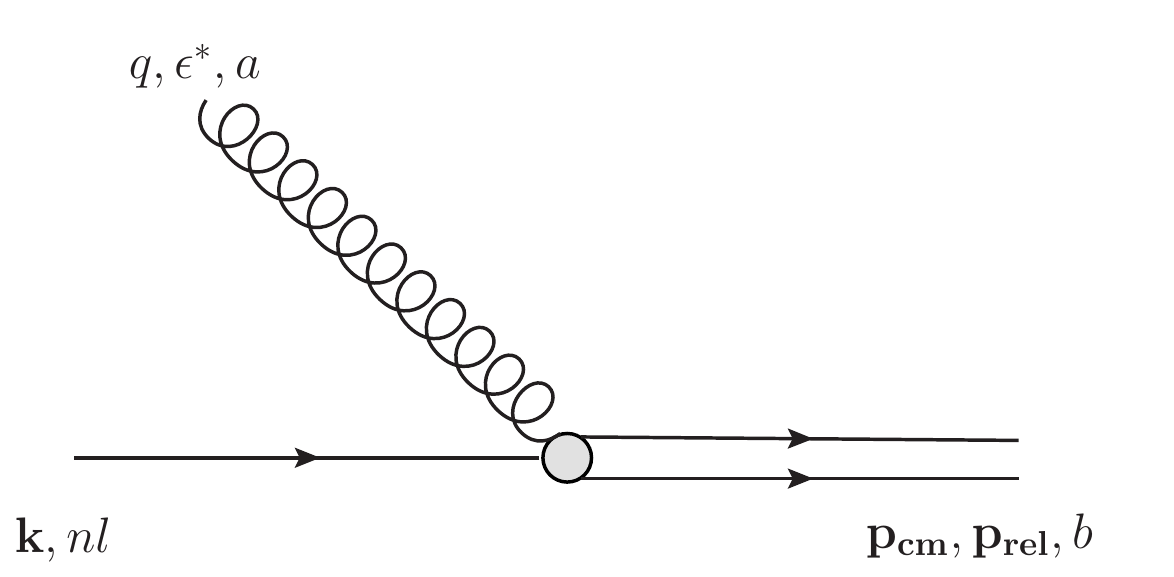}
        \caption{}\label{subfig:a}
    \end{subfigure}%
    ~ 
    \begin{subfigure}[b]{0.33\textwidth}
        \centering
        \includegraphics[height=1.2in]{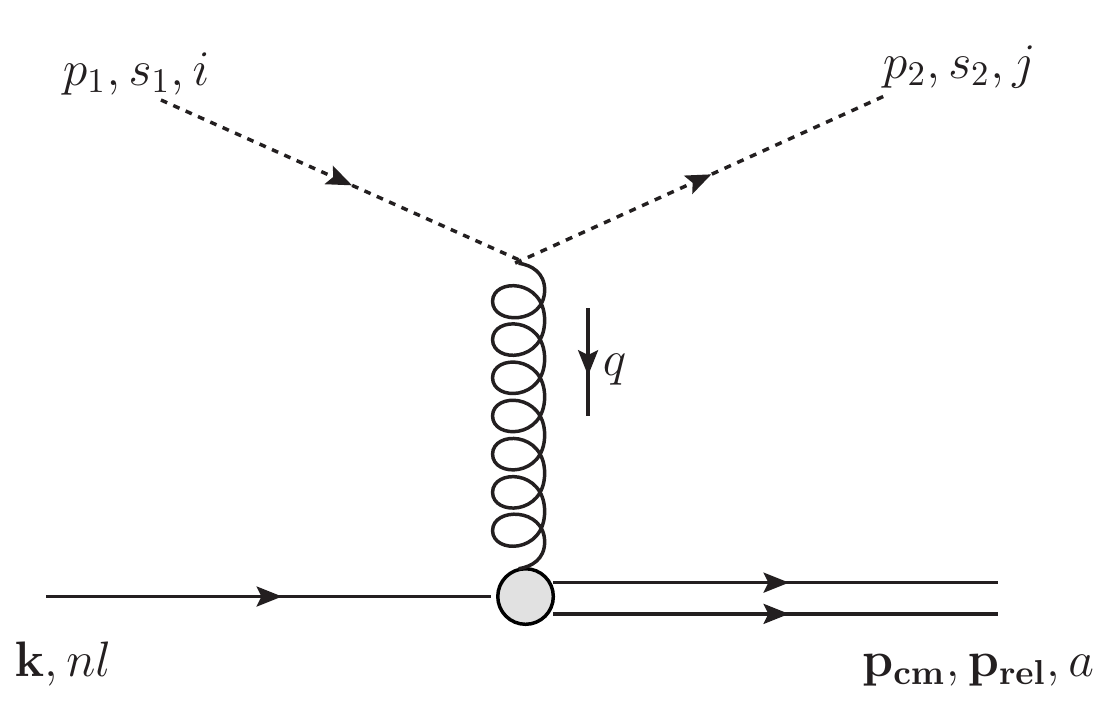}
        \caption{}\label{subfig:b}
    \end{subfigure}%
    ~ 
    \begin{subfigure}[b]{0.33\textwidth}
        \centering
        \includegraphics[height=1.2in]{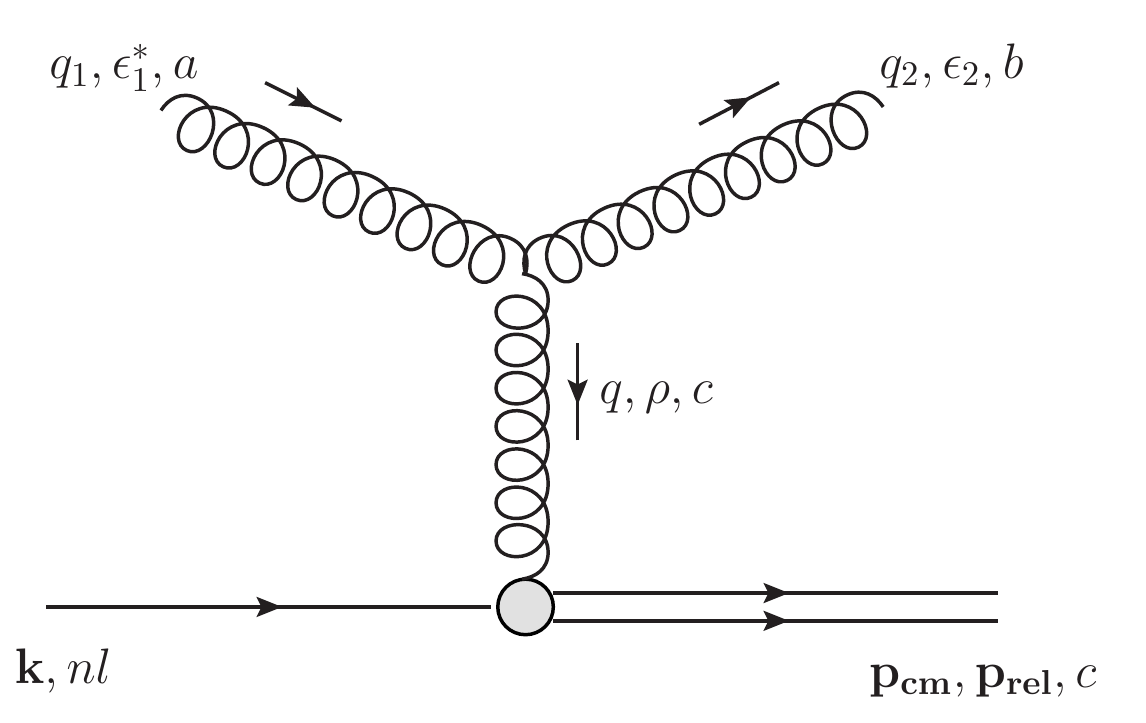}
        \caption{}\label{subfig:c}
    \end{subfigure}%

    \begin{subfigure}[b]{0.33\textwidth}
        \centering
        \includegraphics[height=1.2in]{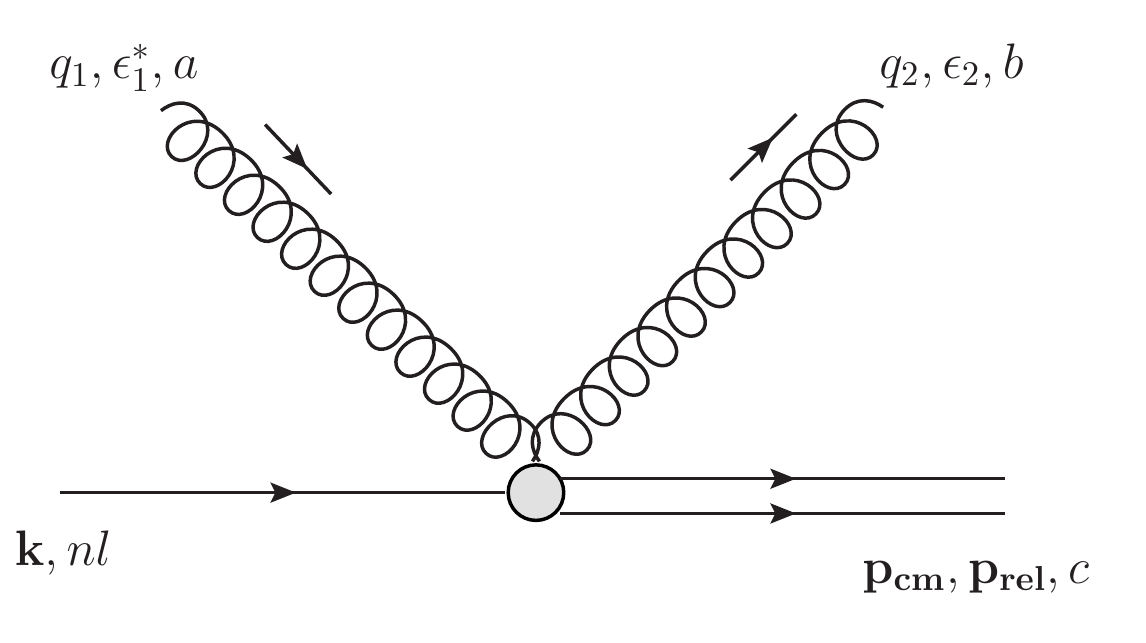}
        \caption{}\label{subfig:d}
    \end{subfigure}%
    ~ 
    \begin{subfigure}[b]{0.33\textwidth}
        \centering
        \includegraphics[height=1.2in]{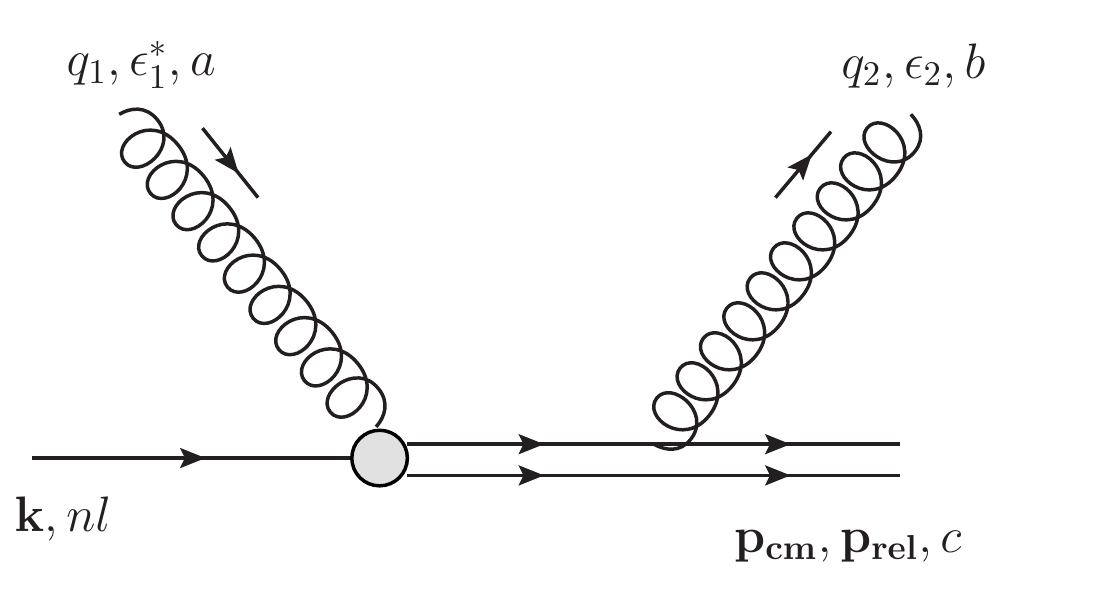}
        \caption{}\label{subfig:e}
    \end{subfigure}%
    ~ 
    \begin{subfigure}[b]{0.33\textwidth}
        \centering
        \includegraphics[height=1.2in]{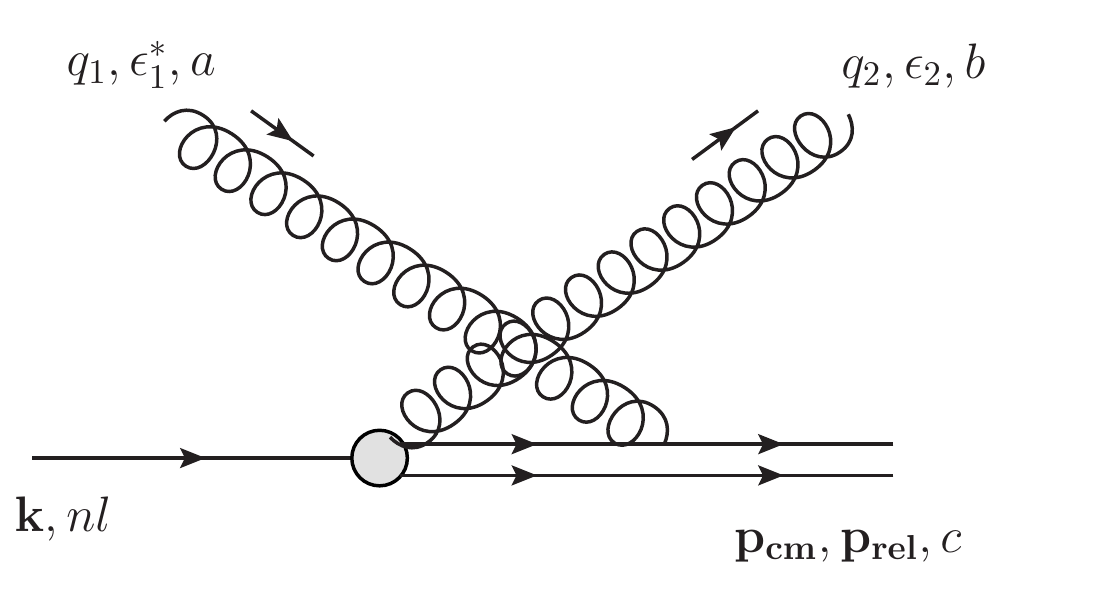}
        \caption{}\label{subfig:f}
    \end{subfigure}%
    
   \begin{subfigure}[b]{0.33\textwidth}
        \centering
        \includegraphics[height=1.2in]{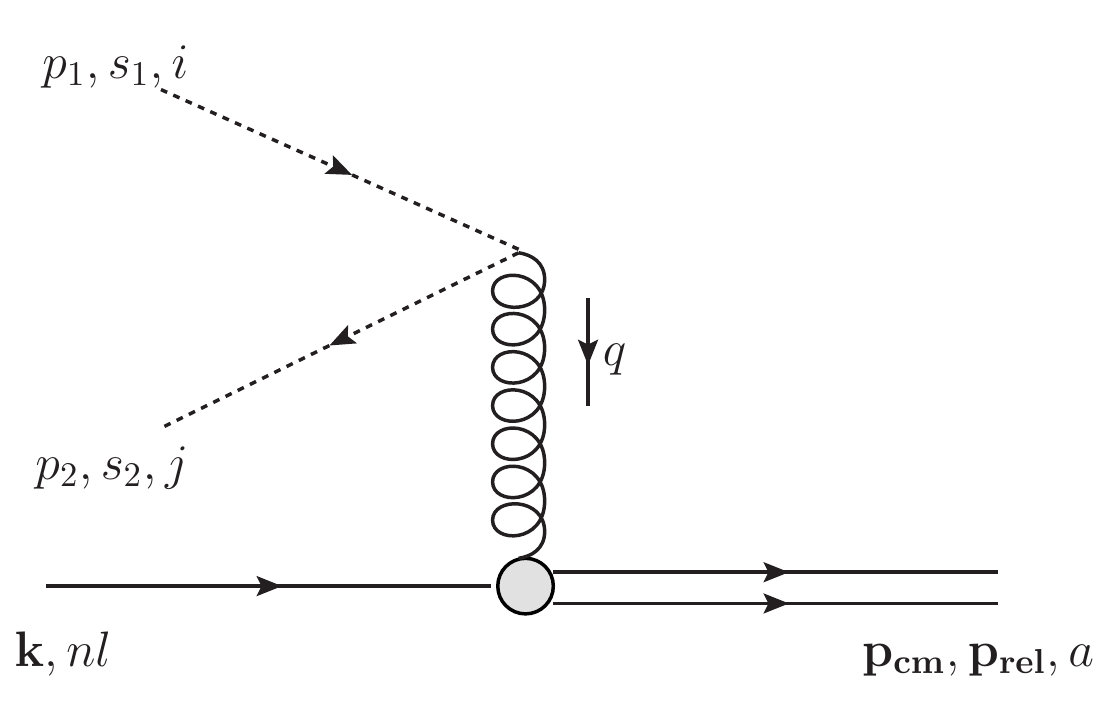}
        \caption{}\label{subfig:g}
    \end{subfigure}%
    ~ 
    \begin{subfigure}[b]{0.33\textwidth}
        \centering
        \includegraphics[height=1.2in]{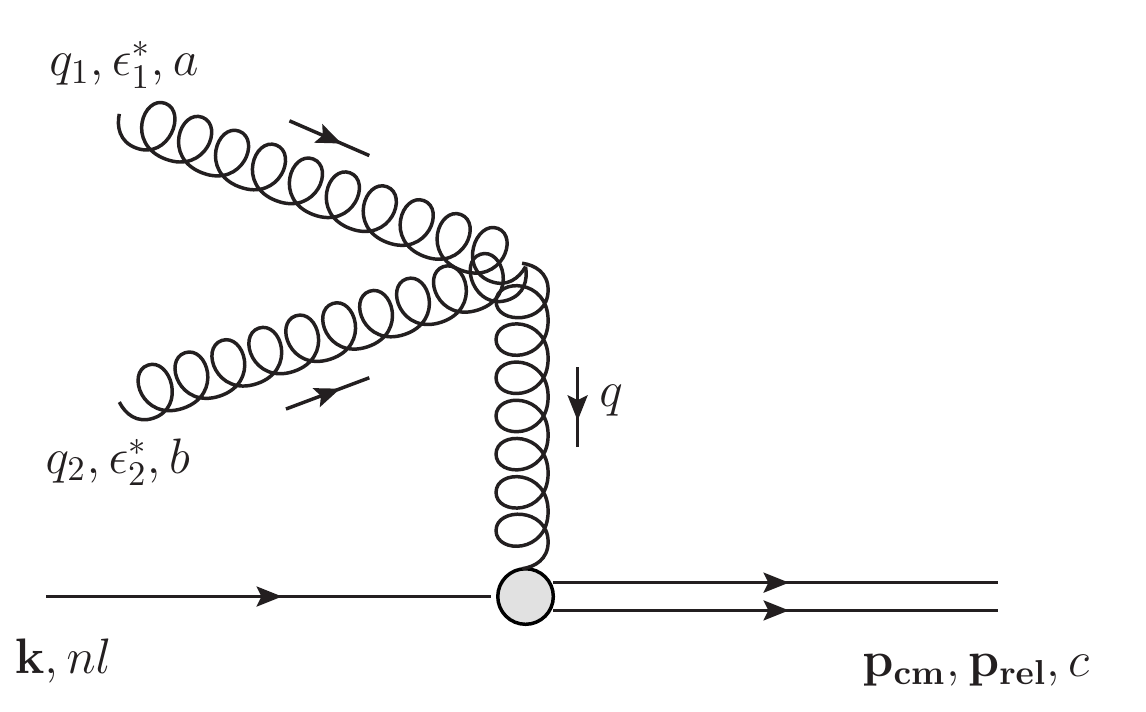}
        \caption{}\label{subfig:h}
    \end{subfigure}%
    ~  
    \begin{subfigure}[b]{0.33\textwidth}
        \centering
        \includegraphics[height=1.2in]{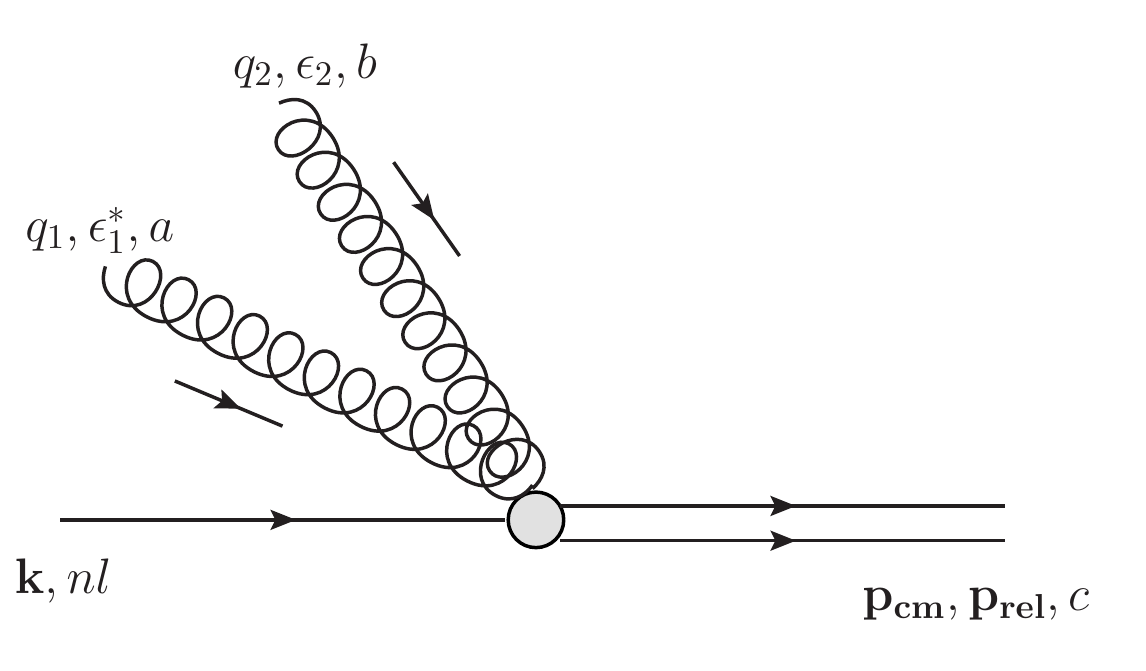}
        \caption{}\label{subfig:i}
    \end{subfigure}%
    
    \begin{subfigure}[b]{0.33\textwidth}
        \centering
        \includegraphics[height=1.2in]{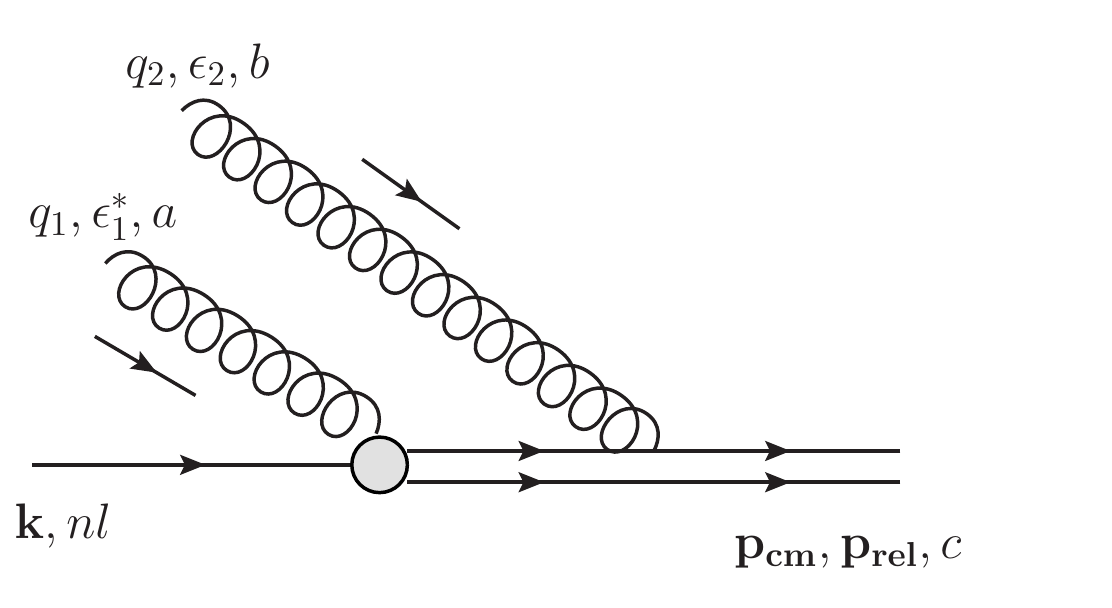}
        \caption{}\label{subfig:j}
    \end{subfigure}%
    ~
    \begin{subfigure}[b]{0.33\textwidth}
        \centering
        \includegraphics[height=1.2in]{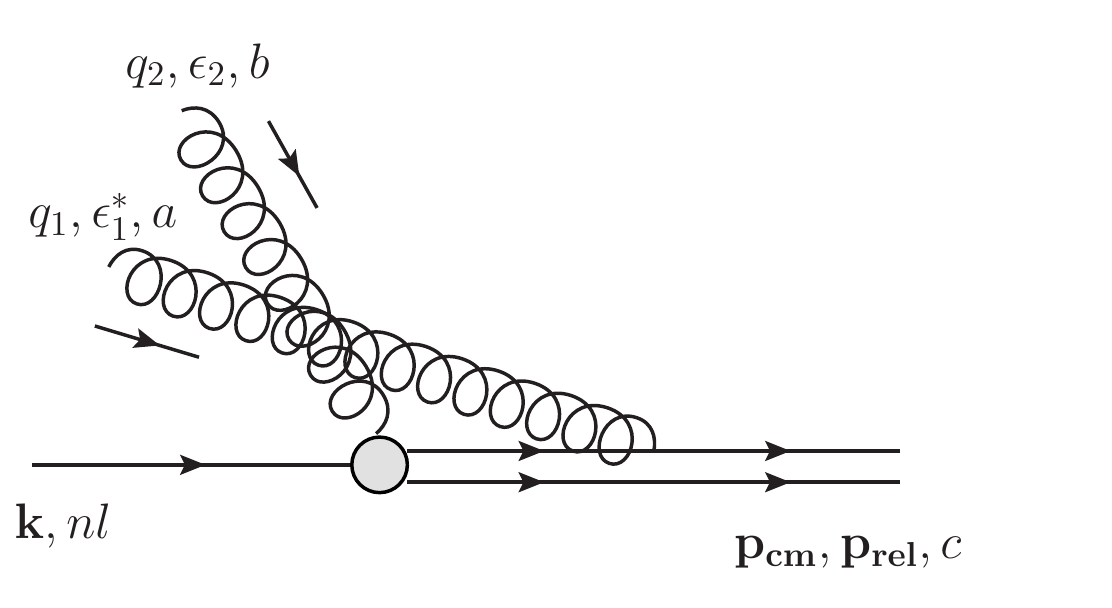}
        \caption{}\label{subfig:k}
    \end{subfigure}%
    ~      
   \begin{subfigure}[b]{0.33\textwidth}
        \centering
        \includegraphics[height=1.2in]{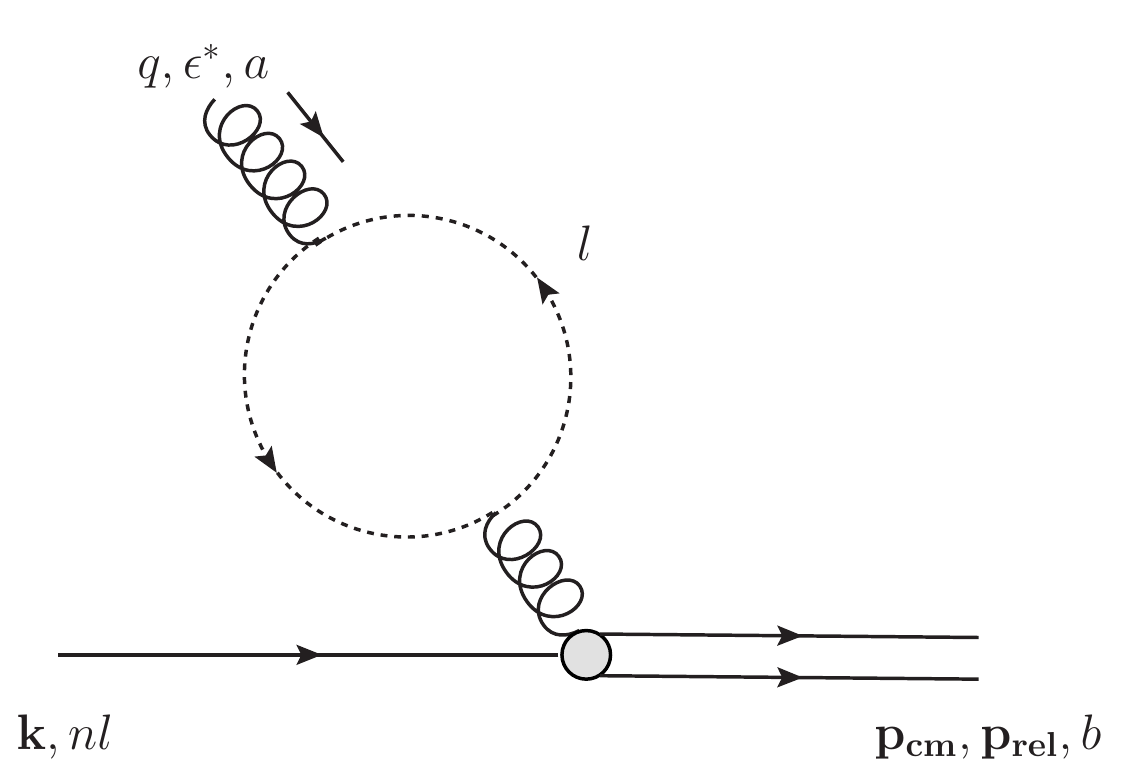}
        \caption{}\label{subfig:l}
    \end{subfigure}%
     
     \begin{subfigure}[b]{0.33\textwidth}
        \centering
        \includegraphics[height=1.2in]{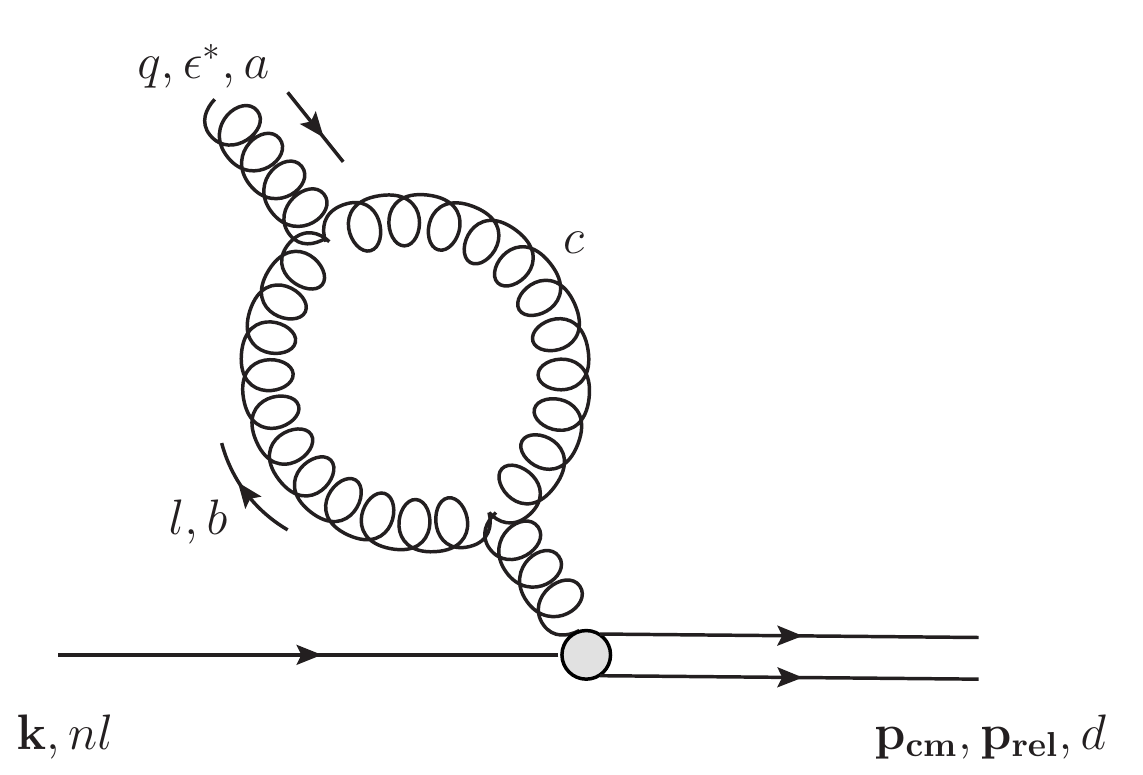}
        \caption{}\label{subfig:m}
    \end{subfigure}%
    ~ 
     \begin{subfigure}[b]{0.33\textwidth}
        \centering
        \includegraphics[height=1.2in]{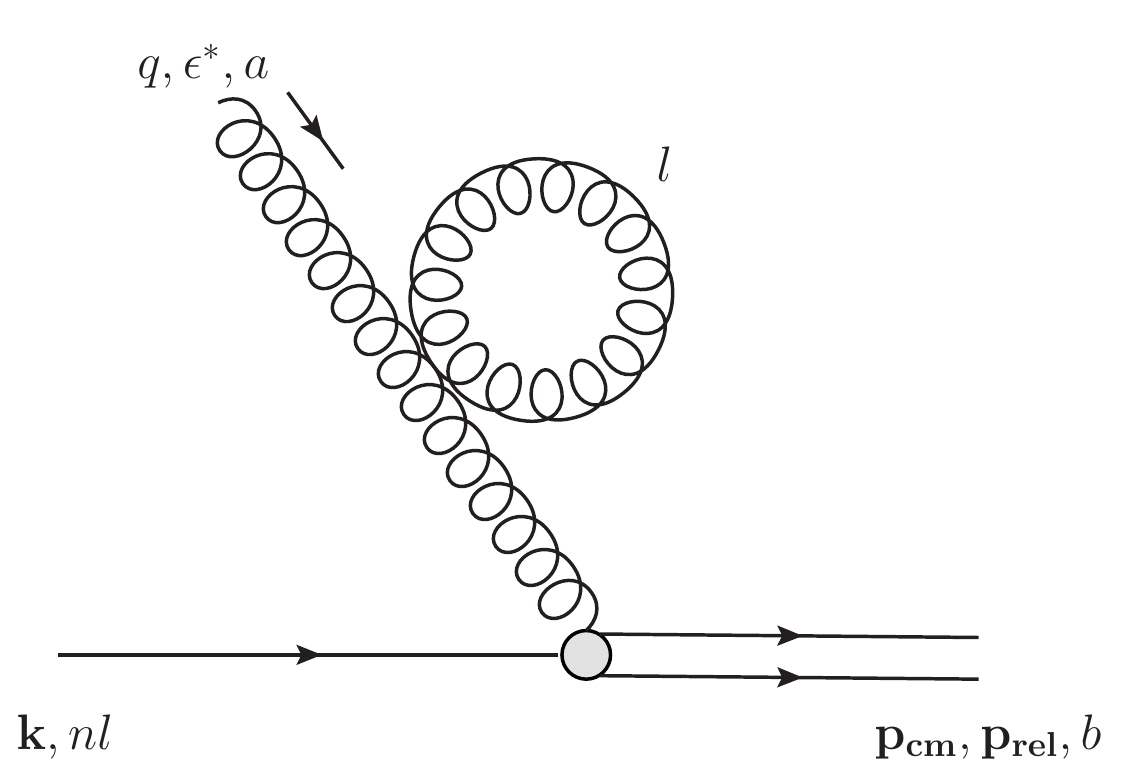}
        \caption{}\label{subfig:n}
    \end{subfigure}%
    ~
    \begin{subfigure}[b]{0.33\textwidth}
        \centering
        \includegraphics[height=1.2in]{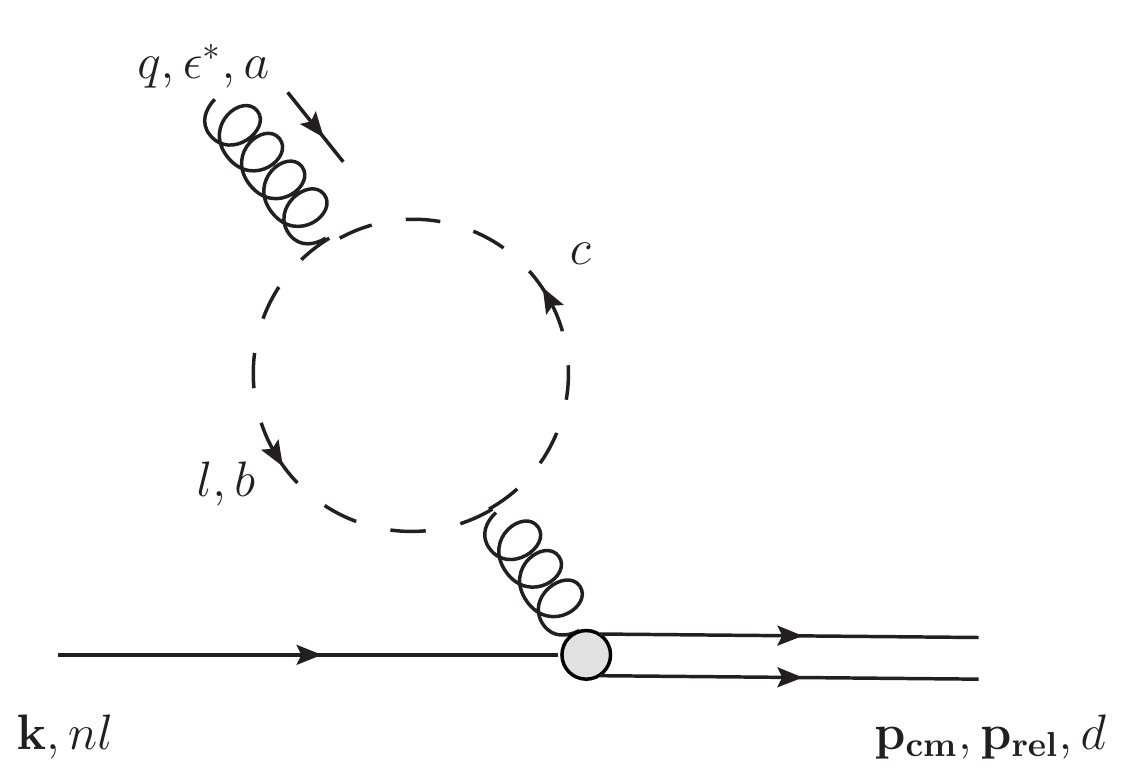}
        \caption{}\label{subfig:o}
    \end{subfigure}%
\end{figure}
\begin{figure}[htb]\ContinuedFloat        
     \begin{subfigure}[b]{0.33\textwidth}
        \centering
        \includegraphics[height=1.2in]{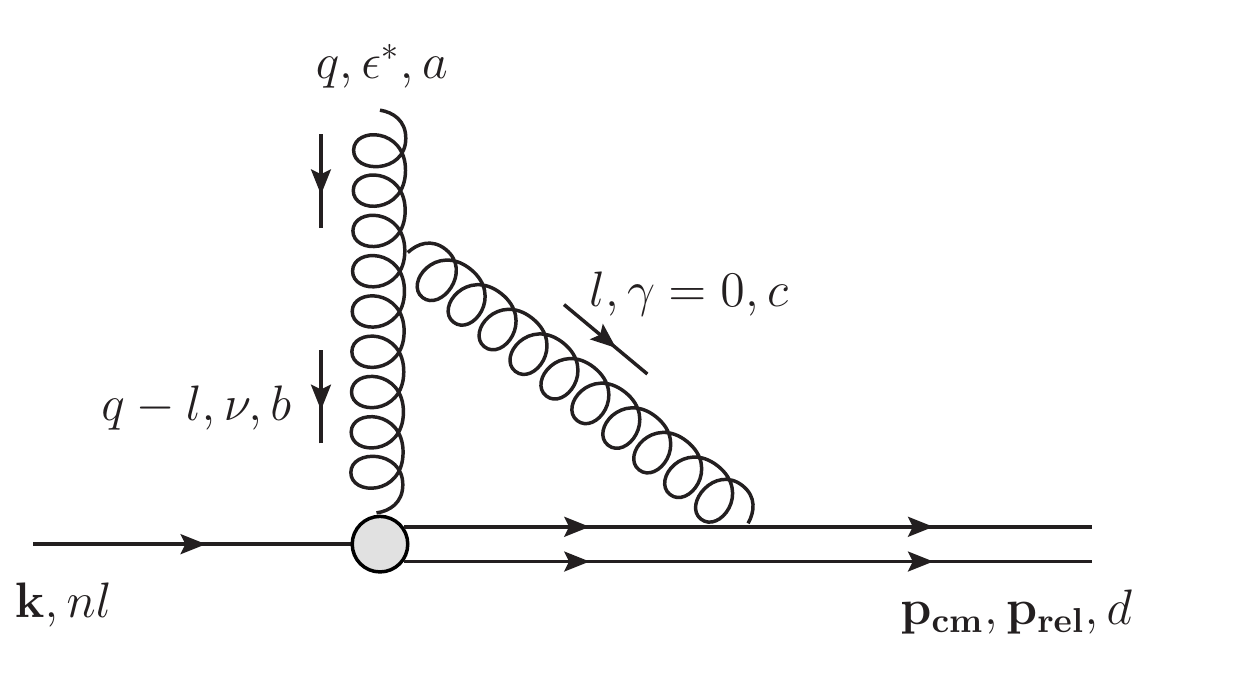}
        \caption{}\label{subfig:p}
    \end{subfigure}%
    ~   
     \begin{subfigure}[b]{0.33\textwidth}
        \centering
        \includegraphics[height=1.2in]{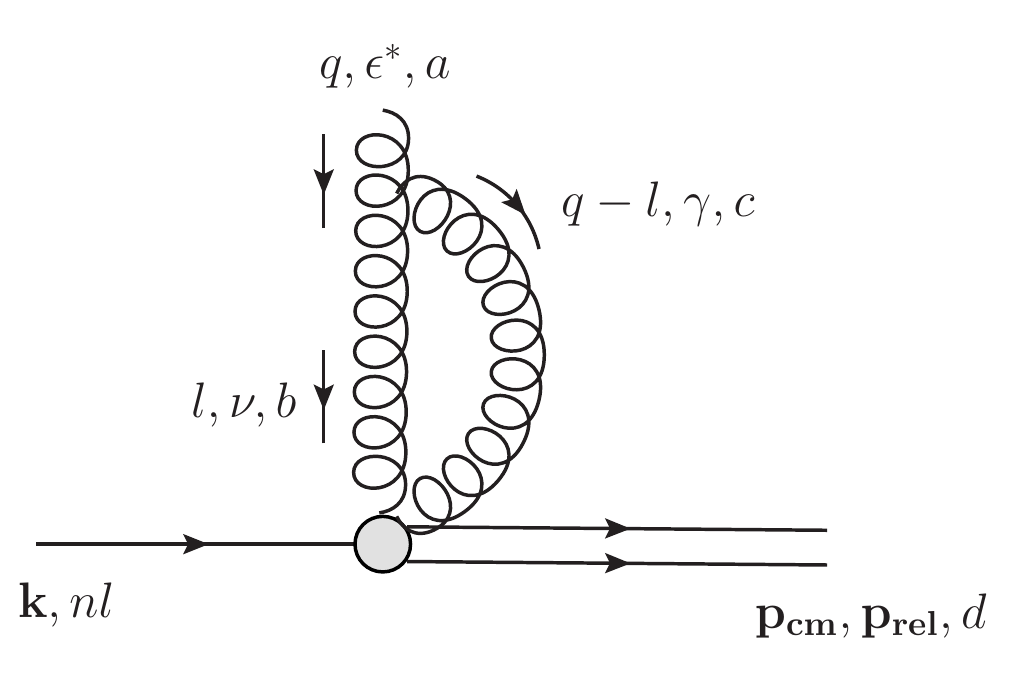}
        \caption{}\label{subfig:q}
    \end{subfigure}%
    ~    
    \begin{subfigure}[b]{0.33\textwidth}
        \centering
        \includegraphics[height=1.2in]{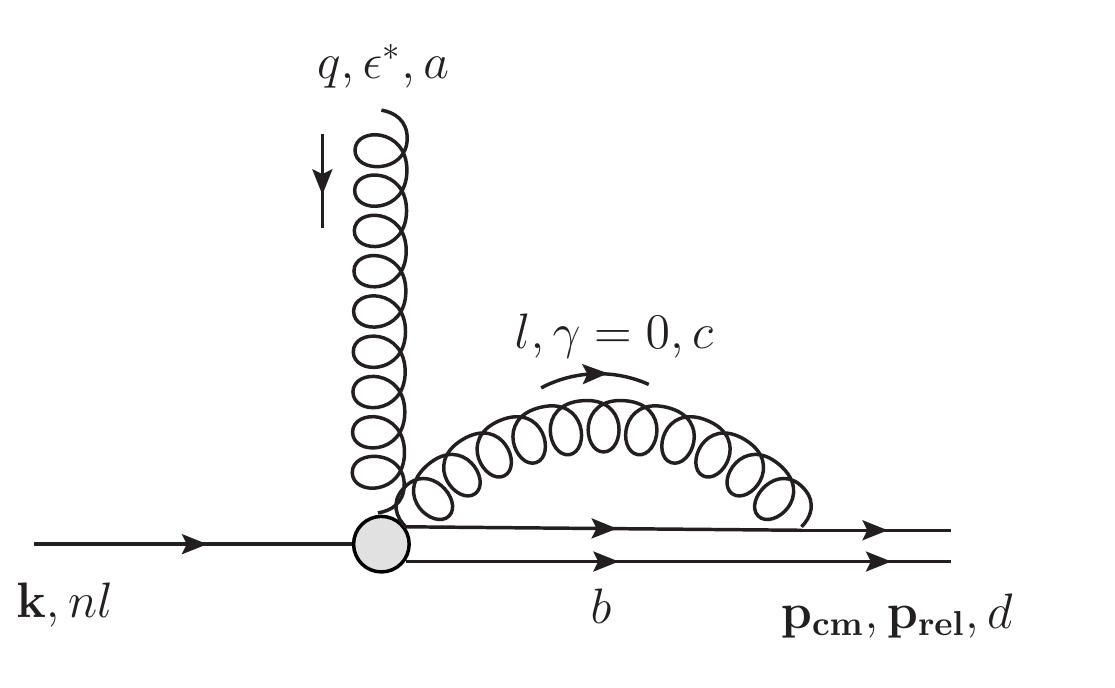}
        \caption{}\label{subfig:r}
    \end{subfigure}%

     \begin{subfigure}[b]{0.33\textwidth}
        \centering
        \includegraphics[height=1.2in]{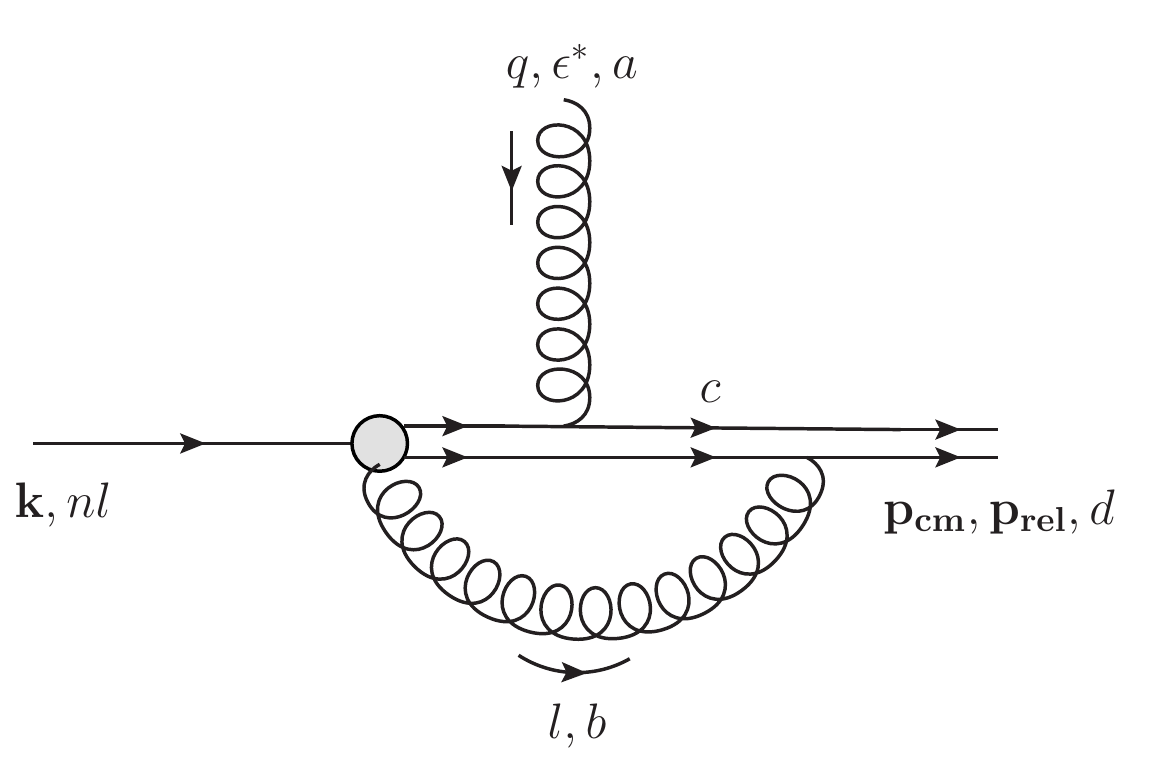}
        \caption{}\label{subfig:s}
    \end{subfigure}%
    ~
    \begin{subfigure}[b]{0.33\textwidth}
        \centering
        \includegraphics[height=1.2in]{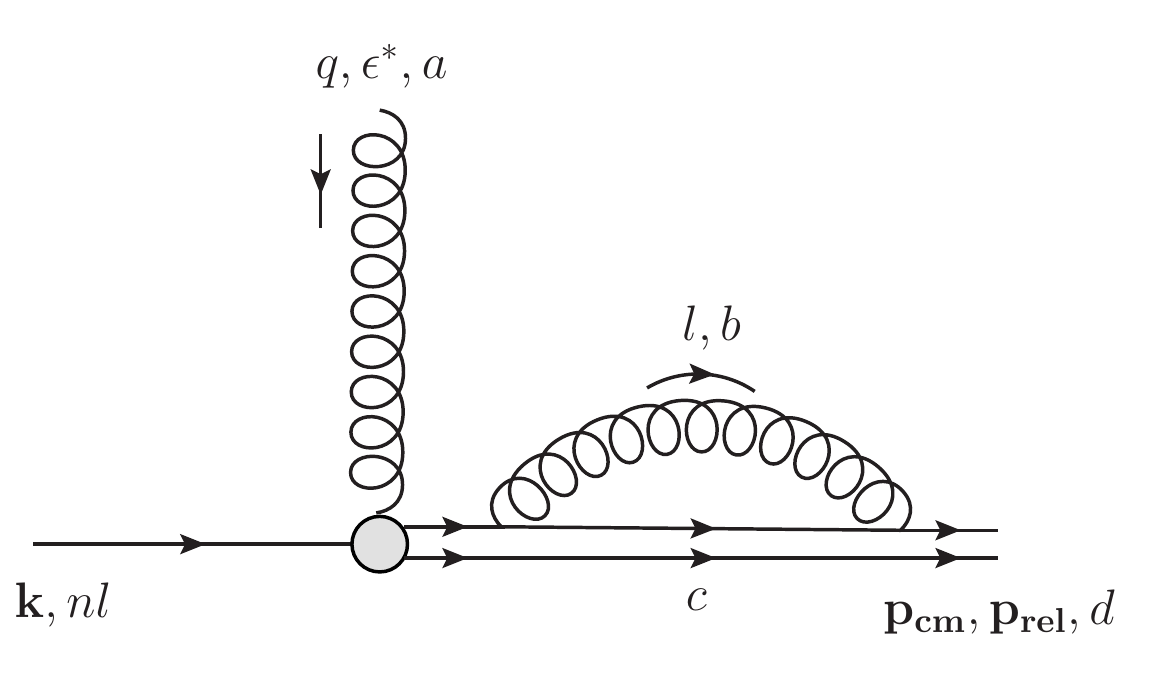}
        \caption{}\label{subfig:t}
    \end{subfigure}%
    \caption[Feynman diagrams contributing to the dissociation and recombination terms in the Boltzmann equation.]{Feynman diagrams contributing to the dissociation and recombination terms in the Boltzmann equation. Single solid line represents the bound color singlet while double solid lines represent the unbound color octet. Short dashed line indicates a light quark (up or down quark) while the long dashed line is the ghost. The grey blob indicates the dipole interaction.}
    \label{chap3_fig:diagrams1}
\end{figure}

\vspace{0.2in}
\subsection{Contributions at the Order $gr$}
\label{subsect: a}
At the order $gr$, only the diagram in Fig.~\ref{subfig:a} contributes. It is a gluon absorption process for dissociation and a gluon emission process for recombination. The scattering amplitude is given by
\be
i\ml{M}_{(a)} &=& g\sqrt{\frac{T_F}{N_c}}(q^0\epsilon^{*i}  - q^i \epsilon^{*0}) \langle  \Psi_{{\bs p}_\ma{rel}}  | r^i |  \psi_{nl}  \rangle \delta^{ab}\\
& \equiv&  i \epsilon^{*\mu} (\ml{M}_{(a)})_\mu \,.
\ee
The gluon is on-shell so $q^0=|\bs q|\equiv q$. The Ward identity can be easily verified:
\be
\label{eqn:ward1}
 q^{\mu}  (\ml{M}_{(a)})_\mu= 0\,.
\ee
So we can compute the amplitude in any gauge we want. In Coulomb gauge, we define the square of the amplitude magnitude, summed over the gluon and octet colors $a,b$ and the gluon polarizations $\epsilon$
\be
\sum | \ml{M}_{(a)} |^2 \equiv \sum_{a,b,\epsilon}  | \ml{M}_{(a)} |^2 = g^2C_Fq^2(\delta^{ij}-\hat{q}^i\hat{q}^j)  \langle \psi_{nl} | r^i | \Psi_{{\bs p}_\ma{rel}} \rangle  \langle \Psi_{{\bs p}_\ma{rel}} | r^j |  \psi_{nl}  \rangle \,,
\ee
where we have used the polarization sum in Coulomb gauge
\be 
\sum_{\epsilon} \epsilon^\mu \epsilon^{*\nu} \equiv \sum_{\lambda = +,-} \epsilon^\mu(\lambda) \epsilon^{*\nu}(\lambda)  
=\begin{cases}
     \delta^{ij}-\hat{q}^i\hat{q}^j & \text{if } \mu=i,\nu=j \\
     0             & \text{otherwise} \,.
\end{cases}
\ee
As mentioned in Chapter 1.4.3, we average over the third component of angular momentum of quarkonium for $l>0$ so we omit the quantum number $m_l$. Here we will show explicitly how the average simplifies the calculation. To this end, we temporarily restore the quantum number $m_l$ in the bound state wave function. When integrating over the relative momentum of the heavy quark-antiquark pair from the dissociation, the average leads to
\be
\nn
&&\frac{1}{2l+1}\sum_{m_l=-l}^l \int \diff^3 p_{\ma{rel}} \langle \psi_{nlm_l} | r^i | \Psi_{{\bs p}_\ma{rel}} \rangle  \langle \Psi_{{\bs p}_\ma{rel}} | r^j |  \psi_{nlm_l}  \rangle \\  \nn
&=& \frac{1}{3}\delta^{ij}  \frac{1}{2l+1}\sum_{m_l=-l}^l \int \diff^3 p_{\ma{rel}}    |  \langle \Psi_{{\bs p}_\ma{rel}}  | {\bs r} |   \psi_{nlm_l}  \rangle | ^2 \\
\label{eqn:average_m} 
&\equiv& \frac{1}{3}\delta^{ij}  \int \diff^3 p_{\ma{rel}} | \langle \Psi_{{\bs p}_\ma{rel}}  | {\bs r} |   \psi_{nl} \rangle |^2\,.
\ee
This allows us to write
\be
\sum | \ml{M}_{(a)} |^2 \equiv  \frac{2}{3}g^2C_Fq^2 |  \langle   \Psi_{{\bs p}_\ma{rel}} | {\bs r} |   \psi_{nl}  \rangle |^2\,.
\ee

To simplify the notation of the dissociation and recombination terms for a quarkonium state $nls$ in the Boltzmann equation, we define \footnote{ In $\ml{F}^+_{nls(a)}$, the positions in the heavy quark and antiquark distributions can be different, as shown in the last section. }
\be\nn
\ml{F}^+_{nls(a)} & \equiv & g_+ \int \frac{\diff^3 k}{(2\pi)^3}  \frac{\diff^3 p_{Q}}{(2\pi)^3}  \frac{\diff^3 p_{\bar{Q}} }{(2\pi)^3} \frac{\diff^3 q}{2q(2\pi)^3} (1+n_B(q)) f_Q({\bs x}_Q, {\bs p}_Q,t) f_{\bar{Q}} ({\bs x}_{\bar{Q}}, {\bs p}_{\bar{Q}},t) \\
\label{chap3_eqn_F+_gluon}
&& (2\pi)^4\delta^3({\bs k} + {\bs q} - {\bs p}_{\ma{cm}}) \delta(-|E_{nl}|+q-\frac{p^2_{\ma{rel}}}{M})  \sum | \ml{M}_{(a)} |^2\\ \nn
\ml{F}^-_{nls(a)} & \equiv &  \int \frac{\diff^3 k}{(2\pi)^3}  \frac{\diff^3 p_{\ma{cm}}}{(2\pi)^3}  \frac{\diff^3 p_{\ma{rel}}}{(2\pi)^3} \frac{\diff^3 q}{2q(2\pi)^3} n_B(q) f_{nls}({\bs x}, {\bs k}, t) \\
&& (2\pi)^4\delta^3({\bs k} + {\bs q} - {\bs p}_{\ma{cm}}) \delta(-|E_{nl}|+q-\frac{p^2_{\ma{rel}}}{M})   \sum | \ml{M}_{(a)} |^2\,,
\ee
where $n_B$ is the Bose-Einstein distribution function, $g_+ = \frac{1}{N_c^2}g_s$ and $g_s$ is the multiplicity factor of spin: $g_s = \frac{3}{4}$ for a quarkonium with spin $S=1$ and $\frac{1}{4}$ for spin $S=0$. In the definition of $\ml{F}^+_{nls(a)}$, ${\bs p}_{\ma{cm}} = {\bs p}_Q+{\bs p}_{\bar{Q}},$ and ${\bs p}_{\ma{rel}} = \frac{{\bs p}_Q - {\bs p}_{\bar{Q}}}{2}$ are the c.m.~and relative momenta of a pair of heavy quark and antiquark with momenta ${\bs p}_Q$ and ${\bs p}_{\bar{Q}}$.

We further define a ``$\delta-$derivative" symbol, first introduced in Ref.~\cite{Yao:2018zze}
\be \nn
&&\frac{\delta }{\delta{{\bs p}_i}} \int \prod_{j=1}^n \frac{\Diff{3}p_j}{(2\pi)^3} h({\bs p}_1, {\bs p}_2, \cdots, {\bs p}_n)\Big|_{{\bs p}_i = {\bs p}}  \\ \nn
\label{chap3_eqn_delta}
&\equiv&  \frac{\delta }{\delta{w({\bs p}})} \int \prod_{j=1}^n \frac{\Diff{3}p_j}{(2\pi)^3} h({\bs p}_1, {\bs p}_2, \cdots, {\bs p}_n)w({\bs p}_i) \\ 
&=& \int \prod_{j=1, j\neq i}^n \frac{\Diff{3}p_j}{(2\pi)^3} h({\bs p}_1, {\bs p}_2, \cdots, {\bs p}_{i-1}, {\bs p}, {\bs p}_{i+1}, \cdots, {\bs p}_n)\,,
\ee
where the second $\delta$ denotes the standard functional variation and $h({\bs p}_1, {\bs p}_2, \cdots, {\bs p}_n)$ and $w({\bs p}_i)$ are arbitrary independent smooth functions. Then the $\ml{C}^{\pm}_{nls}$ terms in the Boltzmann equation (\ref{chap3_eqn_boltz_transport_nls}) can be written as
\be
\ml{C}_{nls(a)}^{\pm}({\bs x}, {\bs p}, t) &=&  \frac{\delta \ml{F}^\pm_{nls(a)}}{\delta{{\bs k}}}  \Big|_{{\bs k}={\bs p}}\,.
\ee
The results $\ml{C}_{nls(a)}^{\pm}$ calculated in this way agree with Eqs.~(\ref{chap3_eqn_Cnls+}) and~(\ref{chap3_eqn_Cnls-}).
For $\ml{C}_{nls(a)}^{+}$, we further require $\frac{1}{2}({\bs x}_Q + {\bs x}_{\bar{Q}}) = {\bs x}$, i.e., the position of a recombined quarkonium is given by the c.m.~position of the heavy quark-antiquark pair. Recombinations from the heavy quark-antiquark pairs with a distance much larger than the Bohr radius $|{\bs x}_Q - {\bs x}_{\bar{Q}} | \gg a_B$ has been shown to be negligible in the last section.

The dissociation rate of the quarkonium state $nls$ from the diagram in Fig.~\ref{subfig:a} is given by
\be \nn
\Gamma_{nls(a)}^{\ma{disso}}({\bs x}, {\bs p}, t) &\equiv& \frac{\ml{C}_{nls(a)}^{-}({\bs x}, {\bs p}, t)}{ f_{nls}({\bs x}, {\bs p}, t) } \\
\label{chap3_eqn_disso_gluon}
&=&  \frac{M}{2\pi} \int \frac{\diff^3 q}{2q(2\pi)^3} n_B(q) \sqrt{M(q-|E_{nl}|)}   \sum | \ml{M}_{(a)} |^2
\ee
The recombination rate of a heavy quark into the quarkonium state $nls$ surrounded by heavy antiquarks with the distribution $f_{\bar{Q}} ({\bs x}_{\bar{Q}}, {\bs p}_{\bar{Q}},t)$ is given by
\be
\label{chap3_eqn_recom_gluon}
\Gamma_{nls(a)}^{\ma{recom}}({\bs x}, {\bs p}, t) \equiv \frac{1}{f_Q({\bs x}_Q, {\bs p}_Q,t) }  \frac{\delta \ml{F}^+_{nls(a)}}{\delta{{\bs p}_Q}} \Big|_{{\bs x}_Q={\bs x},\ {\bs p}_Q={\bs p}}\,.
\ee

\vspace{0.2in}
\subsection{Contributions at the Order $g^2r$}

\subsubsection{Contributions from Diagram \ref{subfig:b}}
\label{subsect: b}
Fig.~\ref{subfig:b} depicts the inelastic scattering with light quarks (up and down) in the medium. The light quarks are assumed massless. First we check the amplitude is independent of the gauge choice. The gauge invariance reflects in the invariance of the amplitude when the $\epsilon_\mu$ in the gluon propagator is replaced with $\epsilon_\mu + q_\mu$. It has been shown in the last section \ref{subsect: a} that the dipole interaction between the singlet and octet is invariant under such a replacement, by virtue of the Ward identity Eq.~(\ref{eqn:ward1}). What remains to show is the invariance of the vertex of the light quark. This is guaranteed by the Dirac equation:
\be
\bar{u}_{s_2}(p_2)\gamma^\mu T^a u_{s_1}(p_1) q_\mu = \bar{u}_{s_2}(p_2)(\slashed{p}_1 - \slashed{p}_2 )T^au_{s_1}(p_1) = 0\,.
\ee
In Coulomb gauge,
\be \nn
i\ml{M}_{(b)} &=&  g^2V_A \sqrt{\frac{T_F}{N_c}} \langle \Psi_{{\bs p}_\ma{rel}} | r^k |  \psi_{nl}  \rangle \\ 
&& \Big[ \frac{-q^0 (\delta^{kl} - \hat{q}^k\hat{q}^l) }{(q^0)^2-{\bs q}^2+i\epsilon}  \bar{u}_{s_2}(p_2) \gamma^l T^a u_{s_1}(p_1)
+ \frac{q^k}{ {\bs q}^2} \bar{u}_{s_2}(p_2)\gamma^0 T^a u_{s_1}(p_1)   \Big]\,.
\ee
We define $|\ml{M}_{(b)}|^2$ summed over the octet color $a$, the spins $s_1,s_2$, colors $i,j$ and flavors of the incoming and outgoing light quarks and include also the contribution from antiquarks as
\be
\label{eqn:collinear_q}
&& \sum |\ml{M}_{(b)}|^2 \equiv \sum_{a,i,j} \sum_{s_1,s_2} \sum_{u,\bar{u},d,\bar{d}} |\ml{M}_{(b)}|^2 \\  \nn
&=& \frac{16}{3}g^4V_A^2T_FC_F  |\langle \Psi_{{\bs p}_\ma{rel}} | {\bs r} |  \psi_{nl}  \rangle|^2 \Big[  \frac{p_1p_2 + {\bs p}_1\cdot {\bs p}_2}{{\bs q}^2} 
 + \frac{2(q^0)^2(p_1p_2-{\bs p}_1\cdot\hat{q} \cdot {\bs p}_2\cdot \hat{q})}{((q^0)^2-{\bs q}^2+i\epsilon)^2} \Big] \,.
\ee

Next we check the infrared sensitivity of the term inside the square bracket. The energy-momentum conservation gives $q^0 = p_1 - p_2 = |{\bs p}_1| - |{\bs p}_2| = |E_{nl}| + p^2_{\ma{rel}}/M$, ${\bs q} = {\bs p}_1 - {\bs p}_2 = {\bs p}_{\ma{cm}}-{\bs k}$. Assuming the angle between ${\bs p}_1$ and ${\bs p}_2$ is $\theta$, we have
\be
\frac{p_1p_2 + {\bs p}_1\cdot {\bs p}_2}{{\bs q}^2} &=& \frac{p_1p_2+p_1p_2\cos\theta}{p_1^2+p_2^2-2p_1p_2\cos\theta} \\
\frac{2(q^0)^2(p_1p_2-{\bs p}_1\cdot\hat{q} \cdot {\bs p}_2\cdot \hat{q})}{((q^0)^2-{\bs q}^2+i\epsilon)^2}  &=& \frac{p_1^2+p_2^2-2p_1p_2}{2p_1p_2(1-\cos\theta)}\times
\frac{p_1^2+p_2^2+p_1p_2(1-\cos\theta)}{p_1^2+p_2^2-2p_1p_2\cos\theta}\,. \ \ \ \ \ \ \ 
\ee
In both terms, there is no soft divergence because the binding energy $|E_{nl}|$ serves as a soft regulator: $p_1^2+p_2^2-2p_1p_2\cos\theta \geq (p_1-p_2)^2 \geq |E_{nl}|^2$. The first term has no collinear divergence either. The collinear divergence happens in the second term when $\cos\theta\rightarrow1$. Physically this occurs when the momenta of both the incoming and outgoing light quarks are in the same direction. The transferred gluon is on-shell. In this case, the inelastic scattering cannot be distinguished from the real gluon process shown in Fig.~\ref{subfig:a}. As we show below in Section~\ref{subsect: lmno}, the interference between the diagram in Fig.~\ref{subfig:a} and its thermal loop correction Fig.~\ref{subfig:l} cancels this collinear divergence.

\subsubsection{Contributions from Diagrams \ref{subfig:c}, \ref{subfig:d}, \ref{subfig:e} and \ref{subfig:f}}
\label{subsect: cdef}
The processes of inelastic scattering with gluons in the medium are depicted in Fig.~\ref{subfig:c}, \ref{subfig:d}, \ref{subfig:e} and \ref{subfig:f}. All the four diagrams are needed for the gauge invariance. First we consider the gauge transformation of the internal gluon line in Fig.~\ref{subfig:c}. If we cut the diagram into two halves by cutting the internal gluon line, we need to show both the upper and lower parts vanish when contracted with $q_\rho$. For the dipole interaction in the lower part, this has been shown by the Ward identity Eq.~(\ref{eqn:ward1}). For the three-gluon vertex in the upper part, it can be shown that
\be
-gf^{abc}(\epsilon^*_1)_\mu (\epsilon_2)_\nu\Big[  g^{\nu\rho}(q_2-q)^\mu + g^{\rho\mu}(q+q_1)^\nu + g^{\mu\nu}(-q_1-q_2)^\rho  \Big] q_{\rho} = 0\,,
\ee
by using $q_1 \cdot \epsilon_1 = q_2 \cdot \epsilon_2 =0$.

Next we consider the gauge transformation of the external gluon line. We fix the internal gluon line to be in the Lorentz gauge.
\be \nn
i\ml{M}_{(c)} &\equiv& i\ml{M}^{\mu\nu}_{(c)} (\epsilon_1^*)_\mu (\epsilon_2)_\nu \\ \nn
&=&-g^2V_A\sqrt{\frac{T_F}{N_c}} f^{abc} \Big[  g^{\nu\rho}(q_2-q)^\mu + g^{\rho\mu}(q+q_1)^\nu + g^{\mu\nu}(-q_1-q_2)^\rho  \Big]  \\
&&\frac{-ig_{\rho\sigma}}{(q_0)^2-{\bs q}^2+i\epsilon} (q^0\delta^{\sigma i} - q^i \delta^{\sigma0})  \langle \Psi_{{\bs p}_\ma{rel}} | r^i |  \psi_{nl}  \rangle (\epsilon_1^*)_\mu (\epsilon_2)_\nu 
\\ \nn
i\ml{M}_{(d)} &\equiv& i\ml{M}^{\mu\nu}_{(d)} (\epsilon_1^*)_\mu (\epsilon_2)_\nu \\ 
&=& ig^2V_A\sqrt{\frac{T_F}{N_c}} f^{abc}\Big[ (\epsilon_1^*)^0 (\epsilon_2)^i - (\epsilon_1^*)^i (\epsilon_2)^0 \Big] \langle \Psi_{{\bs p}_\ma{rel}} | r^i |  \psi_{nl}  \rangle
\\ \nn
i\ml{M}_{(e)} &\equiv& i\ml{M}^{\mu\nu}_{(e)} (\epsilon_1^*)_\mu (\epsilon_2)_\nu \\ \nn
&=& g^2V_A\sqrt{\frac{T_F}{N_c}} f^{abc}(\epsilon_2)^0 \Big[ (q_1)^0 (\epsilon_1^*)^i - (q_1)^i (\epsilon_1^*)^0 \Big]
\langle \Psi_{{\bs p}_\ma{rel}} | r^i |  \psi_{nl}  \rangle \\
&& \frac{i}{E_p-q_2 - \frac{{\bs p}^2_{\ma{rel}}}{M} - \frac{({\bs p}_{\ma{cm}}-{\bs q}_2)^2}{4M} + i\epsilon}
\\ \nn
i\ml{M}_{(f)} &\equiv& i\ml{M}^{\mu\nu}_{(f)} (\epsilon_1^*)_\mu (\epsilon_2)_\nu \\ \nn
&=& g^2V_A\sqrt{\frac{T_F}{N_c}} f^{abc}(\epsilon_1^*)^0 \Big[ (q_2)^0 (\epsilon_2)^i - (q_2)^i (\epsilon_2)^0 \Big]
\langle \Psi_{{\bs p}_\ma{rel}} | r^i |  \psi_{nl}  \rangle \\
&& \frac{i}{E_p-q_1 - \frac{{\bs p}^2_{\ma{rel}}}{M} - \frac{({\bs p}_{\ma{cm}}-{\bs q}_1)^2}{4M} + i\epsilon}\,.
\ee
We show the Ward identity by replacing $(\epsilon_1)_\mu$ with $(q_1)_\mu$:
\be
i(q_1)_\mu \ml{M}^{\mu\nu}_{(c)}  (\epsilon_2)_\nu &=& i g^2V_A\sqrt{\frac{T_F}{N_c}} f^{abc}  \Big[ - q^0 (\epsilon_2)^i + q^i (\epsilon_2)^0 \Big] \langle \Psi_{{\bs p}_\ma{rel}} | r^i |  \psi_{nl}  \rangle\\
i(q_1)_\mu \ml{M}^{\mu\nu}_{(d)}  (\epsilon_2)_\nu &=& i g^2V_A\sqrt{\frac{T_F}{N_c}} f^{abc}  \Big[  (q_1)^0 (\epsilon_2)^i - (q_1)^i (\epsilon_2)^0 \Big] \langle \Psi_{{\bs p}_\ma{rel}} | r^i |  \psi_{nl}  \rangle\\
i(q_1)_\mu \ml{M}^{\mu\nu}_{(e)}  (\epsilon_2)_\nu &=& 0\\ \nn
i(q_1)_\mu \ml{M}^{\mu\nu}_{(f)}  (\epsilon_2)_\nu &=& i g^2V_A\sqrt{\frac{T_F}{N_c}} f^{abc}  \Big[  (q_2)^0 (\epsilon_2)^i - (q_2)^i (\epsilon_2)^0 \Big] \frac{(q_1)^0\langle \Psi_{{\bs p}_\ma{rel}} | r^i |  \psi_{nl}  \rangle}{E_p-q_1 - \frac{{\bs p}^2_{\ma{rel}}}{M} - \frac{({\bs p}_{\ma{cm}}-{\bs q}_1)^2}{4M}} \\ \nn
&=& - i g^2V_A\sqrt{\frac{T_F}{N_c}} f^{abc}  \Big[  (q_2)^0 (\epsilon_2)^i - (q_2)^i (\epsilon_2)^0 \Big] \langle \Psi_{{\bs p}_\ma{rel}} | r^i |  \psi_{nl}  \rangle + \ml{O}(v^2)\,, \\
\ee
where in the last line we have used $E_p = \frac{{\bs p}_{\ma{rel}}^2}{M} + \frac{{\bs p}_{\ma{cm}}^2}{4M}$ and kept only the relative kinetic energy of the octet by neglecting the c.m.~kinetic energy. This is consistent with our power counting. Since $q^\mu = q_1^\mu - q_2^\mu$, the Ward identity is satisfied up to $v^2$-corrections
\be
i(q_1)_\mu \Big[ \ml{M}^{\mu\nu}_{(c)} + \ml{M}^{\mu\nu}_{(d)}+\ml{M}^{\mu\nu}_{(e)}+\ml{M}^{\mu\nu}_{(f)} \Big] (\epsilon_2)_\nu = 0\,.
\ee
Therefore we can compute these diagrams in any gauge we want. In Coulomb gauge, the zero component of the gauge field is not dynamical. So we only need to compute the diagram in Fig.~\ref{subfig:c},
\be \nn
i\ml{M}_{(c)} &=& -g^2V_A\sqrt{\frac{T_F}{N_c}}  (\epsilon_1^*)_\mu (\epsilon_2)_\nu f^{abc}  \Big[  g^{\nu\rho}(q_2-q)^\mu + g^{\rho\mu}(q+q_1)^\nu + g^{\mu\nu}(-q_1-q_2)^\rho  \Big] \\
&&\Big[ \delta_{\rho j} q_0 \frac{i(\delta_{ji} - \hat{q}_j\hat{q}_i)}{(q_0)^2-{\bs q}^2+i\epsilon }  - \delta_{\rho 0} q_i \frac{i}{{\bs q}^2}  \Big]  \langle \Psi_{{\bs p}_\ma{rel}} | r_i |  \psi_{nl}  \rangle\,.
\ee
We define the square of the amplitude magnitude, summed over all color indexes and polarizations:
\be  \nn
&& \sum |\ml{M}_{(c)} |^2 \equiv \sum_{a,b,c}\sum_{\epsilon_1,\epsilon_2}|\ml{M}_{(c)} |^2 \\ \nn
&=& \frac{1}{3}g^4V_A^2C_F | 
\langle \Psi_{{\bs p}_\ma{rel}} | {\bs r} |  \psi_{nl}  \rangle|^2 
\Big[ \frac{1+(\hat{q}_1\cdot\hat{q}_2)^2}{{\bs q}^2} (q_1+q_2)^2 \\ \nn
&& + \frac{1 }{((q_0)^2 - {\bs q}^2 + i\epsilon )^2} P^T_{i_1i_2}({\bs q}_1) P^T_{j_1j_2}({\bs q}_2) P^T_{k_1k_2}({\bs q})\\ \nn
&&  \big( g_{j_1k_1}(q_2-q)_{i_1}  + g_{k_1i_1}(q+q_1)_{j_1} + g_{i_1j_1} (-q_1-q_2)_{k_1}  \big)  \\ 
\label{eqn:collinear_g}
&&  \big( g_{j_2k_2}(q_2-q)_{i_2}  + g_{k_2i_2}(q+q_1)_{j_2} + g_{i_2j_2} (-q_1-q_2)_{k_2}  \big)
 \Big]\,,
\ee
where the transverse polarization tensor is defined as $P^T_{ij}({\bs q}) = \delta_{ij}-\hat{q}_i\hat{q}_j$ and $P^T_{00}=P^T_{0i}=P^T_{i0}=0$. As in the process of inelastic scattering with light quarks, the first term in the square bracket is infrared safe because of the finite binding energy. The second term is collinear divergent when the momenta of the incoming and outgoing gluons are in the same direction. In that case, the transferred gluon is on-shell. As will be shown in Section~\ref{subsect: lmno}, the interference between the diagrams in Figs.~\ref{subfig:a} and~\ref{subfig:m} will cancel this divergence.

\subsubsection{Contributions from Diagrams \ref{subfig:g}, \ref{subfig:h}, \ref{subfig:i}, \ref{subfig:j} and \ref{subfig:k}}
\label{subsect: ghijk}
The diagrams in Figs.~\ref{subfig:g}, \ref{subfig:h}, \ref{subfig:i}, \ref{subfig:j} and \ref{subfig:k} describe the processes of $l+\bar{l}+H\leftrightarrow Q+\bar{Q}$ (here $l$ denotes a light quark and $H$ denotes a quarkonium state) and $g+g+H\leftrightarrow Q+\bar{Q}$. They can be computed similarly as in Sections~\ref {subsect: b} and \ref{subsect: cdef}. However, their contributions to the dissociation and recombination in the Boltzmann equation are much smaller than those from Figs.~\ref{subfig:b}, \ref{subfig:c}, \ref{subfig:d}, \ref{subfig:e} and \ref{subfig:f} because of the limited phase space of the incoming particles. In Coulomb gauge, we only need to consider Figs.~\ref{subfig:g} and \ref{subfig:h}. The energy transferred via the internal gluon is fixed by $q^0=|E_{nl}| + {\bs p}_\ma{rel}^2/M$ and $p_\ma{rel}\sim Mv$ otherwise the dipole transition between wave functions $|\langle \Psi_{{\bs p}_\ma{rel}} | r_i |  \psi_{nl}  \rangle|^2$ vanishes. The phase spaces constrained by $p_1+p_2 = q^0$ in Fig.~\ref{subfig:g} and $q_1+q_2 = q^0$ in Fig.~\ref{subfig:h} are much smaller than those of $p_1-p_2 = q^0$ in Fig.~\ref{subfig:b} and $q_1-q_2 = q^0$ in Fig.~\ref{subfig:c}. The suppression of processes with two incoming light quarks or gluons has been noted before \cite{Song:2012at}.

\subsubsection{Contributions from Diagrams \ref{subfig:l}, \ref{subfig:m}, \ref{subfig:n} and \ref{subfig:o}}
\label{subsect: lmno}
These diagrams are the one-loop corrections of the gluon propagator. If resummed, they will give a thermal mass to the in-medium gluon. The loop correction is at the order $g^2$ so the whole diagram is at the order $g^3r$. Their interference with the diagram in Fig.~\ref{subfig:a} will give contributions equivalent to amplitudes at the order $g^2r$. We will show the interference cancels the collinear divergence in Eqs.~(\ref{eqn:collinear_q}) and (\ref{eqn:collinear_g}). Thus there is no need to resum these diagrams here.

In Coulomb gauge,
\be
i\ml{M}_{(a)} &=& gV_A\sqrt{\frac{T_F}{N_c}}q_0\epsilon^*_i  \langle \Psi_{{\bs p}_\ma{rel}} | r_i |  \psi_{nl}  \rangle \delta^{ab} \\
\label{eqn:M_l}
i\ml{M}_{(l)} &=&  gV_A\sqrt{\frac{T_F}{N_c}}\epsilon^*_i  \langle \Psi_{{\bs p}_\ma{rel}} | r_k |  \psi_{nl}  \rangle \delta^{ab} \Big[ i\Pi^{(l)}_{ij} \frac{iq_0(\delta_{jk}-\hat{q}_j\hat{q}_k)}{q_0^2-{\bs q}^2 + i\epsilon}   - i\Pi^{(l)}_{i0}\frac{iq_k}{{\bs q}^2}    \Big]\,, \ \ \ \ \ 
\ee
where we set $q_0=|\bs q| \equiv q$ for the on-shell massless particle and $\Pi^{(l)}_{\mu\nu}$ is the time-ordered gluon polarization tensor contributed from the fermion loop. The time-ordered gluon propagator and polarization at finite temperature cannot be directly obtained by analytically continuing the imaginary time propagator and polarization. The time-ordered propagators and polarizations can only be obtained from the retarded and advanced ones via the relation
\be \nn
D_{\ml{T}}(q_0,{\bs q})   &=& \frac{1}{2}\big( D_R(q_0,{\bs q}) + D_A(q_0,{\bs q})    \big)  + \big( \frac{1}{2}+n_B(q_0) \big) \big( D_R(q_0,{\bs q}) - D_A(q_0,{\bs q}) \big)\\
\\ \nn
\Pi_{\ml{T}}(q_0,{\bs q}) &=& \frac{1}{2}\big( \Pi_R(q_0,{\bs q}) + \Pi_A(q_0,{\bs q}) \big)  + \big( \frac{1}{2}+n_B(q_0) \big) \big( \Pi_R(q_0,{\bs q}) - \Pi_A(q_0,{\bs q}) \big)\,. \\
\ee
We will focus on the first term $\frac{D_R(q_0,{\bs q}) + D_A(q_0,{\bs q})}{2}$ because this term contributes to the dissociation rate when the gluon is on-shell. The second term contributes to the dissociation rate when the gluon has space-like momentum, which corresponds to the inelastic scattering and has been accounted for above. For a more detailed discussion on this, see Ref.~\cite{Brambilla:2013dpa}. The fermion loop gives:
\be
i\Pi^{(l)}_{\mu\nu}(q_0, {\bs q}) = -g^2T_F\sum_{\ma{flavor}}\int\frac{\diff^4 l}{(2\pi)^4} \frac{\Tr{(\gamma_\mu(\slashed{l}+\slashed{q})\gamma_\nu \slashed{l}})}{ (l_0^2-{\bs l}^2)^2 ((l_0+q_0)^2-({\bs l}+{\bs q})^2)^2}\,.
\ee
In the imaginary formalism of thermal field theory, the integral over $l_0$ is a summation in Matsubara frequency. After the summation
\be \nn
\label{eqn:pi_q}
\Pi_{ij}^{(l)}(q_0=q, {\bs q}) &=& g^2T_F\sum_{\ma{flavor}} \int\frac{\diff^3 l}{(2\pi)^3} \frac{\Tr{(\gamma_i (\slashed{l}+\slashed{q})\gamma_j \slashed{l}})}{4E_1E_2} \\ \nn
&& \Big[  (1-n_F(E_1)-n_F(E_2) ) \Big(\frac{1}{q-E_1-E_2} - \frac{1}{q+E_1+E_2} \Big) \\
&& -  (n_F(E_1)-n_F(E_2) ) \Big(\frac{1}{q+E_1-E_2} - \frac{1}{q-E_1+E_2} \Big)
\Big]\,, \ \ \ \ \ 
\ee
where $E_1 = |{\bs l}+{\bs q}|$, $E_2 = |{\bs l}|$. Here we only need $\frac{\Pi_R(q_0,{\bs q}) + \Pi_A(q_0,{\bs q})}{2} = \Re{\Pi_R(q_0,{\bs q})}$ and we do not need to analytically continue. We can just plug it into (\ref{eqn:M_l}).
To see the cancellation of collinear divergence, we define the interference term summed over colors and gluon polarizations
\be \nn
&& \sum (\ml{M}_{(a)}^*\ml{M}_{(l)} + \ml{M}_{(a)} \ml{M}_{(l)}^*) \equiv \sum_\epsilon \sum_{a,b} (\ml{M}_{(a)}^*\ml{M}_{(l)} + \ml{M}_{(a)} \ml{M}_{(l)}^*) \\
&=& -\frac{2}{3}g^2V_A^2C_F q_0^2  | \langle \Psi_{{\bs p}_\ma{rel}} | {\bs r} |  \psi_{nl}  \rangle |^2  \frac{\delta_{ij}-\hat{q}_i\hat{q}_j}{q_0^2-{\bs q}^2 } \Pi^{(l)}_{ij}\,,
\ee
and consider the following integral that is used in the dissociation in the case of two flavors (up and down quarks)
\be
I_1 \equiv  \int\frac{\diff^3 q}{2q (2\pi)^3} n_B(q)\sum (\ml{M}_{(a)}^*\ml{M}_{(l)} + \ml{M}_{(a)} \ml{M}_{(l)}^*)\,.
\ee
We focus on the term with $(n_F(E_1)-n_F(E_2))$ in the square bracket in Eq. (\ref{eqn:pi_q}). Under a change of variables ${\bs p}_1 = {\bs l}+{\bs q}$, ${\bs p}_2 = {\bs l}$
\be \nn
I_1 &\equiv&  \int\frac{\diff^3 p_1}{2p_1 (2\pi)^3} \int\frac{\diff^3 p_2}{2p_2 (2\pi)^3} n_B(|{\bs p}_1-{\bs p}_2|) (n_F(p_1) - n_F(p_2)) \\ 
&& \frac{16}{3}g^4V_A^2T_FC_F\frac{2(p_1-p_2)|{\bs p}_1-{\bs p}_2|}{((p_1-p_2)^2-({\bs p}_1-{\bs p}_2)^2)^2} (p_1p_2-{\bs p}_1\cdot\hat{q}{\bs p}_2\cdot\hat{q}) + \cdots\,, \ \ \ \ 
\ee
where $\cdots$ means the first term in the square bracket in Eq.~(\ref{eqn:pi_q}). In the collinear limit, ${\bs p}_1$ and ${\bs p}_2$ are in the same direction, so $p_1-p_2 = |{\bs p}_1-{\bs p}_2|$. With $n_B(p_1-p_2) (n_F(p_1) - n_F(p_2)) = -n_F(p_1)(1-n_F(p_2))$, one immediately sees $I_1$ cancels the collinear divergence in
\be
\int \frac{\diff^3 p_1}{2p_1(2\pi)^3} \frac{\diff^3 p_2}{2p_2(2\pi)^3}  n_F(p_1)(1-n_F(p_2)) \sum |\ml{M}_{(b)}|^2\,,
\ee
where $|\ml{M}_{(b)}|^2$ is given in Eq.~(\ref{eqn:collinear_q}). This is for the dissociation contribution from $|\ml{M}_{(b)}|^2$. The cancellation of the collinear divergence in the recombination can be shown similarly.

We next consider the interference between the diagrams in Figs.~\ref{subfig:a} and \ref{subfig:m}. The amplitude $i \ml{M}_{(m)}$ is exactly the same as $i\ml{M}_{(l)}$ under the replacement of $\Pi^{(l)}$ with $\Pi^{(m)}$. $\Pi^{(m)}_{\mu\nu}$ is the gluon polarization tensor in Fig.~\ref{subfig:m}. After summing over the Matsubara frequencies
\be \nn
\Pi^{(m)}_{ij}(q_0=q,{\bs q}) &=& \frac{1}{2}g^2T_A \int\frac{\diff^3l }{(2\pi)^3} \frac{1}{4E_1E_2} P^T_{i_1i_2}({\bs l}+{\bs q}) P^T_{j_1j_2}({\bs l})\\ \nn
&& \Big[ (1+n_B(E_1)+n_B(E_2)) \Big(  \frac{1}{q+E_1+E_2} - \frac{1}{q-E_1-E_2} \Big) \\ \nn
&& -(n_B(E_1)-n_B(E_2)) \Big( \frac{1}{q+E_1-E_2} - \frac{1}{q-E_1+E_2}   \Big)   \Big]  \\ \nn
&& \Big[ g_{j_1i}(l-q)_{i_1}  + g_{ii_1}(2q+l)_{j_1} + g_{i_1j_1} (-q-2l)_{i}   \Big] \\
&& \Big[ g_{j_2j}(l-q)_{i_2}  + g_{ji_2}(2q+l)_{j_2} + g_{i_2j_2} (-q-2l)_{j}  \Big]\,,
\ee
where $T_A=N_c$, $E_1 = |{\bs l}+{\bs q}|$ and $E_2 = |{\bs l}|$. We will focus on the term with $(n_B(E_1)-n_B(E_2))$ and neglect the other term. Under a change of variables ${\bs q}_1={\bs l}+{\bs q}$, ${\bs q}_2={\bs l}$, we can show that the collinear divergence of the integral
\be
I_2 \equiv  \int\frac{\diff^3 q}{2q (2\pi)^3} n_B(q)\sum (\ml{M}_{(a)}^*\ml{M}_{(m)} + \ml{M}_{(a)} \ml{M}_{(m)}^*)\,,
\ee
cancels the collinear divergence in 
\be
\int \frac{\diff^3 q_1}{2q_1(2\pi)^3} \frac{\diff^3 q_2}{2q_2(2\pi)^3}  n_B(q_1)(1+n_B(q_2)) \sum |\ml{M}_{(c)}|^2\,,
\ee
by the virtual of $n_B(q_1-q_2)(n_B(q_1)-n_B(q_2)) = -n_B(q_1)(1+n_B(q_2))$. Cancellation of the divergence in the recombination process can be similarly shown.

\subsubsection{Contributions from Diagrams \ref{subfig:p}, \ref{subfig:q}, \ref{subfig:r} and \ref{subfig:s}}
\label{subsect: pqrs}
These diagrams are the one-loop corrections to the dipole interaction between the singlet and the octet. The correction is at the order $g^3r$ but its interference with the diagram in Fig.~\ref{subfig:a} gives contributions equivalent to amplitudes at the order $g^2r$. We will use the Lorentz gauge for convenience in the calculation (we have shown that the amplitudes satisfy the Ward identity) and dimensional regularization. We will show the detail of computing the diagram \ref{subfig:p} here and list the results of other diagrams. Using the Feynman rules explained in Chapter 1, we can write the diagram \ref{subfig:p} as
\be \nn
i\ml{M}_{(p)} &=& g^3V_A\sqrt{\frac{T_F}{N_c}}\epsilon_\mu^*f^{abc}f^{bcd}   \langle \Psi_{{\bs p}_\ma{rel}}  | r^i |  \psi_{nl}  \rangle \int \frac{\diff^4\ell}{(2\pi)^4} \big[  (q^0 - \ell^0) \delta_\nu^{\,i} - (q^i - \ell^i) \delta_\nu^{\,0} \big] \\  \nn
&& \big[ g^{\mu\nu}(2q-\ell)^0 + g^{\nu0}(-q+2\ell)^\mu + g^{0\mu}(-\ell - q)^\nu \big]  \\
&& \frac{-i}{\ell^2 + i\epsilon} \frac{-i}{(q-l)^2 + i\epsilon} \frac{i}{E_p - \ell^0 - \frac{{\bs p}_\ma{rel}^2}{M} - \frac{({\bs p}_\ma{cm} - {\bs \ell} )^2}{4M} + i\epsilon}  \,.
\ee
Using $f^{abc}f^{bcd} = T_A\delta^{ad}$ where $T_A=N_c$, $E_p = \frac{{\bs p}_\ma{rel}^2}{M} + \frac{{\bs p}_\ma{cm}^2}{4M}$ and the Feynman trick, we find
\be 
i\ml{M}_{(p)} &=&  -ig^3V_A T_A \sqrt{\frac{T_F}{N_c}} \delta^{ad}  \langle \Psi_{{\bs p}_\ma{rel}}  | r^i |  \psi_{nl}  \rangle I \\ 
I &\equiv&   \int \frac{\diff^4\ell}{(2\pi)^4} \int_0^1 \diff x \frac{1}{(\ell^2 - 2x\ell\cdot q + xq^2 + i\epsilon)^2} \frac{1}{-\ell^0 - \frac{{\bs \ell}^2 - 2{\bs \ell}\cdot {\bs p}_\ma{cm}}{4M} + i\epsilon} \\ \nn
&& \Big\{ - \epsilon_0^*(q^0-\ell^0)(q^i+\ell^i) + \epsilon_j^*\big[  g^{ji}(q^0-\ell^0)(2q^0-\ell^0) - (q^i-\ell^i)(2\ell^j - q^j) \big] \Big\} \,.
\ee
Replacing the loop momentum $\ell$ with $\ell \to \ell + xq$ and using the on-shell condition $q^2=0$, we can get
\be \nn
I &=& \int \frac{\diff^4\ell}{(2\pi)^4} \int_0^1 \diff x \frac{1}{(\ell^2 + i\epsilon)^2} \frac{1}{-\ell^0 - xq^0 - \frac{({\bs \ell}+x{\bs q})^2 - 2({\bs \ell}+x{\bs q})\cdot {\bs p}_\ma{cm}}{4M} + i\epsilon} \\ \nn
&& \Big\{ \epsilon_0^*(- q^0 + \ell^0 + xq^0)(q^i+\ell^i + xq^i) + \epsilon_j^*\big[  g^{ji}(q^0-\ell^0-xq^0)(2q^0-\ell^0-xq^0) \\
&& - (q^i-\ell^i-xq^i)(2\ell^j +2xq^j - q^j) \big] \Big\} \,.
\ee
We will do the integration over $\ell^0$ by closing the contour in the lower half plane. We will have a double pole at $\ell^0 = |{\bs \ell}| - i\epsilon = \ell - i\epsilon$. For notational simplicity, we define $D\equiv xq^0 + \frac{({\bs \ell}+x{\bs q})^2 - 2({\bs \ell}+x{\bs q})\cdot {\bs p}_\ma{cm}}{4M}$. The residue theorem gives
\be \nn
I &=& -i \int_0^1 \diff x \int \frac{\diff^3\ell}{(2\pi)^3} \bigg\{ \Big[ \frac{1}{4\ell^3}\frac{1}{\ell + D} + \frac{1}{4\ell^2}\frac{1}{(\ell+D)^2} \Big] \Big[ \epsilon_0^*(- q^0 + \ell^0 + xq^0)(q^i+\ell^i + xq^i) \\ \nn
&& + \epsilon_j^*\big[  g^{ji}(q^0-\ell^0-xq^0)(2q^0-\ell^0-xq^0) - (q^i-\ell^i-xq^i)(2\ell^j +2xq^j - q^j) \big] \Big]
\\
&& - \frac{1}{4\ell^2} \frac{1}{\ell+D} \big[ \epsilon_0^*(q^i + xq^i +\ell^i) + \epsilon_j^*g^{ji} ( -3q^0+2xq^0 + 2\ell )    \big]
\bigg\} \,.
\ee
Now we use the dimensional regularization and regularize the integral $I$ in $d=3-\epsilon$ (the $\epsilon$ in the dimensional regularization should be distinguished from the incoming gluon polarization $\epsilon^{*\mu}$). We will focus on the divergent parts in the integral. So we write the first two terms with $\frac{1}{4\ell^3}\frac{1}{\ell + D} + \frac{1}{4\ell^2}\frac{1}{(\ell+D)^2}$ as
\be \nn
I_{1,2} &\equiv& \int \frac{\diff^d\ell}{(2\pi)^d} \Big[ \frac{1}{4\ell^3}\frac{1}{\ell + D} + \frac{1}{4\ell^2}\frac{1}{(\ell+D)^2} \Big] \Big[   \epsilon_0^*\big[ -q^0\ell^i(1-x) +\ell q^i(1+x) + \ell \ell^i \big] \\
& +&  \epsilon_j^* \big[  -g^{ji} \ell q^0(3-2x) + \ell^iq^j(2x-1) - 2\ell^jq^i(1-x) + g^{ji}\ell^2 + 2\ell^i\ell^j   \big] \Big]\,.
\ee
Using 
\be
\int \diff^d \ell\, \ell^i f(\ell) &=& 0 \\
\int \diff^d \ell\, \ell^i \ell^j f(\ell) &=& \frac{\delta^{ij}}{3} \int \diff^d \ell\, \ell^2f(\ell) \,,
\ee
we can show
\be
I_{1,2} = \frac{1}{8\pi^2\epsilon} \big[   \epsilon^{0*}q^i (2+2x) + \epsilon^{i*}q^0 (-6+3x)   \big] \,.
\ee
The last term with $ - \frac{1}{4\ell^2} \frac{1}{\ell+D}$, up to finite corrections, can be written as 
\be \nn
I_3 &\equiv& - \int \frac{\diff^d\ell}{(2\pi)^d} \frac{1}{4\ell^2} \frac{1}{\ell+D} \big[ \epsilon_0^*(q^i + xq^i +\ell^i) + \epsilon_j^*g^{ji} ( -3q^0+2xq^0 + 2\ell )    \big] \\
&=& -\frac{1}{8\pi^2\epsilon} \big[ \epsilon^{0*}q^i(1+x) - 3 \epsilon^{i*}q^0  \big] \,.
\ee
Therefore the divergent part of $I$ is
\be
I = \frac{1}{8\pi^2\epsilon} \big[ \epsilon^{0*}q^i(1+x)  + \epsilon^{i*}q^0 (-3+3x)  \big] \,.
\ee
After the trivial integration over $x$, we obtain
\be
i\ml{M}_{(p)} =  - \frac{3g^3}{16\pi^2\epsilon} V_A T_A\sqrt{\frac{T_F}{N_c}} \delta^{ad} \langle \Psi_{{\bs p}_\ma{rel}} | r^i |  \psi_{nl}  \rangle 
(\epsilon^{*0} q^i - \epsilon^{*i} q^0 ) +\cdots \,,
\ee
where finite corrections are omitted. 

Similarly we can compute the divergent part of the other three diagrams to obtain
\be
i\ml{M}_{(q)} &=&  \frac{3g^3}{16\pi^2\epsilon}  V_A T_A\sqrt{\frac{T_F}{N_c}} \delta^{ad} \langle \Psi_{{\bs p}_\ma{rel}} | r^i |  \psi_{nl}  \rangle 
(\epsilon^{*0} q^i - \epsilon^{*i} q^0 ) +\cdots\\
i\ml{M}_{(r)} &=& 0 +\cdots\\
i\ml{M}_{(s)} &=& 0 +\cdots\,,
\ee
where the off-shell scheme has been used to extract the logarithmic divergence. Only the terms with the $\epsilon$ poles are shown. Finite terms and corrections at higher orders in $v^2$ are omitted. Therefore
\be
i\ml{M}_{(p)} + i\ml{M}_{(q)} + i\ml{M}_{(r)} + i\ml{M}_{(s)} = \frac{0}{\epsilon} + \cdots\,,
\ee
which means the dipole interaction term between the singlet and octet is independent of scale at the one-loop level
\be
\frac{\diff}{ \diff \mu} V_A(\mu) = 0\,.
\ee
This has already been shown in Ref. \cite{Pineda:2000gza}. From the matching condition, Eq.~(\ref{eqn:match}), we may set $V_A=1$ in the following, no matter the scale involved.

\subsubsection{Contributions from Diagram \ref{subfig:t} }
\label{subsect: t}
The diagram in Fig.~\ref{subfig:t} describes the one-loop correction to the octet propagator at the order $g^2$. The whole diagram is at the order $g^3r$ but its interference with the diagram \ref{subfig:a} is equivalent to an amplitude at the order $g^2r$. The one-loop correction to the octet propagator is given by
\be
L_{o} \equiv g^2f^{abc}f^{cbd} \int\frac{\diff^4l}{(2\pi)^4} \frac{-i}{l_0^2-{\bs l}^2 + i\epsilon} \frac{i}{E_p - l_0 - \frac{{\bs p}_\ma{rel}^2}{M} - \frac{({\bs p}_\ma{cm}-{\bs l})^2}{4M} + i\epsilon}\,,
\ee
where $E_p = \frac{{\bs p}_\ma{rel}^2}{M} + \frac{ {\bs p}_\ma{cm}^2 }{4M} $ is the energy of the external octet field.
We first integrate over $l_0$ by closing the contour in the lower half plane
\be \nn
L_{o} &=& i g^2N_c \delta^{ad} \int\frac{\diff^3l}{(2\pi)^3} \frac{1}{2l} \frac{1}{l+   (l^2 - 2 {\bs l} \cdot {\bs P}_\ma{cm})/(4M) } \\
&=& \frac{ig^2N_c \delta^{ad} M}{4\pi^2 P_\ma{cm}} \int \diff l \ln{\frac{l+4M+2P_\ma{cm}}{l+4M-2P_\ma{cm}}}\,.
\ee
If we expand the integrand in powers of $1/M$ (in our power counting, $v^2 \sim \frac{P_\ma{cm}}{M}$) and use dimensional regularization\footnote{Similar argument has been used in Ref.~\cite{Bodwin:1994jh}. See the appendix therein.}, we find
\be
L_{o} = 0 + \ml{O}(v^2)\,.
\ee
The power divergence is proportional to $v^2$ and neglected here, consistent with the power counting.

\subsubsection{Summary}
To write the dissociation and recombination terms in the Boltzmann equations explicitly, we define $\ml{F}^\pm_{nls(b)}$ and $\ml{F}^\pm_{nls(c)}$ as
\be \nn
\ml{F}^+_{nls(b)} &\equiv&  g_+ \int \frac{\diff^3 k}{(2\pi)^3}  \frac{\diff^3 p_{Q}}{(2\pi)^3}  \frac{\diff^3 p_{\bar{Q}} }{(2\pi)^3} \frac{\diff^3 p_1}{2p_1(2\pi)^3} \frac{\diff^3 p_2}{2p_2(2\pi)^3}  \\ \nn
&& n_F(p_2)(1-n_F(p_1)) f_Q({\bs x}_Q, {\bs p}_Q,t) f_{\bar{Q}} ({\bs x}_{\bar{Q}}, {\bs p}_{\bar{Q}},t) (2\pi)^4 \\
\label{chap3_eqn_F+_ineq}
&& \delta^3({\bs k} + {\bs p}_1 - {\bs p}_{\ma{cm}} - {\bs p}_2) \delta(-|E_{nl}|+p_1-\frac{p^2_{\ma{rel}}}{M}-p_2)  \sum | \ml{M}_{(b)} |^2   \\ \nn
\ml{F}^-_{nls(b)}  &\equiv&  \int \frac{\diff^3 k}{(2\pi)^3}  \frac{\diff^3 p_{\ma{cm}}}{(2\pi)^3}  \frac{\diff^3 p_{\ma{rel}} }{(2\pi)^3} \frac{\diff^3 p_1}{2p_1(2\pi)^3} \frac{\diff^3 p_2}{2p_2(2\pi)^3}  \\ \nn
&& n_F(p_1)(1-n_F(p_2))  f_{nls}({\bs x}, {\bs k}, t) (2\pi)^4 \\
\label{chap3_eqn_F+_ineg}
&& \delta^3({\bs k} + {\bs p}_1 - {\bs p}_{\ma{cm}} - {\bs p}_2) \delta(-|E_{nl}|+p_1-\frac{p^2_{\ma{rel}}}{M}-p_2)  \sum | \ml{M}_{(b)} |^2 \\ \nn
\ml{F}^+_{nls(c)} &\equiv&  g_+ \int \frac{\diff^3 k}{(2\pi)^3}  \frac{\diff^3 p_{Q}}{(2\pi)^3}  \frac{\diff^3 p_{\bar{Q}} }{(2\pi)^3} \frac{\diff^3 q_1}{2q_1(2\pi)^3} \frac{\diff^3 q_2}{2q_2(2\pi)^3}  \\ \nn
&& n_B(q_2)(1+n_B(q_1)) f_Q({\bs x}_Q, {\bs p}_Q,t) f_{\bar{Q}} ({\bs x}_{\bar{Q}}, {\bs p}_{\bar{Q}},t)   (2\pi)^4  \\
&& \delta^3({\bs k} + {\bs q}_1 - {\bs p}_{\ma{cm}} - {\bs q}_2) \delta(-|E_{nl}|+q_1-\frac{p^2_{\ma{rel}}}{M}-q_2)  \sum | \ml{M}_{(c)} |^2   \\ \nn
\ml{F}^-_{nls(c)}  &\equiv&  \int \frac{\diff^3 k}{(2\pi)^3}  \frac{\diff^3 p_{\ma{cm}}}{(2\pi)^3}  \frac{\diff^3 p_{\ma{rel}} }{(2\pi)^3} \frac{\diff^3 q_1}{2q_1(2\pi)^3} \frac{\diff^3 q_2}{2q_2(2\pi)^3}  \\ \nn
&& n_B(q_1)(1+n_B(q_2))  f_{nls}({\bs x}, {\bs k}, t)   (2\pi)^4  \\
&& \delta^3({\bs k} + {\bs q}_1 - {\bs p}_{\ma{cm}} - {\bs q}_2) \delta(-|E_{nl}|+q_1-\frac{p^2_{\ma{rel}}}{M}-q_2)  \sum | \ml{M}_{(c)} |^2 \,, \ \ \ \ 
\ee
The $g_+$ factor and the relation between ${\bs p}_{\ma{cm}}$, ${\bs p}_{\ma{rel}}$ and ${\bs p}_{Q}$, ${\bs p}_{\bar{Q}}$ are defined in Section~\ref{subsect: a}. The collinear divergent parts in the square of amplitudes have been shown to be cancelled by the interference between the tree-level process of gluon absorption/emission and its one-loop corrections. After regularization, we can drop the terms that are originally collinear divergent if they are small, so we can write (by setting $V_A=1$)
\be
\sum | \ml{M}_{(b)} |^2  &=& \frac{16}{3}g^4T_FC_F  |\langle \Psi_{{\bs p}_\ma{rel}} | {\bs r} |  \psi_{nl}  \rangle|^2   \frac{p_1p_2 + {\bs p}_1\cdot {\bs p}_2}{{\bs q}^2} \\
\sum | \ml{M}_{(c)} |^2  &=& \frac{1}{3}g^4C_F | \langle \Psi_{{\bs p}_\ma{rel}} | {\bs r} |  \psi_{nl}  \rangle|^2 
 \frac{1+(\hat{q}_1\cdot\hat{q}_2)^2}{{\bs q}^2} (q_1+q_2)^2\,.
\ee
The dissociation and recombination terms in the Boltzmann equation from the inelastic scattering with light quarks and gluons are given by
\be
\ml{C}_{nls,\ma{inel}}^{\pm}({\bs x}, {\bs p}, t) &=&  \frac{\delta \ml{F}^\pm_{nls(b)}}{\delta{{\bs k}}}  \Big|_{{\bs k}={\bs p}} + \frac{\delta \ml{F}^\pm_{nls(c)}}{\delta{{\bs k}}}  \Big|_{{\bs k}={\bs p}}\,.
\ee
The dissociation rate of the quarkonium state $nls$ from the inelastic scattering is given by
\be
\label{chap3_eqn_disso_inel}
\Gamma_{nls,\ma{inel}}^{\ma{disso}}({\bs x}, {\bs p}, t) \equiv \frac{\ml{C}_{nls,\ma{inel}}^{-}({\bs x}, {\bs p}, t)}{ f_{nls}({\bs x}, {\bs p}, t) }\,.
\ee
The recombination rate of a heavy quark into the quarkonium state $nls$ surrounded by heavy antiquarks with the distribution $f_{\bar{Q}} ({\bs x}_{\bar{Q}}, {\bs p}_{\bar{Q}},t)$ is given by
\be
\label{chap3_eqn_recom_inel}
\Gamma_{nls,\ma{inel}}^{\ma{recom}}({\bs x}, {\bs p}, t) \equiv \frac{1}{f_Q({\bs x}_Q, {\bs p}_Q,t) }  \frac{\delta (\ml{F}^+_{nls(b)}+\ml{F}^+_{nls(c)}) }{\delta{{\bs p}_Q}} \Big|_{{\bs x}_Q={\bs x},\ {\bs p}_Q={\bs p}}\,.
\ee

\vspace{0.2in}
\subsection{Dipole Matrix Elements}
For practical calculations, one needs to calculate the dipole matrix elements given by
\be
|\langle \Psi_{{\bs p}_\ma{rel}} | {\bs r} |  \psi_{nl}  \rangle|^2 \,.
\ee
We will first consider the case where the final $Q\bar{Q}$ wave function is a plane wave $|\Psi_{{\bs p}_\ma{rel}}\rangle = | {\bs p}_\ma{rel} \rangle$. We will demonstrate the use of the spherical tensor operators and then apply this method to the case where the final $Q\bar{Q}$ wave function is a Coulomb scattering wave.

\subsubsection{Plane Wave}
For a Coulomb potential $V_s = -\frac{C_F\alpha_s}{r}$, the 1S bound state wave function is
\be
\psi_{\ma{1S}}({\bs r}) &=& R_{10}(r)Y_{0}^{\,0}(\theta,\phi) = 2 a_B^{-3/2} e^{-r/a_B} \sqrt{ \frac{1}{4\pi} }\,,
\ee
where the Bohr radius is $a_B = \frac{2}{C_F\alpha_sM}$ and the spherical harmonics $Y_{\ell}^{\,m}$ is defined in terms of the associated Legendre polynomial $P_{\ell}^{\,m}$
\be
Y_{\ell}^{\,m}(\theta, \phi)  &=& (-1)^m\sqrt{\frac{(2\ell+1)}{4\pi}\frac{(\ell - m)!}{(\ell+m)!}}  P_{\ell}^{\,m}(\cos\theta) e^{im\phi} \\
P_{\ell}^{\,m}(x)   &=&  \frac{(-1)^m}{2^\ell \ell! } (1-x^2)^{m/2} \frac{\diff^{\ell+m}}{\diff x^{\ell+m} } (x^2-1)^\ell \,.
\ee
We can compute the dipole matrix element directly without using the method of spherical tensor operators
\be
\langle {\bs p}_\ma{rel}  | {\bs r} |  \psi_\ma{1S}  \rangle = \int \diff^3r e^{-ip_\ma{rel}r\cos\theta} (r\sin\theta\cos\phi,\, r\sin\theta\sin\phi,\, r\cos\theta) \frac{e^{-r/a_B} }{\sqrt{\pi}a_B^{3/2}} \,.
\ee
Only the $z$-component will survive after the integration over $\phi$.
\be \nn
\langle {\bs p}_\ma{rel}  | z |  \psi_\ma{1S}  \rangle &=& \frac{2\sqrt{\pi}}{a_B^{3/2}}\int r^2\diff r \, e^{-r/a_B} \frac{ -i p_\ma{rel} r(e^{-i p_\ma{rel} r} + e^{i p_\ma{rel} r})  -(e^{-i p_\ma{rel} r} - e^{i p_\ma{rel} r} )}{(-i p_\ma{rel} r)^2} \\
&=& \frac{32i\sqrt{\pi} a_B^{7/2}  p_\ma{rel}  }{ (1+a_B^2p_\ma{rel}^2)^3  } \,.
\ee
So we have
\be \nn
|\langle {\bs p}_\ma{rel}  | {\bs r} |  \psi_\ma{1S}  \rangle|^2 &=& |\langle {\bs p}_\ma{rel}  | x |  \psi_\ma{1S}  \rangle|^2 +  |\langle {\bs p}_\ma{rel}  | y |  \psi_\ma{1S}  \rangle|^2 +  |\langle {\bs p}_\ma{rel}  | z |  \psi_\ma{1S}  \rangle|^2 \\ \nn
&=&   |\langle {\bs p}_\ma{rel}  | z |  \psi_\ma{1S}  \rangle|^2 \\
\label{chap3_eqn_matrix1S_direct}
&=&  \frac{2^{10}\pi a_B^7 p_\ma{rel}^2 }{(1+a_B^2p_\ma{rel}^2)^6  } \,.
\ee
Similarly for the 2S bound state and unbound plane wave, we have
\be
|\langle {\bs p}_\ma{rel}  | {\bs r} |  \psi_\ma{2S}  \rangle|^2 = \frac{2^{21}\pi a_B^7 p_\ma{rel}^2 (1-2a_B^2p_\ma{rel}^2)^2}{  (1+4a_B^2p_\ma{rel}^2)^8  } \,.
\ee

Now we will use the method of spherical tensor operators and repeat the above calculations. This method will be useful later when we deal with the Coulomb scattering wave as the unbound $Q\bar{Q}$ wave function. First we introduce the spherical tensor operators representing ${\bs r}$
\be
r_+ &\equiv & \sqrt{\frac{4\pi}{3}} r Y_{1}^{\,1} \\
r_- &\equiv & \sqrt{\frac{4\pi}{3}} r Y_{1}^{\,-1} \\
r_0 &\equiv & \sqrt{\frac{4\pi}{3}} r Y_{1}^{\,0} \,,
\ee
such that ${\bs r}^2 = |r_+|^2 + |r_-|^2 + |r_0|^2 $. The Wigner-Eckart theorem states
\be
\langle j\,j_z | T_{qq_z} | j'\, j_z' \rangle = \langle j\,j_z;\, q\,q_z | j'\,j_z' \rangle \langle j || T_q || j' \rangle \,,
\ee
where $T_{qq_z}$ is a spherical tensor operator with the angular momentum $q$ and third component $q_z$. $\langle j\,j_z;\, q\,q_z | j'\,j_z' \rangle$ is the Clebsch-Gordan coefficient of adding two angular momenta $j\,j_z$ and $q\,q_z$ to obtain another angular momentum $j'\,j_z'$. $\langle j || T_q || j' \rangle$ is a reduced matrix element of just radial wave functions. Using spherical harmonics, the plane wave can be written as
\be
\langle {\bs r} | {\bs p}_\ma{rel} \rangle = 4\pi \sum_{\ell,m} i^\ell j_\ell( p_\ma{rel} r) Y_\ell^{\,m*}(\hat{\bs p}_\ma{rel})  Y_\ell^{\,m}(\hat{\bs r})\,, 
\ee
in which $j_\ell(x)$ is the spherical Bessel function that is regular at $x=0$. Using the Wigner-Eckart theorem, we obtain
\be \nn
\langle  \psi_\ma{1S}   | r_+ | {\bs p}_\ma{rel}  \rangle &=& 4\pi \sum_{\ell,m} i^\ell \langle R_{10} || r || j_\ell( p_\ma{rel} ) \rangle \langle \ell \,m ; \, 1\,1 | 0\,0 \rangle Y_\ell^{\,m*}(\hat{\bs p}_\ma{rel}) \\
&=& 4\pi i \langle R_{10} || r || j_1( p_\ma{rel} ) \rangle   Y_1^{\,1*}(\hat{\bs p}_\ma{rel}) \,,
\ee
where $\langle r | j_\ell( p_\ma{rel} ) \rangle = j_\ell( p_\ma{rel}r)$. Similarly,
\be
\langle  \psi_\ma{1S}   | r_- | {\bs p}_\ma{rel}  \rangle &=& 4\pi i \langle R_{10} || r || j_1( p_\ma{rel} ) \rangle   Y_1^{\,-1*}(\hat{\bs p}_\ma{rel}) \\
\langle  \psi_\ma{1S}   | r_0 | {\bs p}_\ma{rel}  \rangle &=& 4\pi i \langle R_{10} || r || j_1( p_\ma{rel} ) \rangle   Y_1^{\,0*}(\hat{\bs p}_\ma{rel}) \,.
\ee
The reduced matrix element can be calculated
\be \nn
\langle R_{10} || r || j_1( p_\ma{rel} ) \rangle  &=&  \frac{1}{\sqrt{3}} \int r^2\diff r\, R_{10}(r) r j_1(p_\ma{rel}r) \\ \nn
&=& \frac{1}{\sqrt{3}} \int r^2\diff r\, \frac{2} {a_B^{3/2}} e^{-r/a_B} r \Big[  \frac{\sin(p_\ma{rel}r)}{(p_\ma{rel}r)^2} - 
\frac{\cos(p_\ma{rel}r)}{p_\ma{rel}r}  \Big] \\
&=& \frac{2}{a_B^{3/2}} \frac{ 2^3a_B^5p_\ma{rel} }{ \sqrt{3}(1+a_B^2p_\ma{rel}^2)^3 }  \,.
\ee
So we have
\be \nn
|\langle \psi_\ma{1S}  | {\bs r} |  {\bs p}_\ma{rel}   \rangle|^2 &=& |\langle  \psi_\ma{1S}   | r_+ | {\bs p}_\ma{rel}  \rangle|^2 + 
| \langle  \psi_\ma{1S}   | r_- | {\bs p}_\ma{rel} \rangle  |^2 + |\langle  \psi_\ma{1S}   | r_0 | {\bs p}_\ma{rel} \rangle |^2  \\
&=& \frac{2^{10}\pi a_B^7 p_\ma{rel}^2 }{(1+a_B^2p_\ma{rel}^2)^6  }\,,
\ee
which agrees with the result (\ref{chap3_eqn_matrix1S_direct}) of the previous method.

\subsubsection{Coulomb Scattering Wave}
Now we apply the method of spherical tensor operators to the case where the unbound $Q\bar{Q}$ wave function is a Coulomb scattering wave, which is the solution to the Schr\"odinger equation with a repulsive Coulomb potential $V_0 = \frac{\alpha_s}{2N_cr}$. The Coulomb wave function is given by
\be
\langle {\bs r} | \Psi_{{\bs p}_\ma{rel}} \rangle = 4\pi \sum_{\ell,m}i^\ell e^{i \sigma_\ell } \frac{F_\ell(\eta,\rho)}{\rho} Y_\ell^{\,m*}(\hat{\bs p}_\ma{rel})  Y_\ell^{\,m}(\hat{\bs r}) \,,
\ee
where $\sigma_\ell=\Arg \Gamma(\ell+1+i\eta)$ is introduced in Chapter 2. Here $\rho = p_\ma{rel} r $ and $\eta = \frac{\alpha_s M}{4N_c} p_\ma{rel}$. The function $F_\ell(\eta,\rho)$ is defined
\be
F_\ell(\eta,\rho) &\equiv& C_\ell(\eta) \rho^{\ell+1} e^{i\rho} \,_1F_1 (\ell+1+i\eta; 2\ell+2; -2i\rho) \\
C_\ell(\eta) &\equiv&  \frac{2^\ell e^{-\pi\eta/2}  |\Gamma(\ell+1+i\eta)|  }{\Gamma(2\ell+2)} \,,
\ee
where $_1F_1(a;b;z)$ is the confluent hypergeometric function of the first kind and $\Gamma(z)$ is the gamma function.
Using the Wigner-Eckart theorem, we obtain
\be
\langle  \psi_\ma{1S}   | r_+ | \Psi_{{\bs p}_\ma{rel}}  \rangle  &=& 4\pi i e^{i\sigma_1} \langle R_{10} || r || \Psi_{ p_\ma{rel}} \rangle   Y_1^{\,1*}(\hat{\bs p}_\ma{rel}) \\
\langle  \psi_\ma{1S}   | r_- |  \Psi_{{\bs p}_\ma{rel}}  \rangle  &=& 4\pi i e^{i\sigma_1} \langle R_{10} || r || \Psi_{ p_\ma{rel}}  \rangle   Y_1^{\,-1*}(\hat{\bs p}_\ma{rel}) \\
\langle  \psi_\ma{1S}   | r_0 | \Psi_{{\bs p}_\ma{rel}}  \rangle  &=& 4\pi i e^{i\sigma_1} \langle R_{10} || r || \Psi_{ p_\ma{rel}}  \rangle   Y_1^{\,0*}(\hat{\bs p}_\ma{rel}) \,,
\ee
where the reduced matrix element is given by
\be 
\langle R_{10} || r || \Psi_{ p_\ma{rel}}  \rangle &=& \frac{1}{\sqrt{3} }\int r^2 \diff r \, R_{10}(r) r \frac{F_1(\eta,\rho)}{\rho} \\ \nn
&=& \frac{2}{\sqrt{3}a_B^{3/2}} C_1(\eta) p_\ma{rel} \int r^4\diff r\, e^{-r/a_B+ip_\ma{rel}r}\,_1F_1 (2 + i\eta; 4 ; -2i\rho) \\ \nn
&=& \frac{2}{\sqrt{3}a_B^{3/2}} C_1(\eta) p_\ma{rel} \Gamma(5) \Big(\frac{1}{a_B}-ip_\ma{rel} \Big)^5\,_2F_1\Big(2+i\eta;5;4;\frac{-2ip_\ma{rel}a_B}{1-ip_\ma{rel}a_B}\Big) \,,
\ee
where
\be
_2F_1(a;5;4;z) = \frac{1}{4}(1-z)^{-1-a}(4-4z+az) \,.
\ee
So we have
\be
_2F_1\Big(2+i\eta;5;4;\frac{-2ip_\ma{rel}a_B}{1-ip_\ma{rel}a_B}\Big) &=& \frac{4+2\eta p_\ma{rel}a_B}{4(1- ip_\ma{rel}a_B)} \Big( 1+\frac{2ip_\ma{rel}a_B}{1 - ip_\ma{rel}a_B} \Big)^{-3-i\eta} \,. \ \ \ \ \ 
\ee
To estimate this, we write
\be
1+\frac{2ip_\ma{rel}a_B}{1 - ip_\ma{rel}a_B} = \frac{1 - p^2_\ma{rel}a_B^2}{1 + p^2_\ma{rel}a^2_B} + \frac{2ip_\ma{rel}a_B}{1+p^2_\ma{rel}a^2_B} = re^{i\theta}\,,
\ee
where $r=1$ and
\be
\label{chap3_eqn_2F1_theta}
\theta = \arctan{\frac{2p_\ma{rel}a_B}{1-p^2_\ma{rel}a^2_B}} \,.
\ee
Using
\be
(a+bi)^{c+di} = (a^2 + b^2)^{c/2} \exp\Big\{  \frac{i}{2}d \ln(a^2+b^2) + ic \arctan{\frac{b}{a}} - d\arctan{\frac{b}{a}} \Big\}\,,
\ee
we find
\be
\Big| \,_2F_1\Big(2+i\eta;5;4;\frac{-2ip_\ma{rel}a_B}{1-ip_\ma{rel}a_B}\Big) \Big|^2 = \frac{(2+\eta p_\ma{rel}a_B)^2 }{4(1+p^2_\ma{rel}a^2_B)^2} e^{2\eta \theta}\,,
\ee
where $\theta$ is given by Eq.~(\ref{chap3_eqn_2F1_theta}). Using $\Gamma(2+i\eta) = (1+i\eta)\Gamma(1+i\eta)$, we can calculate $C_1(\eta)$ to get
\be
|C_1(\eta)|^2 = \Big( \frac{e^{-\pi\eta/2}}{3} | \Gamma(2+i\eta) | \Big)^2 = \frac{1+\eta^2}{9}\frac{2\pi\eta}{e^{2\pi\eta}-1}\,.
\ee 
Putting everything together, we find
\be \nn
|\langle  \psi_\ma{1S}  | {\bs r} |   \Psi_{{\bs p}_\ma{rel}} \rangle|^2 &=& |\langle  \psi_\ma{1S}  | r_+ |   \Psi_{{\bs p}_\ma{rel}} \rangle|^2 
+|\langle  \psi_\ma{1S}  | r_- |   \Psi_{{\bs p}_\ma{rel}} \rangle|^2 + |\langle  \psi_\ma{1S}  | r_0 |   \Psi_{{\bs p}_\ma{rel}} \rangle|^2 \\
&=& \frac{2^9\pi^2\eta  p^2_\ma{rel} a_B^7 (2+\eta p_\ma{rel}a_B)^2 (1+\eta^2)   }{(1+p^2_\ma{rel}a_B^2)^6 (e^{2\pi\eta}-1)} e^{4\eta\arctan(p_\ma{rel}a_B)} \,,
\ee
where we have used $\tan\frac{\theta}{2} = p_\ma{rel}a_B$.

Similarly we can find
\be  \nn
|\langle  \psi_\ma{2S}  | {\bs r} |   \Psi_{{\bs p}_\ma{rel}} \rangle|^2 
&=& \frac{2^{18}\pi^2\eta  p^2_\ma{rel} a_B^7 (1+\eta^2)   }{(1+4p^2_\ma{rel}a_B^2)^8 (e^{2\pi\eta}-1)} e^{4\eta\arctan(2p_\ma{rel}a_B)} \\
&& \frac{1}{E_\ma{2S}^2} \Big[    q(2+\rho_c) - 2|E_\ma{2S}|  ( 2\rho_c^2 + 5\rho_c +3  )      \Big]^2\,,
\ee
where
\be
E_\ma{2S} &=& -\frac{1}{4} \frac{C_F^2\alpha_s^2M}{4}\\
q &=& |E_\ma{2S} | (1 + 4p^2_\ma{rel}a_B^2) \\
\rho_c &=& \frac{1}{N_c^2-1} \,.
\ee

\vspace{0.2in}

\section{Diffusion and In-Medium Energy Loss}
In addition to dissociation and recombination, quarkonium can diffuse inside the QGP. In other words, a quarkonium state can scatter elastically with medium constituents and change its momentum, but not its energy. The elastic scattering can happen because quarkonium still carries color, although the net color vanishes. Quarkonium is of finite size and has an internal color distribution. In the multipole expansion, the leading non-vanishing order is the dipole term. This is similar to a hydrogen atom, which is neutral in charge. But the hydrogen atom has an internal structure and can still respond to an external electromagnetic field. Since the response of quarkonium to an external chromo field is via a dipole, we expect the elastic scattering rate of quarkonium inside QGP to be smaller than the elastic scattering rate of open heavy quarks, which carry a net color. This is true when the medium constituents have a wave length much larger than the size of quarkonium, i.e., $T \ll Mv$, which is indeed the case in our assumed hierarchy of scales. 

\begin{figure}
    \centering
    \begin{subfigure}[b]{0.33\textwidth}
        \centering
        \includegraphics[height=1.25in]{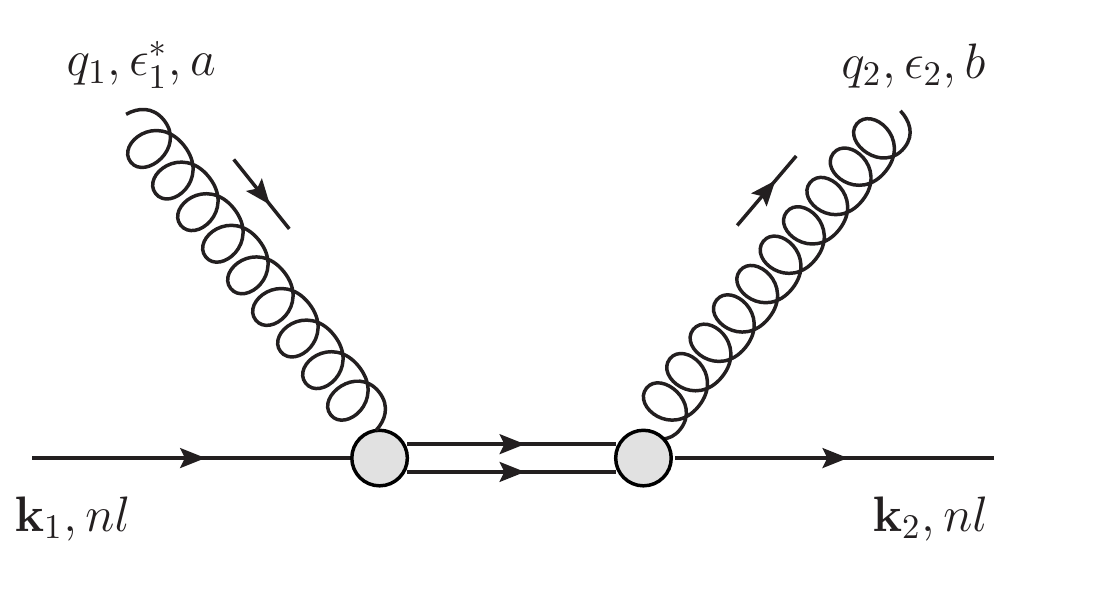}
        \caption{}\label{subfig:elastic1}
    \end{subfigure}%
    ~ 
    \begin{subfigure}[b]{0.33\textwidth}
        \centering
        \includegraphics[height=1.25in]{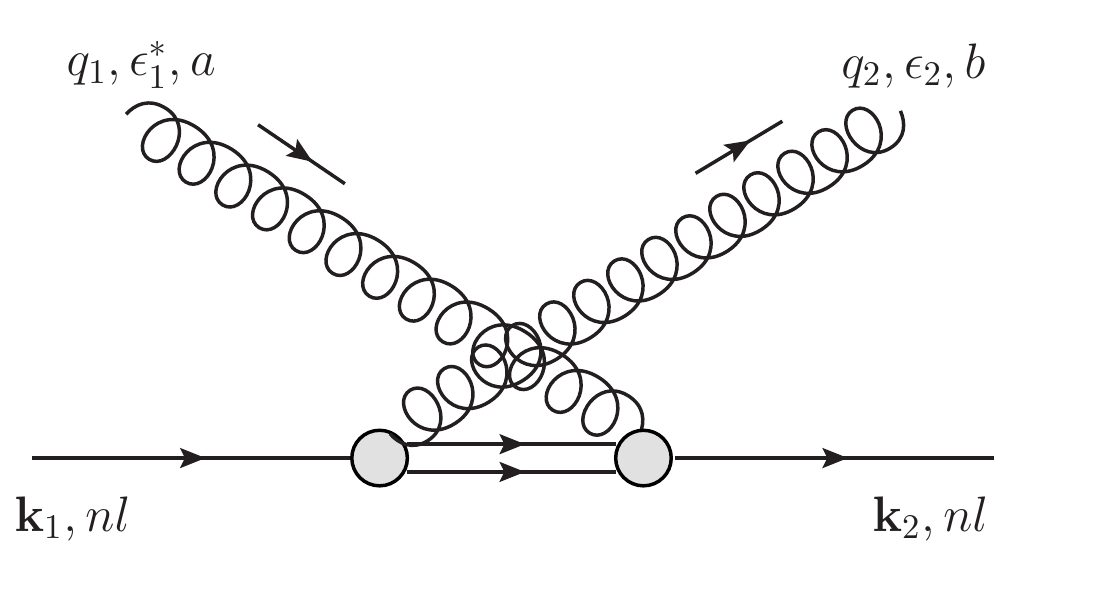}
        \caption{}\label{subfig:elastic2}
    \end{subfigure}%
    ~
    \begin{subfigure}[b]{0.33\textwidth}
        \centering
        \includegraphics[height=1.25in]{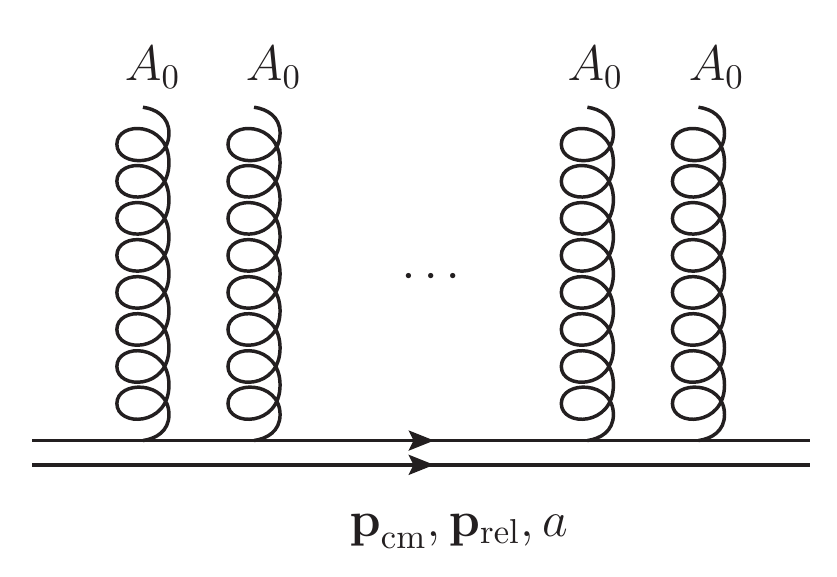}
        \caption{}\label{subfig:diffuse}
    \end{subfigure}%
    \caption[Feynman diagrams contributing to the quarkonium elastic scattering in the QGP.]{Feynman diagrams contributing to the quarkonium elastic scattering in the QGP. First two diagrams are the processes at the order $g^2r^2$. The last diagram is schematic and shows the virtual octet propagator can in principle obtain an infinite series of momentum ``kicks" from the medium.}
    \label{chap3_fig:diffuse}
\end{figure}

The quarkonium diffusion cannot happen at the order $r$ in amplitude, because at this order the color singlet field $S$ only couples to an octet $O^a$, which means quarkonium has to turn to an unbound octet pair. The diffusion process starts to happen at the order $r^2$ because the color singlet can turn to an octet and then become a singlet again. At the order $g^2r^2$, contributing diagrams are shown in Fig.~\ref{subfig:elastic1} and \ref{subfig:elastic2}. They are from the dipole vertex at the second order in perturbation theory. At the order $g^2r^2$, we may need to include quadrupole terms that show up in the Lagrangian in the multipole expansion. A quadrupole term of the form $g^2S^\dagger r_ir_j S$ contracted with $E_{i}$ or $A_{i}$ can contribute to this physical process. However, such a term has a vanishing matching coefficient \cite{Brambilla:2003nt}. The amplitudes satisfy the Ward identity by virtue of Eq.~(\ref{eqn:ward1}). In Coulomb gauge, the amplitudes of diagrams \ref{subfig:elastic1} and \ref{subfig:elastic2} are
\be \nn
i\ml{M} &=& -g^2\frac{T_F}{N_c}\delta^{ab}(\epsilon^*_1)_i(\epsilon_2)_j q_1q_2  \int\frac{\diff^3 p_\ma{rel}}{(2\pi)^3}  \\
\label{chap3_eqn_M_diffuse}
&& \bigg[
\frac{\langle \psi_{nl} | r_j | \Psi_{{\bs p}_\ma{rel}} \rangle   \langle  \Psi_{{\bs p}_\ma{rel}} |r_i | \psi_{nl} \rangle}{q_1-|E_{nl}| - \frac{{\bs p}_\ma{rel}^2}{M} + i\epsilon }  
+ \frac{\langle \psi_{nl} | r_j | \Psi_{{\bs p}_\ma{rel}} \rangle \langle  \Psi_{{\bs p}_\ma{rel}} |r_i | \psi_{nl} \rangle }{-q_1-|E_{nl}| - \frac{{\bs p}_\ma{rel}^2}{M} + i\epsilon }    \bigg]
\,.
\ee
When $q_1 \ge |E_{nl}|$, the first term in the big bracket has a pole. At the pole, the term becomes imaginary. Physically, this happens when the intermediate octet state becomes on-shell so the process becomes the quarkonium dissociation. Therefore we should take the principal value of the integral $\ml{P}\int\diff^3 p_{\ma{rel}}$. We need to show the principal value is well-defined, i.e., the divergent contributions from both sides of the pole cancel out. We abstract the integral of the first term inside the square bracket of (\ref{chap3_eqn_M_diffuse}) as
\be
\ml{P}\int_0^\infty \frac{\diff x\,f(x)}{x^2 - A}\,,
\ee
where $x = p_\ma{rel}$, $A = M(q_1 - |E_{nl}|)$ and pole starts to appear when $A \geq 0$. First we consider the case $A>0$. The pole is at $\sqrt{A}$ and is a single pole. We study an integral in an infinitesimal region centered at $\sqrt{A}$: $[\sqrt{A}-\delta, \sqrt{A}+\delta]$,
\be
I(\delta) \equiv \int_{ \sqrt{A}-\delta } ^{\sqrt{A}} \frac{\diff x\, f(x)}{(x-\sqrt{A})(x+\sqrt{A})} + \int^{\sqrt{A}-\delta}_{\sqrt{A}} \frac{\diff x\, f(x)}{(x-\sqrt{A})(x+\sqrt{A})} \,.
\ee
Changing the variable $y = x - \sqrt{A}$ and defining $g(y) = \frac{f(x)}{x+\sqrt{A}}$, we obtain
\be \nn
I(\delta) &=& \int_{-\delta}^0 \frac{\diff y\, g(y)}{y} + \int^{\delta}_0 \frac{\diff y\, g(y)}{y} \\ \nn
& = & \int_{-\delta}^0 \frac{\diff y\, g(0) + yg'(0) + \cdots }{y} + \int^{\delta}_0 \frac{\diff y\, g(0) + yg'(0) + \cdots }{y} \\
& = & 2\delta g'(0) + \ml{O}(\delta^2) \,.
\ee
So an infinitesimal region around the pole position $\sqrt{A}$ almost does not contribute to the integral. So the principal value can be well-defined. Next we consider the case $A=0$. We have a double pole at $x=0$. We consider the integral
\be
\lim_{\delta\to0} \int_\delta^\infty \frac{\diff x\,f(x)}{x^2} \,.
\ee
This integral is well-defined because we know that $f(x)\sim\ml{O}(x^2)$ where a quadratic term comes from $\diff^3 p_\ma{rel}$. Therefore, the principal value of (\ref{chap3_eqn_M_diffuse}) is well-defined.

During the lifetime of the virtual octet, its momentum may change due to a number of collisions that transfer a small momentum, as depicted in Fig.~\ref{subfig:diffuse}. These processes are at the order $r^0$, so not suppressed by the multipole expansion. The virtual octet diffuses as if it were an open heavy quark. Since the contributions cancel out near the pole of the octet propagator, the octet behaves like a state with lifetime $\Delta\tau\sim\frac{1}{Mv^2}$. The rate of transferring the square of momentum is about $\alpha_s^2T^3$ \cite{Moore:2004tg}. So the square of momentum transferred during its lifetime is about $\frac{\alpha_s^2T^3}{Mv^2}\lesssim\alpha_s^2T^2$ since we assume $T \lesssim Mv^2$. The momentum transferred is about $\alpha_s T$. The c.m.~momentum of the octet is at least $q_1\sim T \gg \alpha_sT$. So the effect from the virtual octet diffusion is small and there is no need to resum $gA_0$ into the virtual octet.

We define the square of the total amplitude, summed over colors and polarizations of gluons,
\be 
\sum |\ml{M}|^2 &\equiv& \sum_{a,b}\sum_{\epsilon_1,\epsilon_2} |\ml{M}|^2 \\ \nn
&=& \frac{4}{9}g^4\frac{T_F}{N_c}C_F |\vec{\epsilon}_1^{\ *}\cdot \vec{\epsilon}_2|^2
q_1^2q_2^2 \bigg[ \ml{P}\int\frac{\diff^3 p_\ma{rel}}{(2\pi)^3} \frac{|\langle  \Psi_{{\bs p}_\ma{rel}}  | {\bs r} | \psi_{nl}  \rangle|^2 ( |E_{nl}|+\frac{{\bs p}^2_\ma{rel}}{M}) }{( |E_{nl}|+\frac{{\bs p}^2_\ma{rel}}{M})^2-q_1^2} \bigg]^2\,.
\ee
Then we can define the diffusion coefficient of quarkonium, which is the square of momentum transferred per unit time
\be \nn
3\kappa &\equiv& \int\frac{\diff^3k_2}{(2\pi)^3} \int\frac{\diff^3q_1}{2q_1(2\pi)^3} \int\frac{\diff^3q_2}{2q_2(2\pi)^3} n_B(q_1)(1+n_B(q_2)) \\
&&({\bs q}_1-{\bs q}_2)^2(2\pi)^4\delta^3({\bs k}_1+{\bs q}_1 - {\bs k}_2 - {\bs q}_2) \delta(q_1-q_2)  \sum |\ml{M}|^2\,.
\ee
After some simplications
\be \nn
\kappa &=& \frac{32}{729\pi^5}\alpha_s^2\int \diff q \,q^8 n_B(q)(1+n_B(q))   \bigg[ \ml{P}\int  \diff p_\ma{rel} \frac{p_\ma{rel}^2 
|\langle \Psi_{{\bs p}_\ma{rel}} | {\bs r} | \psi_{nl} \rangle|^2
( |E_{nl}|+\frac{{\bs p}^2_\ma{rel}}{M})}{(|E_{nl}|+\frac{{\bs p}^2_\ma{rel}}{M})^2-q^2} \bigg]^2\,. \\
&&
\ee
For 1S state, we obtain
\be \nn
\kappa(\ma{1S}) &=& \frac{32}{729\pi^5}\alpha_s^2\int \diff q \,q^8 n_B(q)(1+n_B(q)) \\
 \label{eqn:kappa}
&&  \bigg[ \ml{P}\int  \diff p_\ma{rel} \frac{p_\ma{rel}^2 
|\langle \Psi_{{\bs p}_\ma{rel}} | {\bs r} | \psi_\ma{1S} \rangle|^2
( |E_\ma{1S}|+\frac{{\bs p}^2_\ma{rel}}{M})}{(|E_\ma{1S}|+\frac{{\bs p}^2_\ma{rel}}{M})^2-q^2} \bigg]^2\,.
\ee

\begin{figure}
    \centering
    \includegraphics[height=2.5in]{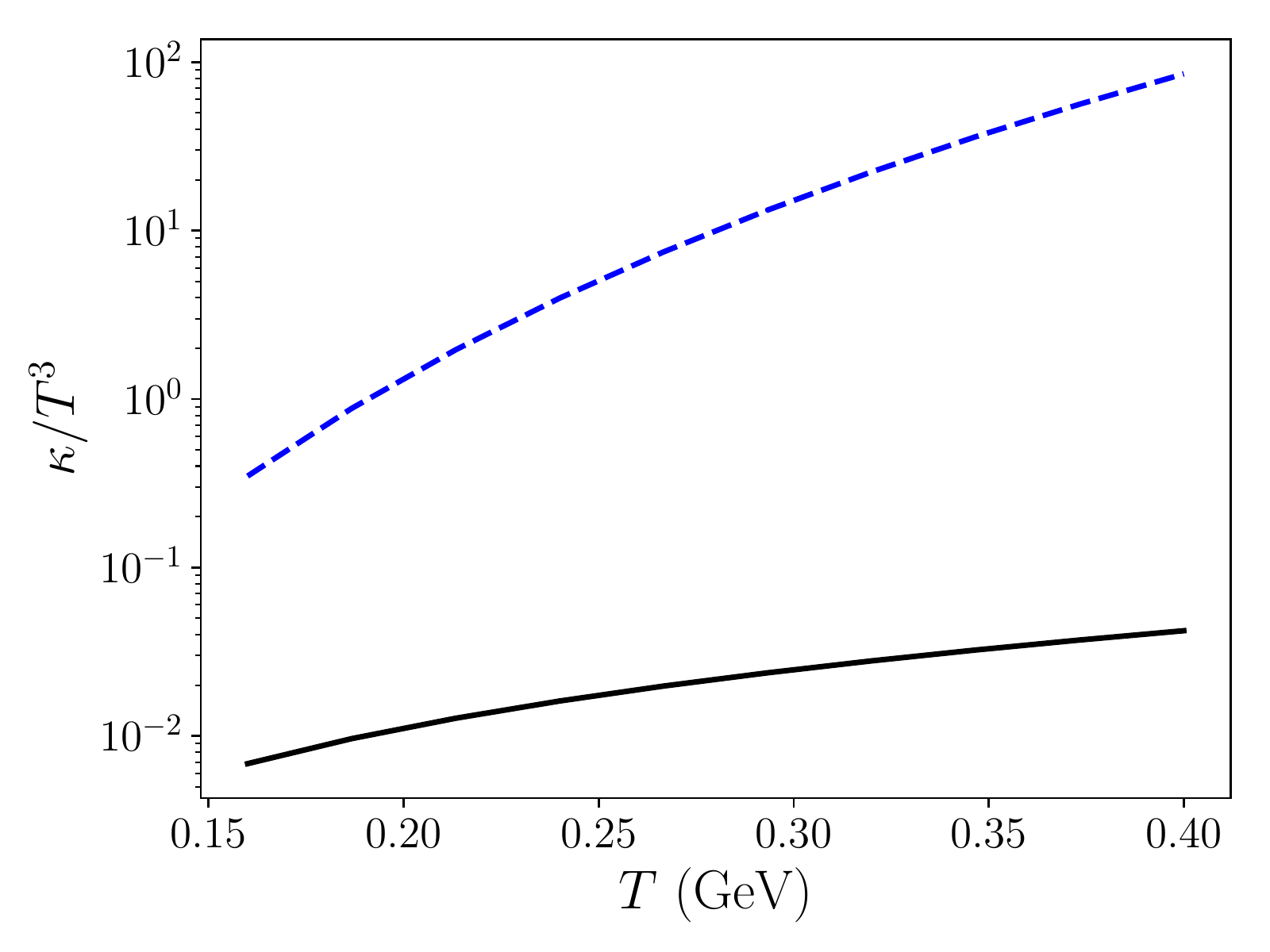}
    \caption[$\Upsilon$(1S) diffusion coefficient $\kappa$ as a function of temperature.]{$\Upsilon$(1S) diffusion coefficient $\kappa$ as a function of temperature, the solid line is the exact result from Eq.~(\ref{eqn:kappa}) while the dashed line is the approximate result from Eq.~(\ref{eqn:kappa_appro}).}
    \label{fig:kappa}
\end{figure}

\begin{figure}
    \centering
    \includegraphics[height=2.5in]{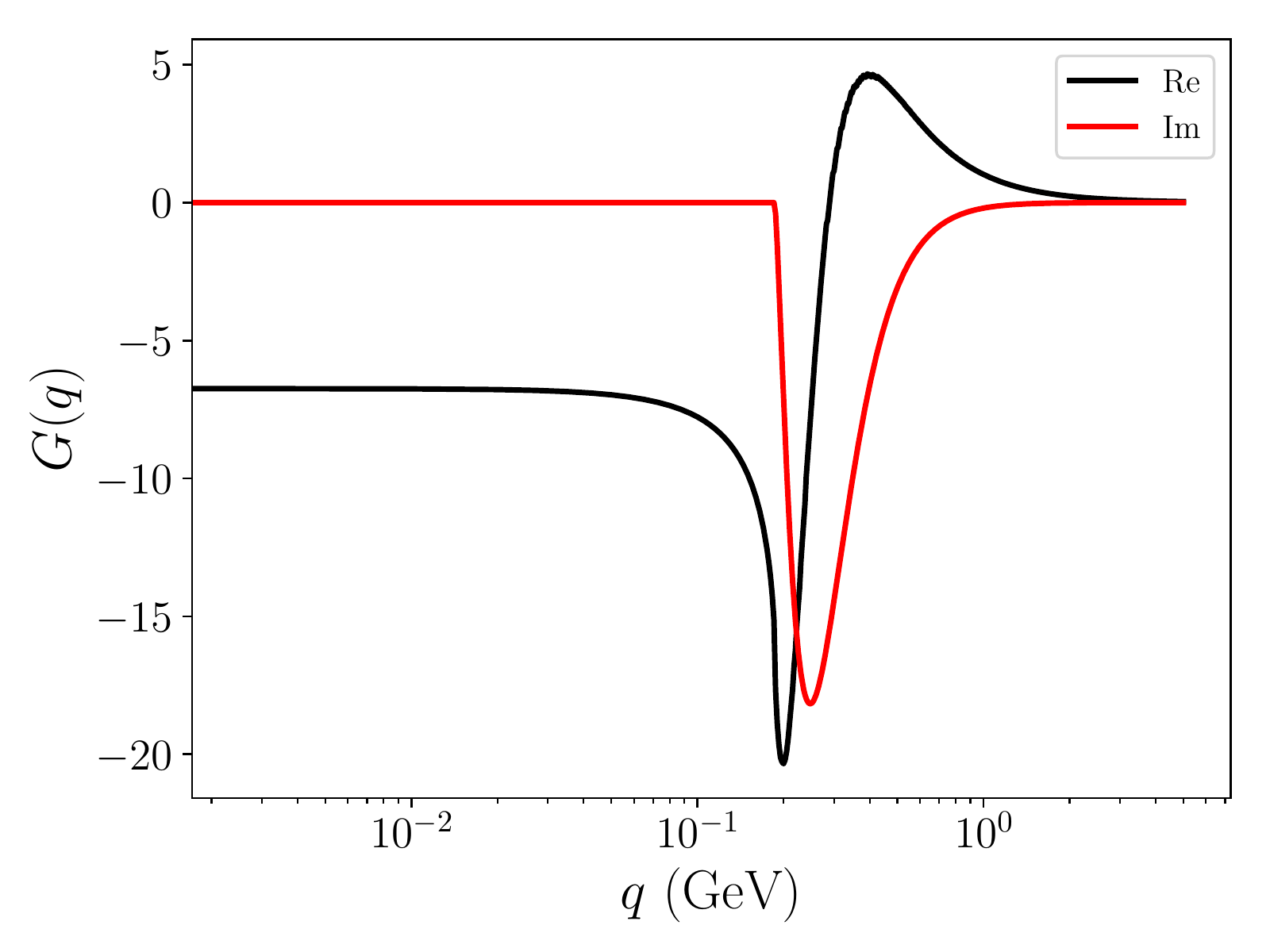}
    \caption{Response function $G(q)$.}
    \label{chap3_fig_response}
\end{figure}

If the $q^2$ in the denominator inside the integral over $p_\ma{rel}$ is neglected and the octet relative wave function is a plane wave, we can show 
\be \nn
\kappa'(\ma{1S}) &=& \frac{32}{729\pi^5}\alpha_s^2\int \diff q \,q^8 n_B(q)(1+n_B(q))  \bigg[ \int  \diff p_\ma{rel} \frac{p_\ma{rel}^2 
|\langle \Psi_{{\bs p}_\ma{rel}} | {\bs r} | \psi_\ma{1S} \rangle|^2}{|E_\ma{1S}|+\frac{{\bs p}^2_\ma{rel}}{M}} \bigg]^2 \\
\label{eqn:kappa_appro}
&=&   \frac{T^3(\pi T a_B)^6}{N_c^2} \frac{50176\pi}{1215} \frac{2}{C_F^2}\,,
\ee
where we have used the result of Eq.~(\ref{chap3_eqn_matrix1S_direct})
\be
|\langle \Psi_{{\bs p}_\ma{rel}} | {\bs r} | \psi_{1S} \rangle|^2 =  \frac{2^{10} \pi a_B^5(p_\ma{rel}a_B)^2}{(1+(p_\ma{rel}a_B)^2)^6}\,.
\ee
If one takes the large-$N_c$ approximation (so we can neglect the repulsive Coulomb potential for the octet), $C_F=N_c/2=3/2$, and multiplies the expression (\ref{eqn:kappa_appro}) by a factor of $9/8$ (because when we sum over colors, there is a factor of $8$, in large-$N_c$, the factor is $9$), then Eq.~(\ref{eqn:kappa_appro}) agrees with a previous estimate using perturbative calculations in an effective field theory where the octet is integrated out \cite{Dusling:2008tg}. The approximate result (\ref{eqn:kappa_appro}) scales as $\kappa' \propto T^9$. Both the exact result, Eq.~(\ref{eqn:kappa}), and the approximate result, Eq.~(\ref{eqn:kappa_appro}), are shown in Fig.~\ref{fig:kappa} for $\Upsilon$(1S) with $M=4.65$ GeV and $\alpha_s = 0.3$. The two results differ by two to three orders of magnitude. To understand the difference, we define a response function of quarkonium to an incoming gluon with an energy $q$
\be
G(q) \equiv \int\frac{\diff^3 p_\ma{rel}}{(2\pi)^3} \bigg[
\frac{ |\langle \Psi_{{\bs p}_\ma{rel}} | {\bs r} | \psi_\ma{1S} \rangle|^2 }{q - |E_{nl}| - \frac{{\bs p}_\ma{rel}^2}{M} + i\epsilon }  
+ \frac{ |\langle \Psi_{{\bs p}_\ma{rel}} | {\bs r} | \psi_\ma{1S} \rangle|^2 }{ - q - |E_{nl}| - \frac{{\bs p}_\ma{rel}^2}{M} + i\epsilon }    \bigg] \,.
\ee
The real part of $G(q)$ corresponds to the elastic scattering of quarkonium while the imaginary part corresponds to the dissociation of quarkonium. The function $G(q)$ is a response function of how the dipole responds to the gradient of gauge fields ${\bs E}^a$. It is plotted in Fig.~\ref{chap3_fig_response}. 
We can see that the imaginary part starts to be nonzero when $q>|E_\ma{1S}| = \frac{C_F^2\alpha_s^2M}{4}$. Both the real and imaginary parts vanish as $q\to\infty$. The approximate result (\ref{eqn:kappa_appro}) corresponds to using $G(q) = G(0)$ for all values of $q$. This is the reason why our result (\ref{eqn:kappa}) is almost three orders of magnitude smaller than (\ref{eqn:kappa_appro}): in (\ref{eqn:kappa}), $G(q)$ quickly goes to zero as $q$ increases. To support this argument, let us check the typical value of $q$ in the integrand of both (\ref{eqn:kappa}) and (\ref{eqn:kappa_appro}). Due to the term $q^8n_B(q)$, we expect $q \lesssim 8T $. So $q$ can be a few GeV, at which $G(q)$ almost vanishes. This is why (\ref{eqn:kappa_appro}) overestimates the diffusion constant. The approximate result Eq.~(\ref{eqn:kappa_appro}) is only valid when $q \ll Mv^2$ so one can neglect the $q^2$ in the denominator. However, for real QGP, $T\gtrsim 160$ MeV, it is not a good approximation even for the bottom quark with $Mv^2\sim 500$ MeV. In fact, even if we take the limit $T \ll Mv^2$, the full result (\ref{eqn:kappa}) does not reduce to be (\ref{eqn:kappa_appro}). In other words, (\ref{eqn:kappa_appro}) is not the correct limit of (\ref{eqn:kappa}) when $T \ll Mv^2$. This can be seen from the expansion of 
\be
\frac{1}{(|E_\ma{1S}|+\frac{{\bs p}^2_\ma{rel}}{M})^2-q^2} \,.
\ee
We write $|E_\ma{1S}|+\frac{{\bs p}^2_\ma{rel}}{M}$ as $Mv^2$ for simplicity and expand assuming $q \ll Mv^2$
\be
\frac{1}{M^2v^4 - q^2} = \frac{1}{M^2v^4} \sum_{n=0}^\infty \Big(\frac{q^2}{M^2v^4}\Big)^n \,.
\ee
But we need to integrate over $q$ with $q^8n_B(q)$. So the actual scaling law of $q$ in the n-th term is given by $q^{8+2n} n_B(q)$ and thus is $q\sim (8+2n)T$. When $n$ becomes large, the expansion condition $q \ll Mv^2$ breaks down even if we have $T \ll Mv^2$. In a nutshell, (\ref{eqn:kappa_appro}) is just the first term of an asymptotic series of (\ref{eqn:kappa}). 

At high temperatures $Mv \gg T \gg Mv^2$ (this is beyond the separation of scales we consider in this dissertation but the first inequality assures our power counting for the dipole interaction), the $q^2$ in the denominator dominates over $Mv^2$ and we expect $\kappa\propto T^5$. Furthermore, if we assume the octet relative wave function is a plane wave (which is true at large-$N_c$) and use Eq.~(\ref{chap3_eqn_matrix1S_direct}), we expect $\kappa(\ma{1S}) \propto (Ma_B^2)^2  \propto M^{-2}$ at high temperatures. 
The mass dependence of $\kappa$ is also plotted for three difference heavy quark masses. At high temperature the mass scaling is approximately valid.

\begin{figure}
    \centering
    \includegraphics[height=2.5in]{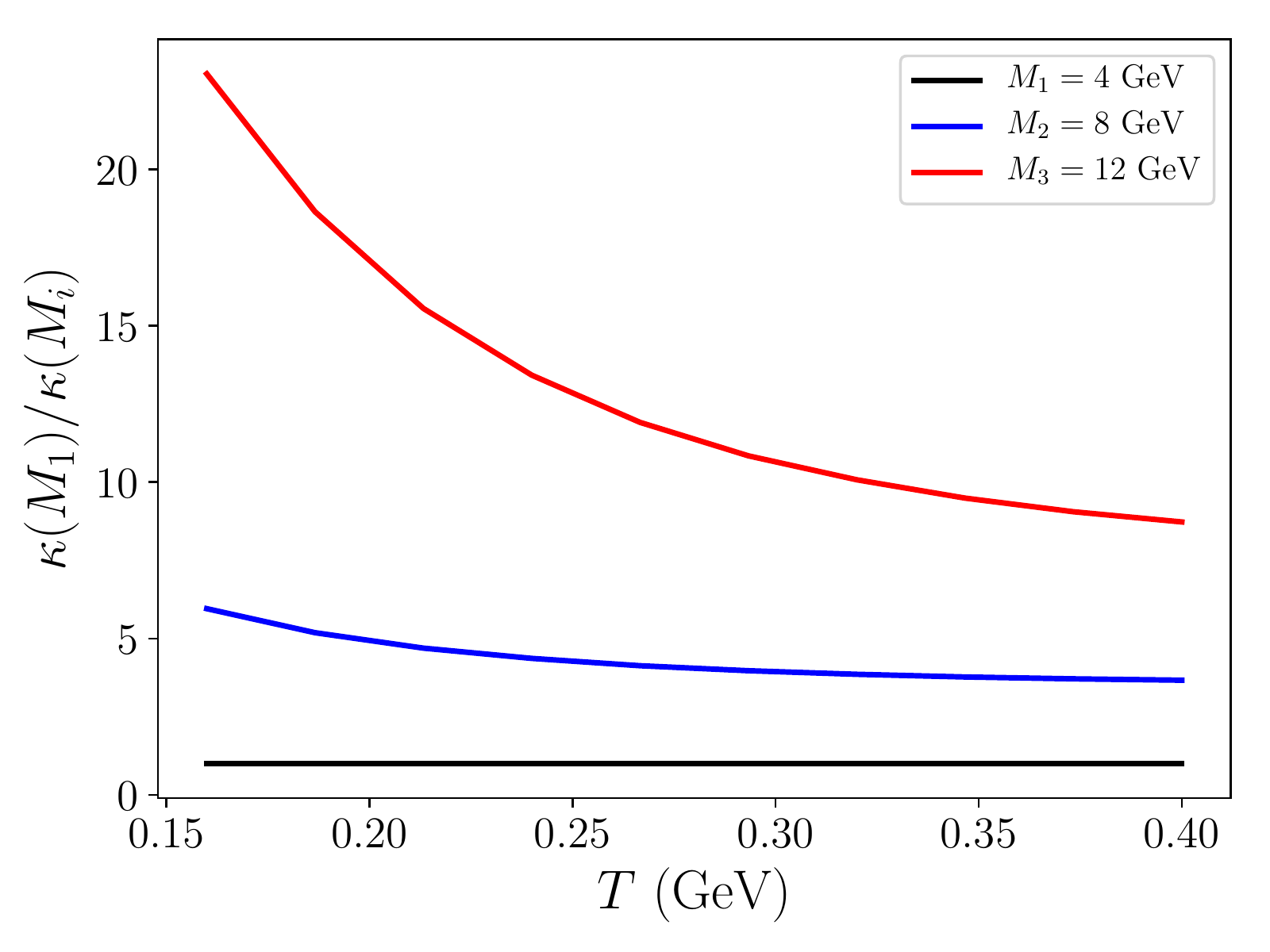}
    \caption[Heavy quark mass dependence of the diffusion coefficient $\kappa$(1S).]{Heavy quark mass dependence of the diffusion coefficient $\kappa$(1S). The lower, middle and upper lines correspond to $\kappa(M_1)/\kappa(M_i)$ with $i=1,2,3$ respectively.}
    \label{fig:kappa_M}
\end{figure}

Finally, we can include a new collision term $\ml{C}_{nls}({\bs x}, {\bs p}, t)$ in the Boltzmann equation (\ref{chap3_eqn_boltz_transport_nls}) to represent the elastic scattering of quarkonium
\be
\label{chap3_eqn_boltz_transport_nls_diffuse}
\frac{\partial}{\partial t} f_{nls}({\bs x}, {\bs p}, t) + {\bs v}\cdot \nabla_{\bs x}f_{nls}({\bs x}, {\bs p}, t) = \ml{C}_{nls}({\bs x}, {\bs p}, t) + \ml{C}_{nls}^{+}({\bs x}, {\bs p}, t) - \ml{C}_{nls}^{-}({\bs x}, {\bs p}, t)\,. \ \ \ \ 
\ee
To write the new collision term $\ml{C}_{nls}({\bs x}, {\bs p}, t)$ out explicitly, we define
\be \nn
\ml{F}_{nls} &\equiv& \int\frac{\diff^3k_1}{(2\pi)^3} \int\frac{\diff^3k_2}{(2\pi)^3} \int\frac{\diff^3q_1}{2q_1(2\pi)^3} \int\frac{\diff^3q_2}{2q_2(2\pi)^3} n_B(q_1)(1+n_B(q_2)) f_{nls}({\bs x}, {\bs k}_1, t)\\
&&(2\pi)^4\delta^3({\bs k}_1+{\bs q}_1 - {\bs k}_2 - {\bs q}_2) \delta(q_1-q_2)  \sum |\ml{M}|^2 \,.
\ee
The diffusion term in the Boltzmann equation (\ref{chap3_eqn_boltz_transport_nls_diffuse}) can be written as
\be
\ml{C}_{nls}({\bs x}, {\bs p}, t) = -\frac{\delta \ml{F}_{nls}}{\delta {\bs k}_1} \Big|_{{\bs k}_1 = {\bs p}} + \frac{\delta \ml{F}_{nls}}{\delta {\bs k}_2} \Big|_{{\bs k}_2 = {\bs p}}\,.
\ee

In principle, a quarkonium has two ways to lose energy inside the QGP (we assume quarkonium has been formed completely before entering the QGP). One way is the elastic scattering or diffusion. The other way is to dissociate first, lose energy as an unbound heavy quark antiquark pair and then recombine later. The former mechanism only works when the quarkonium is a well-defined bound state inside QGP. So it only makes sense when the QGP temperature is below the quarkonium melting temperature. As shown in Fig.~\ref{fig:kappa}, the rate of momentum transfer due to the diffusion is very slow for $\Upsilon$(1S), compared with that of open heavy quarks ($\kappa/T^3$ of heavy quarks is on the order of $1$ or $10$ \cite{Cao:2018ews}). This is also true for $J/\psi$, because we expect its diffusion coefficient is $10$ times larger than that of $\Upsilon$(1S) from the mass scaling. But $J/\psi$ has a lower melting temperature, so the diffusion coefficient would probably only make sense below $300$ MeV. Therefore the latter mechanism (dissociation followed by energy loss and recombination) probably dominates the quarkonium energy loss inside the QGP, even though not every quarkonium finally observed has to go through this sequence of processes. Some of the primordially produced quarkonia may survive the in-medium evolution and lose almost no energy.

\vspace{0.2in}

\section{Quarkonium Annihilation inside QGP}
In this section, we consider the process of quarkonium annihilation inside the QGP. 
As discussed in Chapter 1, the NRQCD Lagrangian has $4$-fermion interactions, which can describe the annihilation of quarkonium (decay into other hadrons or leptons) and are not included in the pNRQCD Lagrangian Eq.~(\ref{eq:lagr}). 
We can add $-\Gamma S^\dagger S \rho_S $ into the Lindblad equation (\ref{eqn:lindblad}) to describe the annihilation. But this would break the conservation of probability $\Tr(\frac{\diff}{\diff t}\rho_S)=0$. So one needs also to include terms of the form $\rho_S S^\dagger S$ and $S \rho_S S^\dagger $. A pedagogical discussion of how to construct an open effective field theory in order to conserve probability can be found in Ref.~\cite{Braaten:2016sja}. The annihilation is too slow to be of much interest for phenomenology but we study it as an interesting example of how Lindblad-type operators enter the time evolution equation for the density matrix. In our case, we first restore the standard pNRQCD notation of singlet field, $S({\bs R}, {\bs r}, t) \equiv \langle {\bs r} | S({\bs R}, t) \rangle$, i.e., we project the wave function of the relative motion onto the relative position space. Then we can add two new terms in the density matrix evolution equation
\be \nn
\rho_S(t) &=& \cdots + \int_0^t dt_1 \int \diff^3R \int \diff^3r \Big( -\frac{\Gamma({\bs r})}{2} \{ S^\dagger({\bs R}, {\bs r}, t_1)S({\bs R}, {\bs r}, t_1), \rho_S(0)  \}
+\\
&& \Gamma({\bs r}) S({\bs R}, {\bs r}, t_1) \rho_S(0) S^\dagger({\bs R}, {\bs r}, t_1)
\Big)\,,
\ee
where the evolution term is explicitly trace-preserving. 

As in Section 3.3, we are interested in the bound state and will sandwich the density matrix between two bound quarkonium states and then do a Wigner transform.

\subsection{$\{ S^\dagger({\bs R}, {\bs r}, t)S({\bs R}, {\bs r}, t), \rho_S(0)  \}$ Term} 
We first compute the $S^\dagger({\bs R}, {\bs r}, t)S({\bs R}, {\bs r}, t) \rho_S(0)$ term sandwiched between $ \langle   {\bs k}_1 , nl, 1 |$ and $|   {\bs k}_2 , nl, 1\rangle$ and insert a complete set of states $|  {\bs k}_3, n_3l_3,1 \rangle$.
\be \nn
&&  \int_0^t dt_1 \int \diff^3R \int \diff^3r \frac{-\Gamma({\bs r})}{2} \int\frac{\diff^3k_3}{(2\pi)^3}\sum_{n_3,l_3}    \\  \nn
&& \langle  {\bs k}_1, nl,1   | S^\dagger({\bs R}, {\bs r}, t_1)S({\bs R}, {\bs r}, t_1) |  {\bs k}_3, n_3l_3,1 \rangle  \langle {\bs k}_3, n_3l_3,1 | \rho_S(0)  |   {\bs k}_2 , nl, 1\rangle \\ \nn
&=& \int_0^t dt_1 \int \diff^3R \int \diff^3r \frac{-\Gamma({\bs r})}{2} \int\frac{\diff^3k_3}{(2\pi)^3}\sum_{n_3,l_3}  \\ \nn
&&\psi_{n_3l_3}({\bs r}) \psi^*_{nl}({\bs r}) 
e^{ i|E_{nl}|t_1-i{\bs k}_1\cdot {\bs R}} e^{ - i|E_{n_3l_3}|t_1+i{\bs k}_3\cdot {\bs R}} \langle {\bs k}_3, n_3l_3,1 | \rho_S(0)  |   {\bs k}_2 , nl, 1\rangle \\
&=& \frac{-1}{2}t  \int \diff^3r \,  \psi^*_{nl}({\bs r}) \Gamma({\bs r}) \psi_{nl}({\bs r}) \langle {\bs k}_1, nl,1 | \rho_S(0)  |   {\bs k}_2 , nl, 1\rangle \,.
\ee
where we have used the Markovian approximation and written the delta function in energy as $t$. In the summation over $n_3$ and $l_3$, only $n_3=n$, $l_3=l$ contributes due to the delta function in energy (we assume no degeneracy in the bound state eigenenergy beyond that implied by rotational invariance). 

We can define the annihilation rate of a quarkonium state $nl$,
\be
\Gamma_{nl}   \equiv \int \diff^3r \,  \psi^*_{nl}({\bs r}) \Gamma({\bs r}) \psi_{nl}({\bs r})\,,
\ee
which is non-zero even in vaccum and should be distinguished from the dissociation rate inside QGP.
For S-wave, one may set $\Gamma({\bs r}) = \Gamma \delta^3(\bs r)$ and then $\Gamma_S =  \Gamma |\psi_{S}(0)|^2$, i.e., the annihilation rate depends on the wave function of the relative motion at the origin.

The other term in the anticommutator will give the same result. Under a Wigner transform, these two terms lead to in the Boltzmann equations
\be
\label{eqn:annhi1}
- \Gamma_{nl}  f_{nl}({\bs x}, {\bs k}, t)\,.
\ee
Typically $\Gamma_{nl} \sim 10 $ keV, so for a QGP with a lifetime $\sim10$ fm $\sim0.05$ MeV$^{-1}$ , the effect from quarkonium annihilations is negligible on the in-medium evolution. It is justified to assume that the total number of heavy quarks is conserved during the in-medium evolution.

\subsection{$S({\bs R}, {\bs r}, t) \rho_S(0) S^\dagger({\bs R}, {\bs r}, t)$ Term}
Then we compute the contribution from the $S({\bs R}, {\bs r}, t) \rho_S(0) S^\dagger({\bs R}, {\bs r}, t)$ term sandwiched between $ \langle   {\bs k}_1 , nl, 1 |$ and $|   {\bs k}_2 , nl, 1\rangle$.
\be  \nn
&&\int_0^t dt_1 \int \diff^3R \int \diff^3r \Gamma({\bs r}) \langle  {\bs k}_1, nl, 1   |   S({\bs R}, {\bs r}, t_1) \rho_S(0)  S^\dagger({\bs R}, {\bs r}, t_1)   |   {\bs k}_2 , nl, 1\rangle \\ \nn
&=& \int \diff^3r \Gamma({\bs r}) \int\frac{\diff^3k_3}{(2\pi)^3}\sum_{n_3,l_3} \int\frac{\diff^3k_4}{(2\pi)^3}\sum_{n_4,l_4} \psi_{n_3l_3}({\bs r})  \psi^*_{n_4l_4}({\bs r}) \\ \nn
&&  (2\pi)^4\delta^3({\bs k}_3 - {\bs k}_4) \delta(E_{k_3}-E_{k_4}) \langle    {\bs k}_1, nl, 1;\ {\bs k}_3, n_3l_3, 1    | \rho_S(0) |       {\bs k}_2, nl, 1;\ {\bs k}_4, n_4l_4, 1     \rangle \\\nn
&=& t \int\frac{\diff^3k_3}{(2\pi)^3}\sum_{n_3,l_3}  \Gamma_{n_3l_3}   \int\frac{\diff^3k'_3}{(2\pi)^3} \int \diff^3x' e^{i{\bs k}_3'\cdot {\bs x}'} \\
&&\langle    {\bs k}_1, nl, 1;\ {\bs k}_3 + \frac{{\bs k}_3'}{2}, n_3l_3, 1    | \rho_S(0) |       {\bs k}_2, nl, 1;\ {\bs k}_3-\frac{{\bs k}_3'}{2}, n_3l_3, 1  \rangle \,,
\ee
where we have inserted an identity $\int\diff^3k'_3\delta^3({\bs k}'_3)=1$ and written $\delta^3({\bs k}'_3)$ as a spatial integral over ${\bs x}'$. It should be noted that the integral over ${\bs k}_3'$ is already a Wigner transform on the density matrix of the second particle with momentum ${\bs k}_3$ and position ${\bs x}'$. If we further apply a Wigner transform on the density matrix of the first particle and properly reshuffle labels, we obtain the contribution of this term in the Boltzmann equation
\be
 \sum_{n',l' } \Gamma_{n'l'}   \int\frac{\diff^3k'}{(2\pi)^3} \int \diff^3x'  f({\bs x}, {\bs k}, nl;\ {\bs x}', {\bs k}', n'l';\ t)\,.
\ee
It involves the two-particle distribution function $ f({\bs x}, {\bs k}, nl;\ {\bs x}', {\bs k}', n'l';\ t)$ of two quarkonium states $nl$ and $n'l'$ with positions ${\bs x}$, ${\bs x}'$ and momenta ${\bs k}$, ${\bs k}'$ respectively. When the second quarkonium with the quantum number $n'l'$ annihilates, it leads to an increase in the one-particle distribution function of quarkonium with quantum number $nl$. Therefore, this term together with the term in Eq.~(\ref{eqn:annhi1}) guarantees the conservation of probability in the one-particle distribution of quarkonium. However, as mentioned earlier,  the annihilation effect is negligible in current heavy ion collision experiments.

\singlespacing
\chapter{Quarkonium Transport inside QGP: Phenomenology}
\vspace{0.2in}
\section{Coupled Transport Equations of Open Heavy Quarks and Quarkonium}

\doublespacing
As we derived the Boltzmann transport equation of quarkonium in the last chapter, we have seen the necessity to keep the heavy quark-antiquark octet as an active degree of freedom. With the octet kept track of, one can include the recombination effect based on first-principle calculations in a consistent way as in dissociation. Inside the QGP, the octet can be generated from quarkonium dissociation, or come from a statistical combination of open heavy quarks and antiquarks. In the latter case, the distance between the unbound heavy quark pair is not necessarily small. Thus, a dipole description of the interaction between the octet and the medium breaks down. We have to resum all the octet-medium interactions in the multipole expansion. If the octet potential is weak, we can approximately treat each open heavy quark and antiquark as independent and assume they individually interact with the medium. To this end, we couple the transport equation of quarkonium with the transport equations of open heavy quarks and antiquarks 
\be
\label{chap4:eqn_LBE}
(\frac{\partial}{\partial t} + \dot{{\bs x}}\cdot \nabla_{\bs x})f_Q({\bs x}, {\bs p}, t) &=& \ml{C}_Q  -  \ml{C}_Q^{+} +  \ml{C}_Q^{-}\\
(\frac{\partial}{\partial t} + \dot{{\bs x}}\cdot \nabla_{\bs x})f_{\bar{Q}}({\bs x}, {\bs p}, t) &=& \ml{C}_{\bar{Q}}  -  \ml{C}_{\bar{Q}}^{+} + \ml{C}_{\bar{Q}}^{-}\\
(\frac{\partial}{\partial t} + \dot{{\bs x}}\cdot \nabla_{\bs x})f_{nls}({\bs x}, {\bs p}, t) &=& \ml{C}_{nls} + \ml{C}_{nls}^{+} - \ml{C}_{nls}^{-}\,.
\ee
Here $f_i({\bs x}, {\bs p}, t)$ is the phase space distribution function of the particle species $i$ at time $t$, with position ${\bs x}$ and momentum $\bs p$, and $i$ can be $Q$ for open heavy quark, $\bar{Q}$ for open heavy antiquark and $nls$ for quarkonium with the quantum number $n$ (radial excitation of the wave function), $l$ (orbital angular momentum) and $s$ (spin). 
The left hand sides of the equations describe the free streaming of distribution functions in phase space while the right hand sides contain interactions between the heavy particles and the medium constituents. If these interactions are weak, the medium constituents can be thought of as light quarks and gluons (abbreviated as $q$ and $g$ from now on). For heavy quarks $Q$ (and similarly for $\bar{Q}$), the collision term $\ml{C}_Q$ includes three types of scattering processes: the elastic ${2\rightarrow2}$ scattering $qQ\rightarrow qQ$ and $gQ\rightarrow gQ$, the inelastic ${2\rightarrow3}$ scattering $qQ\rightarrow qQg$ and $gQ\rightarrow gQg$ and the inelastic ${3\rightarrow2}$ scattering $qQg\rightarrow qQ$ and $gQg\rightarrow gQ$. These processes in the Boltzmann transport equation have been studied both theoretically and numerically \cite{Svetitsky:1987gq,Gossiaux:2008jv,Gossiaux:2009mk,Uphoff:2014hza,Ke:2018tsh}. In this dissertation work, we will use the calculation and numerical implementation of $\ml{C}_{Q(\bar{Q})}$ in Ref.~\cite{Ke:2018tsh}. 

For each quarkonium species $nls$, its distribution function changes due to dissociation and recombination induced by real or virtual gluon absorptions and emissions. Virtual gluon absorption and emission correspond to inelastic scattering between quarkonium and medium constituents. The expressions for $\ml{C}_{nls}^{\pm}$ were given in the Chapter 3. 
The equation for a quarkonium species $nls$ only exists when the local QGP temperature is below its melting temperature. Otherwise the color attraction is too strongly screened to support that specific bound state with the quantum number $nls$. When the QGP temperature is above its melting temperature, the quarkonium is not a well-defined bound state and thus should only be considered as correlated unbound $Q\bar{Q}$. The sign of the interaction between the unbound $Q\bar{Q}$ depends on its color configuration (attraction for a color singlet and repulsion for a color octet). The color configuration may change every time one of the heavy flavors interacts with the medium. Here we will not explicitly keep track of the color degrees of freedom, so we treat the unbound $Q\bar{Q}$ approximately as independent objects traveling through the medium.

The dissociation and recombination of quarkonium also changes the distributions of heavy quarks and antiquarks. The quarkonium dissociation produces a pair of unbound heavy quark and antiquark, thus the corresponding heavy quark (antiquark) distributions increase. Similarly the quarkonium recombination leads to a decrease in the distributions of $Q$ and $\bar{Q}$. This effect has been accounted in the set of coupled Boltzmann equations~(\ref{chap4:eqn_LBE}). It should be emphasized that the signs of $\ml{C}_{nls}^{\pm}$ differ in the equations of open heavy flavor and quarkonium. One also needs to sum over all quarkonium species $nls$ whose melting temperatures are above the local QGP temperature in the equations for $Q$ and $\bar{Q}$.

The collision term $\ml{C}_{nls}$ for quarkonium has some similarity with the $\ml{C}_{Q(\bar{Q})}$. It describes the quarkonium diffusion inside the QGP. It has a contribution from the elastic scattering studied in the last chapter. But there is no contribution from any process where only a single gluon line is attached to the quarkonium, because it changes the color of the bound $Q\bar{Q}$ and thus corresponds to a process of dissociation/recombination.

\vspace{0.2in}

\section{Monte Carlo Method}
In this section, we describe a numerical method to solve the coupled Boltzmann transport equations.

We will use the Monte Carlo method to simulate the evolution event. For each particle species, we sample a certain number of particles according to the distribution function at initial time $t_0$, $f_i({\bs x}, {\bs p}, t=t_0)$. Mathematically,
\be
\tilde{f}_i({\bs x}, {\bs p}, t=t_0) = \sum_{k=1}^{N_i}  \delta^3({\bs x} - {\bs x}_k) \delta^3({\bs p} - {\bs p}_k)\,,
\ee
where the total number $N_i$ is fixed by the initial condition in one collision event:
\be
N_i =  \int \diff^3x \diff^3p   \tilde{f}_i({\bs x}, {\bs p}, t=t_0)\,.
\ee
In practice, the total number may be less than one, $N_i<1$. This means that not every collision event produces such kind of a particle $i$. In this case, we will do a certain number of simulations without any particle $i$, and some other number of simulations with just one particle $i$ such that on average $N_i$ particle $i$ is produced initially. Collision events with more than one particle $i$ produced initially will be extremely rare and not considered here. Each $i$-particle has definite position ${\bs x}_k$ and momentum ${\bs p}_k$. Then we evolve each sampled particle according to the transport equation at discretized time step $\Delta t$,
\be
\tilde{f}_i(t+\Delta t) =  \tilde{f}_i(t) - \dot{\bs x} \cdot \nabla \tilde{f}_i \Delta t + \ml{C}_i \Delta t \,,
\ee
where we omit the dependence on ${\bs x}$ and ${\bs p}$ and simply write the different collision terms as $\ml{C}_i$ for the particle species $i = Q$, $\bar{Q}$ or $nls$. We will discuss each collision term separately later. At each time step, we will consider the following processes: free streaming, diffusion and energy loss of open heavy flavors, dissociation of quaurkonium and recombination of unbound $Q\bar{Q}$ into a specific quarkonium state $nls$. Our basic numerical procedure is to calculate the rate $\Gamma$ of each collision process. Then the probability of some process happening is given by $\Gamma \Delta t$ (we will choose $\Delta t$ such that $\Gamma \Delta t \ll 1$). We will determine whether that specific process happens by sampling a random number between $0$ and $1$ and compare with $\Gamma \Delta t $. If the random number is smaller, then the process occurs and vice versa. If some process happens in a time step, we will change the particle species when it is necessary and update the momentum of the particle according to the energy-momentum conservation in the process. We will iterate over a number of time steps. If we sample enough number of collision events, we expect to reproduce the transport evolution of the phase space distribution from the distribution of the sampled particles,
\be
\tilde{f}_i({\bs x}, {\bs p}, t) \xrightarrow{N_\ma{event}\rightarrow\infty} f_i({\bs x}, {\bs p}, t) \,.
\ee

Now we explain how to implement this for each process.

\subsection{Free Streaming}
At each time step, we will first consider the free streaming. For a particle with position ${\bs x}$ and momentum ${\bs p}$ in the laboratory frame, we will update the position according to
\be
{\bs x}' = {\bs x} + \frac{{\bs p}}{\sqrt{M^2 + {\bs p}^2}} \Delta t \,,
\ee
where $M$ is the particle's mass.

\subsection{Open Heavy Flavor Diffusion and Energy Loss}
The numerical implementation of the open heavy flavor transport is the main topic of the dissertation work of my colleagues Weiyao Ke and Yingru Xu (supervisor: Steffen Bass). In practical numerical simulations, we will use their code for the linearized Boltzmann equation, described in Ref.~\cite{Ke:2018tsh}. The code will determine whether an open heavy quark with a given momentum scatters in a time step and if so, determine whether it is an elastic ${2\rightarrow2}$ scattering, inelastic ${2\rightarrow3}$ scattering or inelastic ${3\rightarrow2}$ scattering. Then the code will give the final momentum of the open heavy quark after the scattering. So we can update the momentum of the open heavy quarks.

\subsection{Dissociation of Quarkonium}
At each time step, for each quarkonium with the quantum number $nls$, position ${\bs x}$ and momentum ${\bs p}_\ma{lab}$, we will obtain the medium information at the position ${\bs x}$ from the hydrodynamics simulation. For our current implementation, we will obtain the hydro-cell temperature $T$ (defined in the hydro-cell rest frame) and velocity ${\bs v}$ with respect to the laboratory frame. We will first boost the quarkonium into the rest frame of the hydro-cell according to
\be
p_\ma{cell} &=& \Lambda({\bs v}) p_\ma{lab} \\
p_\ma{cell} &=& \begin{pmatrix}
E_\ma{cell} \\
p_{x\ma{cell}}\\
p_{y\ma{cell}}\\
p_{z\ma{cell}}
\end{pmatrix} \\
p_\ma{lab} &=& \begin{pmatrix}
E_\ma{lab} = \sqrt{M^2_{nls} + {\bs p}_\ma{lab}^2}\\
p_{x\ma{lab}}\\
p_{y\ma{lab}}\\
p_{z\ma{lab}}
\end{pmatrix} \,,
\ee
where the most general form of the Lorentz transformation is given by
\be
\Lambda({\bs v}) &=&  
\begin{pmatrix}
\gamma                          & -\gamma v_{x} & -\gamma v_{y} & -\gamma v_{z}  \\
-\gamma v_{x}  & 1+(\gamma-1)\frac{v_{x}^2}{{\bs v}^2} &
 (\gamma-1)\frac{v_{x}v_{y}}{{\bs v}^2} & (\gamma-1)\frac{v_{x}v_{z}}{{\bs v}^2}  \\
-\gamma v_{y}  & (\gamma-1)\frac{v_{y}v_{x} }{{\bs v}^2} &
 1+(\gamma-1)\frac{ v^2_{y}}{{\bs v}^2} & (\gamma-1)\frac{v_{y}v_{z}}{{\bs v}^2}  \\
 -\gamma v_{z}  & (\gamma-1)\frac{v_{z}v_{x} }{{\bs v}^2} &
 (\gamma-1)\frac{ v_{z} v_{y}}{{\bs v}^2} & 1+(\gamma-1)\frac{v^2_{z}}{{\bs v}^2} 
\end{pmatrix} \,,
\ee
in which $\gamma = 1/\sqrt{1-{\bs v}^2}$. In the hydro-cell rest frame, the quarkonium velocity is given by
\be
{\bs v}_\ma{cell} = \frac{{\bs p}_\ma{cell}}{E_\ma{cell}} \,.
\ee

Our formula for dissociation rates in the last chapter is derived in the rest frame of the quarkonium, in which the nonrelativistic expansion holds, and by assuming the quarkonium is at rest with respect to the medium. In practice, quarkonium is moving with respect to the medium hydro-cell in general. So we need to modify the rate formula by replacing
\be
n_B(q_0) &\longrightarrow&  n_B(\Lambda^0_{\ \mu}(-{\bs v}_\ma{cell}) q^\mu) \\
n_E(q_0) &\longrightarrow&  n_E(\Lambda^0_{\ \mu}(-{\bs v}_\ma{cell}) q^\mu) \,.
\ee
Then we obtain the rates in the rest frame of the quarkonium. To get the rates in the rest frame of the hydro-cell, we divide by another $\gamma$-factor
\be
\gamma_\ma{cell} = \frac{1}{\sqrt{1-{\bs v}_\ma{cell}^2}} \,.
\ee
We will explain these procedures for each process that results in quarkonium dissociation.

\subsubsection{Real Gluon Absorption}
For the dissociation induced by real gluon absorption, we modify the expression~(\ref{chap3_eqn_disso_gluon}) and obtain the rate in the hydro-cell rest frame
\be 
\label{chap4_eqn_disso_real}
\Gamma^g_{\ma{cell}} &=& \frac{1}{\gamma_\ma{cell}} \frac{M}{2\pi}  \int \frac{\diff^3 q}{2q(2\pi)^3} n_B( \Lambda(-{\bs v}_\ma{cell})q) p_\ma{rel} 
 \frac{2}{3}g^2C_Fq^2 |  \langle   \Psi_{{\bs p}_\ma{rel}} | {\bs r} |   \psi_{nl}  \rangle |^2 \,,
\ee
where $p_\ma{rel} = |{\bs p}_\ma{rel} | =  \sqrt{M(q-|E_{nl}|)}$.
The only angular dependence in the integrand is via the Bose-Einstein distribution. To simplify the expression, we assume ${\bs v}_\ma{cell}$ is along the $z$-axis,
\be
\int \frac{\diff^3 q}{2q(2\pi)^3} n_B( \Lambda(-{\bs v}_\ma{cell})q) = \int \frac{q^2\diff q \diff \cos\theta \diff \phi}{2q(2\pi)^3} \frac{1}{e^{\gamma_\ma{cell}(1+v_\ma{cell}\cos\theta)q/T} - 1}
\ee
We will focus on the angular integral and omit the subscript ``cell" for a moment,
\be \nn
I_{\theta}(v) &\equiv&\int_{-1}^1\diff\cos\theta\frac{1}{e^{\gamma(1+v\cos\theta)q/T}-1}\\ \nn
&=&\frac{T}{\gamma v q}\int_{\gamma(1-v)q/T}^{\gamma(1+v)q/T} \diff x \frac{1}{e^x-1} \\\nn
&=& \frac{T}{\gamma vq}\Big(  \ln{(1-e^{-\gamma(1+v)q/T})} -\ln{(1-e^{-\gamma(1-v)q/T})}   \Big)\,.
\ee
One can also check the $v\rightarrow0$ limit.
\be \nn
\lim_{v\rightarrow0}I_{\theta}(v) 
&=& \frac{T}{q}\frac{\diff}{\diff v}\Big\{\frac{1}{\gamma} \big[ \ln{(1-e^{-\gamma(1+v)q/T})} -\ln{(1-e^{-\gamma(1-v)q/T})} \big] \Big\}_{v=0} \\
&=& \frac{2}{e^{q/T}-1}\,.
\ee
Then we can write Eq.~(\ref{chap4_eqn_disso_real}) as
\be
\label{chap4_eqn_disso_real2}
\Gamma^g_{\ma{cell}} = \frac{2 \alpha_s  MT }{9\pi^2 \gamma_\ma{cell}^2v_\ma{cell}} \int q^2 \diff q  \sqrt{M(q-|E_{nl}|)} \ln{ \frac{1-e^{-\gamma_\ma{cell}(1+v_\ma{cell})q/T}}{1-e^{-\gamma_\ma{cell}(1-v_\ma{cell})q/T}} }
|  \langle   \Psi_{{\bs p}_\ma{rel}} | {\bs r} |   \psi_{nl}  \rangle |^2 \,, \,\,\,\,\,\,
\ee
in which $|{\bs p}_\ma{rel} | = \sqrt{ M(q-|E_{nl}|) }$ and the dissociation process cannot happen unless enough energy is absorbed $q>|E_{nl}|$ to excite the bound state to the continuum.

\subsubsection{Inelastic Scattering with Light Quarks}
The dissociation rate caused by inelastic scattering with light quarks when the quarkonium is at rest with respect to the medium is written in Eq.~(\ref{chap3_eqn_disso_inel}). When the quarkonium is moving at the velocity ${\bs v}_\ma{cell}$ with respect to the medium, we repeat the above procedure and obtain the rate in the hydro-cell rest frame
\be \nn
\Gamma^{\ma{inel},q}_{\ma{cell}} &=& \frac{1}{\gamma_\ma{cell}} \int \frac{\diff^3 p_1}{2p_1(2\pi)^3} \frac{\diff^3 p_2}{2p_2(2\pi)^3} n_F(\Lambda(-{\bs v}_\ma{cell})p_1) \big( 1- n_F(\Lambda(-{\bs v}_\ma{cell})p_2) \big)   \\
&& \frac{Mp_\ma{rel}}{2\pi} \frac{16}{3}g^4T_FC_F  |\langle \Psi_{{\bs p}_\ma{rel}} | {\bs r} |  \psi_{nl}  \rangle|^2   \frac{p_1p_2 + {\bs p}_1\cdot {\bs p}_2}{{\bs q}^2} \,,
\ee
in which $p_\ma{rel} = |{\bs p}_\ma{rel}| = \sqrt{M(p_1 - |E_{nl}| - p_2)}$.
Again we will assume ${\bs v}_\ma{cell}$ as the $z$-axis. Then the momentum of the incoming light quark is specified by the magnitude $p_1 = |{\bs p}_1|$ and the polar angle $\theta_1$. The momentum of the outgoing light quark is specified by the magnitude $p_2 = |{\bs p}_2|$, the polar angle $\theta_2$ and the azimuthal angle $\phi_2$. Then we can simplify $\Gamma^{\ma{inel},q}_{\ma{cell}}$,
\be \nn
\Gamma^{\ma{inel},q}_{\ma{cell}} &=& \frac{2\alpha_s^2 M}{9\pi^4\gamma_\ma{cell}} \int p_1 \diff p_1 \diff c_1 n_F \Big(\frac{p_1(1+v_\ma{cell}c_1)}{\sqrt{1-v_\ma{cell}^2}} \Big) \\  \nn
&& \int p_2 \diff p_2 \diff c_2 \diff \phi_2 \Big[ 1 - n_F\Big(\frac{p_2(1+v_\ma{cell}c_2)}{\sqrt{1-v_\ma{cell}^2}} \Big)  \Big] \sqrt{M(p_1+|E_{nl}| - p_2)}  \\
\label{chap4_eqn_disso_ineq}
&& \frac{p_1p_2(1+s_1s_2\cos\phi_2+c_1c_2)}{p_1^2+p_2^2 - 2p_1p_2(s_1s_2\cos\phi_2 + c_1c_2)} |\langle \Psi_{{\bs p}_\ma{rel}} | {\bs r} |  \psi_{nl}  \rangle|^2  \,,
\ee
where we define $s_i \equiv \sin\theta_i$ and $c_i \equiv \cos\theta_i$.

\subsubsection{Inelastic Scattering with Gluons}
When the quarkonium is at rest in the hydro-cell, the dissociation rate induced by the inelastic scattering with medium gluons is written in Eq.~(\ref{chap3_eqn_disso_inel}). We generalize it to the case in which the quarkonium is moving at the velocity ${\bs v}_\ma{cell}$ with respect to the medium
\be \nn
\Gamma^{\ma{inel},g}_{\ma{cell}} &=& \frac{1}{\gamma_\ma{cell}} \int \frac{\diff^3 q_1}{2q_1(2\pi)^3} \frac{\diff^3 q_2}{2q_2(2\pi)^3} n_B(\Lambda(-{\bs v}_\ma{cell})q_1) \big( 1 + n_B(\Lambda(-{\bs v}_\ma{cell})q_2) \big)  \\
&& \frac{Mp_\ma{rel}}{2\pi}  \frac{1}{3}g^4C_F | \langle \Psi_{{\bs p}_\ma{rel}} | {\bs r} |  \psi_{nl}  \rangle|^2 \frac{1+(\hat{q}_1\cdot\hat{q}_2)^2}{{\bs q}^2} (q_1+q_2)^2 \,,
\ee
in which $p_\ma{rel} = |{\bs p}_\ma{rel}| = \sqrt{M(q_1 - |E_{nl}| - q_2)}$.
Again we will assume ${\bs v}_\ma{cell}$ as the $z$-axis. Then the momentum of the incoming gluon is specified by the magnitude $q_1 = |{\bs q}_1|$ and the polar angle $\theta_1$. The momentum of the outgoing gluon is specified by the magnitude $q_2 = |{\bs q}_2|$, the polar angle $\theta_2$ and the azimuthal angle $\phi_2$. Then we can simplify $\Gamma^{\ma{inel},q}_{\ma{cell}}$,
\be \nn
\Gamma^{\ma{inel},g}_{\ma{cell}} &=& \frac{\alpha_s^2M}{12\pi^4\gamma_\ma{cell}} \int q_1 \diff q_1 \diff c_1 n_B \Big(\frac{q_1(1+v_\ma{cell}c_1)}{\sqrt{1-v_\ma{cell}^2}} \Big) \\  \nn
&& \int q_2 \diff q_2 \diff c_2 \diff \phi_2 \Big[ 1 + n_B\Big(\frac{q_2(1+v_\ma{cell}c_2)}{\sqrt{1-v_\ma{cell}^2}} \Big)  \Big] \sqrt{M(q_1+|E_{nl}| - q_2)}  \\
\label{chap4_eqn_disso_ineg}
&& | \langle \Psi_{{\bs p}_\ma{rel}} | {\bs r} |  \psi_{nl}  \rangle|^2 \frac{(q_1+q_2)^2(1+s_1s_2\cos\phi_2 + c_1c_2)}{q_1^2 + q_2^2 - 2q_1q_2(s_1s_2\cos\phi_2 + c_1c_2)} \,,
\ee
where we define $s_i \equiv \sin\theta_i$ and $c_i \equiv \cos\theta_i$.

\vspace{0.2in}
The total dissociation rate of the quarkonium state $nls$ in the hydro-cell frame is given by
\be
\Gamma_{\ma{cell}} = \Gamma^g_{\ma{cell}} + \Gamma^{\ma{inel},q}_{\ma{cell}}  + \Gamma^{\ma{inel},g}_{\ma{cell}} \,.
\ee
For a time step $\Delta t$ in the laboratory frame, the dissociation probability in the laboratory frame is given by
\be
P_\ma{lab} = \Gamma_{\ma{cell}} \frac{ E_\ma{cell} }{E_\ma{lab} } \Delta t  = P^g_\ma{lab} + P^{\ma{inel},q}_\ma{lab} + P^{\ma{inel},g}_\ma{lab}  \,.
\ee
After calculating the dissociation probability in the laboratory frame, we sample a random number $r$ from a uniform distribution between $0$ and $1$ and decide which process occurs based on
\begin{enumerate}
\item if $r \leq P^g_\ma{lab}$, the quarkonium $nls$ dissociates in the real gluon absorption channel;
\item if $P^g_\ma{lab} < r \leq P^g_\ma{lab}  + P^{\ma{inel},q}_\ma{lab}$, the quarkonium $nls$ dissociates in the channel of inelastic scattering with a light quark;
\item if $P^g_\ma{lab}  + P^{\ma{inel},q}_\ma{lab} < r \leq  P_\ma{lab}$, the quarkonium $nls$ dissociates in the channel of inelastic scattering with a gluon;
\item if $P_\ma{lab} < r$, the quarkonium $nls$ does not dissociate in this time step.
\end{enumerate}

If the quarkonium $nls$ dissociates in a certain process in this time step, we will remove this particle from the list of all quarkonium states with $nls$ and add a heavy quark (antiquark) to the list of open heavy quark (antiquark). To this end, we need to know the position and momentum of the open heavy quark (antiquark) from the dissociation. The positions of the $Q$ and $\bar{Q}$ are both given by ${\bs x}$, the position of the quarkonium just before the dissociation. The position assignment can be improved by sampling a relative position ${\bs r}$ from the bound state wave function $|\psi_{nl}({\bs r})|^2$ and assign ${\bs x} \pm \frac{{\bs r}}{2}$ to the $Q$ and $\bar{Q}$ in the quarkonium rest frame and then boost the positions back into the laboratory frame. The assignment of momenta is more involved and we will explain this in the following for each dissociation process.

\subsubsection{Momentum Sampling in Real Gluon Absorption}
In the rest frame of the quarkonium, the momentum magnitude of the absorbed gluon can be sampled from the integrand of Eq.~(\ref{chap4_eqn_disso_real2})
\be
I_g(q) = q^2 \sqrt{M(q-|E_{nl}|)} \ln{ \frac{1-e^{-\gamma_\ma{cell}(1+v_\ma{cell})q/T}}{1-e^{-\gamma_\ma{cell}(1-v_\ma{cell})q/T}} }
|  \langle   \Psi_{{\bs p}_\ma{rel}} | {\bs r} |   \psi_{nl}  \rangle |^2 \,.
\ee
We will use the acceptance-rejection method to sample the momentum magnitude. We will first sample a random number $q_\ma{try}$ from a uniform distribution between $|E_{nl}|$ (enough energy has to be transferred to the bound state to break it up) and, for example, $55|E_{nl}|$ (above which the integrand is almost vanishing). Then we will sample another random number $r$ from a uniform distribution between $0$ and $1$. If
\be
r \leq \frac{I_g(q_\ma{try})}{ \max{I_g(q)} } \,,
\ee
we will accept this $q_\ma{try}$ as the momentum magnitude of the incoming gluon. If the inequality fails, we will repeat the whole procedure by sampling a new set of $q_\ma{try}$ and $r$ and compare. We will stop repeating when some $q_\ma{try}$ is accepted. 

Once we have the momentum magnitude of the incoming gluon, we can sample its direction with respect to ${\bs v}_\ma{cell}$, the quarkonium velocity in the hydro-cell. We have taken ${\bs v}_\ma{cell}$ to be the $z$-axis in Eq.~(\ref{chap4_eqn_disso_real2}). The part in Eq.~(\ref{chap4_eqn_disso_real}) that is relevant to the $\theta$ sampling is (we omit the subscript ``cell" for simplicity) 
\be
\frac{1}{e^{\gamma(1+v\cos\theta)q/T} - 1} \,.
\ee
We will use the inverse function method to sample $\theta$. We first define
\be \nn
F(x) &\equiv& \int^x_{-1} \diff\cos\theta \frac{1}{e^{\gamma(1+v\cos\theta)q/T} - 1} \\
&=& \frac{T}{\gamma v q}\Big[  \ln{(1-e^{-\gamma(1+vx)q/T})}  -  \ln{(1-e^{-\gamma(1-v)q/T})} \Big] \,.
\ee
Then we generate a random number $r$ from a uniform distribution between $0$ and $1$, and solve $x$ from the equation
\be
r = \frac{F(x) }{F(1)} = \frac{  \ln{(1-e^{-\gamma(1+vx)q/T})}  -  \ln{(1-e^{-\gamma(1-v)q/T})} }{  \ln{(1-e^{-\gamma(1+v)q/T})}  -  \ln{(1-e^{-\gamma(1-v)q/T})} } \,.
\ee
The solution is
\be
x  &=& \frac{1}{v} \Big[ -1-\frac{1}{B}\ln{(1-C)} \Big] \\
B &=& \frac{\gamma q}{T} \\
C &=& r \ln{(1-e^{-B(1+v)})} + (1-r) \ln{(1-e^{-B(1-v)})} \,.
\ee
The solution gives the polar angle of the incoming gluon momentum $\theta$ with respect to the $z$-axis. We can write the incoming gluon four-momentum in the quarkonium rest frame as
\be
p_g = \begin{pmatrix} q \\ q\sin\theta \\ 0 \\ q\cos\theta \end{pmatrix} \,.
\ee
After absorbing this gluon, the quarkonium splits into an unbound $Q\bar{Q}$ pair. The relative kinetic energy between the $Q\bar{Q}$ is given by
\be
\frac{{\bs p}_\ma{rel}^2}{M} = q - |E_{nl}| \,.
\ee
But the direction of ${\bs p}_\ma{rel}$ is not fixed. Since there is no preferred direction in the matrix element squared $| \langle \Psi_{{\bs p}_\ma{rel}} | {\bs r} |  \psi_{nl}  \rangle|^2$ (we have averaged over the third component of the angular momentum), we can sample a $\cos\theta_\ma{rel}$ and a $\phi_\ma{rel}$ uniformly between $-1$ and $1$ and between $0$ and $2\pi$ respectively. Then the relative momentum between the $Q\bar{Q}$ is
\be
\label{chap4_eqn_prel}
{\bs p}_\ma{rel} = \begin{pmatrix}  p_\ma{rel}\sin\theta_\ma{rel}\cos\phi_\ma{rel} \\ p_\ma{rel}\sin\theta_\ma{rel}\sin\phi_\ma{rel} \\ p_\ma{rel}\cos\theta_\ma{rel} \end{pmatrix} \,.
\ee 
So the outgoing open $Q$ and $\bar{Q}$ has the following momentum in the rest frame of the quarkonium,
\be
\label{chap4_eqn_pQpQbar}
{\bs p}_Q &=& {\bs p}_\ma{rel} + \frac{{\bs q}_g}{2} \\
{\bs p}_{\bar{Q}} &=& -{\bs p}_\ma{rel} + \frac{{\bs q}_g}{2} \,.
\ee
The energy of the open heavy flavor can be calculated as $ E_{Q(\bar{Q})} = \sqrt{M^2 + {\bs p}_{Q(\bar{Q})}^2 }$. We need to boost it back to the hydro-cell frame and then boost it back to the laboratory frame. But before we do the Lorentz transformations, we notice that ${\bs v}_\ma{cell}$ is generally not aligned with the $z$-axis in the hydro-cell frame. So we need to rotate $p_{Q(\bar{Q})}$ by a sequence of rotations. We first rotate them around the $y$-axis counterclockwise by $\theta$ and then rotate around the $z$-axis counterclockwise by $\phi$, where $\theta$ and $\phi$ are the polar and azimuthal angles of ${\bs v}_\ma{cell}$ in the hydro-cell frame. In the matrix representation, we do a sequence of matrix multiplications
\be
\label{chap4_eqn_rotate}
{\bs p}^\ma{rot}_{Q(\bar{Q})} = \begin{pmatrix}
  \cos\phi &  - \sin\phi &   \\
  \sin\phi & \cos\phi &   \\
   &  & 1
\end{pmatrix} 
\begin{pmatrix}
  \cos\theta &   & \sin\theta  \\
   & 1 &   \\
  -\sin\theta &  & \cos\theta
\end{pmatrix} {\bs p}_{Q(\bar{Q})} \,.
\ee
Then we will boost it back to the hydro-cell frame and then boost it back to the laboratory frame by two Lorentz transformations
\be
\label{chap4_eqn_boostlab}
p^\ma{lab}_{Q(\bar{Q})} = \Lambda(-{\bs v}) \Lambda(-{\bs v}_\ma{cell})  p^\ma{rot}_{Q(\bar{Q})}\,,
\ee
in which $p^\ma{rot}_{Q(\bar{Q})} = (E_{Q(\bar{Q})}, {\bs p}^\ma{rot}_{Q(\bar{Q})})$.
Eq.~(\ref{chap4_eqn_boostlab}) gives the momenta of the open heavy quark and antiquark from the dissociation.

\subsubsection{Momentum Sampling in Inelastic Scattering with Light Quarks}
The sampling of the momenta of the incoming and outgoing light quarks is more involved in this case because the integrand of Eq.~(\ref{chap4_eqn_disso_ineq}) has complicated dependence on the momenta. A naive rejection sampling that starts with uniform distributions of ${\bs p}_1$ and ${\bs p}_2$ is computationally inefficient. To speed up the sampling procedure, we need to use the importance sampling method: for a function $f(x)$ that is changing rapidly with $x$, we first factorize out a factor $g(x)$ and write 
\be
\int \diff x f(x) = \int \diff x g(x) \frac{f(x)}{g(x)} \,. 
\ee
We hope we can find such a function $g(x)$ that we know how to sample $x$ from the distribution $g(x)$ efficiently and the remaining part $\frac{f(x)}{g(x)}$ is as flat as possible. Then we can sample a $x$ from the the distribution $g(x)$ in a fast way and then do a rejection sampling on the remaining piece. We will apply the importance sampling method to the integrand of Eq.~(\ref{chap4_eqn_disso_ineq}). We will first change the variable from $p_2$ to $p_\ma{rel}$ (it turns out that sampling $p_\ma{rel}$ is much faster than sampling $p_2$),
\be
p_2 = p_1 - |E_{nl}| - \frac{p_\ma{rel}^2}{M} \,.
\ee
Then we write Eq.~(\ref{chap4_eqn_disso_ineq}) as, up to a multiplication of some constant
\be
\int \diff p_1 \diff p_\ma{rel} \diff c_1 \diff c_2 \diff \phi_2 \,g_1(p_1,c_1) g_2(p_\ma{rel}) h(p_1, p_\ma{rel}, c_1, c_2, \phi_2)  \,,
\ee
in which the functions $g_i$ are the factors we factorize out and $h$ is the remaining part. They are defined as (again, we omit the subscript ``cell" for simplicity)
\be
g_1(p_1, c_1) &=& p_1  \frac{1}{e^{\gamma(1+vc_1)p_1/T}+1}    \frac{1}{\frac{ p_2 }{p_1} + \frac{p_1}{p_2} - 2} \\
g_2(p_\ma{rel}) &=& p_\ma{rel}^2 | \langle \Psi_{{\bs p}_\ma{rel}} | {\bs r} |  \psi_{nl}  \rangle|^2 \\ \nn
h(p_1, p_\ma{rel}, c_1, c_2, \phi_2) &=& \frac{p_2}{p_1}\big[ 1-n_F(\gamma(1+vc_2)p_2) \big]  \frac{1+s_1s_2\cos\phi_2+c_1c_2}{2} \\
&& \frac{\frac{ p_2 }{p_1} + \frac{p_1}{p_2 } - 2}{\frac{p_1}{p_2} + \frac{p_2}{p_1} - 2(s_1s_2\cos\phi_2 + c_1c_2)} \,.
\ee
The sampling of $p_1$ and $c_1$ according to the distribution function $g_1(p_1, c_1)$ is very similar to the sampling procedure described for the real gluon absorption process. We will sample $p_1$ using the rejection method on the function
\be
\int_{-1}^1 \diff c_1 g_1(p_1, c_1) = \frac{T}{\gamma v}   \ln{\frac{1+e^{-\gamma(1-v)p_1/T}}{1+e^{-\gamma(1+v)p_1/T}}}   \frac{1}{\frac{ p_2 }{p_1} + \frac{p_1}{p_2 } - 2} \,.
\ee
The range of the trial uniform distribution is taken as $[|E_{nl}|, 15T/\sqrt{1-v_\ma{cell}}]$, where $v_\ma{cell}$ is the velocity magnitude of the quarkonium in the rest frame of the hydro-cell.
We sample $c_1$ using the inverse function method once we have sampled a $p_1$. The function we need to invert is defined as
\be \nn
G(x) &\equiv&   \int_{-1}^x \diff c_1 \frac{1}{e^{\gamma(1+vc_1)p_1/T}+1} \\
&=& -\frac{T}{\gamma v p_1} \Big[  \ln{ (1+e^{-\gamma(1+vx)p_1/T})} -  \ln{ (1+e^{-\gamma(1-v)p_1/T} )}    \Big]  \,.
\ee
The $x = \cos\theta$ can be solved from the equation $r = G(x)$ where $r$ is a random number generated from a uniform distribution between $0$ and $1$.

The magnitude of the relative momentum $p_\ma{rel}$ between the unbound $Q$ and $\bar{Q}$ is sampled using the rejection method on the function $g_2(p_\ma{rel})$. The lower and upper limits of $p_\ma{rel}$ used in the rejection method are taken as $0$ and $3.5/a_B$ for 1S state and $0$ and $1.5/a_B$ for 2S state.

Then we will do a rejection sampling on the remaining part $h(p_1, p_\ma{rel}, c_1, c_2, \phi_2)$. We will further generate a random number $c_2$ uniformly between $-1$ and $1$ and $\phi_2$ uniformly between $0$ and $2\pi$. Together with the sampled $p_1$, $p_\ma{rel}$ and $ c_1$ as described above, we can calculate the function value $h(p_1, p_\ma{rel}, c_1, c_2, \phi_2)$. We can further generate a random number $r$ uniformly between $0$ and $1$ and decide to accept this sampling if $r ( \max{h}) \leq h(p_1, p_\ma{rel}, c_1, c_2, \phi_2)$. The maximum of $h$ is not easy to obtain. In practice, we notice that $h\leq1$. So we simply set $\max{h} = 1$ in the acceptance criterial. We may lose some efficiency in this step of the sampling procedure but the overall efficiency in the sampling is good enough for our purpose.

\begin{figure}
    \centering
    \begin{subfigure}[b]{0.48\textwidth}
        \centering
        \includegraphics[height=2.1in]{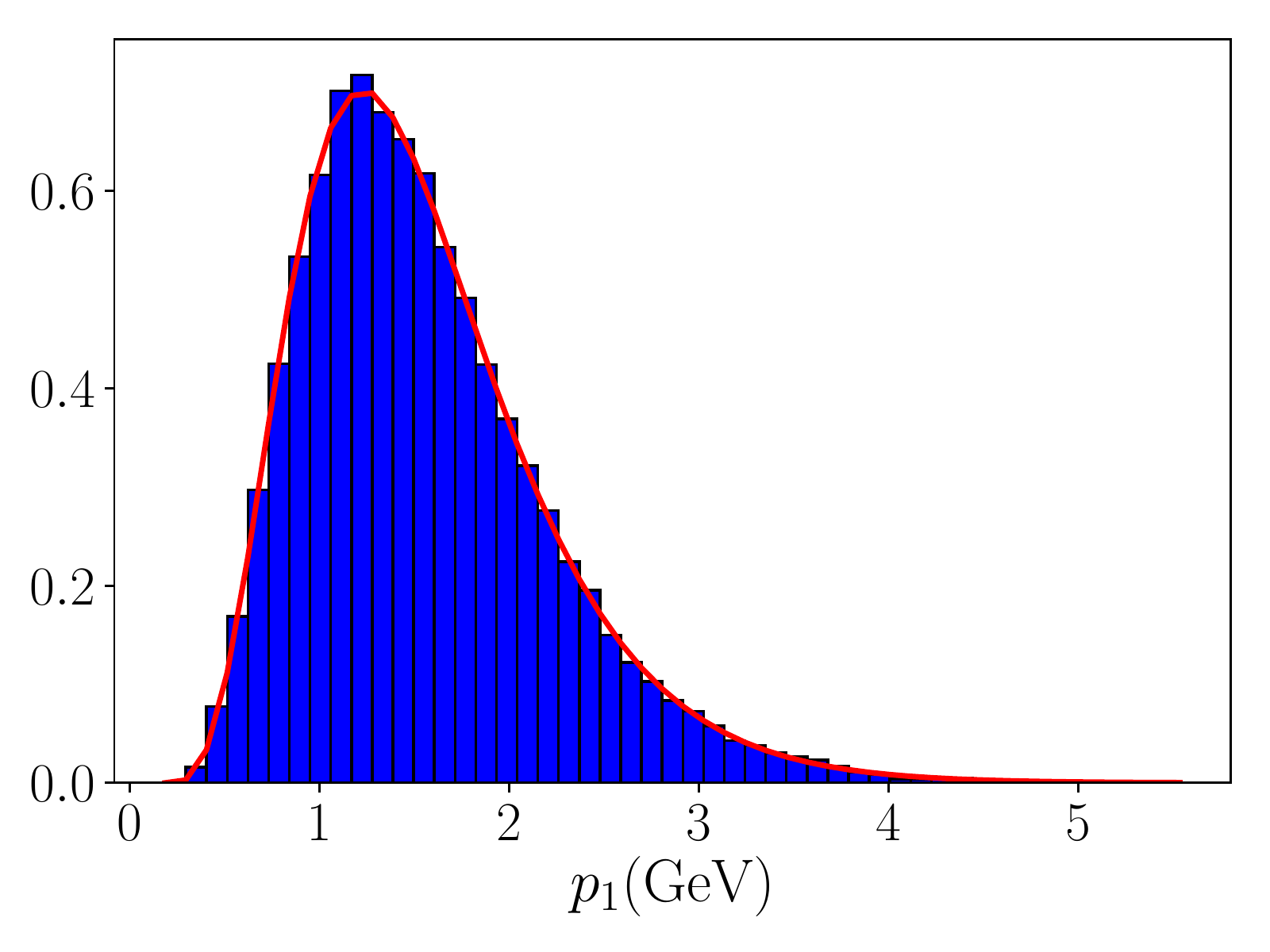}
        \caption{$p_1$}\label{}
    \end{subfigure}%
    ~
    \begin{subfigure}[b]{0.48\textwidth}
        \centering
        \includegraphics[height=2.1in]{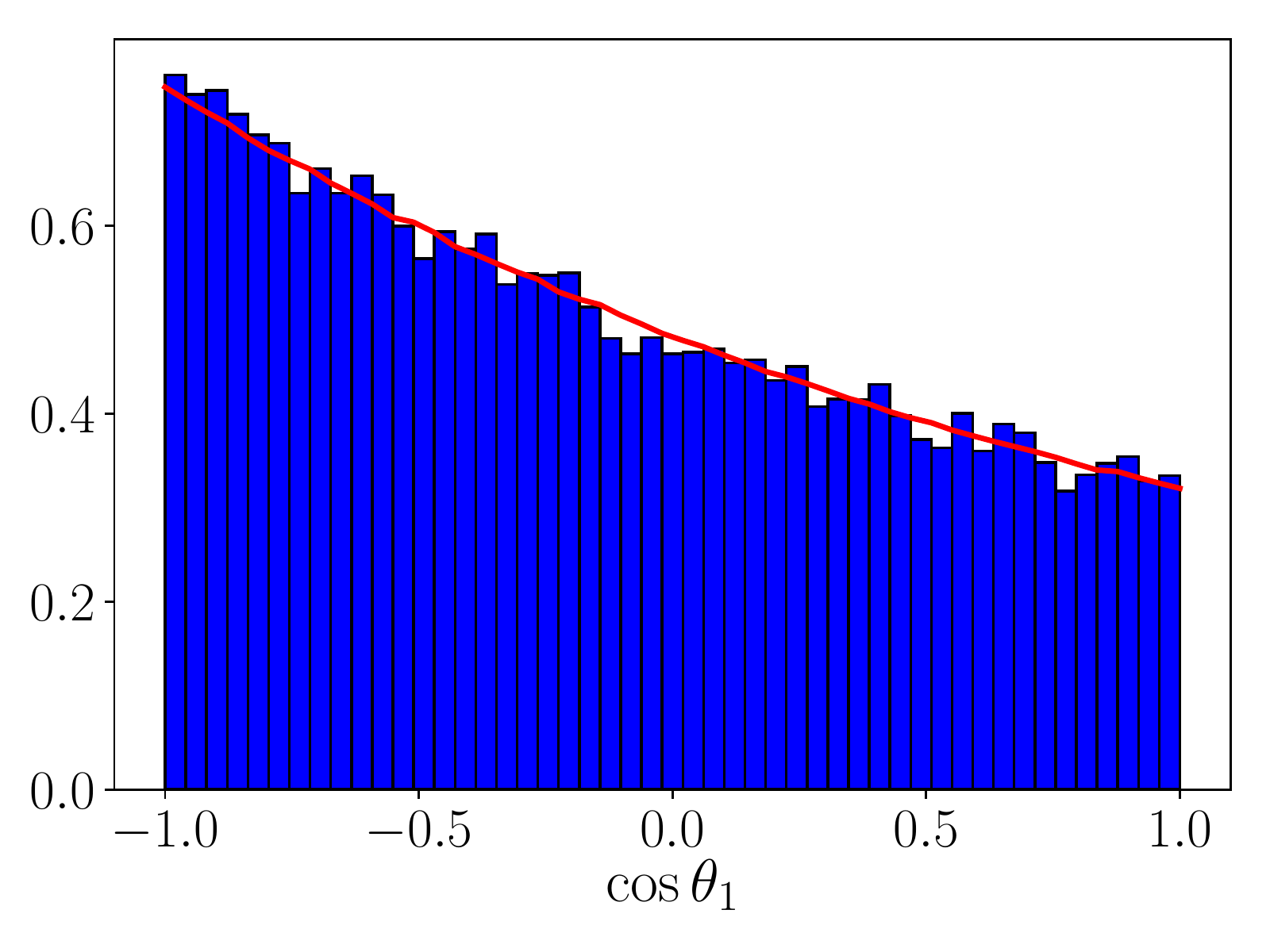}
        \caption{$c_1$}\label{}
    \end{subfigure}%
    
    \begin{subfigure}[b]{0.48\textwidth}
        \centering
        \includegraphics[height=2.1in]{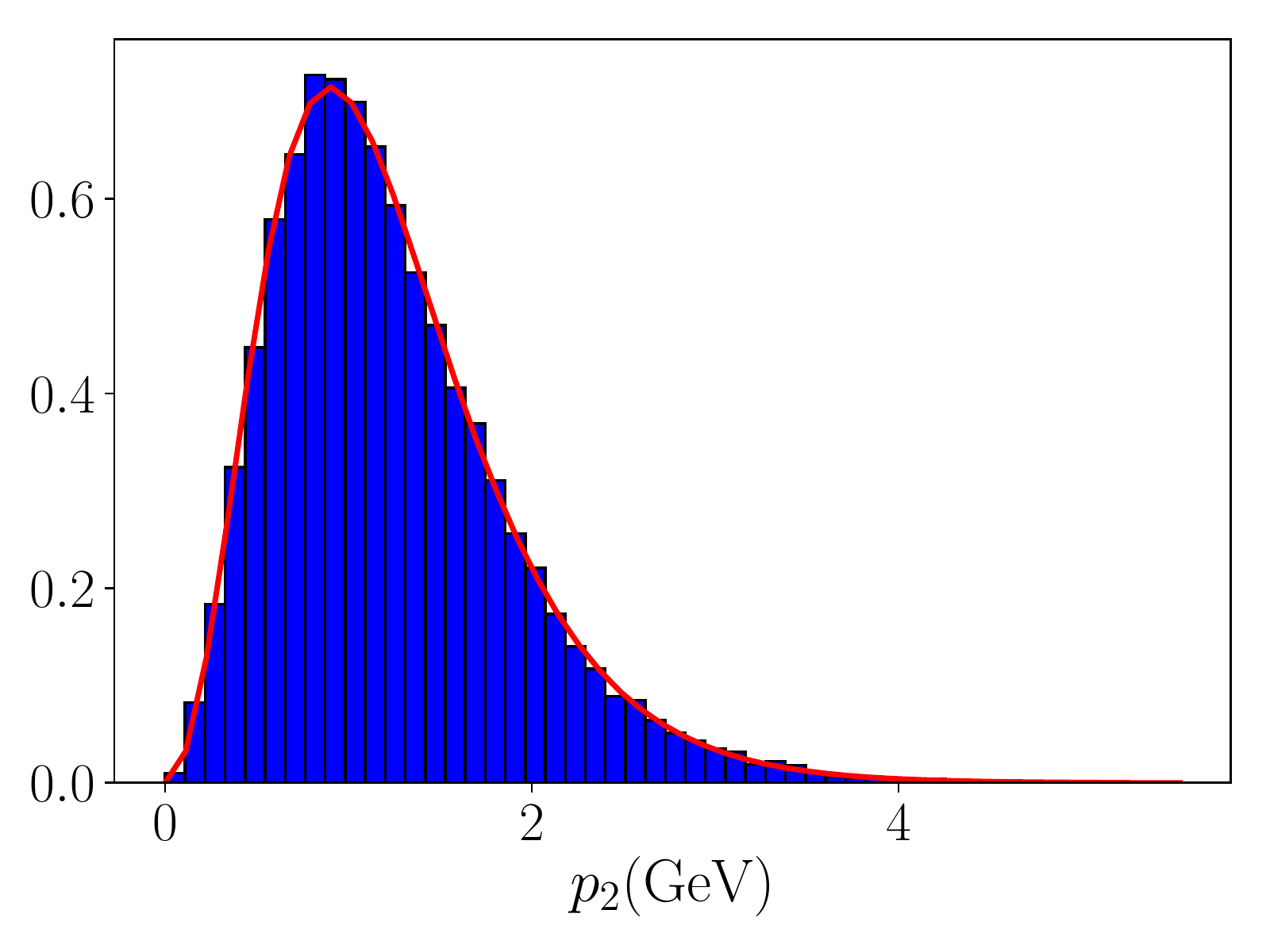}
        \caption{$p_2$}\label{}
    \end{subfigure}%
    ~
    \begin{subfigure}[b]{0.48\textwidth}
        \centering
        \includegraphics[height=2.1in]{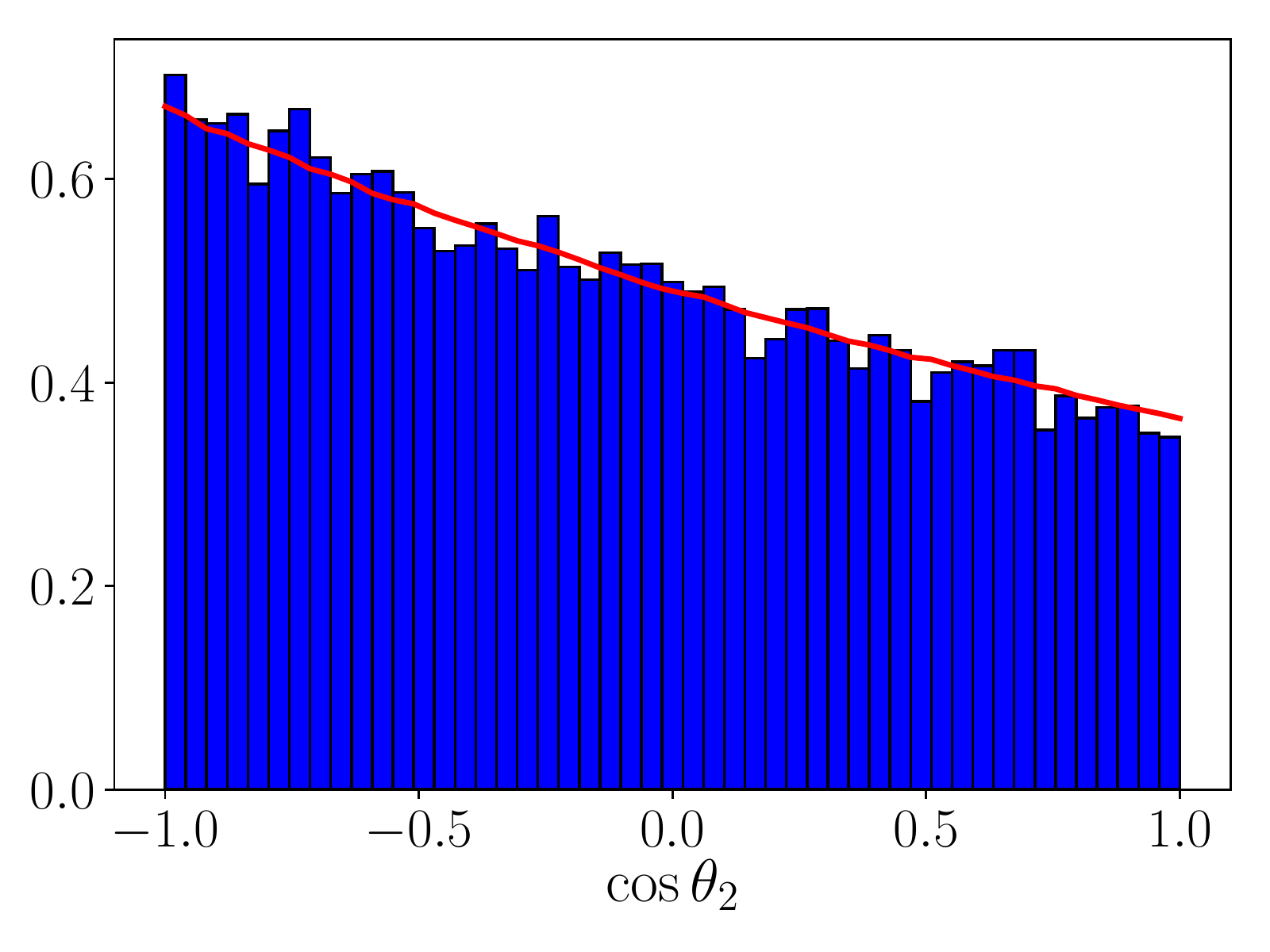}
        \caption{$c_2$}\label{}
    \end{subfigure}%

    \begin{subfigure}[b]{0.48\textwidth}
        \centering
        \includegraphics[height=2.1in]{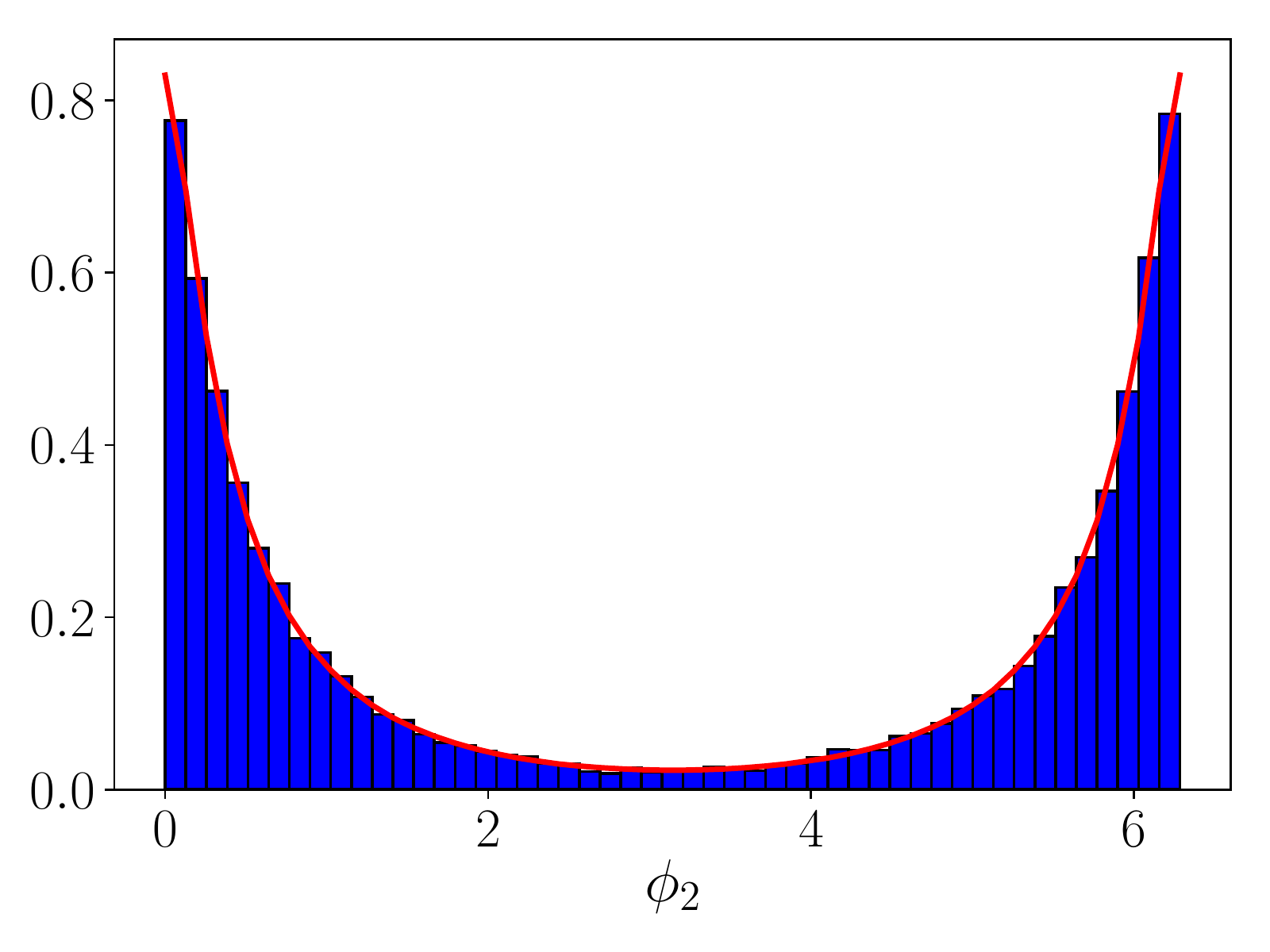}
        \caption{$\phi_2$}\label{}
    \end{subfigure}%
    ~
             
    \caption[Histograms of sampled momenta of the light quarks compared with the marginal distributions when $v_\ma{rel}=0.1$ and $T=0.35$ GeV.]{Histograms of sampled momenta of the incoming and outgoing light quarks (blue) compared with the marginal distributions (red) when $v_\ma{rel}=0.1$ and $T=0.35$ GeV. Normalization is unity.}
    \label{chap4_fig_disso_ineq_sample_compare1}
\end{figure}

\begin{figure}
    \centering
    \begin{subfigure}[b]{0.48\textwidth}
        \centering
        \includegraphics[height=2.1in]{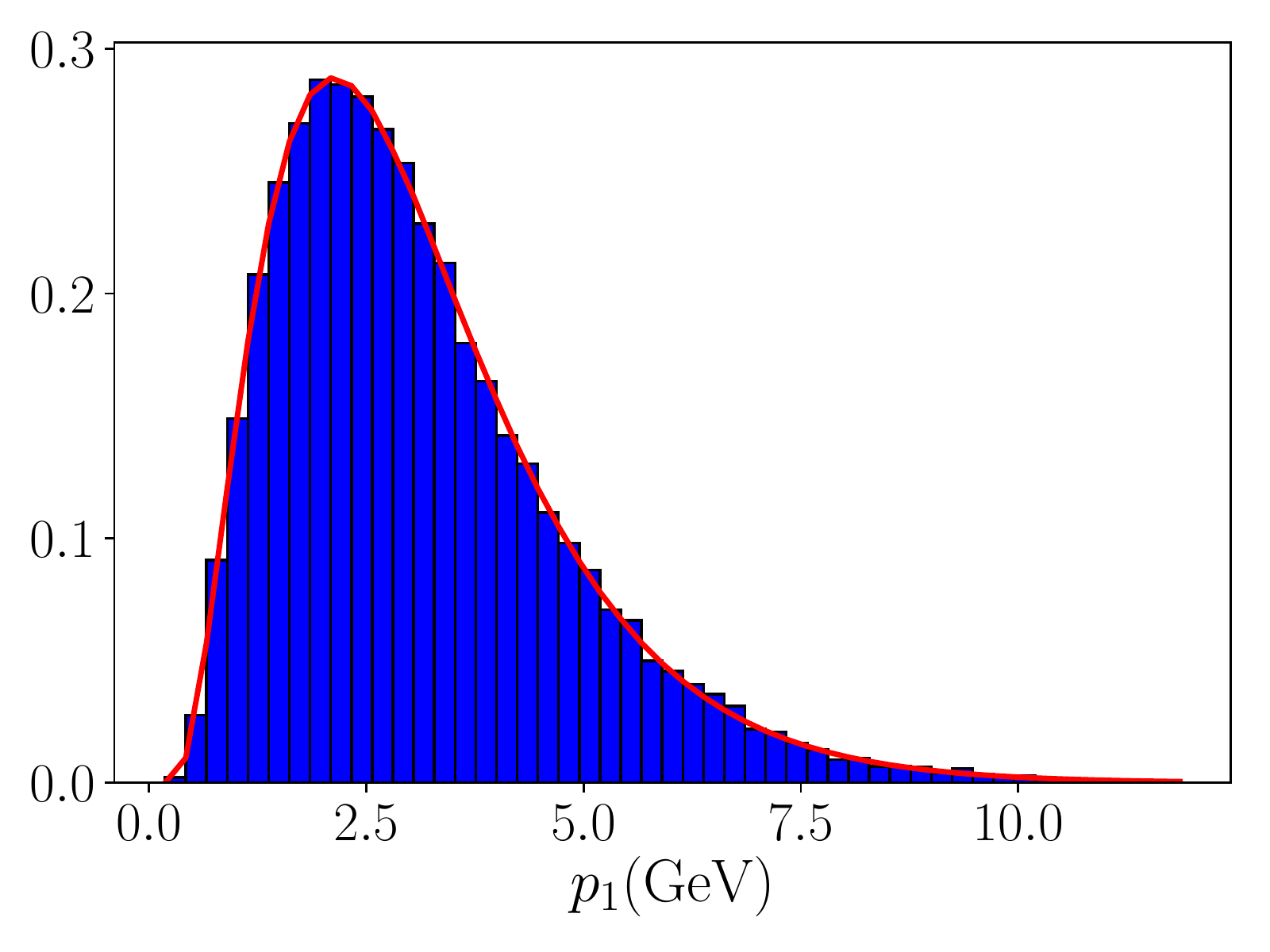}
        \caption{$p_1$}\label{}
    \end{subfigure}%
    ~
    \begin{subfigure}[b]{0.48\textwidth}
        \centering
        \includegraphics[height=2.1in]{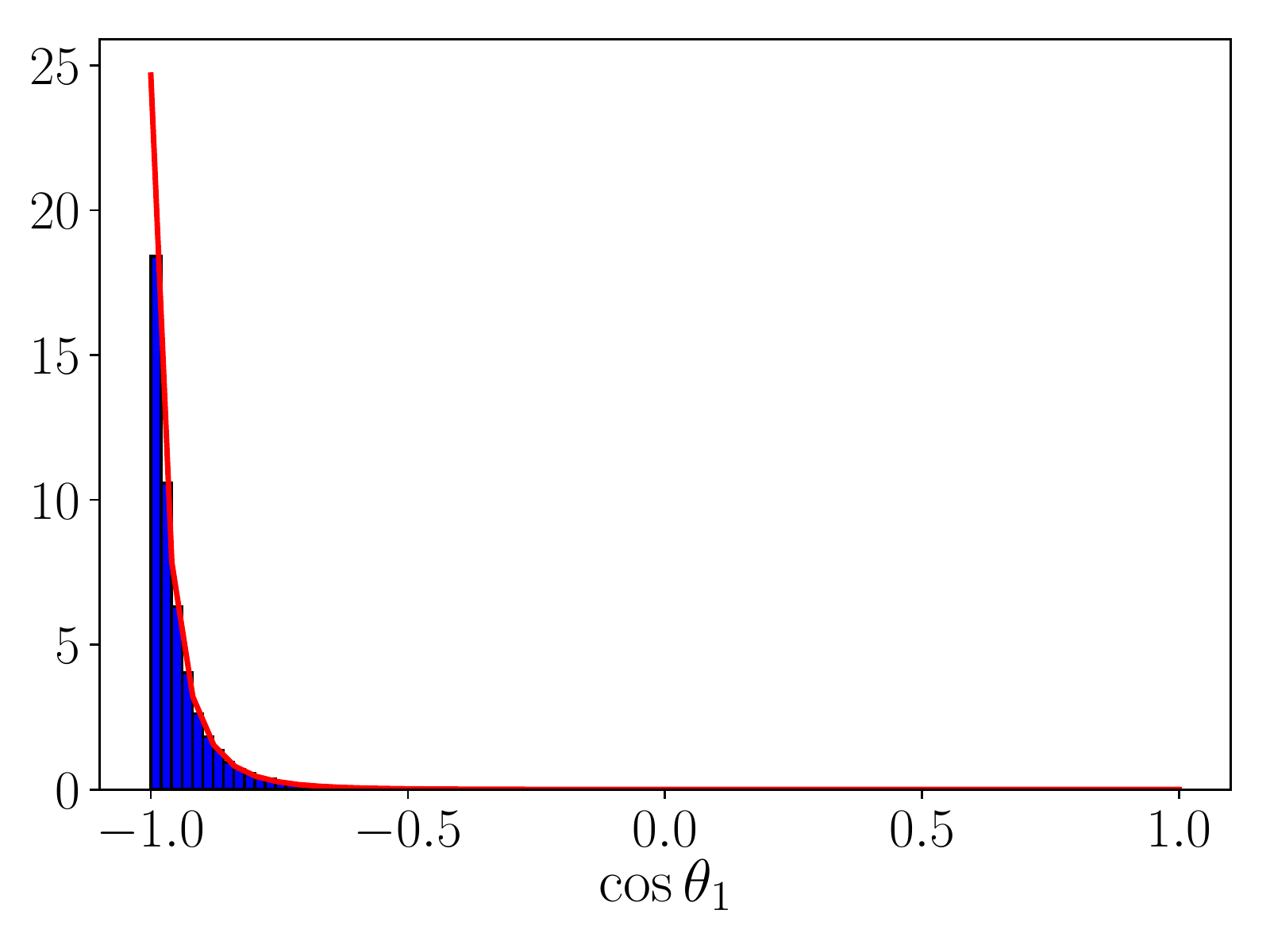}
        \caption{$c_1$}\label{}
    \end{subfigure}%
    
    \begin{subfigure}[b]{0.48\textwidth}
        \centering
        \includegraphics[height=2.1in]{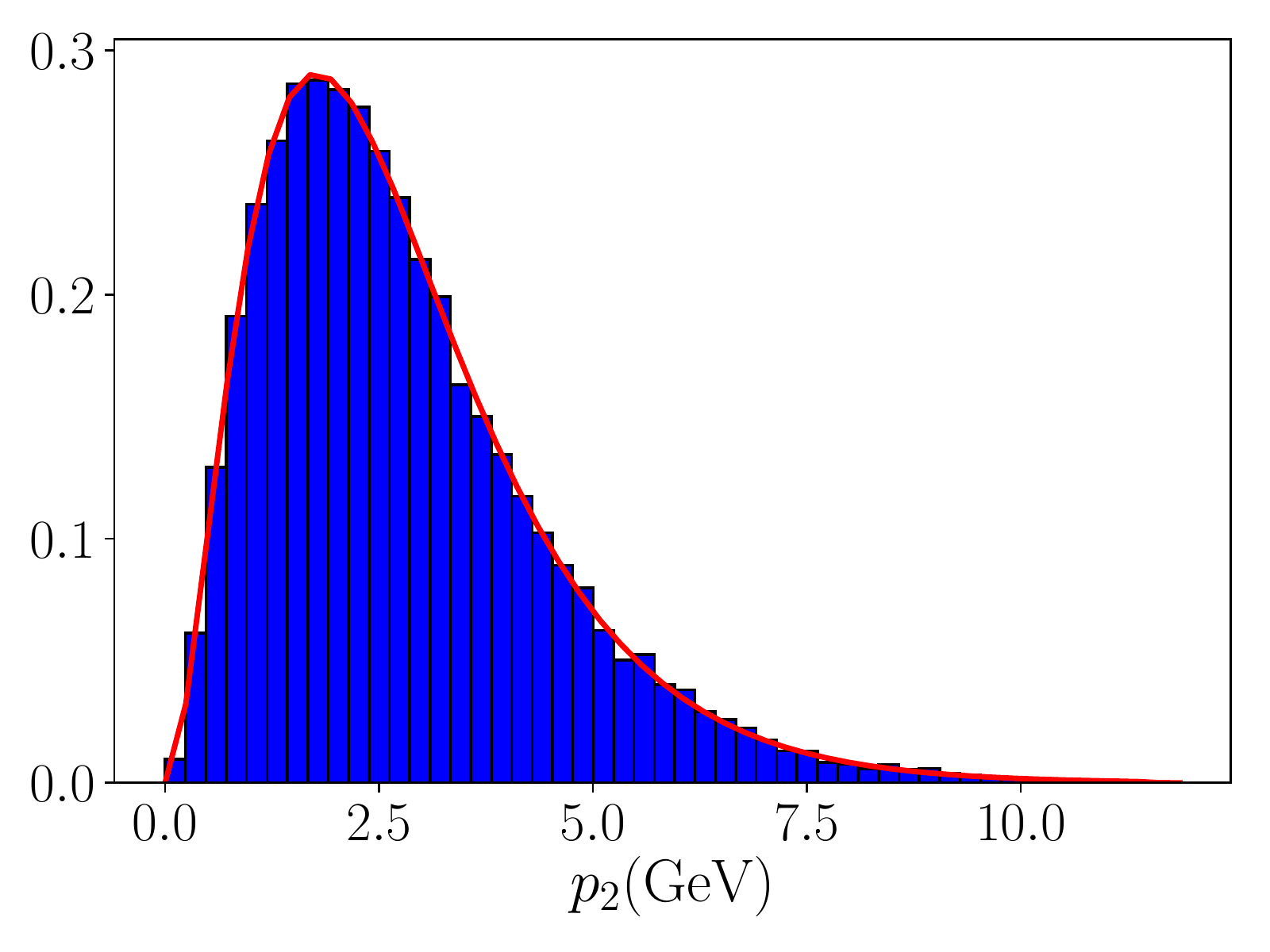}
        \caption{$p_2$}\label{}
    \end{subfigure}%
    ~
    \begin{subfigure}[b]{0.48\textwidth}
        \centering
        \includegraphics[height=2.1in]{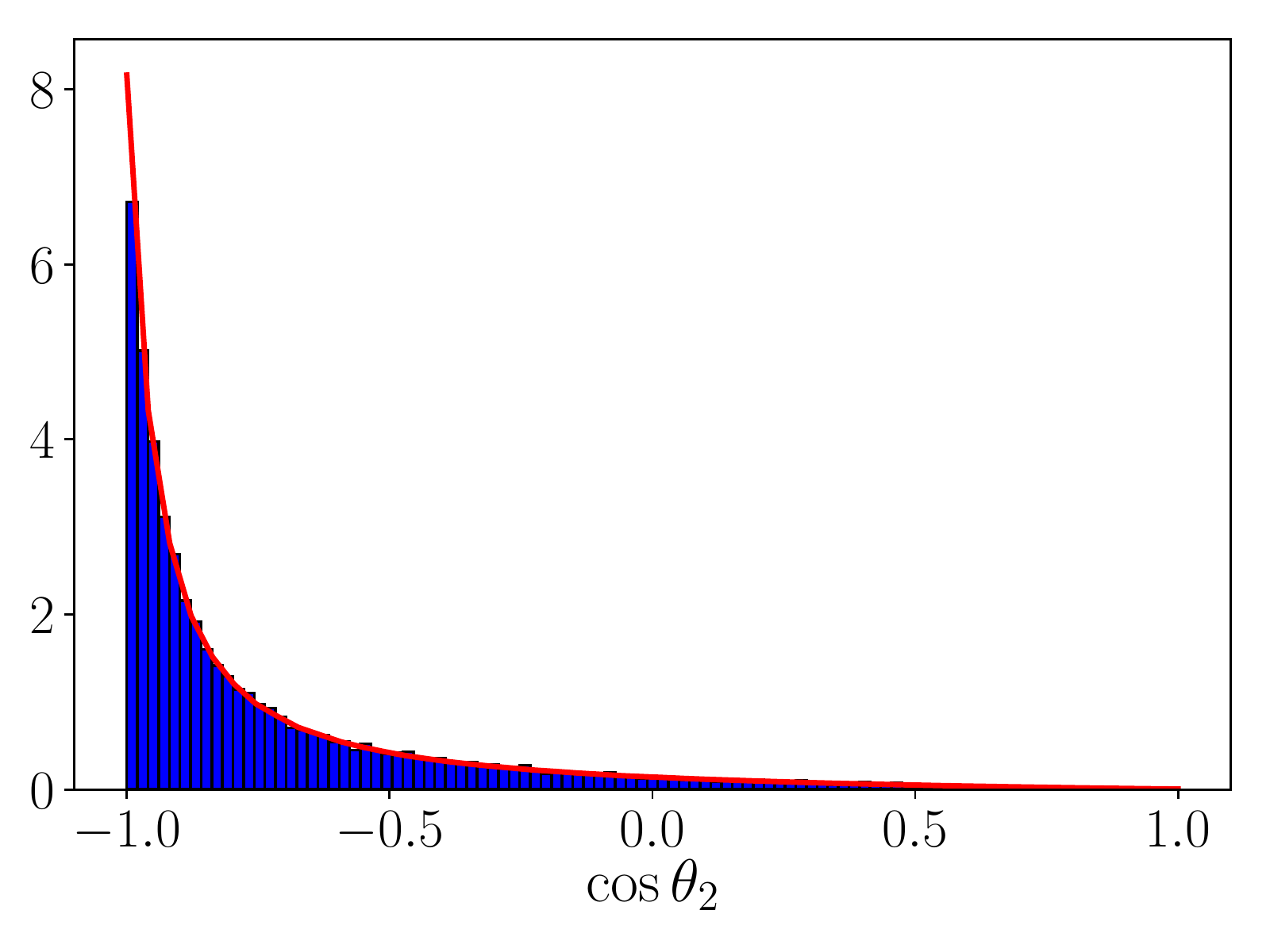}
        \caption{$c_2$}\label{}
    \end{subfigure}%

    \begin{subfigure}[b]{0.48\textwidth}
        \centering
        \includegraphics[height=2.1in]{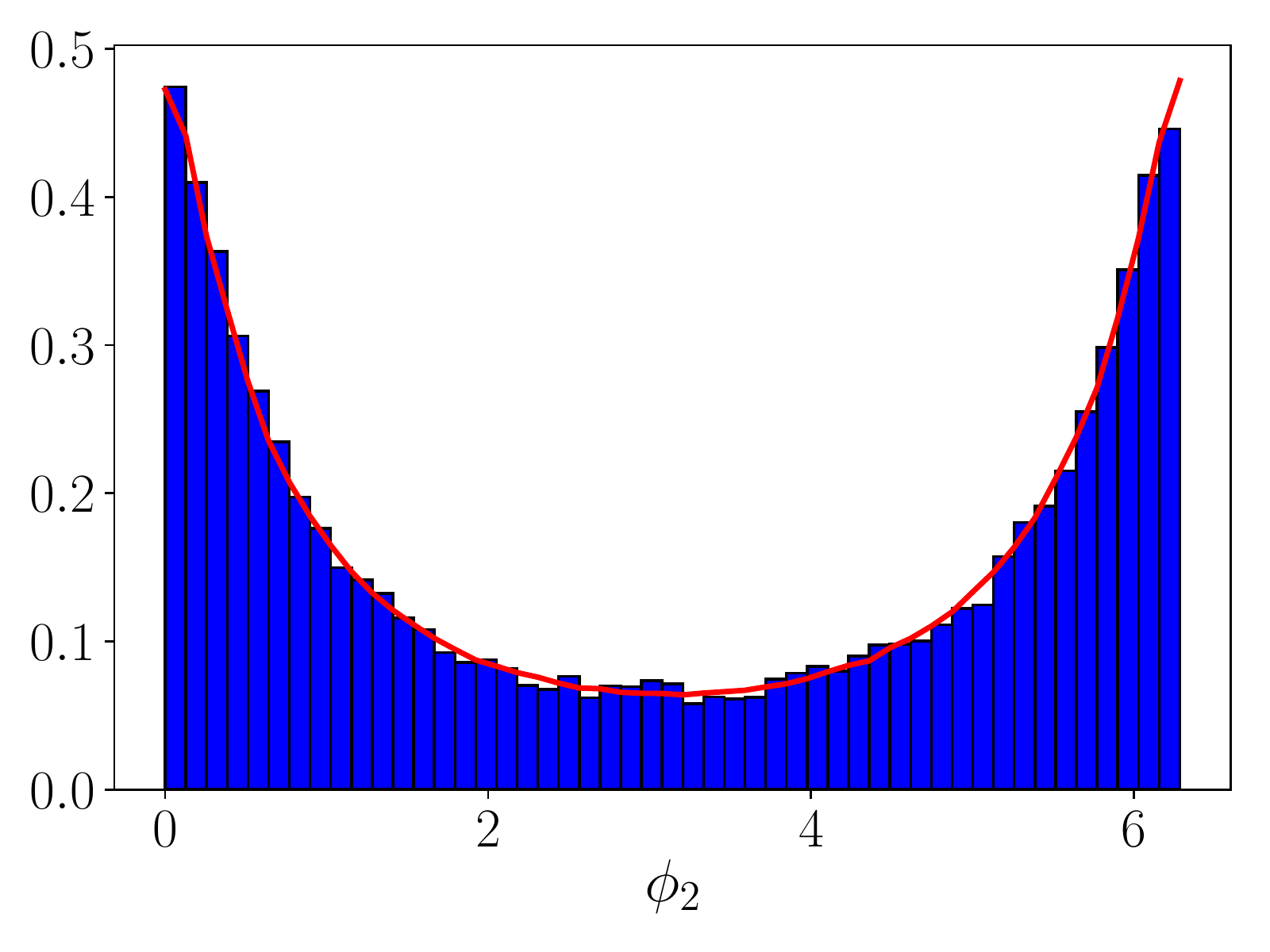}
        \caption{$\phi_2$}\label{}
    \end{subfigure}%
    ~
             
    \caption[Histograms of sampled momenta of the light quarks compared with the marginal distributions when $v_\ma{rel}=0.9$ and $T=0.25$ GeV.]{Histograms of sampled momenta of the incoming and outgoing light quarks (blue) compared with the marginal distributions (red) when $v_\ma{rel}=0.9$ and $T=0.25$ GeV. Normalization is unity.}
    \label{chap4_fig_disso_ineq_sample_compare2}
\end{figure}

To test this sampling procedure, we compare the sampling algorithm described above with the marginal distributions. We assume $\alpha_s=0.3$ and $M=4.65$ GeV. The marginal distribution in a variable $X$ is obtained by integrating the integrand in Eq.~(\ref{chap4_eqn_disso_ineq}) over all the variables except the $X$. Here $X$ can be $p_1$, $c_1$, $p_2$, $c_2$ and $\phi_2$. The comparisons are shown in Figs.~\ref{chap4_fig_disso_ineq_sample_compare1} and~\ref{chap4_fig_disso_ineq_sample_compare2} for two cases: $v_\ma{cell} = 0.1$, $T=0.35$ GeV and $v_\ma{cell} = 0.9$, $T=0.25$ GeV. We sampled $30000$ dissociation events for each case and plot the histograms in blue. Once we have sampled $p_1$ and $p_\ma{rel}$, we can compute $p_2$. It can be seen that the sampled distributions agree well with the marginal distributions drawn in red.

After we have determined the momenta of the incoming and outgoing light quarks, we can calculate the four-momentum of the transferred gluon (Here $p_1$ and $p_2$ indicate the four-momentum of the light quarks. Previously we use them to denote the magnitude of the three-momentum. The meaning should be clear from the context.)
\be
p_g = p_1 - p_2 \,,
\ee
where
\be
p_1 = \begin{pmatrix}  p_1 \\ p_1\sin\theta_1  \\ 0 \\ p_1\cos\theta_1 \end{pmatrix} \,\,\,\,\,\,\,\,
p_2 = \begin{pmatrix}  p_2 \\ p_2\sin\theta_2\cos\phi_2  \\ p_2\sin\theta_2\sin\phi_2 \\ p_2\cos\theta_2 \end{pmatrix} \,.
\ee
The energy transferred to the quarkonium in its rest frame is $p_1-p_2$. We can use this to calculate the relative kinetic energy of the final $Q\bar{Q}$ and sample their relative momentum as in Eq.~(\ref{chap4_eqn_prel}), by replacing $p_\ma{rel}$ with $\sqrt{M(p_1-p_2-|E_{nl}|)}$. Then the momenta of the unbound $Q\bar{Q}$ in the quarkonium rest frame can be written as, similar to Eq.~(\ref{chap4_eqn_pQpQbar}),
\be
{\bs p}_{Q(\bar{Q})} = \pm {\bs p}_\ma{rel}  + \frac{{\bs p}_1 - {\bs p}_2}{2} \,.
\ee
The next step would be to rotate them as in Eq.~(\ref{chap4_eqn_rotate}) and boost them back to the laboratory frame as in Eq.~(\ref{chap4_eqn_boostlab}).

\subsubsection{Momentum Sampling in Inelastic Scattering with Gluons}
We will use the importance sampling method to sample the momenta of the incoming and outgoing gluons according to the integrand of Eq.~(\ref{chap4_eqn_disso_ineg}). We first do a change of variable from $q_2$ to $p_\ma{rel}$,
\be
\frac{p^2_\ma{rel}}{M} = q_1 - q_2 - |E_{nl}| \,.
\ee
Then we can write Eq.~(\ref{chap4_eqn_disso_ineg}) as, up to a multiplication of some constant
\be
\int \diff q_1 \diff p_\ma{rel} \diff c_1 \diff c_2 \diff \phi_2 \,g_1(q_1,c_1) g_2(p_\ma{rel}) h(q_1, p_\ma{rel}, c_1, c_2, \phi_2)\,,
\ee
in which the functions $g_i$ are the factors we factorize out in the importance sampling and $h$ is the remaining part. They are defined as (again, we omit the subscript ``cell" for simplicity)
\be
g_1(q_1, c_1) &=& q_1  \frac{1}{e^{\gamma(1+vc_1)q_1/T} - 1}    \frac{\frac{ q_2 }{q_1} + \frac{q_1}{ q_2 } + 2}{\frac{ q_2 }{q_1} + \frac{q_1}{ q_2 } - 2} \\
g_2(p_\ma{rel}) &=& p_\ma{rel}^2 | \langle \Psi_{{\bs p}_\ma{rel}} | {\bs r} |  \psi_{nl}  \rangle|^2 \\ \nn
h(q_1, p_\ma{rel}, c_1, c_2, \phi_2) &=& \frac{q_2}{q_1}\big[ 1+n_B(\gamma(1+vc_2)q_2) \big] \frac{1+s_1s_2\cos\phi_2+c_1c_2}{2} \\
&& \frac{ \frac{q_1}{q_2} + \frac{q_2}{q_1} -2  }{\frac{q_1}{q_2} + \frac{q_2}{q_1} - 2(s_1s_2\cos\phi_2 + c_1c_2)} \,.
\ee
The sampling of $q_1$ and $c_1$ according to the distribution function $g_1(q_1, c_1)$ is very similar to the sampling procedure described for the inelastic scattering with light quarks. We will sample $q_1$ using the rejection method on the function
\be
\int_{-1}^1 \diff c_1 g_1(q_1, c_1) = \frac{T}{\gamma v}  \ln{\frac{1-e^{-\gamma(1+v)q_1/T}}{1-e^{-\gamma(1-v)q_1/T}}}   \frac{\frac{ q_2 }{q_1} + \frac{q_1}{q_2 } + 2}{\frac{ q_2 }{q_1} + \frac{q_1}{ q_2 } - 2} \,.
\ee
The range of the trial uniform distribution is taken as $[|E_{nl}|, 15T/\sqrt{1-v_\ma{cell}}]$, where $v_\ma{cell}$ is the velocity magnitude of the quarkonium in the rest frame of the hydro-cell.
We sample $c_1$ using the inverse function method once we have sampled a $q_1$. The function we need to invert is defined as
\be \nn
G(x) &\equiv&   \int_{-1}^x \diff c_1 \frac{1}{e^{\gamma(1+vc_1)q_1/T}-1} \\
&=& \frac{T}{\gamma v q_1} \Big[  \ln{ (1 - e^{-\gamma(1+vx)q_1/T})} -  \ln{ (1 - e^{-\gamma(1-v)q_1/T} )}    \Big]  \,.
\ee
The $x = \cos\theta$ can be solved from the equation $r = G(x)$ where $r$ is a random number generated from a uniform distribution between $0$ and $1$.

The magnitude of the relative momentum $p_\ma{rel}$ between the open $Q$ and $\bar{Q}$ is sampled using the rejection method on the function $g_2(p_\ma{rel})$. The lower and upper limits of $p_\ma{rel}$ used in the rejection method are taken as $0$ and $3.5/a_B$ for 1S state and $0$ and $1.5/a_B$ for 2S state. 

Then we will do rejection sampling on the remaining part $h(q_1, p_\ma{rel}, c_1, c_2, \phi_2)$. We will further generate a random number $c_2$ uniformly between $-1$ and $1$ and $\phi_2$ uniformly between $0$ and $2\pi$. Together with the sampled $q_1$, $p_\ma{rel}$ and $c_1$ as described above, we can calculation the function value $h(q_1, p_\ma{rel}, c_1, c_2, \phi_2)$. We can further generate a random number $r$ uniformly between $0$ and $1$ and decide to accept this sampling if $r  (\max{h}) \leq h(q_1, p_\ma{rel}, c_1, c_2, \phi_2)$. The maximum of $h$ is not easy to obtain. In practice, we can show that
\be
h \leq \frac{q_2}{q_1} \frac{1}{\gamma(1+vc_2)q_2/T } \leq \frac{T}{ |E_{nl}| } \frac{1}{\gamma(1-v)} \,.
\ee
So we simply set $\max{h}$ to be $\frac{T}{ |E_{nl}| } \frac{1}{\gamma(1-v)}$ in the acceptance criterion. We may lose some efficiency in this step of the sampling procedure but the overall efficiency in the sampling is good enough for our purpose.

\begin{figure}
    \centering
    \begin{subfigure}[b]{0.48\textwidth}
        \centering
        \includegraphics[height=2.1in]{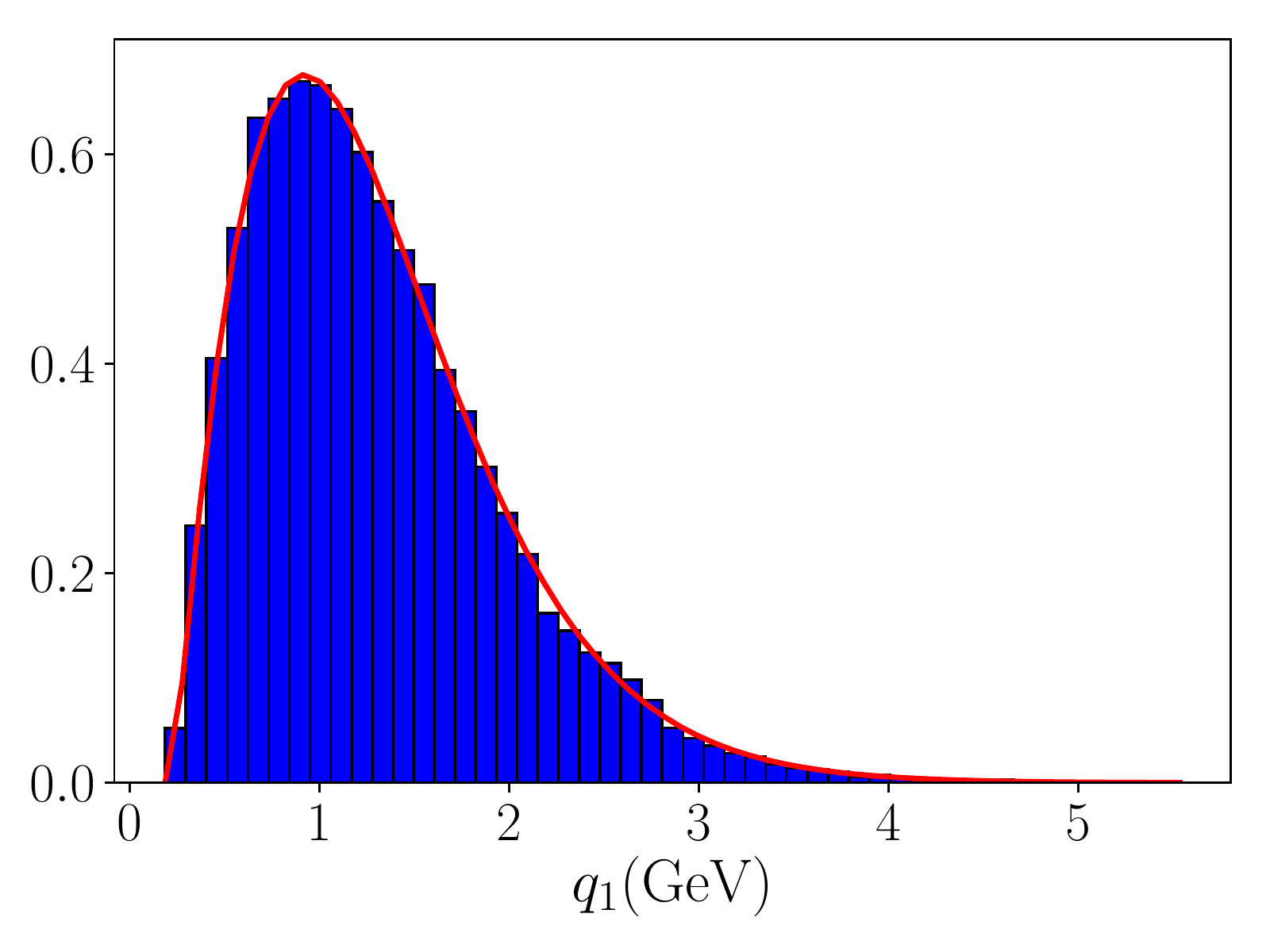}
        \caption{$q_1$}\label{}
    \end{subfigure}%
    ~
    \begin{subfigure}[b]{0.48\textwidth}
        \centering
        \includegraphics[height=2.1in]{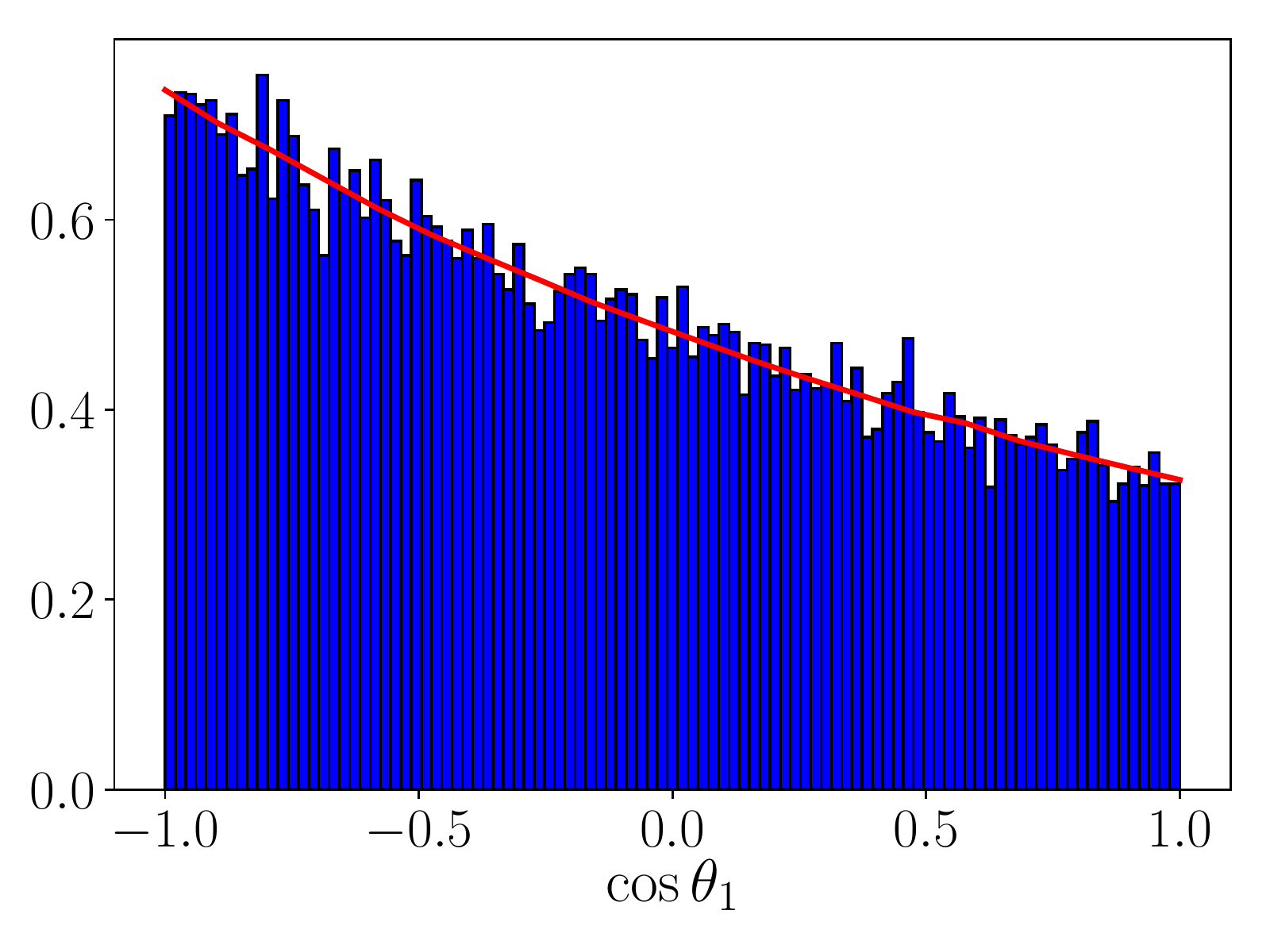}
        \caption{$c_1$}\label{}
    \end{subfigure}%
    
    \begin{subfigure}[b]{0.48\textwidth}
        \centering
        \includegraphics[height=2.1in]{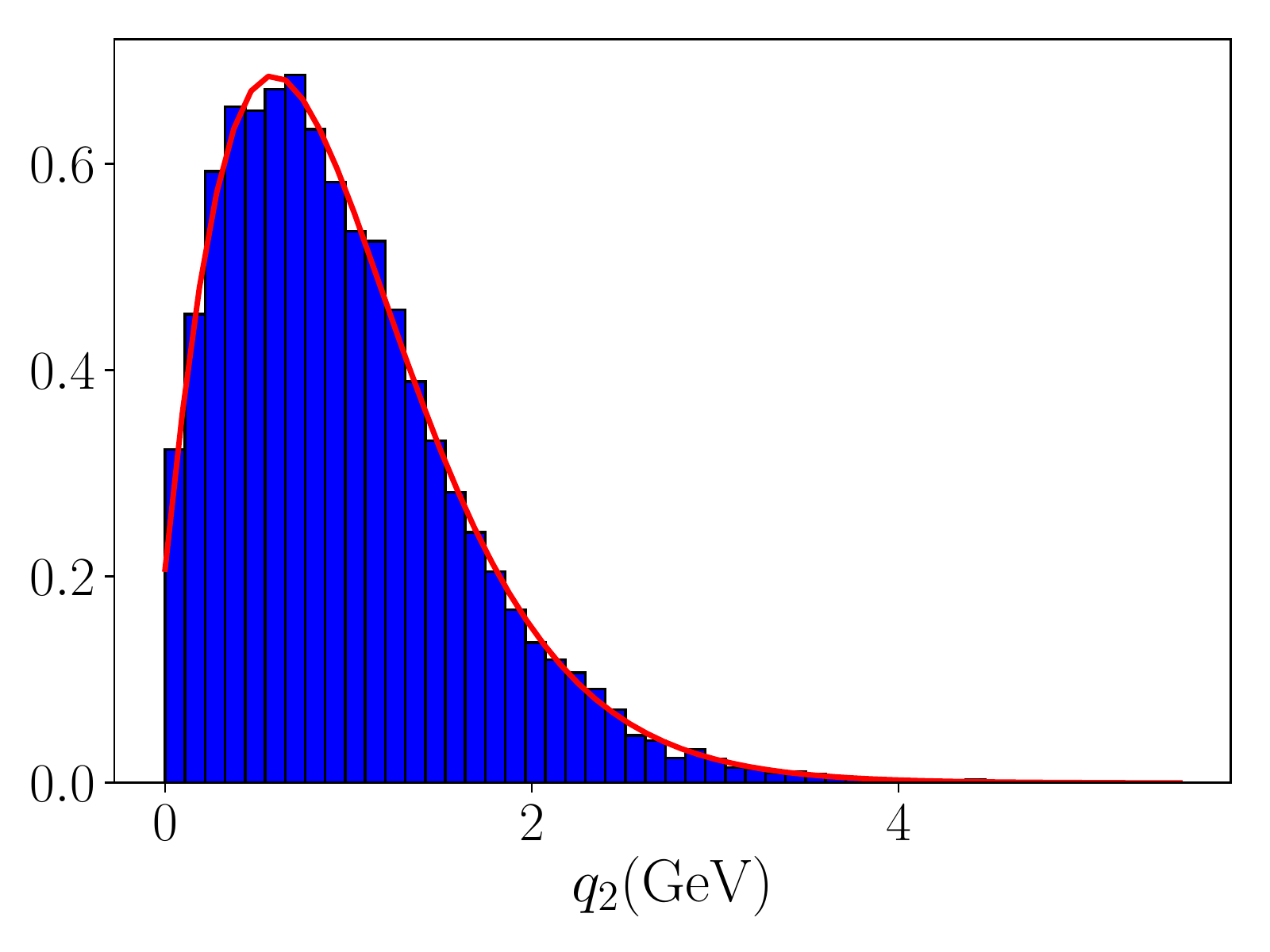}
        \caption{$q_2$}\label{}
    \end{subfigure}%
    ~
    \begin{subfigure}[b]{0.48\textwidth}
        \centering
        \includegraphics[height=2.1in]{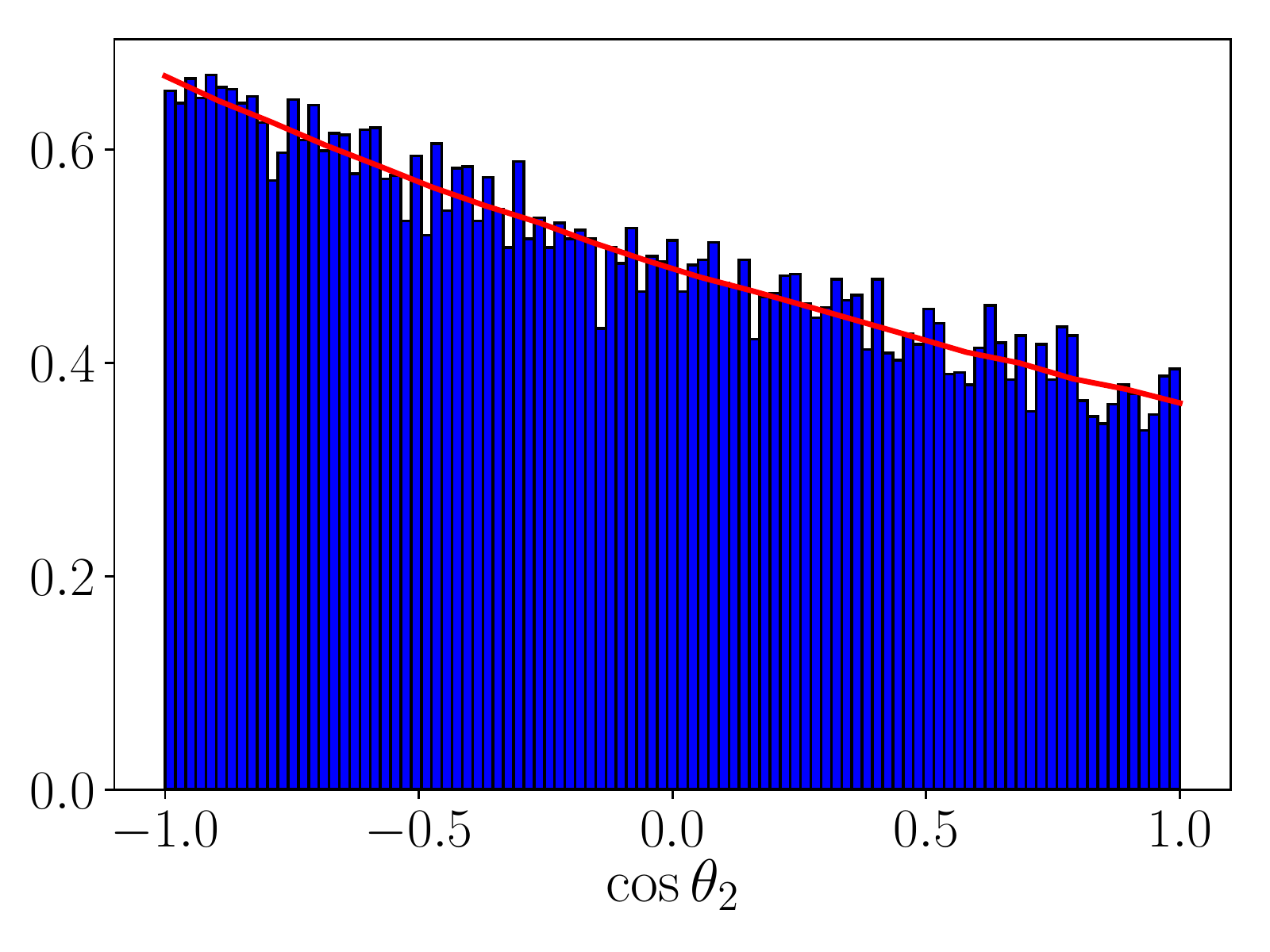}
        \caption{$c_2$}\label{}
    \end{subfigure}%

    \begin{subfigure}[b]{0.48\textwidth}
        \centering
        \includegraphics[height=2.1in]{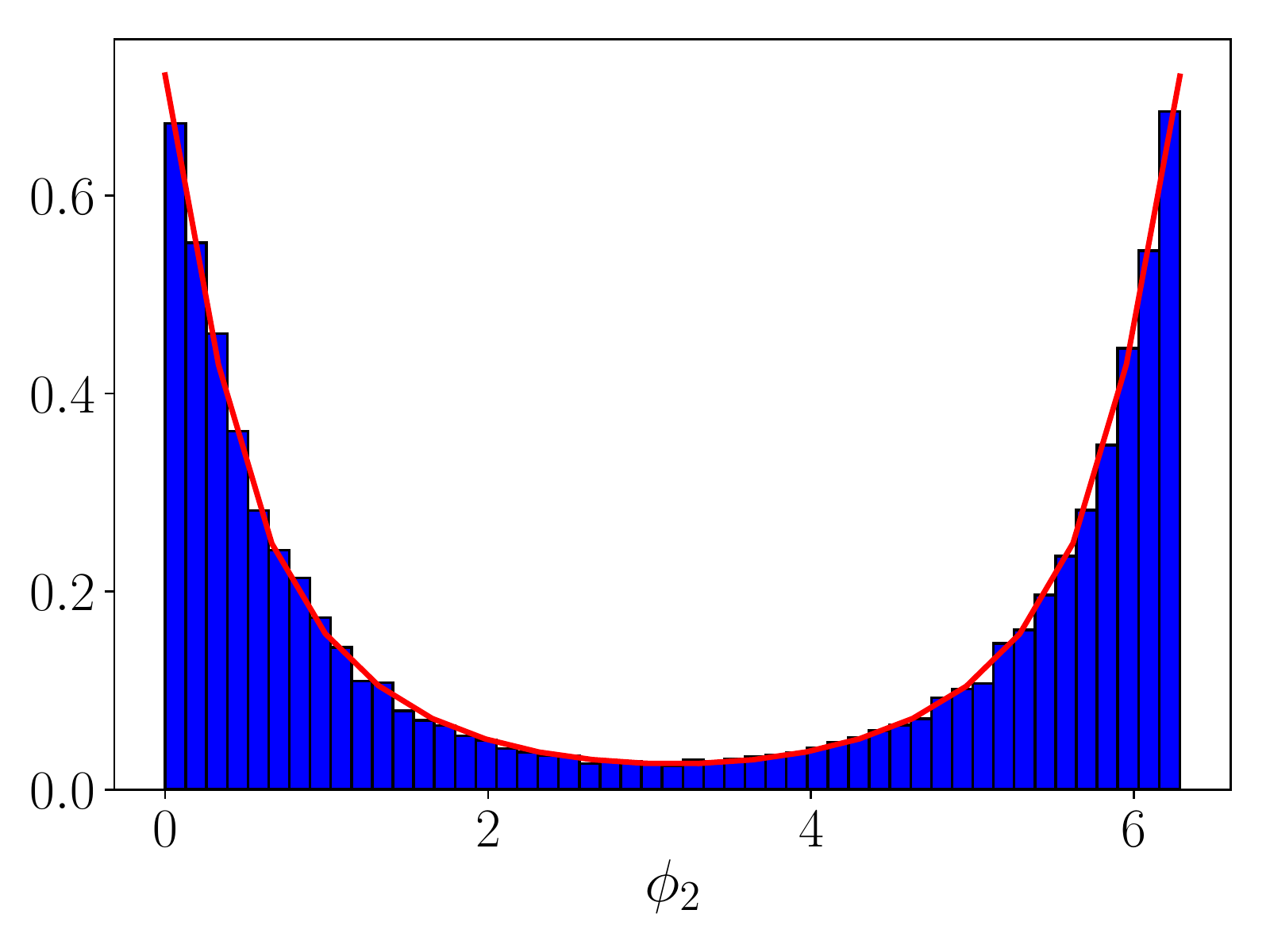}
        \caption{$\phi_2$}\label{}
    \end{subfigure}%
    ~
             
    \caption[Histograms of sampled momenta of the gluons compared with the marginal distributions when $v_\ma{rel}=0.1$ and $T=0.35$ GeV.]{Histograms of sampled momenta of the incoming and outgoing gluons (blue) compared with the marginal distributions (red) when $v_\ma{rel}=0.1$ and $T=0.35$ GeV. Normalization is unity.}
    \label{chap4_fig_disso_ineg_sample_compare1}
\end{figure}

\begin{figure}
    \centering
    \begin{subfigure}[b]{0.48\textwidth}
        \centering
        \includegraphics[height=2.1in]{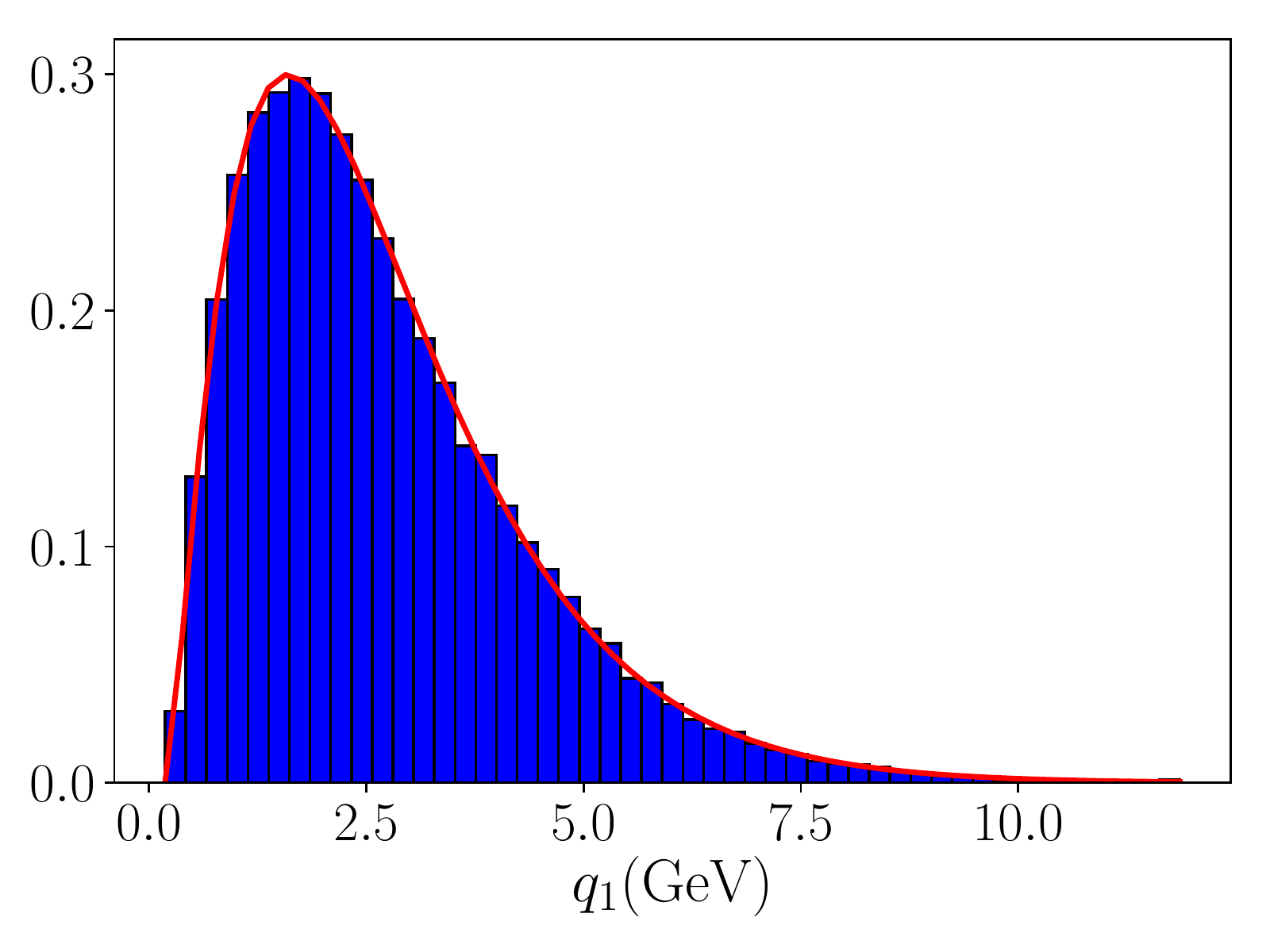}
        \caption{$q_1$}\label{}
    \end{subfigure}%
    ~
    \begin{subfigure}[b]{0.48\textwidth}
        \centering
        \includegraphics[height=2.1in]{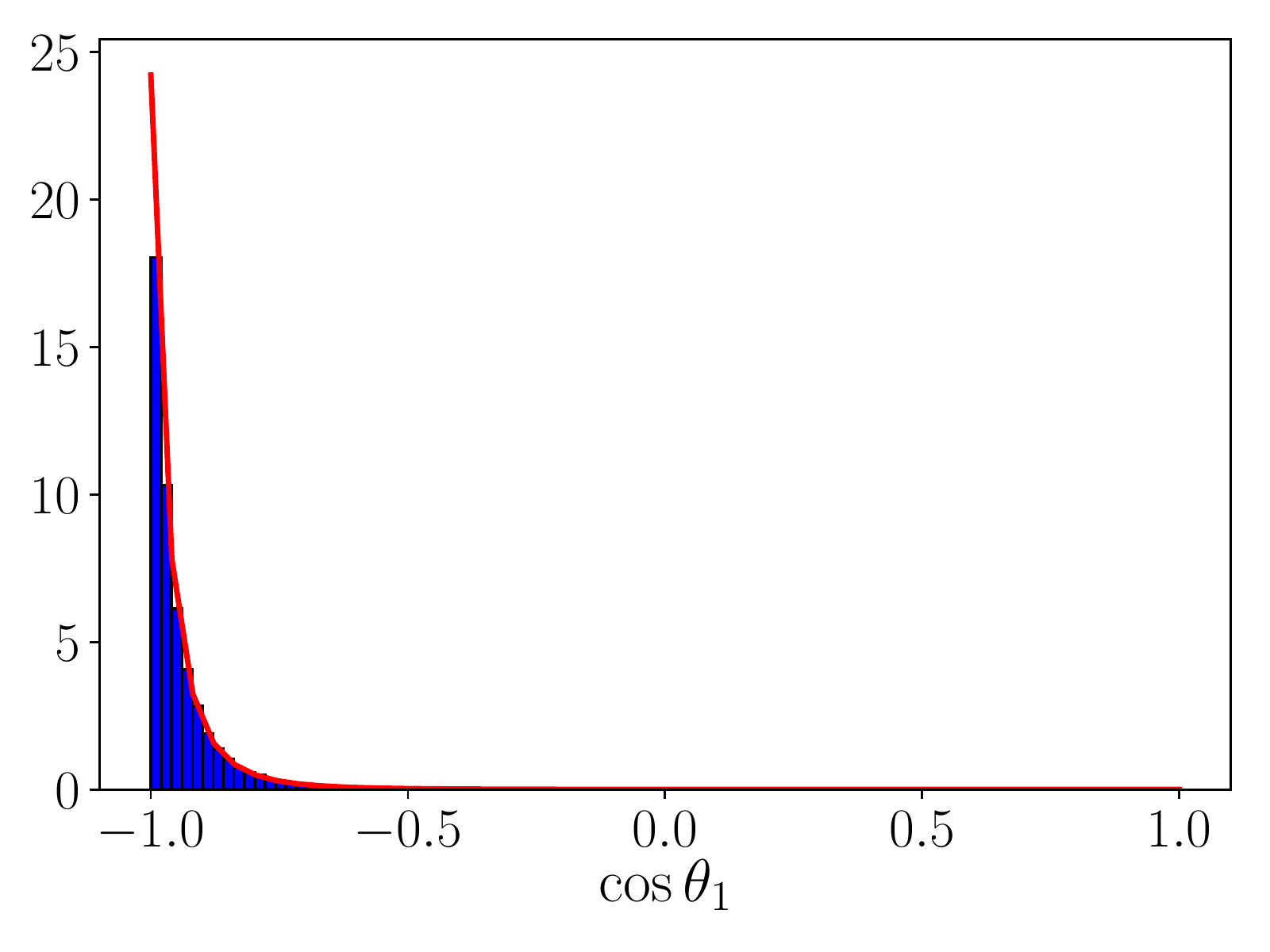}
        \caption{$c_1$}\label{}
    \end{subfigure}%
    
    \begin{subfigure}[b]{0.48\textwidth}
        \centering
        \includegraphics[height=2.1in]{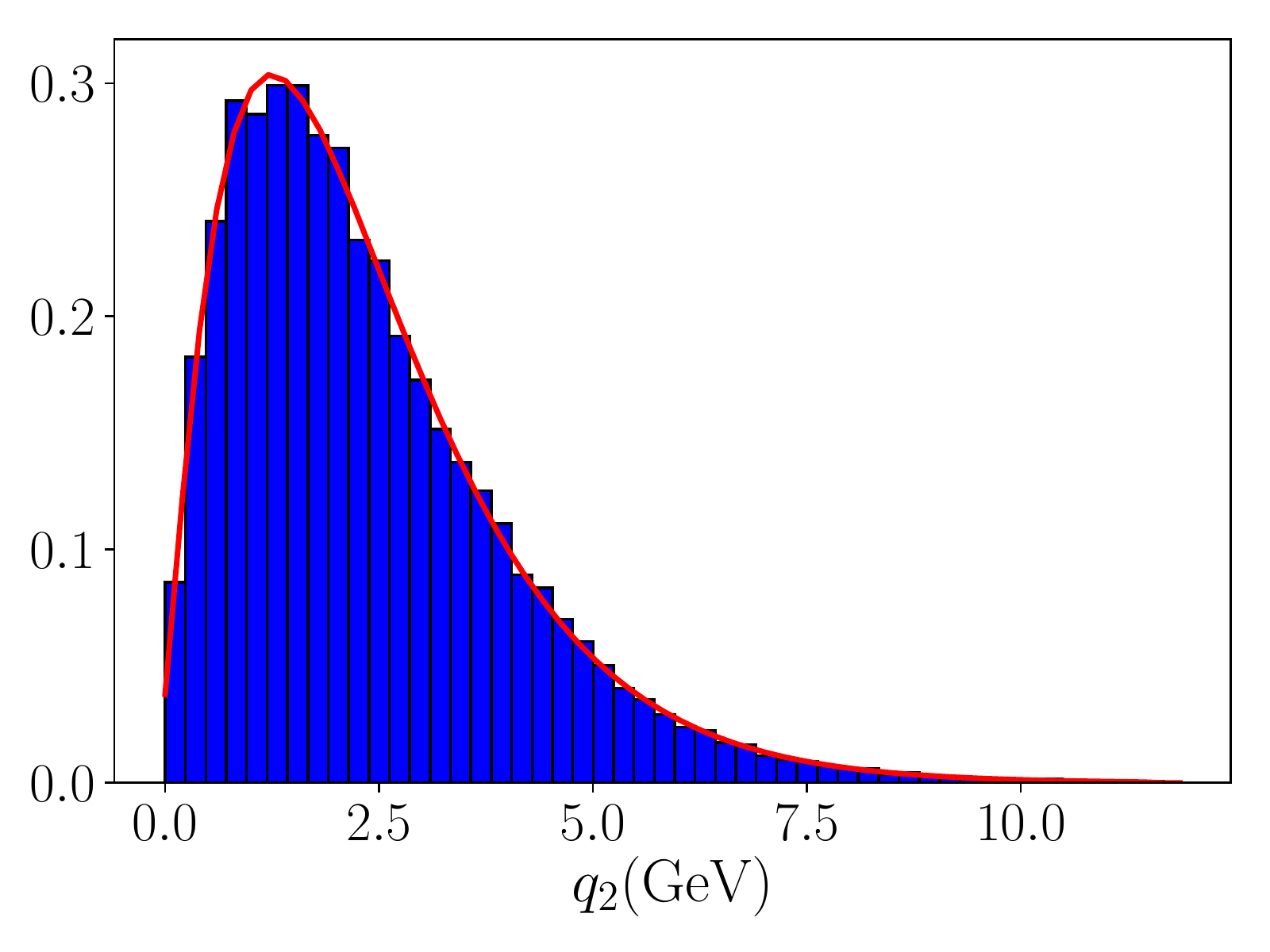}
        \caption{$q_2$}\label{}
    \end{subfigure}%
    ~
    \begin{subfigure}[b]{0.48\textwidth}
        \centering
        \includegraphics[height=2.1in]{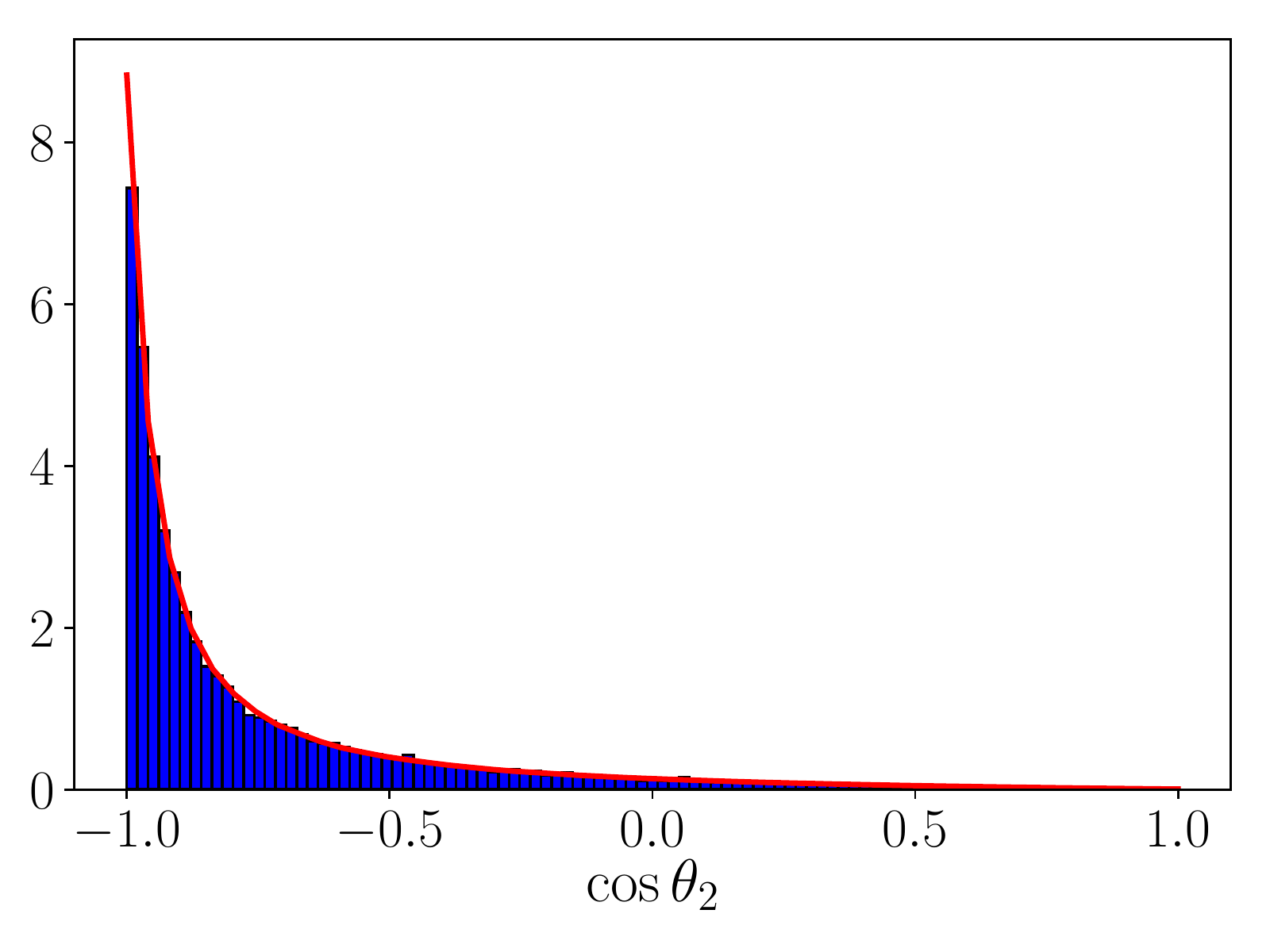}
        \caption{$c_2$}\label{}
    \end{subfigure}%

    \begin{subfigure}[b]{0.48\textwidth}
        \centering
        \includegraphics[height=2.1in]{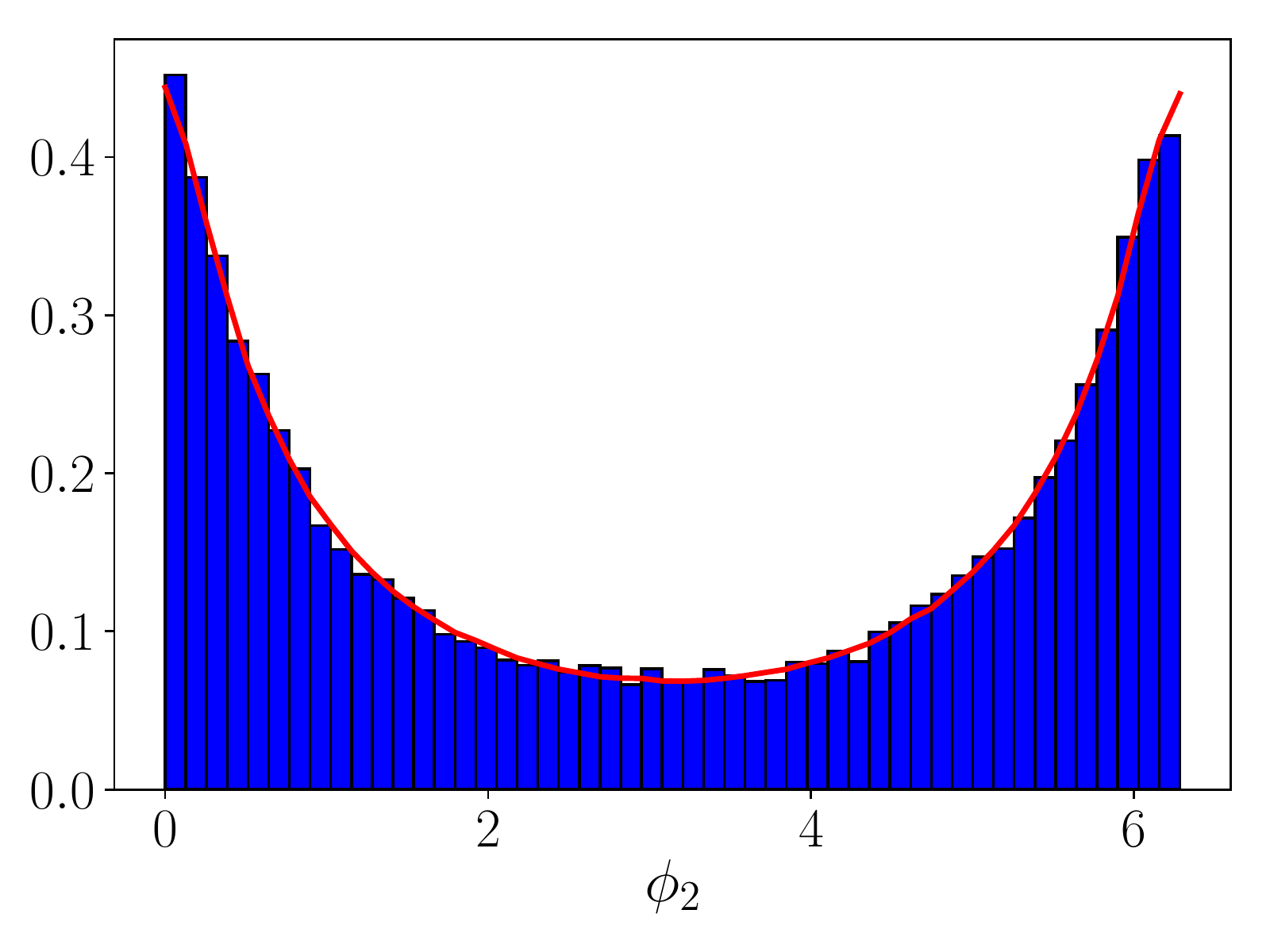}
        \caption{$\phi_2$}\label{}
    \end{subfigure}%
    ~
             
    \caption[Histograms of sampled momenta of the gluons compared with the marginal distributions when $v_\ma{rel}=0.9$ and $T=0.25$ GeV.]{Histograms of sampled momenta of the incoming and outgoing gluons (blue) compared with the marginal distributions (red) when $v_\ma{rel}=0.9$ and $T=0.25$ GeV. Normalization is unity.}
    \label{chap4_fig_disso_ineg_sample_compare2}
\end{figure}

We test the sampling of the momenta with the marginal distributions. We assume $\alpha_s=0.3$ and $M=4.65$ GeV. The marginal distribution in a variable $X$ is obtained by integrating the integrand in Eq.~(\ref{chap4_eqn_disso_ineg}) over all the variables except the $X$. Here $X$ can be $q_1$, $c_1$, $q_2$, $c_2$ and $\phi_2$. The comparisons are shown in Figs.~\ref{chap4_fig_disso_ineg_sample_compare1} and~\ref{chap4_fig_disso_ineg_sample_compare2} for two cases: $v_\ma{cell} = 0.1$, $T=0.35$ GeV and $v_\ma{cell} = 0.9$, $T=0.25$ GeV. We sampled $30000$ dissociation events for each case and plot the histograms in blue. Once we have sampled $q_1$ and $p_\ma{rel}$, we can compute $q_2$. It can be seen that the sampled distributions agree well with the marginal distributions drawn in red.

\vspace{0.2in}

\subsection{Recombination of Open Heavy Flavors}
The recombination rate of a heavy quark into the quarkonium state $nls$ surrounded by heavy antiquarks with the distribution $f_{\bar{Q}} ({\bs x}_{\bar{Q}}, {\bs p}_{\bar{Q}},t)$ is defined in Eq.~(\ref{chap3_eqn_recom_gluon}) for gluon radiation and Eq.~(\ref{chap3_eqn_recom_inel}) for inelastic scattering. To compute the recombination rate, we first need to evaluate the expressions of $\ml{F}^+$ defined in Eqs.~(\ref{chap3_eqn_F+_gluon}), (\ref{chap3_eqn_F+_ineq}) and (\ref{chap3_eqn_F+_ineg}). In our Monte Carlo sampling method, we need to do the following replacement,
\be 
&& f_Q({\bs x}_Q, {\bs p}_Q,t) f_{\bar{Q}} ({\bs x}_{\bar{Q}}, {\bs p}_{\bar{Q}},t) \\ \nn
&\longrightarrow& (2\pi)^6 \sum_{i,j} \delta^3( {\bs x}_Q - {\bs x}_i(t) ) \delta^3( {\bs x}_{\bar{Q}} - \widetilde{\bs x}_j(t) )  \delta^3( {\bs p}_Q - {\bs p}_i(t) ) \delta^3( {\bs p}_{\bar{Q}} - \widetilde{\bs p}_j(t) ) \,,
\ee
where ${\bs x}_i$ and ${\bs p}_i$ are the positions and momenta of open heavy quarks in the system while $\widetilde{\bs x}_j$ and $\widetilde{\bs p}_j$ are the positions and momenta of open heavy antiquarks. However, with this simple replacement, a $Q\bar{Q}$ far away from each other has the same probability to recombine as a $Q\bar{Q}$ close to each other does (assuming they have the same c.m.~and relative momenta). This fact disagrees with the original expression (\ref{chap3_eqn:reco_boltz}) derived in Chapter 3 and it is also counterintuitive. As will be discussed in detail later, we will modify the replacement by using a Gaussian function for the relative positions
\be  
\label{chap4_eqn_gauss_dist}
&& f_Q({\bs x}_Q, {\bs p}_Q,t) f_{\bar{Q}} ({\bs x}_{\bar{Q}}, {\bs p}_{\bar{Q}},t) \\  \nn
&\longrightarrow& (2\pi)^6 \sum_{i,j}  \delta^3 \Big( \frac{  {\bs x}_Q + {\bs x}_{\bar{Q}}  }{2}  -  \frac{{\bs x}_i + \widetilde{\bs x}_j  }{2}  \Big) \frac{e^{- (  {\bs x}_i - \widetilde{\bs x}_j   )^2/(2\sigma^2)}}{(2\pi\sigma^2)^{3/2}}  \delta^3( {\bs p}_Q - {\bs p}_i ) \delta^3( {\bs p}_{\bar{Q}} - \widetilde{\bs p}_j ) \,, \,\,\,\,\,\,\,\,
\ee
where the c.m.~position is still a $\delta$-function. We have omitted the time dependence of ${\bs x}_i $, $\widetilde{\bs x}_j $, ${\bs p}_i$ and $\widetilde{\bs p}_j$. The width of the Gaussian function can be chosen as the Bohr radius of the quarkonium bound state $\sigma = a_B$. We now are ready to write down the recombination rate of a heavy quark into the quarkonium state $nls$ surrounded by heavy antiquarks with the distribution $f_{\bar{Q}} ({\bs x}_{\bar{Q}}, {\bs p}_{\bar{Q}},t)$.

\subsubsection{Real Gluon Radiation}
An unbound $Q\bar{Q}$ pair in the color octet can form a bound state by radiating out a gluon. The induced recombination rate of a heavy quark with position ${\bs x}_i$ and momentum ${\bs p}_i$ into the quarkonium state $nls$ surrounded by heavy antiquarks with the distribution $f_{\bar{Q}} ({\bs x}_{\bar{Q}}, {\bs p}_{\bar{Q}},t)$, is given by
\be \nn
\Gamma^g &=& g_+ \int\frac{ \diff^3 p_{\bar{Q}} }{(2\pi)^3} f_{\bar{Q}}({\bs x}_{\bar{Q}}, {\bs p}_{\bar{Q}}, t)  \int\frac{\diff^3 q}{2q(2\pi)^3} (1+n_B(q))  \\
&& (2\pi) \delta(-|E_{nl}| + q - \frac{{\bs p}^2_\ma{rel}}{M}) \frac{2}{3}C_F g^2q^2  | \langle \Psi_{{\bs p}_\ma{rel}} | {\bs r} | \psi_{nl} \rangle |^2 \,.
\ee
Our recombination formula is derived in the rest frame of each $Q\bar{Q}$ pair and we also assume the $Q\bar{Q}$ pair is at rest with respect to the medium.
In practical calculations, each $Q\bar{Q}$ pair moves at a different c.m.~velocity relative to the medium. So we need to deal with the integral $\int \diff^3 p_{\bar{Q}}  f_{\bar{Q}}({\bs x}_{\bar{Q}}, {\bs p}_{\bar{Q}}, t)$ carefully. We first look at the remaining term
\be
\label{chap4_eqn_reco_real}
 \int\frac{\diff^3 q}{2q(2\pi)^3} (1+n_B(q)) 
 (2\pi) \delta(-|E_{nl}| + q - \frac{{\bs p}^2_\ma{rel}}{M}) \frac{2}{3}C_F g^2q^2  | \langle \Psi_{{\bs p}_\ma{rel}} | {\bs r} | \psi_{nl} \rangle |^2 \,.
\ee
For a pair of $Q\bar{Q}$ with positions ${\bs x}_i$, $\widetilde{\bs x}_j$ and four-momenta $p_i = (E_i, {\bs p}_i)$,\, $\widetilde{p}_j = ( \widetilde{E}_j,\, \widetilde{\bs p}_j)$ in the laboratory frame, their c.m.~momentum and total energy is given by
\be
{\bs p}_\ma{cm}^\ma{lab} &=& {\bs p}_i + \widetilde{\bs p}_j\\
\label{chap4_eqn_cmE}
E^\ma{lab}_\ma{cm}  &=& \sqrt{(2M)^2 + ({\bs p}_\ma{cm}^\ma{lab})^2} \,.
\ee
The four-momenta in the hydro-cell frame (which is moving at ${\bs v}$ in the laboratory frame) are given by
\be
p_i^\ma{cell} &=& \Lambda({\bs v})  p_i \\
\widetilde{p}_j^\ma{cell} &=& \Lambda({\bs v}) \widetilde{ p}_j \,.
\ee
Then the c.m.~momentum and velocity in the hydro-cell frame are given by
\be
{\bs p}_\ma{cm}^\ma{cell} &=& {\bs p}_i^\ma{cell} + \widetilde{\bs p}_j^\ma{cell} \\
{\bs v}_\ma{cm}^\ma{cell} &=& \frac{{\bs p}_\ma{cm}^\ma{cell}}{ \sqrt{(2M)^2+ ({\bs p}_\ma{cm}^\ma{cell})^2}} \,.
\ee
Then we boost the four-momenta to the rest frame of the $Q\bar{Q}$ pair,
\be
p_i^\ma{rest} &=&  \Lambda({\bs v}_\ma{cm}^\ma{cell})  {p}_i^\ma{cell} \\
\widetilde{p}_j^\ma{rest} &=&  \Lambda({\bs v}_\ma{cm}^\ma{cell})  \widetilde{p}_j^\ma{cell} \,.
\ee
In this frame we can define the relative momentum that will be used in the calculation of expression (\ref{chap4_eqn_reco_real})
\be
\label{chap4_eqn_p_relative}
{\bs p}_\ma{rel}  &=&  \frac{1}{2} ({\bs p}_i^\ma{rest}  -  \widetilde{\bs p}_j^\ma{rest}  ) \,.
\ee
We want to mention that the frame in the calculation of recombination (in the rest frame of the unbound $Q\bar{Q}$ pair) is different from that in the calculation of dissociation (in the rest frame of the quarkonium). But the difference in this frame choice is suppressed by powers of $T/M$ and neglected throughout the dissertation work. 

Then we can compute the expression (\ref{chap4_eqn_reco_real}) in the $Q\bar{Q}$ rest frame (we have discussed in the dissociation part how to boost the gluon distribution)
\be \nn
&& \int\frac{\diff^3 q}{2q(2\pi)^3} \Big[1+n_B\big(\Lambda^0_{\ \mu}(-{\bs v}_\ma{cm}^\ma{cell} )q^\mu \big) \Big] (2\pi) \delta(-|E_{nl}| + q - \frac{ ({\bs p}_\ma{rel})^2}{M})  \\
&& \frac{2}{3}C_F g^2q^2  | \langle \Psi_{  {\bs p}_\ma{rel}  } | {\bs r} | \psi_{nl} \rangle |^2 \,,
\ee
where the relative momentum ${\bs p}_\ma{rel}$ is defined in (\ref{chap4_eqn_p_relative}). 
We will first assume ${\bs v}_\ma{cm}^\ma{cell}$ as the $z$-axis, so we can simplify the above integral ($\gamma_\ma{cm}^\ma{cell} = 1/\sqrt{1-(v_\ma{cm}^\ma{cell})^2}$)
\be
\label{chap4_recom_gluon_sample}
&&\int \frac{q\diff \cos\theta }{ 4\pi }  \frac{1}{ 1 - e^{-\gamma_\ma{cm}^\ma{cell}(1 + v_\ma{cm}^\ma{cell} \cos\theta)q/T }  } \frac{2}{3}C_F g^2q^2  | \langle \Psi_{  {\bs p}_\ma{rel}  } | {\bs r} | \psi_{nl} \rangle |^2   \Big|_{q = |E_{nl}| + \frac{ ({\bs p}_\ma{rel})^2}{M}} \\
\label{chap4_eqn_recom_gluon_GV}
&=& \frac{8}{9}\alpha_s q^3 \Big( 2 + \frac{T}{\gamma^\ma{cell}_\ma{cm} v^\ma{cell}_\ma{cm} q}   \ln{\frac{1-e^{-\gamma^\ma{cell}_\ma{cm}(1+v^\ma{cell}_\ma{cm})q/T}}{ 1-e^{-\gamma^\ma{cell}_\ma{cm}(1-v^\ma{cell}_\ma{cm})q/T} }}  \Big) | \langle \Psi_{  {\bs p}_\ma{rel}  } | {\bs r} | \psi_{nl} \rangle |^2   \Big|_{q = |E_{nl}| + \frac{ ({\bs p}_\ma{rel})^2}{M}} \,, \ \ \ \ \ \ \ \ \ 
\ee
where the angular integral has been done similarly as in the last section. The recombination rate of this $Q\bar{Q}$ pair in their rest frame is given by
\be
\label{chap4_eqn_reco_real_rest_1pair}
\Gamma^g_\ma{rest} &=& g_+ \frac{ \diff^3 p^\ma{rest}_{\bar{Q}} }{(2\pi)^3} f^\ma{rest}_{\bar{Q}}({\bs x}^\ma{rest}_{\bar{Q}}, {\bs p}^\ma{rest}_{\bar{Q}}, t) \\ \nn
&& \frac{8}{9}\alpha_s q^3 \Big( 2 + \frac{T}{\gamma^\ma{cell}_\ma{cm} v^\ma{cell}_\ma{cm} q}   \ln{\frac{1-e^{-\gamma^\ma{cell}_\ma{cm}(1+v^\ma{cell}_\ma{cm})q/T}}{ 1-e^{-\gamma^\ma{cell}_\ma{cm}(1-v^\ma{cell}_\ma{cm})q/T} }}  \Big) | \langle \Psi_{  {\bs p}_\ma{rel}  } | {\bs r} | \psi_{nl} \rangle |^2   \Big|_{q = |E_{nl}| + \frac{ ({\bs p}_\ma{rel})^2}{M}} \,,
\ee
where we have not integrated over ${\bs p}^\ma{rest}_{\bar{Q}}$. To get the recombination rate in the laboratory frame, we multiply (\ref{chap4_eqn_reco_real_rest_1pair}) by $1/\gamma = \frac{2M}{E_\ma{cm}^\ma{lab}}$ where $E_\ma{cm}^\ma{lab}$ is given in Eq.~(\ref{chap4_eqn_cmE}). To obtain the total recombination rate of that specific $Q$ in the laboratory frame, we integrate over the distribution of $\bar{Q}$ to get
\be 
\Gamma^g_\ma{lab} &=& g_+ \int \frac{ \diff^3 p^\ma{rest}_{\bar{Q}} }{(2\pi)^3} f^\ma{rest}_{\bar{Q}}({\bs x}^\ma{rest}_{\bar{Q}}, {\bs p}^\ma{rest}_{\bar{Q}}, t)  \frac{2M}{E_\ma{cm}^\ma{lab}}  \\ \nn
&& \frac{8}{9}\alpha_s q^3 \Big( 2 + \frac{T}{\gamma^\ma{cell}_\ma{cm} v^\ma{cell}_\ma{cm} q}   \ln{\frac{1-e^{-\gamma^\ma{cell}_\ma{cm}(1+v^\ma{cell}_\ma{cm})q/T}}{ 1-e^{-\gamma^\ma{cell}_\ma{cm}(1-v^\ma{cell}_\ma{cm})q/T} }}  \Big) | \langle \Psi_{  {\bs p}_\ma{rel}  } | {\bs r} | \psi_{nl} \rangle |^2   \Big|_{q = |E_{nl}| + \frac{ ({\bs p}_\ma{rel})^2}{M}} \,.
\ee
The physical meaning of $\frac{ \diff^3 p_{\bar{Q}} }{(2\pi)^3} f_{\bar{Q}}({\bs x}_{\bar{Q}}, {\bs p}_{\bar{Q}}, t)$ is the number density of $\bar{Q}$ with a given ${\bs p}_{\bar{Q}}$, which is given by $N_{\bar{Q}}({\bs p}_{\bar{Q}}) / V$ where $V$ is the volume. The proper frame to define the volume of the QGP is the laboratory frame. So we have
\be
V_\ma{lab} = \gamma V_\ma{rest} \,,
\ee
where ``rest" means the rest frame of a $Q\bar{Q}$ pair and the associated $\gamma$ factor is given by $\frac{E_\ma{cm}^\ma{lab}}{2M}$ where $E_\ma{cm}^\ma{lab}$ is given in Eq.~(\ref{chap4_eqn_cmE}). So we have
\be
\frac{ \diff^3 p^\ma{rest}_{\bar{Q}} }{(2\pi)^3} f^\ma{rest}_{\bar{Q}}({\bs x}^\ma{rest}_{\bar{Q}}, {\bs p}^\ma{rest}_{\bar{Q}}, t)  \frac{2M}{E_\ma{cm}^\ma{lab}} = \frac{ \diff^3 p^\ma{lab}_{\bar{Q}} }{(2\pi)^3} f^\ma{lab}_{\bar{Q}}({\bs x}^\ma{lab}_{\bar{Q}}, {\bs p}^\ma{lab}_{\bar{Q}}, t) \,.
\ee
Finally, the total recombination rate of that specific $Q$ in the laboratory frame is given by
\be
\Gamma^g_\ma{lab} &=& g_+ \int \frac{ \diff^3 p^\ma{lab}_{\bar{Q}} }{(2\pi)^3} f^\ma{lab}_{\bar{Q}}({\bs x}^\ma{lab}_{\bar{Q}}, {\bs p}^\ma{lab}_{\bar{Q}}, t)   \\ \nn
&& \frac{8}{9}\alpha_s q^3 \Big( 2 + \frac{T}{\gamma^\ma{cell}_\ma{cm} v^\ma{cell}_\ma{cm} q}   \ln{\frac{1-e^{-\gamma^\ma{cell}_\ma{cm}(1+v^\ma{cell}_\ma{cm})q/T}}{ 1-e^{-\gamma^\ma{cell}_\ma{cm}(1-v^\ma{cell}_\ma{cm})q/T} }}  \Big) | \langle \Psi_{  {\bs p}_\ma{rel}  } | {\bs r} | \psi_{nl} \rangle |^2   \Big|_{q = |E_{nl}| + \frac{ ({\bs p}_\ma{rel})^2}{M}} \,.
\ee
Using the replacement in (\ref{chap4_eqn_gauss_dist}), we can write
\be
\label{chap4_eqn_reco_lab_g}
\Gamma^g_\ma{lab} &=& g_+ \sum_j \frac{e^{- (  {\bs x}_i - \widetilde{\bs x}_j   )^2/(2a_B^2)}}{(2\pi a_B^2)^{3/2}} 
\\ \nn
&& \frac{8}{9}\alpha_s q^3 \Big( 2 + \frac{T}{\gamma^\ma{cell}_\ma{cm} v^\ma{cell}_\ma{cm} q}   \ln{\frac{1-e^{-\gamma^\ma{cell}_\ma{cm}(1+v^\ma{cell}_\ma{cm})q/T}}{ 1-e^{-\gamma^\ma{cell}_\ma{cm}(1-v^\ma{cell}_\ma{cm})q/T} }}  \Big) | \langle \Psi_{  {\bs p}_\ma{rel}  } | {\bs r} | \psi_{nl} \rangle |^2   \Big|_{q = |E_{nl}| + \frac{ ({\bs p}_\ma{rel})^2}{M}} \,,
\ee
where ${\bs x}_i$ is the position of the heavy quark $Q$ whose recombination rate into the quarkonium state $nls$ is under consideration. $\widetilde{\bs x}_j $ is the position of each heavy antiquark $\bar{Q}$. The relative momentum ${\bs p}_\ma{rel}$ is calculated from the momenta ${\bs p}_i$ of $Q$ and $\overline{\bs p}_j$ of $\bar{Q}$ as in Eq.~(\ref{chap4_eqn_p_relative}).

\subsubsection{Inelastic Scattering with Light Quarks}
The recombination rate caused by inelastic scattering with light quarks when the $Q\bar{Q}$ is at rest with respect to the medium is written in Eq.~(\ref{chap3_eqn_recom_inel}), which is
\be 
\Gamma^{\ma{inel},q} &=& g_+ \int\frac{ \diff^3 p_{\bar{Q}} }{(2\pi)^3} f_{\bar{Q}}({\bs x}_{\bar{Q}}, {\bs p}_{\bar{Q}},t)
\int \frac{\diff^3 p_1}{2p_1(2\pi)^3} \frac{\diff^3 p_2}{2p_2(2\pi)^3} n_F(p_1) \big( 1- n_F(p_2) \big) \\ \nn
&&  (2\pi)\delta( p_1 + \frac{{\bs p}^2_\ma{rel}}{M} - p_2 + |E_{nl}| ) \frac{16}{3}g^4T_FC_F  |\langle \Psi_{{\bs p}_\ma{rel}} | {\bs r} |  \psi_{nl}  \rangle|^2   \frac{p_1p_2 + {\bs p}_1\cdot {\bs p}_2}{{\bs q}^2} \,.
\ee
In general the $Q\bar{Q}$ can move at a velocity ${\bs v}_\ma{cm}^\ma{cell}$ with respect to the hydro-cell. The hydro-cell may also move at a velocity relative to the laboratory frame. For a pair of $Q\bar{Q}$ with positions ${\bs x}_i$, $\widetilde{\bs x}_j$ and four-momenta $p_i = (E_i, {\bs p}_i)$,\, $\widetilde{p}_j = ( \widetilde{E}_j,\, \widetilde{\bs p}_j)$ in the laboratory frame, we can repeat the analysis shown in the real gluon absorption case. The total recombination rate of a heavy quark $Q$ surrounded by a certain number of $\bar{Q}$'s in the laboratory frame is
\be \nn
\Gamma^{\ma{inel},q}_\ma{lab} &=& g_+ \sum_j \frac{e^{- (  {\bs x}_i - \widetilde{\bs x}_j   )^2/(2a_B^2)}}{(2\pi a_B^2)^{3/2}} \int \frac{\diff^3 p_1}{2p_1(2\pi)^3} \frac{\diff^3 p_2}{2p_2(2\pi)^3} \\ \nn
&& n_F( \Lambda(-{\bs v}_\ma{cm}^\ma{cell} )p_1) \big( 1- n_F( \Lambda(-{\bs v}_\ma{cm}^\ma{cell}) p_2) \big) (2\pi)\delta( p_1 + \frac{{\bs p}^2_\ma{rel}}{M} - p_2 + |E_{nl}| )  \\
&&  \frac{16}{3}g^4T_FC_F  |\langle \Psi_{{\bs p}_\ma{rel}} | {\bs r} |  \psi_{nl}  \rangle|^2   \frac{p_1p_2 + {\bs p}_1\cdot {\bs p}_2}{{\bs q}^2} \,.
\ee
If we assume the velocity ${\bs v}_\ma{cm}^\ma{cell}$ as the $z$-axis, we can write the above expression as (by omitting the subscript ``cm" and the superscript ``cell" for simplicity)
\be \nn
\Gamma^{\ma{inel},q}_{\ma{lab}} &=& g_+ \sum_j \frac{e^{- (  {\bs x}_i - \widetilde{\bs x}_j   )^2/(2a_B^2)}}{(2\pi a_B^2)^{3/2}} \\ \nn
&&  \frac{8\alpha_s^2}{9\pi^2}    \int p_1 \diff p_1 \diff c_1 n_F(\gamma(1+vc_1)p_1) \int p_2 \diff c_2 \diff \phi_2 
\big[  1 - n_F(\gamma(1+vc_2)p_2)  \big] \\ 
\label{chap4_eqn_reco_lab_ineq}
&&  \frac{p_1p_2(1+s_1s_2\cos\phi_2+c_1c_2)}{p_1^2 + p_2^2 - 2p_1p_2(s_1s_2\cos\phi_2 + c_1c_2)} |\langle \Psi_{{\bs p}_\ma{rel}} | {\bs r} |  \psi_{nl}  \rangle|^2 \,.
\ee

\subsubsection{Inelastic Scattering with Gluons}
The recombination rate caused by inelastic scattering with gluons when the $Q\bar{Q}$ is at rest with respect to the medium is written in Eq.~(\ref{chap3_eqn_recom_inel}), which is
\be 
\Gamma^{\ma{inel},g}_{\ma{cell}} &=& g_+ \int\frac{ \diff^3 p_{\bar{Q}} }{(2\pi)^3}  f_{\bar{Q}}({\bs x}_{\bar{Q}}, {\bs p}_{\bar{Q}},t)
\int \frac{\diff^3 q_1}{2q_1(2\pi)^3} \frac{\diff^3 q_2}{2q_2(2\pi)^3} n_B(q_1) \big( 1 + n_B(q_2) \big) \\ \nn
&& (2\pi)\delta( q_1 + \frac{{\bs p}^2_\ma{rel}}{M} - q_2 + |E_{nl}| )  \frac{1}{3}g^4C_F | \langle \Psi_{{\bs p}_\ma{rel}} | {\bs r} |  \psi_{nl}  \rangle|^2 \frac{1+(\hat{q}_1\cdot\hat{q}_2)^2}{{\bs q}^2} (q_1+q_2)^2 \,,
\ee
The delta function in energy gives $p_\ma{rel} = |{\bs p}_\ma{rel}| = \sqrt{M(q_2 - |E_{nl}| - q_1)}$. In general the $Q\bar{Q}$ can move at a velocity ${\bs v}_\ma{cm}^\ma{cell}$ with respect to the hydro-cell. The hydro-cell may also move at a velocity relative to the laboratory frame. For a pair of $Q\bar{Q}$ with positions ${\bs x}_i$, $\widetilde{\bs x}_j$ and four-momenta $p_i = (E_i, {\bs p}_i)$,\, $\widetilde{p}_j = ( \widetilde{E}_j,\, \widetilde{\bs p}_j)$ in the laboratory frame, we can repeat the analysis shown in the real gluon absorption case. The total recombination rate of a heavy quark $Q$ surrounded by a certain number of $\bar{Q}$'s in the laboratory frame is
\be \nn
\Gamma^{\ma{inel},g}_\ma{lab} &=& g_+ \sum_j \frac{e^{- (  {\bs x}_i - \widetilde{\bs x}_j   )^2/(2a_B^2)}}{(2\pi a_B^2)^{3/2}} \int \frac{\diff^3 q_1}{2q_1(2\pi)^3} \frac{\diff^3 q_2}{2q_2(2\pi)^3} \\ \nn
&& n_B( \Lambda(-{\bs v}_\ma{cm}^\ma{cell} )q_1) \big( 1 + n_B( \Lambda(-{\bs v}_\ma{cm}^\ma{cell}) q_2) \big) (2\pi)\delta( q_1 + \frac{{\bs p}^2_\ma{rel}}{M} - q_2 + |E_{nl}| )  \\
&&  \frac{1}{3}g^4C_F | \langle \Psi_{{\bs p}_\ma{rel}} | {\bs r} |  \psi_{nl}  \rangle|^2 \frac{1+(\hat{q}_1\cdot\hat{q}_2)^2}{{\bs q}^2} (q_1+q_2)^2 \,.
\ee
If we assume the velocity ${\bs v}_\ma{cm}^\ma{cell}$ as the $z$-axis, we can write the above expression as (by omitting the subscript ``cm" and the superscript ``cell" for simplicity)
\be \nn
\Gamma^{\ma{inel},g}_{\ma{lab}} &=& g_+ \sum_j \frac{e^{- (  {\bs x}_i - \widetilde{\bs x}_j   )^2/(2a_B^2)}}{(2\pi a_B^2)^{3/2}} \\ \nn
&&  \frac{\alpha_s^2}{3\pi^2}    \int q_1 \diff q_1 \diff c_1 n_B(\gamma(1+vc_1)q_1) \int q_2 \diff c_2 \diff \phi_2 \big[  1 + n_B(\gamma(1+vc_2)q_2)  \big]  \\
\label{chap4_eqn_reco_lab_ineg}
&& \frac{ (q_1+q_2)^2(1+s_1s_2\cos\phi_2 +c_1c_2) }{q_1^2 + q_2^2 - 2q_1q_2(s_1s_2\cos\phi_2 + c_1c_2)} | \langle \Psi_{{\bs p}_\ma{rel}} | {\bs r} |  \psi_{nl}  \rangle|^2 \,.
\ee

\vspace{0.2in}

Then the $nls$-recombination probability of a given heavy quark with position ${\bs x}_i$ and momentum ${\bs p}_i$ in the laboratory frame, surrounded by a certain number of heavy antiquarks, is given by 
\be
P_\ma{lab} = (\Gamma^{g}_\ma{lab} + \Gamma^{\ma{inel},q}_\ma{lab} + \Gamma^{\ma{inel},g}_\ma{lab}) \Delta t =  P^g_\ma{lab} + P^{\ma{inel},q}_\ma{lab} + P^{\ma{inel},g}_\ma{lab} \,.
\ee
After calculating the recombination probability in the laboratory frame, we sample a random number $r$ from a uniform distribution between $0$ and $1$ and decide which process occurs based on
\begin{enumerate}
\item if $r \leq P^g_\ma{lab}$, the recombination into a quarkonium $nls$ occurs in the real gluon radiation channel;
\item if $P^g_\ma{lab} < r \leq P^g_\ma{lab}  + P^{\ma{inel},q}_\ma{lab}$, the recombination into a quarkonium $nls$ occurs in the channel of inelastic scattering with a light quark;
\item if $P^g_\ma{lab}  + P^{\ma{inel},q}_\ma{lab} < r \leq  P_\ma{lab}$, the recombination into a quarkonium $nls$ occurs in the channel of inelastic scattering with a gluon;
\item if $P_\ma{lab} < r$, the recombination into a quarkonium $nls$ does not happen in this time step.
\end{enumerate}

If the recombination into the quarkonium state $nls$ happens in a certain process in this time step, we will further determine which $\bar{Q}$ the $Q$ recombines with. This can be done in a similar Monte Carlo way by writing the recombination probability in that process as a sum over all the contributions from different $\bar{Q}$'s (see the summation in expressions (\ref{chap4_eqn_reco_lab_g}), (\ref{chap4_eqn_reco_lab_ineq}) and (\ref{chap4_eqn_reco_lab_ineg})) and check where the random number $r$ fits into the increasing series. Once we determine which $Q\bar{Q}$ pair recombines, we will remove the $Q\bar{Q}$ from the lists of open heavy quarks and antiquarks. We will add a new particle to the list of the quarkonium state $nls$. The position of the new quarkonium state $nls$ is given by the c.m.~position of the recombining $Q\bar{Q}$. 
The assignment of momenta is more involved and we will explain this in the following for each recombination process.

\subsubsection{Momentum Sampling in Real Gluon Radiation}
We will first determine the quarkonium momentum in the rest frame of the recombining $Q\bar{Q}$ and then boost it back to the laboratory frame. Once we determine the relative kinetic energy ${\bs p}_\ma{rel}^2/M$ of the recombining $Q\bar{Q}$ by using Eq.~(\ref{chap4_eqn_p_relative}), the energy of the radiated gluon is fixed by
\be
q = \frac{{\bs p}_\ma{rel}^2}{M} + |E_{nl}| \,.
\ee
The only thing we need to determine is the polar $\theta_g$ and azimuthal $\phi_g$ angles of the radiated gluon with respect to the $Q\bar{Q}$ c.m.~velocity ${\bs v}_\ma{cm}^\ma{cell}$, which is defined in the hydro-cell frame. The angles can be sampled from the angular part of the integrand of Eq.~(\ref{chap4_recom_gluon_sample})
\be
f(\cos\theta) \equiv \frac{1}{ e^{ \gamma_\ma{cm}^\ma{cell}(1 + v_\ma{cm}^\ma{cell} \cos\theta)q/T }  -1} \,.
\ee
We can use the acceptance-rejection method to sample $\cos\theta$. The maximum of $f(\cos\theta)$ is obtained at $\cos\theta = 1$. We will first sample a random number $c$ from a uniform distribution between $-1$ and $1$ and another random number $r$ from a uniform distribution between $0$ and $1$, and if
\be
rf(1) \leq f(c) \,,
\ee
we will accept this sampling and use $c$ as the $\cos\theta_g$ we want to sample. Otherwise we just repeat the process till we obtain the first $c$ satisfying the condition. The azimuthal angle $\phi_g$ is sampled from a uniform distribution between $0$ and $2\pi$ because the integrand is independent of $\phi_g$. Then the gluon momentum in the $Q\bar{Q}$ rest frame can be written as
\be
{\bs q} = \begin{pmatrix}  q \sin\theta_g \cos\phi_g  \\ q\sin\theta_g\sin\phi_g \\ q\cos\theta_g \end{pmatrix} \,.
\ee
The momentum of the quarkonium in the $Q\bar{Q}$ rest frame is
\be
{\bs k}_\ma{rest} = - {\bs q}\,.
\ee
The next step is to rotate ${\bs k}_\ma{rest}$ in the same way as we rotate the $z$-axis to the direction of ${\bs v}_\ma{cm}^\ma{cell}$, because when we sample the momentum of the radiated gluon, we assume ${\bs v}_\ma{cm}^\ma{cell}$ as the $z$-axis. Assuming the polar and azimuthal angles of ${\bs v}_\ma{cm}^\ma{cell}$ are $\theta$ and $\phi$, we get
\be
\label{chap4_eqn_recom_rotate}
{\bs k}_\ma{rest}^\ma{rot} = \begin{pmatrix}
  \cos\phi &  - \sin\phi &   \\
  \sin\phi & \cos\phi &   \\
   &  & 1
\end{pmatrix} 
\begin{pmatrix}
  \cos\theta &   & \sin\theta  \\
   & 1 &   \\
  -\sin\theta &  & \cos\theta
\end{pmatrix} {\bs k}_\ma{rest} \,.
\ee
Then we boost it back to the hydro-cell frame and to the laboratory frame
\be
\label{chap4_eqn_recom_boostlab}
k_\ma{lab} =  \Lambda(-{\bs v}) \Lambda(- {\bs v}_\ma{cm}^\ma{cell} ) k_\ma{rest}^\ma{rot}  \,,
\ee
in which
\be    
k_\ma{rest}^\ma{rot}  =  \begin{pmatrix} \sqrt{M_{nls}^2 + {\bs k}_\ma{rest}^2} \\ {\bs k}_\ma{rest} \end{pmatrix} \,,
\ee
where the mass of the quarkonium state $nls$ is given by $M_{nls} = 2M - |E_{nl}|$. The binding energy $|E_{nl}|$ is independent of the spin $s$ because we work at lowest order in the nonrelativistic expansion.

\subsubsection{Momentum Sampling in Inelastic Scattering with Light Quarks}
We only need the following part of the integrand of Eq.~(\ref{chap4_eqn_reco_lab_ineq}) to sample the momenta of the incoming $p_1$ and outgoing $p_2$ light quarks
\be \nn
\label{chap4_eqn_recom_ineq_sample}
&&\int p_1 \diff p_1 \diff c_1 n_F(\gamma(1+vc_1)p_1) 
\int p_2 \diff c_2 \diff \phi_2 \big[  1 - n_F(\gamma(1+vc_2)p_2)  \big] \\
&& \frac{p_1p_2(1+s_1s_2\cos\phi_2+c_1c_2)}{p_1^2 + p_2^2 - 2p_1p_2(s_1s_2\cos\phi_2 + c_1c_2)}  \,.
\ee
We neglect $| \langle \Psi_{{\bs p}_\ma{rel}} | {\bs r} |  \psi_{nl}  \rangle|^2$ because in recombination, ${\bs p}_\ma{rel}$ is fixed by the $Q\bar{Q}$ pair in the initial state and thus $| \langle \Psi_{{\bs p}_\ma{rel}} | {\bs r} |  \psi_{nl}  \rangle|^2$ is just a constant.
Again we will use the combination of importance sampling, rejection sampling and inverse function sampling methods. We rewrite the above integral (\ref{chap4_eqn_recom_ineq_sample}) as
\be
\int \diff p_1 \diff c_1 \diff c_2 \diff \phi_2 \,g(p_1,c_1)  h(p_1, c_1, c_2, \phi_2) \,,
\ee
where the value of $p_2$ is fixed in the recombination process by the initial relative kinetic energy ${\bs p}_\ma{rel}^2  /M$
\be
p_2 = p_1 + |E_{nl}| + \frac{ {\bs p}_\ma{rel}^2 }{M} \,.
\ee
The functions $g$ and $h$ are defined as
\be
g(p_1,c_1) &=& p_1p_2 \frac{1}{e^{\gamma(1+vc_1)p_1/T} + 1} \frac{1}{\frac{p_1}{p_2} + \frac{p_2}{p_1} - 2} \\ \nn
h(p_1, c_1, c_2, \phi_2) &=& \Big(  1-\frac{1}{e^{\gamma(1+vc_2)p_2/T} + 1} \Big) \frac{1+s_1s_2\cos\phi_2+c_1c_2}{2}  \\ 
&&  \frac{\frac{p_1}{p_2} + \frac{p_2}{p_1} - 2}{ \frac{p_1}{p_2} + \frac{p_2}{p_1} - 2(s_1s_2\cos\phi_2+c_1c_2) } \,.
\ee

The sampling of $p_1$ and $c_1$ according to the distribution function $g(p_1, c_1)$ is very similar to the sampling procedure described for the dissociation process induced by inelastic scattering with light quarks. We will sample $p_1$ using the rejection method on the function
\be
\int_{-1}^1 \diff c_1 g(p_1, c_1) = \frac{T}{\gamma v}  p_2 \ln{\frac{1+e^{-\gamma(1-v)p_1/T}}{1+e^{-\gamma(1+v)p_1/T}}}   \frac{1}{ \frac{p_1}{p_2 } + \frac{ p_2 }{p_1} - 2} \,,
\ee
and sample $c_1$ using the inverse function method once we have sampled a $p_1$. The function we need to inverse is defined as
\be
G(x) &\equiv&   \int_{-1}^x \diff c_1 \frac{1}{e^{\gamma(1+vc_1)p_1/T}+1} \\
&=& -\frac{T}{\gamma v p_1} \Big[  \ln{ (1+e^{-\gamma(1+vx)p_1/T})} -  \ln{ (1+e^{-\gamma(1-v)p_1/T} )}    \Big]  \,.
\ee
The $x = \cos\theta$ can be solved from the equation $r = G(x)$ where $r$ is a random number generated from a uniform distribution between $0$ and $1$.

Then we sample a $c_2$ and a $\phi_2$ uniformly from $-1$ to $1$ and from $0$ to $2\pi$ respectively. Together with the sampled $p_1$ and $c_1$, we can compute the value of $h(p_1, c_1, c_2, \phi_2)$. We also notice that $h \leq 1$. So we sample another random number uniformly from $0$ to $1$ and compare $h(p_1, c_1, c_2, \phi_2)$ with $r$. If $r \leq h(p_1, c_1, c_2, \phi_2)$, we will accept this sampling. Otherwise we just repeat the whole sampling procedure till we can find a set of variables $p_1, c_1, c_2, \phi_2$ such that the condition is satisfied. Once we have $p_1$, $c_1$, $c_2$ and $\phi_2$, we can determine the momenta of the incoming and outgoing light quarks in the rest frame of the $Q\bar{Q}$,
\be
p_1 = \begin{pmatrix}  p_1 \\ p_1\sin\theta_1  \\ 0 \\ p_1\cos\theta_1 \end{pmatrix} \,,\ \ \ \ \ \ \ 
p_2 = \begin{pmatrix}  p_2 \\ p_2\sin\theta_2\cos\phi_2  \\ p_2\sin\theta_2\sin\phi_2 \\ p_2\cos\theta_2 \end{pmatrix} \,.
\ee
The four-momentum of the transferred gluon in the rest frame of the $Q\bar{Q}$ is
\be
p_g = p_1 - p_2 \,.
\ee
Then the momentum of the quarkonium in the rest frame of the $Q\bar{Q}$ is
\be
{\bs k}_\ma{rest} = - {\bs p}_g \,.
\ee
The next step would be to rotate the quarkonium momentum as in Eq.~(\ref{chap4_eqn_recom_rotate}) and boost it back to the laboratory frame as in Eq.~(\ref{chap4_eqn_recom_boostlab}).

\begin{figure}
    \centering
    \begin{subfigure}[b]{0.48\textwidth}
        \centering
        \includegraphics[height=2.1in]{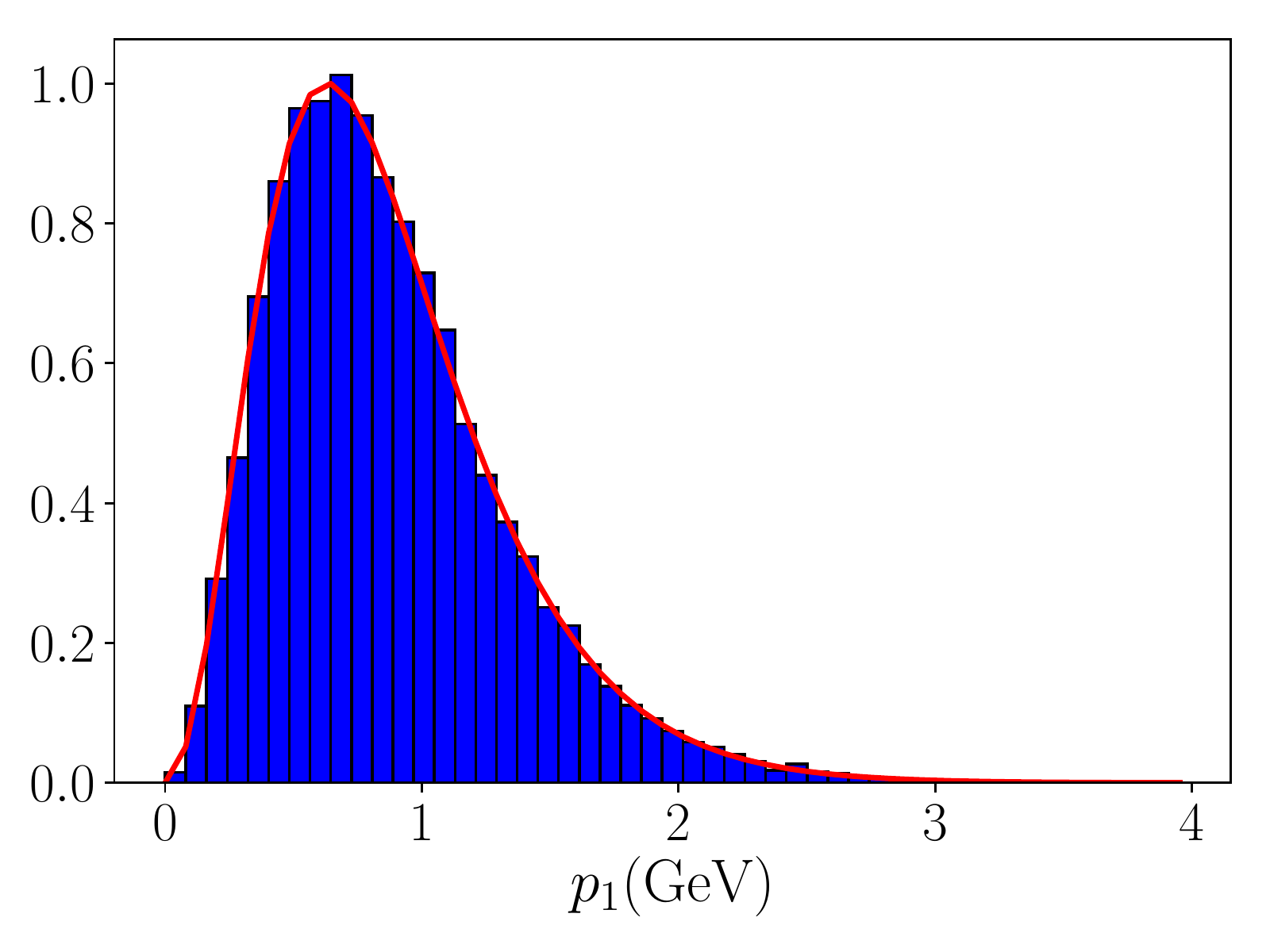}
        \caption{$p_1$}\label{}
    \end{subfigure}%
    ~
    \begin{subfigure}[b]{0.48\textwidth}
        \centering
        \includegraphics[height=2.1in]{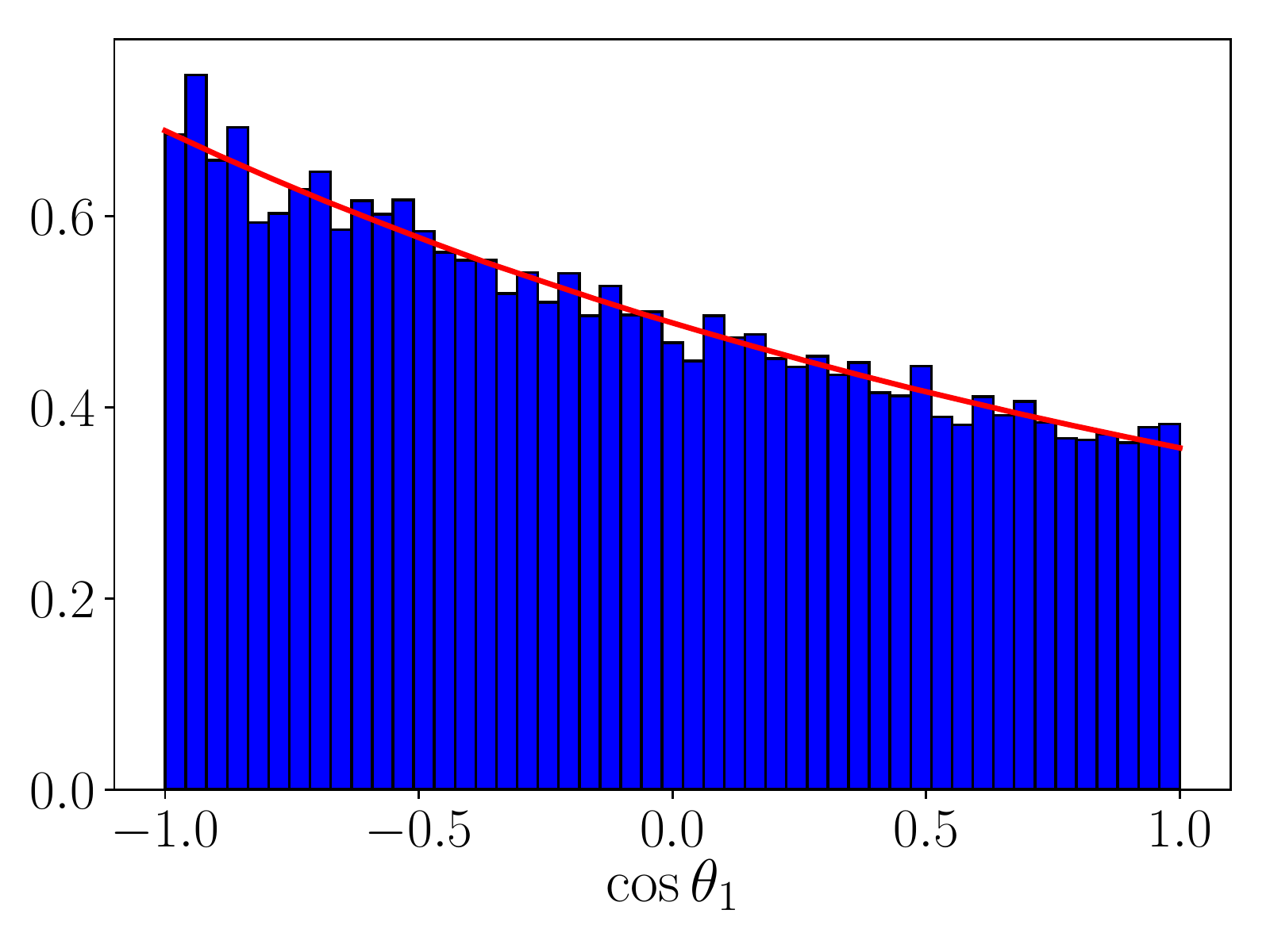}
        \caption{$c_1$}\label{}
    \end{subfigure}%
    
    \begin{subfigure}[b]{0.48\textwidth}
        \centering
        \includegraphics[height=2.1in]{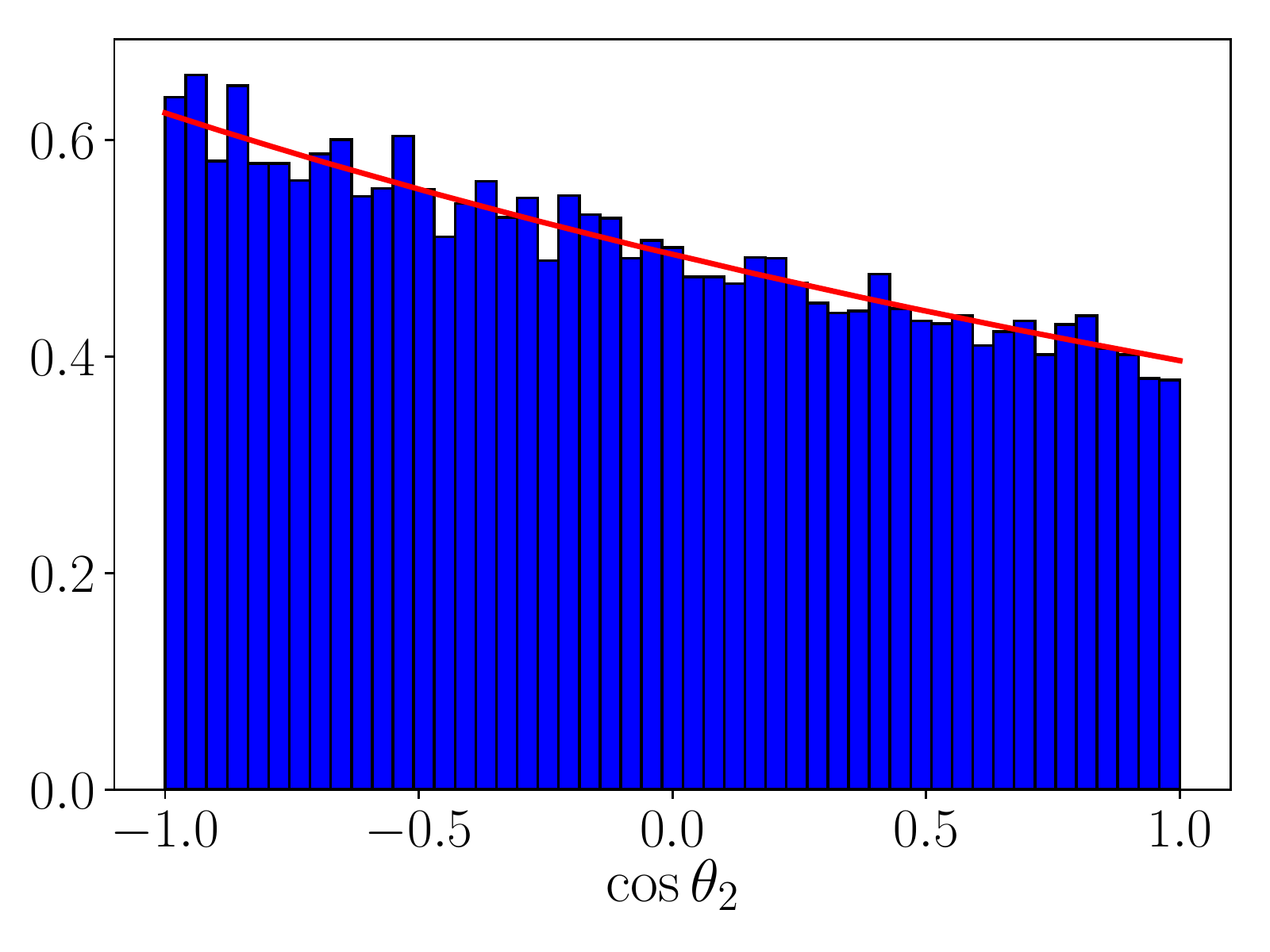}
        \caption{$c_2$}\label{}
    \end{subfigure}%
    ~
    \begin{subfigure}[b]{0.48\textwidth}
        \centering
        \includegraphics[height=2.1in]{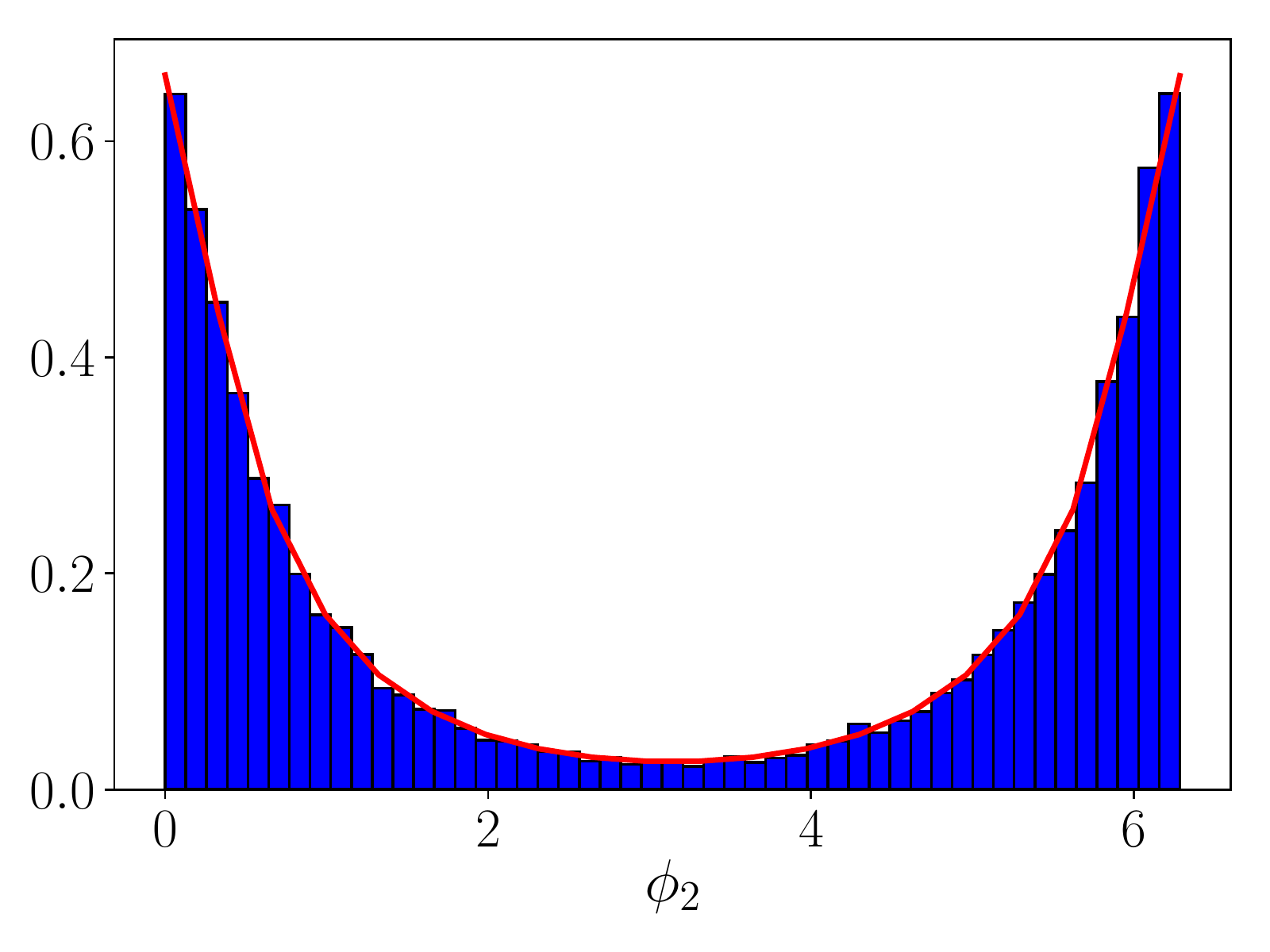}
        \caption{$\phi_2$}\label{}
    \end{subfigure}%
    \caption[Histograms of sampled momenta of the light quarks compared with the marginal distributions when $v_\ma{rel}=0.1$ and $T=0.25$ GeV.]{Histograms of sampled momenta of the incoming and outgoing light quarks (blue) compared with the marginal distributions (red) when $v_\ma{rel}=0.1$ and $T=0.25$ GeV. Normalization is unity.}
    \label{chap4_fig_recom_ineq_sample_compare1}
\end{figure}

\begin{figure}
    \centering
    \begin{subfigure}[b]{0.48\textwidth}
        \centering
        \includegraphics[height=2.1in]{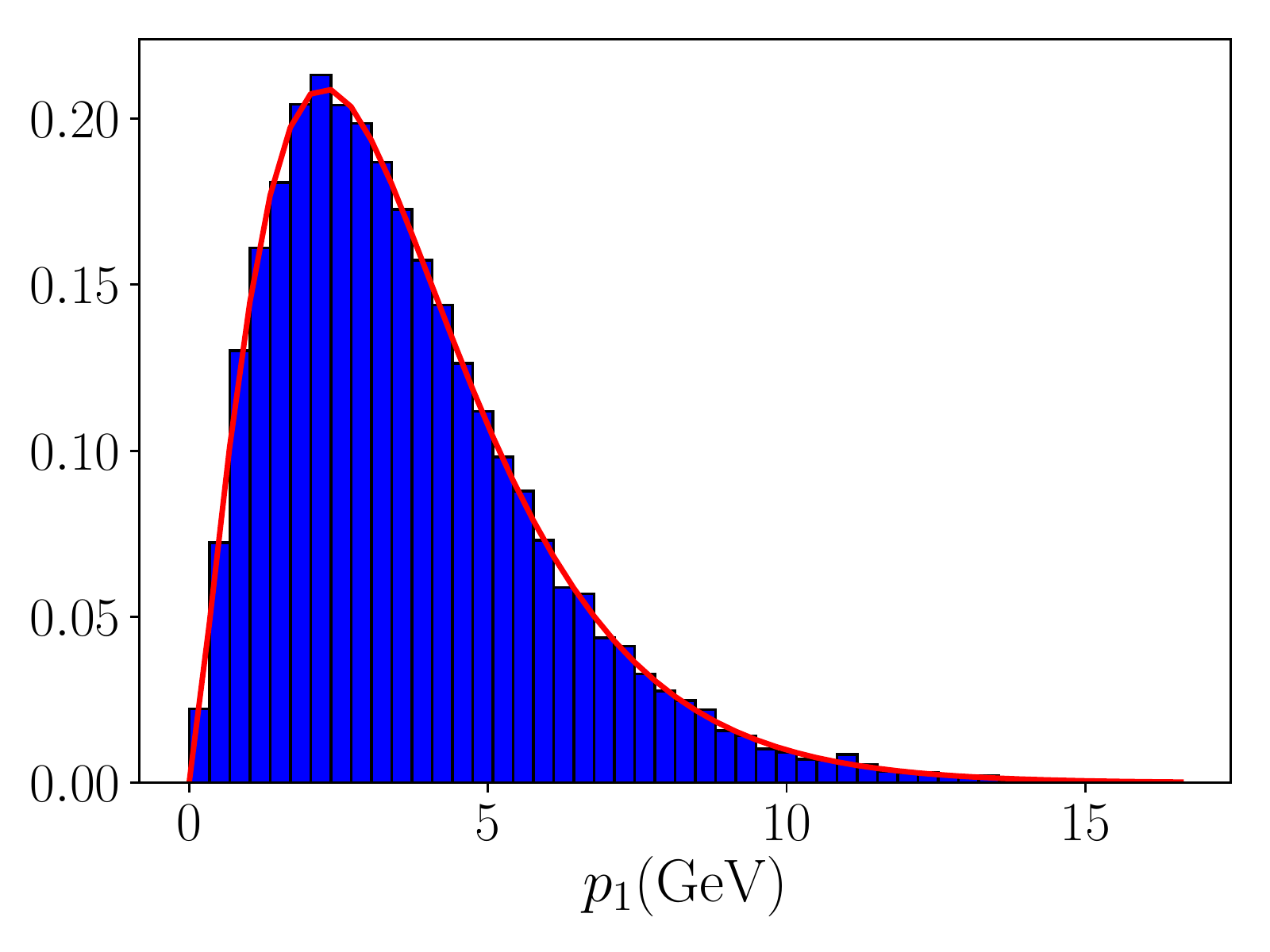}
        \caption{$p_1$}\label{}
    \end{subfigure}%
    ~
    \begin{subfigure}[b]{0.48\textwidth}
        \centering
        \includegraphics[height=2.1in]{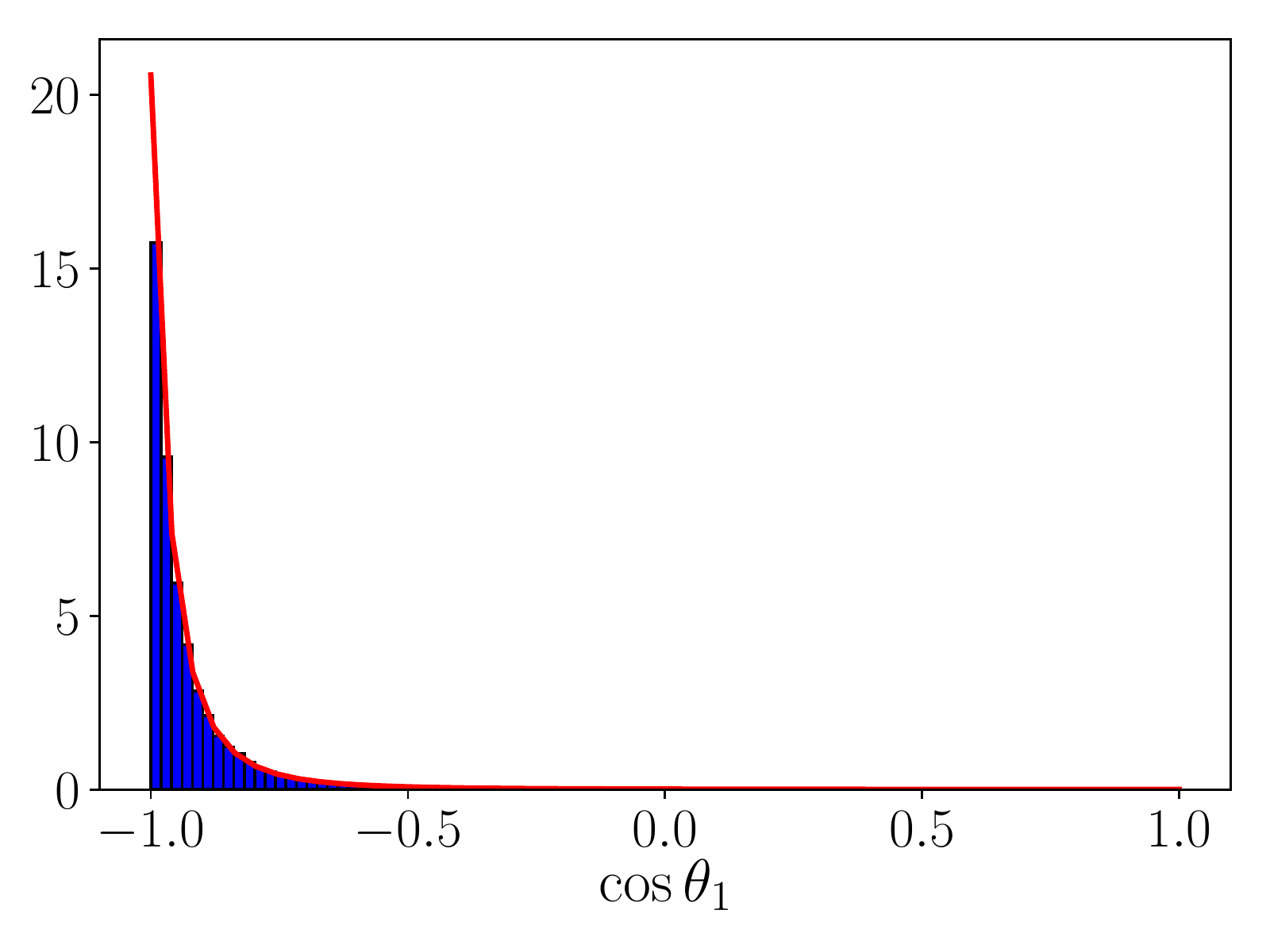}
        \caption{$c_1$}\label{}
    \end{subfigure}%
    
    \begin{subfigure}[b]{0.48\textwidth}
        \centering
        \includegraphics[height=2.1in]{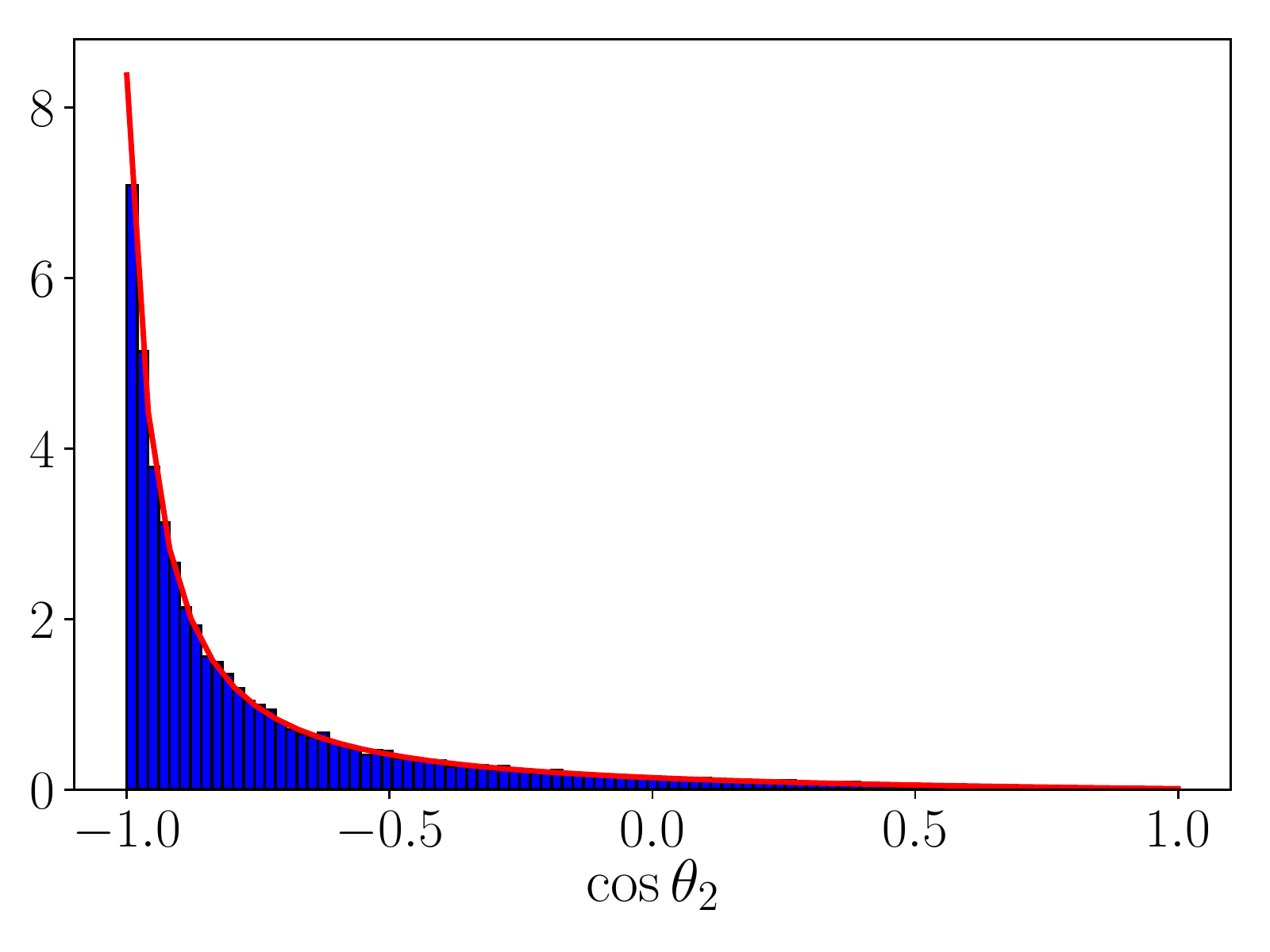}
        \caption{$c_2$}\label{}
    \end{subfigure}%
    ~
    \begin{subfigure}[b]{0.48\textwidth}
        \centering
        \includegraphics[height=2.1in]{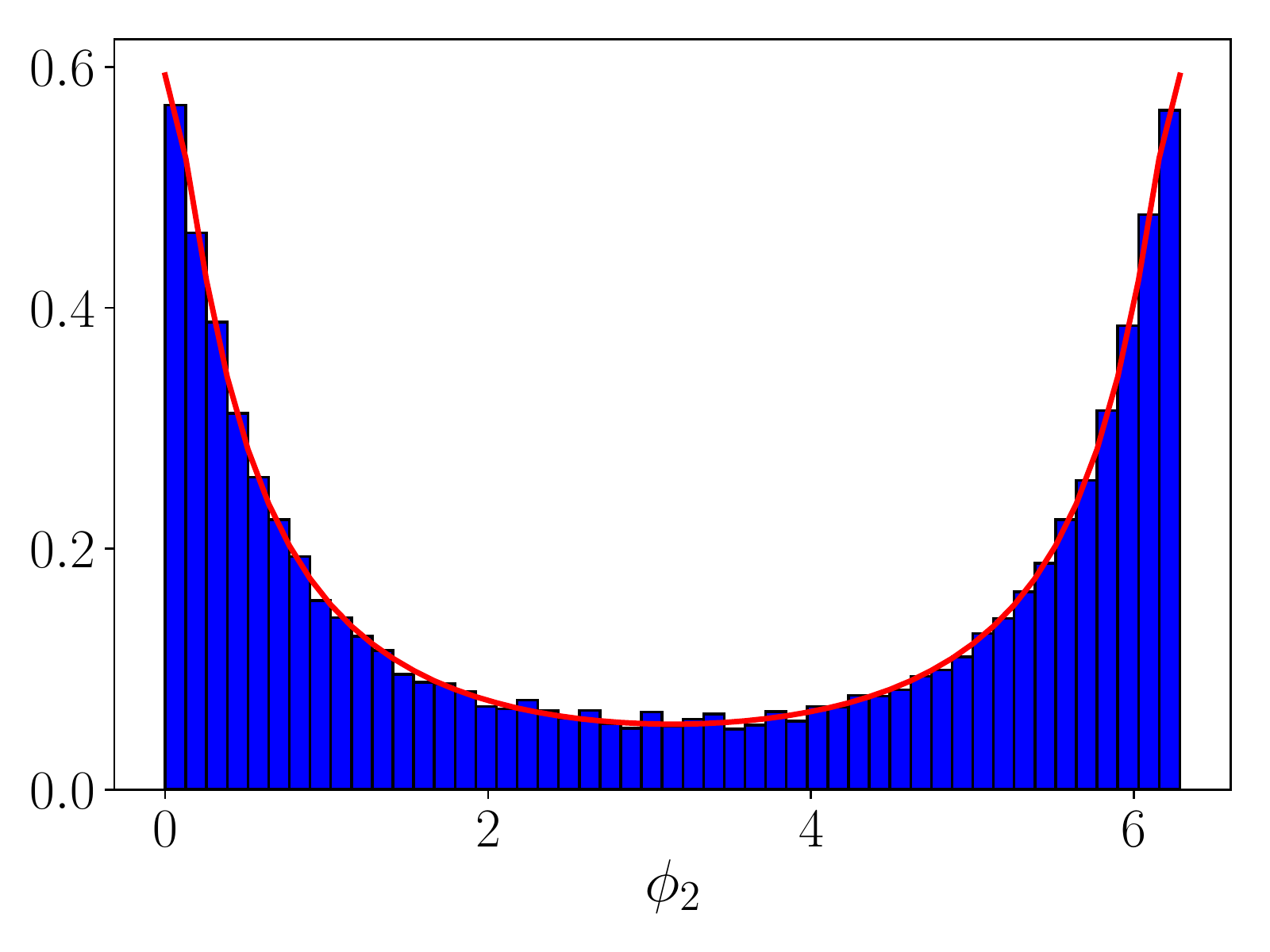}
        \caption{$\phi3_2$}\label{}
    \end{subfigure}%
    \caption[Histograms of sampled momenta of the light quarks compared with the marginal distributions when $v_\ma{rel}=0.9$ and $T=0.35$ GeV.]{Histograms of sampled momenta of the incoming and outgoing light quarks (blue) compared with the marginal distributions (red) when $v_\ma{rel}=0.9$ and $T=0.35$ GeV. Normalization is unity.}
    \label{chap4_fig_recom_ineq_sample_compare2}
\end{figure}

We test the momentum sampling by sampling a large number of events and comparing with the marginal distributions. We assume $\alpha_s=0.3$, $M=4.65$ GeV and ${\bs p}_\ma{rel}=0.8$ GeV. The marginal distribution in a variable $X$ is obtained by integrating the integrand in Eq.~(\ref{chap4_eqn_recom_ineq_sample}) over all the variables except the $X$. Here $X$ can be $p_1$, $c_1$, $c_2$ and $\phi_2$. The comparisons are shown in Figs.~\ref{chap4_fig_recom_ineq_sample_compare1} and~\ref{chap4_fig_recom_ineq_sample_compare2} for two cases: $v_\ma{cell} = 0.1$, $T=0.25$ GeV and $v_\ma{cell} = 0.9$, $T=0.35$ GeV. We sampled $30000$ recombination events for each case and plot the histograms in blue. It can be seen that the sampled distributions agree well with the marginal distributions drawn in red.

\subsubsection{Momentum Sampling in Inelastic Scattering with Gluons}
We neglect $| \langle \Psi_{{\bs p}_\ma{rel}} | {\bs r} |  \psi_{nl}  \rangle|^2$ because in recombination, ${\bs p}_\ma{rel}$ is fixed by the $Q\bar{Q}$ pair in the initial state and thus $| \langle \Psi_{{\bs p}_\ma{rel}} | {\bs r} |  \psi_{nl}  \rangle|^2$ is just a constant.
We only need the following part of the integrand of Eq.~(\ref{chap4_eqn_reco_lab_ineg}) to sample the momenta of the incoming $q_1$ and outgoing $q_2$ gluons
\be \nn
&& \int q_1 \diff p_1 \diff c_1 n_B(\gamma(1+vc_1)q_1)  \int q_2 \diff c_2 \diff \phi_2 \big[  1 + n_B(\gamma(1+vc_2)q_2)  \big] \\
\label{chap4_eqn_recom_ineg_sample}
&& \frac{(q_1+q_2)^2(1+s_1s_2\cos\phi_2+c_1c_2)}{q_1^2 + q_2^2 - 2q_1q_2(s_1s_2\cos\phi_2 + c_1c_2)}  \,.
\ee
Again we will use the combination of importance sampling, rejection sampling and inverse function sampling methods. We rewrite the above integral (\ref{chap4_eqn_recom_ineg_sample}) as
\be
\int \diff q_1 \diff c_1 \diff c_2 \diff \phi_2\, g(q_1,c_1)  h(q_1, c_1, c_2, \phi_2) \,,
\ee
where the value of $q_2$ is fixed in the recombination process by the initial relative kinetic energy ${\bs p}_\ma{rel}^2  /M$
\be
q_2 = q_1 + |E_{nl}| + \frac{ {\bs p}_\ma{rel}^2 }{M} \,.
\ee
The functions $g$ and $h$ are defined as
\be
g(q_1,c_1) &=& q_1q_2 \frac{1}{e^{\gamma(1+vc_1)q_1/T} - 1} \frac{ \frac{q_1}{q_2} + \frac{q_2}{q_1} + 2 }{\frac{q_1}{q_2} + \frac{q_2}{q_1} - 2} \\ \nn
h(q_1, c_1, c_2, \phi_2) &=& \big[  1 + \frac{1}{e^{\gamma(1+vc_2)q_2/T} - 1} \big] \frac{1+s_1s_2\cos\phi_2+c_1c_2}{2} \\
&& \frac{\frac{q_1}{q_2} + \frac{q_2}{q_1} - 2}{ \frac{q_1}{q_2} + \frac{q_2}{q_1} - 2(s_1s_2\cos\phi_2+c_1c_2) }  \,.
\ee

The sampling of $q_1$ and $c_1$ according to the distribution function $g(q_1, c_1)$ is very similar to the sampling procedure described for the dissociation process induced by inelastic scattering with gluons. We will sample $q_1$ using the rejection method on the function
\be
\int_{-1}^1 \diff c_1 g(q_1, c_1) = \frac{T}{\gamma v}  q_2 \ln{\frac{1 - e^{-\gamma(1+v)q_1/T}}{1 - e^{-\gamma(1-v)q_1/T}}}   \frac{ \frac{q_1}{q_2} + \frac{q_2}{q_1} + 2 }{\frac{q_1}{q_2} + \frac{q_2}{q_1} - 2} \,,
\ee
and sample $c_1$ using the inverse function method once we have sampled a $q_1$. The function we need to inverse is defined as
\be \nn
G(x) &\equiv&   \int_{-1}^x \diff c_1 \frac{1}{e^{\gamma(1+vc_1)q_1/T} - 1} \\
&=& \frac{T}{\gamma v q_1} \Big[  \ln{ (1 - e^{-\gamma(1+vx)p_1/T})} -  \ln{ (1 - e^{-\gamma(1-v)p_1/T} )}    \Big]  \,.
\ee
The $x = \cos\theta$ can be solved from the equation $r = G(x)$ where $r$ is a random number generated from a uniform distribution between $0$ and $1$.

Then we sample a $c_2$ and a $\phi_2$ uniformly from $-1$ to $1$ and from $0$ to $2\pi$ respectively. Together with the sampled $q_1$ and $c_1$, we can compute the value of $h(q_1, c_1, c_2, \phi_2)$. We also notice that 
\be
h \leq 1 + \frac{1}{ e^{\gamma(1-v)(|E_{nl}| + \frac{{\bs p}_\ma{rel}^2}{M})/T} - 1 } \equiv h_\ma{max} \,.
\ee
So we sample another random number uniformly from $0$ to $1$ and compare $h(q_1, c_1, c_2, \phi_2)$ with $r h_\ma{max}$. If $ r h_\ma{max} \leq h(q_1, c_1, c_2, \phi_2) $, we will accept this sampling. Otherwise we just repeat the whole sampling procedure till we find a set of variables $q_1, c_1, c_2, \phi_2$ such that the condition is satisfied. Once we have $q_1$, $c_1$, $c_2$ and $\phi_2$, we can determine the momenta of the incoming and outgoing gluon in the rest frame of the $Q\bar{Q}$,
\be
q_1 = \begin{pmatrix}  q_1 \\ q_1\sin\theta_1  \\ 0 \\ q_1\cos\theta_1 \end{pmatrix} \,,\ \ \ \ \ \ \ 
q_2 = \begin{pmatrix}  q_2 \\ q_2\sin\theta_2\cos\phi_2  \\ q_2\sin\theta_2\sin\phi_2 \\ q_2\cos\theta_2 \end{pmatrix} \,.
\ee
The four-momentum of the transferred gluon is
\be
p_g = q_1 - q_2 \,.
\ee
Then the momentum of the produced quarkonium in the rest frame of the $Q\bar{Q}$ is
\be
{\bs k}_\ma{rest} = - {\bs p}_g \,.
\ee
The next step would be to rotate the quarkonium momentum as in Eq.~(\ref{chap4_eqn_recom_rotate}) and boost it back to the laboratory frame as in Eq.~(\ref{chap4_eqn_recom_boostlab}).

\begin{figure}
    \centering
    \begin{subfigure}[b]{0.48\textwidth}
        \centering
        \includegraphics[height=2.1in]{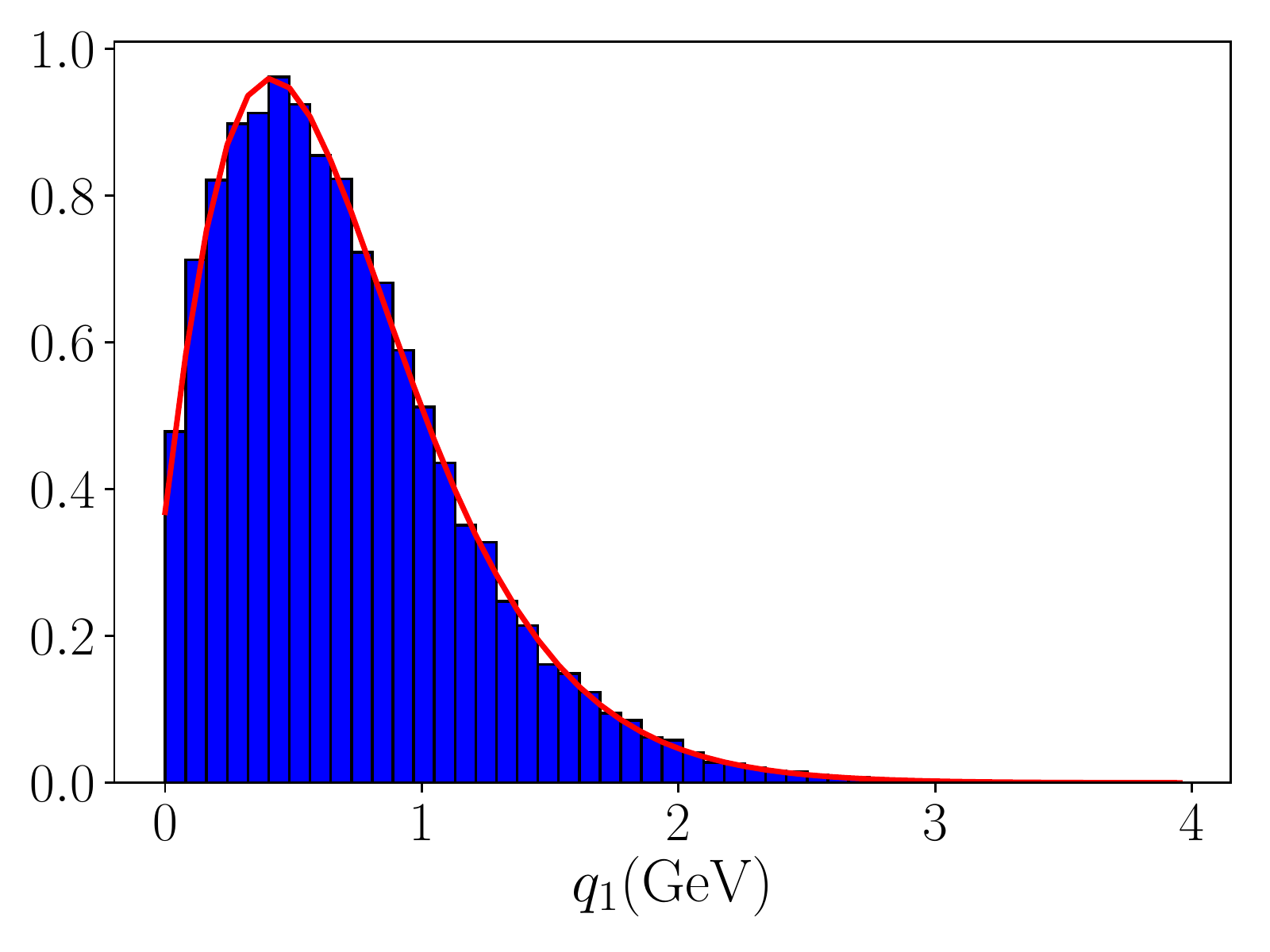}
        \caption{$q_1$}\label{}
    \end{subfigure}%
    ~
    \begin{subfigure}[b]{0.48\textwidth}
        \centering
        \includegraphics[height=2.1in]{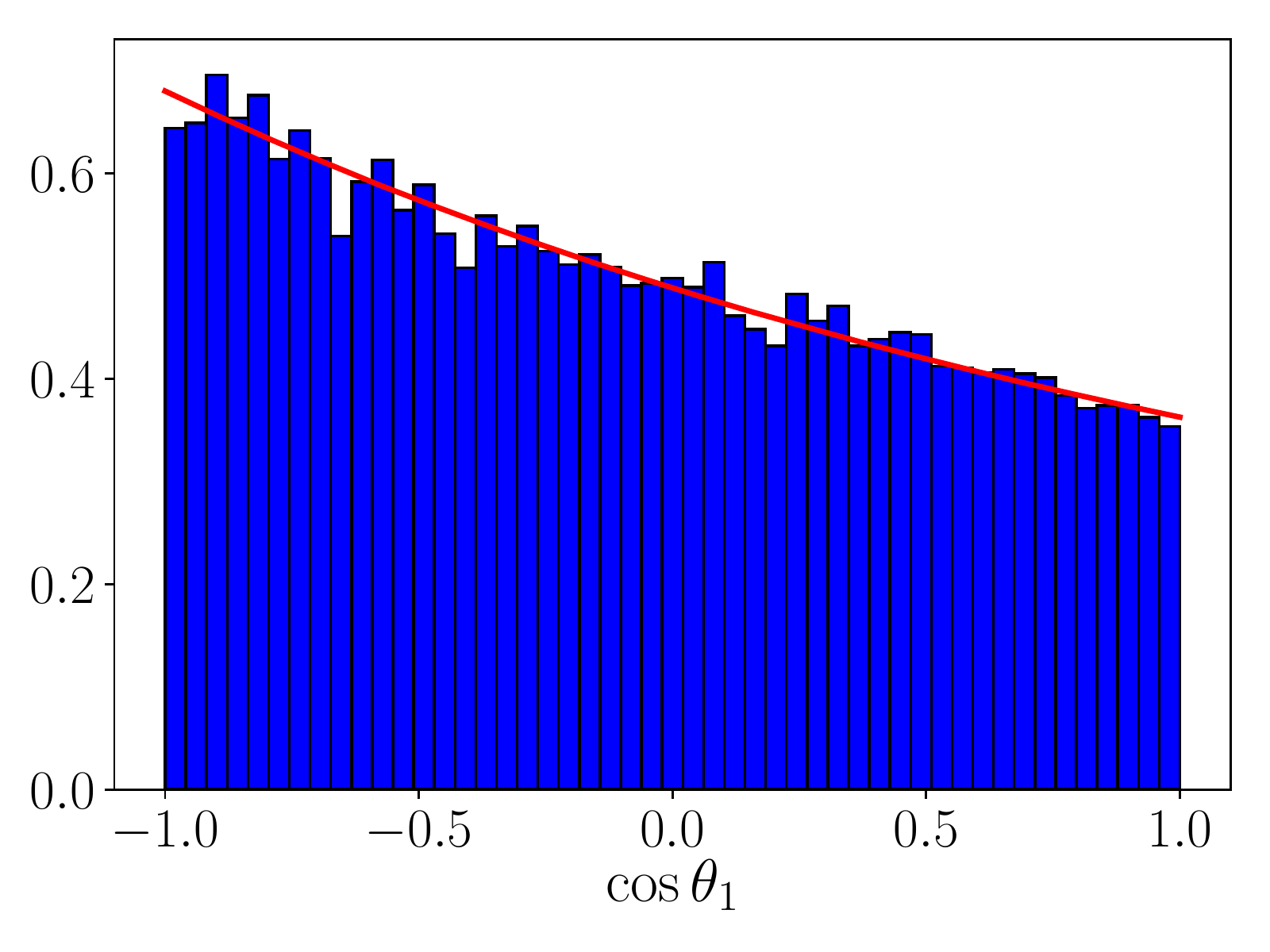}
        \caption{$c_1$}\label{}
    \end{subfigure}%
    
    \begin{subfigure}[b]{0.48\textwidth}
        \centering
        \includegraphics[height=2.1in]{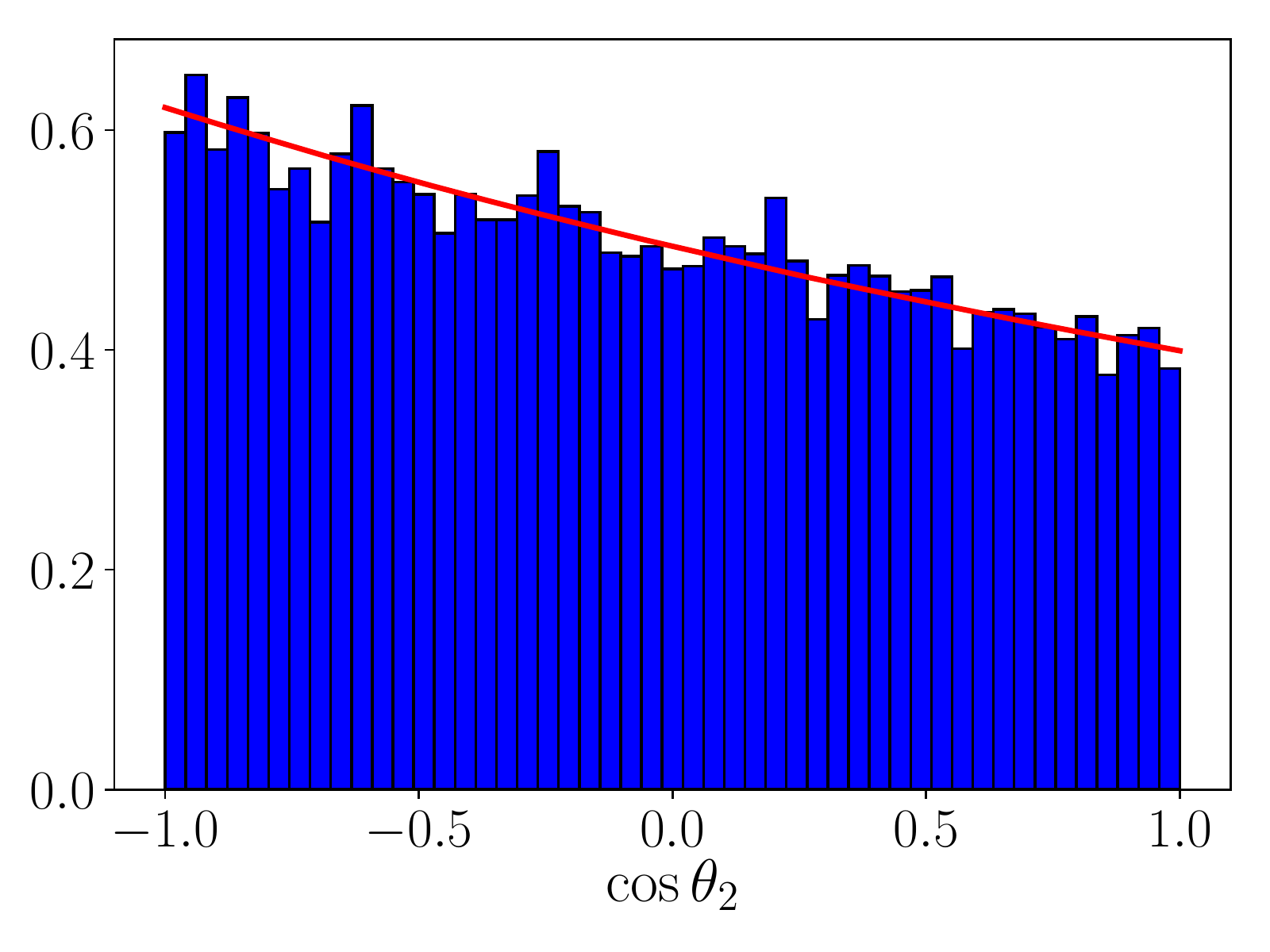}
        \caption{$c_2$}\label{}
    \end{subfigure}%
    ~
    \begin{subfigure}[b]{0.48\textwidth}
        \centering
        \includegraphics[height=2.1in]{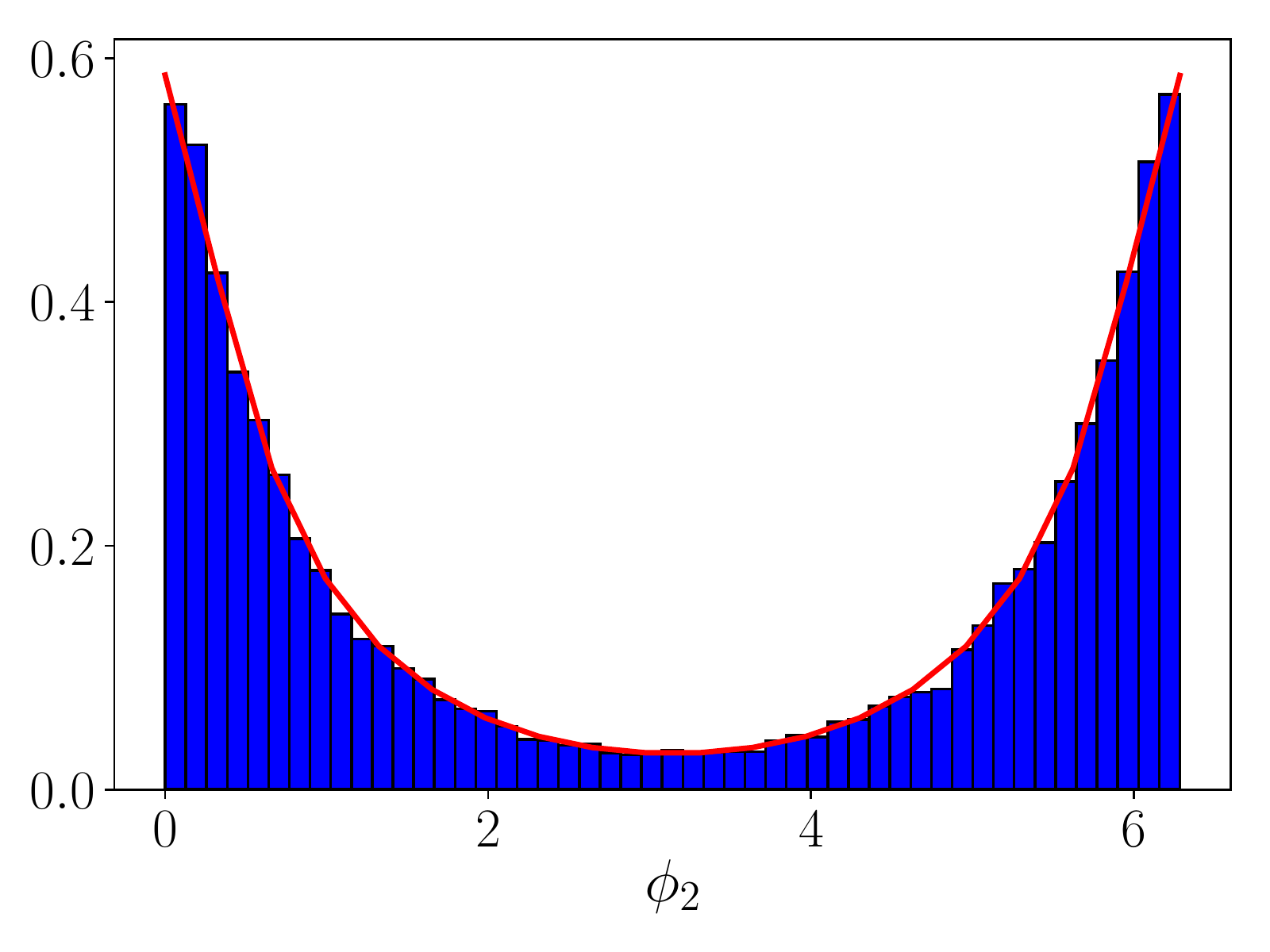}
        \caption{$\phi_2$}\label{}
    \end{subfigure}%
    \caption[Histograms of sampled momenta of the gluons compared with the marginal distributions when $v_\ma{rel}=0.1$ and $T=0.25$ GeV.]{Histograms of sampled momenta of the incoming and outgoing gluons (blue) compared with the marginal distributions (red) when $v_\ma{rel}=0.1$ and $T=0.25$ GeV. Normalization is unity.}
    \label{chap4_fig_recom_ineg_sample_compare1}
\end{figure}

\begin{figure}
    \centering
    \begin{subfigure}[b]{0.48\textwidth}
        \centering
        \includegraphics[height=2.1in]{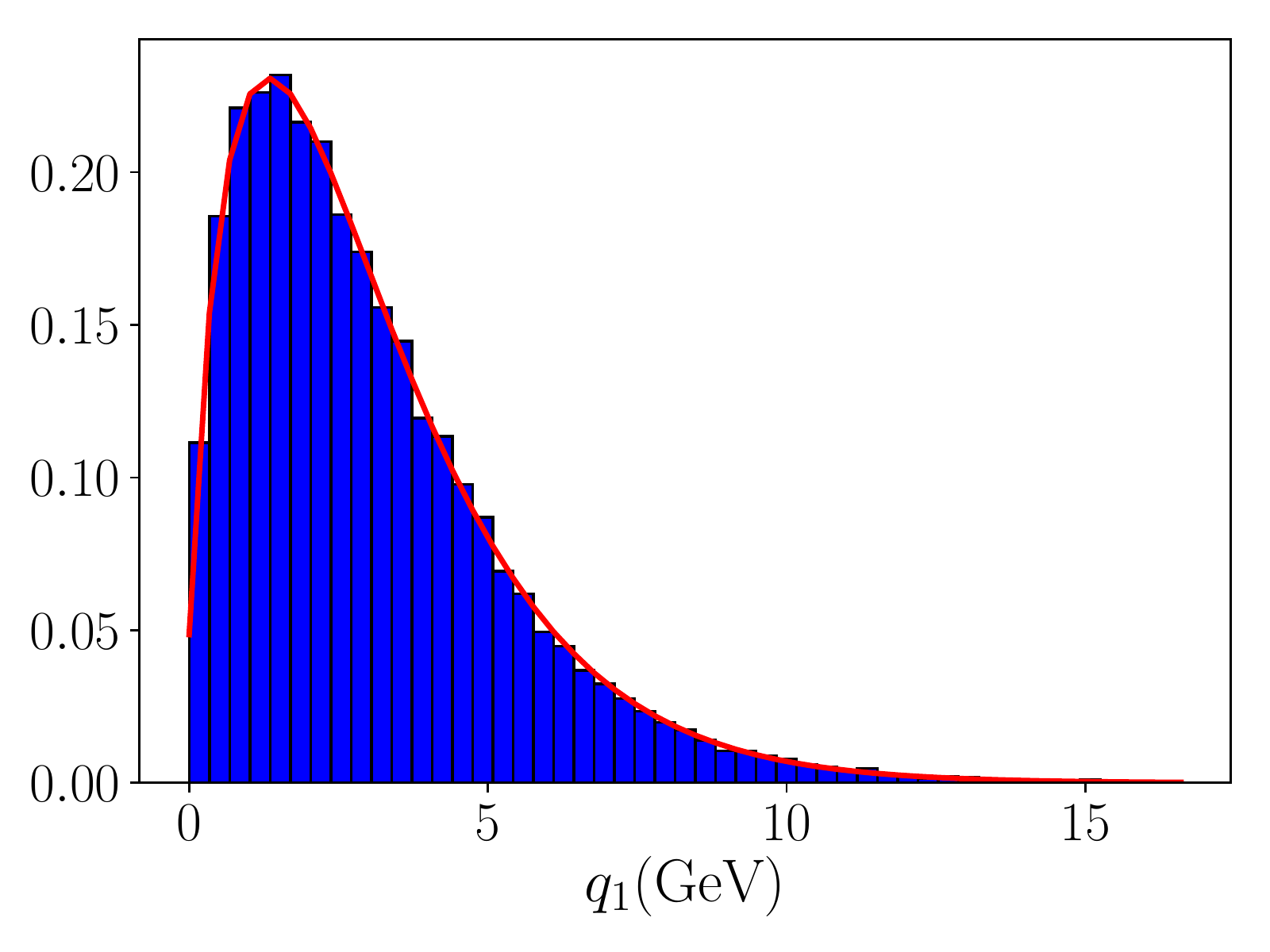}
        \caption{$q_1$}\label{}
    \end{subfigure}%
    ~
    \begin{subfigure}[b]{0.48\textwidth}
        \centering
        \includegraphics[height=2.1in]{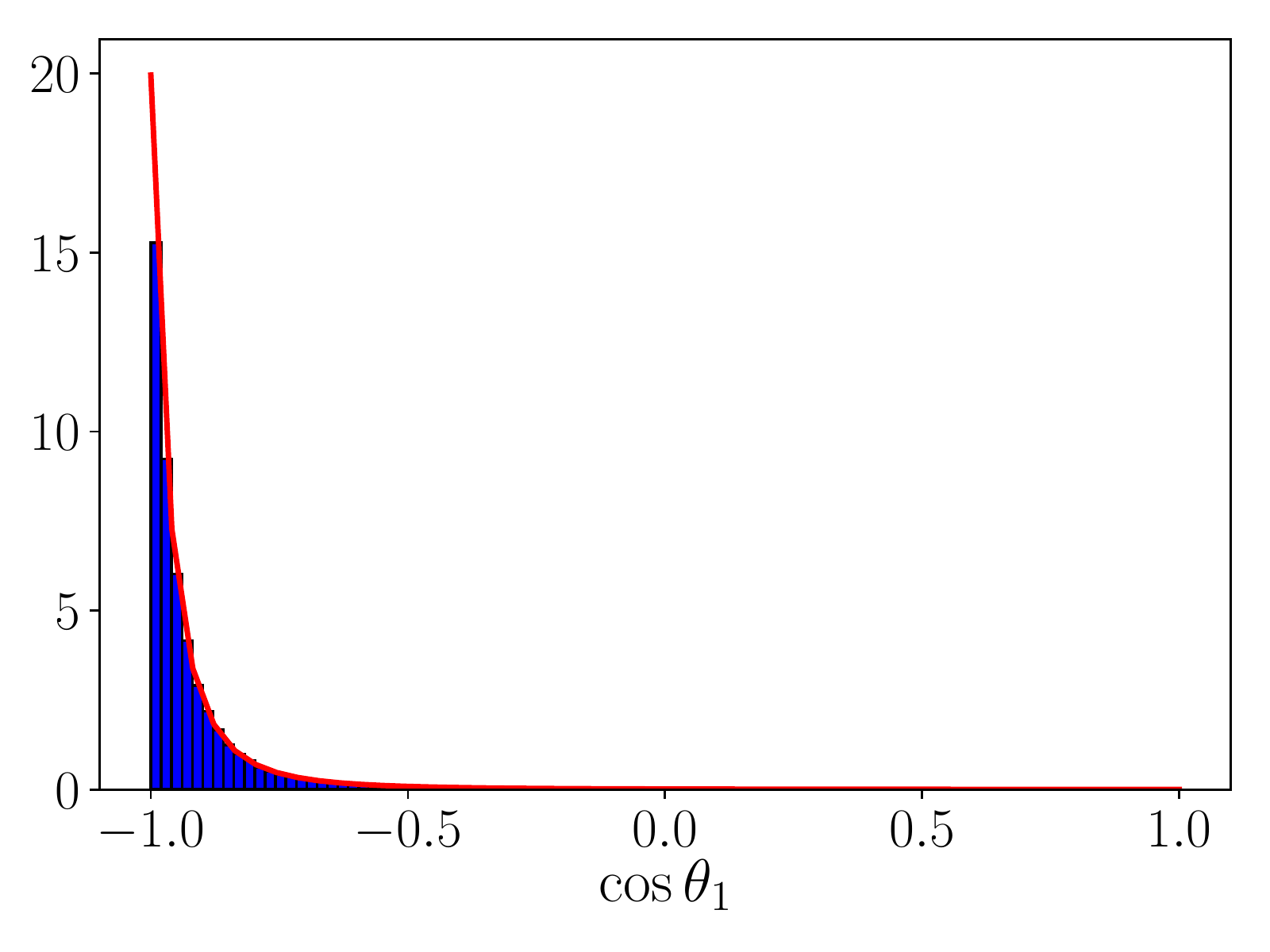}
        \caption{$c_1$}\label{}
    \end{subfigure}%
    
    \begin{subfigure}[b]{0.48\textwidth}
        \centering
        \includegraphics[height=2.1in]{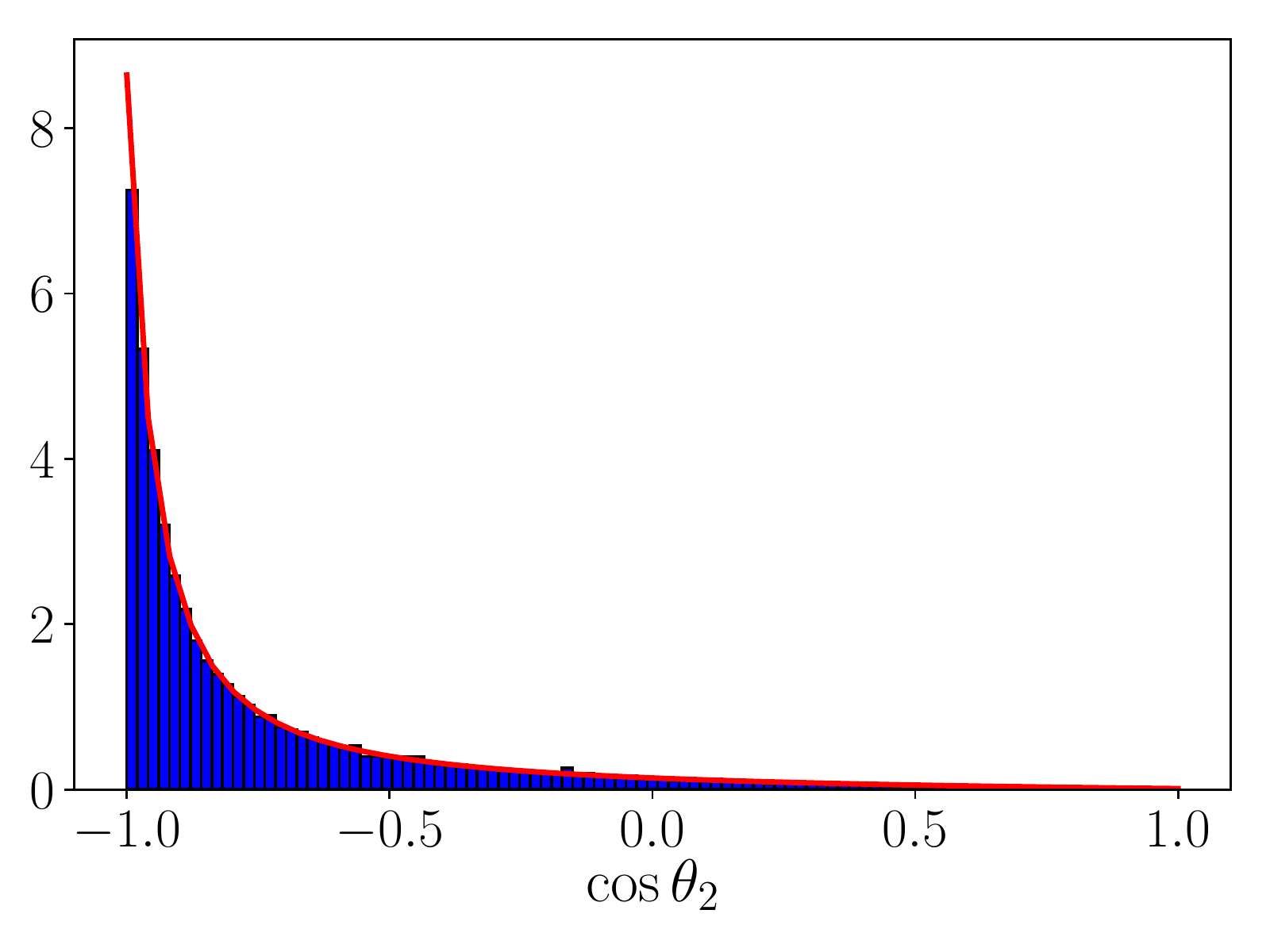}
        \caption{$c_2$}\label{}
    \end{subfigure}%
    ~
    \begin{subfigure}[b]{0.48\textwidth}
        \centering
        \includegraphics[height=2.1in]{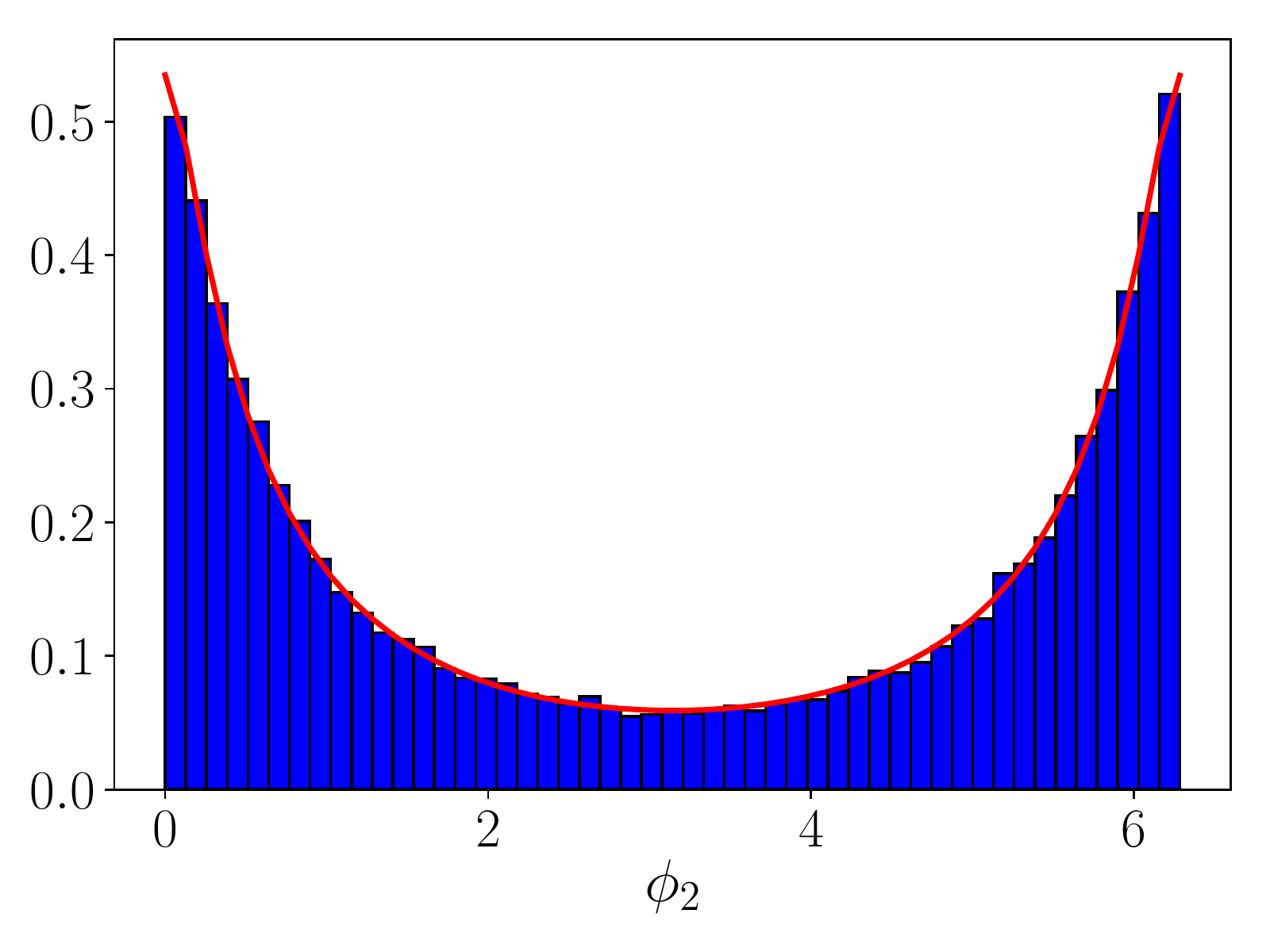}
        \caption{$c_2$}\label{}
    \end{subfigure}%
    \caption[Histograms of sampled momenta of the gluons compared with the marginal distributions when $v_\ma{rel}=0.9$ and $T=0.35$ GeV.]{Histograms of sampled momenta of the incoming and outgoing gluons (blue) compared with the marginal distributions (red) when $v_\ma{rel}=0.9$ and $T=0.35$ GeV. Normalization is unity.}
    \label{chap4_fig_recom_ineg_sample_compare2}
\end{figure}

We test the momentum sampling by sampling a large number of events and comparing with the marginal distributions. We assume $\alpha_s=0.3$, $M=4.65$ GeV and ${\bs p}_\ma{rel}=0.8$ GeV. The marginal distribution in a variable $X$ is obtained by integrating the integrand in Eq.~(\ref{chap4_eqn_recom_ineg_sample}) over all the variables except the $X$. Here $X$ can be $q_1$, $c_1$, $c_2$ and $\phi_2$. The comparisons are shown in Figs.~\ref{chap4_fig_recom_ineg_sample_compare1} and~\ref{chap4_fig_recom_ineg_sample_compare2} for two cases: $v_\ma{cell} = 0.1$, $T=0.25$ GeV and $v_\ma{cell} = 0.9$, $T=0.35$ GeV. We sampled $30000$ recombination events for each case and plot the histograms in blue. It can be seen that the sampled distributions agree well with the marginal distributions drawn in red.

\vspace{0.2in}

\section{Detailed Balance and Thermalization}
Before we move on to the phenomenological studies of quarkonium production in heavy ion collisions, we test the proposed coupled Boltzmann equations in a cubic box of a static QGP. The medium has a constant temperature and no flows. We will apply a periodic boundary condition to the sampled particles, i.e., when they just cross one facet of the cubic box, they will appear on the other side of the box. We must implement the boundary condition correctly when searching pairs of $Q\bar{Q}$ in the simulation of recombination. In the searching algorithm, we can make a bigger cubic box by attaching the same box along the faces. The bigger box can be divided into $3\times3\times3$ cubic boxes each of which is the same as the cubic box we start with. Then for each $Q$ in the central small cubic box, we will search for $\bar{Q}$'s in the bigger cubic box that is within a certain radius $R$. The contribution to recombination of the $Q$ from the outside is negligible because $R\gg a_B$, where the Bohr radius $a_B$ is the Gaussian width used in the calculation of recombination rates (\ref{chap4_eqn_reco_lab_g}), (\ref{chap4_eqn_reco_lab_ineq}) and (\ref{chap4_eqn_reco_lab_ineg}). There is no double counting if we choose $2R<L$ where $L$ is the side length of the box. We choose $L=10$ fm for the QGP box.

We will focus our studies on the bottom quark and bottomonium. Because they have a larger mass than the charm quark and charmonium, the assumed separation of scales works better. We will assume the bottom quark mass is $M=4.65$ GeV and the coupling constant is $\alpha_s=0.3$ throughout this section. We will test the numerical implementation from three perspectives: dissociation, recombination and their interplays.

\begin{figure}
    \centering
    \begin{subfigure}[t]{0.48\textwidth}
        \centering
        \includegraphics[height=2.1in]{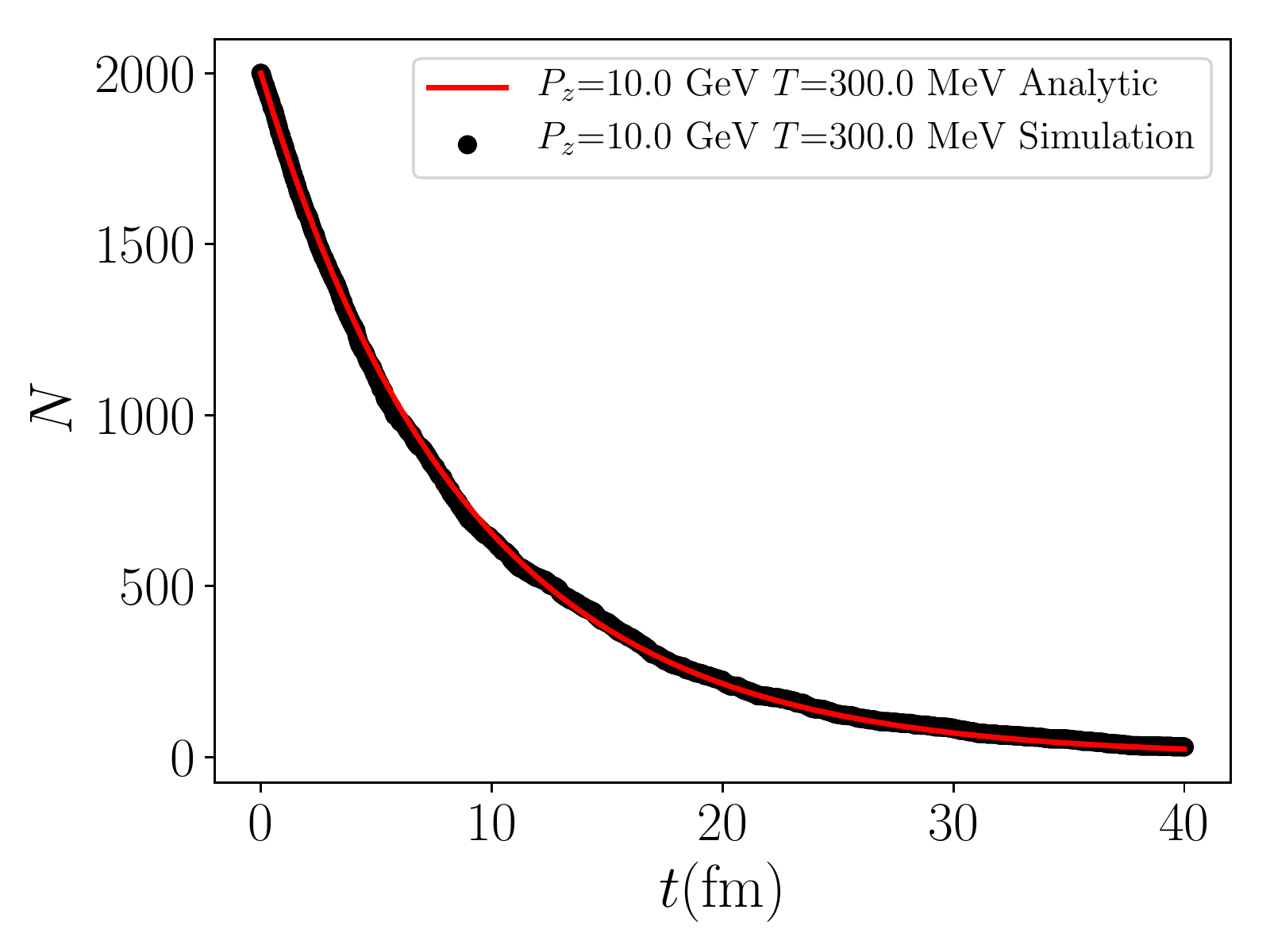}
        \caption{Gluon absorption when $T=300$ MeV, ${\bs P} =  P_z \hat{z} $ and $P_z=10$ GeV.}\label{}
    \end{subfigure}%
    ~
    \begin{subfigure}[t]{0.48\textwidth}
        \centering
        \includegraphics[height=2.1in]{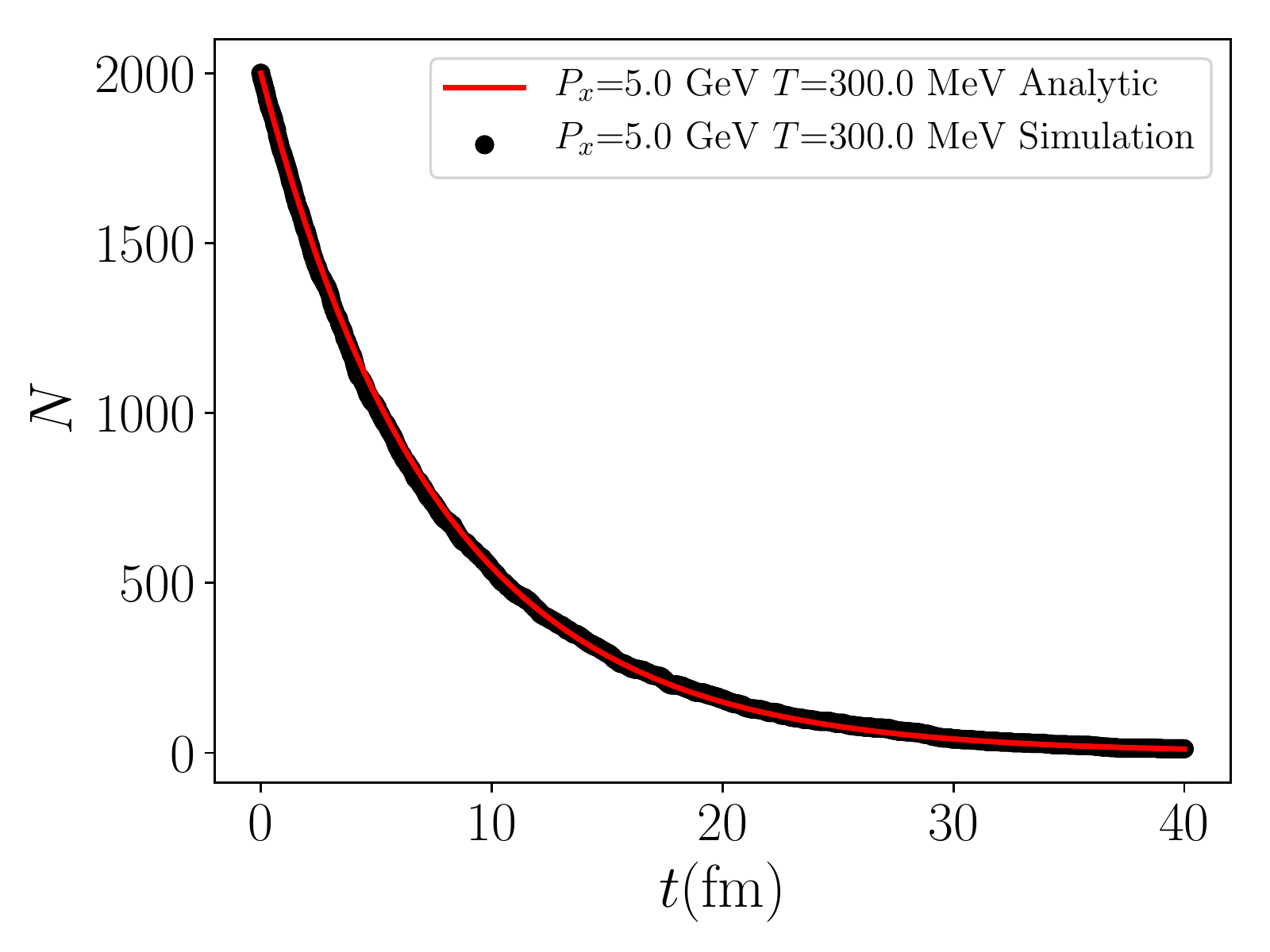}
        \caption{Inelastic scattering with light quarks when $T=300$ MeV, ${\bs P} =  P_x \hat{x} $ and $P_x=5$ GeV.}\label{}
    \end{subfigure}%
    
    \begin{subfigure}[t]{0.48\textwidth}
        \centering
        \includegraphics[height=2.1in]{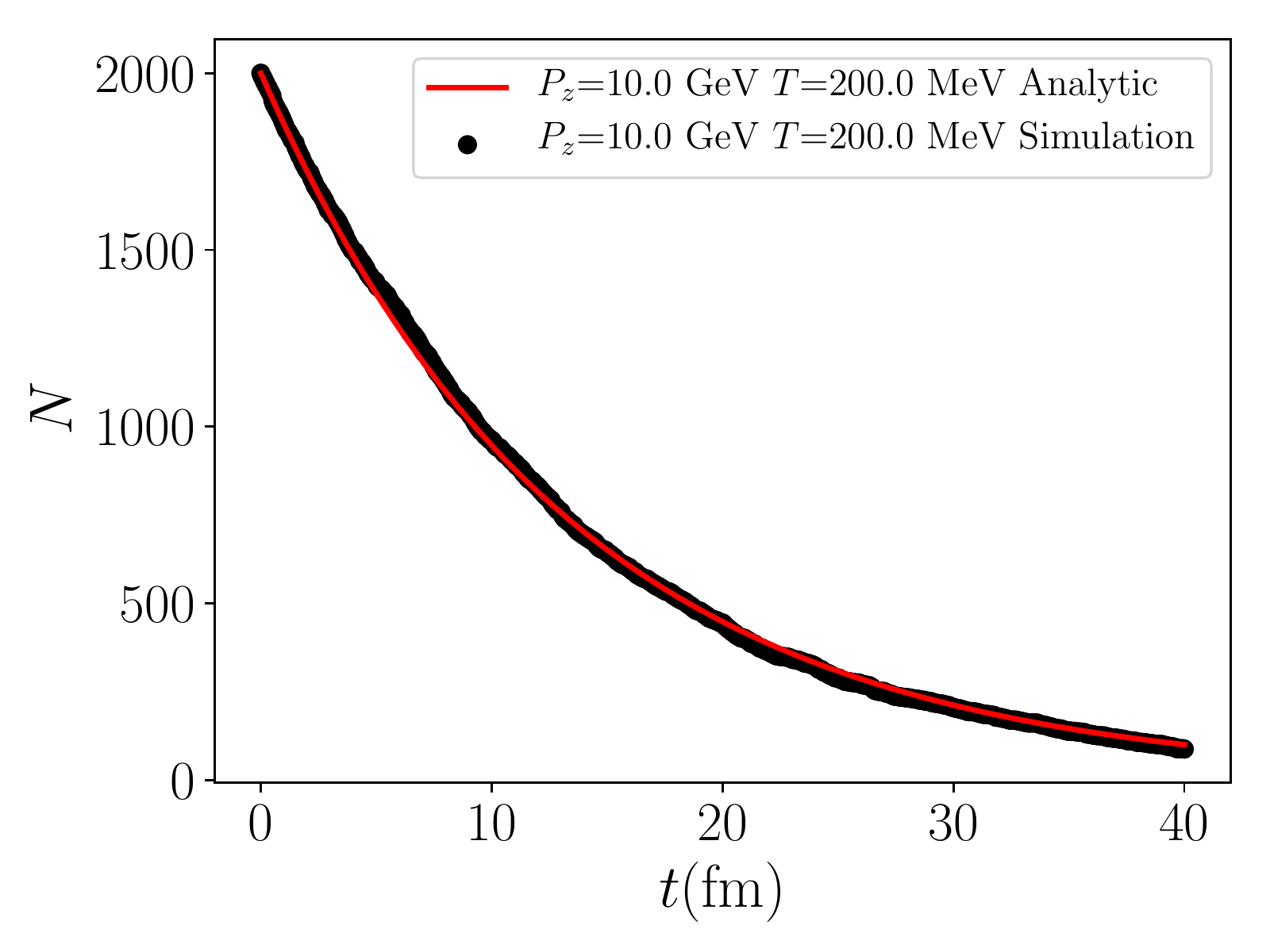}
        \caption{Inelastic scattering with gluons when $T=200$ MeV, ${\bs P} =  P_z \hat{z} $ and $P_z=10$ GeV.}\label{}
    \end{subfigure}%
    ~
    \begin{subfigure}[t]{0.48\textwidth}
        \centering
        \includegraphics[height=2.1in]{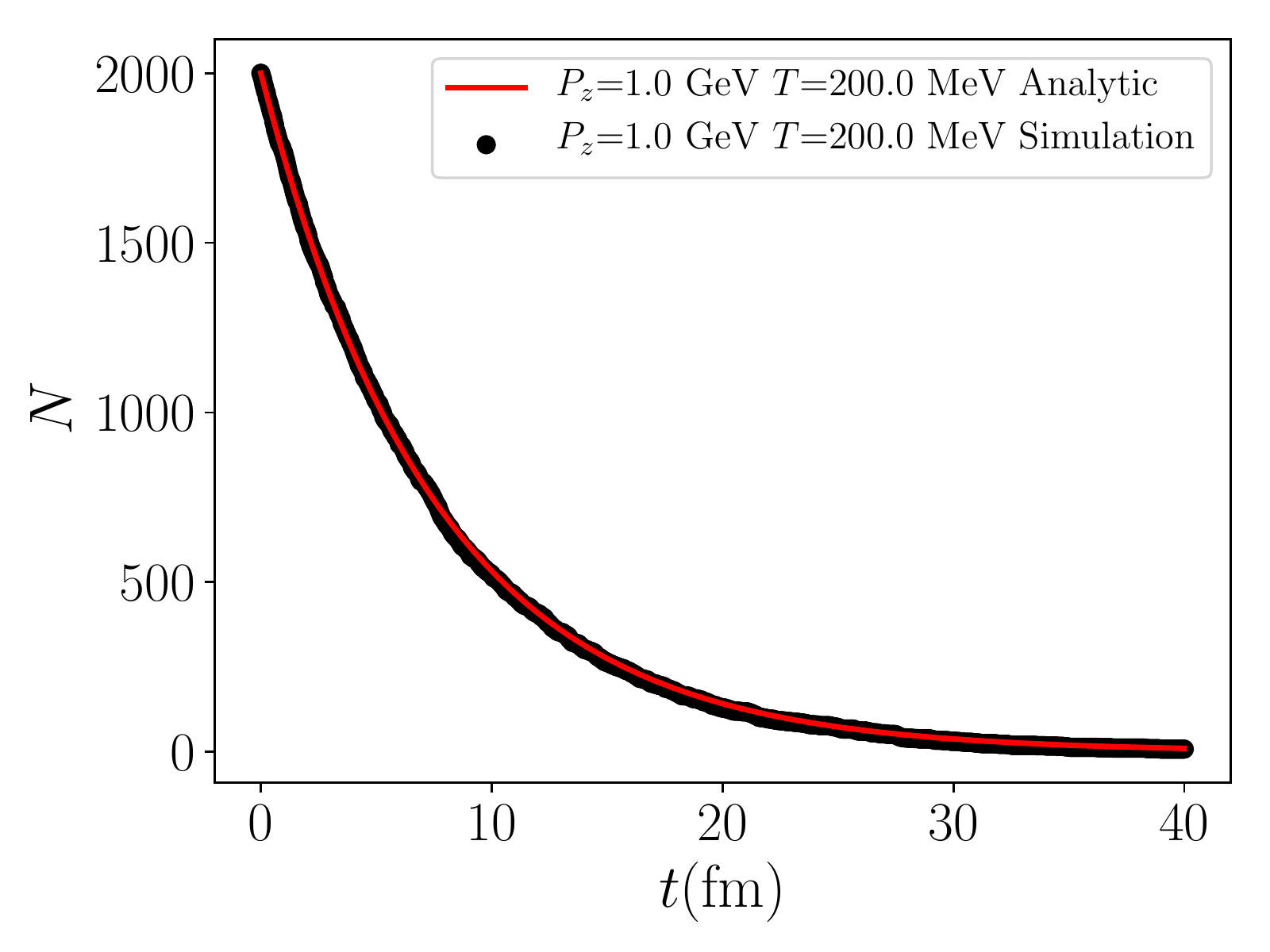}
        \caption{All three processes when $T=200$ MeV, ${\bs P} =  P_z \hat{z} $ and $P_z=1$ GeV.}\label{}
    \end{subfigure}%
    \caption[Comparison between the Monte Carlo simulation and the analytic expression for different dissociation channels at different $v$ and $T$.]{Comparison between the Monte Carlo simulation and the analytic expression (\ref{chap4_eqn_analytic_decayN}) for different dissociation channels at different $v$ and $T$.}
    \label{chap4_fig_box_disso}
\end{figure}

\vspace{0.2in}
\subsection{Dissociation Test}
The first thing we test is the dissociation in our numerical implementation. We sample $N(t=0) = 2000$ particles of $\Upsilon$(1S) inside the QGP box. Their positions are randomly sampled. Their momenta are the same constant ${\bs P}$. Then we keep track of the evolution of the number of $\Upsilon$(1S) and compare with the analytic expression
\be
\label{chap4_eqn_analytic_decayN}
N(t) = N(0) e^{-\Gamma^d t}\,,
\ee
where $\Gamma^d = \Gamma^d(v, T)$ is the dissociation rate calculated in the last section. Here $v$ is the speed of $\Upsilon$(1S) with respect to the QGP box and $T$ is the temperature of the QGP box. The comparisons between the Monte Carlo simulation and the analytic expression (\ref{chap4_eqn_analytic_decayN}) for different dissociation channels at different $v$ and $T$ are shown in Fig.~\ref{chap4_fig_box_disso}. It can be seen that the numerical implementation of the dissociation rate is consistent with the analytic calculation.

\vspace{0.2in}
\subsection{Recombination Test}
We next test the recombination implementation. We sample $N_b=50$ bottom quarks and $N_{\bar{b}}=50$ antibottom quarks inside the QGP box. Their positions inside the box are randomly sampled. Their momenta are sampled from the Boltzmann thermal distribution
\be
\label{chap4_eqn_boltzmann_distribution}
f_B({\bs p}) \equiv C e^{-\sqrt{M^2+{\bs p}^2}/T} \,,
\ee
where $C$ is the normalization constant for the condition $\int \frac{\diff^3 p}{(2\pi)^3} f_B({\bs p}) =1$.
We will study the process of forming $\Upsilon$(1S) via recombination.

On one hand, we can calculate the average recombination rate of bottom quarks via
\be
\label{chap4_eqn_average_recom}
\langle \Gamma^r \rangle &\equiv& \frac{ g_+ \int  \frac{\diff^3 p_b}{(2\pi)^3} \frac{\diff^3 p_{\bar{b}}}{(2\pi)^3} f_b({\bs x}_b, {\bs p}_b, t) f_{\bar{b}}({\bs x}_{\bar{b}}, {\bs p}_{\bar{b}}, t) (A_g + A_{\ma{inel},q} + A_{\ma{inel},g}) }{ \int  \frac{\diff^3 p_b}{(2\pi)^3}f_b({\bs x}_b, {\bs p}_b, t) } \\
A_g  &=&   \frac{8}{9}\alpha_s q^3 \Big( 2 + \frac{T}{\gamma v q}   \ln{\frac{1-e^{-\gamma (1+v)q/T}}{ 1-e^{-\gamma(1-v)q/T} }}  \Big) | \langle \Psi_{  {\bs p}_\ma{rel}  } | {\bs r} | \psi_{nl} \rangle |^2   \Big|_{q = |E_{nl}| + \frac{ ({\bs p}_\ma{rel})^2}{M}}  \\ \nn
A_{\ma{inel},q} &=&  \frac{8\alpha_s^2}{9\pi^2}    \int p_1 \diff p_1 \diff c_1 n_F(\gamma(1+vc_1)p_1) \int p_2 \diff c_2 \diff \phi_2 
\big[  1 - n_F(\gamma(1+vc_2)p_2)  \big] \\ 
&&  \frac{p_1p_2(1+s_1s_2\cos\phi_2+c_1c_2)}{p_1^2 + p_2^2 - 2p_1p_2(s_1s_2\cos\phi_2 + c_1c_2)} |\langle \Psi_{{\bs p}_\ma{rel}} | {\bs r} |  \psi_{nl}  \rangle|^2  \\ \nn
A_{\ma{inel},g} &=&  \frac{\alpha_s^2}{3\pi^2}    \int q_1 \diff q_1 \diff c_1 n_B(\gamma(1+vc_1)q_1) \int q_2 \diff c_2 \diff \phi_2 \big[  1 + n_B(\gamma(1+vc_2)q_2)  \big]  \\
&& \frac{ (q_1+q_2)^2(1+s_1s_2\cos\phi_2 +c_1c_2) }{q_1^2 + q_2^2 - 2q_1q_2(s_1s_2\cos\phi_2 + c_1c_2)} | \langle \Psi_{{\bs p}_\ma{rel}} | {\bs r} |  \psi_{nl}  \rangle|^2  \,,
\ee
where the $A$'s are taken from the recombination rate defined in expressions (\ref{chap4_eqn_reco_lab_g}), (\ref{chap4_eqn_reco_lab_ineq}) and (\ref{chap4_eqn_reco_lab_ineg}). $v$ is the velocity of a $b\bar{b}$ pair relative to the QGP box and $\gamma = 1/\sqrt{1-v^2}$. The relative momentum ${\bs p}_\ma{rel}$ between the $b\bar{b}$ pair is calculated in rest frame of their c.m.~motion, as in Eq.~(\ref{chap4_eqn_p_relative}). The distribution functions of the bottom and antibottom quarks are
\be
f_b({\bs x}_b, {\bs p}_b, t) &=& \frac{N_b}{L^3} f_B({\bs p}_b)   \\
f_{\bar{b}}({\bs x}_{\bar{b}}, {\bs p}_{\bar{b}}, t) &=& \frac{ N_{\bar{b}} }{L^3} f_B({\bs p}_{\bar{b}})   \,.
\ee
We will use the Monte Carlo method to numerically integrate the expression, using the package ``Vegas". 

\begin{figure}
    \centering
    \begin{subfigure}[t]{0.48\textwidth}
        \centering
        \includegraphics[height=2.1in]{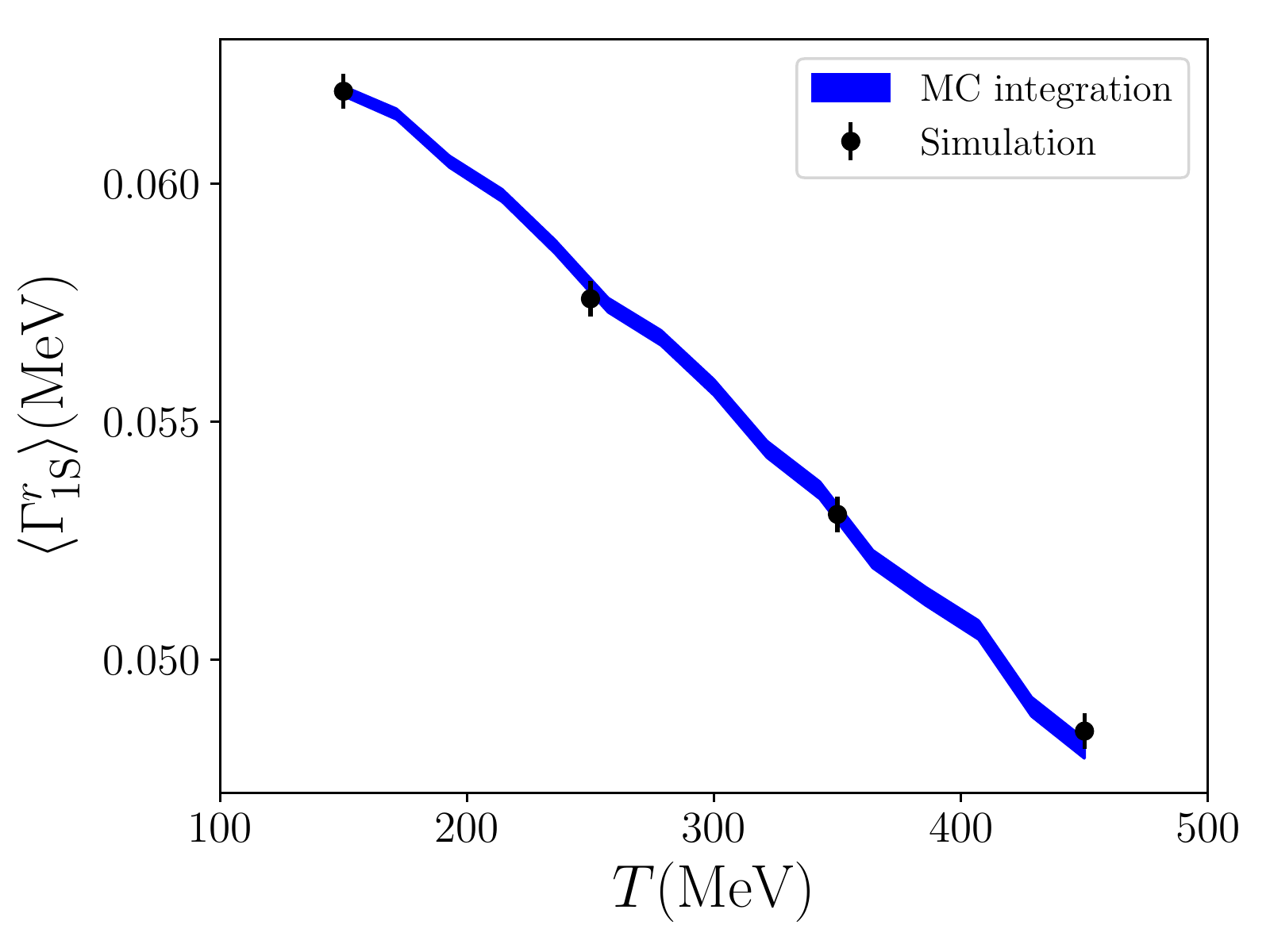}
        \caption{Gluon radiation.}\label{}
    \end{subfigure}%
    ~
    \begin{subfigure}[t]{0.48\textwidth}
        \centering
        \includegraphics[height=2.1in]{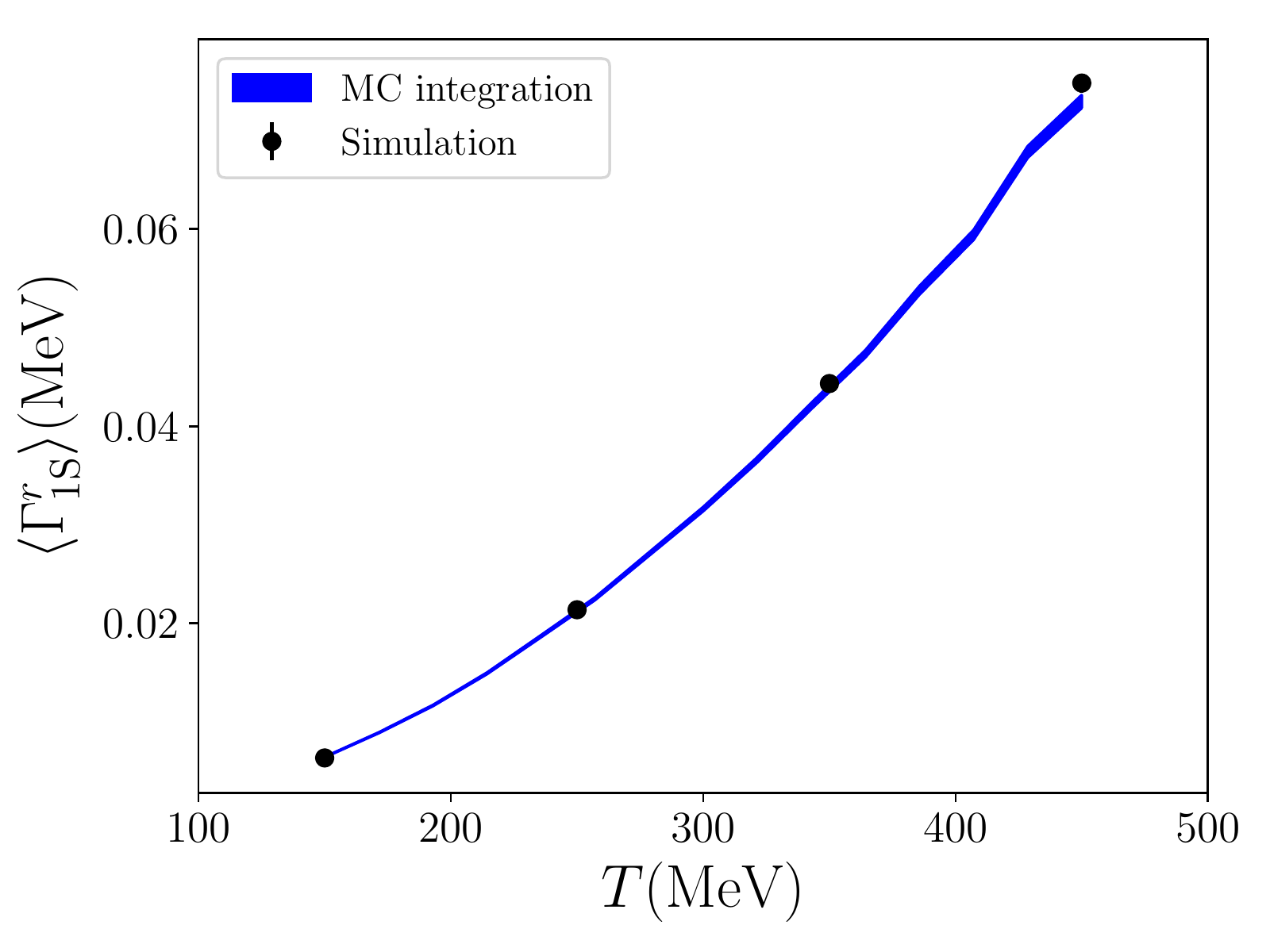}
        \caption{Inelastic scattering with light quarks.}\label{}
    \end{subfigure}%
    
    \begin{subfigure}[t]{0.48\textwidth}
        \centering
        \includegraphics[height=2.1in]{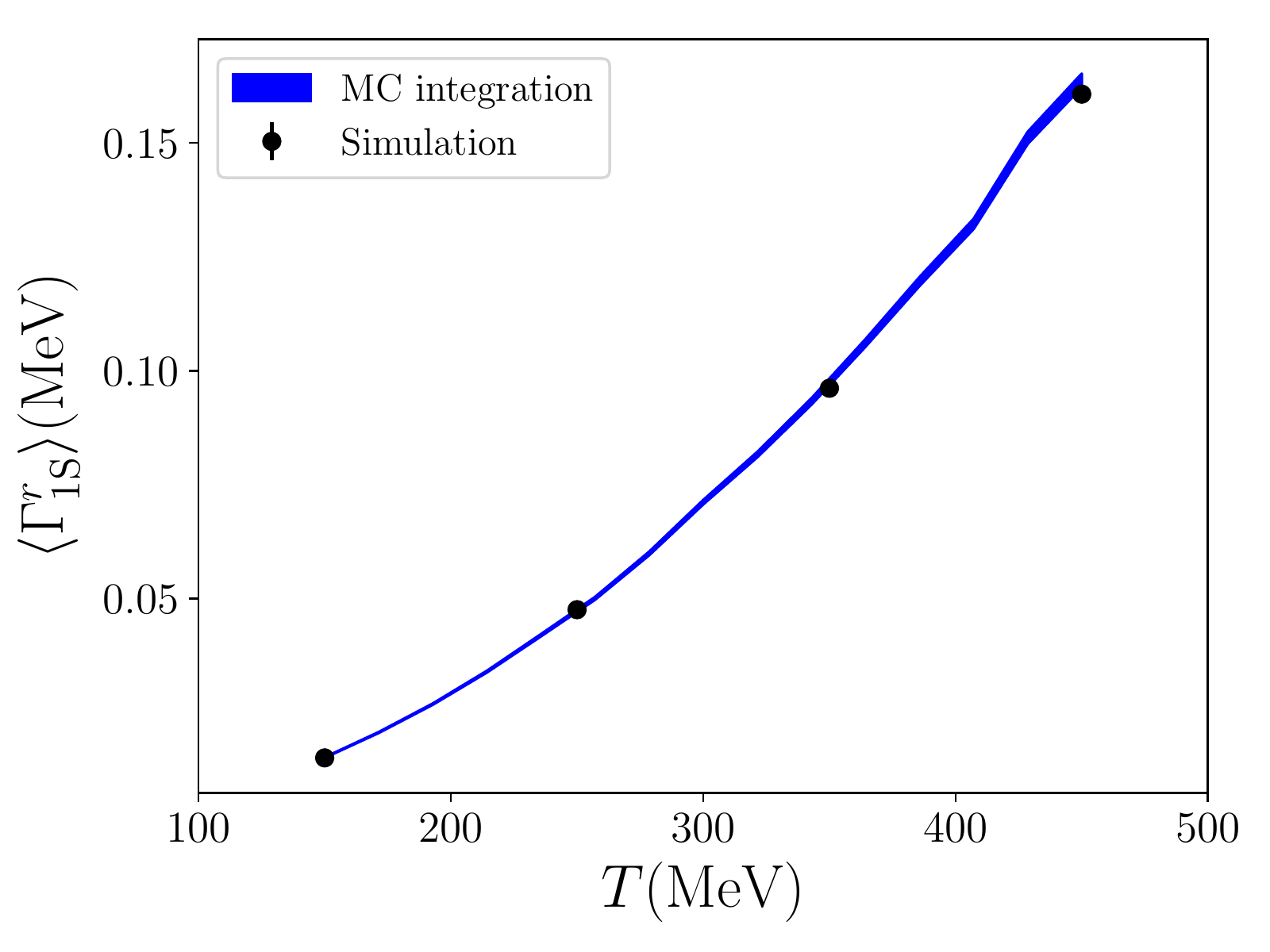}
        \caption{Inelastic scattering with gluons.}\label{}
    \end{subfigure}%
    ~         
    \caption[Comparison between the numerical simulation and the analytic expression for different recombination channels.]{Comparison between the numerical simulation and the analytic expression (\ref{chap4_eqn_average_recom}) computed using the Monte Carlo integration method for different recombination channels.}
    \label{chap4_fig_box_recom}
\end{figure}

On the other hand, we can estimate the average recombination rate in the Monte Carlo simulation. For each bottom quark inside the box, we search for $\bar{b}$ within a radius equal to $R=5a_B$. For each $\bar{b}$ within, our numerical implementation can give the recombination rate of this $b\bar{b}$ pair. If we sum the rate over all ${\bar{b}}$'s that are within the search radius, we get the recombination rate of that specific bottom quark. If we average over all bottom quarks, we obtain the average recombination rate of bottom quarks. We sample $30000$ events to estimate the average recombination rate of bottom quarks. The comparison between the numerical simulation and the analytic expression (\ref{chap4_eqn_average_recom}) computed using the Monte Carlo integration method for different recombination channels is shown in Fig.~\ref{chap4_fig_box_recom}. The band in the analytic expression is due to the statistical uncertainty when we numerically integrate the expression. The statistical uncertainty is also shown for the Monte Carlo simulations.

\vspace{0.2in}
\subsection{Interplay between Dissociation and Recombination}
Finally we test the interplay between the dissociation and recombination. We will sample events in several different cases:
\begin{enumerate}
\item We sample a certain number of bottom and antibottom quarks and no $\Upsilon$(1S) states. Their positions are randomly sampled inside the box while their momenta are sampled from the Boltzmann thermal distribution. We turn off the transport of open heavy flavors.
\item We sample a certain number of bottom and antibottom quarks and no $\Upsilon$(1S) states. Their positions are randomly sampled inside the box while their momenta are sampled from a uniform distribution
\be
\label{chap4_eqn_uniform}
f_\ma{uniform} ({\bs p}) = \ma{const.} \ \ \ \ \ \ \ma{if}\ \ 0 \leq p_x,p_y,p_z \leq P_{\max} \,.
\ee
We turn off the transport of open heavy flavors so that the momentum spectra of open bottom and antibottom quarks will not thermalize.
\item The same as in case 2 except that we turn on the transport of open heavy flavors so that the momentum spectra of open bottom and antibottom quarks will thermalize eventually.
\item We sample a certain number of $\Upsilon$(1S) states and no bottom and antibottom quarks. Their positions are randomly sampled inside the box while their momenta are sampled from a uniform distribution (\ref{chap4_eqn_uniform}). We turn on the transport of open heavy flavors.
\end{enumerate}

We will simulate $N_\ma{event}=10000$ events in each case. In each event simulation, we keep track of how the hidden bottom flavor percentage changes with time. The hidden bottom flavor percentage is defined as
\be
\frac{N_{b,\ma{hidden}}}{N_{b,\ma{tot}}} = \frac{N_{\Upsilon \ma(1S)}}{N_b + N_{\Upsilon \ma(1S)}} \,.
\ee
The total number of bottom flavor $N_{b,\ma{tot}}$ is equal to the sum of the number of open bottom quarks and the number of $\Upsilon \ma(1S)$, because $\Upsilon \ma(1S)$ contains one bottom flavor and one antibottom flavor. 
We will compare the simulation results with the hidden bottom flavor percentage at thermal equilibrium. At equilibrium
\be
\label{chap4_eqn_Neq}
N^\ma{eq}_{i}=g_i V \int\frac{\diff^3p}{(2\pi)^3}\lambda_{i}e^{-E_i(p)/T}\,,
\ee 
with $V$ is the volume of the QGP box. Relativistically $E_i(p)=\sqrt{M_i^2+p^2}$ and nonrelativistically $M_i+\frac{p^2}{2M_i}$ for $i=b,\bar{b}$ or $\Upsilon$(1S). The degeneracy factors are $g_b=g_{\bar{b}}=6$ for spin and color and $g_{\Upsilon\ma{(1S)}}=3$ (because the hyperfine splitting between $\eta_b$ and $\Upsilon$(1S) has been considered in the recombination rate, see the definition of $g_+$). The fugacities are related by $\lambda_{\Upsilon}=\lambda_{b}\lambda_{\bar{b}}=\lambda_{b}^2$ and solved from $N^\ma{eq}_{b}+N^\ma{eq}_{\Upsilon}=N_{b,\ma{tot}}$. Once we determine the fugacities, we can compute the hidden bottom flavor percentage at equilibrium. 

\begin{figure}
    \centering
    \begin{subfigure}[t]{0.48\textwidth}
        \centering
        \includegraphics[height=2.1in]{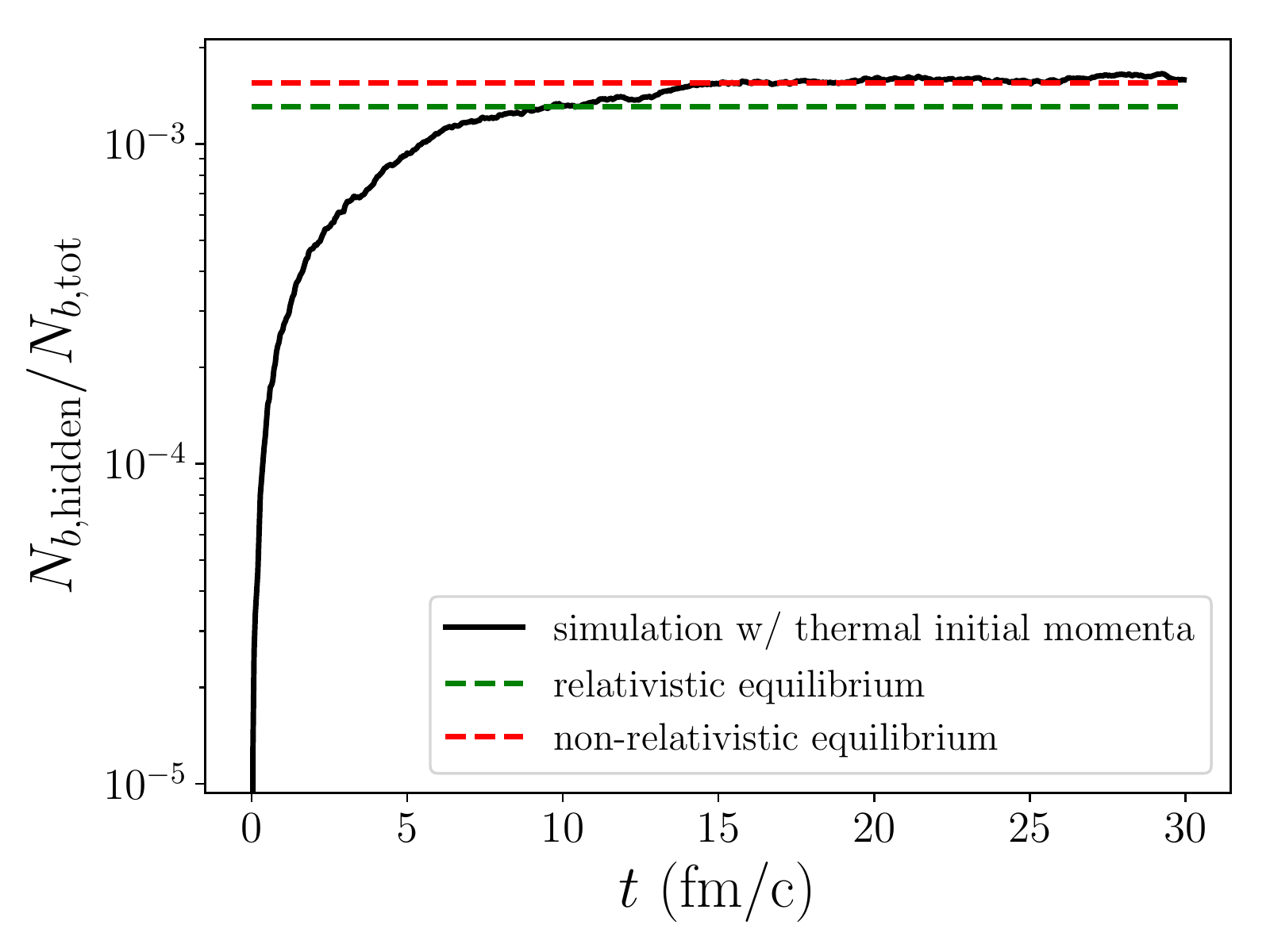}
        \caption{Only gluon absorption and radiation turned on at $T=0.3$ GeV.}\label{}
    \end{subfigure}%
    ~
    \begin{subfigure}[t]{0.48\textwidth}
        \centering
        \includegraphics[height=2.1in]{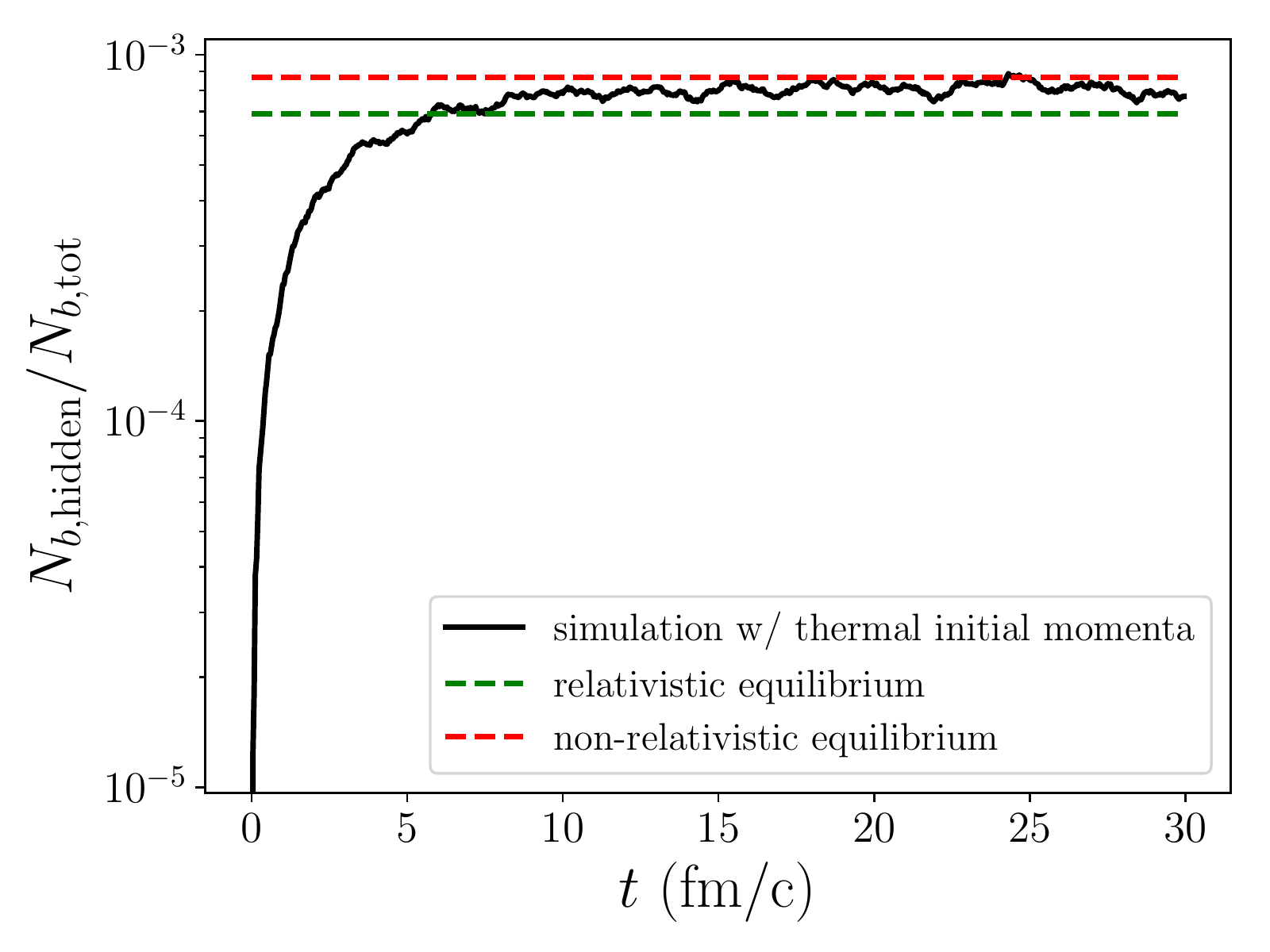}
        \caption{Only inelastic scattering with light quarks turned on at $T=0.4$ GeV.}\label{}
    \end{subfigure}%
    
    \begin{subfigure}[t]{0.48\textwidth}
        \centering
        \includegraphics[height=2.1in]{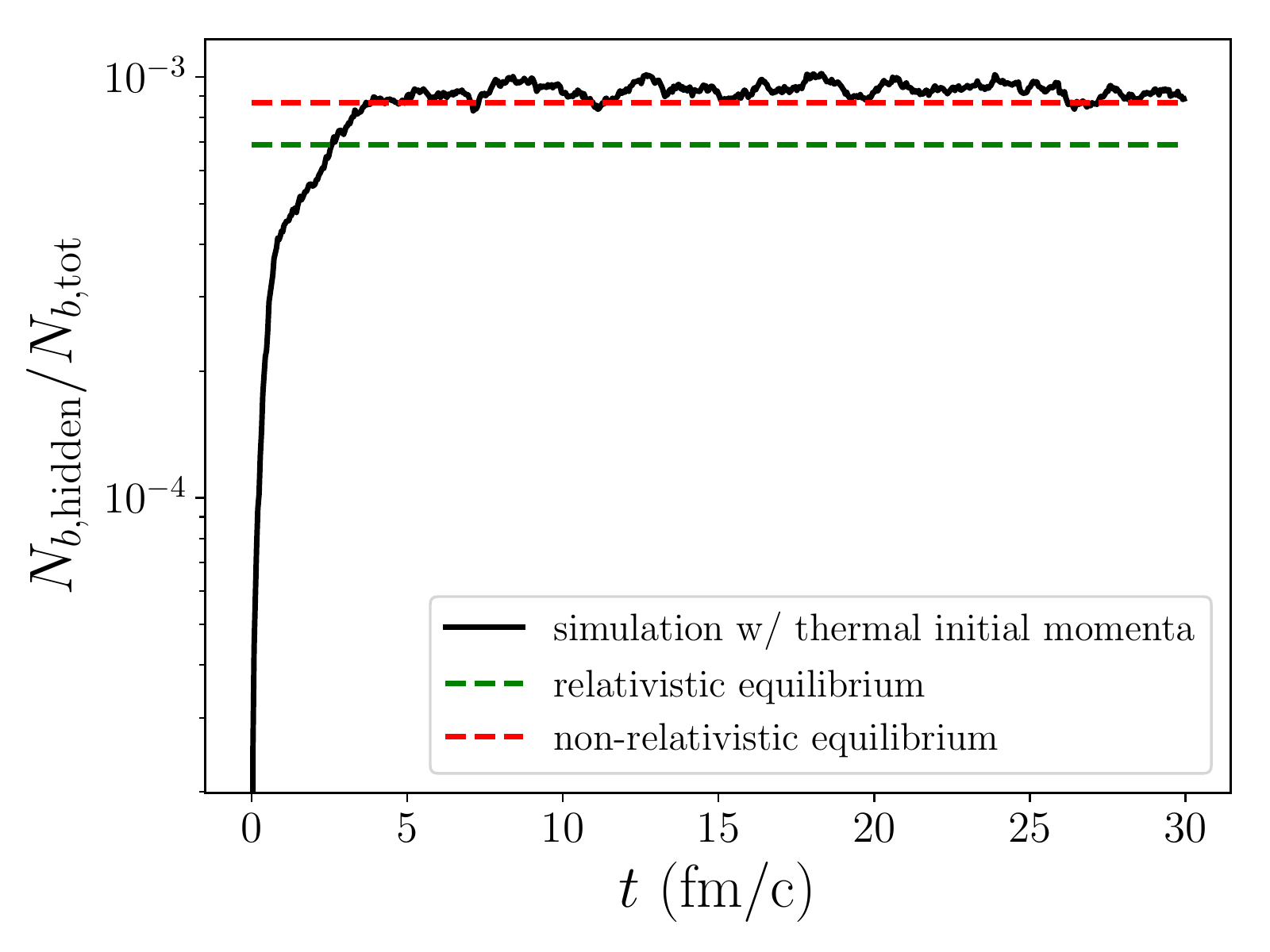}
        \caption{Only inelastic scattering with gluons turned on at $T=0.4$ GeV.}\label{}
    \end{subfigure}%
    ~
    \begin{subfigure}[t]{0.48\textwidth}
        \centering
        \includegraphics[height=2.1in]{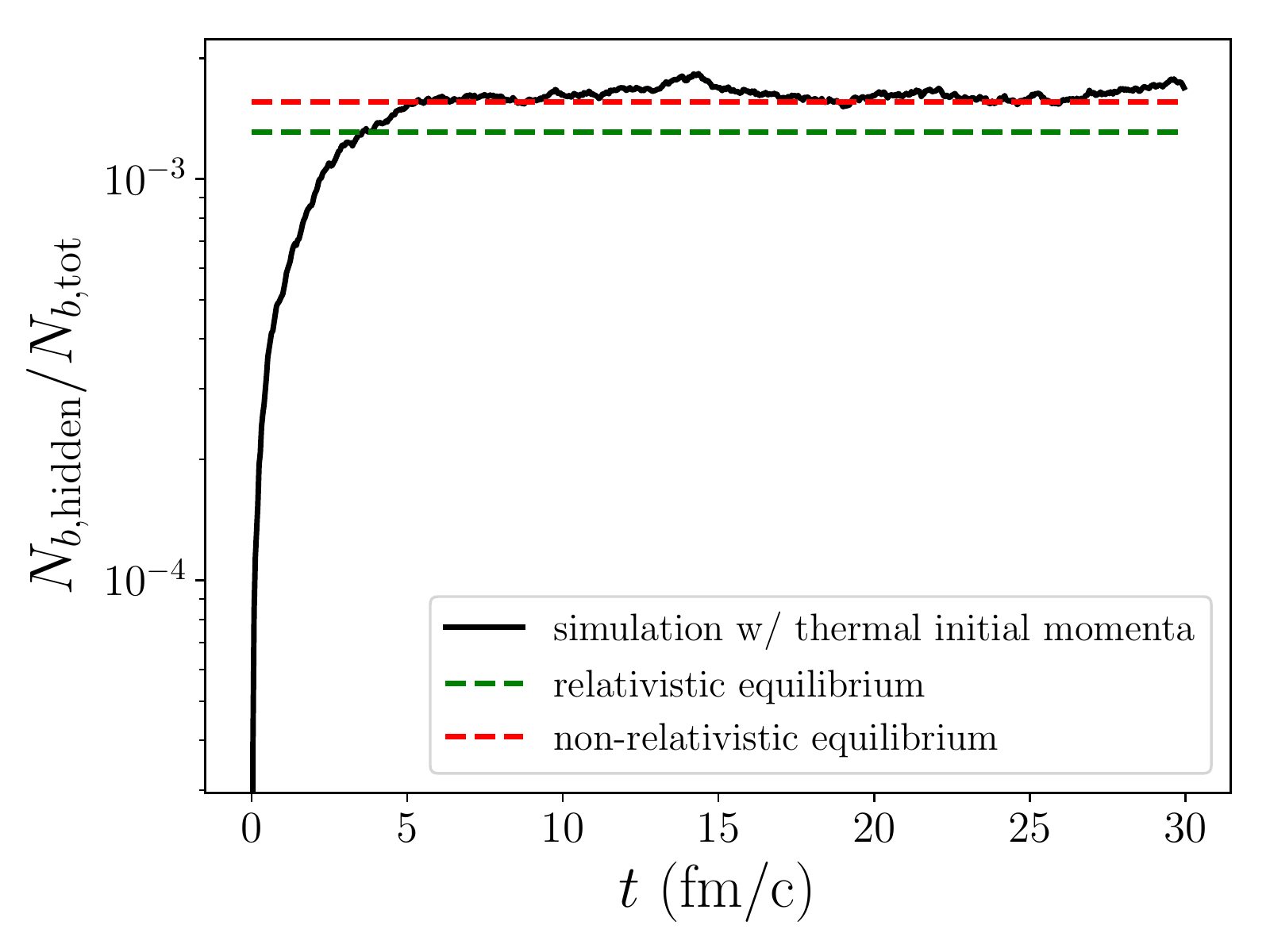}
        \caption{All three processes turned on at $T=0.3$ GeV.}\label{}
    \end{subfigure}%
                
    \caption{Comparison of the hidden bottom flavor percentage calculated via the numerical simulation in case 1 and that at equilibrium.}
    \label{chap4_fig_case1}
\end{figure}

The simulation results of the hidden bottom flavor percentage in case 1 and the comparison with that at thermal equilibrium are shown in Fig.~\ref{chap4_fig_case1}. We show the results for four different scenarios: only gluon absorption and radiation turned, only inelastic scattering with light quarks turned on, only inelastic scattering with gluons turned on and all three processes turned on. We can see from the plots that the interplay between dissociation and recombination drives the system to detailed balance.  At detailed balance, the number of quarkonia dissociating in a time step is equal to the number of quarkonia generated from recombination. We see that the detailed balance can be reached via each scattering process. At detailed balance, the hidden bottom flavor percentage agrees with that at thermal equilibrium, as expected because the initial momenta of all the particles are sampled from a thermal distribution. The agreement is achieved when we use the nonrelativistic dispersion relation in Eq.~(\ref{chap4_eqn_Neq}). This is because the scattering amplitudes are calculated in a nonrelativistic theory. The energy-momentum conservation in the calculation has the nonrelativistic dispersion relation.

\begin{figure}
    \centering
    \begin{subfigure}[t]{0.48\textwidth}
        \centering
        \includegraphics[height=2.1in]{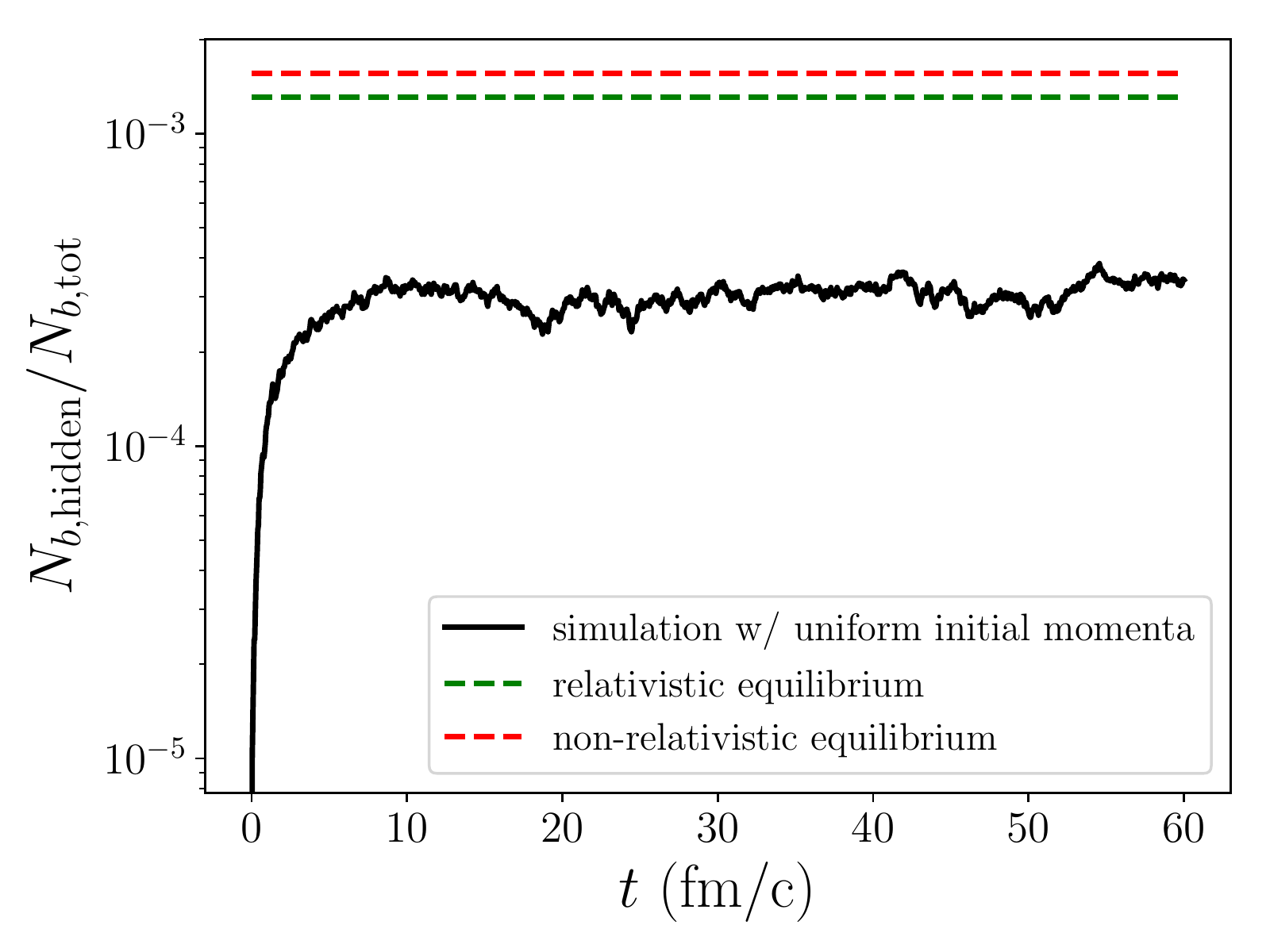}
        \caption{Simulation result in case 2 at $T=0.3$ GeV.}\label{}
    \end{subfigure}%
    ~
    \begin{subfigure}[t]{0.48\textwidth}
        \centering
        \includegraphics[height=2.1in]{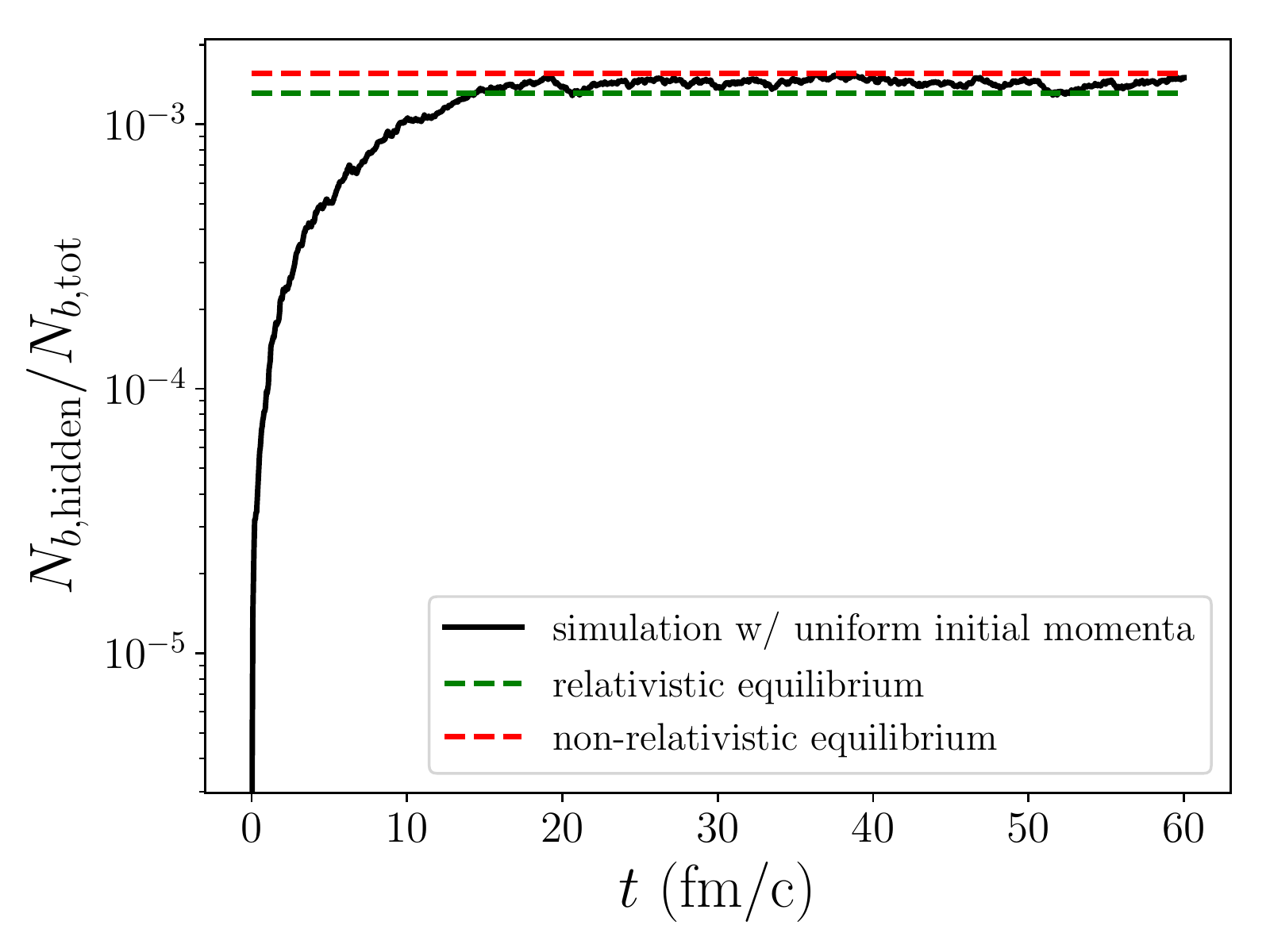}
        \caption{Simulation result in case 3 at $T=0.3$ GeV.}\label{}
    \end{subfigure}%
                
    \caption{Hidden bottom flavor percentage calculated via the numerical simulation in case 2 and 3.}
    \label{chap4_fig_case23}
\end{figure}

The simulation results of the hidden bottom flavor percentage in case 2 and 3 are shown in Fig.~\ref{chap4_fig_case23}. In these cases, the initial momenta of the open bottom and antibottom quarks are not thermal. They are sampled from a uniform distribution. The system can still reach detailed balance. The hidden bottom flavor percentage in case 2 differs from that at thermal equilibrium. This is because in case 2, we turn off the transport of open heavy flavors. As a result, the initial uniform momentum distribution will not be thermalized during the evolution. On the contrary, in case 3, we turn on the transport of open heavy flavors, which can thermalize the spectra of open heavy flavors. Consequently, the hidden bottom flavor percentage at detailed balance in case 3 agrees with that at thermal equilibrium. From this comparison we learn that the interplay between dissociation and recombination only drives the system to chemical equilibrium. The kinetic equilibrium is only achieved via interactions between the open heavy flavors and the medium. At the order of expansions we are working currently, interactions between quarkonium and the medium that can drive the kinetic thermalization of quarkonium do not occur.

\begin{figure}
    \centering
    \begin{subfigure}[t]{0.48\textwidth}
        \centering
        \includegraphics[height=2.1in]{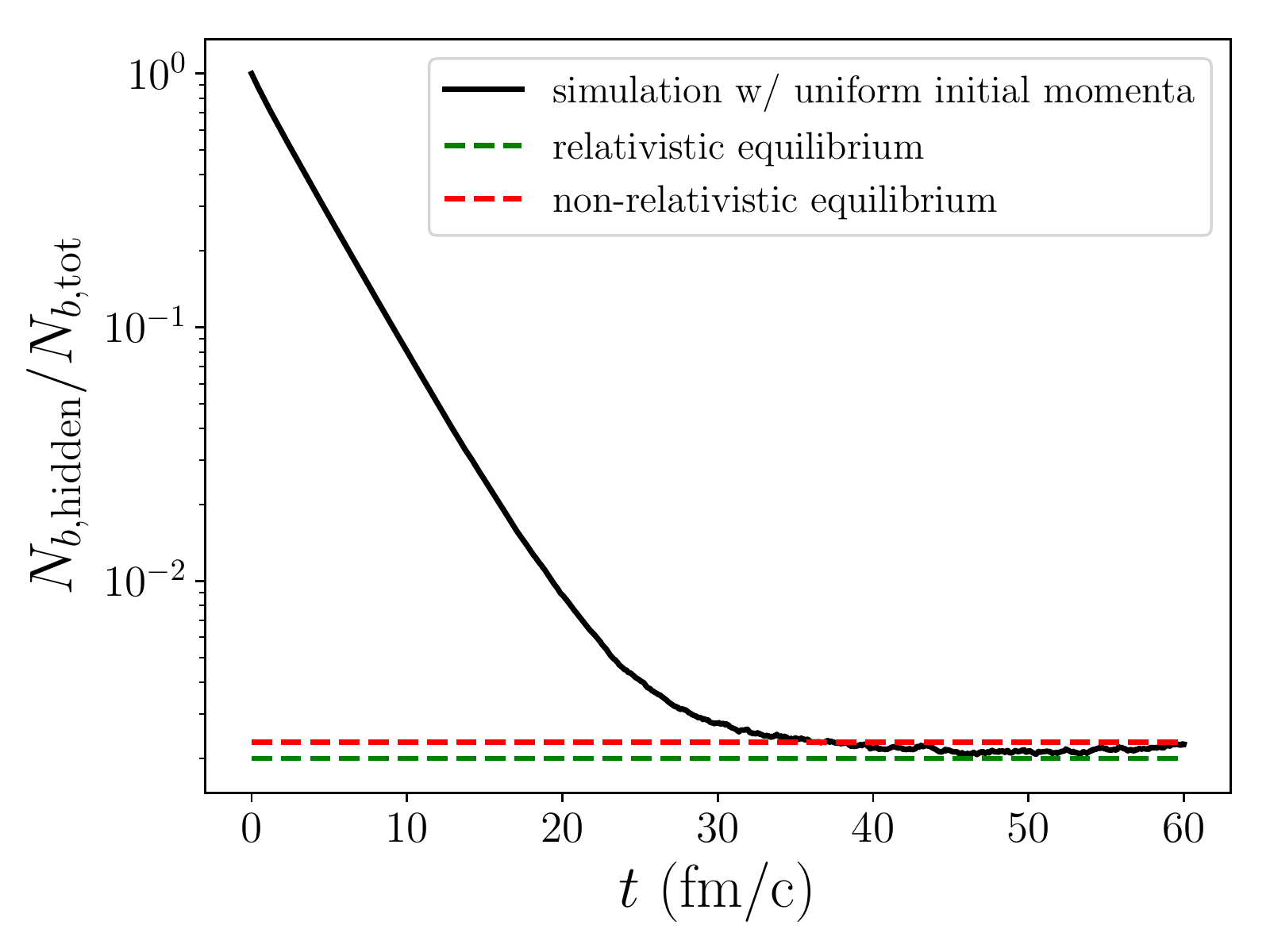}
        \caption{$N_{\Upsilon\ma{(1S)}} = 50$ initially at $T=0.25$ GeV.}\label{}
    \end{subfigure}%
    ~
    \begin{subfigure}[t]{0.48\textwidth}
        \centering
        \includegraphics[height=2.1in]{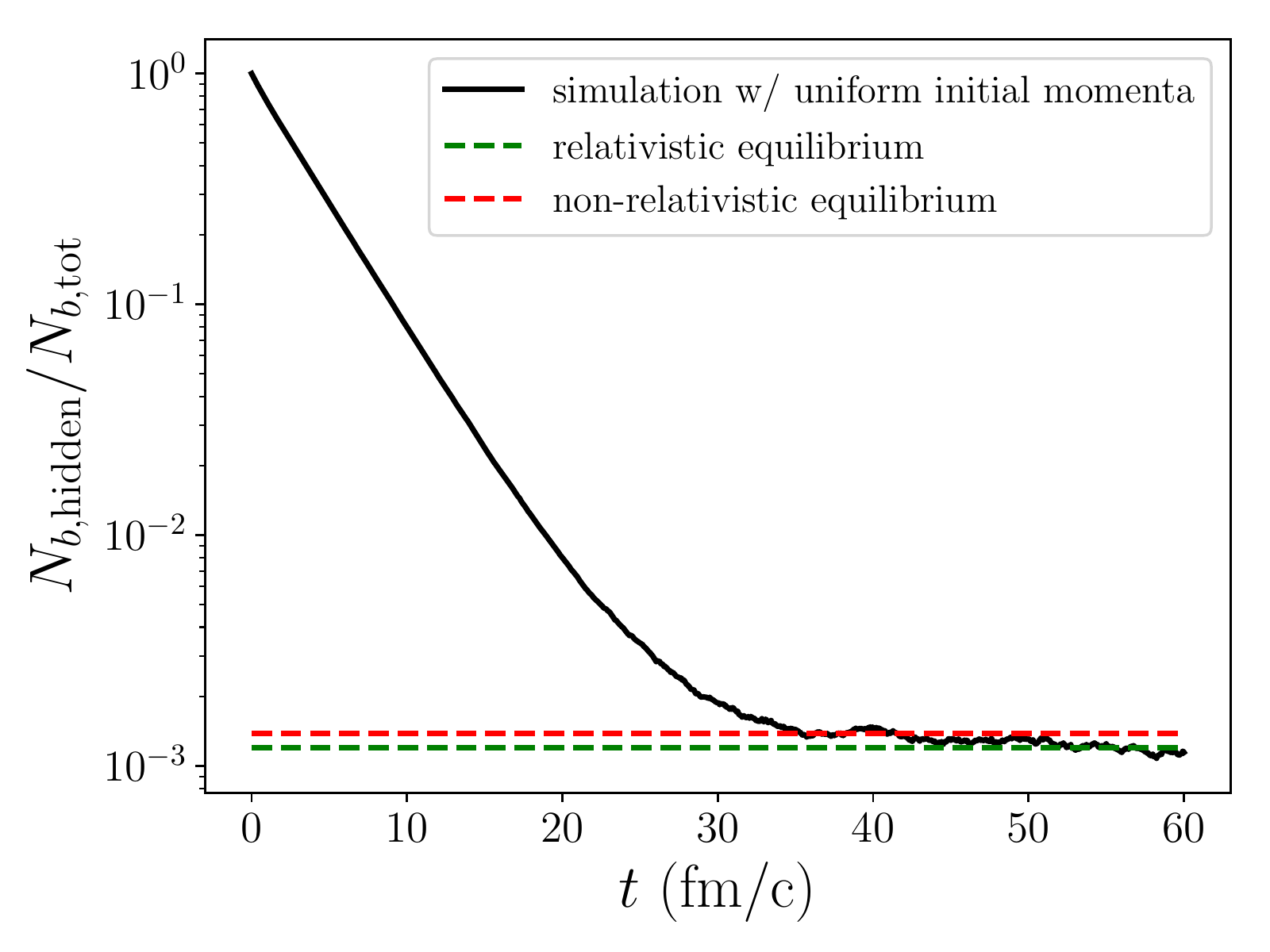}
        \caption{$N_{\Upsilon\ma{(1S)}} = 30$ initially at $T=0.25$ GeV.}\label{}
    \end{subfigure}%
                
    \caption{Hidden bottom flavor percentage calculated via the numerical simulation in case 4.}
    \label{chap4_fig_case4}
\end{figure}

Finally the simulation results of the hidden bottom flavor percentage in case 4 are shown in Fig.~\ref{chap4_fig_case4}. We show the results for two different initial $\Upsilon\ma{(1S)}$ numbers. Both can reach the detailed balance at thermal equilibrium. The thermalization process is as follows: in the early stage, the dominant process is the dissociation of $\Upsilon\ma{(1S)}$. After the dissociation, the unbound $b$ and $\bar{b}$ can thermalize by interacting with the medium. As more $\Upsilon\ma{(1S)}$'s dissociate, there are more unbound $b\bar{b}$ pairs and recombination starts to be manifest. When the number of dissociating $\Upsilon\ma{(1S)}$ equals the number of recombining $\Upsilon\ma{(1S)}$, the system dynamically reaches the detailed balance and thermal equilibrium.

We can understand the reason why the system reaches detailed balance from the perspective of random matrix theory or eigenstate thermalization hypothesis. The interaction between the bound singlet and the unbound octet in pNRQCD is given by
\be
\Tr(\ma{S}^\dagger g{\bs r}\cdot {\bs E} \ma{O} ) + h.c. \,,
\ee
where the chromo-electric field ${\bs E}$ comes from the medium and obeys some distribution. 
When computing the dissociation and recombination rates, we take average over the medium configurations. This introduces randomness in the interaction part of the theory, which drives the system to thermalization. Our numerical implementation demonstrates this.

\vspace{0.2in}
In this section, we tested our numerical implementation of the coupled Boltzmann transport equations. We demonstrated how the system dynamically reaches detailed balance and thermal equilibrium. Now we are ready to study quarkonium production in heavy ion collisions.

\vspace{0.2in}

\section{Bottomonium in Heavy Ion Collisions}
To study bottomonium production in heavy ion collisions, we need to study the evolution of $b\bar{b}$ pairs inside the QGP produced in the collisions. The dynamical evolution is described by the coupled Boltzmann equations. To solve the transport equations, we need an initial condition and a description of the medium.

\vspace{0.2in}
\subsection{Initial Conditions}
The initial momentum distributions of open bottom (antibottom) quarks and bottomonium are generated from the event generator \textsc{Pythia} \cite{Sjostrand:2014zea}. The cross section of each process is calculated in the framework of collinear factorization in \textsc{Pythia}. For a particle species $A$, the inclusive cross section in proton-proton collision is given by
\be
\label{chap4_eqn_collinear_factor}
\diff \sigma^{p+p\to A+X} = \sum_{i,j} \int_0^1 \diff x_i \int_0^1 \diff x_j f_i(x_i,Q^2) f_j(x_j, Q^2) \diff \sigma^{i+j\to A+X} \,.
\ee
In the collinear factorization formula (\ref{chap4_eqn_collinear_factor}), $i$ and $j$ indicate partons which can be either quarks or gluons. $f_i(x_i,Q^2)$ is the parton distribution function (PDF) of the parton species $i$ inside the proton at the scale $Q^2$. $\diff \sigma^{i+j\to A+X}$ is the differential cross section of the partonic process $i+j\to A+X$. For open bottom quarks, the partonic process $i+j\to b+X$ can be calculated in QCD perturbation. For a bottomonium state $H$, the cross section of the process $i+j\to H+X$ is calculated in the framework of NRQCD factorization in \textsc{Pythia}. The NRQCD factorization was explained in Chapter 1, which states that the cross section factorizes into a short-distance part of producing a $b\bar{b}$ pair in certain quantum numbers and a long-distance part in which the $b\bar{b}$ pair forms the bottomonium $H$. Mathematically
\be
\label{chap4_eqn_NRQCD_factor}
\diff \sigma^{i+j\to H+X} = \sum_n \diff \sigma^{i+j\to (b\bar{b})_n+X} \langle 0 | \ml{O}^H_n | 0 \rangle \,,
\ee
where $n$ indicates the quantum numbers (color, spin, orbital angular momentum) of the four-fermion operators $\ml{O}_n$ introduced in Chapter 1. 

For heavy ion collisions, we assume the collinear factorization (\ref{chap4_eqn_collinear_factor}) and the NRQCD factorization (\ref{chap4_eqn_NRQCD_factor}) are still valid for the calculation of bottom quark and quarkonium production in each initial nucleon-nucleon binary collision. The only difference is the PDF. The PDF of a nucleon inside a nuclei is generally different from that of a proton. So we need to use the nuclear PDF in the factorization formula. The nuclear PDF can be studied and fitted by using measurements in proton-nucleus collisions. We will apply the parametrization EPS09 \cite{Eskola:2009uj} to \textsc{Pythia}. The initial production suppression due to the cold nuclear matter effect is accounted by the nuclear PDF.
 
The number of particles of a certain species $i$ produced in all the initial nucleon-nucleon binary collisions of one heavy ion collision is given by
\be
N_i = \sigma^{n+n\to i+X} T_{AA}({\bs b}) \,,
\ee
where $T_{AA}({\bs b})$ is the nuclear overlap function of the A-A collision at the impact parameter ${\bs b}$, defined in Eq.~(\ref{chap1_eqn_overlap}). We will use the binary collision model \trento\ \cite{Moreland:2014oya} to calculate $T_{AA}({\bs b})$ in each centrality class.

The \trento\ model can also calculate the density distribution $\rho_{AA}({\bs b}, {\bs r})$ of the nuclear overlap function in the transverse plane, perpendicular to the beam axis. According to Eq.~(\ref{chap1_eqn_overlap}), it is given by
\be
\rho_{AA}({\bs b}, {\bs r}) = T_A({\bs r}) T_A({\bs b}-{\bs r}) \,.
\ee
$\rho_{AA}({\bs b}, {\bs r})$ gives the density of nucleon-nucleon binary collisions in the transverse plane. We can treat each nucleon-nucleon binary collision as a proton-proton collision with a nuclear PDF effectively. So we can sample the positions of the produced bottom quarks and bottomonia based on the density $\rho_{AA}({\bs b}, {\bs r})$.

So now we have a method to sample both the position and the momentum of each particle produced in the initial binary collisions. We assume they are produced at time $t=0$ in the laboratory frame.

\vspace{0.2in}
\subsection{Medium Description}
The medium background is given by a boost-invariant viscous hydrodynamic simulation. We will use the simulation package VISHNU \cite{Song:2007ux,Shen:2014vra}. The initial condition of the hydrodynamics, i.e., the initial averaged entropy density, can be calculated in the \trento\ model. With given initial conditions, the simulation package numerically solves the hydrodynamic equation
\be
\partial_{\mu} T^{\mu\nu} &=& 0
\ee
with the energy-momentum tensor
\be
T^{\mu\nu} &=& eu^{\mu}u^{\nu} - (p+\Pi)(g^{\mu\nu}-u^{\mu}u^{\nu}) + \pi^{\mu\nu} \\
\Pi &=& -\zeta\nabla\cdot u \\
\pi^{\mu\nu} &=& 2\eta\nabla^{\langle\mu}u^{\nu\rangle} \,.
\ee
Here $e$ and $p$ are the local energy density and pressure, and $u^{\mu}$ is the local four-velocity of the QGP. $\Pi$ is the bulk stress with the bulk viscosity $\zeta$, and $\pi^{\mu\nu}$ is the shear stress tensor with the shear viscosity $\eta$. Here the angle bracket means traceless symmetrization.

We will conduct a large number of hydrodynamic simulations at each collision energy and impact parameter, and obtain the event-averaged QGP profile at each collision energy and impact parameter. The event-averaged QGP profile contains information such as the space-time dependence of the QGP temperature and hydro-cell velocity. We will use the information to calculate the scattering rate of open bottom quarks, the dissociation and the recombination rates of bottomonia at a given time step. 

The parameters of the \trento\ model and the medium properties used in the hydrodynamic simulation have been calibrated with experimental observables (such as yields and flow parameters $v_2$) of light particles with small transverse momenta \cite{Bernhard:2016tnd}.

\vspace{0.2in}
\subsection{Results}
We will focus on studying the production of $\Upsilon$(1S) and $\Upsilon$(2S) here. We include open bottom and antibottom quarks, $\Upsilon$(1S) and $\Upsilon$(2S) in the coupled Boltzmann transport equations. We set the bottom quark mass to be $M=4.65$ GeV. We fix the coupling constant of the dipole vertex of pNRQCD to be $\alpha_s=0.3$ (since it is determined at the scale $Mv\approx1.5$ GeV). But we leave the coupling constant $\alpha_s^\ma{pot}$ in the potentials
\be
V_s(r) = -C_F\frac{\alpha_s^\ma{pot}}{r}\,, \ \ \ \ \ \ \ \ \ \ V_o(r) = \frac{1}{2N_c}\frac{\alpha_s^\ma{pot}}{r} \,,
\ee
as a parameter. Since we use Coulomb potentials here, solutions of $\Upsilon$(nS) states in the Schr\"ordinger equation exist at an arbitrary high temperature, which contradicts the current understanding. So we include the melting temperature by hand. Since the initial temperature of our hydrodynamic simulation is about $\sim 450$ MeV, we will not assign a melting temperature to the $\Upsilon$(1S) state. But we will assign a melting parameter $T_\ma{2S}$ to the $\Upsilon$(2S) state. Above $T_\ma{2S}$, no $\Upsilon$(2S) state can be produced from recombination. 

We assume the QGP is formed at the co-moving time $\tau=0.6$ fm/c where $\tau = \sqrt{t^2-z^2}$ and $t$ and $z$ are observed in the laboratory frame. So we will start the simulation of the coupled Boltzmann transport equations when $\tau=0.6$ fm/c. For the time period between $t=0$ and $\tau=0.6$ fm/c, we will assume all the bottom, antibottom quarks and bottomonia produced at $t=0$ are free streaming. Their momenta do not change during this time period. 

We will stop the simulation of transport equations when the local QGP temperature drops to $T=154$ MeV. We will assume no more dissociation and recombination in the hadronic gas stage. But we include the feed-down process of $\Upsilon$(2S) to $\Upsilon$(1S). The branching ratio of $\Upsilon$(2S) to $\Upsilon$(1S) in the hadronic gas stage is $0.26$ \cite{Tanabashi:2018oca}. 

\begin{figure}[h]
    \centering
    \begin{subfigure}[t]{0.48\textwidth}
        \centering
        \includegraphics[height=2.2in]{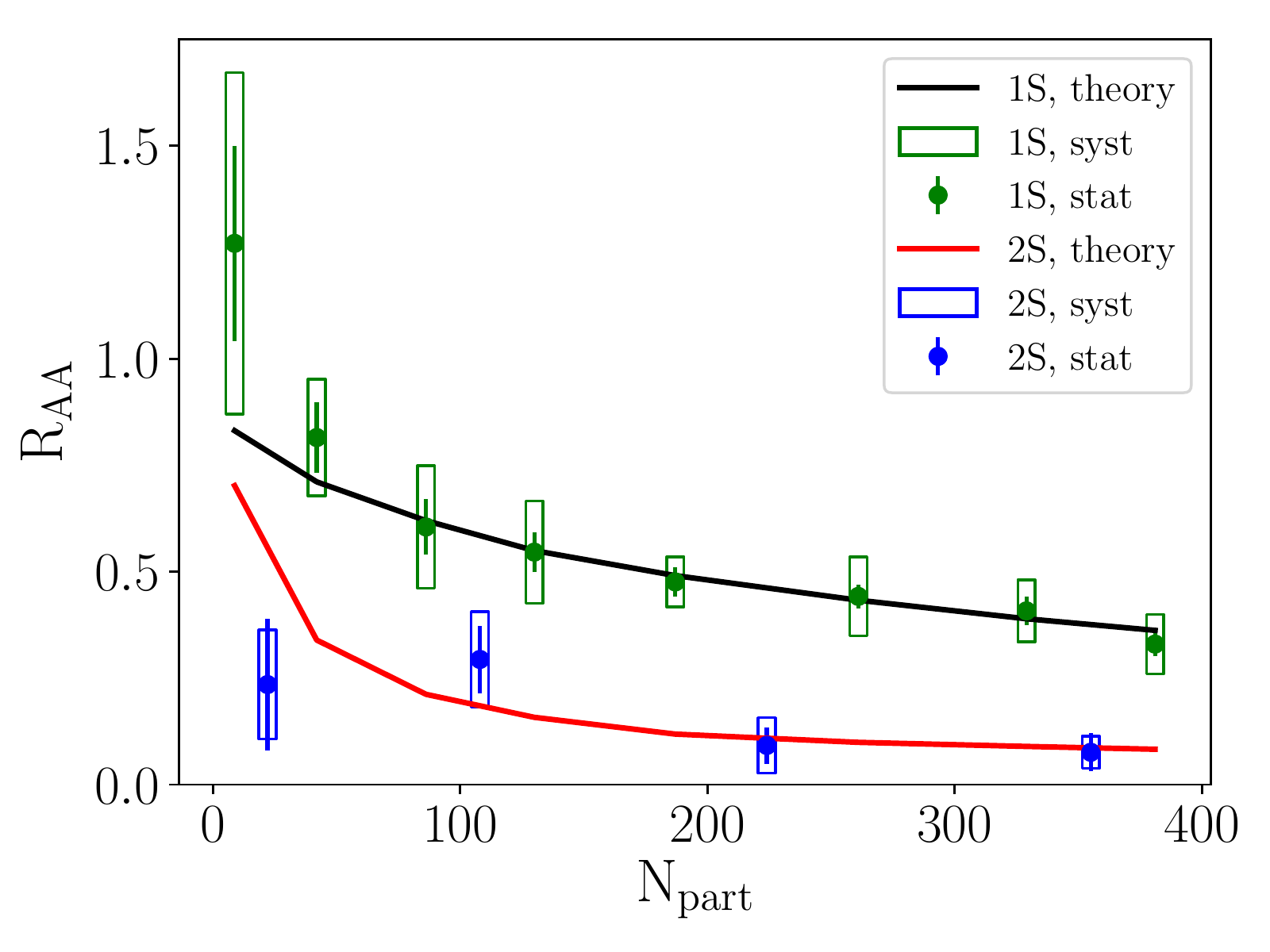}
        \caption{$R_{AA}$ as a function of centrality.}
    \end{subfigure}
    ~ 
    \begin{subfigure}[t]{0.48\textwidth}
        \centering
        \includegraphics[height=2.2in]{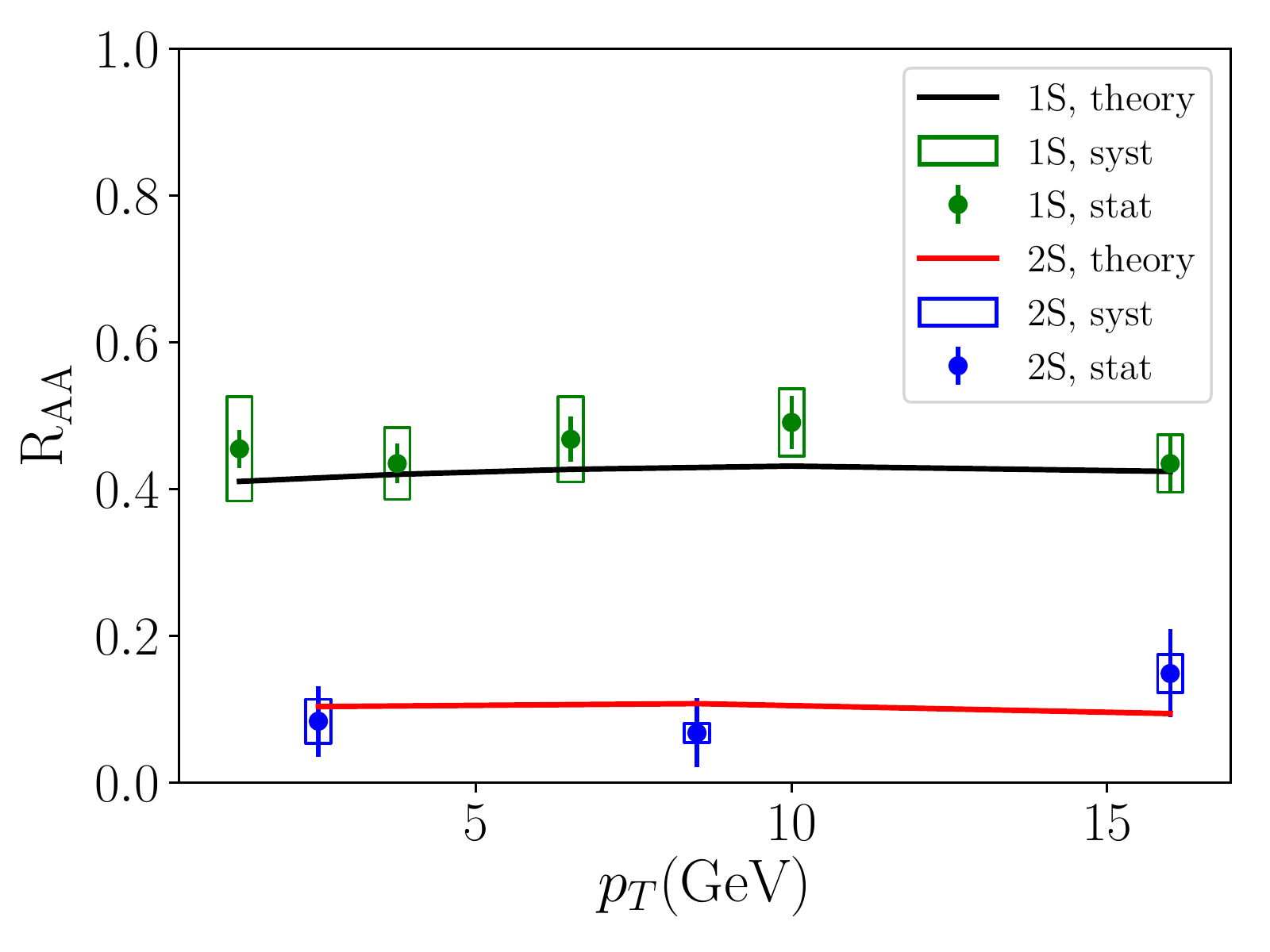}
        \caption{$R_{AA}$ as a function of transverse momentum in $0\%-100\%$ centrality.}
    \end{subfigure}%
    
    \begin{subfigure}[t]{0.48\textwidth}
        \centering
        \includegraphics[height=2.2in]{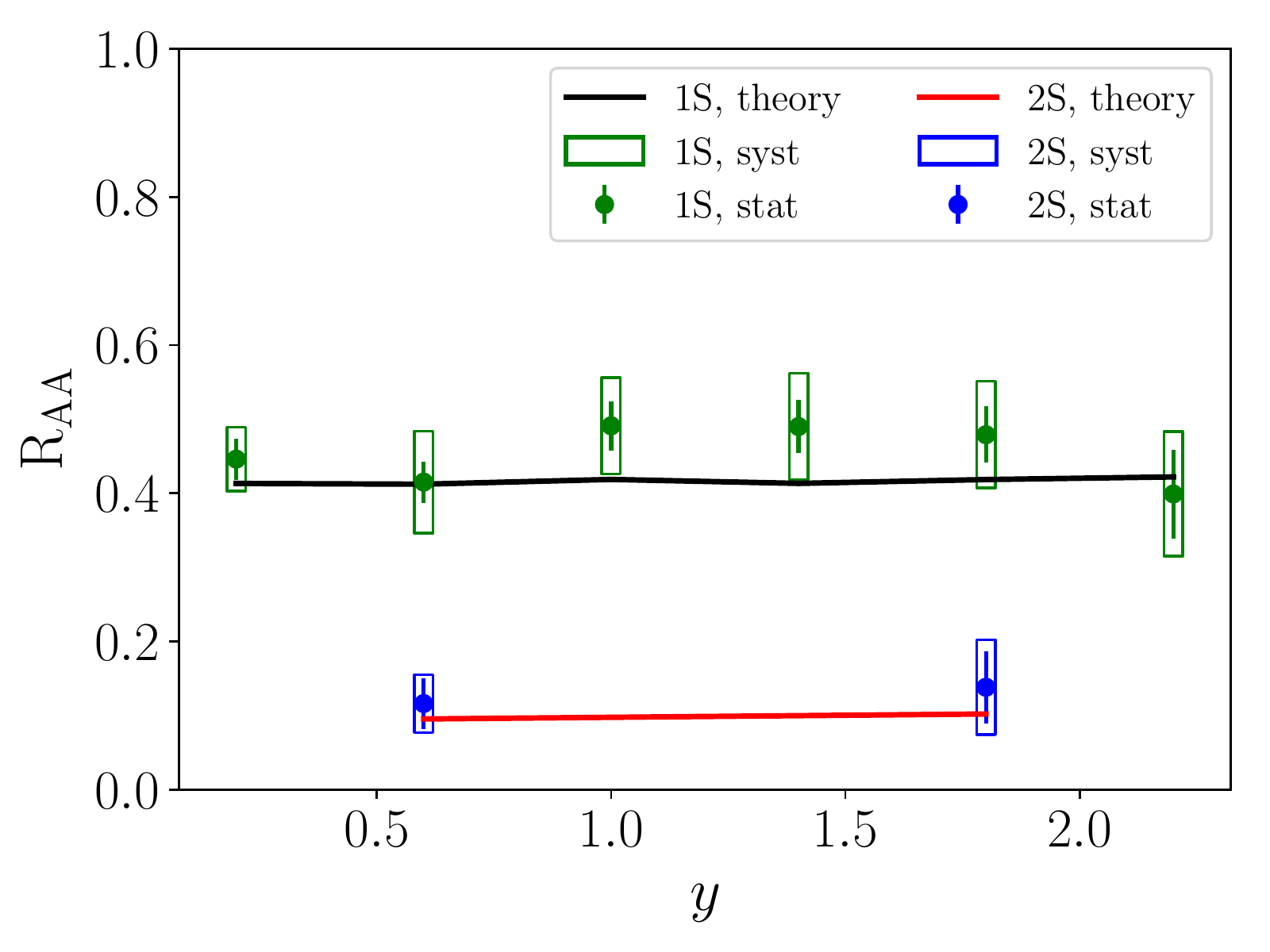}
        \caption{$R_{AA}$ as a function of rapidity in $0\%-100\%$ centrality.}
    \end{subfigure}%
    \caption[Results of $R_{AA}$ of $\Upsilon$(nS) in $2.76$ TeV Pb-Pb collisions in $|y|<2.4$ and the CMS measurements.]{Results of $R_{AA}$ of $\Upsilon$(nS) in $2.76$ TeV Pb-Pb collisions in $|y|<2.4$ and the CMS measurements. Data are taken from Ref.~\cite{Khachatryan:2016xxp}.}
    \label{chap4_fig:cms}
\end{figure}

After the event simulation, we can estimate the nuclear modification factor $R_{AA}$. The cold nuclear matter suppression factor is estimated by \textsc{Pythia} with a nuclear PDF. For the $2.76$ TeV Pb-Pb and $5.02$ TeV Pb-Pb collisions at LHC, the estimates give $0.87$ and $0.85$ respectively. For the $200$ GeV Au-Au collisions at RHIC, the estimate given by \textsc{Pythia} is around or even larger than $1$ (anti-shadowing effect), which is inconsistent with the measurements in p-Au collisions \cite{Ye:2017fwv}. So we fix this by using the input of the experimental measurements. We simply use $R_{pAu}^2\approx0.72$ \cite{Ye:2017fwv} as the cold nuclear matter suppression factor in Au-Au collisions.

We have two parameters in the calculation: the coupling constant in the potential $\alpha_s^\ma{pot}$ and the melting temperature of $\Upsilon$(2S) $T_{\ma{2S}}$. We will use the data at $2.76$ TeV Pb-Pb collision to fix these two parameters. We find $\alpha_s^\ma{pot} = 0.42$ and $T_{\ma{2S}}=210$ MeV can well describe the data. The comparisons between our calculation and the data are shown in Fig.~\ref{chap4_fig:cms}. We compare $R_{AA}$ as functions of the centrality, the transverse momentum $p_T$ and the rapidity $y$. The $R_{AA}(y)$ is flat. In our calculations, since we use a rapidity independent cold nuclear matter suppression, the only rapidity dependence can come from the in-medium evolution. We use a boost-invariant $2+1D$ hydrodynamics. So the medium looks the same in any frame that is boosted along the beam axis. As a result, the in-medium evolution is rapidity independent and we obtain a flat $R_{AA}(y)$, which agrees with the experimental data. This may also indicate that a boost-invariant medium is enough to describe bottomonium production up to $|y|=2.4$.

\begin{figure}[h]
    \centering
        \begin{subfigure}[t]{0.48\textwidth}
        \centering
        \includegraphics[height=2.2in]{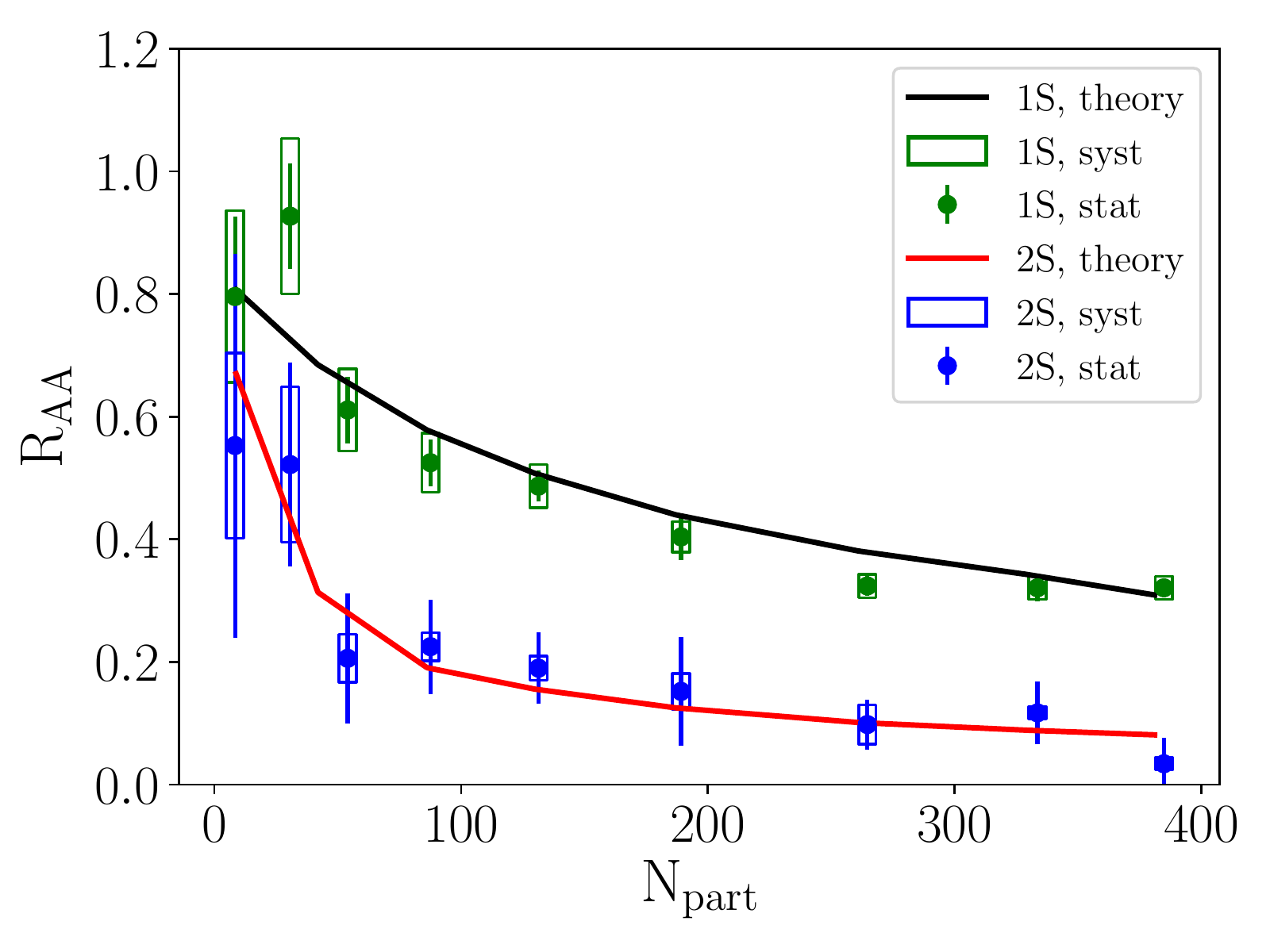}
        \caption{$R_{AA}$ as a function of centrality.}
    \end{subfigure}%
    ~
    \begin{subfigure}[t]{0.48\textwidth}
        \centering
        \includegraphics[height=2.2in]{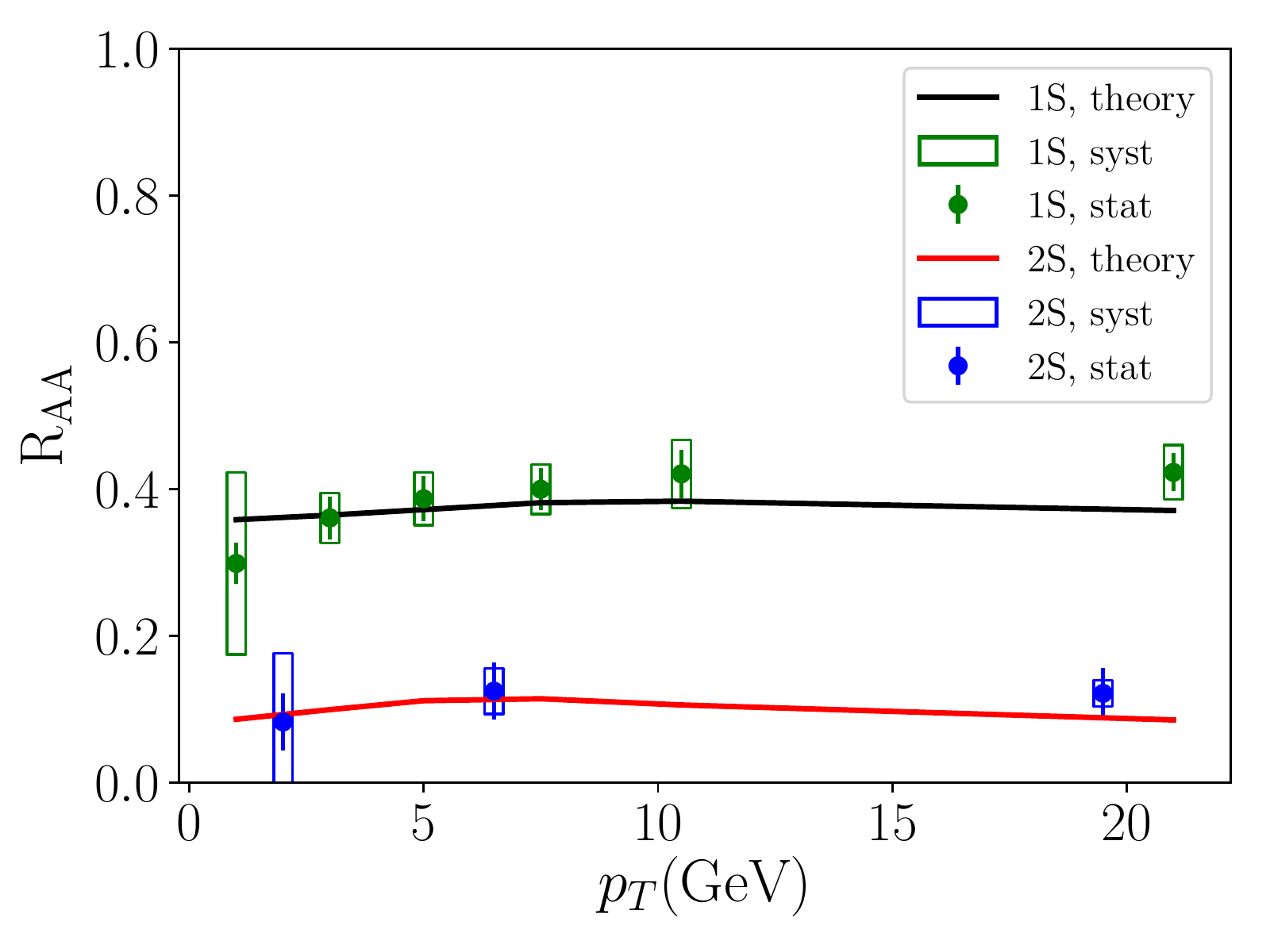}
        \caption{$R_{AA}$ v.s. $p_T$ in $0\%-100\%$ centrality.}
    \end{subfigure}%
    
    \begin{subfigure}[t]{0.48\textwidth}
        \centering
        \includegraphics[height=2.2in]{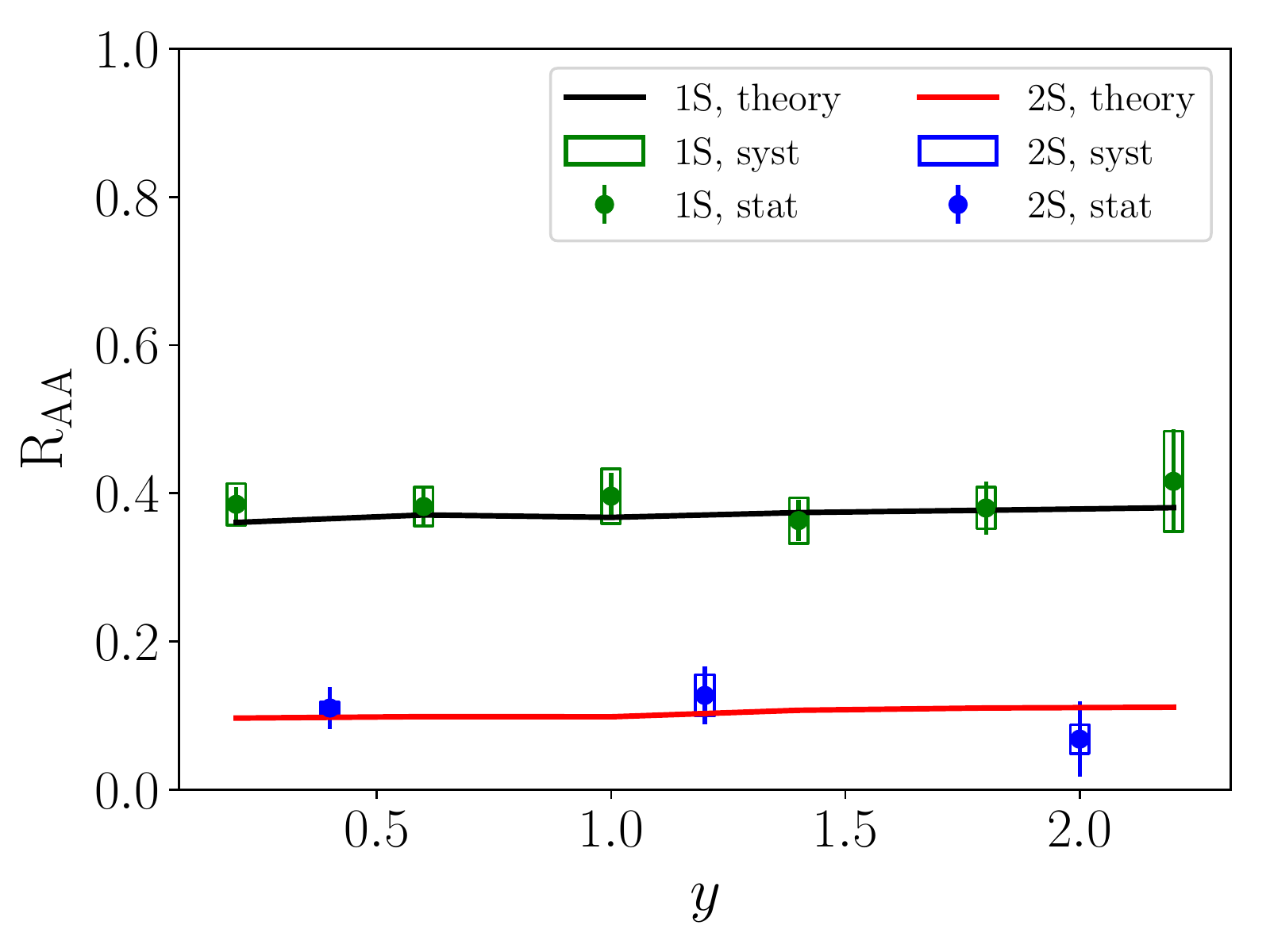}
        \caption{$R_{AA}$ v.s. $y$ in $0\%-100\%$ centrality.}
    \end{subfigure}%
    \caption[Results of $R_{AA}$ of $\Upsilon$(nS) in $5.02$ TeV Pb-Pb collisions in $|y|<2.4$ and the CMS measurements.]{Results of $R_{AA}$ of $\Upsilon$(nS) in $5.02$ TeV Pb-Pb collisions in $|y|<2.4$ and the CMS measurements. Data are taken from Ref.~\cite{Sirunyan:2018nsz}}
    \label{chap4_fig:cms5}
\end{figure}

\begin{figure}[h]
    \centering
    \begin{subfigure}[t]{0.48\textwidth}
        \centering
        \includegraphics[height=2.2in]{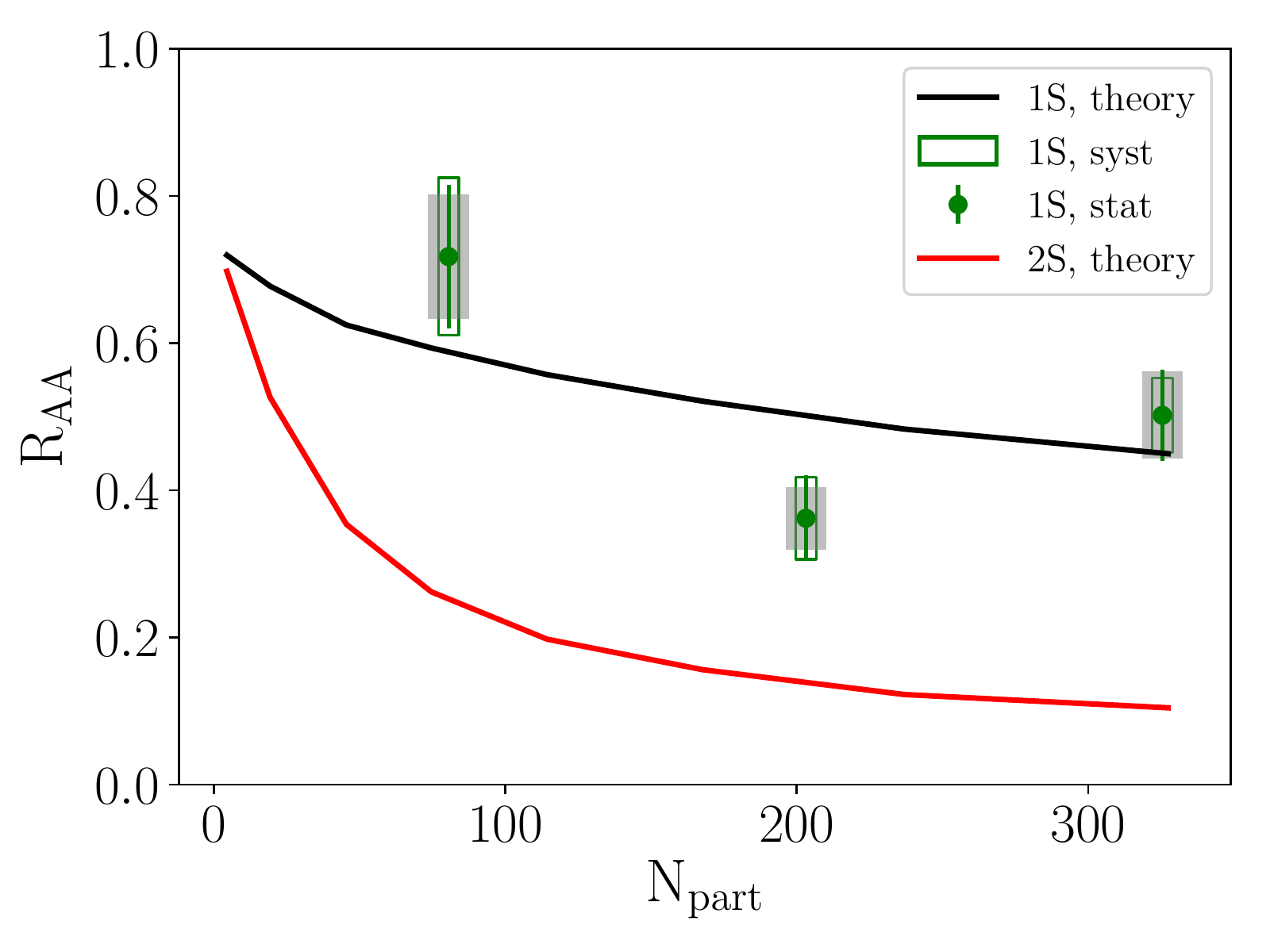}
        \caption{$R_{AA}$ as a function of centrality.}
    \end{subfigure}%
    ~ 
    \begin{subfigure}[t]{0.48\textwidth}
        \centering
        \includegraphics[height=2.2in]{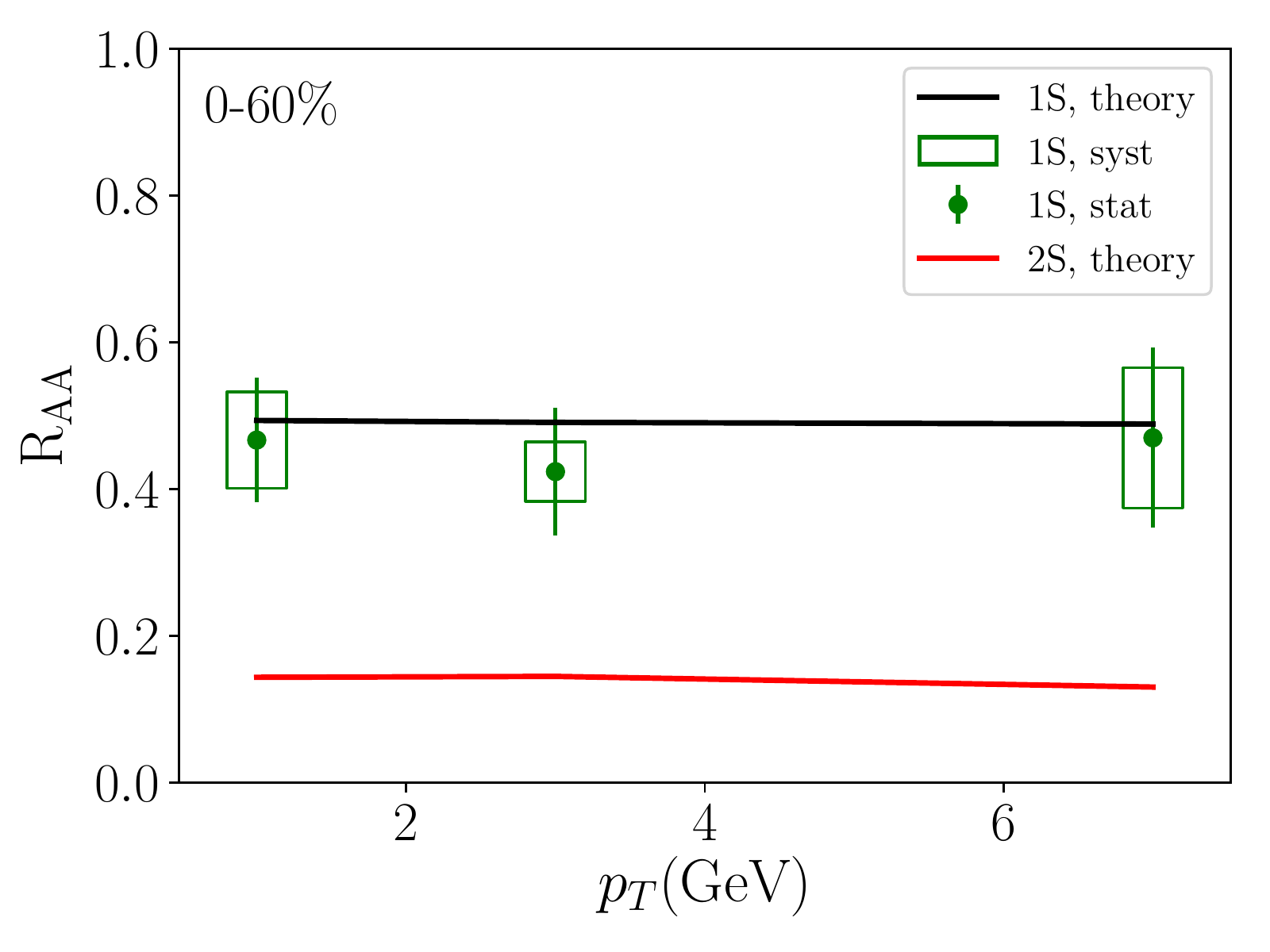}
        \caption{$R_{AA}$ as a function of transverse momentum in $0\%-60\%$ centrality.}
    \end{subfigure}%
    \caption[Results of $R_{AA}$ of $\Upsilon$(nS) in $200$ GeV Au-Au collisions in $|y|<0.5$ and the STAR measurements.]{Results of $R_{AA}$ of $\Upsilon$(nS) in $200$ GeV Au-Au collisions in $|y|<0.5$ and the STAR measurements. Data are taken from Ref.~\cite{Ye:2017fwv}. The data points of $\Upsilon$(2S) are not available since Ref.~\cite{Ye:2017fwv} only reports $R_{AA}$ of $\Upsilon\ma{(2S)}+\Upsilon\ma{(3S)}$.}
    \label{chap4_fig:star}
\end{figure}

\begin{figure}[h]
    \centering
    \begin{subfigure}[t]{0.48\textwidth}
        \centering
        \includegraphics[height=2.2in]{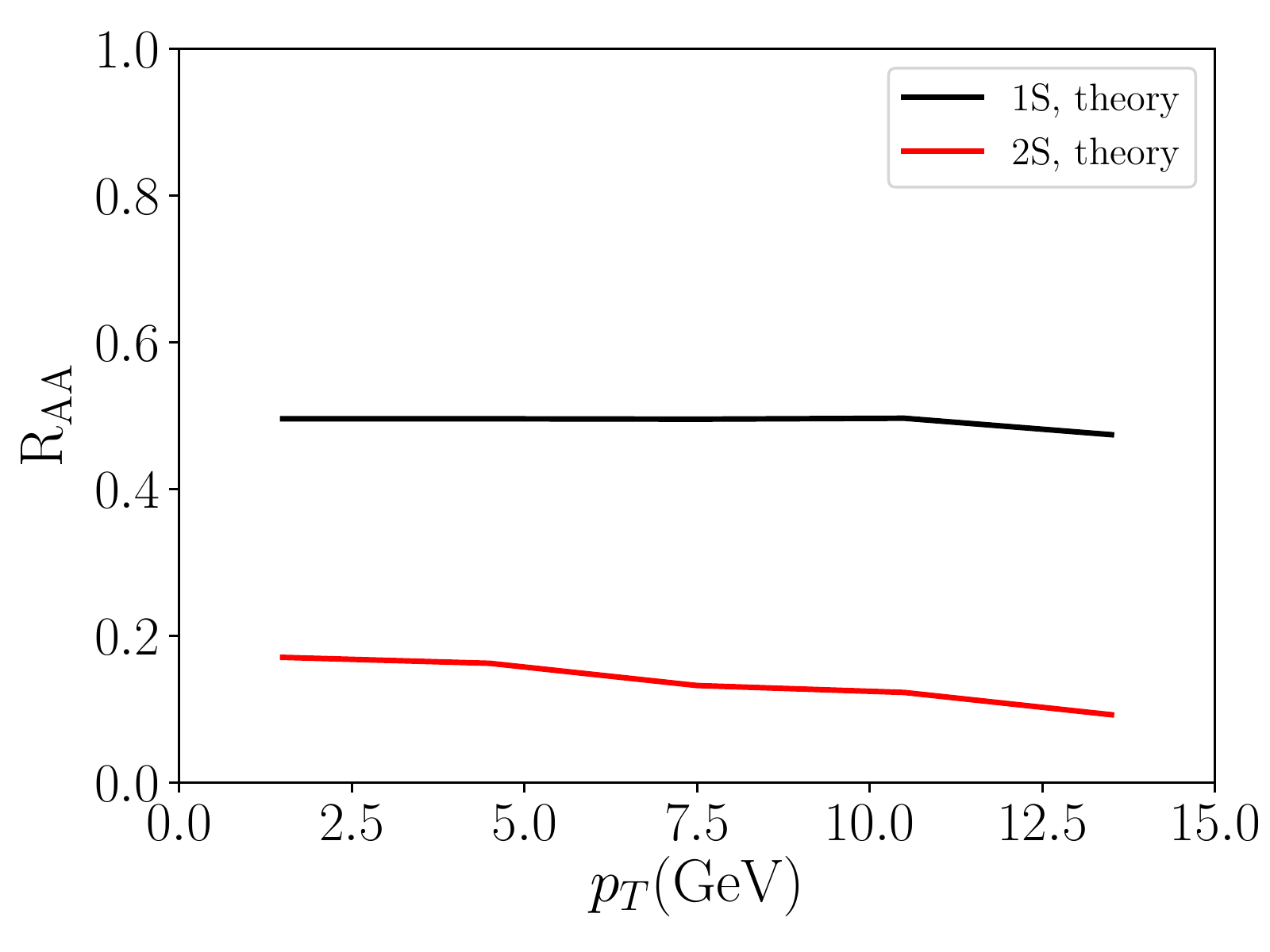}
        \caption{$R_{AA}$ as a function of transverse momentum in $0\%-100\%$ centrality.}
    \end{subfigure}%
    ~ 
    \begin{subfigure}[t]{0.48\textwidth}
        \centering
        \includegraphics[height=2.2in]{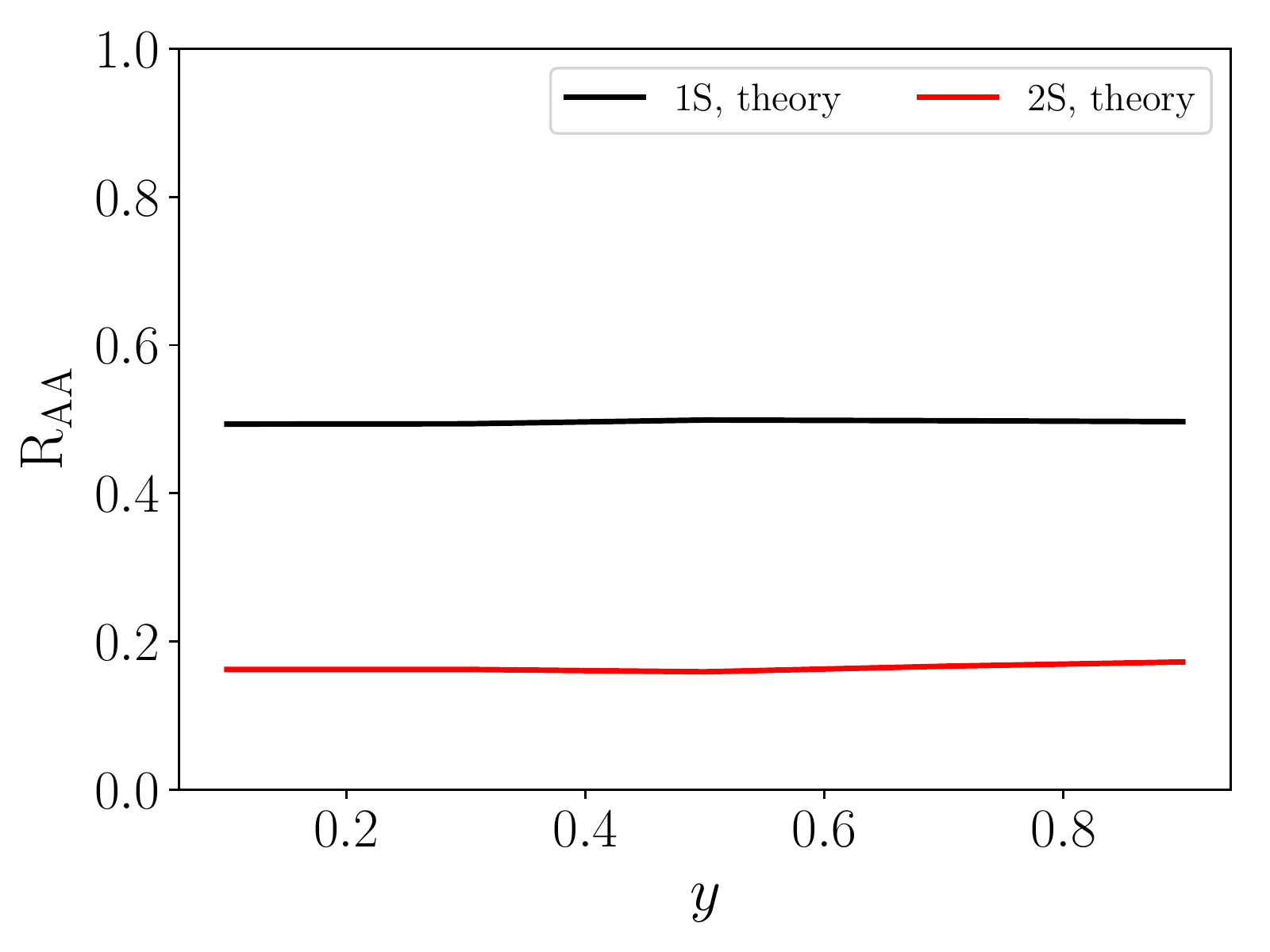}
        \caption{$R_{AA}$ as a function of rapidity in $0\%-100\%$ centrality.}
    \end{subfigure}%
    \caption{Predictions of $R_{AA}$ of $\Upsilon$(nS) in $200$ GeV Au-Au collisions for the kinetic range ($|y|<1$, $0<p_T<15$ GeV) covered by the sPHENIX detector.}
    \label{chap4_fig:sphenix}
\end{figure}

We use the same set of parameters $\alpha_s^\ma{pot} = 0.42$ and $T_{\ma{2S}}=210$ MeV to calculate $R_{AA}$ of $\Upsilon$(nS) in the $5.02$ TeV Pb-Pb and $200$ GeV Au-Au collisions. The results of our calculations are shown in Figs.~\ref{chap4_fig:cms5} and \ref{chap4_fig:star}. For the $200$ GeV Au-Au collisions, we only compare the results with measurements of $\Upsilon$(1S). Experimental results of $\Upsilon$(2S) are not available since Ref.~\cite{Ye:2017fwv} only reports $R_{AA}$ of $\Upsilon\ma{(2S)}+\Upsilon\ma{(3S)}$. The good agreement indicates that our theoretically well-justified description works reasonably in phenomenology. For the $R_{AA}$ of $\Upsilon$(nS) in $200$ GeV Au-Au collisions, we also give predictions for the kinetic range covered by the forthcoming sPHENIX detector, shown in Fig.~\ref{chap4_fig:sphenix}. The rapidity range is $|y|<1$ and the transverse momentum can be measured up to $15$ GeV.

\begin{figure}[h]
    \centering
	\includegraphics[height=2.5in]{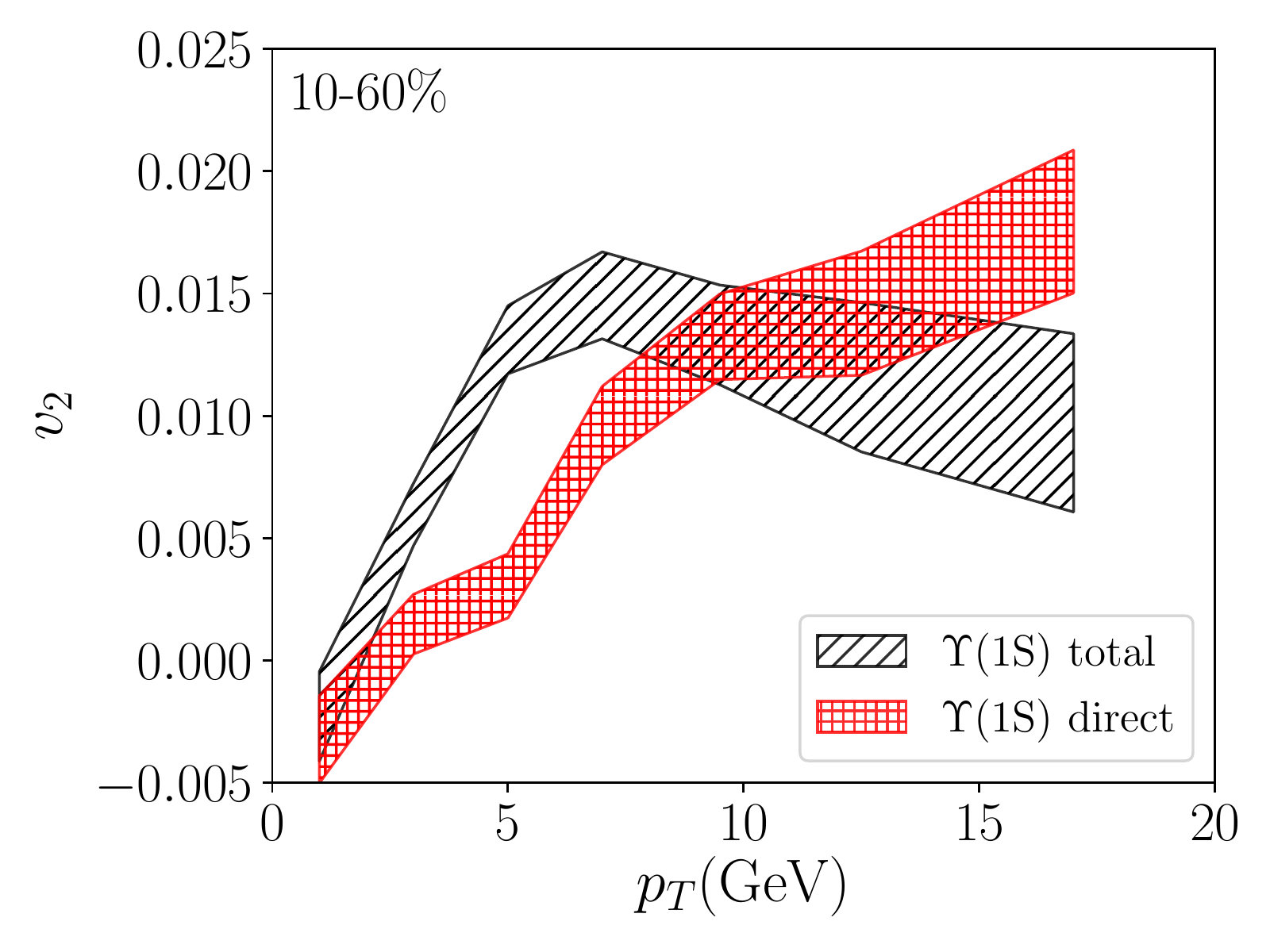}
    \caption{Predictions of $v_2$ of $\Upsilon$(1S) in $5.02$ TeV Pb-Pb collisions in the centrality range $10\%-60\%$ and in the rapidity range $|y|<2.4$.}
    \label{chap4_fig:v2}
\end{figure}

Finally, we estimate the azimuthal angular anisotropy parameter $v_2$ of $\Upsilon$(1S). The prediction of $v_2$ in the $5.02$ TeV Pb-Pb collisions in the centrality range $10\%-60\%$ and in the rapidity range $|y|<2.4$ is shown in Fig.~\ref{chap4_fig:v2}. Here we decompose the $v_2$ of $\Upsilon$(1S) into two parts: $v_2$ of those $\Upsilon$(1S) that are produced in the initial nucleon-nucleon binary collisions and survive the in-medium evolution; and $v_2$ of those $\Upsilon$(1S) that are produced from recombination. Besides the total $v_2$, we also plot the contribution from those $\Upsilon$(1S) that are produced in the initial binary collision and survive the in-medium evolution (we label this $v_2$ contribution as ``direct").

An experimental measurement of $v_2$ of $\Upsilon$ will be very interesting. We can use the $v_2$ measurements to test different theoretical calculations and constrain the parameters. It may be possible that the correct description of both $R_{AA}$ and $v_2$ require including quantum effect in the evolution that is beyond the semi-classical Boltzmann transport equation. As more precision data become available with high statistics, an exciting era of theoretical and phenomenological studies of quarkonium in-medium transport will come. 

\singlespacing
\chapter{Doubly Heavy Baryon Production in Heavy Ion Collisions}
\doublespacing
\vspace{0.2in}
\section{Physical Motivation}
Recently the LHCb Collaboration reported the observation of a doubly charmed baryon carrying two units of positive charge, $\Xi_{cc}^{++}$, with a mass $m(\Xi_{cc})\approx 3621$ MeV \cite{Aaij:2017ueg}, which is consistent with a previous theoretical estimate \cite{Karliner:2014gca}. Though it is still unclear why the observed mass differs from the previous SELEX result \cite{Mattson:2002vu}, the existence of hadrons with more than one heavy quark is now on a more solid ground. We will first explain the structure of hadrons with multiple heavy quarks.

\vspace{0.2in}

\subsection{Multi-Heavy Hadrons}
The quark model description of $\Xi_{cc}^{++}$ is $(ccu)$. The color and spin wave functions of these three valence quarks are expected to be simpler than those of a proton $(uud)$ or neutron $(udd)$ because of the large charm quark mass. A pair of two light quarks $qq$ (we will call it {\it diquark} from now on), can be in either a color anti-triplet or a color sextet. Two quarks in the color anti-triplet (sextet) generally interact attractively (repulsively) with each other. If the pair is in a color anti-triplet (sextet), the spin will be (anti-)parallel for a S-wave state because the wave function has to be antisymmetric for identical fermions. Since the light quark mass is small, the energy difference between the color anti-triplet and the sextet due to the difference in the attractive and repulsive potential is probably on the same order as the hyperfine structure splitting of different spins and the fine structure splitting of different combinations of spins and orbital angular momenta. Therefore it is not easy to tell the quantum numbers of the light diquark in the ground state of $(uud)$ or $(udd)$. In reality, the structure of the light diquark could be a mixture of both the color anti-triplet and sextet. The quantum number structure is much simpler for the $(QQq)$ ground state. Since the heavy quark mass is large, the fine and hyperfine structure splittings are suppressed relative to the energy difference between the anti-triplet and the sextet. Therefore, the heavy diquark $QQ$ inside the $(QQq)$ ground state is probably in the anti-triplet state and forms a bound state. For the ground state, we expect all the relative orbital angular momenta to be S-wave. So the heavy diquark is a spin triplet. For a spin-$\frac{1}{2}$ $(QQq)$ ground state, the spin of the light quark will point to the opposite direction of the spin of the heavy quarks.

Since the experimentally determined mass of $\Xi_{cc}^{++}$ is below the threshold of one D meson and one $\Lambda_c$ baryon, the particle is stable under strong interactions and only decays weakly. Experimental measurements rely on the reconstruction from decay products of $\Xi_{cc}^{++}$. The decay properties of doubly heavy baryons have been intensely studied \cite{SanchisLozano:1993kh,Kiselev:1998sy,Guberina:1999mx,Egolf:2002nk,Ebert:2004ck,Hu:2005gf,Albertus:2011xz,Karliner:2014gca,Wang:2017mqp,Li:2017pxa,Lu:2017meb}.

According to the heavy quark diquark symmetry, in the infinitely large mass limit, a heavy diquark in the anti-triplet $(QQ)_{\bar{3}}$ can be thought of as a single heavy antiquark. In this picture, a doubly heavy baryon can be thought of as a singly heavy meson. More interestingly, a singly heavy baryon should have a corresponding partner, which is a doubly heavy tetraquark $(QQqq)$. The stability of heavy tetraquarks has been investigated previously in Ref.~\cite{Ader:1981db}. The $bb\bar{u}\bar{d}$ ground state with $J^P=1^+$ is predicted to be stable \cite{Karliner:2017qjm,Eichten:2017ffp,Francis:2016hui,Bicudo:2016ooe,Czarnecki:2017vco}. An experimental confirmation of a stable $bb\bar{u}\bar{d}$ tetraquark will be exciting.

\vspace{0.2in}
\subsection{$\Xi_{cc}^{++}$ in Heavy Ion Collisions}
Here we consider the production of $\Xi_{cc}^{++}$ in high energy heavy ion collisions and study the hot medium effect on its production. Previous work was based on quark coalescence at hadronization and assumed that heavy quarks are thermally distributed \cite{Becattini:2005hb, Zhao:2016ccp}. Here we pursue out a more dynamical approach considering the formation of bound heavy diquarks within the QGP and the incomplete equilibration of the heavy quark spectrum.

In hadron-hadron collisions, it is difficult to produce a pair of heavy quarks in the color anti-triplet at leading order in a fragmentation process. On the other hand, the coalescence process involving two independently produced charm quarks is sensitive to the relative momentum between the heavy quark pair. In proton-proton collisions, the relative momentum is uncontrolled and likely large, suppressing the coalescence. Heavy ion collisions have two advantages for $\Xi_{cc}^{++}$ production: First, the rapidity density of charm quarks produced in a single collision is higher. Second, the deconfined QGP medium lasts roughly $10$ fm/c, during which time the charm quarks can diffuse in the QGP via interactions with light quarks and gluons. This is confirmed by recent measurements from the STAR Collaboration, which shows that charm quarks participate in the collective flow of the QGP \cite{Adamczyk:2017xur}. 
As a result, the relative momentum of a charm quark pair can be on the order of the QGP temperature. The coalescence probability into a charm diquark bound state is thus enhanced if the temperature of the QGP is not too high. 

After its formation the charm diquark also diffuses in the QGP because it carries color charge. At the same time, the charm diquark may dissociate by absorbing a real or virtual gluon. So the whole process is a dynamical in-medium evolution involving charm diquark formation, diffusion and dissociation. This is similar to the in-medium evolution of heavy quarkonium described in Chapters 3 and 4, except that the heavy diquark carries color while quarkonium is color neutral. As a result, physical processes that change the momenta of heavy diquarks must be included in the evolution. At the transition from the deconfined QGP phase to the hadronic gas phase, the charm diquarks hadronize into doubly charmed baryons by absorbing an up or down quark from the medium.

We will describe the in-medium dynamical evolution of charm quarks and diquarks by generalizing the coupled Boltzmann equations that we have used in Chapter 4 to study the bottomonium production.

\vspace{0.2in}
\section{Coupled Boltzmann Transport Equations}

The set of coupled Boltzmann transport equations for the charm quark and diquark distribution functions $f({\bs x}, {\bs p}, t)$ is given by
\be \nn
(\frac{\partial}{\partial t} + \dot{{\bs x}}\cdot \nabla_{\bs x})f_c({\bs x}, {\bs p}, t) &=& \ml{C}_c  - \ml{C}_c^+ + \ml{C}_c^-\\
\label{eq:LBE}
(\frac{\partial}{\partial t} + \dot{{\bs x}}\cdot \nabla_{\bs x})f_{cc}({\bs x}, {\bs p}, t) &=& \ml{C}_{cc} + \ml{C}_{cc}^+ - \ml{C}_{cc}^-\,,
\ee
where all the collision terms $\ml{C}$, $\ml{C}^{\pm}$ depend on ${\bs x}, {\bs p}, t$.
Here we will focus on the ground charm diquark state $(cc)_{\bar{3}}$(1S) because excited states are loosely bound and cannot survive at high temperature. In the following, by charm diquark we mean the $(cc)_{\bar{3}}$(1S) state.
The collision terms $\ml{C}_c$ and $\ml{C}_{cc}$ describe their collisions with thermal constituents of the QGP. These collisions will change the momenta of charm quarks and diquarks. For $\ml{C}_{cc}$, we simply treat the diquark as a single heavy quark. We will use the calculation and implementation of Ref.~\cite{Ke:2018tsh} as in Chapter 4. The diquark gain term $\ml{C}_{cc}^+$ is from the combination of a charm quark pair by gluon emission and the loss term $\ml{C}_{cc}^-$ is from dissociation by gluon absorption. The formation and dissociation of diquarks also change the charm quark distribution function, which are represented by $\ml{C}_{c}^{\pm}$, as in the case of quarkonium transport. The transport equation of diquarks can be similarly derived from the formalism of open quantum system, as done for quarkonium in Chapter 3. Here we will not repeat the derivation. We will focus on calculating the combination and dissociation terms by using an EFT for heavy diquarks.

\vspace{0.2in}
\section{Potential NRQCD: Heavy Diquark}
We will use a pNRQCD for the diquark sector \cite{Brambilla:2005yk,Fleming:2005pd}. The EFT can be derived from QCD under the hierarchy of scales $M \gg Mv \gg Mv^2 \gtrsim T \gtrsim m_D$ where $M=1.3$ GeV is the charm quark mass, $v\sim0.4$ is the relative velocity of the $cc$ pair inside the diquark, $T$ is the QGP temperature, and $m_D$ is the Debye mass. If $T$ or $m_D$ scales as $Mv$, the Debye static screening of the color attraction is so strong that no diquark bound states can be formed inside the QGP. So the above hierarchy of scales is relevant to the diquark formation. 

The construction starts with two heavy quark fields of NRQCD $\psi_i({\bs x}_1,t)\psi_j({\bs x}_2,t)$ where $i$ and $j$ are color indexes. We want to map it onto two composite fields which depend on the c.m.~position ${\bs R} = {\bs x}_1 + {\bs x}_2$ and the relative position ${\bs r} = \frac{{\bs x}_1-{\bs x}_2}{2}$:
\be
\psi_i({\bs x}_1,t)\psi_j({\bs x}_2,t) \sim t^l_{ij}T^l({\bs R}, {\bs r}, t) + \sigma^{\nu}_{ij}\Sigma^{\nu}({\bs R}, {\bs r}, t)\,,
\ee
where $T^l$ and $\Sigma^{\nu}$ are the anti-triplet and sextet fields. The generators of the anti-triplet $t^l$ and the sextet $\sigma^{\nu}$ representations are given by
\be
 t^l_{ij} &=& \frac{1}{\sqrt{2}}\epsilon_{ijl}\\
 \label{eqn:six}
 \sigma^{1}_{11} &=& \sigma^{4}_{22}= \sigma^{6}_{33} = 1 \ \ \ \ \ \sigma^{2}_{12} =  \sigma^{2}_{21} =  \sigma^{3}_{13} =  \sigma^{3}_{31} =  \sigma^{5}_{23} =  \sigma^{5}_{32} = \frac{1}{\sqrt2}\,.
\ee
But we need to make sure that both sides of the mapping have the same gauge transformation properties. So we add Wilson lines 
\be
\Psi_{ij}({\bs x}_1, {\bs x}_2, t) &\equiv& \psi_i({\bs x}_1,t) \psi_j({\bs x}_2,t) \\
\Psi_{ij}({\bs x}_1, {\bs x}_2, t) &=&  W_{ii'}({\bs x}_1, {\bs R}, t) W_{jj'}({\bs x}_2, {\bs R}, t) \big[ t^l_{i'j'}T^l({\bs R}, {\bs r}, t) + \sigma^{\nu}_{i'j'}\Sigma^{\nu}({\bs R}, {\bs r}, t) \big] \ \ \ \ \ \  \\
W_{ij}({\bs y}, {\bs z}, t) &=& \bigg( \exp\Big\{ig\int_{\bs z}^{{\bs y}}\diff{\bs r} \cdot {\bs A}({\bs r},t) \Big\} \bigg)_{ij} \,,
\ee 
where the gauge field is $A = T^aA^a$. The Lagrangian of the fields $\Psi_{ij}({\bs x}_1, {\bs x}_2, t) $ can be written down from the Lagrangian of NRQCD
\be
\label{chap5_eqn_Lpre}
\ml{L}({\bs x}_1, {\bs x}_2, t) =  \Tr\Big\{ \Psi^{\dagger}({\bs x}_1, {\bs x}_2, t) \Big( iD_0 + \frac{{\bs D}^2_{{\bs x}_1}}{2M}   + \frac{{\bs D}^2_{{\bs x}_2}}{2M} +\cdots   \Big) \Psi ({\bs x}_1, {\bs x}_2, t) \Big\}\,.
\ee
Expanding Eq.~(\ref{chap5_eqn_Lpre}) in terms of the coupling constant and the relative distance $r$, one can show the Lagrangian of pNRQCD is given by
\be \nn
\ml{L}_{\ma{pNRQCD}}({\bs R}, t) &=& \int \Diff{3}r \Tr \Big\{ \ma{T}^{\dagger} (iD_0 - H_T) \ma{T} 
+ \Sigma^{\dagger} (iD_0 - H_{\Sigma}) \Sigma  \\
&+& \ma{T}^{\dagger} {\bs r} \cdot g {\bs E} \Sigma 
+ \Sigma^{\dagger} {\bs r} \cdot g {\bs E} \ma{T}\Big\} + \cdots\,,
\ee
where higher order interaction terms in $v^2$ and $r$ are omitted.
The Lagrangian of light quarks and gluons is just QCD with momenta $\lesssim Mv$. The degrees of freedom are the anti-triplet $\ma{T}(\bs R, \bs r, t)$ and sextet $\Sigma(\bs R, \bs r, t)$. They are defined as
\be
\ma{T} = t^l T^l \ \ \ \ \ \ \ \ \Sigma = \sigma^{\nu} \Sigma^{\nu} \,.
\ee

The equations of motion of the anti-triplet and sextet are Schr\"odinger equations with the Hamiltonians expanded in powers of $1/M$ or $v^2$
\be
H_{T,\Sigma} &=& -\frac{{\bs D}_{\bs R} ^2 }{4M} - \frac{\nabla_{\bs r}^2}{M}  + V_{T,\Sigma}^{(0)} + \frac{V_{T,\Sigma}^{(1)}}{M} + \frac{V_{T,\Sigma}^{(2)}}{M^2} + \cdots\,,
\ee
where ${\bs D}_{\bs R}$ is the covariant derivative associated with the c.m.~position. By the virial theorem, $-\nabla_{\bs r}^2/M \sim V_{T,\Sigma}^{(0)}$. So the order of the relative kinetic term is accounted as leading order $\sim Mv^2$. The c.m.~kinetic term is suppressed because momenta $\sim Mv$ have been integrated out in the construction and then ${\bs D}_{\bs R}\ll Mv$ as argued similarly in Chapter 1. Higher-order terms of the potentials are also suppressed by $v^2$ which include relativistic corrections, spin-orbital and spin-spin interactions. We only work to the order $v^2$ since the charm quark mass is large. At this order, the Hamiltonians only contain the relative kinetic term and $V_{T,\Sigma}^{(0)}$. The leading order potentials in the Lagrangian are given by Coulomb interactions
\be
V_{T}^{(0)} = -\frac{2}{3}\frac{\alpha_s}{r}\ \ \ \ \ \ \ \ V_{\Sigma}^{(0)} = \frac{1}{3}\frac{\alpha_s}{r}\,,
\ee
which are approximately valid inside the QGP because the confining part is significantly screened. 

We will assume the medium is translationally invariant, as in the quarkonium case. Then the interaction between the anti-triplet diquark and the medium can be decomposed into two parts: a part that only changes the c.m.~motion and leaves the bound state intact and the other part that only modifies the relative motion and can destroy the bound state. The decomposition is explicit in the pNRQCD Lagrangian by the multipole expansion. At the order we are working, the c.m.~motion part is fully described by the gauged kinetic term of the anti-triplet field, in the same way as the interaction between the open heavy quarks and the medium. Through the coupling of this term, the c.m.~motion of a diquark can change when it scatters with medium constituents, similar to the scattering between open heavy quarks and the constituents (see $\ml{C}_{c}$ and $\ml{C}_{cc}$ in expression (\ref{eq:LBE})). The change of the relative motion is described by terms of at least linear order in $r$. For example, the anti-triplet can interact with the sextet via a color dipole interaction where the chromoelectric field is given by
\be
{\bs E} = T^a {\bs E}^a\,,
\ee
and $T^a$ is the generator of the fundamental representation.

\begin{figure}
\centering
\vspace{0.1in}
\includegraphics[width=2.5in]{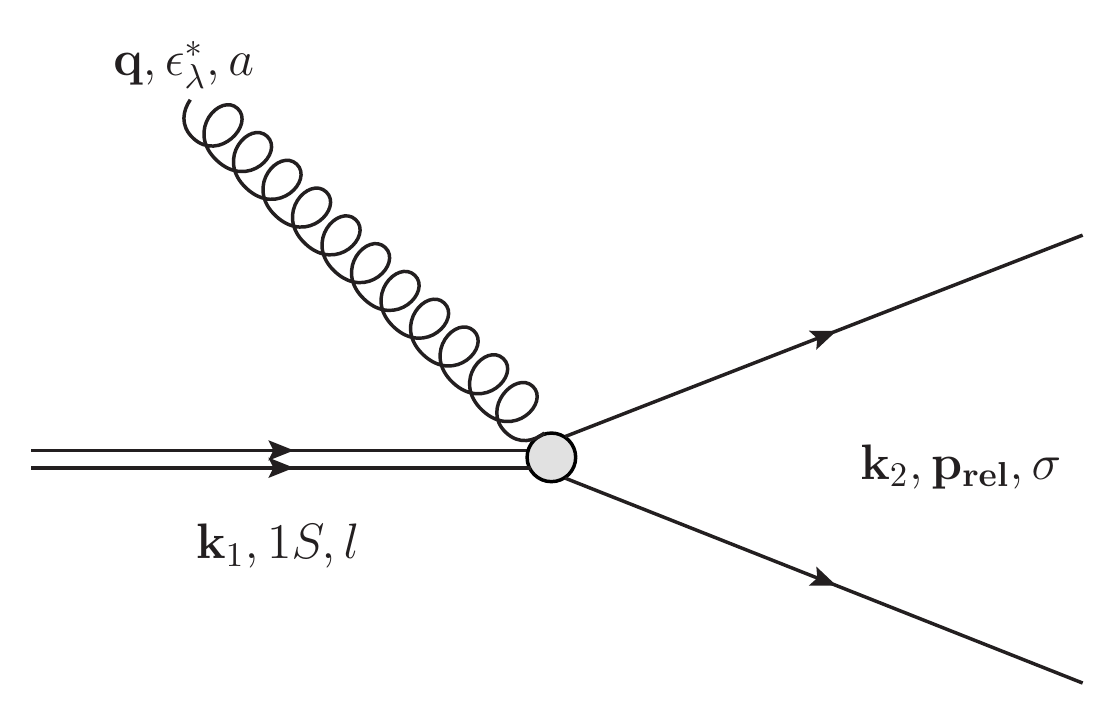}
\caption[Transition between a bound charm diquark in the anti-triplet and an unbound charm quark pair in the sextet.]{Transition between a bound charm diquark in the anti-triplet and an unbound charm quark pair in the sextet by absorbing or emitting an on-shell gluon. Narrow double lines indicate the diquark while widely open double lines represent the unbound pair.}\label{fig:t_sigma}
\end{figure}

At leading order in $r$, the transition between unbound charm quark pairs and bound diquarks can only occur between an unbound sextet and a bound anti-triplet. 
The Feynman diagram of the transition via gluon absorption or emission is shown in Fig.~\ref{fig:t_sigma}. For simplicity, we only consider the interaction with on-shell gluons in the QGP. Transitions caused by virtual gluons (inelastic scattering with medium constitutes) are at next order in $\alpha_s$ and neglected here. They can be easily included as in Chapter 3. The scattering amplitude in Coulomb gauge is given by 
\be
\ml{T}^{\nu l a}_\lambda &=& (2\pi)^4 \delta^3({\bs k}_1 + {\bs q} - {\bs k}_2) \delta(E_{1S} +q -\frac{p_\ma{rel}^2}{M}) \ml{M}^{\nu l a}_\lambda\\
\ml{M}^{\nu l a}_\lambda &=& -igq \Tr( \sigma^{\nu}  T^a t^l ) (\epsilon_{\lambda}^*)_i \langle  \Psi_{{\bs p}_\ma{rel}}| r_i | \psi_\ma{1S} \rangle \,,
\ee
where ${\bs k}_{1,2}$ are the c.m.~momenta, ${\bs p}_{\ma{rel}}$ is the relative momentum between the unbound quark pair and $q=|{\bs q}|$ is the gluon energy. In the matrix element, $|\psi_{1S}\rangle$ is the hydrogen-like $1S$ wave function for the bound diquark in the anti-triplet, and $| \Psi_{{\bs p}_\ma{rel}} \rangle$ is the Coulomb wave function for the unbound sextet. The dipole matrix element $\langle  \Psi_{{\bs p}_\ma{rel}}| r_i | \psi_\ma{1S} \rangle$ has been calculated in Chapter 3. The $1S$ binding energy is given by $E_{1S} = - \alpha_s^2M/9$. According to the power counting explained above, the c.m.~kinetic energies have been neglected. Throughout this section we will set $\alpha_s = g^2/(4\pi)= 0.4$. 

To calculate rates, we need to average and sum over certain quantum numbers. For convenience, we define
\be
\overline{|\ml{M}|^2} & \equiv & \sum_{a=1}^8 \sum_{l=1}^3 \sum_{\nu=1}^6 \sum_{\lambda = \pm}|\ml{M}^{\nu l a}_\lambda|^2 
= 2g^2q^2 |   \langle  \Psi_{{\bs p}_\ma{rel}}  | {\bs r} |\psi_\ma{1S}\rangle |^2\,.
\ee
Most calculations are the same as in the quarkonium case. We will explain the summations over the color indexes here
\be \nn
&&\sum_{a=1}^8 \sum_{l=1}^3 \sum_{\nu=1}^6 | {\sigma}^\nu_{km}  T^a_{mn} t^l_{nk} |^2 = \sum_{a=1}^8 \sum_{l=1}^3 \sum_{\sigma=1}^6 \big(   {\sigma}_{km}^\nu T^a_{mn} t^l_{nk}   \big) \big(   \sigma^{\nu*}_{k'm'}  T^{a*}_{m'n'} t^{l*}_{n'k'}   \big) \\  \nn
&=&  \sum_{\nu=1}^6  {\sigma}^\nu_{km}  {\sigma}^\nu_{k'm'} \Big( \sum_{l=1}^3 t^l_{nk}  t^l_{n'k'} \Big)\Big( \sum_{a=1}^8 T^a_{mn} T^a_{n'm'} \Big) \\ \nn
&=&  \sum_{\nu=1}^6  {\sigma}^\nu_{km}  {\sigma}^\nu_{k'm'} \Big(\sum_{l=1}^3\frac{1}{2}\epsilon_{lnk}\epsilon_{ln'k'} \Big)\Big( \frac{1}{2}\delta_{mm'}\delta_{nn'} - \frac{1}{6}\delta_{mn}\delta_{m'n'} \Big) \\ \nn
&=&  \sum_{\nu=1}^6  {\sigma}^\nu_{km}  {\sigma}^\nu_{k'm'} \frac{1}{2}   \Big(\delta_{nn'}\delta_{kk'} - \delta_{nk'}\delta_{n'k} \Big)\Big( \frac{1}{2}\delta_{mm'}\delta_{nn'} - \frac{1}{6}\delta_{mn}\delta_{m'n'} \Big)\\ \nn
&=& \sum_{\nu=1}^6 \Big[ \big(\frac{1}{2} - \frac{1}{12}\big)  {\sigma}^\nu_{mn} {\sigma}^\nu_{mn} +\frac{1}{12} {\sigma}^\nu_{mn} {\sigma}^\nu_{nm} \Big] \\ \nn
&=& \sum_{\nu=1}^6 \frac{1}{2} {\sigma}^\nu_{mn} {\sigma}^\nu_{mn} = \frac{1}{2} \times 6 = 3\,,
\ee
where we have used
\be
(T^a)^*_{ij} &=& T^a_{ji}\\
{\sigma}_{mn}^\nu &=&{\sigma}^\nu_{nm} \\
{\sigma}_{mn}^\nu{\sigma}^{\nu'}_{mn} &=& \delta^{\nu\nu'}\,.
\ee

To write out the dissociation and combination rates out explicitly, we first define
\be \nn
\ml{F}^+ &\equiv& \frac{1}{2} g_+ \int\frac{\Diff{3}p_1}{(2\pi)^3}\frac{\Diff{3}p_2}{(2\pi)^3} \frac{\Diff{3}k_1}{(2\pi)^3} \frac{\Diff{3}q}{(2\pi)^32q} 
\big(1+n_B{(q)}\big)  f_c({\bs x}, {\bs p}_1, t)f_c({\bs x}, {\bs p}_2, t) \\
\label{chap5_eqn:reco}
&& (2\pi)^4 \delta^3({\bs k}_1 + {\bs q} - {\bs k}_2) \delta(E_{1S} +q -\frac{p_\ma{rel}^2}{M}) \overline{|\ml{M}|^2}  \\ \nn
\ml{F}^- &\equiv& \frac{1}{2} g_- \int \frac{\Diff{3}k_1}{(2\pi)^3}  \frac{\Diff{3}k_2}{(2\pi)^3}  \frac{\Diff{3}p_{\ma{rel}}}{(2\pi)^3} \frac{\Diff{3}q}{(2\pi)^32q}n_B{(q)}  f_{cc}({\bs x}, {\bs k}_1, t)  \\
&&  (2\pi)^4 \delta^3({\bs k}_1 + {\bs q} - {\bs k}_2) \delta(E_{1S} +q -\frac{p_\ma{rel}^2}{M}) \overline{|\ml{M}|^2} \, ,
\ee
where ${\bs p}_{\ma{rel}}$ and ${\bs k}_2$ are the relative and c.m.~momenta of the unbound charm quark pair with momenta ${\bs p}_1$ and ${\bs p}_2$. The pre-factor $\frac{1}{2}$ avoids double counting in the phase space of two charm quarks. The $g$-factors are given by 
\be
g_+ &=& \frac{2J+1}{(2S+1)^2}\frac{d_6}{N_c^2}\frac{1}{d_6} = \frac{1}{12} \\
g_- &=& \frac{1}{d_{\bar{3}}} = \frac{1}{3}\,,
\ee
where $J=1$ is the diquark spin, $S=\frac{1}{2}$ is the heavy quark spin, $d_6 = 6$ is the sextet multiplicity and $d_{\bar{3}}=3$ is the anti-triplet multiplicity. For the formation process, one needs to average over the initial sextet multiplicity and only a fraction ${d_6}/{N_c^2}$ of unbound charm quark pairs are in the sextet, which can form a diquark by radiating out a gluon at the order of $r$. The formed 1S diquark is a color anti-triplet and thus has to be in the spin triplet because of the antisymmetric nature of fermions. So another spin factor $\frac{2J+1}{(2S+1)^2} =\frac{3}{4}$ is inserted. For the dissociation process, one needs to average over the initial anti-triplet multiplicity.
The phase space measure is relativistic for gluons and nonrelativistic for charm quarks and diquarks, which is consistent with our field definitions. 
Formation from unbound anti-triplet pairs only happens at higher orders in $r$ and $v^2$.

The gain and loss collision terms in the Boltzmann transport equations can be written as
\be
\label{eq:formterm}
\ml{C}_{c}^{\pm} &=& \frac{\delta \ml{F}^{\pm} }{\delta{{\bs p}_1}}  \bigg|_{{\bs p}_1 = {\bs p}} + \frac{\delta \ml{F}^{\pm} }{\delta{{\bs p}_2}}  \bigg|_{{\bs p}_2 = {\bs p}} \\
\ml{C}_{cc}^{\pm} &=& \frac{\delta \ml{F}^{\pm} }{\delta{{\bs k}_1}}  \bigg|_{{\bs k}_1 = {\bs p}} \,,
\ee
where the ``$\delta-$derivative" symbol has been defined in Eq.~(\ref{chap3_eqn_delta}) in Chapter 3.

The rate of charm quarks combining $\Gamma_f$ and the dissociation rate of a diquark $\Gamma_d$ can be defined as
\be
\label{eq:formrate}
\ml{C}_{c}^{+} &\equiv& \Gamma_f({\bs x}, {\bs p}, t) f_{c}({\bs x}, {\bs p}, t) \\
\ml{C}_{cc}^{-} &\equiv& \Gamma_d({\bs x}, {\bs p}, t) f_{cc}({\bs x}, {\bs p}, t) \, .
\ee

The scattering amplitude and the rate are calculated in the rest frame of the diquark for dissociation and that of the unbound quark pair for formation, where the pNRQCD is valid. When the diquark or the unbound pair moves relatively to the medium, the Bose distribution of medium gluons $n_B{(q)}$ needs to be boosted into the rest frames, respectively, as demonstrated in Chapter 4. The two frames are not equivalent but since the gluon energy is small compared to $M$ ($T\ll M$), the difference is suppressed by $T/M$.

\vspace{0.2in}
\section{Estimate of Production Rate}
\begin{figure}[h]
    \centering
    \begin{subfigure}[t]{0.48\textwidth}
        \centering
        \includegraphics[height=2.2in]{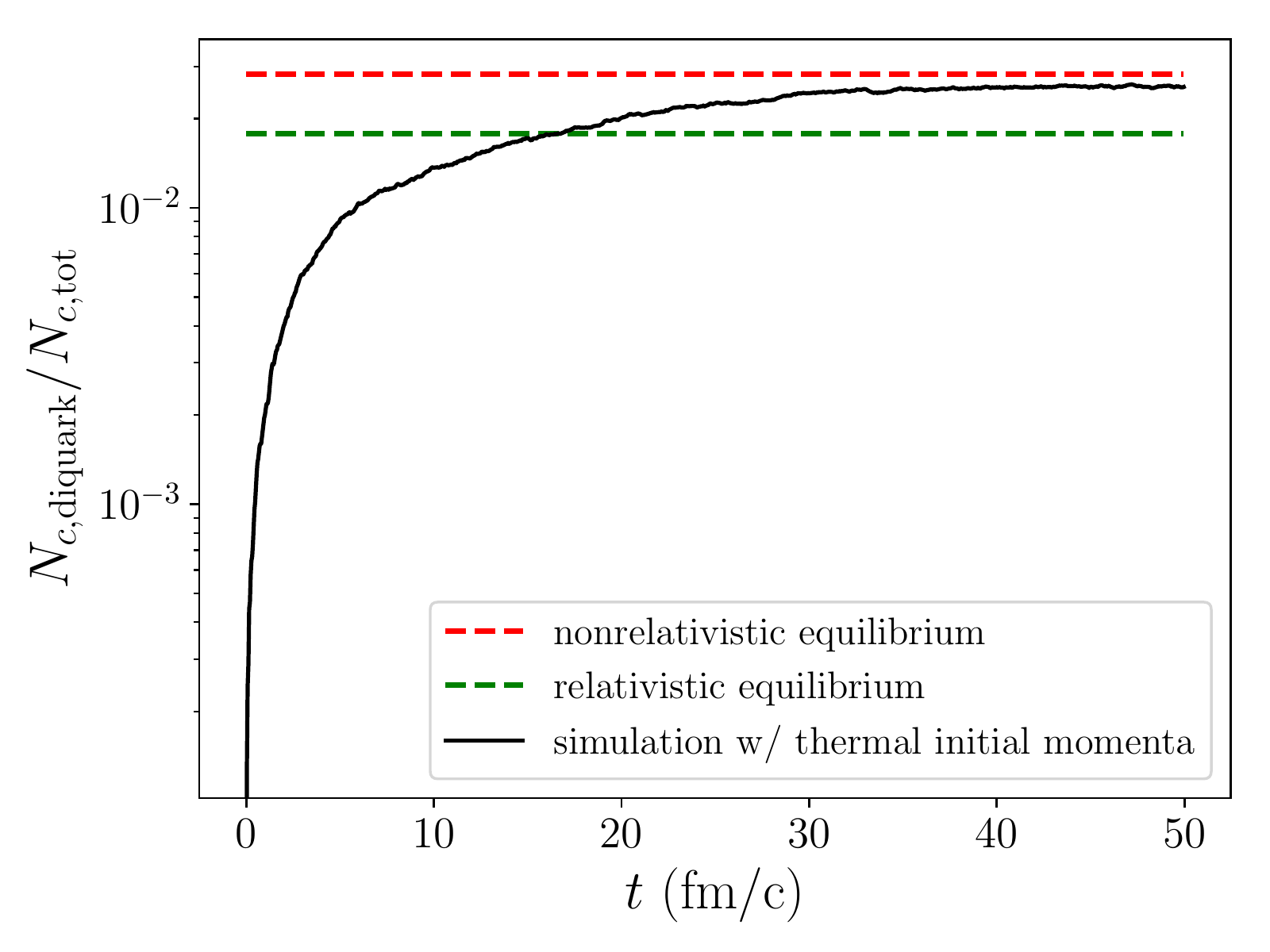}
        \caption{$T=250$ MeV.}
    \end{subfigure}%
    ~ 
    \begin{subfigure}[t]{0.48\textwidth}
        \centering
        \includegraphics[height=2.2in]{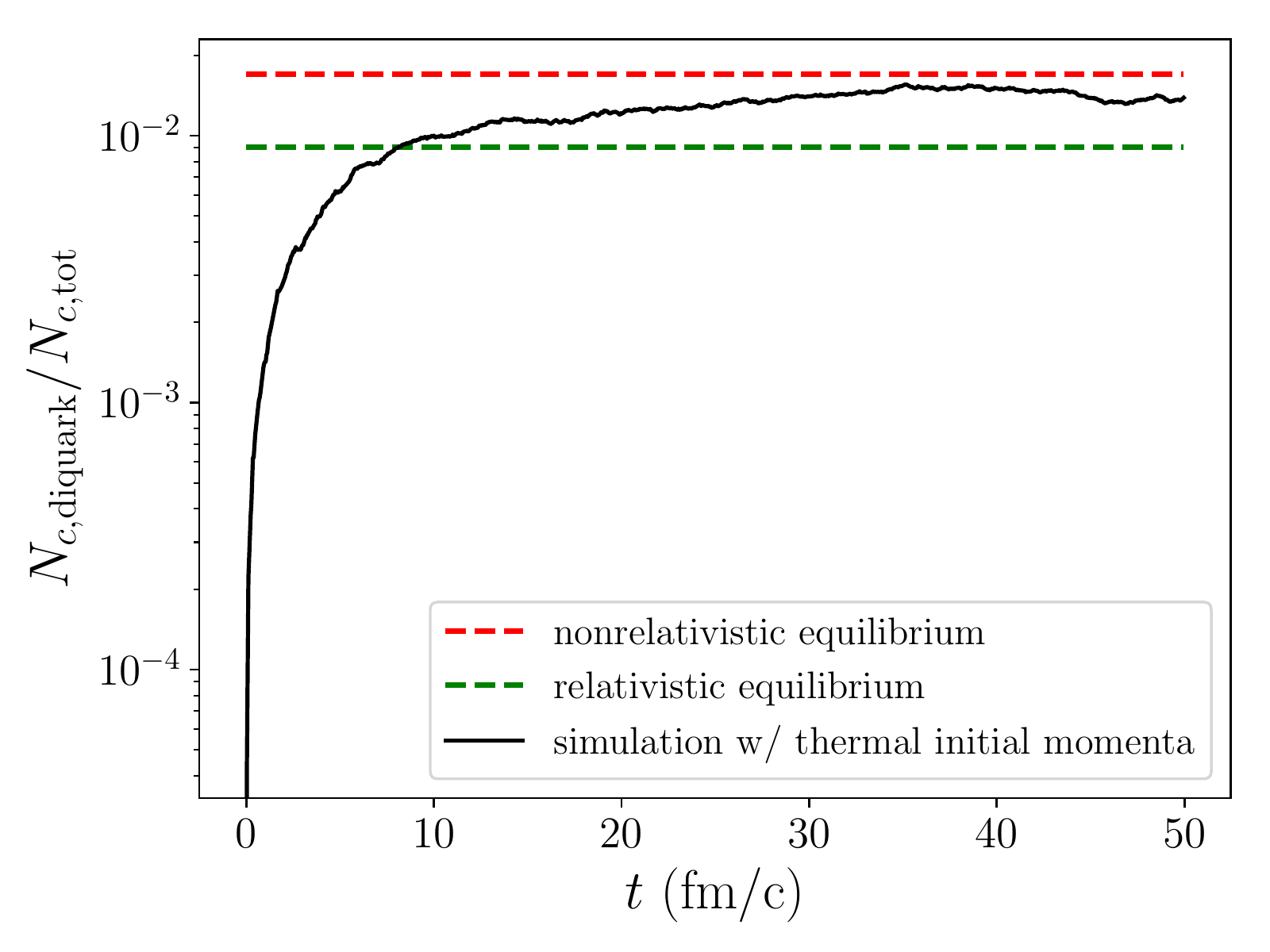}
        \caption{$T=350$ MeV.}
    \end{subfigure}%
    \caption{Simulation results on the diquark percentage compared with that at thermal equilibrium. }
    \label{chap5_fig:eq}
\end{figure}

We will solve the coupled Boltzmann equations in Monte Carlo methods, as explained in Chapter 4 in detail. Before we show phenomenological results, we will show some test studies. As explained in Chapter 4, we will test the evolution of the system consisting of charm quarks and diquarks inside a QGP box with a side length $L=10$ fm and a constant temperature. We sample $N_{c,\ma{tot}} = N_{\bar{c},\ma{tot}} = 30$ charm and anti-charm quarks initially. Their momenta are sampled from the Boltzmann distribution (\ref{chap4_eqn_boltzmann_distribution}) and their positions are sampled randomly inside the QGP box. We keep track of the diquark percentage defined by
\be
\frac{N_{c,\ma{diquark}}}{N_{c,\ma{tot}}} = \frac{N_{c,\ma{diquark}}}{N_{c,\ma{open}}+N_{c,\ma{diquark}}} \,,
\ee
and compare it with the quantity at thermal equilibrium. The diquark percentage at thermal equilibrium can be calculated by using Eq~(\ref{chap4_eqn_Neq}). The fugacities are related by $\lambda_c^2 = \lambda_{cc}$ and can be solved from the number conservation: $N_{c}^\ma{eq} + 2N_{cc}^\ma{eq} = N_{c,\ma{tot}}$. The results and the comparison are shown in Fig.~\ref{chap5_fig:eq}. We see that the interplay between combination and dissociation can drive the system to a detailed balance, as in the quarkonium case. If the momentum spectrum is thermal, the diquark percentage at detailed balance is the same as that in thermal equilibrium. The simulation result of the diquark percentage agrees better with the equilibrium property with a nonrelativistic dispersion relation because the energy-momentum conservation in the pNRQCD scattering amplitude is nonrelativistic.

Next we move on to study the production in real collisions. We will assume only open charm quarks are produced in the initial hard binary collisions. No charm diquarks are produced in the beginning. The initial transverse momentum and rapidity distribution from the hard scattering is calculated from FONLL \cite{FONLL} with the nuclear PDF EPS$09$ \cite{Eskola:2009uj}, as in our study on bottomonium production in Chapter 4. The FONLL calculation is done with the renormalization and factorization scale $m_T=\sqrt{M^2+p_T^2}$. The number of charm quarks produced in one collision event is determined by $\sigma^{p+p\to c+X} T_{AA}$, the product of the cross section $\sigma$ per binary collision calculated in FONLL, and the nuclear thickness function $T_{AA}$ calculated from binary collision models. Here we will focus on collisions with $0-10\%$ centrality, which corresponds to impact parameters from $0$ to $5$ fm roughly and $T_{AA}\approx23$ mb \cite{Abelev:2013qoq}.

The initial position of the charm quark produced is sampled using the \trento\, model, as in our study on bottomonium production in Chapter 4. The charm quark production is a short-distance process due to its large mass, implying that its initial position is roughly the same as the location where the two parent nucleons scatter.

The medium we will use is provided by the $2+1$ dimensional viscous hydrodynamic simulation VISHNU, as in Chapter 4. 
With the initial condition and hydrodynamical background given, we solve the transport equations by test particles Monte Carlo simulations. The hydrodynamic medium is assumed to be formed at the co-moving time $\tau=0.6$ fm/c where $\tau = \sqrt{t^2-z^2}$ and $t$ and $z$ are observed in the laboratory frame. Before this, we assume the charm quarks are just free-streaming without interactions. After $\tau=0.6$ fm/c, we consider three types of processes at each time step $\Delta t=0.04$ fm/c in the laboratory frame: momentum change of open charm quarks and diquarks, formation and dissociation of charm diquarks.

First, for each charm quark and diquark, we determine their thermal scattering rate with medium constituents. The product of the rate and time step $\Delta t$ gives the scattering probability. Then we use random numbers to determine whether a certain process occurs. If so, we sample the momenta of the incoming medium constituent from a thermal distribution and obtain the momenta of outgoing particles by energy-momentum conservation. Finally, we update both particles' momenta and positions after one time step.

Second, for each diquark, we calculate its dissociation rate and probability within a time step as above. If the diquark is determined to dissociate, we replace it by two unbound charm quarks whose momenta are determined from energy-momentum conservation and whose positions are given by that of the diquark just before the dissociation. Here we just briefly explain the implementation of each process. Detailed descriptions have been given in Chapter 4.

Finally, for each charm quark with position ${\bs y}_i$ and momentum $\tilde{{\bs p}}_i$, whose neighboring charm quarks have positions ${\bs y}_j$ and momenta $\tilde{{\bs p}}_j$, we need to determine the diquark formation rate by using expressions (\ref{chap5_eqn:reco}), (\ref{eq:formterm}), (\ref{eq:formrate}). We will replace the product of two delta functions in positions by the product of a Gaussian function in the relative position and a delta function in the c.m.~position, as done in Chapter 4.
The width of the Gaussian function is chosen as the diquark Bohr radius $a_B=\alpha_sM/3$. This ensures that the combination rate for a widely separated charm quark pair vanishes. The product of the local distributions in (\ref{chap5_eqn:reco}) is thus replaced with
\be \nn
&& f_c({\bs x}, {\bs p}_1, t)f_{c}({\bs x}, {\bs p}_2, t) \rightarrow  \\
\label{eq:ff}
&&\sum_{i,j}  \frac{e^{-({\bs y}_i - {\bs y}_j)^2/2a_B^2}}{(2\pi a_B^2)^{3/2}} \delta^3\left({\bs x}-\frac{{\bs y}_i+{\bs y}_j}{2}\right)   
\delta^3({\bs p}_1-\tilde{{\bs p}}_i) \delta^3({\bs p}_2-\tilde{{\bs p}}_j)\,,
\ee
where the sum runs over all unbound charm quark pairs. For each charm quark $i$, the diquark formation rate in expression (\ref{eq:formrate}) involves a sum over $j$. If a diquark is formed, we replace the unbound charm quark pair by a diquark whose momentum is determined by momentum conservation and whose position is given by the c.m.~position of the quark pair as indicated in (\ref{eq:ff}).

When particles reach the hadronization hypersurface determined by the local transition temperature $T_c\approx 154$ MeV, each diquark combines with a thermal up or down quark to form a doubly charmed baryon. Here we use a simple hadronization model: a massless up or down quark is sampled from a Fermi-Dirac distribution with the temperature $T_c$, and its momentum is added to the diquark momentum to determine the baryon momentum. The baryon energy is fixed by the momentum and vacuum mass $m(\Xi_{cc})$.  We assume all diquarks end up as the ground $\Xi_{cc}$ states because excited states decay to the ground state much faster than the weak decay of the ground state \cite{Li:2017pxa,Lu:2017meb}. In this way, roughly half the diquarks end up as $\Xi_{cc}^{++}$. A more realistic hadronization model would include the effect of the baryon wave function on the coalescence probability.

\begin{figure}
\centering
\vspace{0.1in}
\includegraphics[width=3.0in]{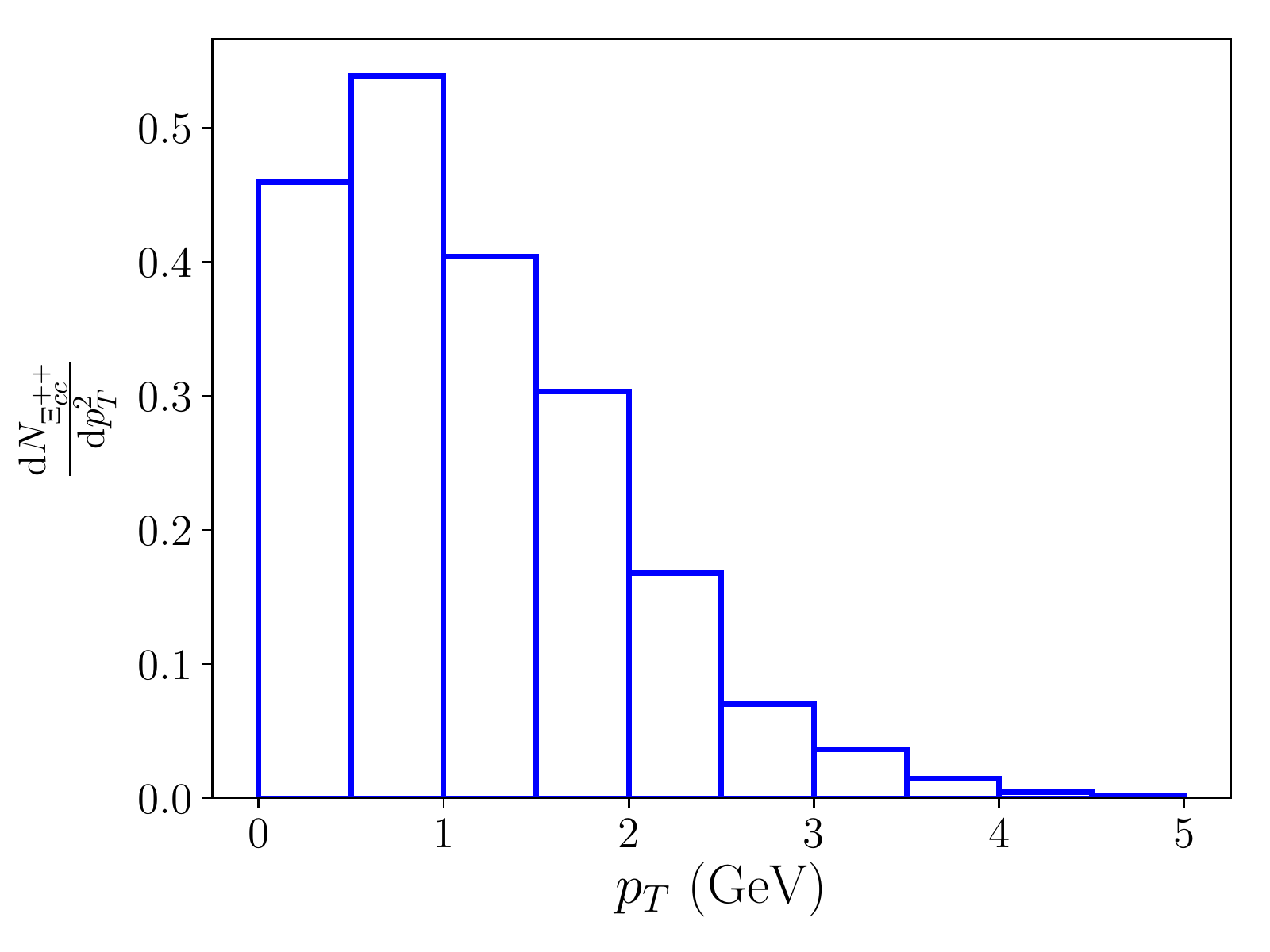}
\caption[$p_T$ spectra of emitted $\Xi_{cc}^{++}$ integrated over the rapidity window $-1 \leq y \leq 1$.]{$p_T$ spectra of emitted $\Xi_{cc}^{++}$ integrated over the rapidity window $-1 \leq y \leq 1$. The normalization is arbitrary.}
\label{fig:pt}
\end{figure}

We have simulated 40,000 nuclear collision events. In each event, the initial charm quark momentum is sampled over the range $p_T \in [0,\,30]$ GeV and $y \in [-8,\,8]$. At the end of each calculation, we accept $\Xi_{cc}^{++}$ in the kinematic range $p_T \in [0,\,5]$ GeV and $y \in [-1,\,1]$. The $p_T$ spectra integrated over this rapidity range are shown in Fig.~\ref{fig:pt}. The yield within this kinematic range is $N(\Xi_{cc}^{++})\approx0.02$ per collision.

So far, we have assumed that the diquark can be formed at any temperature. This cannot be true due to the Debye screening of the attractive color force inside the QGP. To understand the influence of Debye screening on $\Xi_{cc}^{++}$ production, we repeat the calculation but assume a melting temperature $T_{cc\ma{(1S)}}=250$ MeV above which the charm diquark cannot be formed inside the QGP. The yield in the same kinematic range is then reduced to $N(\Xi_{cc}^{++})\approx0.0125$ per collision.

The calculation can be extended to the production of other doubly heavy baryons, such as $\Xi_{bb}$ and $\Xi_{bc}$, and doubly heavy tetraquarks. For $\Xi_{bb}$, the only difference is that fewer bottom quarks are produced than charm quarks. This implies that the probability of having two bottom quarks come close and form a bottom diquark is much smaller. Thus, one expects a correspondingly smaller yield of $\Xi_{bb}$. For $\Xi_{bc}$, there exist extra dipole terms in the pNRQCD Lagrangian for transitions among anti-triplets (or sextets) \cite{Brambilla:2005yk}, which means that an unbound pair of bottom and charm quarks in the color anti-triplet can form a bound $bc$ diquark via a dipole transition.

For tetraquarks, the in-medium evolution of heavy quarks and diquarks proceeds in the same way, but the anti-triplet diquark hadronizes by coalescing with two light antiquarks. This process is analogous to the formation of an antibaryon containing a single heavy antiquark, while the formation of a doubly heavy baryon is analogous to the creation of a heavy meson. Heavy baryon ($\Lambda_c$) emission is known to be enhanced relative to heavy meson ($D^0$) emission in relativistic heavy ion collisions \cite{Zhou:2017ikn} as a consequence of quark recombination from the thermal quark-gluon plasma \cite{Oh:2009zj}, compared with proton-proton collisions. A similar enhancement of the production of doubly heavy tetraquarks, relative to the production of doubly heavy (anti-)baryons, can be expected. The measured ratio $\Lambda_c/D^0 \approx 1$ in Au+Au collisions at RHIC suggests that the yield of doubly heavy baryons and tetraquarks should also be approximately equal.

\vspace{0.2in}
\section{Free Energy of Heavy Diquark}

The melting temperature of heavy diquarks can be studied from their free energies, in a similar way as quarkonia melting temperatures \cite{Mocsy:2013syh}. The free energy of a heavy quark pair could be studied on a lattice by calculating the correlations of two Polyakov loops at different lattice locations, where each Polyakov loop corresponds to a static thermal heavy quark \cite{McLerran:1981pb}. The free energy projected onto the color anti-triplet state can be used to study the binding energies and spectral functions of diquarks, from which one can obtain the melting temperature. The projections onto the anti-triplet and sextet states were first studied in Ref.~\cite{Nadkarni:1986as}. In this section, we explain how to project onto the anti-triplet in a gauge invariant but path dependent way. We also show that under a weak coupling expansion, the free energy of a pair of heavy quarks in the anti-triplet is the sum of the free energies of two individual heavy quarks and their attractive potential energy. A previous gauge dependent lattice study can be found in Ref.~\cite{Doring:2007uh}.

The anti-triplet and sextet states of a heavy quark pair at different lattice locations can be defined as
\be
| QQ_{\bar{3}}({\bs 0}, {\bs r}, \tau)\rangle^{l} &\equiv& \frac{1 }{\sqrt{2}}\epsilon_{ikl}\psi_i^{\dagger}({\bs 0},\tau)  \psi_j^{\dagger}({\bs r},\tau) W^{\dagger}_{jk} ( ({\bs 0},\tau), ({\bs r},\tau))|s\rangle \\
| QQ_{6}({\bs 0}, {\bs r}, \tau)\rangle^{\nu} &\equiv& \sigma^{\nu}_{ik} \psi_i^{\dagger}({\bs 0},\tau)  \psi_j^{\dagger}({\bs r},\tau) W^{\dagger}_{jk} ( ({\bs 0},\tau), ({\bs r},\tau))|s\rangle\,,
\ee
where $\tau$ is the Euclidean time and $|s\rangle$ can be any state with no heavy quarks. The symbol $\sigma^{\nu}_{ik}$ is defined in the expression (\ref{eqn:six}) and satisfies $\sigma^{\nu}_{ik}\sigma^{\nu}_{i'k'} = (\delta_{ii'}\delta_{kk'} + \delta_{ik'}\delta_{i'k})/2$. The symbol $W(y,z)$ denotes a Wilson line from lattice site $z$ to site $y$. The definitions depend on the spatial path of the Wilson line. The heavy quark annihilation $\psi$ and creation $\psi^{\dagger}$ operators satisfy the anti-commutation relation on the lattice
\be
\{  \psi_{i}({\bs r},\tau) , \psi_j^{\dagger}({\bs r'},\tau)   \} = \delta_{{\bs r}{\bs r'}} \delta_{ij}\,.
\ee

\begin{figure}[h]
     \begin{subfigure}[t]{0.5\textwidth}
        \centering
        \includegraphics[height=2.2in]{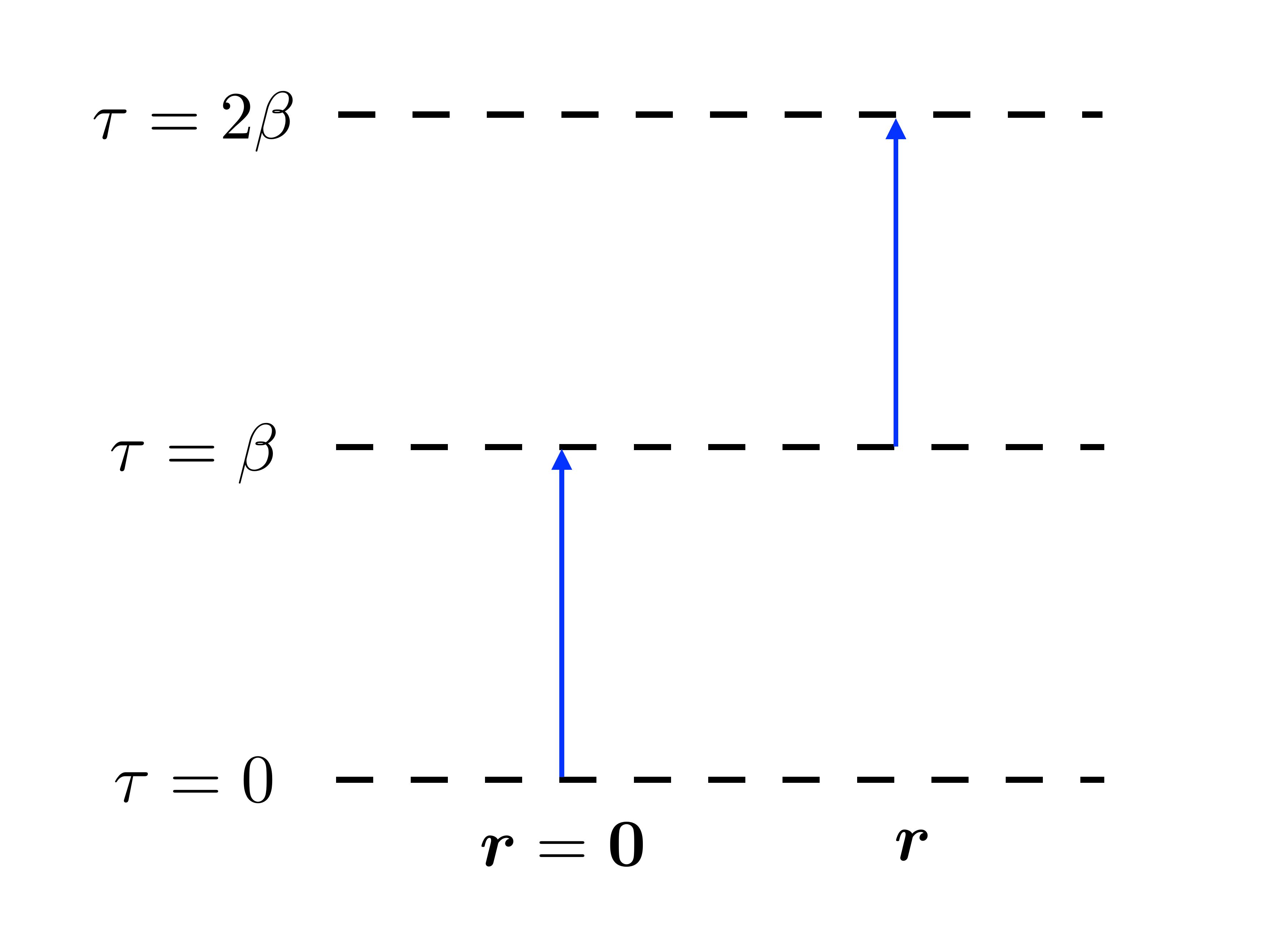}
        \caption{$ \langle  \Tr L({\bs 0}) \Tr L({\bs r})  \rangle_T $.}
    \end{subfigure}%
    ~
    \begin{subfigure}[t]{0.5\textwidth}
        \centering
        \includegraphics[height=2.2in]{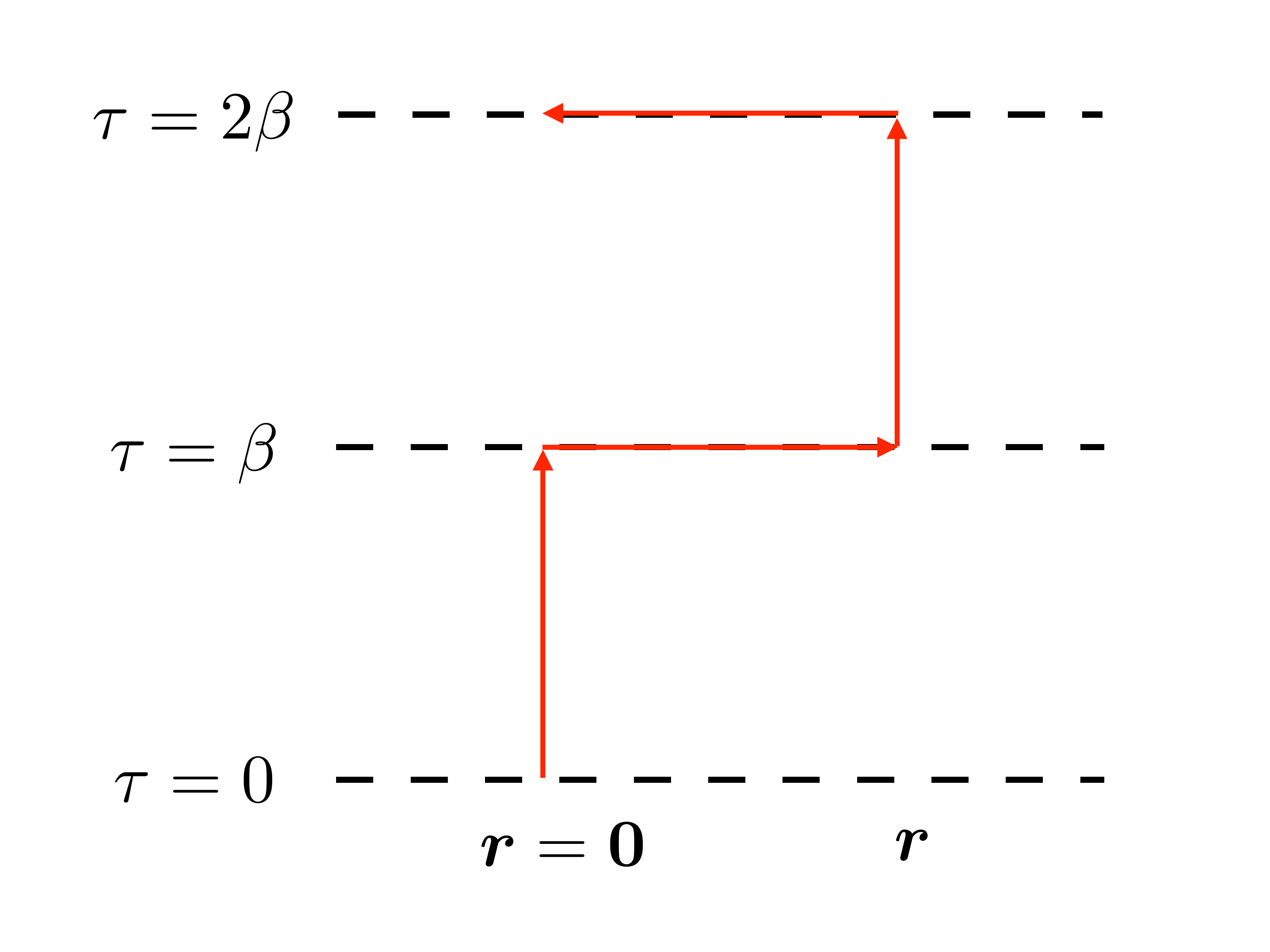}
        \caption{$ \langle   \Tr[W  L({\bs r})  W^{\dagger}  L({\bs 0})      ]    \rangle_T $.}
    \end{subfigure}%
\caption[Schematic diagrams for the correlation terms in $\exp(-F_{QQ_{\bar{3}}}(\bs r)/T)$.]{Schematic diagrams for the correlation terms in $\exp(-F_{QQ_{\bar{3}}}(\bs r)/T)$. The sub-plots (a) and (b) correspond to the first and second terms separately. The three black dashed lines label the same Euclidean time due to the periodicity at finite temperature. The region from $\tau=\beta$ to $\tau =2\beta$ is just a duplicate of the region from $\tau=0$ to $\tau = \beta$. In (a), the two blue arrows indicate the two Polyakov loops which are located at $\bs r = 0$ and $\bs r$. In (b), the four red arrows indicate the trace in the second term. It consists of a Polyakov line at $\bs r =0$, followed by a Wilson line from $\bs r = 0$ to $\bs r$, then another Polyakov line at $\bs r$ and finally a Wilson line from $\bs r$ to $\bs r = 0$. Though straight lines are used to denote the Wilson lines, they can be any spatial paths connecting the two ends. Due to the periodicity of gauge fields, the four red arrows form a loop.}
\label{fig:lattice}
\end{figure}

The free energy of a heavy quark pair in the anti-triplet can be defined as
\be \nn
e^{-F_{QQ_{\bar{3}} }(\bs r)/T} &=& \frac{1}{N_c}\sum_{|s\rangle}    \langle QQ_{\bar{3}}({\bs 0}, {\bs r}, 0)|^{l}  e^{-\beta H}  | QQ_{\bar{3}}({\bs 0}, {\bs r}, 0)\rangle^{l} \\ \nn
&=&\frac{1}{2N_c}\epsilon_{i'k'l}\epsilon_{ikl}\sum_{|s\rangle} \langle s | W_{k'j'}( ({\bs 0},0), ({\bs r},0) )\psi_{j'}({\bs r},0) \psi_{i'}({\bs 0},0) \\ \nn
&&e^{- \beta H} \psi_i^{\dagger}({\bs 0},0)  \psi_j^{\dagger}({\bs r},0) W^{\dagger}_{jk} ( ({\bs 0},0), ({\bs r},0) )|s\rangle \\\nn
&=&\frac{1}{6} (\delta_{ii'}\delta_{kk'} - \delta_{ik'}\delta_{i'k}) \sum_{|s\rangle} \langle s |  e^{- \beta H}  W_{k'j'}( ({\bs 0},\beta), ({\bs r},\beta) )  \\
&& \psi_{j'}({\bs r},\beta) \psi_{i'}({\bs 0},\beta) \psi_i^{\dagger}({\bs 0},0)  \psi_j^{\dagger}({\bs r},0) W^{\dagger}_{jk} ( ({\bs 0},0), ({\bs r},0) )   |s\rangle \,.
\ee 
In the static heavy quark limit,
\be
\psi_i ({\bs r},\beta) = \ml{T}(e^{ig\int_0^{\beta} \diff\tau A_0(\bs r, \tau)})_{ij} \psi_j ({\bs r},0) \equiv L(\bs r)_{ij} \psi_j ({\bs r},0)\,,
\ee 
where $\ml{T}$ is the time ordering operator. The starting and ending points of the Wilson line along the Euclidean time direction are the same due to the periodicity of gauge fields at finite temperature and is denoted as the Polyakov line $L(\bs r)$. Then using the anti-commutation relation of heavy quark operators it can be shown
\be \nn
e^{-F_{QQ_{\bar{3}} }(\bs r)/T} &=& \frac{1}{6} \langle  \Tr L({\bs 0}) \Tr L({\bs r})  \rangle_T \\
&& -\frac{1}{6} \langle   \Tr[W( ({\bs 0},\beta), ({\bs r},\beta)) L({\bs r})  W^{\dagger}( ({\bs 0},0), ({\bs r},0) )   L({\bs 0})      ]    \rangle_T\,,
\ee
where $\langle \hat{O} \rangle_T \equiv \sum_{|s\rangle} \langle s |e^{-\beta H} \hat{O} |s\rangle $ and $\Tr L$ is the Polyakov loop. Both the correlation terms in the above expression are gauge invariant because of the cyclic property of the trace and the periodicity of gauge fields. Schematic diagrams for the two correlation terms are shown in Fig.~\ref{fig:lattice}. 

In a similar way, the sextet free energy can be defined as
\be \nn
e^{-F_{QQ_6}(\bs r)/T} &=& \frac{1}{6}\sum_{|s\rangle}    \langle QQ_{6}({\bs 0}, {\bs r}, 0)|^{\nu}  e^{-\beta H}  | QQ_{6}({\bs 0}, {\bs r}, 0)\rangle^{\nu} \\  \nn
&=&\frac{1}{12} \langle  \Tr L({\bs 0}) \Tr L({\bs r})  \rangle_T \\
&& + \frac{1}{12} \langle   \Tr[W( ({\bs 0},\beta), ({\bs r},\beta)) L({\bs r})  W^{\dagger}( ({\bs 0},0), ({\bs r},0) )   L({\bs 0})      ]    \rangle_T\,,
\ee
which is also gauge invariant. Both definitions depend on the spatial paths of the Wilson lines.

Under a weak coupling expansion in powers of $g$, we obtain in the static gauge $\dot{A}_0=0$ (where $A_0 $ is a constant matrix)
\be \nn
e^{-F_{QQ_{\bar{3}}}({\bs r})/T} &=& 1 + \frac{g^2\beta^2}{12}\delta^{ab} \langle  A_0^{a}(\bs r)A_0^{b}(\bs 0)\rangle_T - \frac{g^2\beta^2}{12}\delta^{ab} \langle  A_0^{a}(\bs 0)A_0^{b}(\bs 0)\rangle_T \\
&&- \frac{g^2\beta^2}{12}\delta^{ab} \langle  A_0^{a}(\bs r)A_0^{b}(\bs r)\rangle_T +\ml{O}(g^3)\,.
\ee
The last two terms are independent of the positions and are just the free energies of two individual heavy quarks at order $g^2$. The free energy of a single heavy quark can be calculated from
\be
e^{-F_Q/T} = \frac{1}{3} \langle \Tr L  \rangle_T \,.
\ee
Therefore,
\be
F_{QQ_{\bar{3}} }({\bs r})  = 2F_Q - \frac{g^2\beta}{12}\delta^{ab} \langle  A_0^{a}(\bs r)A_0^{b}(\bs 0)\rangle_T+ \ml{O}(g^3)\,.
\ee
In the static gauge and under the hard thermal loop approximation
\be
\langle  A_0^{a}(\bs r)A_0^{b}(\bs 0)\rangle_T = T\sum_n\int \frac{\diff^3q}{(2\pi)^3}\frac{e^{i{\bs q}\cdot{\bs r} } }{{\bs q}^2+m_D^2}\delta_{n0} \delta^{ab}
= T\delta^{ab} \frac{1}{4\pi r}e^{-m_Dr}\,.
\ee
So finally,
\be
F_{QQ_{\bar{3}}}({\bs r})  = 2F_Q - \frac{2}{3} \frac{g^2}{4\pi r}e^{-m_Dr} + \ml{O}(g^3)\,.
\ee
The free energy of an anti-triplet heavy quark pair is the sum of the free energies of two individual heavy quarks and their color attractive potential energy.

In a similar way,
\be
F_{QQ_6}({\bs r})  = 2F_Q + \frac{1}{3} \frac{g^2}{4\pi r}e^{-m_Dr} + \ml{O}(g^3)\,.
\ee
The free energy of a sextet is the sum of the free energies of two individual heavy quarks and their color repulsive potential energy. Though up to order $g^2$ the anti-triplet and sextet free energies are independent of the Wilson line paths in the definition, they are generally dependent on the paths beyond the order $g^2$ \cite{Nadkarni:1986cz}.

\singlespacing
\chapter{Conclusions}
\doublespacing
\vspace{0.2in}

In this dissertation, we showed three studies of applying effective field theory to research problems in nuclear physics.

We first applied pionless EFT and hard thermal loop effective field theory to study the plasma screening effect on the $\alpha$-$\alpha$ scattering at the $^8$Be resonance. Because of well-separated scales in the $\alpha$ mass and the resonance energy, the EFT provides a systematic description of the low-energy scattering between the $\alpha$ particles. The parameters of the EFT satisfy a manifest power counting and can be fitted from the experimental data of the phase shift. When the $\alpha$ particles scatter inside an $e^-e^+\gamma$ plasma, the plasma screening effect will modify the Coulomb repulsion between the $\alpha$ particles. The static screening effect induces an effective Debye mass of the photon in propagators and suppresses the Coulomb repulsion. The dynamical screening effect induces an imaginary part in the Coulomb potential, which originates from the collisions between the $\alpha$ particles and the medium constituents. We demonstrated that if only static screening is considered, the $^8$Be resonance becomes more stable as the plasma temperature increases: the resonance energy is lowered and the width shrinks. When $m_D\gtrsim 0.3$ MeV, the $^8$Be resonance becomes a bound state. When both the static and dynamical screening effects are taken into account, the cross section observed becomes broader and has a higher peak location as the plasma temperature increases. We argued that it is extremely difficult to extract the formation of the $^8$Be bound state if we only measure the asymptotic final states. The energy and momentum of the $\alpha$ particles from the dissociation of the $^8$Be bound state change significantly when they travel through the plasma. 

Then we applied pNRQCD to study the quarkonium evolution inside a QGP under the hierarchy of scales $M\gg Mv \gg Mv^2 \gtrsim T \gtrsim m_D$. We started from the open quantum system formalism and wrote down the Lindblad equation to describe the non-unitary and time-irreversible evolution of the $Q\bar{Q}$ subsystem inside the QGP. Under a Wigner transform and Markovian approximation, we showed that the Lindblad equation leads to a Boltzmann transport equation. The Markovian approximation is a separation of time scales of the thermal bath and the subsystem relaxation. It can be justified in our assumed separation of scales. The key of the argument is that the interaction between the bound quarkonium and the medium is a dipole interaction, which can be treated as a perturbation in our assumed hierarchy of scales.

The Boltzmann equation has three kinds of collision terms: dissociation, recombination and diffusion. We calculated the scattering amplitudes of dissociation and recombination to the order $g^2r$. We demonstrated that these amplitudes satisfy the Ward identity and are gauge invariant. We also verified that the dipole interaction vertex does not run at the one-loop level. The collinear singularity in the inelastic scattering processes (which are at the order $g^2r$) are cancelled by the interference terms of the amplitude of order $gr$ and its thermal loop corrections. The finite binding energy of the bound state serves as a soft regulator in the inelastic scattering processes. We also computed the elastic scattering amplitudes at the order $g^2r^2$ which contribute to the diffusion of quarkonium. We estimated the diffusion coefficient of $\Upsilon$(1S) and showed it is much smaller than that of open bottom quarks because it is a process at $r^2$. In our practical calculations, we work to the order $r$ and neglect the quarkonium diffusion.

The recombination term in the Boltzmann equation of quarkonium needs distribution functions of open heavy flavors. So we couple the transport equations of open heavy flavors and quarkonia. We solved the coupled transport equations using Monte Carlo simulations. We tested our simulations inside a QGP box with constant temperature. We showed that the dissociation and recombination rates in the numerical implementation are consistent with the analytic expressions. We demonstrated that the interplay between the dissociation and recombination drives the subsystem of bound and unbound $Q\bar{Q}$ to detailed balance. The transport of open heavy flavors is necessary for the subsystem to reach kinetic thermalization. We solved the transport equations for real collision situations and can successfully describe the experimentally measured $R_{AA}$ of $\Upsilon$(1S) and $\Upsilon$(2S). We also predicted the $v_2$ parameter of $\Upsilon$(1S) in the azimuthal angular anisotropy.

Finally we applied a pNRQCD of the diquark sector to study the evolution of diquarks inside a QGP and estimate the production rate of $\Xi^{++}_{cc}$ in heavy ion collisions. The in-medium transport of diquarks is similar to that of quarkonium, except that diquarks carry color and thus their momenta can change in the same way as an open heavy quark. We gave two estimates of the production rate of $\Xi^{++}_{cc}$ based on two different melting temperatures of the charm diquark. We showed how the melting temperature of heavy diquarks can be studied on a lattice by calculating their free energies. We explained the lattice formulation of a gauge invariant but path dependent free energy of the color anti-triplet.

The theoretical and phenomenological studies of the quarkonium transport inside the QGP would be improved in several aspects. First one could work to higher orders in the weak-coupling and nonrelativistic expansions. This could be done in two parts in the pNRQCD calculation. One part is the interaction between bound singlet and unbound octet. One could extend the study to include the quadrupole interaction. The other part is the gauge field correlation $ \langle {\bs E}^a_i {\bs E}^b_j \rangle$. Currently we use weakly-coupled thermal field theory (by assuming the QGP is weakly-coupled) to calculate it to the next-leading order. In reality, we know that the QGP produced in current heavy ion collisions is strongly-coupled. So we need to consider higher order effects in the $ \langle {\bs E}^a_i {\bs E}^b_j \rangle$ correlation. Alternatively, one can reformulate the construction such that a gauge invariant correlation $ \langle {\bs E}(x_1) W(x_1,x_2) {\bs E}(x_2) \rangle$ shows up in the Lindblad and transport equations where $W(x_1,x_2)$ is a Wilson line. Then we could use the AdS/CFT correspondence and map the correlation in our strongly-coupled case into a quantity in a weakly-coupled Einsteinian gravity theory. In this way, we could formulate the transport equation in a strongly-coupled plasma. It would be interesting to see how much the phenomenological results for quarkonium differ in such a strongly-coupled scenario.

Recently, it was shown how to formulate and calculate the complex potential of $Q\bar{Q}$ on a lattice \cite{Rothkopf:2011db}. It would be interesting to explore if one could formulate the Lindblad and transport equations in terms of the complex potential. This would provide a bridge between lattice calculations, which are in Euclidean time generally, and phenomenology, which is in Minkowski time.

Finally for the numerical implementation, we could improve the calculation by using temperature-dependent potentials. Then using these potentials, we could calculate the melting temperature of different quarkonium states, thus avoiding the need to treat the melting temperature as an independent parameter. Furthermore, we could parametrize the potentials and use a systematic theory-experiment comparison (for example, the Bayesian analysis and machine learning) to extract the potential parameters. After this we could compare the potentials extracted from experiments with the lattice results. We could also improve by including other quarkonium states (1P, 2P, 3S) in the Boltzmann equations and using event-by-event hydrodynamic simulations.

As more precision data come, these improvements will probably be necessary. The future data may even imply that we have to go beyond the semi-classical approximation and include some quantum effects in the evolution. EFT will be useful to tell which quantum effects are more important. With both theory and experiment improved in the following years, we will gain a deeper and more complete understanding of quarkonium transport inside the QGP and have a better use of quarkonium as a probe of the QGP.

\singlespacing

\nocite{*}
\bibliographystyle{alpha}

\end{document}